\pdfoutput=1

\documentclass[11pt,
a4paper,
twoside=semi,%
openright, %
final
]{scrbook}

\usepackage[automark]{scrlayer-scrpage}
\clearpairofpagestyles
\usepackage{subfiles}

\providecommand{\MainFolder}{.} %
\providecommand{\MyPackagesFolder}{\MainFolder/LatexPackages} %
\providecommand{\GraphicsFolder}{\MainFolder/Graphics} %

\providecommand{\BibliographyFile}{\MainFolder/Bibliography/bibliography.bib} %

\usepackage[T1]{fontenc}
\usepackage[utf8]{inputenc}

\usepackage[ngerman,english]{babel}
\usepackage{csquotes} %

\usepackage{lmodern} %

\usepackage[
style=alphabetic, %
giveninits=true, %
backend=biber %
]{biblatex}

\DeclareSourcemap{
 \maps[datatype=bibtex]{ %
 \map{\step[fieldsource=doi,final] %
      \step[fieldset=isbn,null]}   %
 \map{\step[fieldsource=arxivId,final] %
      \step[fieldset=url,null]}        %
 \map[overwrite=true]{ %
      \step[fieldsource=arxivId] %
      \step[fieldset=eprint, origfieldval] %
      \step[fieldsource=eprint, match={/}, final] %
      \step[fieldset=primaryClass, null]} %
 }}

\DefineBibliographyExtras{english}{\let\finalandcomma=\empty} %

\usepackage{microtype} 

\usepackage{hyperref}

\usepackage{setspace}

\usepackage{\MyPackagesFolder/my_math_external_packages}

\usepackage{\MyPackagesFolder/my_tools}

\usepackage{\MyPackagesFolder/my_math_1}

\usepackage{\MyPackagesFolder/my_lists}

\usepackage{\MyPackagesFolder/my_amsthm}

\usepackage{\MyPackagesFolder/my_todo}

\pgfplotsset{compat=1.6} %

\begin{document}

\ToDoList

\frontmatter

\title{IBL-Infinity Model of String Topology from Perturbative Chern-Simons Theory}
\author{Pavel H\'ajek}
\begingroup
\newlength\SmallSkip
\newlength\BigSkip
\newlength\BBigSkip %
\setlength{\SmallSkip}{0.3cm}
\setlength{\BigSkip}{1cm}
\setlength{\BBigSkip}{2cm}

\makeatletter %
\renewcommand*{\maketitle}{
\begin{titlepage}
    
    \centering
    {\linespread{1.1}\Huge\bfseries\@title\par}
    \vskip\BigSkip
\begin{otherlanguage}{ngerman}
	{\Large
   	{\LARGE\bfseries Dissertation}
	\vskip\BigSkip
	zur Erlangung des akademischen Grades
	\vskip\SmallSkip
    Dr.~rer.~nat.
	\vskip\BigSkip
    eingereicht an der
    \vskip\SmallSkip
    Mathematisch-Naturwissenschaftlich-Technischen Fakult\"at
    \vskip\SmallSkip
	der Universit\"at Augsburg
	\vskip\BigSkip
	von
	\vskip\SmallSkip
	{\bfseries\LARGE\@author} 
	\vskip\BBigSkip
	Augsburg, Oktober 2019
	\vskip\BigSkip
	\includegraphics[scale=0.3]{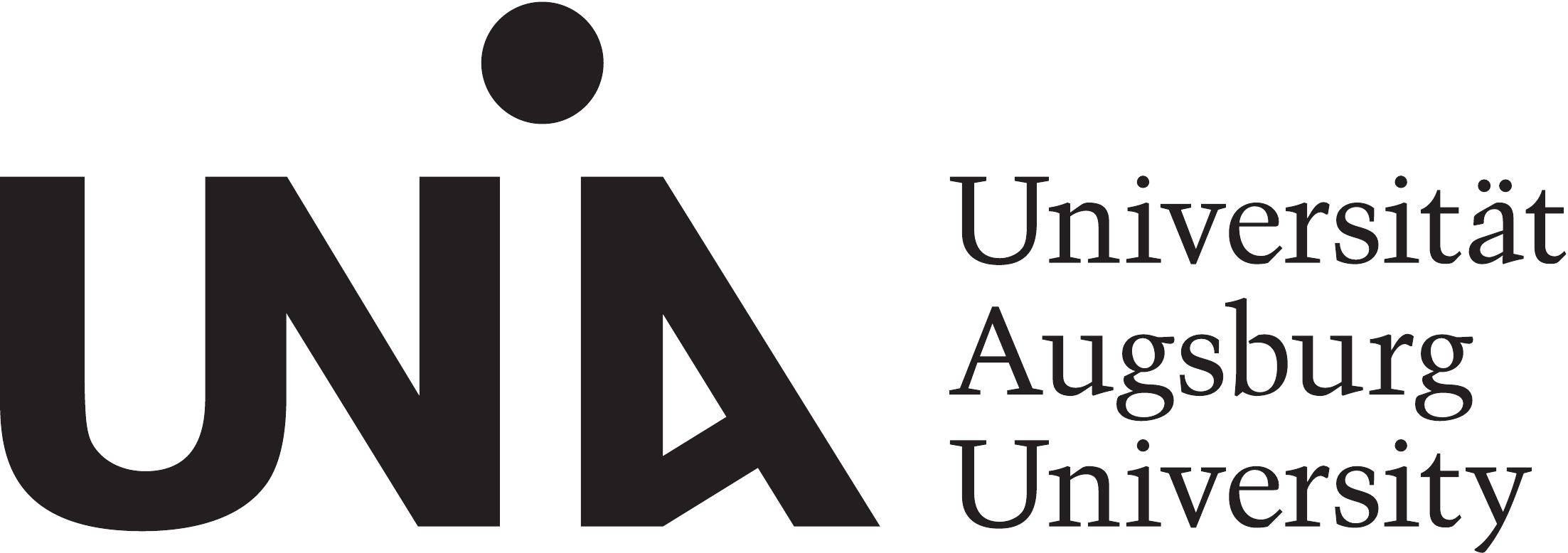}
	\clearpage
    \thispagestyle{empty}
	\null\vfill	
	\begin{flushleft}
  	\begin{tabular}{ll}
   		1.~Gutachter:  & Prof.~Dr.~Kai Cieliebak, Universit\"at Augsburg \\
    	2.~Gutachter: & Prof.~Dr.~Janko Latschev, Universit\"at Hamburg 
    \end{tabular}
    
\vspace{1cm}    
\begin{tabular}{ll}
Tag der mündlichen Prüfung: & Freitag, 20.~Dezember, 2019
\end{tabular}    	%

    \end{flushleft}
    }\end{otherlanguage}
\end{titlepage}
}
\maketitle  
\makeatother
\endgroup

\onehalfspacing

\begin{center}
  {\LARGE Abstract}\\[0.5cm]
  Pavel H\'ajek: {\itshape IBL-Infinity Model of String Topology from Perturbative Chern-Simons Theory}
\end{center}

Following Cieliebak, Fukaya, Latschev and Volkov, we construct an IBL-infinity chain model for equivariant string topology on cyclic Hochschild cochains of de Rham cohomology. We study its properties and perform explicit computations.\vspace{.8cm}
\clearpage 
{\null\vfill
\begin{flushleft}
\makeatletter
\copyright\ \the\year\ \@author
\makeatother \\[\baselineskip]

Published online via OPUS:  \url{https://opus.bibliothek.uni-augsburg.de/}

Personal online version: \url{https://github.com/p135246/phd-thesis}.\\[\baselineskip]

The thesis was typeset in \LaTeX\ using the \textsf{scrbook} class of \Komaname. \\
The bibliography was created using \Biblatexname\ and checked against \textsf{MathSciNet}. \\
Figures were drawn in \Tikzname.\\
Font: \Libertinename, $11$\,pt.\\
Computations were done in \Mathematicaname.\\[\baselineskip]
\end{flushleft}
}

\addchap*{Acknowledgements}
\Modify[noline]{DONE Check who has Dr. in name!}
First and foremost, I thank my parents, \emph{Dana H\'ajkov\'a} and \emph{Libor H\'ajek,} for supporting me throughout my studies, both emotionally and materially, for discussing various decisions I had to make, for listening when I needed to talk and for wishing me all the best in whatever I wanted to~do.
D\'iky moc mami a tati!

I thank my grandparents,  And\v{e}la H\'ajkov\'a, Anna Uhl\'ikov\'a and Jaroslav Uhl\'ik, for their love and support, although they would rather see me being happy and having a family than pursuing a doctorate. D\'iky moc babi a d\v{e}do! \Add[noline,caption={DONE Add cross}]{DO NOT Add cross to Anna Uhlikova}

Next, I thank my supervisor, Prof.~Dr.~Kai Cieliebak, for giving me time and support to get through the graduate studies, to develop academically and to finish the thesis. I~enjoyed discussions with him very much and learned from his insights and methods. I~appreciate and admire his skill to explain everything anytime starting from elementary facts, his passion for mathematics and physics, scientific productivity and lecturing skills. Danke sch\"on, Kai!

I thank Dr.~Evgeny Volkov for discussing his work on Chern-Simons theory, string topology, cyclic homology and Chen's iterated integrals, for providing me his preprints in progress and for supportive chats at the university. Thank you, Evgeny, good luck fighting with lions in Siberia!

I~thank Prof.~Robert Bryant for proposing an alternative notation for the Hodge propagator for spheres in an online discussion. I thank Prof.~John Rognes for answering my question about spectral sequences in an online discussion. I thank Dr.~Najib Idrissi for answering my question about formality in an online discussion. I also thank the online discussion --- Mathoverflow and Math Stackexchange --- for its existence.

I~thank Dr.~Alexandru Doicu for checking a tedious sign computation and for friendly talks. I~thank Dr.~Andreas Hermann for discussing the standard Hodge propagator, for sending me his notes on the Green kernel for spheres and for his interest in my work. I~thank Prof.~Dr.~Hông Vân Lê for discussing her work on formality and Poincar\'e $\DGA$'s and for her interest in my work. I thank J\'an Pulmann and Lada Peksov\'a for discussing $\BV$-formalism and properads. I thank Ond\v{r}ej~Hul\'ik for a Skype discussion about physics and for friendly talks. I thank Dr.~Oliver Lindblad Petersen for bringing up the name Melrose blow-up for our blow-up construction. I thank Ji\v{r}\'i Zeman for pointing out that absolute convergence is equivalent to invariance on resummations and for friendly talks. I thank Dr.~Kyler Siegel for a discussion about A-infinity in Stony Brook. I thank Dr.~Raymond Puzio for his interest in my work, for thinking about an action for our propagator and for friendly talks. I~thank Prof.~Alberto Cattaneo and Dr.~Pavel Mnev for discussing Feynman integrals and an action for our propagator in Stony Brook and for sending me a list of relevant literature later. I~thank Prof.~Dr.~Bra\v{n}o Jur\v{c}o for discussing $\BV$-formalism and for offering me a position in Prague.

I thank Prof.~Dr.~Christian~B\"ar, Prof.~Dr.~Kaoru Ono and Prof.~Dr.~Janko Latschev for inviting me to their institutes to discuss and give a talk, for their hospitality and for valuable feedback and insights. I am grateful to Prof.~Dr.~Janko Latschev for giving me a position in Hamburg and for waiting until I finish the thesis.

Although everything mentioned in the last paragraphs occurred in the final fifth year of my Ph.D., it inspired me and motivated me to complete the work.

I thank members of the department for mathematics at the University of Augsburg for creating and maintaining a rich and inspiring academic environment. I thank Prof.~Dr.~Urs Frauenfelder for various discussions and for always coming up with a highly interesting topic for the symplectic seminar, which, at the beginning, I often preferred to my own work. I~thank Prof.~Dr.~Edward Belbruno, a~visiting professor from Princeton, for his series of talks about chaotic dynamics in celestial mechanics, for the explanation of the weak stability boundary, for organizing exciting events related to space travel and for stopping by our office to chat about various topics.

I thank my colleagues at the university; those who I liked for being friendly and encouraging, and those who I did not for putting up with me and for not fueling conflicts. Specifically, I thank Alexei Kudryashov and Thorsten Hertl, who hanged out with me in the final two years when I was spending the most of my time at the university, and who had to listen to my laments about lost years, starting to work on concrete problems too late, terrible topic which is not geometric at all, internet overuse and screwed up personal life. I can not forget to highlight their top-level expertise in model categories and algebraic topology, respectively.

Unrelated to academia, I thank my tennis trainer, Thorsten Moser, alias Mr.~T, for his university tennis courses in a cheerful atmosphere, him being dressed in one of his crazy costumes and constantly making fun of my tennis skills. There were times when the regular tennis training was the only thing I was looking forward to and the only order I could stick to. I thank my sparring partners and friends, Lukas Moosbrugger, Thomas Gumpinger and Tobias Watzka, for playing tennis and drinking beer with me. I~thank Surf Club Augsburg, R\"ugen Piraten and Windsurfing F4 for the great time I had windsurfing on Mandichosee, Ostsee and Bol, respectively, during my holiday. Memories of planing along the coast when a storm was coming helped me to overcome times when I~felt really down. I thank Cuban Salsa Power in Augsburg, where I learned to dance Salsa properly, and where I spent almost every other evening in the beginnings of my studies. I~thank my dance partner, Beatrix Fertl, for visiting me in Augsburg and for going out with me. I thank Ushi Haller for her interest in my well-being and for baking cookies and cakes for me. Finally, I want to mention an old friend, Milan Va\v nk\'at, and thank him for the occasional obscene phone calls we had and for his few visits when we made fun of every possible social norm. He also left me the two African snails (die Schnecken) who accompanied me from the beginning to the end; I~never felt too slow with them by my side.

\Add[inline,caption={DONE Add more names}]{Add Ushi for giving me cookies and cakes. Tennis partners: Thomas, Lukas, Tobi}
\Add[inline,caption={DONE Add picture}]{Add the picture of Mikesch with snails and Augsburg}
\begin{figure}[!b]
\centering
\includegraphics[trim=130 28 75 57, clip, scale=.64]{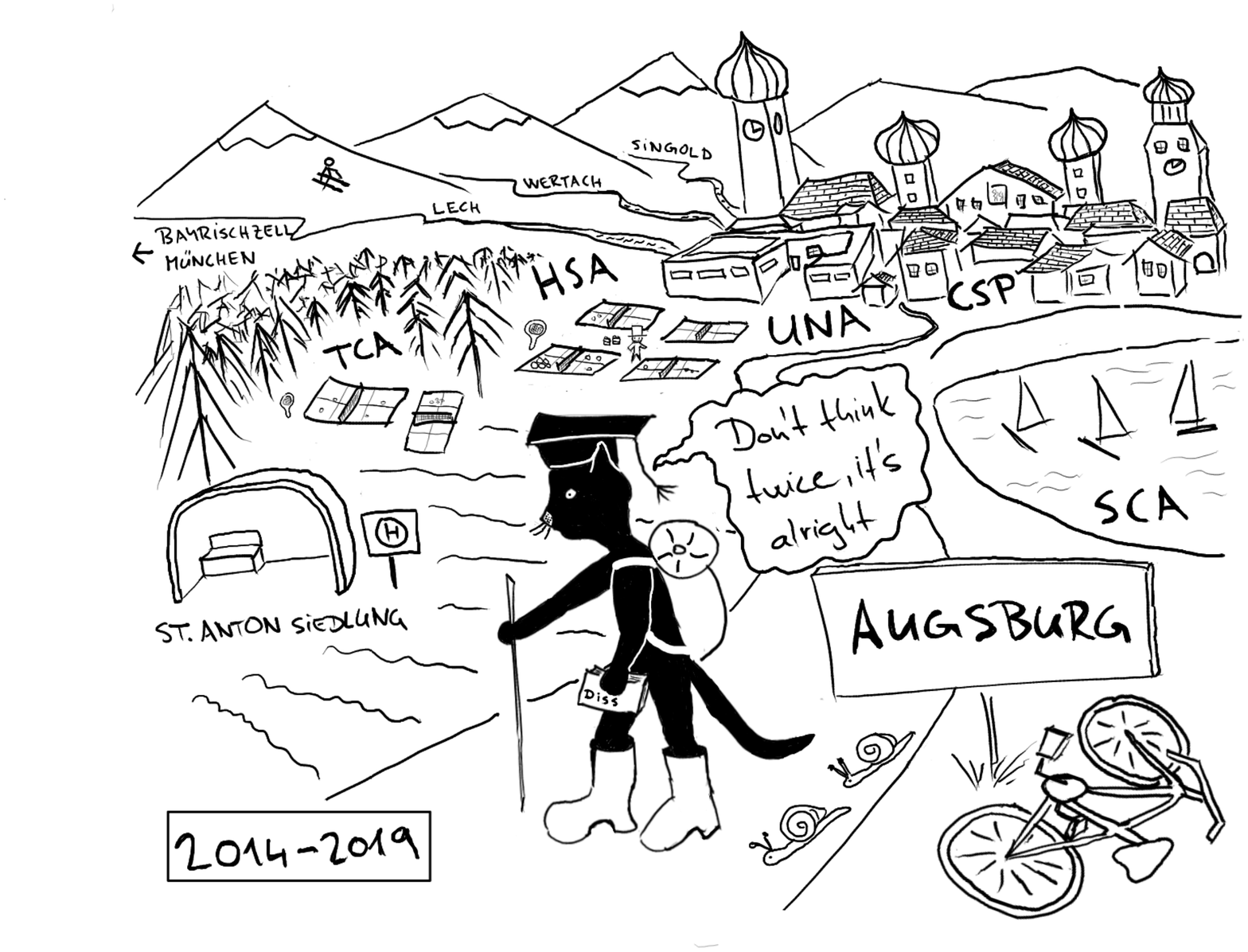}
\end{figure}

\addchap*{Structure of the thesis}

The thesis consists of the \emph{Introduction} and the following three parts:
\begin{itemize}
 \item \emph{Part I---IBL-infinity chain model:} the goal is to compute the twisted $\IBLInfty$-algebra on de Rham cohomology of $\Sph{n}$ explicitly and compare it to string topology (the original task). A version of Part~I was previously made available online and can be found under the following reference: 
\begin{center}
\fullcite{Hajek2018}
\end{center}
 \item \emph{Part~II---Follow-up topics:} additional topics discovered while working on Part~I.
 \item \emph{Part III---Appendices:} algebraic details of Part~I and related questions.
\end{itemize}

\noindent The three parts were written chronologically and reflect how author's view of the theory developed. The mathematical notation is consistent but the terminology and understanding might vary slightly from part to part. All in all, Part~I is more about ``getting things done'' while Parts~II and III are more about ``optimizing''.

\newcommand\partentrynumberformat[1]{\partname\ #1}
\RedeclareSectionCommand[
  tocentrynumberformat=\partentrynumberformat,
  tocnumwidth=4em
]{part}
\renewcommand*{\contentsname}{Contents}
\KOMAoptions{toc=chapterentrydotfill}
\tableofcontents
\listoffigures

\mainmatter
\Correct[inline,caption={DONE Fitlration index}]{Standardize the position of the filtration index!}
\Correct[inline,caption={DONE(only in PI and PIII) Add Part reference}]{Add Part I to the references to other parts.}
\Correct[inline,caption={DONE Completed tensor product}]{Replace completed tensor product with COtimes because of the correct spacing of binary operator.}
\Add[inline,caption={Completed tensor product}]{Add a remark that Koszul convention does not matter for $\odot$ because tensor products of more morphisms appear only in morphisms which have even degree. Polemize once more that this pairing is important in order to make Chapter 10 work, ie, the cyclic bar complex. One might change the definitions of canonical operations and the evaluation of ribbon graphs and redo do the proofs in those section. However, the author failed. But maybe just because of his newbie. Now it seems more plausible. }
\Modify[inline,caption={DONE Replace overline with widebar}]{}
\Correct[inline,caption={DONE The correct definition equality :=}]{Replace := with $\coloneqq$}
\Modify[inline,caption={DONE Chern-Simons Maurer-Cartan element}]{Change formal pushforward Maurer-Cartan element to Chern-Simons Maurer-Cartan element}
\Correct[inline,caption={DONE Filtration by weights}]{The filtration on the dual cyclic bar complex is the dual to the filtration by weights on the bar complex.}
\Correct[inline,caption={DONE $\circ_{h_1,\dotsc,h_r}$ not defined for bialgebra}]{In fact, $\OPQ_{klg}$ and the operations $\circ_{h_1,\dotsc,h_r}$ can not be defined on an arbitrary bialgebra. We still need the filtration by weights. It might be possible to extend this to weight-graded bialgebras (operations are continuous with respect to the grading and weights). Also resolve the notion of "continuous extension" and what is, in fact, continuous. Filtration preserving versus non-negative filtration degree. Maybe call all guys with finite-filtration degree continuous.}
\Correct[inline,caption={DONE Hodge htpy}]{Change Green operator to Hodge homotopy. Also GOp to HOp or something like that. Add admissible to symmetric Hodge propagator which extends smoothly to the blow-up. Add special is it satisfies the additional properties.}
\Add[inline,caption={Melrose Blowup}]{I was told by Oliver that we do Melrose blowups.}
\Add[inline,caption={Name of MaurerCarta}]{We should call $\int$ Chern-Simons Maurer-Cartan element and $\PMC$ effective Chern-Simons element. We previously used also formal pushforward Maurer-Cartan element. However, in fact, $\int$ is not a Maurer-Cartan element of the Fr\'echet algebra.
}
\Correct[inline,caption={Questions}]{Make questions with ending QED Sign}

\chapter{Introduction}

\section{String topology and Chen's iterated integrals}

String topology of a manifold~$M$ is the study of the \emph{free loop space} 
\[ \Loop M = \{\gamma: \Sph{1}\rightarrow M\text{ continuous}\}, \]
which is equipped with the compact-open topology, and of natural structures on it.
Each loop~$\gamma$ is parametrized, with base-point~$1$, and there is a natural $\Sph{1}$-action changing the base-point.
Therefore, we can distinguish the following two homology theories:
\begin{center}
\begin{tabular}{rl}
 $\H(\Loop M)\quad\dotsc$& the \emph{singular homology} and \\[1ex]
 $\H^{\Sph{1}}\!(\Loop M)\quad\dotsc$ & \parbox[t]{10cm}{the \emph{equivariant homology} --- ``the singular homology of the space of parametrized loops with the base-point forgotten.''}
\end{tabular}
\end{center}
In this thesis, we consider $\H^{\Sph{1}}\!(\Loop M)$ with coefficients in $\R$ only.

If $M=\Sigma$ is an oriented surface, we consider immersed loops with transverse double points and define a bracket and cobracket by Figure~\ref{Fig:ConstrLoop}.
\begin{figure}[t]
\begin{equation*}
\begin{aligned}
\StringOp_2\left(
\parbox[c]{2.85cm}{
\begin{tikzpicture}
	\def\rad{.8cm}
	\draw[green,dashed,thick,decoration={markings, mark=at position 0.25 with {\arrow{>}}},postaction={decorate}] ([shift=(0:\rad)]0,0) arc (0:360:\rad);  
	\draw[red,thick,decoration={markings, mark=at position 0.25 with {\arrow{>}}},postaction={decorate}] (1.5*\rad,0) circle (\rad);
	\end{tikzpicture}}
\right)
&=\parbox[c]{2.85cm}{
\begin{tikzpicture}
	\def\rad{.8cm}
	\draw[blue,thick] ([shift=(90:\rad)]0,0) arc (90:360:\rad); %
    \draw[blue,thick] ([shift=(0:\rad)]0,0) arc (0:180:.25*\rad); %
	\draw[blue,thick] ([shift=(180:\rad)]1.5*\rad,0) arc (180:450:\rad); %
	\draw[blue,thick,decoration={markings, mark=at position 0.5 with {\arrow{<}}},
        postaction={decorate}] ([shift=(90:\rad)]0,0) to ([shift=(90:\rad)]1.5*\rad,0); %
\end{tikzpicture}}
\ -\ 
\parbox[c]{2.85cm}{
\begin{tikzpicture}
	\def\rad{.8cm}
	\draw[blue,thick] ([shift=(0:\rad)]0,0) arc (0:270:\rad); %
	\draw[blue,thick] ([shift=(180:\rad)]1.5*\rad,0) arc (180:360:.25*\rad); %
	\draw[blue,thick] ([shift=(270:\rad)]1.5*\rad,0) arc (270:540:\rad); %
	\draw[blue,thick,decoration={markings, mark=at position 0.5 with {\arrow{>}}},
        postaction={decorate}] ([shift=(270:\rad)]0,0) to ([shift=(270:\rad)]1.5*\rad,0);
\end{tikzpicture}}\\
\StringCoOp_2\left(\hspace{-.4em}
\parbox[c]{3.3cm}{
\begin{tikzpicture}
	\def\rad{.8cm}
	\draw[blue,thick,decoration={markings, mark=at position 0.25 with {\arrow{>}}},
        postaction={decorate}] ([shift=(45:\rad)]0,0) arc (45:315:\rad);
	\draw[blue,thick,decoration={markings, mark=at position 0.25 with {\arrow{<}}},
        postaction={decorate}] ([shift=(-135:\rad)]2*\rad,0) arc (-135:135:\rad);
	\draw[blue,thick] (45:\rad) to[out=-45,in=135] ($(-135:\rad)+(2*\rad,0)$);
	\draw[blue,thick] (-45:\rad) to[out=45,in=225] ($(135:\rad)+(2*\rad,0)$);
\end{tikzpicture}}\right)
&=\parbox[c]{1.64cm}{
\begin{tikzpicture}
\def\rad{.8cm}
\draw[green,thick,dashed,decoration={markings, mark=at position 0.25 with {\arrow{>}}},
        postaction={decorate}] (0,0) circle (\rad);
\end{tikzpicture}}\otimes
\parbox[c]{1.64cm}{
\begin{tikzpicture}
\def\rad{.8cm}
\draw[red,thick,decoration={markings, mark=at position 0.25 with {\arrow{<}}},
        postaction={decorate}] (0,0) circle (\rad);
\end{tikzpicture}}\ -\ 
\parbox[c]{1.64cm}{
\begin{tikzpicture}
\def\rad{.8cm}
\draw[red,thick,decoration={markings, mark=at position 0.25 with {\arrow{<}}},
        postaction={decorate}] (0,0) circle (\rad);
\end{tikzpicture}} \otimes
\parbox[c]{1.64cm}{
\begin{tikzpicture}
\def\rad{.8cm}
\draw[green,dashed,thick,decoration={markings, mark=at position 0.25 with {\arrow{>}}},
        postaction={decorate}] (0,0) circle (\rad);
\end{tikzpicture}}
\end{aligned}
\end{equation*}
\caption[Bracket and cobracket in equivariant string topology.]{The bracket and cobracket in equivariant string topology. Imagine some genus so that the operations are homotopically non-trivial.}
\label{Fig:ConstrLoop}
\end{figure}
In words:\ToDo[noline,caption={How is it with loop parametrization}]{How is it precisely with the speeds?}
\begin{description}[leftmargin=*]
 \item[$\StringOp_2$:] Imagine putting one base-point $b_1$ on the first loop $\gamma_1$ and another base-point $b_2$ on the second loop $\gamma_2$ in all possible positions. Whenever $\gamma_1(b_1) = \gamma_2(b_2) = p$, construct a new loop~$\gamma = \gamma_1 \Star_p \gamma_2$ by running first along $\gamma_1$ with double speed starting and ending at~$b_1$ and continuing along $\gamma_2$ starting and ending at $b_2$. Forget the base-points and multiply $\gamma$ with the sign of the intersection $\varepsilon(p;\gamma_1, \gamma_2)$. Should more intersections occur, take the sum.
 \item[$\StringCoOp_2$:] Imagine putting the basepoints $b_1$ and $b_2$ on $\gamma$ in all possible positions such that $b_1 \neq b_2$. Whenever $\gamma(b_1) = \gamma(b_2)=p$, split $\gamma$ into~$\gamma_1$ and $\gamma_2$ as follows. The first loop~$\gamma_1$ is the portion of~$\gamma$ from $b_1$ to $b_2$ and the second loop $\gamma_2$ is the portion from $b_2$ to $b_1$ ran along with the correspondingly scaled speed. Forget the base-points, form the tensor product $\gamma_1 \otimes \gamma_2$ and multiply it with $\varepsilon(p; \gamma)$. Should more self-intersections occur, take the sum of the tensors.
\end{description}
The operations $\StringOp_2$ and $\StringCoOp_2$ are known as the \emph{Goldman bracket} and the \emph{Turaev cobracket} and were defined and studied in \cite{Goldman1986} and \cite{Turaev1991}, respectively.

In \cite{Sullivan1999}, it was demonstrated that the construction of~$\StringOp_2$ extends to families of loops and to an arbitrary dimension~$n$ of an oriented manifold~$M$; it produces a Lie bracket on the equivariant homology.
The construction of $\StringCoOp_2$ generalizes too and gives a Lie cobracket; this is explained for instance in~\cite{Cieliebak2007}. 
The picture is always Figure~\ref{Fig:ConstrLoop}, just the intersection points come from transverse intersections of smooth parameter spaces of points on loops in $M$; consequently, both~$\StringOp_2$ and~$\StringCoOp_2$ have degrees $2-n$.
Another definition of the coproduct is used in \cite{Basu2011}, where loops in~$M$ are viewed as open strings in $M\times M$ with endpoints at the diagonal.

In order to make these geometric constructions rigorous, the most straightforward way (which would work over $\Z$) is to use a \emph{geometric homology theory} of $M$ based on smooth chains such that the transverse intersection of two smooth chains is again a smooth chain.
Such theory for smooth manifolds was constructed in \cite{Lipyanskiy2014}.
A~version for general topological spaces is proposed in \cite{Cieliebak2013}, and some details regarding triangulations are addressed in~\cite{Hajek2014} (see also the discussion at \cite{MO157762}).
Note that $\StringOp_2$ and $\StringCoOp_2$ are only ``transversally defined'' on the chain level, and in order to define them on homology, it is important that we can homotop to a generic situation within the homology class.

The main theorem of \cite{Sullivan2002} asserts that $\StringOp_2$ and $\StringCoOp_2$ induce the structure of an \emph{involutive bi-Lie algebra}, abbreviated $\IBL$, of degree $2-n$ on the equivariant homology $H^{\Sph{1}}\!(\Loop M,M)$ relative to constant loops $M\xhookrightarrow{}\Loop M$.
Modding out constant loops is necessary for $\StringCoOp_2$ to be well defined because of the phenomenon of \emph{``vanishing of small loops''} illustrated for instance in~\cite{Cieliebak2007}:  Let $\sigma\in C_1(\Loop M)$ be a $1$-chain supported on $[0,1]$ which for $t=0$ agrees with the loop in the argument of $\StringCoOp_2$ in Figure~\ref{Fig:ConstrLoop}, next, for $t\in (0,1)$, the left knot~$L$ contracts to the mid-point $p$, and for $t=1$, only the right knot $R$ remains (thus the name ``vanishing of small loops'').
It is then easy to see that 
\[ 0 \neq \Bdd \StringCoOp_2(\sigma) - \StringCoOp_2(\Bdd \sigma) = p\otimes R - L \otimes p \in C(M)\otimes C(\Loop M) + C(\Loop M)\otimes C(M). \]
Since the string bracket $\StringOp_2$ applied to a chain of constant loops gives a degenerate chain, it restricts to a Lie bracket on the relative homology.\footnote{This contrasts with the situation on the non-equivariant homology $\H(\Loop M)$, where constant loops do not always form an ideal for the associative loop product; this is easy to see in the case of torus~$\T^2$. They do form an ideal, however, provided that the Euler characteristics $\chi(M)$ is non-zero, see \cite{Tamanoi2010}.
On the other hand, if $\chi(M)=0$, then the loop coproduct admits an extension to $\H(\Loop M)$; this is possibly dependent on the choice of a non-vanishing vector field, see \cite{Basu2011}.}
In work in progress~\cite{CieliebakHingston2018}, Poincar\'e duality on the Rabinowitz-Floer homology of the unit cotangent bundle of $M$ is introduced and related to the non-equivariant string topology via a long exact sequence.
This shall give a canonical extension of $\StringCoOp_2$ to $\H^{\Sph{1}}\!(\Loop M,\mathrm{pt})$, the relative homology modulo one point, which together with $\StringOp_2$ would make $\H^{\Sph{1}}\!(\Loop M,\mathrm{pt})$ into a bi-Lie algebra.
In fact, this is what our chain model is supposed to compute.
Being aware of this context, we will use the symbol $\RedEquivHom(\Loop M)$ as an avatar for either $\H^{\Sph{1}}\!(\Loop M,M)$ or $\H^{\Sph{1}}\!(\Loop M,\mathrm{pt})$.

It is expected that the $\IBL$-structure on $\RedEquivHom(\Loop M)$ is induced from a much richer and in some sense natural algebraic structure on the chain level, whose homotopy type is an invariant of $M$.
In fact, there is a notion of \emph{strong homotopy involutive bi-Lie algebra}, abbreviated $\IBLInfty$, which was developed in~\cite{Cieliebak2015}.
An $\IBLInfty$-algebra consists of operations $(\OPQ_{klg})$ for $k$, $l \ge 1$ and $g\ge 0$, where $\OPQ_{110}$ is a boundary operator, $\OPQ_{210}$ a bracket and $\OPQ_{120}$ a cobracket which satisfy the $\IBL$-relations up to a coherent system of higher homotopies~$(\OPQ_{klg})$.

Consider the string space 
\[ \StringSpace M \coloneqq (\EG\Sph{1}\times \Loop M)/\Sph{1}, \]
i.e., the homotopically correct version of the quotient $\Loop M/\Sph{1}$, and let $(C(\StringSpace M),\Bdd)$ be the singular chain complex of $\StringSpace M$.
Recall that~$\StringOp_2$ and~$\StringCoOp_2$ are partially defined on transverse smooth chains therein.
An \emph{$\IBLInfty$-chain model for equivariant string topology} is an $\IBLInfty$-algebra $(\Model,(\OPQ_{klg}))$ ($\Model$ stands for ``model'') together with a weak homotopy equivalence  ($\coloneqq$~zig-zag of quasi-isomorphisms) of $(C(\StringSpace M),\Bdd)$ and $(\Model,\OPQ_{110})$ which induces an isomorphism of $\IBL$-algebras
\[ (\RedEquivHom(\Loop M),\StringOp_2,\StringCoOp_2) \simeq (\H(\Model,\OPQ_{110}), \OPQ_{210}, \OPQ_{120}). \]
Note that there can be various non-homotopically equivalent $\IBLInfty$-chain models.
On the other hand, the properad $\IBLInfty$ is a quasi-free resolution of the properad $\IBL$ and as such has convenient homotopy theoretical properties; for example, homotopy inverses of quasi-isomorphisms exist.
These properties imply that any two weakly homotopy equivalent $\IBLInfty$-algebras are homotopy equivalent.

In this thesis, we use a version of \emph{perturbative Chern-Simons theory} for an oriented compact Riemannian manifold~$M$ to construct an $\IBLInfty$-chain model for the equivariant string topology of $M$.
The chains are cyclic Hochschild cochains of the de Rham cohomology $\HDR\coloneqq\HDR(M)$, and the homotopy type of the model is supposed to be an invariant of~$M$ (perhaps topological).
The construction involves a version of Feynman integrals, and the proof that it is well-defined relies on the theory of integration on certain compactifications of configuration spaces, which is currently being developed in~\cite{Cieliebak2018}.
The concrete form of this chain model was sketched in \cite{Cieliebak2015}.
For the sake of the big picture we remark that it is expected that evaluations at boundaries of pseudo-holomorphic curves in the symplectization of the unit cotangent bundle of $M$ induce an $\IBLInfty$-quasi-isomorphism of the corresponding symplectic field theory and the $\IBLInfty$-chain model of string topology; see~\cite{Cieliebak2007}. 

We now describe the underlying chain complex of our $\IBLInfty$-chain model and the quasi-isomorphism to string topology in more details.
Let $\DR\coloneqq\DR(M)$ be the space of smooth de Rham forms on $M$, and let $\BCyc \DR$ be the graded vector space generated by cyclic words 
\[ \omega_1\dotsb\omega_k = (-1)^{\Abs{\omega_k}(\Abs{\omega_1}+\dotsb+\Abs{\omega_k})} \omega_k\omega_1 \dotsb \omega_{k-1} \]
with homogenous components $\omega_1$, $\dotsc$, $\omega_k \in \DR$ for $k\ge 1$.
The grading satisfies $\Abs{\omega_i} = \Deg(\omega_i) - 1$, where $\Deg(\omega_i)$ denotes the form-degree of $\omega_i$. 
We call $\BCyc \DR$ the \emph{cyclic bar complex of $\DR$} (it might be described as ``reduced'' because we omit $k=0$).
On $\BCyc \DR$, we consider the \emph{Hochschild differential}
\[ \Hd(\omega_1 \dotsb \omega_k) = \begin{aligned}[t]
&\sum_{i=1}^k (-1)^{\Abs{\omega_1} + \dotsb + \Abs{\omega_{i-1}}}\omega_1 \dotsb\Dd\omega_i\dotsb \omega_k \\
+&\sum_{i=1}^{k-1} (-1)^{\Abs{\omega_1}+\dotsb + \Abs{\omega_{i-1}} + \Abs{\omega_i} + 1}\omega_1 \dotsb \omega_i \wedge \omega_{i+1} \dotsb \omega_k\\
+&(-1)^{\Abs{\omega_k}(\Abs{\omega_1} + \dotsb + \Abs{\omega_{k-1}}) + \Abs{\omega_k} + 1}\omega_k \wedge \omega_1 \dotsb \omega_{k-1}.
\end{aligned}\]
It descends from the Hochschild differential on the bar construction $\B \DR$, which is defined as the sum of the unique extensions of degree shifts of $\Dd$ and $\wedge$ to coderivatives of $\B \DR$ plus the wrap-around term.

\emph{Chen's iterated integral} is the map
\[ I: \begin{aligned}[t]
    \BCyc \DR& \longrightarrow C^*(\StringSpace M)\\    
    \omega_1 \dotsb \omega_k & \longmapsto  \Bigl(\sigma \mapsto \varepsilon(\omega) \int_{K_\sigma \times \Delta^k} \omega_1(\tilde{\sigma}(x,t_1)) \dotsb \omega_k(\tilde{\sigma}(x,t_k))\Bigr),
   \end{aligned}\]
where $\varepsilon(\omega)$ is the sign
\[ \varepsilon(\omega) = (-1)^{(k-1)(\Abs{\omega_1} + 1) + (k-2)(\Abs{\omega_2} + 1) + \dotsb + \Abs{\omega_{k-1}} + 1}, \]
$K_\sigma$ is a smooth chain in $M$, i.e., a manifold with corners, and $\tilde{\sigma}: K_\sigma\times\Sph{1} \rightarrow M$ is the projection of a lift $K_\sigma \xrightarrow{\hspace{.2em}\sigma\hspace{.2em}}\StringSpace M \dasharrow \EG \Sph{1} \times \Loop M$ to the second factor.
We also identify $\Sph{1} = \R/\Z$.
The map $I$ is a chain map with respect to the grading of $\BCyc\DR$ by $\Abs{\omega_1 \dotsc \omega_k} = \Abs{\omega_1} + \dotsb + \Abs{\omega_k}$.

If $\pi_1(M) = \{1\}$, then $I$ induces an isomorphism 
\begin{equation}\label{Eq:IsomKai}
\H(\BCyc \DR,\Hd)/\Span\{[1^{2k-1}]\mid k\in\N\} \simeq \H^*(\StringSpace M)/\R[u],
\end{equation}
where $\Span\{[1^{2k-1}]\mid k\in\N\} = \H(\BCyc \R,\Hd)$ and the polynomial ring $\R[u]$ with $\Abs{u} = 2$ comes from the module structure on $\H^*(\StringSpace M)$ induced from $\StringSpace M = (\EG \Sph{1} \times \Loop M)/\Sph{1} \rightarrow \EG\Sph{1}/\Sph{1} = \CP^{\infty}$.
Note that the quotient on the left hand side agrees with the homology of $\coker(\BCyc \R \xhookrightarrow{} \BCyc \DR)$.

Proving \eqref{Eq:IsomKai} is the goal of \cite{Cieliebak2018b}.
They study different totalizations of the Connes' cyclic bicomplex of $\DR$ and identify the one which $(\BCyc \DR,\Hd)$ is weakly equivalent to.
Then they use the isomorphism from \cite{Getzler}.

\section{IBL-infinity chain model and Chern-Simons theory}

We consider the \emph{dual cyclic bar complex}
\[ \CDBCyc \DR = \bigoplus_{d\in\Z} \prod_{k=1}^\infty (\BCyc_k \DR)^{d*}, \]
where $(\BCyc_k \DR)^{d*}$ denotes the linear dual to the degree $d$ component of the weight $k$ component $\BCyc_k \DR$ of $\BCyc \DR$ ($k$ is the number of letters in the generating word).
We equip $\CDBCyc \DR$ with the dual Hochschild differential~$\Hd^*$.
By taking~$\prod_k$, i.e., the completion of~$\bigoplus_k$ with respect to the filtration by weights, we allow ``bubbling off'' of forms $\omega_i$ of form-degree~$1$ with $\Abs{\omega_i}=0$ and constants $1$ with $\Abs{1} = -1$.
Notice that if we take $\HDR$ instead of $\DR$, then the completion is relevant only if $\HDR^1 \neq 0$.
The dual of Chen's iterated integral map provides a quasi-isomorphism
\[I^*: (C(\StringSpace M),\Bdd) \rightarrow (\CDBCyc\DR,\Hd^*) \]
for simply-connected $M$.

\begin{figure}[t]
 \centering
 \def\rad{2}
 \def\len{.4}
 \def\smalllen{.1}
 \def\num{6} 
 \begin{tikzpicture}
 \tikzset{point/.style = {draw, circle, fill=black, minimum size=2pt,inner sep=0pt}}
   \coordinate (C) at (0,0);
   \draw([shift=(0:\rad)]C) arc (0:360:\rad);
   \foreach \x in {1,...,\num} {
   	\node at ([shift=(\x*360/\num-360/\num:\rad+\len)]C) {$\omega_\x$};
   	\node[point,style={fill=white}] at ([shift=(\x*360/\num-360/\num:\rad)]C) {};
    }
 \end{tikzpicture}
 \caption{Inserting fields on strings.}
 \label{Fig:GeomStr}
\end{figure}
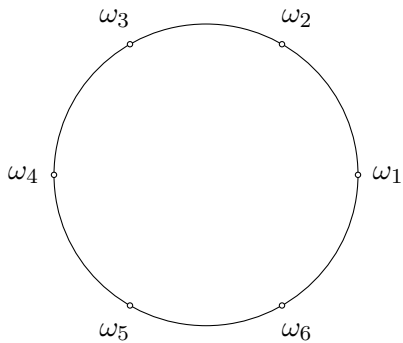

In the following discussion, which does not aim to be rigorous at all, we introduce a \emph{physical interpretation.}
We think of
\begin{itemize}
\item elements of $\DR$ as \emph{fields,}
\item elements of $\BCyc \DR$ as \emph{field strings} (not to confuse with string fields :-)) and
\item elements of the space
\[ \Fun(\BCyc \DR[1]) \coloneqq \hat{\Sym}(\DBCyc\DR[1])\COtimes \R((\hbar)) \]
as \emph{observables on field strings.}
Here, $\R((\hbar))$ is the ring of Laurent series in Planck's constant $\hbar$, $\hat{\Sym}$ the completed symmetric algebra and $\hat{\otimes}$ the completed tensor product.
\end{itemize}
If $\sigma:\Sph{1}\rightarrow M$ is a string, the observable $I^*(\sigma)$ ``localizes'' on field strings which approximate $\sigma$; for instance, it holds $I^*(\omega) = \int_{\sigma} \omega$, and thus $I^*(\omega)$ ``localizes'' at fundamental forms of $\sigma(\Sph{1})$. We imagine that we decorate $\sigma$ with a field string $\omega_1 \dotsc \omega_k$ as in Figure~\ref{Fig:GeomStr} and get a number $I(\sigma)(\omega_1\dotsc\omega_k)$. The isomorphism \eqref{Eq:IsomKai} guarantees that the observable~$I(\sigma)$ determines $\sigma$ up to a boundary term.

We will be dealing with \emph{two ``dynamical'' theories:} one is the theory of fields $\omega\in \DR$ and one of field strings $\omega_1 \dotsb \omega_k \in \BCyc \DR$.

The field theory is at hand.
We know that $(\DR,\Dd,\wedge,\langle\cdot,\cdot\rangle)$, where $\langle\omega_1,\omega_2\rangle = \int_M \omega_1 \wedge \omega_2$ for $\omega_1$, $\omega_2\in \DR$, is a \emph{symmetric dg-Frobenius algebra}.
It is well-known that finite-dimensional symmetric Frobenius algebras $V$ are equivalent to $2d$ topological quantum field theories (TQFT).
A finite dimension is necessary so that one can write the identity as $\Id = \sum \langle \cdot,e^i\rangle e_i$, or, in other words, that the identity propagator 
\[
T = \sum \pm e^i\otimes e_i
\]
is well defined.
We will ignore this issue and substitute $V=\DR$ for now, although we will soon transfer to $\HDR$, where everything works just fine.
We will represent interactions of fields via Feynman graphs drawn on surfaces --- the trivial cylinder for fields, i.e., the free propagation, will be a line, and the pair of pants, i.e., the interaction via the intersection~$\wedge$, will be a point with $3$ segments emanating from it (we do not have to distinguish inputs and outputs by cyclic symmetry).

\begin{figure}[t]
\centering
 \input{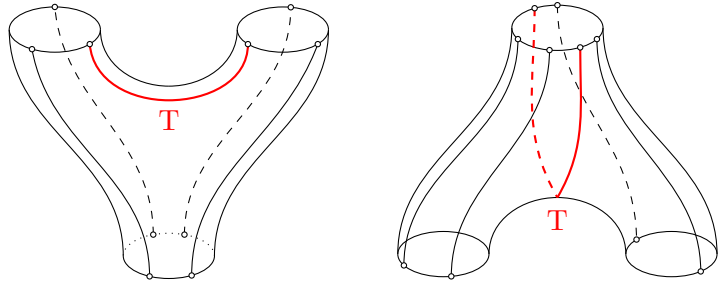}
 \caption[Operations $\OPQ_{210}$ and $\OPQ_{120}$.]{Operations $\OPQ_{210}$ and $\OPQ_{120}$. Fields propagate from the bottom to the top. An additional dualization is required when two outputs/inputs are connected; hence the emergence of the identity propagator.}
 \label{Fig:OpCoOpDiag}
\end{figure}

Let us now consider field strings. Figure \ref{Fig:OpCoOpDiag} defines the operations
\[ \OPQ_{210}: \Ext_2 \DBCyc V \longrightarrow \DBCyc V \quad\text{and}\quad \OPQ_{120}: \DBCyc V \longrightarrow \Ext_2\DBCyc V,
\]
where $\Ext_k \DBCyc V$ denotes the $k$-th exterior power of $\DBCyc V$ seen as the $k$-th symmetric power of the degree shift $(\DBCyc V)[1]$.
We read the diagram from the top to the bottom but imagine fields $\omega\in \BCyc V$ being fed into $\psi\in \DBCyc V$ from the bottom to the top.
We might think of these digrams as of \emph{string interaction diagrams} for strings freely moving in a topological space $M$, connecting and disconnecting.
This suggests that $\OPQ_{210}$ and $\OPQ_{120}$ are related to $\StringOp_2$ and $\StringCoOp_2$.
Formulas for~$\OPQ_{210}$ and~$\OPQ_{120}$ were written down in \cite{Cieliebak2015}; they are also clear from the figure by decorating the world-lines with the identity (or $T$) and evaluating in a straightforward way.
It was proven that $\OPQ_{210}$ and $\OPQ_{120}$ indeed constitute an $\IBL$-algebra on~$\DBCyc V$ (note that this is not a TQFT for strings!). 
\ToDo[noline,caption={Which degree shift}]{Need to sort out which degree shift for bislgebra one needs!
How is it with $2-n$.}

As a mathematical remark, we will show that $\OPQ_{210}$ is obtained from the Gerstenhaber bracket on Hochschild cochains via cyclization by $\langle\cdot,\cdot\rangle$ and that $\OPQ_{120}$ is a factorization of an extension of the canonical Schwarz's $\BV$-operator on $\Fun(V[1])$ to cyclic invariants with respect to the cyclic shuffle product.
This makes sense because odd degree shift of a finite-dimensional symmetric dg-Frobenius algebra is an odd symplectic vector space. 

Since $\DBCyc V [1]$ is not naturally an odd symplectic vector space, there is no Schwarz's $\BV$-operator on $\Fun(\DBCyc V [1])$. However, the following canonical operator
\[ \BVOp_{\mathrm{s}} \coloneqq  \hat{\OPQ}_{120} + \hbar \hat{\OPQ}_{210}: \Fun(\DBCyc V [1]) \rightarrow \Fun(\DBCyc V [1]), \]
where $\hat{\cdot}$ denotes the canonical extension to (co)derivatives, is a $\BV$-operator with respect to the function multiplication; we call it the \emph{string $\BV$-operator.}

In physics, a $\BV$-operator $\BVOp$ on $\Fun(U)$, where $U$ is the space of fields (typically an odd cotangent bundle with classical fields in the base and ghost fields in the fibers), is related to the measure in the path integral $\int \mu$.
An action $S\in \Fun(U)$ satisfying the \emph{quantum master equation} (QME)
\[ \BVOp S + \frac{1}{2}\{S,S\} = 0 \]
defines a new measure $e^{-S} \mu$, and the corresponding twisted $\BV$-operator (or rather $\BVInfty$-operator) satisfies $\BVOp^S = e^{-S} \BVOp e^{S}$.

For field strings, we define the following \emph{actions} $S_{\mathrm{free}}$, $S_{\mathrm{int}}\in \Fun(\BCyc V[1])$, which remind us of the \emph{Chern-Simons functional:}
\[
S_{\mathrm{free}}(\omega_1 \omega_2) \coloneqq \pm \hbar^{-1}\int_M \omega_1 \wedge \Dd \omega_2 \quad\text{and}\quad S_{\mathrm{int}}(\omega_1 \omega_2 \omega_3) \coloneqq \pm \hbar^{-1}\int_M \omega_1 \wedge \omega_2 \wedge \omega_3.
\]
More precisely, $S_{\mathrm{free}}$ and $S_{\mathrm{int}}$ are linear functions on $\BCyc V[1]$ which vanish everywhere but on field strings of lengths $2$ and $3$, respectively.
It turns out that $S_{\mathrm{free}}$ and $S\coloneqq S_{\mathrm{free}} + S_{\mathrm{int}}$ satisfy the QME for $\BVOp_{s}$.
The twisted $\BV$-operators look like
\[
\BVOp^{S_{\mathrm{free}}}_{s} = \hat{\OPQ}_{110} + \BVOp_{\mathrm{s}} \quad \text{and}\quad \BVOp^{S}_{s} = \hat{\OPQ}_{110} + \reallywidehat{\OPQ_{210}(S_{\mathrm{int}},\cdot)} + \BVOp_{\mathrm{s}}.
\]
Figure~\ref{Fig:NewTerm} depicts the new terms $\OPQ_{110}$ and $\OPQ_{210}(S_{\mathrm{int}},\cdot)$ in $\BVOp^\Action_s$.
\begin{figure}[t]
 \centering
 \def\caphght{.6}
\def\BddMin{.2}
\def\BddMaj{.4}
\def\HorLen{2}
\def\PMCVert{1}
\def\PantsVert{2}
\def\PantsPlunge{.5}
\newcommand{\BddSurf}[6][0]{
\draw[#5,rotate=#1] ([shift=(0:{#3} and {#4})]#2) arc (0:180:{#3} and {#4});
\draw[#6,rotate=#1] ([shift=(180:{#3} and {#4})]#2) arc (180:360:{#3} and {#4});
}
\vcenterline{
\begin{tikzpicture}[scale=1.5]
\tikzset{point/.style = {draw, circle, fill=black, minimum size=2pt,inner sep=0pt}}
\node at (0,.9) {};
\coordinate (CT) at (0,0);
\coordinate (CB) at (0,-\PantsVert);
\draw ([shift={(-\BddMaj,0)}]CB) -- ([shift={(-\BddMaj,0)}]CT);
\draw ([shift={(\BddMaj,0)}]CB) -- ([shift={(\BddMaj,0)}]CT);
\BddSurf{CB}{\BddMaj}{\BddMin}{dotted}{}
\BddSurf{CT}{\BddMaj}{\BddMin}{}{}
\node[point,style={fill=white}] (NT1) at ([shift=(-40:{\BddMaj} and {\BddMin})]CT) {};
\node[point,style={fill=white}] (NT2) at ([shift=(-100:{\BddMaj} and {\BddMin})]CT) {};
\node[point,style={fill=white}] (NT3) at ([shift=(140:{\BddMaj} and {\BddMin})]CT) {};
\node[point,style={fill=white}] (NT4) at ([shift=(70:{\BddMaj} and {\BddMin})]CT) {};
\node[point,style={fill=white}] (NB1) at ([shift=(-40:{\BddMaj} and {\BddMin})]CB) {};
\node[point,style={fill=white}] (NB2) at ([shift=(-100:{\BddMaj} and {\BddMin})]CB) {};
\node[point,style={fill=white}] (NB3) at ([shift=(140:{\BddMaj} and {\BddMin})]CB) {};
\node[point,style={fill=white}] (NB4) at ([shift=(70:{\BddMaj} and {\BddMin})]CB) {};
\draw (NT1) -- (NB1);
\draw (NT2) -- (NB2);
\draw[dashed] (NT3) -- (NB3);
\draw[dashed] (NT4) -- (NB4);
\node[fill=white] at ($.5*(NT2)+.5*(NB2)$) {$\color{green}\Dd$};
\end{tikzpicture}}
\qquad + \qquad\vcenterline{
\begin{tikzpicture}[scale=1.5]
\tikzset{point/.style = {draw, circle, fill=black, minimum size=2pt,inner sep=0pt}}
\coordinate (C1) at (0,0); %
\coordinate (CC) at ($(C1) + (\HorLen,0)$); %
\coordinate (CV) at ($(C1) + (.5*\HorLen,\PMCVert)$); %

\coordinate (C2) at ($(CC) + (.5*\HorLen,-\PantsVert)$); %
\coordinate (CP) at ($(CC) + (.5*\HorLen,-\PantsPlunge)$); %
\coordinate (C3) at ($(CC) + (\HorLen,0)$); %

\coordinate (CT) at ($(CC) + (0,\caphght)$);
\draw ($(CC) + (-\BddMaj,0)$) to[out=90,in=180] (CT);
\draw (CT) to[out=0,in=90] ($(CC)+(\BddMaj,0)$);

\BddSurf{C2}{\BddMaj}{\BddMin}{dotted}{}
\BddSurf{C3}{\BddMaj}{\BddMin}{}{}
\BddSurf{CC}{\BddMaj}{\BddMin}{dotted}{dotted}

\draw ([shift={(-\BddMaj,0)}]CC) to[out=-90,in=90] ([shift={(-\BddMaj,0)}]C2);
\draw ([shift={(\BddMaj,0)}]C2) to[out=90,in=-90] ([shift={(\BddMaj,0)}]C3);

\draw ([shift={(\BddMaj,0)}]CC) to[out=-90,in=180] (CP);
\draw (CP) to[out=0,in=-90] ([shift={(-\BddMaj,0)}]C3);

\draw[thick,red] (CT) to[out=-50,in=90] ([shift=(-40:{\BddMaj} and {\BddMin})]CC) to[out=-85,in=180] ([shift={(0,.-.25*\PantsPlunge)}]CP) to[out=0,in=-90] ([shift=(-140:{\BddMaj} and {\BddMin})]C3);
\draw ([shift=(245:{\BddMaj} and {\BddMin})]C2) to[out=90,in=-80] ([shift=(240:{\BddMaj} and {\BddMin})]CC) to[out=90,in=-120] (CT);

\draw[dashed] ([shift=(110:{\BddMaj} and {\BddMin})]C2) to[out=90, in=-85] ([shift=(90:{\BddMaj} and {\BddMin})]CC) to[out=90,in=-90] (CT);

\draw ([shift=(-60:{\BddMaj} and {\BddMin})]C2) to[out=90, in=-110] ([shift=(-40:{\BddMaj} and {\BddMin})]C3);

\draw[dashed] ([shift=(70:{\BddMaj} and {\BddMin})]C2) to[out=90, in=-90] ([shift=(80:{\BddMaj} and {\BddMin})]C3);
\node[label={[yshift=-.8cm] $\color{red}\mathrm{T}$}] at (CP) {};

\node[point,style={fill=white}] at ([shift=(80:{\BddMaj} and {\BddMin})]C3) {};

\node[point,style={fill=white}] at ([shift=(-40:{\BddMaj} and {\BddMin})]C3) {};

\node[point,style={fill=white}] at ([shift=(-140:{\BddMaj} and {\BddMin})]C3) {};
\node[point,style={fill=white}] at ([shift=(245:{\BddMaj} and {\BddMin})]C2) {};
\node[point,style={fill=white}] at ([shift=(110:{\BddMaj} and {\BddMin})]C2) {};
\node[point,style={fill=white}] at ([shift=(-60:{\BddMaj} and {\BddMin})]C2) {};
\node[point,style={fill=white}] at ([shift=(70:{\BddMaj} and {\BddMin})]C2) {};

\node[point,label={[above]$\color{green}\wedge$}] at (CT) {};

\end{tikzpicture}}
 \caption{Adding $\Dd$ and $\wedge$ via $\Action$.}
 \label{Fig:NewTerm}
\end{figure}
The corresponding $\dIBL$-algebra reads
\[ \Bigl(\DBCyc V,\OPQ_{110}^\MC\coloneqq \OPQ_{110} + \OPQ_{210}(S_{\mathrm{int}},\cdot),\OPQ_{210},\OPQ_{120}\Bigr). \]
One can show that $\OPQ_{110}^\MC$ is the Hochschild differential.
If this was well-defined for $V = \DR$, then it would surely be a model of string topology.

As in quantum field theories, we are going to ``formally'' \emph{integrate out redundant degrees of freedom in the path integral} of our ill-defined theory and obtain a well-defined theory on $\DBCyc\HDR$, which is ``formally'' homotopy equivalent to the original one.
We pick a Riemannian metric on $M$ and consider the Hodge decomposition
\[
\DR = \Harm \oplus \Dd\DR \oplus \CoDd\DR,
\]
where $\Harm \simeq \HDR$ is the space of harmonic forms defined by $\Dd \omega = \CoDd \omega = 0$.
One may interpret $\Dd \omega = 0$ as the Euler-Lagrange equation and $\CoDd \omega = 0$ as the Lorentz gauge.
The inverse of $\Dd: \CoDd\DR \rightarrow \Dd\DR$ extended by $0$ to $\Harm$ and $\CoDd\DR$ is called the \emph{standard Hodge homotopy $\HtpStd$;} equivalently, it is the unique coexact solution of
\[
\Dd \Htp + \Htp \Dd = \pi_\Harm - \Id,
\]
where $\pi_\Harm: \DR \rightarrow \Harm$ is the orthogonal projection.
The Schwartz kernel of $\HtpStd$ is the \emph{standard Hodge propagator} $\PrpgStd$.

A formula for the \emph{effective action} $W\in\Fun(\BCyc\HDR[1])$ was given in \cite{Cieliebak2015}; in their terminology, $W$ is equivalent to the \emph{(formal) pushforward Maurer-Cartan element}.
We have
\[ W = \hbar^{-1}\sum_{l\ge 1, g\ge 0} \PMC_{lg} \hbar^{g}, \]
where $\PMC_{lg} \in \hat{\Ext}_l \DBCyc\HDR$ is computed by summing over $(l+g-1)$-loop Feynman diagrams with interaction vertices $\wedge$ and propagator $\StdPrpg$.
We remark that the Feynman diagrams in~$W$ have at least one external vertex.
One might try to construct a refinement $(W^0_{lg})_{l\ge 1, g\ge 0}$ of the Chern-Simons invariant by summing over diagrams with no external vertex, but it seems to be unrelated to the $\IBLInfty$-theory so far.

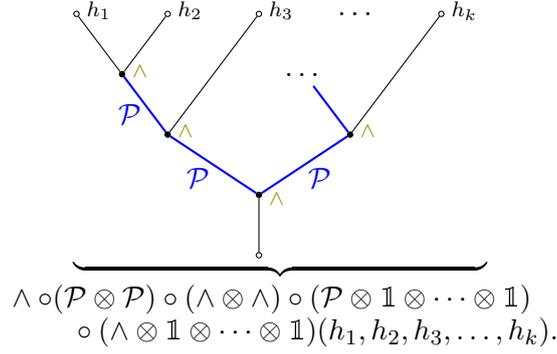
\begin{figure}[t]
\centering
\[\underbrace{\vcenterline{\begin{tikzpicture}[scale=1,
every label/.append style={font=\scriptsize},
point/.style = {draw, circle, fill=black, minimum size=2pt,inner sep=0pt},
leaf/.style = {draw, circle, fill=white, minimum size=2pt,inner sep=0pt},
]
\def\vertdist{.8}
\def\hordist{.6}
\node[leaf] (R) at (0,0) {};
\node[point, label={[right,yshift=-1mm] $\color{olive}\wedge$}] (RU) at ($(R) + (0,\vertdist)$) {};
\node[point, label={[right] $\color{olive}\wedge$}] (RUL) at ($(RU) + (-2*\hordist,\vertdist)$) {};
\node[point, label={[right] $\color{olive}\wedge$}] (RUR) at ($(RU) + (2*\hordist,\vertdist)$) {};
\node[point, label={[right] $\color{olive}\wedge$}] (RULL) at ($(RUL) + (-1*\hordist,1*\vertdist)$) {};
\coordinate (RULR) at ($(RUL) + (\hordist,\vertdist)$);
\coordinate (RURR) at ($(RUR) + (\hordist,\vertdist)$);
\node[leaf, label={[right] $h_1$}] (RULLL) at ($(RULL) + (-1*\hordist,1*\vertdist)$){};
\node[leaf, label={[right] $h_2$}] (RULLR) at ($(RULL) + (1*\hordist,1*\vertdist)$) {};
\node[leaf, label={[right] $h_3$}] (RULRR) at ($(RULR) + (1*\hordist,1*\vertdist)$) {};
\node[leaf, label={[right] $h_k$}] (RURRR) at ($(RURR) + (1*\hordist,1*\vertdist)$) {};
\node[] (RURRL) at ($(RURR) + (-1*\hordist,1*\vertdist)$) {\ \,\dots};
\node (RURL) at ($(RUR) + (-\hordist,\vertdist)$) {\dots};
\draw (R) edge (RU); 
\draw[thick,blue] (RU) -- (RUL) node[below,midway,shift={(-2mm,1mm)}] {$\Htp$}; 
\draw[thick,blue] (RUL) -- (RULL) node[below,midway,shift={(-2mm,1.5mm)}] {$\Htp$}; 
\draw (RULL) edge (RULLL);
\draw (RULL) edge (RULLR);
\draw (RUL) edge (RULRR);
\draw[thick,blue] (RU) -- (RUR) node[below,midway,shift={(2mm,1mm)}] {$\Htp$};
\draw[thick,blue] (RUR) edge (RURL);
\draw (RUR) edge (RURRR);
\end{tikzpicture}}}_{\begin{multlined} \wedge \circ (\Htp \otimes \Htp)\circ (\wedge \otimes \wedge)\circ (\Htp \otimes \Id \otimes \dotsb  \otimes \Id)\\
 \circ (\wedge \otimes \Id \otimes \dotsb \otimes \Id)(h_1, h_2, h_3, \dots, h_k).
\end{multlined}}\]
\caption{Kontsevich-Soibelman evaluation of a decorated tree.}
\label{Fig:KSTree}
\end{figure}

The twisted string $\BV$-operator on $\Fun(\BCyc\HDR[1])$ reads
\[ \BVOp^W_s = \hat{\OPQ}^{\PMC}_{110}  +  \hbar{\OPQ}_{210} + \sum_{l\ge 2, g\ge 0} \hat{\OPQ}^{\PMC}_{1lg} \hbar^{g}, \]
where $\OPQ_{110}^\PMC = \OPQ_{210}(\PMC_{10},\cdot) = \OPQ_{210} \circ_1 \PMC_{10}$ and $\OPQ_{1lg}^\PMC = \OPQ_{210}\circ_1 \PMC_{lg}$, where $\circ_1$ means that precisely one output of the first operation is connected to precisely one input of the following operation.
The resulting $\IBLInfty$-structure on $\CDBCyc\HDR$ has lots of vanishing operations.
It is in fact a \emph{quantum $\CoLInfty$-algebra $(\OPQ_{1lg}^\PMC)$ with Drinfeld-compatible Lie bracket $\OPQ_{210}$.}
The boundary operator $\OPQ_{110}^\PMC$ is precisely the Hochschild differential of the homotopy transfered $\AInfty$-structure $(m_k)$ on $\Harm$.
The quasi-isomorphism of $(\CDBCyc\HDR,\OPQ_{110}^\PMC)$ and $(C(\StringSpace M),\Bdd)$ inducing an isomorphism of the $\IBL$-structure on homology is given by the composition $F\circ I^*: C(\StringSpace M) \rightarrow \CDBCyc \HDR$ for
\[ F = \HTP_{110} + \HTP_{210}\circ_1 \MC_{10} + \frac{1}{2!} \HTP_{310}\circ_{1,1}(\MC_{10},\MC_{10}) + \dotsb, \]
where $\HTP_{110} = \iota^*$ for $\iota: \HDR\simeq \Harm \xhookrightarrow{} \DR$ and $\HTP_{k10}\circ_{1,\dotsc,1}(\MC_{10},\dotsc,\MC_{10})$ is obtained by summing over trivalent trees as in Figure~\ref{Fig:KSTree}.

Note that in order to evaluate trees, we do not need the Schwartz kernel $\Prpg$ and hence any pairing $\langle\cdot,\cdot\rangle$.
The homotopy $\Htp$ is enough because the graph is directed and we can distinguish inputs and outputs.
On the other hand, an evaluation of the $1$-loop Feynman graph in Figure~\ref{Fig:OneLoopDiag} contributing to $\OPQ_{120}^\PMC$ requires $\Prpg$, and hence also $\langle\cdot,\cdot\rangle$.

It is well-known from Sullivan's minimal model theory of a simply-connected manifold~$M$ that the homotopy type of the homotopy transferred $\AInfty$-structure $(m_k)$ on $\Harm$ is a topological invariant which encodes the rational homotopy theory of $M$.
The $\IBLInfty$-construction is associated to the Poincar\'e $\DGA$ $(\DR,\Dd,\wedge, \langle\cdot,\cdot\rangle)$, i.e, a $\DGA$ whose homology is a Poincar\'e duality algebra.
It is not clear yet to which extent it depends on the pairing and what kind of invariant of $M$ it is. 
However, if $M$ is formal in the sense of $\DGA$'s, then $M$ is formal in the sense of Poincar\'e $\DGA$'s, and we conjecture that it is formal also in the sense of $\IBLInfty$-algebras; by this we mean that the twisted $\IBLInfty$-algebra on $\CDBCyc \HDR$ is homotopy equivalent to the canonical $\IBL$-algebra on $\CDBCyc \HDR$.
\begin{figure}[t]
\centering
\def\BddMin{.2}
\def\BddMaj{.4}
\def\HorLen{2}
\def\PMCVert{1}
\def\PantsVert{2}
\def\PantsPlunge{.5}
\newcommand{\BddSurf}[6][0]{
\draw[#5,rotate=#1] ([shift=(0:{#3} and {#4})]#2) arc (0:180:{#3} and {#4});
\draw[#6,rotate=#1] ([shift=(180:{#3} and {#4})]#2) arc (180:360:{#3} and {#4});
}
\begin{tikzpicture}[scale=1.5]
\tikzset{point/.style = {draw, circle, fill=black, minimum size=2pt,inner sep=0pt}}
\coordinate (C1) at (0,0);
\coordinate (CC) at ($(C1) + (\HorLen,0)$);
\coordinate (CV) at ($(C1) + (.5*\HorLen,\PMCVert)$);

\coordinate (C2) at ($(CC) + (.5*\HorLen,-\PantsVert)$);
\coordinate (CP) at ($(CC) + (.5*\HorLen,-\PantsPlunge)$);
\coordinate (C3) at ($(CC) + (\HorLen,0)$);
 
\BddSurf{C1}{\BddMaj}{\BddMin}{dotted}{}
\BddSurf{C2}{\BddMaj}{\BddMin}{dotted}{}
\BddSurf{C3}{\BddMaj}{\BddMin}{}{}
\BddSurf{CC}{\BddMaj}{\BddMin}{dotted}{dotted}

\draw ([shift={(\BddMaj,0)}]C1) to[out=90,in=180] ([shift={(0,-\BddMaj)}]CV);
\draw ([shift={(0,-\BddMaj)}]CV) to[out=0,in=90] ([shift={(-\BddMaj,0)}]CC);
\draw ([shift={(-\BddMaj,0)}]CC) to[out=-90,in=90] ([shift={(-\BddMaj,0)}]C2);
\draw ([shift={(\BddMaj,0)}]C2) to[out=90,in=-90] ([shift={(\BddMaj,0)}]C3);

\draw ([shift={(-\BddMaj,0)}]C1) to[out=90,in=180] ([shift={(0,\BddMaj)}]CV);
\draw ([shift={(0,\BddMaj)}]CV) to[out=0,in=90] ([shift={(\BddMaj,0)}]CC);
\draw ([shift={(\BddMaj,0)}]CC) to[out=-90,in=180] (CP);
\draw (CP) to[out=0,in=-90] ([shift={(-\BddMaj,0)}]C3);

\BddSurf[90]{CV}{\BddMaj}{\BddMin}{dashed,thick,blue}{thick,blue}

\draw[red,thick] ([shift=(45:{\BddMin} and {\BddMaj})]CV) to[out=0,in=95] ([shift=(-40:{\BddMaj} and {\BddMin})]CC) to[out=-85,in=180] ([shift={(0,.-.25*\PantsPlunge)}]CP) to[out=0,in=-90] ([shift=(-140:{\BddMaj} and {\BddMin})]C3);
\draw ([shift=(-60:{\BddMaj} and {\BddMin})]C1) to[out=90,in=180] ([shift=(-20:{\BddMin} and {\BddMaj})]CV);

\draw ([shift=(-110:{\BddMaj} and {\BddMin})]C1) to[out=90,in=180] ([shift=(20:{\BddMin} and {\BddMaj})]CV);

\draw[dashed] ([shift=(125:{\BddMaj} and {\BddMin})]C1) to[out=80,in=180] ([shift=(145:{\BddMin} and {\BddMaj})]CV);
\draw ([shift=(245:{\BddMaj} and {\BddMin})]C2) to[out=90,in=-80] ([shift=(240:{\BddMaj} and {\BddMin})]CC) to[out=100,in=0] ([shift=(-50:{\BddMin} and {\BddMaj})]CV);

\draw[dashed] ([shift=(110:{\BddMaj} and {\BddMin})]C2) to[out=90, in=-85] ([shift=(90:{\BddMaj} and {\BddMin})]CC) to[out=95, in=15] ([shift=(190:{\BddMin} and {\BddMaj})]CV);

\draw ([shift=(-60:{\BddMaj} and {\BddMin})]C2) to[out=90, in=-110] ([shift=(-40:{\BddMaj} and {\BddMin})]C3);

\draw[dashed] ([shift=(70:{\BddMaj} and {\BddMin})]C2) to[out=90, in=-90] ([shift=(80:{\BddMaj} and {\BddMin})]C3);

\node[point,style={fill=white}] at ([shift=(-140:{\BddMaj} and {\BddMin})]C3) {};
\node[point] at ([shift=(190:{\BddMin} and {\BddMaj})]CV) {};

\node[point] at ([shift=(45:{\BddMin} and {\BddMaj})]CV) {};
\node[point] at ([shift=(-50:{\BddMin} and {\BddMaj})]CV) {};
\node[point] at([shift=(20:{\BddMin} and {\BddMaj})]CV) {};
\node[point] at ([shift=(-20:{\BddMin} and {\BddMaj})]CV) {};
\node[point] at ([shift=(145:{\BddMin} and {\BddMaj})]CV) {};

\node[label={[yshift=.2cm] $\psi$}] at (C3) {};
\node[label={[yshift=-.8cm] $\color{red}\mathrm{T}$}] at (CP) {};
\node at ([shift={(0,-\BddMin)}]CV) {};

\node[point,style={fill=white},label={[below,yshift=-.1cm,xshift=.1cm] $\scriptstyle h_{13}$}] at ([shift=(-60:{\BddMaj} and {\BddMin})]C1) {};
\node[point,style={fill=white},label={[below,yshift=-.1cm,xshift=-.1cm] $\scriptstyle h_{12}$}] at ([shift=(-110:{\BddMaj} and {\BddMin})]C1) {};
\node[point,style={fill=white},label={[below,yshift=+.1cm,xshift=-.6cm] $\scriptstyle h_{11}$}] at ([shift=(125:{\BddMaj} and {\BddMin})]C1) {};

\node[point,style={fill=white},label={[below,xshift=-.1cm,yshift=-.1cm] $\scriptstyle h_{22}$}] at ([shift=(245:{\BddMaj} and {\BddMin})]C2) {};
\node[point,style={fill=white},label={[left,xshift=-.4cm] $\scriptstyle h_{21}$}] at ([shift=(110:{\BddMaj} and {\BddMin})]C2) {};
\node[point,style={fill=white},label={[below,xshift=.1cm,yshift=-.1cm] $\scriptstyle h_{23}$}] at ([shift=(-60:{\BddMaj} and {\BddMin})]C2) {};
\node[point,style={fill=white},label={[right,xshift=.4cm] $\scriptstyle h_{24}$}] at ([shift=(70:{\BddMaj} and {\BddMin})]C2) {};

\node[label={[above,yshift=.03cm] $\scriptstyle x_1$}] at ([shift=(145:{\BddMin} and {\BddMaj})]CV) {};
\node[label={[above,yshift=-.05cm,xshift=.1cm] $\scriptstyle x_2$}] at ([shift=(45:{\BddMin} and {\BddMaj})]CV) {};
\node[label={[right,xshift=-.04cm,yshift=-.18cm] $\scriptstyle x_3$}] at ([shift=(20:{\BddMin} and {\BddMaj})]CV) {};
\node[label={[right,yshift=-.25cm,xshift=-.05cm] $\scriptstyle x_4$}] at ([shift=(-20:{\BddMin} and {\BddMaj})]CV) {};
\node[label={[below,yshift=-.27cm] $\scriptstyle x_5$}] at ([shift=(-50:{\BddMin} and {\BddMaj})]CV) {};
\node[label={[left,yshift=-.24cm,xshift=.05cm] $\scriptstyle x_6$}] at ([shift=(190:{\BddMin} and {\BddMaj})]CV) {};
\node[font=\footnotesize] (ZZ) at ([shift={(0,-4.5ex)}]CV) {$\color{blue}\Prpg$};
\end{tikzpicture}
\[\begin{aligned}
=&\sum_{a,b}\sum_{c=1}^4 \pm  \mathrm{T}^{ab} \psi(e_a  h_{2,c+2}  h_{2,c+3})  \Bigl(\int_{x_1 x_2 x_3 x_4 x_5 x_6} \Prpg(x_1,x_2)\Prpg(x_2,x_3)\Prpg(x_3,x_4)\\
&\Prpg(x_4,x_5)\Prpg(x_5,x_6)\Prpg(x_6,x_1)\bigl( h_{11}(x_1) h_{12}(x_3)
h_{13}(x_4)\bigr)\bigl(e_b(x_2)  h_{2,c}(x_6)  h_{2,c+1}(x_5)\bigr)\Bigr)
\end{aligned}
\]
\caption{A $1$-loop diagram and its contribution to the twisted cobracket.}
\label{Fig:OneLoopDiag}
\end{figure}

As a final remark, it is well-known from the theory of Koszul (pr)operads that $\IBL$ is Koszul dual to $\Frob$, i.e., $\IBL^! = \Frob$ and $\Frob^! = \IBL$. 
Here, $\Frob$ is the properad of Frobenius bialgebras.
It follows that $\IBLInfty = \Omega(\Frob^*)$, where $\Omega$ denotes the cobar construction and ${}^*$ the linear dual coproperad.
This precisely reflects our situation of having a Frobenius bialgebra structure on $\HDR$, where the coproduct is obtained from $\wedge$ via dualization, and an $\IBLInfty$-structure on $\DBCyc \HDR$ if $\BCyc$ is understood as a cyclic version of the cobar construction.

\section{Other relevant work}

In \cite{Cohen2001}, a homotopy theoretical realization of the Chas-Sullivan loop product on $\H(\Loop M)$ by constructing the ``wrong way map'' using the Thom-Pontryagin construction was described.
Note that having the loop product, one constructs $\StringOp_2$ on $\H^{\Sph{1}}(\Loop M)$ via the Gysin sequence for the Borel construction.

In \cite{Chen2012}, the Chas-Sullivan $\BV$-algebra on $\H(\Loop M)$ and the gravity algebra on $\H^{\Sph{1}}(\Loop M)$ were constructed using an algebraic model based on Whitney polynomial forms with coefficients in $\Q$.
The advantage of Whitney forms $A$ over de Rham forms $\DR$ is that the dualization of the product gives a complete coproduct with values in the currents $C = A^*$, which together with the product on $A$ constitutes a dg Frobenius-like algebra.
It is shown that $A\hat{\otimes}\hat{\Omega}(C)$, where $\hat{\Omega}$ denotes the complete cobar construction, carries a natural dg-algebra structure which corresponds to the loop product on $\H(\Loop M)$ under the Jones, et al., quasi-isomorphism from the singular chain complex of $\Loop M$.
The equivariant case is handled with methods of Connes' cyclic homology.

In \cite{Irie2014}, de Rham chains of marked Moore loops and their fiber products and concatenations are used to construct a non-symmetric dg operad $\mathcal{O}$ with a cyclic structure, multiplication and unit together with a morphism $\mathcal{O}\to \End_\DR$. The cyclic Deligne conjecture is applied to obtain an algebra $\widetilde{\mathcal{O}}$ over a chain model of the framed little disk operad (whose homology is the $\BV$-operad) such that the induced quasi-isomorphism from $\widetilde{\mathcal{O}}$ to Hochschild cochains induces an isomorphism of the $\BV$-structures on homology. The latter is known to be isomorphic to the $\BV$-structure on $\H(\Loop M)$ via iterated integrals.
In addition, they use the homotopy transfer from $\widetilde{\mathcal{O}}$ to obtain $\AInfty$- and $\LInfty$-structures on $\H(LM)$ whose operations with $2$ inputs are the loop product and the Gerstenhaber bracket, respectively.
Interestingly, their chain model works for non-simply connected $M$.

In~\cite{DrummondCole2015}, they use diffuse intersection and short geodesic segments to associate to metric chord diagrams operations on the singular chain complex of $\Loop M$.
They should recover the full positive boundary TQFT on $\H(\Loop M)$ described in \cite{Cohen2009}.

In \cite{Sullivan2005}, a rich structure of operations on equivariant chains of $\Loop M$ parametrized by chains in a certain compactification of the moduli space of Riemann surfaces is discussed. 
A part of this structure is an $\IBLInfty$-chain model on the reduced chains.

In \cite{ViterboThm}, it is proven that the $\BV$-algebra of symplectic homology of the cotangent bundle of an oriented manifold $M$ is isomorphic to the $\BV$-algebra~$\H(\Loop M)$.

In \cite{Cieliebak2007}, they sketch a proof that symplectic field theory of the unit cotangent bundle of $M$ and chain level equivariant string topology of $M$ are $\IBLInfty$-quasi-isomorphic via a map induced from evaluations at boundaries of holomorphic curves in symplectizations.
From this point of view, string topology operations arise naturally from the structure of codimension $1$ boundary stratas of the moduli space of holomorphic curves.
In fact, a precise formulation of this correspondence was perhaps the main reason for Cieliebak \& Latschev to think about an $\IBLInfty$-chain model of string topology.\ToDo[caption={Precise formulation of SFT},noline]{How is it precisely, what kind of boundary conditions and punctures?}

In \cite{Fukaya2006}, it is argued that the compactified moduli space $\widehat{\Model}$ of holomorphic discs with boundaries on a Lagrangian submanifold $L$ of a symplectic manifold $M$ gives rise to a filtered $\AInfty$-structure on $\H^*(L)$.
Evaluation at the boundary allows to interpret~$\widehat{\Model}$ as a chain in the free loop space.
The structure of codimension 1 boundary strata of~$\widehat{\Model}$ implies the relation $\Bdd \widehat{\Model} + \frac{1}{2}\{\widehat{\Model},\widehat{\Model}\} = 0$, where~$\{\cdot,\cdot\}$ is the chain level string bracket.
Under iterated integrals, this translates to the Maurer-Cartan equation on the cyclic bar construction of~$\DR(L)$ with the Gerstenhaber bracket.
The twisted coderivation gives the $\AInfty$-structure on $\DR(L)$ which is then homotopy transferred to $\H^*(L)$. 
\ToDo[caption={DONE What paper?},noline]{In what paper is this? Application of Floer homology of Langrangian submanifolds to symplectic topology}

\section{Summary of results}

\begin{enumerate}[label=\arabic*)]
\item The starting point was setting up a formalism and deducing signs for a definition of the formal-pushforward Maurer-Cartan element, aka Chern-Simons Maurer-Cartan element, and the $\IBLInfty$-chain model in the de Rham setting.
A big part of the work was about trying to understand what is happening and discovering and formulating the structure and possible claims.

\item We compute the $\IBLInfty$-chain model for $\Sph{n}$ with $n\neq 2$ by finding an explicit Hodge propagator and computing Feynman integrals.
In fact, for $n\ge 3$, all integrals which are relevant for the $\IBLInfty$-theory vanish.
 
A trick from \cite{Mnev2009} is based on modifying an abstract Hodge propagator to obtain special properties implying vanishing of the integrals.
The author of this thesis was not aware of this trick and tried to compute integrals with an explicit propagator in spherical coordinates for around 3 years until he rediscovered a part of this trick himself. 
The interesting thing is that the discovery was made via explicit computations, and it was a coincidence that the constructed Hodge propagator satisfied the special properties.
 
\item Using the trick from \cite{Mnev2009}, we generalize the previous computation to geometrically formal manifolds and show that the Feynman integrals vanish provided that $\HDR^1(M) = 0$.
For a general manifold, all higher coproducts vanish unless $M$ is a surface or a $3$-manifold with $\HDR^1(M) \neq 0$.
In fact, the homotopy type of the $\IBLInfty$-chain model for a manifold with $\HDR^1(M) = 0$ is determined by the tree-level perturbative Chern-Simons theory for a special Hodge propagator.

\item We conjecture that the $\IBLInfty$-chain model for formal manifolds with $\HDR^1(M) = 0$ is homotopy equivalent to the canonical $\dIBL$-structure.

\item There are two approaches of associating an $\IBLInfty$-homotopy type to a Poincar\'e $\DGA$ like $\DR(M)$.
One is using the homotopy transfer and integrals as explained in the previous section (geometric approach) and one is by taking a Poincar\'e duality model $\Model$ and constructing the canonical $\dIBL$ structure on cyclic cochains of $\Model$ (algebraic approach).
We study both and conjecture that they are equivalent. 

\item We study $\DGA$'s of Hodge type and give an alternative proof of the existence of a Poincar\'e duality model in the category of $\PDGA$'s.
In the $\DGA$ category, this is originally due to Lambrechts \& Stanley.
The new method is based on adding exact partners to non-degenerates rather than adding killers of orphans.
 
\item We prove a proposition that the cyclic homology of a strictly unital $\AInfty$-algebra can be computed from its reduced version.
We do it by extending Loday's cyclic homology theory for $\DGA$'s to $\AInfty$-algebras.

\item We relate $\OPQ_{210}$ to Gerstenhaber bracket and its cyclization and $\OPQ_{120}$ to the Schwarz's $\BV$-operator and cyclic shuffle product.

\item We extend the $\MV$-formalism to filtered $\MV$-formalism and use it to construct a $\BV$-formulation of the weak $\IBLInfty$-theory.
This has the advantage that the exponentials are honest exponentials and honest maps.
This will be useful for studying $\BV$-chain complexes.

\item We formulate the composition at $k$-common channels $\circ_k$ using ``heart with veins'' which appears in the iterated bialgebra compatibility condition.

\item We propose a $\BV$-formulation of the $\IBLInfty$-theory with an action, effective action and quantum master equation.

\item We find the standard Hodge propagator for $\Sph{2}$ up to a constant and prove that it smoothly extends to spherical blow-up.
\end{enumerate}

\part{IBL-infinity chain model}
\chapter{Overview}

An \emph{$\IBLInfty$-algebra} is essentially a collection of multilinear operations $\OPQ_{klg}$ with~$k$ inputs,~$l$ outputs and ``genus''~$g$ satisfying certain relations; in particular,~$\OPQ_{110}$ is a boundary operator, and the pair $\OPQ_{210}$, $\OPQ_{120}$ induces the structure of an involutive Lie bialgebra on the homology of $\OPQ_{110}$.
It was introduced in \cite{Cieliebak2015} and applications to string topology, symplectic field theory and higher genus Lagrangian Floer theory were proposed.

This part of the thesis is an attempt to understand its application to \emph{string topology}.
The idea was to carry out some explicit computations according to the plan sketched in~\cite[Section~13]{Cieliebak2015} and test the string topology conjecture (see below).

The following results from~\cite[Corollary~11.9]{Cieliebak2015} are our starting point.
Precise definitions of all the notions will be given in the next chapters.
$\IBLInfty$-algebras in Part~I will be strict and filtered in the terminology of \cite{Cieliebak2015}.
We denote by $[d]$ the degree shift by $d$; we handle the additional signs in \cite{Cieliebak2015} by working with degree shifted objects.
\begin{enumerate}[listparindent=\parindent,label=\textbf{(\Alph*)}]
\item For a finite-dimensional cyclic cochain complex $(V,\Pair,m_1)$ of degree $2-n$, where~$\Pair: V[1]\otimes V[1]\rightarrow\R$ is a pairing of degree $2-n$ and $m_1: V[1] \rightarrow V[1]$ a differential,
there is a canonical $\dIBL$-structure $\OPP_{110}$, $\OPP_{210}$, $\OPP_{120}$ of bidegree $(n-3,2)$ on the degree shifted dual cyclic bar complex
\[ \CycC(V)\coloneqq \DBCyc V[2-n] \simeq \Bigl(\bigoplus_{k\ge 1} \bigl(V[1]^{\otimes k} / \text{cyc}\bigr)^{\GD} \Bigr)[2-n], \]
where $\mathrm{cyc}$ stands for cyclic permutations with the Koszul sign and $\GD$ denotes the graded dual.
This structure is denoted by $\dIBL(\CycC(V))$.
\item Let $(\Harm,\Pair,m_1) \subset (V,\Pair,m_1)$ be a subcomplex such that the restriction $\Pair: \Harm[1]\otimes\Harm[1]\rightarrow\R$ is non-degenerate. We apply (A) to $(\Harm,\Pair,m_1)$ to get the canonical $\dIBL$-algebra $\dIBL(\CycC(\Harm)) = (\CycC(\Harm),\OPQ_{110},\OPQ_{210},\OPQ_{120})$. 
Suppose that $\pi: V[1] \rightarrow V[1]$ is a projection to $\Harm[1]$ which satisfies
\begin{equation*}
 \begin{aligned}
 \pi \circ m_1 &= m_1 \circ \pi \quad\text{and}\\ \Pair(\pi(v_1),v_2) &= \Pair(v_1,\pi(v_2))
\end{aligned}
\end{equation*}
for all $v_1$, $v_2 \in V[1]$, and let $\iota: \Harm[1]\rightarrow V[1]$ be the inclusion. A linear map $\Htp: V[1]\rightarrow V[1]$ of degree $-1$ such that
\begin{equation}\label{Eq:ConditionOnG}
\begin{aligned}
m_1 \circ \Htp + \Htp \circ m_1 &= \iota\circ \pi - \Id_{V[1]}\quad\text{and} \\ 
\Pair(\Htp(v_1),v_2) &= (-1)^{\Abs{v_1}}\Pair(v_1,\Htp(v_2)) 
\end{aligned}
\end{equation}
for all $v_1$, $v_2 \in V[1]$ induces an $\IBLInfty$-homotopy equivalence 
\[\HTP=(\HTP_{klg}): \dIBL(\CycC(V)) \longrightarrow \dIBL(\CycC(\Harm)) \]
such that $\HTP_{110}: \CycC(V)[1] \rightarrow \CycC(\Harm)[1]$ is the map given by the precomposition with~$\iota$ in every component.
We recall from \cite{Cieliebak2015} that $\HTP_{klg}: \Ext_k \CycC(V) \rightarrow \Ext_l \CycC(\Harm)$ is a linear map between exterior powers of $\CycC$ which are realized as symmetric powers of the degree shift $\CycC[1]$.
\end{enumerate}

The map $\HTP_{klg}$ is constructed as a sum of contributions coming from isomorphism classes of \emph{ribbon graphs} (=:\,multigraphs with a cyclic ordering of half-edges at every internal vertex) with~$k$ internal vertices, $l$ boundary components and genus $g$.
To compute the contribution of a labeled ribbon graph~$\Gamma$ to the value 
\[ \HTP_{klg}(\Psi_1\otimes \dotsb \otimes \Psi_k)(\W_1 \otimes \dotsb\otimes \W_l)\] 
for $\Psi_1$, $\dotsc$, $\Psi_k \in \DBCyc V[3-n]$ and  $\W_1$, $\dotsc$, $\W_l\in \BCyc \Harm [3-n]$, we decorate the~$i$-th internal vertex of $\Gamma$ with $\Psi_i$, external vertices lying on the $i$-th boundary component with components $v_{i1}$, $\dotsc$, $v_{i s_i}\in V[1]$ of $\W_i = \Susp (v_{i 1} \otimes \dotsb \otimes v_{i s_i} / \text{cyc})$, where $\Susp$ is a formal symbol of degree $n-3$, and internal edges with the Schwartz kernel $\Prpg$ of $\Htp$ with respect to $\Pair$.
Decorated ribbon graphs are then evaluated in a consistent way to obtain real numbers (see Appendix \ref{Section:Appendix} for an invariant formalism or \cite[Section 10]{Cieliebak2015} for the original definition in coordinates).

We will also use the following results from \cite[Proposition~12.5 and Theorem~12.9]{Cieliebak2015} about deformations of $\IBLInfty$-algebras:
\begin{enumerate}[resume,listparindent=\parindent,label=\textbf{(\Alph*)}]
 \item If in addition to (A) there is a product $m_2 : V[1]\otimes V[1] \rightarrow V[1]$ making $(V,m_1,m_2)$ into a cyclic dga, then $(-1)^{n-2} m_2^+\in \CycC(V)$, where $m_2^+ = \Pair(m_2(\cdot,\cdot),\cdot)$, defines a canonical Maurer-Cartan element $\MC\coloneqq(\MC_{10})$ for $\dIBL(\CycC(V))$.
The twisted $\IBLInfty$-algebra is again a $\dIBL$-algebra of bidegree $(n-3,2)$; it is denoted by $\dIBL^\MC(\CycC(V))$ and satisfies
\begin{equation} \label{Eq:CanonMCTwist}
\begin{aligned}
&\dIBL^\MC(\CycC(V)) \\ 
&\qquad = (\CycC(V),\OPP^\MC_{110}=\OPP_{110}+\OPP_{210}\circ_1\MC_{10},\ \OPP^\MC_{210}=\OPP_{210},\ \OPP^\MC_{120} = \OPP_{120}).
\end{aligned}
\end{equation}
\item The $\IBLInfty$-morphism $\HTP$ from (B) can be used to pushforward $\MC$ and obtain a Maurer-Cartan element $\PMC = (\PMC_{lg})$ for $\dIBL(\CycC(\Harm))$.
The twist by $\PMC$ is an $\IBLInfty$-algebra of bidegree $(n-3,2)$; it is denoted by $\dIBL^\PMC(\CycC(\Harm))$ and satisfies
\[\begin{aligned}
& \dIBL^\PMC(\CycC(\Harm)) \\
& \quad = \begin{multlined}[t]
\bigl(\CycC(\Harm), \OPQ_{110}^\PMC = \OPQ_{110} + \OPQ_{210}\circ_1 \PMC_{10},\ \OPQ_{210}^\PMC=\OPQ_{210},\ \OPQ_{120}^\PMC = \OPQ_{120} \\+ \OPQ_{210} \circ_1 \PMC_{20},
\text{ plus higher operations }\OPQ_{1lg}^\PMC = \OPQ_{210}\circ_1 \PMC_{lg} \bigr).
\end{multlined} \end{aligned} \]
This $\IBLInfty$-algebra is $\IBLInfty$-homotopy equivalent to $\dIBL^\MC(\CycC(V))$ via a twisted $\IBLInfty$-morphism 
\[\HTP^\MC=(\HTP^\MC_{klg}): \dIBL^\MC(\CycC(V)) \longrightarrow \dIBL^\PMC(\CycC(\Harm)). \]
\end{enumerate}

The pushforward Maurer-Cartan element $\PMC = \HTP_* \MC$ can be expressed as a sum of contributions of isomorphism classes of \emph{trivalent ribbon graphs} ($m_2^+$ has namely three inputs), where a labeled ribbon graph $\Gamma$ is decorated with $m_2^+$ at internal vertices, with the components of the $i$-th argument of $\PMC_{lg}$, i.e., elements of $\Harm[1]$, at the $i$-th boundary component and with $\Prpg$ at internal edges.

The application to string topology of an oriented closed manifold $M$ of dimension~$n$ comes from studying generalizations of (A) -- (D) to the infinite-dimensional oriented dga $(\DR,\allowbreak \Pair,\allowbreak m_1,\allowbreak m_2)$. Here, $\DR = \DR(M)$ is the de Rham complex of $M$ and the maps $\Pair: \DR[1]^{\otimes 2} \rightarrow \R$, $m_1: \DR[1]\rightarrow \DR[1]$ and $m_2: \DR[1]^{\otimes 2} \rightarrow \DR[1]$ are defined for all $\eta$, $\eta_1$, $\eta_2 \in \DR$ as follows:
\begin{equation} \label{Eq:DeRhamDGA}
 \qquad \mathllap{\text{de Rham cyc.~dga}}\left\{ \begin{aligned}
 \Pair(\SuspU\eta_1,\SuspU\eta_2) &\coloneqq (-1)^{\deg \eta_1} \int_M \eta_1\wedge \eta_2, \\ 
 m_1(\SuspU\eta) &\coloneqq \SuspU \Dd \eta,\\[\jot] 
 \quad m_2(\SuspU\eta_1,\SuspU\eta_2) &\coloneqq (-1)^{\deg \eta_1} \SuspU(\eta_1\wedge\eta_2), \end{aligned}\right.
\end{equation}
where $\Dd$ is the de Rham differential, $\wedge$ the wedge product, $\SuspU$ a formal symbol of degree $-1$ and $\deg \eta_1$ is the form-degree of $\eta_1$.

By picking a Riemannian metric on $M$, we obtain the subcomplex of harmonic forms $(\Harm, \Pair, m_1 \equiv 0)$ with the projection $\pi_{\Harm}: \DR \rightarrow \Harm$ coming from the Hodge decomposition.
This cyclic cochain complex shall be taken as the subcomplex in (B).

Because the intersection pairing on $\DR[1]$ is not perfect, one has to restrict the construction in (A) to the subspace $\DBCyc \DR_\infty$ of elements with smooth Schwartz kernel.
Then (A) and (B) work in the setting of the so called Fr\'echet $\IBLInfty$-algebras introduced in \cite[Section 13]{Cieliebak2015}.
However, the element $\MC_{10} \in \DBCyc \DR[3-n]$, which translates into the \emph{Chern-Simons term}
\begin{equation} \label{Eq:ChernSimons}
 m_2^+(\SuspU\eta_1,\SuspU\eta_2,\SuspU\eta_3) \coloneqq (-1)^{\deg \eta_2}  \int_M \eta_1\wedge \eta_2\wedge\eta_3\quad\text{for all }\eta_1, \eta_2, \eta_3 \in \DR(M),
\end{equation}
does not define the canonical Maurer-Cartan element $\MC$ in (C) directly because $m_2^+ \not\in \CDBCyc \DR_\infty$.
This also means that one cannot use (D) to conclude the existence of the pushforward Maurer-Cartan element $\PMC$.

Nevertheless, it was proposed to define $\PMC$ formally using a summation over trivalent ribbon graphs as in the finite-dimensional case.
We call such $\PMC$ the \emph{Chern-Simons} or \emph{formal pushforward Maurer-Cartan element.} In order to compute the contribution of a labeled trivalent ribbon graph $\Gamma$ with $k$ internal vertices, $l$ boundary components and genus $g$ to the value
\[ \PMC_{lg}(\Omega_1\otimes \dotsb \otimes \Omega_l), \]
where $\Omega_i = \Susp \omega_i$ for $\omega_1$,~$\dotsc$, $\omega_l \in \BCyc\Harm$, one starts by decorating internal vertices with integration variables $x_1$, $\dotsc$, $x_k$ on the $k$-fold product $M\times \dotsb \times M$, external vertices on the $i$-th boundary component with the components $\alpha_{i1}$,~$\dotsc$, $\alpha_{i s_i} \in\Harm[1]$ of $\omega_i$ and internal edges with the Hodge propagator $\Prpg$.
In the setting of $\DR$ and $\Harm$, $\Prpg$ shall be the Schwartz kernel of a homotopy $\Htp$ in the sense of pseudo-differential operators (it is necessarily singular at the diagonal and smooth outside of it, i.e., $\Prpg \in \DR^{n-1}(M\times M \backslash \Diag)$).
One then takes the wedge product of all forms in the decorated graph in the order and with the sign deduced from the labeling of~$\Gamma$ and computes the integral over $x_1$, $\dotsc$, $x_k$.
Similar integrals appear in \emph{perturbative Chern-Simons quantum field theory}.

Because of the singularity of $\Prpg$ at $\Diag$, the integrand described above is smooth only on the $k$-th configuration space of $M$.
It is not clear that all the integrals converge and that the resulting $(\PMC_{lg})$ are well-defined and satisfy the Maurer-Cartan equation.
The idea of work in progress~\cite{Cieliebak2018} of K.~Cieliebak and E.~Volkov is to use iterated spherical blow-ups of multiple diagonals to resolve the singularities and obtain integrals of smooth forms on compact manifolds with corners; this guarantees integrability. The Maurer-Cartan equation for $\PMC = (\PMC_{lg})$ is then proven with the help of a Stokes' formula on stratified spaces; the key part is to show that contributions of hidden codimension-$1$ faces cancel. This method is similar to the method from~\cite{Axelrod1991} and~\cite{Axelrod1993}, where Feynman integrals of perturbative Chern-Simons theory were considered.

Having $\PMC$, the twisted $\IBLInfty$-algebra $\dIBL^\PMC(\CycC(\Harm))$, which can be equivalently written as $\dIBL^\PMC(\CycC(\H_{\mathrm{dR}}))$ using the Hodge isomorphism $\Harm\simeq \H_{\mathrm{dR}}$, should satisfy the following conjecture:

\newtheorem*{Conject}{String topology conjecture}
\begin{Conject}[{Quotation of \cite[Conjecture~1.12]{Cieliebak2015}}]
Let $M$ be a closed oriented manifold of dimension $n$ and $\H_{\mathrm{dR}}$ its de Rham cohomology. Then there exists an $\IBLInfty$-structure on (a suitable version of) $\DBCyc \H_{\mathrm{dR}}[2-n]$ whose homology equals the cyclic cohomology of the de Rham complex of $M$.
\end{Conject}

The idea is as follows. The \emph{$\Sph{1}$-equivariant homology of the free loop space} $\StringH(\Loop M)$ is for simply-connected $M$ isomorphic to a version of \emph{Connes' cyclic cohomology of the de Rham algebra} $\CycCoH^*(\DR)$. The precise relation will be established in yet another work in progress~\cite{Cieliebak2018} of K.~Cieliebak and E.~Volkov using a chain-map coming from a cyclic version of Chen's iterated integrals. Now, a suitable degree shift of $\CycCoH^*(\DR)$ is isomorphic to the homology of the boundary operator~$\OPQ_{110}^\MC$ of the hypothetical $\dIBL$-algebra $\dIBL^\MC(\CycC(\DR))$. This hypothetical $\dIBL$-algebra is then by (D) hypothetically quasi-isomorphic to the $\IBLInfty$-algebra $\dIBL^\PMC(\CycC(\Harm))$ via the twisted $\IBLInfty$-morphism $\HTP^\MC$.

The space $\StringH(\Loop M, M)$ carries an $\IBL$-structure consisting of the Chas-Sullivan string bracket $\StringOp_2$ and string cobracket $\StringCoOp_2$; \Add[caption={DONE Reduced loop space}]{Write that they are actually defined on the homology relative to constant loops. Actually, write here modulo problems with modding out constant loops or one constant loop.} these operations were defined geometrically on suitably transverse smooth chains in \cite{Sullivan1999} and \cite{Sullivan2002}, respectively. A natural question is: How is the $\IBL$-structure $\StringOp_2$, $\StringCoOp_2$ related to the $\IBL$-structure $\OPQ_{210}^\PMC$, $\OPQ_{120}^\PMC$ induced on $\StringH(\Loop M)$ via the isomorphism from the string topology conjecture? An extended string topology conjecture asserts that these structures agree, and hence the operations $\OPQ_{210}^\PMC$, $\OPQ_{120}^\PMC$ defined on cyclic cochains provide a \emph{chain model} for~$\StringOp_2$, $\StringCoOp_2$. Based on our observations and explicit computations, we formulate an up-to-date version of the string topology conjecture for simply-connected manifolds (see Conjecture~\ref{Conj:StringTopology}).

Our first result is an explicit computation of $\dIBL^\PMC(\CycC(\HDR({\Sph{n}})))$ by finding a particular Hodge propagator and showing that all integrals which contribute to $\PMC$ vanish for $n\ge 3$; for $n=1$, there is a non-vanishing integral whose value we compute (see Section~\ref{Sec:GreenSphere}); for $n=2$, the existence of a non-vanishing integral remains open.

\begin{IntroThm}[Explicit computation for $\Sph{n}$]\label{IntroThm:A}
Consider the round sphere $\Sph{n}\subset \R^{n+1}$. Define $\NOne\coloneqq \SuspU 1$, $\NVol\coloneqq \SuspU \Vol \in \HDR(\Sph{n})[1]$, where $\Vol$ is the volume form, $1$ the constant one and $\SuspU$ a formal symbol of degree $-1$. There exists a Hodge propagator $\Prpg$ such that the following holds. For the homology of the twisted boundary operator $\OPQ_{110}^\PMC$, we have:
\begin{equation*}
\begin{aligned}
& \HIBL^\PMC(\CycC(\HDR(\Sph{n})))[1] \coloneqq \H(\CDBCyc \HDR(\Sph{n})[3-n],\OPQ_{110}^\PMC) \\
& \qquad =\begin{cases}
\langle \Susp \NVol^{i*}, \Susp \NOne^{2j-1*} \mid i, j \ge 1\rangle & \text{for }n\ge 3 \text{ odd}, \\
\langle \Susp \NVol^{2 i-1*}, \Susp \NOne^{2j-1*} \mid i, j \ge 1\rangle &\text{for }n\text{ even}, \\
 \bigl\langle \Susp\sum_{k=1}^\infty c_k \NVol^{k*}, \Susp \NOne^{2j-1*}\mid c_k\in \R, j \ge 1\bigr\rangle & \text{for }n=1. 
\end{cases}
\end{aligned}
\end{equation*}
Here $\langle \cdot \rangle$ denotes the linear span over $\R$, $*$ the dual and $\Susp$ is a formal symbol of degree $n-3$. The product $\OPQ_{210}^\PMC$ vanishes on $\HIBL^\PMC$ except for the following relations for $n\ge 3$ odd
\[ \OPQ_{210}^\PMC(\Susp \NOne^* \otimes \Susp \NVol^{k*}) = \OPQ_{210}^\PMC(\Susp \NVol^{k*} \otimes \Susp \NOne^*) = -(k-1) \NVol^{k-1*} \]
and the following relations for $n=1$:
\[ \OPQ_{210}^\PMC\Bigl(\Susp \NOne^* \otimes \Susp\sum_{k=1}^\infty c_k \NVol^{k*}\Bigr) = - \Susp \sum_{k=1}^\infty k c_{k+1} \NVol^{k*}.  \]
The coproduct $\OPQ_{120}^\PMC$ as well as all higher operations $\OPQ_{1lg}^\PMC$ vanish on $\HIBL^\PMC$ in every  dimension $n$. For $\Sph{1}$, we have $\OPQ_{120}^\PMC \neq \OPQ_{120}$ on the chain level; i.e., the twisting is non-trivial. For $n\neq 2$, all higher operations vanish on the chain level.

\end{IntroThm}

If we mod out $\Susp \NOne^{2j-1*}$, i.e., if we consider the point-reduced version\Modify[caption={DONE Point reduced}]{Write Point reduced version.}, then, after dropping~$\Susp$, the results agree with the string topology of $M$ relative to one constant loop and with Chas-Sullivan operations. The only exception is $M=\Sph{1}$. This supports the string topology conjecture for simply-connected manifolds and provides a counterexample for non-simply connected manifolds.

\Correct[caption={DONE Degree shift}]{Write here that we have to drop $\Susp$!}

Our second result generalizes the previous explicit computation and shows that in many cases, the twists with $\PMC$ and $\MC$ coincide. Its proof is a combination of facts from Section~\ref{Sec:Vanishing}.

\begin{IntroThm}[Triviality of the twist with $\PMC$ on the chain level] \label{IntroThm:B}
Let $M$ be a closed oriented $n$-manifold. There exists a Hodge propagator $\Prpg$, the so called special Hodge propagator, such that the following holds for the twisted $\IBLInfty$-structure $\dIBL^\PMC(\CycC(\HDR(M)))$:
\begin{enumerate}[label=(\arabic*)]
\item For the basic operations $\OPQ_{110}^\PMC = \OPQ_{210}\circ_1 \PMC_{10}$, $\OPQ_{210}^\PMC = \OPQ_{210}$, $\OPQ_{120}^\PMC = \OPQ_{120} + \OPQ_{210}\circ_1 \PMC_{20}$, we have:
\begin{enumerate}
\item If $\HDR^1(M)=0$, then $\PMC_{20}=0$, and hence $\OPQ_{120}^\PMC = \OPQ_{120}$.
\item If $M$ is geometrically formal, then $\PMC_{10} = \MC_{10}$, and hence $\OPQ_{110}^\PMC = \OPQ_{110}^\MC$. (In fact, if in addition $\HDR^1(M)=0$, then $\PMC = \MC$, at least for $n\ne 2$.)
\end{enumerate}
\item For the higher operations $\OPQ_{1lg}^\PMC=\OPQ_{210}\circ_1 \PMC_{20}$ with $(l,g)\neq (1,0)$, $(2,0)$, we have $\PMC_{lg}=0$, and hence $\OPQ_{1lg}^\PMC=0$ with the possible exception of surfaces and $3$-manifolds with $\HDR^1(M)\neq 0$.
\end{enumerate}
\end{IntroThm}

The following are some interesting directions we plan to work on sorted from the most concrete to the most unclear:

\begin{enumerate}[label=(\arabic*)]
\item Improve Theorem~\ref{IntroThm:B} by showing that the higher operations for $\Sph{2}$ vanish. If this is the case, then the statement that for every manifold~$M$ with $\HDR^1(M)=0$ there is a propagator such that all higher operations vanish is true.
\item Let $M$ be a formal and simply-connected manifold. Are  $\dIBL^\PMC(\CycC(\HDR(M)))$ and $\dIBL^\MC(\CycC(\HDR(M)))$ homotopy equivalent as $\IBLInfty$-algebras? If not, we would like to understand the obstruction. 
\item Does the Schwartz kernel $\PrpgStd$ of $\HtpStd=-\CoDd \Delta^{-1}$, the so called standard Hodge propagator, where $\CoDd$ is the codifferential and~$\Delta$ the Hodge-de~Rham Laplacian, extends smoothly to a blow-up? If yes, then it is a canonical special Hodge propagator for which the statement of Theorem~\ref{IntroThm:B} holds.
\item Compute $\dIBL^\PMC(\CycC(\HDR(M)))$ for surfaces $\Sigma_g$ with $g\ge 1$ and formulate and proof a version of string topology conjecture for non-simply connected manifolds. One can consider differential forms with values in a Lie algebra and obtain an $\IBLInfty$-theory with gauge group. How is this structure related to \cite{Goldman1986} and \cite{Andersen1996}?
\item The $\IBLInfty$-theory uses only graphs with at least one external vertex to construct~$\PMC_{lg}$. Let $W_{lg}$ be the real number obtained by summing over graphs with $l$ boundary components, genus $g$ and with no external vertex. How is this related to the partition function and Chern-Simons invariants and does it fit into a ``weak non-reduced $\IBLInfty$-formalism'' and possibly satisfy some relations?
\item What physical field theory has $\Prpg$ as a propagator and trivalent ribbon graphs with harmonic forms as Feynman graphs?
Due to the polarization $\Prpg = \sum_{k=0}^{n-1}\Prpg^{(k)}(x,y)$, such theory must have a field in every degree (the standard $3d$ Chern-Simons theory before gauge fixing has just one $1$-form).
\item Can the $\IBLInfty$-theory be interpreted from the point of view of string field theory or topological strings? Together with symplectic field theory, what is the global picture?
\end{enumerate}

\chapter{Algebraic structures}

\label{Sec:Alg0}

In Section~\ref{Sec:Alg1a}, we recall weight-grading (Definition \ref{Def:Grading}), Koszul sign (Definition~\ref{Def:Koszul}), degree shift (Definition~\ref{Def:DegreeShift}), filtrations (Definition~\ref{Def:Filtrations}) and completions (Definition~\ref{Def:Completion}).
We prove the K\"unneth formula for completed symmetric (co)homology (Proposition~\ref{Prop:Kuenneth}).

In Section~\ref{Sec:Alg1}, we review basics of $\IBLInfty$-algebras from \cite{Cieliebak2015}.
We define the exterior algebra $\Ext C$ over a graded vector space $C$ as the symmetric algebra~$\Sym$ over~$C[1]$ (Definition~\ref{Def:ExtAlg}) and use the operations $\mu$ and $\Delta$ from the structure of an associative bialgebra on $\Sym(C[1])$ to give explicit formulas for the partial compositions $\circ_{h_1, \dotsc, h_k}$  (Definition~\ref{Def:CircS}).
We use the compositions to define the notion of an $\IBLInfty$-algebra $(\OPQ_{klg})$ on $C$ (Definition~\ref{Def:IBLInfty}), a~Maurer-Cartan element~$(\PMC_{lg})$ (Definition~\ref{Def:MaurerCartan}) and twisted operations $(\OPQ_{klg}^\PMC)$ (Definition~\ref{Def:TwistedOperations}).
We mention that an $\IBL$-algebra according to our definition is an odd degree shift of a classical $\IBL$-algebra (Proposition~\ref{Prop:ClasModIBL}).
We define the induced $\IBL$-structure on homology (Definition~\ref{Def:HomIBL}) and briefly discuss the $\mathrm{BV}$-formalism and mention weak $\IBLInfty$-algebras (Remark~\ref{Rem:BVForm}).
Finally, we summarize the situation for twisted $\dIBL$-algebras (Proposition~\ref{Prop:dIBL}) and briefly discuss higher operations (Remark~\ref{Rem:Higher}).

In Section~\ref{Sec:Alg2}, we define the (weight-reduced) dual cyclic bar-complex $\DBCyc V$ of a graded vector space $V$ (Definition~\ref{Def:BarComplex}) and introduce some notation (Notation~\ref{Def:Notation}).
We then summarize some facts about the completions $\CDBCyc V$ and $\hat{\Ext}_k \DBCyc V$ (Proposition~\ref{Prop:Compl}).
We define the notion of a cyclic $\AInfty$-structure on~$V$ (Definition~\ref{Def:CyclicAinfty}) and its Hochschild and cyclic homology (Definition~\ref{Def:CycHom}).
We recall strict units and strict augmentations (Definition~\ref{Def:AugUnit}), define the reduced dual cyclic bar complex $\RedDBCyc V$ (Definition~\ref{Def:ReducedDual}) and sketch a proof of the fact that the cyclic homology is a direct sum of the reduced cyclic homology and the cyclic homology of the ground field (Proposition~\ref{Prop:Reduced}).
Our version of cyclic homology of a dga $V$ is based on the degree shift $V[1]$.
We undo the degree shift and obtain a version based on $V$, which can be compared to the standard version from~\cite{LodayCyclic} (Proposition~\ref{Prop:DGA}).
We also show that the reduced spaces for a simply connected and connected $V$ are complete (Proposition~\ref{Prop:SimplCon}).

In Section~\ref{Sec:Alg3}, we review the construction of the canonical $\dIBL$-structure $\dIBL(\CycC(V))$ (Definition~\ref{Def:CanonicaldIBL}) and the canonical Maurer-Cartan element $\MC$ (Definition~\ref{Def:CanonMC}) starting from a cyclic dga $(V,\Pair,m_1,m_2)$.
We give formulas for the operations $(\OPQ_{1lg}^\PMC)$ of the $\IBLInfty$-algebra $\dIBL^\PMC(\CycC(V))$ twisted by a Maurer-Cartan element $\PMC$ (Proposition~\ref{Prop:Formulafortwisted}).
We consider the $\AInfty$-structure induced on $V$ by $\PMC_{10}$ (Definition~\ref{Def:MukDef}) and relate its cyclic homology to the homology of $\OPQ_{110}^\PMC$ (Proposition~\ref{Prop:CyclicHom}).
We define the reduced canonical $\dIBL$-algebra $\dIBL(\RedCycC(V))$ (Definition~\ref{Def:ReduceddIBL}) and the notion of a strictly reduced Maurer-Cartan element (Definition~\ref{Def:StrictlyReduced}).
The twisted $\IBLInfty$-structure then splits into the reduced part and the part generated by $\NOne^{i*}$, which we can explicitly compute (Proposition~\ref{Prop:Ones}).

\section{Gradings, degree shifts and completions
}
\allowdisplaybreaks
\label{Sec:Alg1a}

We will work with vector spaces over $\R$, possibly infinite-dimensional, graded by the \emph{degree} $d\in \Z$ and the \emph{weight} $k\in \N_0$.

\begin{Definition}[Weight-graded vector spaces]\label{Def:Grading}
A \emph{graded vector space} is a vector space~$W$ together with a collection  of subspaces $W^d \subset W$ for all $d\in \Z$ such that
\[ W=\bigoplus_{d\in \Z} W^d. \]
Elements of $W^d$ are called \emph{homogenous} of degree $d$; given $w\in W^d$, we denote the degree of $w$ by $\Abs{w} \coloneqq d$. 

A linear map of graded vector spaces $f: W_1 \rightarrow W_2$ is called homogenous of degree $\Abs{f} \in \Z$ if it holds 
\begin{equation}\label{Eq:StdGrading}
f(W_1^d)\subset W_2^{d+\Abs{f}}\quad\text{for all }d\in \Z.
\end{equation}

A \emph{weight-graded vector space} is a graded vector space $W$ together with a collection of subspaces $W_k^d \subset W^d$ for all $k\in \N_0$ and $d\in \Z$ such that
\[ W^d=\bigoplus_{k\in\N_0} W_k^d\quad\text{for all }d\in \Z. \]
We define the \emph{weight-$k$ component} of $W$ by
\[ W_k \coloneqq \bigoplus_{d\in \Z} W_k^d\quad\text{for all }k\in \N_0. \]
If $W_0^d = 0$ for all $d\in \Z$, we say that $W$ is \emph{weight-reduced.} We define the \emph{weight-reduced subspace} $\widebar{W}$ of a weight-graded vector space $W$ by
\[ \widebar{W} \coloneqq \bigoplus_{d\in \Z} \bigoplus_{k\in\N} W_k^d. \]

The superscript $\,{}^*$ usually denotes the dual in the category we work in, e.g., chain complexes. For a (weight-)graded vector space $W$, we also introduce the following notation to avoid confusion:%
\begin{equation}\label{Eq:Duals}\begin{aligned}
W^{\LD} \coloneqq \{\psi: W \rightarrow \R \text{ linear}\} &\;\,\dots&& \text{\emph{linear dual,}}\\[2\jot]
W^{\GD}  \coloneqq \bigoplus_{d\in  \Z} \prod_{k\in \N_0} W_{k}^{d*} &\;\,\dots&& \text{\emph{graded dual,}}\\
W^{\WGD} \coloneqq \bigoplus_{d\in \Z} \bigoplus_{k\in \N_0} W_k^{d*} &\;\,\dots&& \text{\emph{weight-graded dual.}}
\end{aligned}
\end{equation}
Here, $W^{\LD}$ is a vector space, $W^{\GD}$ a graded vector space and $W^{\WGD}$ a weight-graded vector space. The grading convention from \eqref{Eq:Duals} is the \emph{cohomological grading convention}, which differs from the convention \eqref{Eq:StdGrading} for maps $f: W \rightarrow \R$ by degree reversal (for this, see Definition~\ref{Def:DegreeShift}).

We identify $W^{\GD}$ with the subspace of $W^{\LD}$ generated by homogenous maps and~$W^{\WGD}$ with the subspace of $W^{\LD}$ generated by maps which are non-zero only on finitely many~$W_k^d$; hence, under this identification, we have the tower of inclusions 
\[ W^{\WGD} \subset W^{\GD} \subset W^{\LD}.\]
\end{Definition}

\begin{Definition}[Koszul sign] \label{Def:Koszul}
Let $k\ge 1$, and let $\sigma\in \Perm_k$ be a permutation on~$k$ elements, i.e., a bijection of the set $\{1,\dotsc,k\}$. For $i=1$, $\dotsc$, $k$,  let $a_i$ and $b_i$ be graded symbols of degrees~$\Abs{a_i}$ and~$\Abs{b_i}$, respectively. We denote by 
\[\varepsilon(\sigma,a) \quad\text{and}\quad\varepsilon(a,b)\]
the \emph{Koszul signs} of the transformations
 \[ a_1 \dots a_k \longmapsto a_{\sigma_1^{-1}} \dots a_{\sigma_k^{-1}} \quad \text{and} \quad a_1 \dots a_k b_1 \dots b_k \longmapsto a_1 b_1 \dots a_k b_k, \]
respectively. Here $\sigma_i^{-1} \coloneqq \sigma^{-1}(i)$. The Koszul sign is computed by permuting the left-hand side to the right-hand side via transpositions of two adjacent elements such that whenever we transpose two graded symbols, e.g., $a_i \longleftrightarrow a_{j}$, we multiply with $(-1)^{\Abs{a_i}\Abs{a_{j}}}$.
\end{Definition}

We emphasize that the Koszul sign depends only on the initial and the final order of the graded symbols and not on the sequence of transpositions; this is easy to show.

\begin{Definition}[Degree shift and grading reversal]\label{Def:DegreeShift}
Let $A\in \Z$. The \emph{degree shift by $A$} is a functor which associates to a graded vector space $W$ the graded vector space $W[A]$ with degree-$d$ components
\[ W[A]^d \coloneqq W^{d+A}\quad \text{for all }d\in \Z.\]%
There is the canonical degree shift morphism 
\begin{equation}\label{Eq:DegreeShift}
 W\longrightarrow W[A]
\end{equation}
of degree $-A$ mapping $W^d$ identically to $W[A]^{d-A}$. We view this morphism as multiplication from the left with a formal symbol $\Susp_A$ of degree $\Abs{\Susp_A}=-A$, so that~\eqref{Eq:DegreeShift} can be written as $w\in W \longmapsto \Susp_A w\in W[A]$.

Given graded vector spaces $W_1$, $W_2$ and constants $A_1$, $A_2\in \Z$, we associate to a morphism $f: W_1 \rightarrow W_2$ its \emph{degree shift} $f: W_1[A_1] \rightarrow W_2[A_2]$ by defining\footnote{This convention is not optimal; see Remark~\ref{Rem:BadConvention}.}
\begin{equation}\label{Eq:DegreeShiftConv}
f(\Susp_{A_1} w) \coloneqq \Susp_{A_2} f(w)\quad\text{for all } w \in W_1.
\end{equation} 
Notice that if $f:W_1 \rightarrow W_2$ has degree $\Abs{f}$, then $f: W_1[A_1] \rightarrow W_2[A_2]$ has degree $\Abs{f} + A_1 - A_2$.

The \emph{grading reversal} $r$ is a functor which associates to a graded vector space~$W$ the graded vector space $r(W)$ with
\[ r(W)^d \coloneqq W^{-d}\quad\text{for all }d\in \Z. \]
There is the canonical morphism $W\rightarrow r(W)$ mapping $W^d$ identically to $W^{-d}$ for every $d\in \Z$. The degree reversal of a morphism $f: W_1 \rightarrow W_2$ is the morphism $f: r(W_1) \rightarrow  r(W_2)$ defined by conjugating $f$ with the canonical morphism. If $\Abs{f}$ is the degree of $f: W_1 \rightarrow W_2$, then $-\Abs{f}$ is the degree of $f: r(W_1) \rightarrow  r(W_2)$.
\end{Definition}

In our main reference \cite{Cieliebak2015}, they view $W$ and $W[A]$ as one vector space with two different gradings $\Deg(\cdot)$ and $\Abs{\cdot}$, respectively; these are related by
\begin{equation*}
\Abs{w} = \Deg(w) - A \quad\text{for all homogenous }w\in W. 
\end{equation*}
On the other hand, we think of $W$ and $W[A]$ as of two different graded vector spaces and never use the same symbol for an element $w\in W$ and its degree shift $\Susp_A w \in W[A]$. %
It allows us to use just one notation~$\Abs{\cdot}$ for the gradings on both~$W$ and~$W[A]$. However, in order to preserve compatibility with~\cite{Cieliebak2015}, we will also sometimes use the notation $\Deg(w)$ (in the exponent just $(-1)^w$) for the degrees on $W$.

For graded vector spaces $W_1$, $\dotsc$, $W_k$ and constants $A_1$, $\dotsc$, $A_k\in \Z$, we identify 
\[ W_1[A_1]\otimes \dotsb \otimes W_k[A_k] \simeq (W_1\otimes \dotsb \otimes W_k)[A_1+\dotsb+A_k] \] using the \emph{Koszul convention for the tensor product}; for homogenous elements $w_1 \in W_1$,~$\dotsc$, $w_k \in W_k$, it reads
\begin{equation} \label{Eq:KoszulTensor}
\Susp_{A_1} w_1 \otimes \dotsb \otimes \Susp_{A_k} w_k = \varepsilon(\Susp_A,w) \underbrace{\Susp_{A_1}\dots\Susp_{A_k}}_{\displaystyle \mathclap{\eqqcolon \Susp_{A_1 + \dotsb + A_k}}\rule{0ex}{2ex}} w_1 \otimes \dotsb \otimes w_k.
\end{equation}
If $A_1 = \dotsb = A_k \eqqcolon A$ is fixed in the context, which is our usual case, we omit the subscript~$A$ and write just $\Susp$.

In the case of the multilinear map $f: W_1\otimes \dotsb \otimes W_k \rightarrow V_1\otimes \dotsb \otimes V_l$, the combination of~\eqref{Eq:DegreeShiftConv} and~\eqref{Eq:KoszulTensor} gives for $f: W_1[A_1]\otimes \dotsb\otimes W_k[A_k] \rightarrow V_1[B_1]\otimes \dotsb\otimes V_l[B_l]$ the following:
\begin{equation}\label{Eq:DegreeShiftConvII}
 f(\Susp_{A_1 +\dotsb + A_k} w_1 \otimes \dotsb \otimes w_k) = \Susp_{B_1 + \dotsb + B_l} f(w_1 \otimes \dotsb \otimes w_k).
\end{equation}

\begin{Remark}[Why is this sign convention bad?]\label{Rem:BadConvention}
Let us illustrate that \eqref{Eq:DegreeShiftConvII} is not compatible with the following standard Koszul rule:
\begin{equation*}
(K):\qquad (f_1 \otimes f_2)(w_1 \otimes w_2) = (-1)^{\Abs{f_2}\Abs{w_1}} f_1(w_1) \otimes f_2(w_2).
\end{equation*}
On one hand, we get 
\[\begin{aligned}
(f_1 \otimes f_2)(\Susp^2 w_1 \otimes w_2) &\overset{\eqref{Eq:DegreeShiftConvII}}{=} \Susp^2 (f_1\otimes f_2)(w_1 \otimes w_2) \\
& \overset{(K)}{=} (-1)^{\Abs{f_2}\Abs{w_1}} \Susp^2 f_1(w_1) \otimes f_2(w_2)  \\
& \overset{\eqref{Eq:KoszulTensor}}{=} (-1)^{\Abs{f_2}\Abs{w_1} + A(\Abs{f_1} + \Abs{w_1})} \Susp f_1(w_1) \otimes \Susp f_2(w_2).
\end{aligned}\]
On the other hand, we get
\[\begin{aligned}
(f_1 \otimes f_2)(\Susp^2 w_1 \otimes w_2) &\overset{\eqref{Eq:KoszulTensor}}{=} (-1)^{A\Abs{w_1}}(f_1 \otimes f_2)(\Susp w_1 \otimes \Susp w_2)\\
&\overset{(K)}{=} (-1)^{A\Abs{w_1} + \Abs{f_2}(A+\Abs{w_1})} f_1(\Susp w_1) \otimes f_2(\Susp w_2)\\ &\overset{\eqref{Eq:DegreeShiftConvII}}{=}(-1)^{A\Abs{w_1}+ \Abs{f_2}(A+\Abs{w_1})} \Susp f_1(w_1) \otimes \Susp f_2(w_2).
\end{aligned}\]
The results differ by $(-1)^{A(\Abs{f_1}+\Abs{f_2})}$. Therefore, we can not use (K) to identify the tensor product $\Hom(W_1,V_1)\otimes \Hom(W_2,V_2)$ with a subspace of $\Hom(W_1\otimes W_2, V_1\otimes V_2)$ in general. We will rather define an ad-hoc pairing in the case where we need it (see Definition~\ref{Def:Pairings}).

Another caveat is that in the case of the tensor product, the degree shift by $A_1$ followed by the degree shift by~$A_2$ is not the same as the degree shift by $A_1 + A_2$. Indeed, we compute
\[\begin{aligned}
(\Susp_{A_1 + A_2}w_1) \otimes (\Susp_{A_1 + A_2}w_2) &=
(\Susp_{A_2} \Susp_{A_1}w_1) \otimes (\Susp_{A_2}\Susp_{A_1} w_2) \\ &= (-1)^{A_2(A_1 + \Abs{w_1})}\Susp_{A_2}^2 (\Susp_{A_1} w_1) \otimes (\Susp_{A_1} w_2) \\
&= (-1)^{A_2 A_1 + (A_1 + A_2)\Abs{w_1}} \Susp_{A_2}^2 \Susp_{A_1}^2 (w_1 \otimes w_2) \\ &= (-1)^{A_2 A_1 + (A_1 + A_2)\Abs{w_1}} \Susp_{2(A_1 + A_2)} (w_1 \otimes w_2),
\end{aligned}\]
which differs by $(-1)^{A_1 A_2}$ from the direct degree shift by $A_1 + A_2$. Therefore, we have to always remember the vector spaces which we started with and the sequence of degree shifts. 

Note that we also have the unnatural ``$\Susp_{A_1} \Susp_{A_2} = \Susp_{A_1 + A_2} = \Susp_{A_2} \Susp_{A_1}$'' due to \eqref{Eq:KoszulTensor}.
\end{Remark}

\begin{Remark}[Is there a better sign convention?]
The author originally respected the Koszul rule for the algebra with formal symbols and considered the following map $\Susp^{l}_*\widebar{\Susp}^{k*} f: W[A]^{\otimes k} \rightarrow V[A]^{\otimes l}$ as the degree shift of $f: W^{\otimes k} \rightarrow V^{\otimes l}$:
\begin{equation}\label{Eq:AltConv}
(\Susp^{l}_*\widebar{\Susp}^{k*} f)(\Susp^{k} w_1 \otimes \dotsb \otimes w_k ) = (-1)^{k \Abs{f}A + \frac{1}{2}k(k-1) A} \Susp^l f(w_1 \otimes \dotsb \otimes w_k).
\end{equation}
Here $\DeSusp$ denotes the ``inverse'' of $\Susp$ with $\Abs{\widebar{\Susp}} = - \Abs{\Susp}$, $\Susp_*^l f = \Susp^l \circ f$ is the post-composition, $\DeSusp^{k*}f = (-1)^{k A \Abs{f}} f\circ \DeSusp^k$ the pre-composition, and the sign $ \varepsilon(\Susp, \DeSusp) = (-1)^{\frac{1}{2}k(k-1)A}$ comes from the ``collision'' $\DeSusp_1\dots \DeSusp_k \Susp_1 \dots \Susp_k \mapsto \DeSusp_1 \Susp_1 \dots \DeSusp_k \Susp_k$. 

However, the author did not manage to reprove the theory in~\cite{Cieliebak2015} using~\eqref{Eq:AltConv} (because of too many ``external'' signs appearing and a problem with disconnected graphs). A~motivation to try a different sign convention was to explain some artificial signs in~\cite{Cieliebak2015} and formulate their coordinate constructions invariantly in order to generalize them to the ``continuous'' de Rham case.

It might be possible to deduce a ``universal'' sign convention ``respecting'' the Koszul rules by considering the category of chain complexes and graded morphisms $\mathcal{C}$ as the category enriched in the closed monoidal category of chain complexes and chain maps of degree~$0$. One can then define the enriched degree shift functor $\Susp_{A}: \mathcal{C} \rightarrow \mathcal{C}$, embed $\mathcal{C}^{\otimes k} \subset \mathcal{C}$ using~$(K)$ and study enriched natural transformations in the algebra of functors consisting of tensor products and compositions of $\Susp_{A}$, $\Hom(\cdot, \cdot)$ and the dual $*$. The question is whether the sign rules for degree shifts, in particular  \eqref{Eq:AltConv}, are uniquely determined by the requirements on enriched functors and enriched natural transformations.

Another idea for how to specify the degree shift convention~\eqref{Eq:AltConv} is the observation that maps $f: W^{\otimes k}\rightarrow W^{\otimes l}$ under consideration usually form an algebra $\Prop \rightarrow \End_W$ of a certain PROP $\Prop$ over $W$; here $\End_W$ denotes the endomorphism PROP. Suppose that there is a natural notion of the degree shift $\Prop[A]$ of a PROP $\Prop$ and a canonical degree shift morphism $\Prop \rightarrow \Prop[A]$. The PROPs $(\End_W)[A]$ and $\End_{W[A]}$ are better to be isomorphic, and any intelligent degree shift convention $\psi_{k,l}: \Hom(W^{\otimes k},W^{\otimes l})[A(k-l)] \rightarrow \Hom(W[A]^{\otimes k},W[A]^{\otimes l})$ should induce an isomorphism of the PROPs $(\End_W)[A]$ and $\End_{W[A]}$.
\end{Remark}

\begin{Definition}[Standard action of permutations]\label{Def:Permutations}
For $k\ge 1$ and $\sigma\in \Perm_k$ ($\coloneqq$\,the group of permutations on $k$ elements), we define the \emph{standard action of $\sigma$} on $W^{\otimes k}$ by 
\begin{equation}\label{Eq:Perm}
\sigma(w_1 \otimes \dotsb \otimes w_k) \coloneqq  \varepsilon(\sigma,w) w_{\sigma_1^{-1}}\otimes \dotsb \otimes w_{\sigma_k^{-1}}
\end{equation}
for all homogenous $w_1$, $\dotsc$, $w_k\in W$.
\end{Definition}

Notice that the $i$-th vector is permuted to the $\sigma_i$-th place --- this is the ``active'' convention for permutations.

\begin{Definition}[Symmetric algebra]\label{Def:SymAlgebra}
Let $\Ten(V)\coloneqq \bigoplus_{k\ge 0} V^{\otimes k}$ be the tensor algebra over a graded vector space~$V$. The \emph{symmetric algebra} over $V$ is defined by $\Sym(V)\coloneqq \bigoplus_{k\ge 0} \Sym_k(V)$, where
\[ \Sym_k(V) \coloneqq V^{\otimes k} \bigl/ \sum_{\sigma\in \Perm_k} \Im(\Id-\sigma)\quad(\eqqcolon \Perm_k\text{-coinvariants}). \]
It is a weight-graded vector space with components denoted by $(\Sym_k V)^d$ for all $d\in \Z$ and $k\in \N_0$. Note that $\Sym_0 V = \R$ has degree $0$ by definition. Consider the canonical projection
\[\begin{aligned}
\pi : \Ten(V) &\longrightarrow \Sym(V) \\
v_1\otimes \dotsb \otimes v_k &\longmapsto v_1\dotsb v_k.
\end{aligned}\]
If $v_i\in V$ are homogenous, we call $v_1 \dotsb v_k$ a \emph{generating word}; we have
\[ v_1 \dotsb  v_k = \varepsilon(\sigma,v) v_{\sigma_1^{-1}} \dotsb v_{\sigma_k^{-1}}\quad \text{for every }\sigma\in \Perm_k. \]
Let $\iota: \Sym(V) \rightarrow \Ten(V)$ be the section of $\pi$ defined by 
\[ \iota(v_1\dotsb v_k) \coloneqq \frac{1}{k!}\sum_{\sigma\in\Perm_k} \varepsilon(\sigma,v) v_{\sigma_1^{-1}}\otimes \ldots \otimes v_{\sigma_k^{-1}}. \]
We use it to identify $\Sym(V)$ with the subspace of symmetric tensors
\[ \iota(\Sym_k(V)) = \bigcap_{\sigma\in \Perm_k} \Ker(\Id - \sigma) \subset \Ten_k(V)\quad(\eqqcolon\Perm_k\text{-invariants}). \]
\end{Definition}

\begin{Definition}[Filtrations] \label{Def:Filtrations}
Let $W$ be a graded vector space. A filtration of~$W$ is a collection of linear subspaces $\Filtr^\lambda W \subset W$ for $\lambda\in \R$ such that we have either
\begin{itemize}
\item $\Filtr^{\lambda_1}W\subset \Filtr^{\lambda_2}W$ for all $\lambda_1 \le \lambda_2$\quad$\Longleftrightarrow:$\quad\emph{increasing filtration}, or
\item  $\Filtr^{\lambda_1}W\supset \Filtr^{\lambda_2}W$ for all $\lambda_1 \le \lambda_2$\quad$\Longleftrightarrow:$\quad\emph{decreasing filtration.}
\end{itemize}
We will assume that our filtrations are \emph{graded} in the following sense:
\[ \forall \lambda\in \R:\quad \Filtr^\lambda W = \bigoplus_{d\in \Z} \Filtr^\lambda W^d,\quad \text{where}\quad\Filtr^\lambda W^d \coloneqq \Filtr^\lambda W \cap W^d.\]
A filtration $\Filtr^\lambda W$ is called:
\begin{align*}
 \text{\emph{exhaustive}}&\quad:\Equiv\quad\bigcup_{\lambda\in \R} \Filtr^\lambda W = W;\\
 \text{\emph{Hausdorff}} &\quad:\Equiv\quad\bigcap_{\lambda\in \R} \Filtr^\lambda W = 0;\\
 \text{\emph{$\Z$-gapped}} &\quad:\Equiv\quad\Filtr^\lambda W = \Filtr^{\lfloor\lambda\rfloor} W\text{ for all }\lambda \in \R;\\[3\jot]
 \text{\emph{bounded from below}}&\quad:\Equiv\quad\exists \lambda \in \R: \Filtr^\lambda W =0;\\[3\jot]
 \text{\emph{bounded from above}}&\quad:\Equiv\quad\exists \lambda\in \R: \Filtr^\lambda W =W.
\end{align*} 

Given a graded vector space $W$ filtered by a $\Z$-gapped filtration $\Filtr^\lambda W$, we associate to it the bi-graded vector space 
\[ \Gr(W) = \bigoplus_{d\in\Z}\bigoplus_{\lambda\in \Z} \Gr(W)_\lambda^d \]
called the \emph{graded module} whose components are given as follows:\footnote{The definition is made in such a way that if $r$ is the functor which reverses $\lambda$, i.e., $r(\Filtr)^\lambda = \Filtr^{-\lambda}$, then it holds $r(\Gr(\Filtr)) = \Gr(r(\Filtr))$. }\Modify[caption={DONE Graded module},noline]{Should not the grading of $\Gr$ be shitfted? I propose $gr_k = F_k/F_{k\pm 1}$ because $r(gr(\tilde{F})) = gr(F)$ for the degree reversed filtration.}
\[ \forall d, \lambda\in \Z: \quad \Gr(W)_\lambda^d \coloneqq \begin{cases}
                   \Filtr^\lambda W^d /\Filtr^{\lambda-1} W^d & \text{for increasing }\Filtr^\lambda W,\\
                   \Filtr^{\lambda} W^d / \Filtr^{\lambda+1} W^d & \text{for decreasing } \Filtr^\lambda W.
                 \end{cases}\]
We naturally extend a filtration over degree shifts, graded duals, direct sums, tensor products and symmetric products as follows: \allowdisplaybreaks
\begin{align*}
\Filtr^\lambda W[A]^d &\coloneqq \Filtr^\lambda W^{d+A},\\[3.5\jot]
\Filtr^\lambda (W^{\GD})^d &\coloneqq \{\psi\in W^{d*} \mid \Restr{\psi}{\Filtr^\lambda W} = 0\}, \\[3.5\jot]
\Filtr^\lambda\bigl(\bigoplus_{i\in I} W_i\bigr)^d &\coloneqq \bigoplus_{i\in I} \Filtr^{\lambda} W_i^d, \\[2\jot]
\Filtr^{\lambda}(W_1 \otimes \dotsb \otimes W_k)^d &\coloneqq \bigoplus_{\substack{d_1, \dotsc, d_k\in \Z \\ d_1 + \dotsb + d_k = d}}\ \sum_{\substack{\lambda_1, \dotsc, \lambda_k \in \R \\ \lambda_1 + \dotsb + \lambda_k = \lambda}} \Filtr^{\lambda_1} W_1^{d_1} \otimes \dotsb \otimes \Filtr^{\lambda_k} W_k^{d_k},\\
\Filtr^\lambda (\Sym_k V)^d &\coloneqq \pi(\Filtr^\lambda(V^{\otimes k})^d),
\end{align*}
where $\pi: \Ten(V) \rightarrow \Sym(V)$ is the canonical projection. The ground field $\R$ is filtered by the trivial filtration:
\begin{equation}\label{Eq:TrivFiltr}
\Filtr^\lambda \R \coloneqq \begin{cases} \R & \lambda\le 0, \\ 0 & \lambda >0. \end{cases}
\end{equation}
If $(W,\Bdd)$ is a filtered chain complex, we filter the homology as follows:
\[ \forall \lambda \in \R, d\in \Z: \quad \Filtr^\lambda \H_d(W,\Bdd) \coloneqq \{\alpha \in \H_d(C,\Bdd) \mid \exists w\in \alpha: w\in \Filtr^\lambda W^d \}. \]
\end{Definition}

\begin{Definition}[Completions]\label{Def:Completion}
Let $W$ be a graded vector space filtered by a decreasing filtration $\Filtr^\lambda W$. The \emph{filtration degree} of $w\in W$ is defined by
\[ \Norm{w} \coloneqq \sup\{ \lambda\in \R \mid w\in \Filtr^\lambda W\}. \]
The filtration degree of a linear map $f: W_1 \rightarrow W_2$ is defined by
\[ \Norm{f} \coloneqq \sup \{ \lambda\in \R \mid \Norm{f(w)} \ge \Norm{w} + \lambda\ \forall w\in W_1\}. \]
We say that the \emph{filtration degree is finite} if $\Norm{f}>-\infty$. Note that $\Norm{0} = \infty$.

The \emph{completion} of $W$ is the graded vector space
\[ \hat{W}\coloneqq \bigoplus_{d\in \Z} \hat{W}^d, \]
where for all $d\in \Z$  we define
\[ \hat{W}^d \coloneqq \Bigl\{ \sum_{i=0}^\infty w_i\ \Bigl|\   \forall i\in \N_0: w_i\in W^d; \Norm{w_i} \to \infty \text{ as }i\to \infty \Bigl\}\Bigl/\sim.
\]
Here $\sum_{i=0}^\infty w_i \sim \sum_{i=0}^\infty w_i'$ if and only if $\Norm{\sum_{i=0}^n (w_i - w_i')}\to \infty$ as $n\to \infty$.\footnote{\label{Footnote:Compl}In fact, $\hat{W}$ is the inverse limit $\varprojlim_\lambda^{\mathrm{gr}}(W/\Filtr^\lambda W)$ in the category of graded vector spaces and~$\hat{W}^d$ the inverse limit $\varprojlim_\lambda(W^d/\Filtr^\lambda W^d)$ in the category of vector spaces. As a side-remark, if we forget the grading on $W$, we might also consider $\varprojlim_\lambda (W/\Filtr^\lambda W)$, which would be a vector space containing~$\hat{W}$ as a subspace\vphantom{$W^d$}.}
The completion~$\hat{W}$ is canonically filtered by the filtration $\Filtr^\lambda \hat{W}$ defined as follows:
\[ \forall \lambda\in \R, d\in \Z: \quad \Filtr^\lambda\hat{W}^d\coloneqq \Bigl\{\sum_{i=0}^\infty w_i \in \hat{W}^d \ \Bigr|\ \forall i\in \N_0: w_i \in \Filtr^\lambda W^d \Bigr\}. \]
We denote the completion of $W_1 \otimes \dotsb \otimes W_k$ by $W_1 \COtimes \dotsb \COtimes W_k$ and the completion of $\Sym_k V$ by $\hat{\Sym}_k V$.

A map $f: W_1 \rightarrow W_2$ of finite filtration degree
\emph{extends continuously} to a linear map $f: \hat{W}_1 \rightarrow \hat{W}_2$; this continuous extension is defined by
\[ f\Bigl(\sum_{i=0}^\infty w_i\Bigr)\coloneqq \sum_{i=0}^\infty f(w_i)\quad\text{for all }\sum_{i=0}^\infty w_i \in \hat{W}. \]
\end{Definition}

\begin{Remark}[Completed tensor product]\label{Rem:ComplTens}
Using Proposition~\ref{Prop:IsoCrit} below, one can show that the \emph{completed tensor product} $\COtimes$ is associative and that $W_1 \COtimes W_2 \simeq \hat{W}_1 \COtimes \hat{W}_2$. By refining this argument, one can show that $\hat{\Sym}_k V \simeq \hat{\Sym}_k\hat{V}$ for any $k\in \N$.
\end{Remark}

A weight-graded vector space $W$ is canonically \emph{filtered by weights:}
\begin{equation}\label{Eq:FiltrWeights}
\forall \lambda\in \R, d\in \Z:\quad\Filtr^\lambda W^d \coloneqq \bigoplus_{k\le \lambda} W_k^d.
\end{equation}
This filtration is $\Z$-gapped, exhaustive, Hausdorff, increasing and bounded from below. The induced filtration on the graded dual $W^{\GD}$ is $\Z$-gapped, Hausdorff, decreasing and bounded from above (and thus automatically exhaustive). It holds $\Gr(W) \simeq W$, and it is easy to see from \eqref{Eq:Duals} that the canonical map $W^{\WGD} \rightarrow W^{\GD}$ induces the isomorphism
\[ \widehat{W^{\WGD}} \simeq W^{\GD}. \]
We also see that the condition
\[ (WG0): \quad \forall d\in \Z\ \exists J\subset \N_0, \Abs{J}<\infty\ \forall k\in \N_0\backslash J: \quad W_k^d = 0 \]
is equivalent to $W^{\WGD}= W^{\GD}$.

A useful tool to compare completions is the following proposition:

\begin{Proposition}[{\cite[Proposition 7.3.7]{Fresse}}, Isomorphism criterion]\label{Prop:IsoCrit}
Let $W_1$ and~$W_2$ be graded vector spaces filtered by $\Z$-gapped filtrations which are decreasing and bounded from above. Suppose that $f: W_1 \rightarrow W_2$ is a filtration preserving homogenous linear map. Then the continuous extension $f: \hat{W}_1 \rightarrow \hat{W}_2$ is an isomorphism if and only if the induced map $f: \Gr(W_1) \rightarrow \Gr(W_2)$ is an isomorphism.
\end{Proposition}
\begin{proof}
The implication from the right to the left is obtained from the diagram 
\[\begin{tikzcd}
0 \arrow{r} & \Gr(W_1)_\lambda \arrow[hook]{r} \arrow{d}{f} & W_1/\Filtr^{\lambda + 1}W_1 \arrow[two heads]{r} \arrow{d}{f} & W_1/\Filtr^{\lambda} W_1 \arrow{r}   \arrow{d}{f} & 0 \\
0 \arrow{r} & \Gr(W_2)_\lambda \arrow[hook]{r} & W_2/\Filtr^{\lambda+1}W_2 \arrow[two heads]{r} & \arrow{r}W_2/\Filtr^{\lambda} W_2   & 0
\end{tikzcd}\]
by induction using the definition of $\hat{W}$ as the inverse limit of $W/\Filtr^\lambda W$ (see Footnote \ref{Footnote:Compl} on page \pageref{Footnote:Compl}). The other implication follows from $\Filtr^\lambda\hat{W}/\Filtr^{\lambda-1}\hat{W}\simeq\Filtr^\lambda W/\Filtr^{\lambda-1}W$.
\end{proof}

For a graded vector space $W$ filtered by a $\Z$-gapped filtration, consider the following conditions:
\begin{equation}\label{Eq:WGs}
\begin{aligned}
(WG1):\quad& \forall \lambda\in \Z\ \exists I \subset \Z, \Abs{I}<\infty\ \forall d\in \Z\backslash I:& \Gr(W)_\lambda^d &= 0, \\
(WG2):\quad & \forall d, \lambda\in \Z:& \dim(\Gr(W)_\lambda^d) &< \infty.
\end{aligned}
\end{equation}

\begin{Lemma}[Completion of symmetric powers of the graded dual]\label{Lem:Terrible}
Let $W$ be a graded vector space filtered by an exhaustive $\Z$-gapped filtration~$\Filtr^\lambda W$ which is increasing and bounded from below. If (WG1) \& (WG2) are satisfied, then the natural map $\Sym_k(W^{\GD}) \rightarrow (\Sym_k W)^{\GD}$ induces the isomorphism 
\[ \hat{\Sym}_k(W^{\GD}) \simeq  (\Sym_k W)^{\GD} \quad\text{for every }k\in \N.\]
Note that we filter graded duals by the induced filtration from Definition~\ref{Def:Filtrations}.
\end{Lemma}
\begin{proof}
The natural map $\Sym_k(W^{\GD}) \rightarrow (\Sym_k W)^{\GD}$ is clearly filtration preserving, and hence it extends continuously to a map of completions. The target space $(\Sym_k W)^{\GD}$ is already complete (the dual space $W^{\GD}$ is complete, provided that the filtration of~$W$ is exhaustive), and thus we obtain the map $\hat{\Sym}_k(W^{\GD}) \rightarrow (\Sym_k W)^{\GD}$. According to Proposition~\ref{Prop:IsoCrit}, this map is an isomorphism if and only if the induced map $\Gr(\Sym_k(W^{\GD})) \rightarrow \Gr((\Sym_k W)^{\GD})$ is. This is shown by the following computation (the maps involved are natural in at least one direction):
\allowdisplaybreaks
\begin{align*} 
\frac{\Filtr^\lambda ({W^{\otimes k}}^{\GD})^d}{\Filtr^{\lambda+1} ({W^{\otimes k}}^{\GD})^d} &\simeq \frac{\Filtr^\lambda (W^{\otimes k})^{d*}}{\Filtr^{\lambda+1} (W^{\otimes k})^{d*}} \simeq  \Bigl(\frac{\Filtr^{\lambda+1}(W^{\otimes k})^d}{\Filtr^{\lambda}(W^{\otimes k})^d}\Bigr)^* \\ 
 &\simeq \Biggl( \frac{\bigoplus_{\Abs{\vec{d}}= d} \sum_{\Abs{\vec{\lambda}}= \lambda + 1} \Filtr^{\lambda_1} W^{d_1} \otimes \dotsb \otimes \Filtr^{\lambda_k} W^{d_k}}{\bigoplus_{\Abs{\vec{d}} = d}\sum_{\Abs{\vec{\lambda}} = \lambda} \Filtr^{\lambda_1} W^{d_1} \otimes \dotsb \otimes \Filtr^{\lambda_k} W^{d_k}}\Biggr)^* \\
&\simeq \Biggl( \bigoplus_{\Abs{\vec{d}}= d} \frac{\sum_{\Abs{\vec{\lambda}} = \lambda + 1} \Filtr^{\lambda_1} W^{d_1} \otimes \dotsb \otimes \Filtr^{\lambda_k} W^{d_k}}{\sum_{\Abs{\vec{\lambda}} = \lambda} \Filtr^{\lambda_1} W^{d_1} \otimes \dotsb \otimes \Filtr^{\lambda_k} W^{d_k}}\Biggr)^* \\
&\simeq\Bigl( \bigoplus_{\Abs{\vec{d}} = d} \bigoplus_{\Abs{\vec{\lambda}}=\lambda} \frac{\Filtr^{\lambda_1 + 1}W^{d_1}}{\Filtr^{\lambda_1} W^{d_1}} \otimes \dotsb \otimes \frac{\Filtr^{\lambda_k+1}W^{d_k}}{\Filtr^{\lambda_k} W^{d_k}}\Bigr)^*\\
&\simeq\Bigl( \bigoplus_{\Abs{\vec{\lambda}}=\lambda} \bigoplus_{\Abs{\vec{d}} = d} \frac{\Filtr^{\lambda_1 + 1} W^{d_1}}{\Filtr^{\lambda_1} W^{d_1}} \otimes \dotsb \otimes \frac{\Filtr^{\lambda_k + 1} W^{d_k}}{\Filtr^{\lambda_k} W^{d_k}}\Bigr)^* \\
&\mathclap{\substack{\Z-\text{gapped}  \\[1ex] \&\ \text{bounded below}  \\[1ex] \&\ (WG1)}\rightarrow\rule{7.7em}{0pt}}\simeq  \bigoplus_{\Abs{\vec{\lambda}}= \lambda} \bigoplus_{\Abs{\vec{d}} = d} \Bigl( \frac{\Filtr^{\lambda_1 + 1} W^{d_1}}{\Filtr^{\lambda_1} W^{d_1}} \otimes \dotsb \otimes \frac{\Filtr^{\lambda_k + 1} W^{d_k}}{\Filtr^{\lambda_k} W^{d_k}}\Bigr)^* \\
&\mathclap{{\scriptstyle (WG2)}\rightarrow\rule{4em}{0pt}}\simeq  \bigoplus_{\Abs{\vec{\lambda}}= \lambda} \bigoplus_{\Abs{\vec{d}} = d}  \Bigl( \frac{\Filtr^{\lambda_1 + 1} W^{d_1}}{\Filtr^{\lambda_1} W^{d_1}}\Bigr)^* \otimes \dotsb \otimes \Bigl( \frac{\Filtr^{\lambda_k + 1} W^{d_k}}{\Filtr^{\lambda_k} W^{d_k}}\Bigr)^* \\
&\simeq \bigoplus_{\Abs{\vec{d}} = d} \bigoplus_{\Abs{\vec{\lambda}}=\lambda} \frac{\Filtr^{\lambda_1}(W^{\GD})^{d_1}}{\Filtr^{\lambda_1 + 1}(W^{\GD})^{d_1}}\otimes \dotsb \otimes \frac{\Filtr^{\lambda_k}(W^{\GD})^{d_k}}{\Filtr^{\lambda_k + 1}(W^{\GD})^{d_k}} \\
&\simeq \frac{\Filtr^{\lambda} ({W^{\GD}}^{\otimes k})^d}{\Filtr^{\lambda + 1} ({W^{\GD}}^{\otimes k})^d}.
\end{align*}
In fact, this computation shows that $\hat{\Ten}_k(W^{\GD}) \simeq (\Ten_k W)^{\GD}$. The conclusion for $\Sym_k$ follows by checking that the maps above are $\Perm_k$-equivariant.
\end{proof}

Given a chain complex $(W,\Bdd)$, the boundary operator $\Bdd$ induces the boundary operator $\Bdd_k : W^{\otimes k} \rightarrow W^{\otimes k}$ for all $k\in \N$; for all $w_1$, $\dotsc$, $w_k\in W$, it is defined~by
\begin{equation}\label{Eq:BddExt}
\Bdd_k(w_1 \otimes \dotsb \otimes w_k)\coloneqq \sum_{i=1}^k (-1)^{\Abs{w_1} + \dotsb + \Abs{w_{i-1}}} w_1 \otimes \dotsb \otimes \Bdd w_i \otimes \dotsb \otimes w_k.
\end{equation}
The map $\Bdd_k$ is clearly $\Perm_k$-equivariant, and thus induces the boundary operator $\Bdd_k : \Sym_k W \rightarrow \Sym_k W$.

\begin{Proposition}[K\"unneth formula for completed symmetric cohomology]\label{Prop:Kuenneth}
Let $(W,\Bdd)$ be a $\Z$-graded chain complex over $\R$ filtered by an exhaustive $\Z$-gapped filtration~$\Filtr^\lambda W$ which is increasing and bounded from below. Consider the dual cochain complex $(W^{\GD},\Dd\coloneqq \Bdd^*)$. Suppose that $\Dd$ has finite filtration degree, so that $\Dd_k: \Sym_k(W^{\GD}) \rightarrow \Sym_k(W^{\GD})$ extends continuously to $\Dd_k : \hat{\Sym}_k(W^{\GD}) \rightarrow \hat{\Sym}_k(W^{\GD})$ for every $k\in \N$. If (WG1) \& (WG2) are satisfied, then the natural map $\Sym_k \H(W^{\GD},\Dd) \rightarrow \H(\hat{\Sym}_k(W^{\GD}),\Dd_k)$ induces the isomorphism
\begin{equation*}
\hat{\Sym}_k \H(W^{\GD},\Dd) \simeq \H(\hat{\Sym}_k(W^{\GD}), \Dd_k)\quad \text{for all }k\in \N.
\end{equation*}
\end{Proposition}
\begin{proof}
The natural map $\Sym_k \H(W^{\GD},\Dd) \rightarrow \H(\hat{\Sym}_k W^{\GD},\Dd_k)$ is clearly filtration preserving, and hence it extends continuously to a map of completions. The target space $\H(\hat{\Sym}_k W^{\GD},\Dd_k)$ is already complete (the homology of a complete space is complete), and hence we obtain the map $\hat{\Sym}_k \H(W^{\GD},\Dd) \rightarrow \H(\hat{\Sym}_k W^{\GD},\Dd_k)$. The following facts are easy to verify:
\begin{enumerate}[label=(\arabic*)]
 \item The isomorphism from Lemma~\ref{Lem:Terrible} is an isomorphism of cochain complexes 
 \[ (\hat{\Sym}_k W^{\GD}, \Dd_k) \simeq ((\Sym_k W)^{\GD}, \Bdd_k^*). \]
\item If the filtration on~$W$ satisfies (WG1) and (WG2), then the filtration on $\H(W)$ also satisfies (WG1) and (WG2), respectively. Consequently, Lemma~\ref{Lem:Terrible} holds for symmetric powers of $\H(W,\Bdd)^{\GD}$ as well.
\item The Künneth formula $\H(W^{\otimes k}) \simeq \H(W)^{\otimes k}$ implies $\H(\Sym_k W) \simeq \Sym_k \H(W)$ for any $\Z$-graded chain complex $W$ over $\R$.
\item We have $(\H(W))^{\GD} \simeq \H(W^{\GD})$ over $\R$ by the universal coefficient theorem.
\end{enumerate}
Now, we compute
\begin{align*}
\H(\hat{\Sym}_k W^{\GD},\Dd_k ) &
\underset{\substack{\uparrow\rule{0pt}{1.5ex} \\ (1)}}{\simeq}
\H((\Sym_k W)^{\GD}, \Bdd_k^*)
\underset{\substack{\uparrow\rule{0pt}{1.5ex} \\ (4)}}{\simeq}
\H(\Sym_k W, \Bdd_k)^{\GD} 
\underset{\substack{\uparrow\rule{0pt}{1.5ex} \\ (3)}}{\simeq}
(\Sym_k \H(W,\Bdd))^{\GD} \\ 
&\underset{\substack{\uparrow\rule{0pt}{1.5ex} \\ (2)}}{\simeq}
\hat{\Sym}_k (\H(W,\Bdd)^{\GD}) 
\underset{\substack{\uparrow\rule{0pt}{1.5ex} \\ (4)}}{\simeq}
\hat{\Sym}_k \H(W^{\GD},\Dd).
\end{align*}
This proves the proposition.
\end{proof}

\section{Basics of IBL-infinity-algebras
}
\allowdisplaybreaks
\label{Sec:Alg1}

\Correct[caption={DONE Bad picture},noline]{There migh be wrong dimensions on one side of the maurer cartan element. Compare to the section about filtrations. REMARK THAT THE DEFINITION NEEDS THE WEIGHT STRUCTURE OF THE BIALGEBRA. No Problem.}
\Add[caption={DONE New algebraic setting in Appendix},noline]{There is a new algebraic description of the iterated compatibility condition in the appendix.}
\begin{Definition}[Exterior algebra]\label{Def:ExtAlg}
Given a graded vector space $C$ over $\R$, we define the \emph{exterior algebra} over $C$ by
\[ \Ext C \coloneqq \Sym(C[1]). \]
The weight-$k$ component is denoted by $\Ext_k C$ and the weight-reduced part by~$\RExt C$. If $C$ is in addition filtered, then $\Ext_k C$ is filtered by the induced filtration and its completion is denoted by $\hat{\Ext}_k C$.
\end{Definition}

We have the concatenation product $\Prod : \Ext C \otimes \Ext C \rightarrow \Ext C$ and the shuffle coproduct $\CoProd: \Ext C \rightarrow \Ext C\otimes \Ext C$ defined~by
\begin{align*}
&\Prod(c_{11}\dots c_{1k} \otimes c_{21} \dots c_{2k'}) \coloneqq c_{11} \dots c_{1k} c_{21} \dots c_{2k'}\quad\text{and}\\[\jot]
&\CoProd(c_1 \dots c_k) \coloneqq \sum_{\substack{k_1,\,k_2 \ge 0\\ k_1 + k_2 = k}} \sum_{\sigma\in \Perm_{k_1, k_2}} \varepsilon(\sigma,c) c_{\sigma^{-1}_1} \dots c_{\sigma_{k_1}^{-1}}\otimes c_{\sigma_{k_1+1}^{-1}}\dots c_{\sigma_{k_1 + k_2}^{-1}}
\end{align*}
for all homogenous $c_{ij}$, $c_i \in C[1]$ and $k$, $k'\ge 0$, respectively, where $\Perm_{k_1, k_2}\subset \Perm_{k_1+k_2}$ denotes the set of shuffle permutations with blocks of lengths $k_1$ and $k_2$. These operations satisfy the relations of an \emph{associative bialgebra} (see \cite{Loday2012}):
\begin{equation}\label{Eq:Bialgebra}
\text{Assoc.~bialg.}\quad \left\{
\begin{aligned}
\Prod\circ (\Id\otimes \Prod) &= \Prod\circ (\Prod \otimes \Id), \\ 
(\Id\otimes \CoProd)\circ\CoProd &= (\CoProd\otimes \Id)\circ \CoProd, \\
\CoProd \circ \Prod &= (\Prod\otimes \Prod) \circ (\Id \otimes \tau\otimes \Id) \circ (\CoProd\otimes \CoProd).
\end{aligned} \right.
\end{equation}
Here $\tau: C_1 \otimes C_2 \rightarrow C_2 \otimes C_1$, $c_1\otimes c_2 \mapsto (-1)^{\Abs{c_1}\Abs{c_2}}c_2\otimes c_1$ denotes the \emph{twist map}.

We will use the bialgebra calculus ($\coloneqq$\,relations \eqref{Eq:Bialgebra}) to write down explicit formulas for the operations $\circ_{h_1, \dotsc, h_r}$ which were briefly introduced in \cite{Cieliebak2015}; these operations take symmetric maps $f_1$, $\dotsc$, $f_r$ and connect $h_1$, $\dotsc$, $h_r$ of their outputs to the inputs of a symmetric map $f$ in all possible ways, so that the result, which we denote by $f\circ_{h_1, \dotsc, h_r}(f_1, \dotsc,f_r)$, becomes a symmetric map again.

\begin{Definition}[Partial compositions] \label{Def:CircS}
Let $C$ be a graded vector space. For $i$, $j\ge 0$, we denote by 
\begin{align*} \pi_i : \Ext C \longrightarrow \Ext_i C,& \quad \iota_i : \Ext_i C \longrightarrow \Ext C, \\
 \Id_i : \Ext_i C \longrightarrow \Ext_i C,& \quad \begin{aligned}[t]\tau_{i,j}: \Ext_i C\otimes \Ext_j C &\longrightarrow \Ext_j C \otimes \Ext_i C \end{aligned}
 \end{align*}
the components of the canonical projection $\pi$, the canonical inclusion $\iota$, the identity $\Id$ and the twist map $\tau$, respectively. We also set 
\[ \CoProd_{i,j} \coloneqq (\pi_i \otimes \pi_j)\circ \CoProd\circ \iota_{i+j} \quad\text{and}\quad \Prod_{i,j}\coloneqq \pi_{i+j}\circ \Prod\circ (\iota_i \otimes \iota_j). \]
For $k'$, $k_1$, $l'$, $l_1\ge 0$, let $f: \Ext_{k'}C \rightarrow \Ext_{l'} C$ and $f_1: \Ext_{k_1} C \rightarrow \Ext_{l_1} C$ be linear maps, and let $0 \le h \le \min(k', l_1)$. We set  
\[k \coloneqq k' + k_1 - h \quad\text{and}\quad l \coloneqq l' + l_1 - h \]
and define the \emph{composition of $f$ and $f_1$ at $h$ common outputs} to be the linear map $f \circ_h f_1: \Ext_k C \rightarrow \Ext_l C$ given by 
\begin{equation}\label{Eq:CompositionSimple}
f \circ_h f_1 \coloneqq \begin{multlined}[t]\Prod_{l', l_1 - h} \circ (f\otimes \Id_{l_1-h})\circ (\Prod_{h, k'-h}\otimes \Id_{l_1-h})\circ (\Id_{h} \otimes \tau_{\rule{0pt}{7pt}l_1-h,k'-h}) \\[\jot] \circ (\CoProd_{h,l_1-h}\otimes \Id_{k'-h}) \circ (f_1 \otimes \Id_{k'-h} ) \circ \CoProd_{k_1, k'-h}. \end{multlined}
\end{equation}
More generally, we define the composition of $f: E_{k'}\rightarrow E_{l'}$ with $r\ge 1$ linear maps $f_{i}: E_{k_i} \rightarrow E_{l_i}$ with $k_i$, $l_i \ge 0$ for $i=1$,~$\dotsc$, $r$ at $0 \le h_i \le l_i$ common outputs such that $h\coloneqq h_1 + \dotsb + h_r \le k'$ as follows. We set 
\[ k\coloneqq k' + k_1 + \dotsb + k_r-h\quad\text{and}\quad l\coloneqq l' +  l_1 + \dotsb + l_r  - h \]
and define $f\circ_{h_1, \dotsc, h_r}(f_1, \dotsc, f_r): \Ext_k C \rightarrow \Ext_l C$ by\ToDo[caption={DONE Add the other composition},noline]{Add the definition of $(f_1,\dotsc,f_r)\circ_{h_1,\dotsc,h_r} f$. IT IS DONE IN THE APPENDIX.}
\begin{equation} \label{Eq:CompositionFormula}
\begin{aligned}
&f\circ_{h_1, \dotsc, h_r}(f_1, \dotsc, f_r)\\
&\qquad \coloneqq \begin{multlined}[t]
\Prod\circ (f\otimes \Id)\circ(\Prod\otimes\Id)\circ(\Id\otimes \tau) \\[\jot] \circ \big(\big[(\Prod^{(r)}\otimes \Prod^{(r)})\circ (F_{h_1,\dotsc,h_r} \otimes \Id^{\otimes r})\circ \sigma_r\circ \CoProd^{\otimes r}\big] \otimes \Id \big) \\[\jot] \circ (f_1\otimes \dotsb \otimes f_r\otimes \Id)\circ\CoProd^{(r+1)},
\end{multlined}\end{aligned}
\end{equation}
where we have:
\begin{itemize}
\item The operation $\Prod^{(r)}$ is the ``product with $r$ inputs'' and the operation $\CoProd^{(r)}$ the ``coproduct with $r$ outputs''; they are defined~by
\begin{align*}
\Prod^{(r)} &\coloneqq \Prod(\Id \otimes \Prod)\dotsb(\Id^{\otimes r-2} \otimes \Prod), & \Prod^{(1)}&\coloneqq \Id,\\
\CoProd^{(r)} &\coloneqq (\Id^{\otimes r-2} \otimes \CoProd)\dotsb(\Id \otimes \CoProd)\CoProd, &\CoProd^{(1)}&\coloneqq \Id.
\end{align*}

\item $F_{h_1,\dotsc, h_r} \coloneqq (\iota_{h_1}\pi_{h_1}) \otimes \dotsb \otimes (\iota_{h_r}\pi_{h_r})$.%
\item The permutation $\sigma_r \in \Perm_{2r}$ is given by  \Add[caption={DONE Better definition!!!}]{This needs a better definition of $\sigma_r$!!!}
\[\sigma_r: (1,2,3,4\dotsc, 2r-1, 2r) \longmapsto (1,r+1,2,r+2, \dotsc, r, 2r).\]
\item The symbols $f$ and $f_i$ inside the formula denote the \emph{trivial extensions} of $f$ and $f_i$, respectively; we extend a linear map $f: E_{k'} C \rightarrow E_{l'} C$ trivially to $f: \Ext C \rightarrow \Ext C$ by defining $f(\Ext_i C)=0$ for $i\neq k'$.
\end{itemize}
\end{Definition}

\begin{Remark}[On partial compositions]\phantomsection\label{Rem:Compositions}
\begin{RemarkList}
\item Defining $f\circ_{h_1, \dotsc, h_r}(f_1, \dotsc, f_r): \Ext_k C \rightarrow \Ext_l C$ using~\eqref{Eq:CompositionFormula} makes sense because the right hand side is a trivial extension of its component $\Ext_k C \rightarrow \Ext_l C$. In fact, all $\Prod$, $\CoProd$, $\pi$, $\iota$ in \eqref{Eq:CompositionFormula} can be replaced with $\Prod_{i,j}$, $\CoProd_{i,j}$, $\pi_i$, $\iota_i$ for unique $i$, $j$, so that trivial extensions do not have to be used at all. In this way, it can be seen that \eqref{Eq:CompositionSimple} is indeed a special case of~\eqref{Eq:CompositionFormula}. 
\item If $h = k' = l_1$, then $f \circ_{h} f_1 = f\circ f_1$.
\item It holds $f \circ_0 f_1 = (-1)^{\Abs{f}\Abs{f_1}} f_1 \circ_0 f$ and 
\[ f\circ_{h_1,\dotsc,h_r}(f_1,\dotsc,f_r) = \varepsilon(\sigma,f) f\circ_{h_{\sigma_1^{-1}},\dotsc,h_{\sigma_r^{-1}}}(f_{\sigma_1^{-1}},\dotsc,f_{\sigma_r^{-1}}). \]
\item Consider the (``non-trivial'') extension $\hat{f}\coloneqq \Prod(f\otimes \Id)\CoProd: \Ext C \rightarrow \Ext C$ and the symmetric product $f_1 \odot \dotsb \odot f_r \coloneqq \Prod^{(r)}(f_1 \otimes \dotsb \otimes f_r)\CoProd^{(r)}: \Ext C \rightarrow \Ext C$. The following formulas from~\cite{Cieliebak2015} hold:
\begin{equation} \label{Eq:Mix}
\begin{aligned}
f\circ_{h_1,\dotsc,h_{r-1},0}(f_1,\dotsc, f_r) &= f\circ_{h_1,\dotsc, h_{r-1}}(f_1,\dotsc, f_{r-1}) \odot f_r, \\
\hat{f} \circ \hat{f}_1 &= \sum_{h = 0}^{\min(k',l_1)} \widehat{f\circ_h f_1}, \\ 
   \hat{f} \circ (f_1 \odot \dotsb \odot f_r) &= \sum_{\substack{h_1, \dotsc, h_r \ge 0 \\ h_1 + \dotsb + h_r = k'}} f\circ_{h_1,\dotsc, h_r}(f_1,\dotsc, f_r).
\end{aligned}
\end{equation}
We also have the ``weak associativity''
\begin{equation}\label{Eq:WeakAssoc}
\qquad\qquad \mathclap{\sum_{\substack{0 \le h_2 \le \min(f_3^-, f_2^+) \\
0 \le h_1 \le \min(f_1^+,f_2^- + f_3^- - h_2) \\
h_1 + h_2 = h}}}\quad\qquad f_1 \circ_{h_1} (f_2 \circ_{h_2} f_3) = \qquad\quad \mathclap{\sum_{\substack{0 \le h_1 \le \min(f_1^+,f_2^-) \\ 0 \le h_2 \le \min(f_1^+ + f_2^+ - h_1, f_3^-) \\ h_1 + h_2 = h}}}\qquad\quad (f_1 \circ_{h_1} f_2) \circ_{h_2} f_3
\end{equation}
for every $0\le h \le \min(k_1 + k_2 + k_3, l_1 + l_2 + l_3)$, where $f^+$ denotes the number of inputs and $f^-$ the number of outputs of $f$. The weak associativity of $\circ_h$ can be proven using the associativity of $\,\hat{\cdot}$ and the second relation of \eqref{Eq:Mix}.
\item We refer to Section~\ref{Sec:CompConvA} of Appendix~\ref{App:IBLMV} for a thorough treatment of partial compositions. We show there that $\circ_{h_1,\dotsc,h_r}$ can be defined on maps on any connected weight-graded bialgebra using natural bilinear operations $\SquareComp_A$ on polynomials in the convolution product. In Proposition~\ref{Prop:PartCompositions} in the appendix, we prove the relations above.\qedhere
\end{RemarkList}
\end{Remark}

If $C$ is filtered by a decreasing filtration, then the bialgebra operations extend continuously to 
\begin{align*}
\Prod: \hat{\Ext}_{k_1} C \COtimes \hat{\Ext}_{k_2} C &\longrightarrow \hat{\Ext}_{k_1+k_2} C \quad \text{and}\\ 
\CoProd: \hat{\Ext}_k C &\longrightarrow \bigoplus_{\substack{l_1, l_2 \ge 0 \\ l_1 + l_2 = k}} \hat{\Ext}_{l_1}C\COtimes\hat{\Ext}_{l_2}C
\end{align*}
for all $k_1$, $k_2$, $k\in \N_0$ because they preserve the filtration degree (see~\cite{Fresse} for a similar construction). Next, if $f_1: \hat{\Ext}_{k_1}C \rightarrow \hat{\Ext}_{l_1}C$ and $f_2: \hat{\Ext}_{k_2}C\rightarrow \hat{\Ext}_{l_2}C$ have finite filtration degrees, then $f_1 \otimes f_2: \hat{\Ext}_{k_1} C \otimes \hat{\Ext}_{k_2} C \rightarrow \hat{\Ext}_{l_1}C\otimes\hat{\Ext}_{l_2}C$ has finite filtration degree too, and hence it extends continuously to $f_1 \otimes f_2: \hat{\Ext}_{k_1} C\COtimes \hat{\Ext}_{k_2} C \rightarrow \hat{\Ext}_{l_1}C\COtimes\hat{\Ext}_{l_2}C$.
Using these facts, we can canonically extend Definition~\ref{Def:CircS} to maps $f: \hat{\Ext}_{k'} C \rightarrow \hat{\Ext}_{l'} C$ and $f_i: \hat{\Ext}_{k_i}C \rightarrow \hat{\Ext}_{l_i}C$ of finite filtration degrees. The resulting map $f\circ_{h_1,\dotsc,h_r}(f_1,\dotsc,f_r): \hat{\Ext}_k C \rightarrow \hat{\Ext}_l C$ will have finite filtration degree too. Moreover, the formulas in Remark~\ref{Rem:Compositions} will still hold.%

We will now rephrase the definitions of an $\IBLInfty$-algebra, a Maurer-Cartan element and twisted operations from~\cite{Cieliebak2015} in terms of~$\circ_{h_1, \dotsc, h_r}$.
\begin{Def}[$\IBLInfty$-algebra] \label{Def:IBLInfty} Let $C$ be a graded vector space equipped with a decreasing filtration, and let $d\in \Z$ and $\gamma\ge 0$ be fixed constants. An \emph{$\IBLInfty$-algebra of bidegree $(d,\gamma)$} on~$C$ is a collection of linear maps $\OPQ_{klg}: \hat{\Ext}_k C \rightarrow \hat{\Ext}_l C$ for all $k,l\ge 1$, $g\ge 0$ which are homogenous, of finite filtration degree and satisfy the following conditions: 
\begin{enumerate}[label=\arabic*)]
\item $\Abs{\OPQ_{klg}} = - 2d(k+g-1) - 1$.
\item $\Norm{\OPQ_{klg}} \ge \gamma \chi_{klg}$,
where $\chi_{klg}\coloneqq2-2g-k-l$. %
\item The \emph{$\IBLInfty$-relations} hold: for all $k,l\ge 1$, $g\ge 0$, we have
\begin{equation} \label{Eq:IBLInfRel}
\sum_{h=1}^{g+1} \sum_{\substack{k_1, k_2, l_1, l_2 \ge 1 \\ g_1, g_2 \ge 0 \\k_1 + k_2 = k + h \\ l_1 + l_2 = l+ h\\ g_1 + g_2 = g+ 1 -h }} \OPQ_{k_2 l_2 g_2} \circ_h \OPQ_{k_1 l_1 g_1} = 0.
\end{equation}
\end{enumerate}
We denote a given $\IBLInfty$-algebra structure on $C$ by $\IBLInfty(C)$; i.e., we write $\IBLInfty(C)=(C,(\OPQ_{klg}))$.

If $\OPQ_{klg} \equiv 0$ for all $(k,l,g)\neq (1,1,0)$, $(2,1,0)$, $(1,2,0)$, then we call $\IBLInfty(C)$ a \emph{$\dIBL$-algebra} and denote it by $\dIBL(C)$. If in addition $\OPQ_{110} \equiv 0$, then we have an \emph{$\IBL$-algebra} $\IBL(C)$.

If the operations on the completed exterior powers~$\hat{\Ext}_k C$ arise as continuous extensions of operations $\OPQ_{klg}: \Ext_k C \rightarrow \Ext_l C$, then we call the $\IBLInfty$-algebra \emph{completion-free} and denote $C$ together with the operations $\OPQ_{klg}: \Ext_k C \rightarrow \Ext_l C$ by $\ShortIBLInfty(C)$.
\end{Def}

The acronym $\IBL$ stands for an \emph{involutive Lie bialgebra.} It follows namely from the $\IBLInfty$-relations \eqref{Eq:IBLInfRel} that for $\IBL(C) = (C,\OPQ_{210},\OPQ_{120})$ the following holds: 
\begin{equation*}
\raisebox{2ex}{$\text{Lie bialg.}\;\left\{\rule{0pt}{5ex}\right.$}\;
\begin{aligned}   
   0&= \OPQ_{210}\circ_1 \OPQ_{210} &&\leftarrow\text{Jacobi id.}\\
   0 &= \OPQ_{120} \circ_1 \OPQ_{120} &&\leftarrow\text{co-Jacobi id.}\\
   0 &= \OPQ_{120}\circ_1 \OPQ_{210} + \OPQ_{210}\circ_1 \OPQ_{120}&&\leftarrow\text{Drinfeld id.} \\
   0 &= \OPQ_{210} \circ_2 \OPQ_{120} &&\leftarrow\text{Involutivity}
\end{aligned}
\end{equation*}
The acronym $\dIBL$ stands for a \emph{differential involutive Lie bialgebra} --- an involutive Lie bialgebra together with a differential (a boundary operator in our case) such that the bracket and cobracket are chain maps.
 
\begin{Proposition}[Odd degree shift of an $\IBL$-algebra]\label{Prop:ClasModIBL}
Let $(C,\OPQ_{210}, \OPQ_{120})$ be an $\IBL$-algebra of degree $d$ from Definition~\ref{Def:IBLInfty}, and let $\tilde{\OPQ}_{210} : C^{\otimes 2} \rightarrow C$ and $\tilde{\OPQ}_{120}: C \rightarrow C^{\otimes 2}$ be the linear maps defined by
\begin{equation}\label{Eq:ClasModIBL}
\begin{aligned}
\SuspU \tilde{\OPQ}_{210}(x_1 \otimes x_2) &\coloneqq \OPQ_{210}(\pi(\SuspU^2 x_1 \otimes x_2)) \quad\text{and} \\
\SuspU^2 \tilde{\OPQ}_{120}(x) &\coloneqq \iota(\OPQ_{120}(\SuspU x))
\end{aligned}
\end{equation}
for all $x_1$, $x_2$, $x\in C$, where $\iota: \Sym_2(C[1]) \rightarrow C[1]^{\otimes 2}$ is the section of $\pi: C[1]^{\otimes 2} \rightarrow \Sym_2(C[1])$ from Definition~\ref{Def:SymAlgebra} and~$\SuspU$ is a formal symbol of degree $\Abs{\SuspU} = -1$. Then the degrees satisfy
\[ \Abs{\tilde{\OPQ}_{210}} = \Abs{\OPQ_{210}} - 1 = -2d - 2\quad\text{and}\quad\Abs{\tilde{\OPQ}_{120}} = \Abs{\OPQ_{120}} + 1 = 0, \]
the operations $\tilde{\OPQ}_{210}$ and $\tilde{\OPQ}_{120}$ are graded antisymmetric, i.e., we have
\[ \tilde{\OPQ}_{210} \circ\tau = - \tilde{\OPQ}_{210}\quad\text{and}\quad\tau \circ \tilde{\OPQ}_{120} = - \tilde{\OPQ}_{120} \]
for the twist map $\tau$, and the relations
\begin{align*}
0&=\tilde{\OPQ}_{210}\circ (\tilde{\OPQ}_{210}\otimes \Id)\circ (\Id^{\otimes 3}+ t_3 + t_3^2), \\
0&=(\Id^{\otimes 3}+t_3 + t_3^2)\circ (\tilde{\OPQ}_{120}\otimes\Id)\circ \tilde{\OPQ}_{120}, \\
0&= x_1 \cdot \tilde{\OPQ}_{120}(x_2) - (-1)^{ x_1 x_2} x_2 \cdot \tilde{\OPQ}_{120}(x_1) - \tilde{\OPQ}_{120}(\tilde{\OPQ}_{210}(x_1,x_2)), \\
0& = \tilde{\OPQ}_{210} \circ \tilde{\OPQ}_{120},
\end{align*}
hold for all $x_1$, $x_2\in C$. Here, $t_3 \in \Perm_3$ denotes the cyclic permutation with $t_3(1) = 2$ acting on $C^{\otimes 3}$, and we define
\[ x\cdot (y_1 \otimes y_2) \coloneqq \tilde{\OPQ}_{210}(x,y_1)\otimes y_2 + (-1)^{ x y_1} y_1 \otimes \tilde{\OPQ}_{210}(x,y_2) \]
for all $x$, $y_1$, $y_2 \in C$.
\end{Proposition}

\begin{proof}
The proof is a lengthy but straightforward computation.
\end{proof}

Consider the \emph{sign-action of $\Perm_k$} on $C^{\otimes k}$ given by $\sigma \mapsto \bar{\sigma}$, where 
\[ \bar{\sigma}(c_1\otimes\dotsb\otimes c_k)\coloneqq(-1)^\sigma\varepsilon(\sigma,c)c_{\sigma_1^{-1}}\otimes\dotsb\otimes c_{\sigma_k^{-1}} \]
for all $c_1$, $\dotsc$, $c_k\in C$ and $\sigma\in \Perm_k$. We define 
\[ \Lambda C \coloneqq \bigoplus_{k=0}^\infty \Lambda_k C\quad\text{with}\quad \Lambda_k C \coloneqq C^{\otimes k}/\sim, \]
where $c\sim \bar{\sigma}(c)$ for all $c\in C^{\otimes k}$ and $\sigma\in \Perm_k$. It is easy to see that the degree-shift map $\theta^{\otimes k}: c_1\otimes \dotsb \otimes c_k \in C^{\otimes k} \mapsto \varepsilon(c,\theta)(\theta c_1)\otimes\dotsb\otimes(\theta c_k)\in C[1]^{\otimes k}$ is equivariant with respect to the sign-action of $\Perm_k$ on $C^{\otimes k}$ and the standard action of $\Perm_k$ on $C[1]^{\otimes k}$ for all $k$, and thus it induces an isomorphism of vector spaces $\Lambda C$ and $\Ext C$. We use the following notation:
\[\begin{tikzcd}
 \OPQ_{klg}: \hat{\Ext}_k C \arrow{r} \arrow{d}{\theta^{\otimes k}} & \hat{\Ext}_l C \arrow{d}{\theta^{\otimes l}} \\
 \tilde{\OPQ}_{klg}: \hat{\Lambda}_k C \arrow{r} & \hat{\Lambda}_l C.
\end{tikzcd}\]
In fact, $\OPQ_{klg}$ and $\tilde{\OPQ}_{klg}$ are related precisely by the degree shift \eqref{Eq:DegreeShiftConvII}. 

\begin{figure}[t]
\centering
\begin{subfigure}{\textwidth}
\centering
\input{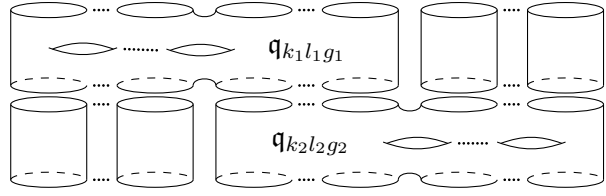}
\caption{The term $\OPQ_{k_2 l_2 g_2} \circ_h \OPQ_{k_1 l_1 g_1}$ in the $\IBLInfty$-equation \eqref{Eq:IBLInfRel}.}
\end{subfigure}
\par\bigskip
\begin{subfigure}{\textwidth}
\centering
\input{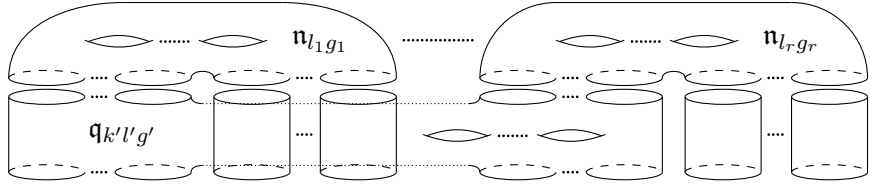}
\caption{The term $\OPQ_{k' l' g'}\circ_{h_1,\dotsc,h_r} (\PMC_{l_1 g_1},\dotsc,\PMC_{l_r g_r})$ in the Maurer-Cartan equation \eqref{Eq:MaurerCartanEquation}. We remark that the contour of the surface corresponding to $\OPQ_{k'l'g'}$ starts on the left and continues to the right along the dotted line behind the two trivial cylinders.}
\end{subfigure}
\par\bigskip
\begin{subfigure}{\textwidth}
\centering
\input{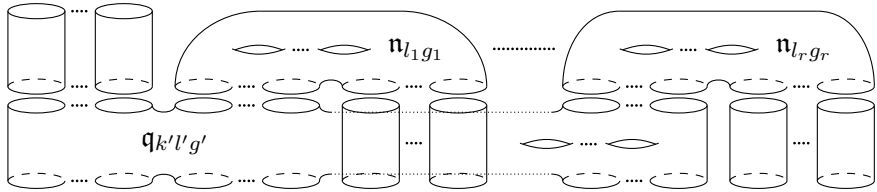}
\caption{The term $\OPQ_{k' l' g'}\circ_{l_1,\dotsc, l_r} (\PMC_{l_1 g_1},\dotsc,\PMC_{l_r g_r})$ in the twisted operation \eqref{Eq:TwistedOperations}. The remark to Figure (b) applies too.}
\end{subfigure}
\caption[$\IBLInfty$-relations, Maurer-Cartan equation and twisting graphically.]{Graphical representation of compositions appearing in Definitions \ref{Def:IBLInfty}, \ref{Def:MaurerCartan} and \ref{Def:TwistedOperations} as gluing of connected Riemannian surfaces. The figure is to be read from the top  to the bottom, the empty cylinder represents the identity, and the resulting surface must be connected. We emphasize that the gluing is not associative (c.f., weak associativity \eqref{Eq:WeakAssoc}).}
\label{Fig:Surfaces}
\end{figure}
\begin{Def}[Maurer-Cartan element] \label{Def:MaurerCartan}
A \emph{Maurer-Cartan element} for an $\IBLInfty$-algebra $\IBLInfty(C)$ from Definition~\ref{Def:IBLInfty} is a collection $\PMC \coloneqq (\PMC_{lg})_{l\ge 1, g\ge 0}$ of elements $\PMC_{lg}\in \hat{\Ext}_l C$ which are homogenous, of finite filtration degree and satisfy the following conditions:
\begin{enumerate}[label=\arabic*)]
\item $\Abs{\PMC_{lg}} = - 2d(g-1)$.
\item $\Norm{\PMC_{lg}}\ge\gamma \chi_{0lg}$ with $>$ for $(l,g)=(1,0)$, $(2,0)$ (see Definition~\ref{Def:IBLInfty} for~$\chi_{klg}$).
\item The \emph{Maurer-Cartan equation} holds: for all $l\ge 1$, $g\ge 0$, we have
\begin{equation} \label{Eq:MaurerCartanEquation}
\begin{aligned}
\sum_{r\ge 1}\frac{1}{r!}  \sum_{\substack{l', k', l_1, \dotsc, l_r\ge 1 \\ g', g_1, \dotsc, g_r \ge 0 \\ h_1, \dotsc, h_r \ge 1 \\ 
l_1 + \dotsb + l_r + l' - k'= l \\ g_1 + \dotsb + g_r + g' +  k' = g + r \\ h_1 + \dotsb + h_r - k' =0 } } \OPQ_{k' l' g'}\circ_{h_1,\dotsc,h_r} (\PMC_{l_1 g_1},\dotsc,\PMC_{l_r g_r}) = 0,
\end{aligned} 
\end{equation}
where we view $\PMC_{lg}$ as a linear map $\PMC_{lg}: \hat{\Ext}_0 C = \R \rightarrow \hat{\Ext}_l C$ with $\PMC_{lg}(1) = \PMC_{lg}$.

\end{enumerate}
\end{Def}

\begin{Def}[Twisted operations] \label{Def:TwistedOperations}
In the setting of Definition~\ref{Def:MaurerCartan}, the \emph{twisted operations} $\OPQ_{klg}^\PMC: \hat{\Ext}_k C\rightarrow \hat{\Ext}_l C$ for $k,l\ge 1$, $g\ge 0$ are defined by
\begin{equation}\label{Eq:TwistedOperations}
 \OPQ_{klg}^\PMC =\sum_{r\ge 0} \frac{1}{r!} \sum_{\substack{k', l', l_1, \dotsc, l_r \ge 1 \\ g', g_1, \dotsc, g_r \ge 0\\ h_1, \dotsc, h_r \ge 1 \\
l_1 + \dotsb + l_r + l' - k' = l-k \\ g_1 + \dotsb +g_r + g' + k' = g + r + k \\ h_1 + \dotsb + h_r - k' = -k}} \OPQ_{k' l' g'}\circ_{h_1,\dotsc, h_r} (\PMC_{l_1 g_1},\dotsc,\PMC_{l_r g_r}).
\end{equation}
In \cite[Proposition~9.3]{Cieliebak2015}, they prove that $(\OPQ_{klg}^\PMC)_{k,l\ge 1, g\ge 0}$ is again an $\IBLInfty$-algebra of bidegree $(d,\gamma)$ on $C$ --- \emph{the twisted $\IBLInfty$-algebra}.  We denote it by $\IBLInfty^\PMC(C)$.
\end{Def}

Let $(\OPQ_{klg})$ be an $\IBLInfty$-algebra on $C$. The boundary operator $\OPQ_{110}: C[1] \rightarrow C[1]$ induces the boundary operator $\Bdd_k : \Ext_k C \rightarrow \Ext_k C$ for every $k\in \N$ (see \eqref{Eq:BddExt}). Because of the finite filtration degree, $\Bdd_k$ continuously extends to $\Bdd_k: \hat{\Ext}_k C \rightarrow \hat{\Ext}_k C$.
The following is easy to see using \eqref{Eq:CompositionSimple}:
\[\begin{aligned}
 \OPQ_{klg} \circ_1 \OPQ_{110} &= \OPQ_{klg} \circ \Bdd_k, \\
 \OPQ_{110} \circ_1 \OPQ_{klg} &= \Bdd_l \circ \OPQ_{klg}.
\end{aligned}\]
Because $\OPQ_{klg}$ are odd ($\coloneqq$\,have odd degree), we have
\[ \begin{aligned}
  [\Bdd,\OPQ_{klg}] &\coloneqq \Bdd_l \circ \OPQ_{klg} - (-1)^{\Abs{\Bdd}\Abs{\OPQ_{klg}}} \OPQ_{klg}\circ \Bdd_k \\
   &= \Bdd_l \circ \OPQ_{klg} + \OPQ_{klg}\circ \Bdd_k \\
   &= \OPQ_{110}\circ_1 \OPQ_{klg} + \OPQ_{klg}\circ_1 \OPQ_{110}.
  \end{aligned}\]
With this notation, the $\IBLInfty$-relations \eqref{Eq:IBLInfRel} for 
$\OPQ_{210}: \hat{\Ext}_2 C \rightarrow \hat{\Ext}_1 C$ and $\OPQ_{120}: \hat{\Ext}_1 C \rightarrow \hat{\Ext}_2 C$ become $[\Bdd,\OPQ_{210}] = 0$ and $[\Bdd,\OPQ_{120}] = 0$, respectively. If moreover the canonical maps $\Ext_k \H(\hat{C},\tilde{\OPQ}_{110}) \rightarrow \H(\hat{\Ext}_k C, \Bdd_k)$ induce the isomorphisms $\hat{\Ext}_k \H(\hat{C},\tilde{\OPQ}_{110})  \simeq \H(\hat{\Ext}_k C, \Bdd_k)$ for $k=1$, $2$, e.g., when Proposition~\ref{Prop:Kuenneth} holds, then we obtain the maps
\[ \OPQ_{210}: \hat{\Ext}_2\H(\hat{C},\tilde{\OPQ}_{110})  \rightarrow \hat{\Ext}_1\H(\hat{C},\tilde{\OPQ}_{110}) \quad \text{and}\quad \OPQ_{120}: \hat{\Ext}_1\H(\hat{C},\tilde{\OPQ}_{110})  \rightarrow \hat{\Ext}_2\H(\hat{C},\tilde{\OPQ}_{110}), \]
and $(\H(\hat{C},\tilde{\OPQ}_{110}),\OPQ_{210},\OPQ_{120})$ becomes an $\IBL$-algebra according to Definition~\ref{Def:IBLInfty} --- the \emph{induced $\IBL$-algebra on homology}.

\begin{Definition}[Homology]\label{Def:HomIBL}
We define the homology of an $\IBLInfty$-algebra $\IBLInfty(C)$ by
\[ \HIBL(C)[1] \coloneqq \H(\hat{C}[1], \OPQ_{110}). \]
It is a graded vector space with the induced filtration. If $\PMC$ is a Maurer-Cartan element for $\IBLInfty(C)$, we denote by $\HIBL^\PMC(C)$ the homology of $\IBLInfty^\PMC(C)$.
\end{Definition}

\begin{Remark}[Weak $\IBLInfty$-algebras and $\mathrm{BV}$-formalism]\phantomsection\label{Rem:BVForm}
\begin{RemarkList}
\item A possible generalization of the $\IBLInfty$-theory is to allow $k=0$ and $l=0$.
Such structures would be called \emph{weak $\IBLInfty$-algebras} while the structures from this section \emph{strict $\IBLInfty$-algebras.}
In fact, one does not need filtrations and completions to deal with the category of strict $\IBLInfty$-algebras unless deformations (twisting) are considered.
On the other hand, one needs filtrations and completions for the definition of a morphism of weak $\IBLInfty$-algebras already.
We refer to Appendix~\ref{App:IBLMV} for more details. 
\item Let $\CExt C[[\hbar]]$, resp.~$\CExt C((\hbar))$ be the spaces of formal power, resp.~Laurent series in the variable~$\hbar$ of degree $\Abs{\hbar} = 2d$ with coefficients in $\Ext C$ completed with respect to a suitable completion.
Operations of an $\IBLInfty$-algebra on $C$ can be encoded in a degree~$-1$ operator $\BVOp: \CExt C[[\hbar]] \rightarrow \CExt C[[\hbar]]$ called the \emph{$\mathrm{BV}$-operator,} while the data of a Maurer-Cartan element~$(\PMC_{lg})$ give rise to an element $e^{\PMC}\in \CExt C((\hbar))$. The prescriptions are
\[ \BVOp \coloneqq \sum_{i\ge 0}\BVOp_{i+1} \hbar^{i}\quad\text{and}\quad e^{\PMC} \coloneqq \sum_{j\in \Z} (e^{\PMC})_j \hbar^{j}, \]
where $\BVOp_i$ and $(e^\PMC)_j$ for $i\ge 1$, $j\in \Z$ are defined by 
\begin{align*}
\BVOp_i & \coloneqq \sum_{\substack{k\ge 1, g\ge 0 \\k+g=i}} \sum_{l\ge 1} \hat{\OPQ}_{klg}\quad\text{and} \\
(e^{\PMC})_j &\coloneqq \sum_{r=0}^\infty \frac{1}{r!} \sum_{\substack{g_1, \dotsc, g_r \ge 0 \\ g_1 + \dotsb +g_r - r= j }} \sum_{l_1, \dotsc, l_r\ge 1} \PMC_{l_1 g_1} \odot \dotsb \odot \PMC_{l_r g_r}.
\end{align*}
It can be shown that the $\IBLInfty$-relations~\eqref{Eq:IBLInfRel} and the Maurer-Cartan equation~\eqref{Eq:MaurerCartanEquation} are equivalent to  
\begin{equation}\label{Eq:BVEquat}
 \BVOp\circ \BVOp = 0\quad\text{and}\quad \BVOp(e^\PMC) = 0,
\end{equation}
respectively, and that the $\BVInfty$-operator $\BVOp^\PMC$ for the twisted $\IBLInfty$-structure $(\OPQ_{klg}^\PMC)$ satisfies
\begin{equation} \label{Eq:TwistBV}
\BVOp^\PMC(\bullet)= e^{-\PMC}\BVOp(e^\PMC\bullet),
\end{equation}
where we multiply with $e^{-\PMC}$ and $e^{\PMC}$, respectively.
These facts were shown in~\cite{Cieliebak2015} using~\eqref{Eq:Mix}.
We refer to Appendix~\ref{App:IBLMV} to the precise formulation of the $\BV$-formalism using a filtered version of the $\MV$-formalism from \cite{Markl2015}.\qedhere
\end{RemarkList}
\end{Remark}

In our applications in string topology, a canonical $\dIBL$-algebra $\dIBL(C)$ with a natural Maurer-Cartan element $\PMC$ coming from the Chern-Simons theory is given, and we want to study $\dIBL^\PMC(C)$, which will be a chain model of string topology.
We are also interested in the homology $\HIBL^\PMC(C)$, the $\IBL$-structure $\IBL(\HIBL^\PMC(C))$ and possible higher operations on $\HIBL^\PMC(C)$ induced by $\OPQ_{klg}^\PMC$; however, these higher maps are not chain maps in general.
The following proposition summarizes some observations in this situation:

\begin{Proposition}[Twist of a $\dIBL$-algebra]\label{Prop:dIBL}
Let $\dIBL(C) = (C,\OPQ_{110},\OPQ_{210},\OPQ_{120})$ be a $\dIBL$-algebra, and let $\PMC = (\PMC_{lg})$ be a Maurer-Cartan element. The Maurer-Cartan equation~\eqref{Eq:MaurerCartanEquation} reduces to the following:
\[ 
\begin{multlined}[b] 0 = \OPQ_{110}\circ_1 \PMC_{lg} + \OPQ_{120} \circ_1 \PMC_{l-1,g} +  \OPQ_{210}\circ_2 \PMC_{l+1,g-1} \\[\jot]+  \frac{1}{2}\sum_{\substack{l_1, l_2\ge 1 \\ g_1, g_2 \ge 0 \\ l_1 + l_2 = l + 1 \\ g_1 + g_2 = g}} \OPQ_{210}\circ_{1,1}(\PMC_{l_1 g_1}, \PMC_{l_2 g_2}) \end{multlined}\quad \forall l\ge1 , g\ge 0.
\]
In particular, the ``lowest'' equation is given by\footnote{In \cite[Definition 2.4.]{Cieliebak2015}, they define a partial ordering on the signatures $(k,l,g)$.}
\begin{equation} \label{Eq:MCEq}
(l,g) = (1,0): \qquad \OPQ_{110}(\PMC_{10}) + \frac{1}{2}\OPQ_{210}(\PMC_{10}, \PMC_{10}) = 0.
\end{equation}
This can be visualized as
{\begingroup \def\dist{0.25} %
  \def\rad{0.5} %
  \def\ecc{0.1} %
  \def\hght{1} %
  \def\dif{1.5} %
  \def\radO{\rad} %
  \def\eccO{\ecc} %
  \def\hghtO{2*\hght+\dist} %
  \def\difO{\dif} %
  \def\gencanc{0.05} %
  \def\genecc{20} %
  \def\genrad{0.45} %
\[0 =\quad \vcenterline{\input{\GraphicsFolder/p110n10.tex}}\; + \frac{1}{2} \quad \vcenterline{\input{\GraphicsFolder/p210n10n10.tex}}. \]
\endgroup}

The twisted $\IBLInfty$-algebra $\dIBL^\PMC(C)$ consists of the operations $\OPQ_{110}^\PMC$, $\OPQ_{210}^\PMC$ and~$\OPQ_{120}^\PMC$, which we call the \emph{basic operations}, and of the operations $\OPQ_{1lg}^\PMC$ for the pairs $(l,g)\in \N \times \N_0 \backslash \{(1,0),(2,0)\}$, which we call the \emph{higher operations}. These operations are given by 
\[ \begin{aligned}
\OPQ_{110}^\PMC &= \OPQ_{110} + \OPQ_{210}\circ_1 \PMC_{10},\\
\OPQ_{210}^\PMC &= \OPQ_{210}, \\
\OPQ_{120}^\PMC & = \OPQ_{120} + \OPQ_{210}\circ_1 \PMC_{20},\\
\OPQ_{1lg}^\PMC & = \OPQ_{210}\circ_1 \PMC_{lg}.
\end{aligned}\]
This can be visualized as
{ \begingroup \allowdisplaybreaks
\def\dist{0.25} %
  \def\rad{0.5} %
  \def\ecc{0.1} %
  \def\hght{1} %
  \def\dif{1.5} %
  \def\radO{\rad} %
  \def\eccO{\ecc} %
  \def\hghtO{2*\hght+\dist} %
  \def\difO{\dif} %
  \def\gencanc{0.05} %
  \def\genecc{20} %
  \def\genrad{0.45} %
\begin{align*}
\OPQ_{110}^\PMC & =\quad\vcenterline{\input{\GraphicsFolder/p110.tex}}\quad +\quad \vcenterline{\input{\GraphicsFolder/twist2.tex}}, \\[1ex] 
\OPQ_{210}^\PMC &=\quad \vcenterline{\input{\GraphicsFolder/p210.tex}}, \\[1ex] 
\OPQ_{120}^\PMC &=\quad \vcenterline{\input{\GraphicsFolder/p120.tex}}+\quad\vcenterline{\input{\GraphicsFolder/twist1.tex}}, \\[1ex]
\OPQ^{\PMC}_{1lg} & =\quad \vcenterline{\input{\GraphicsFolder/twistn.tex}}.
\end{align*}
\endgroup}
The $\IBLInfty$-relations satisfied by $(\OPQ_{klg}^\PMC)$ read for all $l\ge 1$, $g\ge 0$ as follows:
\begin{equation}\label{Eq:IBLInftydIBL}
\begin{aligned}
(3,1,0):\quad 0& = \OPQ_{210}^\PMC \circ_1 \OPQ_{210}^\PMC, \\[\jot]
(2,l,g):\quad 0&=\OPQ^\PMC_{1lg}\circ_1 \OPQ^\PMC_{210} + \OPQ^\PMC_{210}\circ_1\OPQ^\PMC_{1lg}, \\[\jot]
(1,l,g):\quad 0&= \begin{multlined}[t] \sum_{\substack{l_1, l_2 \ge 1 \\ g_1, g_2 \ge 0 \\ l_1 + l_2 = l+1 \\ g_1 + g_2 = g}} \OPQ^\PMC_{1l_1 g_1}\circ_1 \OPQ^\PMC_{1 l_2 g_2}+\OPQ^\PMC_{210} \circ_2 \OPQ^\PMC_{1, l+1, g-1}.\end{multlined}
\end{aligned}
\end{equation}
We call the relations for $(k,l,g) = (1,1,0)$, $(2,1,0)$, $(1,2,0)$, $(3,1,0)$, $(1,3,0)$, $(2,2,0)$, $(1,1,1)$ \emph{basic relations} because they contain all compositions of basic operations. In the order above, they read:
\[\begin{aligned} 
 0 & =\OPQ^\PMC_{110} \circ_1 \OPQ^\PMC_{110}, && \\
0 &=\OPQ^\PMC_{110}\circ_1\OPQ^\PMC_{210} + \OPQ^\PMC_{210}\circ_1 \OPQ^\PMC_{110}, && \\
0 &= \OPQ^\PMC_{110}\circ_1 \OPQ^\PMC_{120} + \OPQ^\PMC_{120}\circ_1 \OPQ^\PMC_{110}, && \\
0 &= \OPQ^\PMC_{210} \circ_1 \OPQ^\PMC_{210}, && \leftarrow\text{Jacobi identity} \\
0 &=\OPQ^\PMC_{120} \circ_1 \OPQ^\PMC_{120} + \OPQ^\PMC_{110}\circ_1 \OPQ^\PMC_{130} + \OPQ^\PMC_{130}\circ_1 \OPQ^\PMC_{110}, && \leftarrow\text{co-Jacobi id. up to htpy.} \\
0 & = \OPQ^\PMC_{120}\circ_1 \OPQ^\PMC_{210} + \OPQ^\PMC_{210}\circ_1 \OPQ^\PMC_{120}, && \leftarrow\text{Drinfeld identity} \\
0 & = \OPQ^\PMC_{210}\circ_2 \OPQ^\PMC_{120} + \OPQ^\PMC_{111}\circ_1 \OPQ^\PMC_{110} + \OPQ^\PMC_{110}\circ_1 \OPQ^\PMC_{111}. && \leftarrow\text{Involutivity up to htpy.}
\end{aligned}\]
The last four equations can be visualized as
{ \begingroup \allowdisplaybreaks
\def\dist{0.25} %
  \def\rad{0.4} %
  \def\ecc{0.1} %
  \def\hght{1} %
  \def\dif{1.1} %
  \def\difbig{1.5*\dif} %
  \def\radO{\rad} %
  \def\eccO{\ecc} %
  \def\hghtO{2*\hght+\dist} %
  \def\difO{\dif} %
  \def\gencanc{0.05} %
  \def\genecc{20} %
  \def\genrad{0.3} %
\begin{align*}
0 & =\quad\vcenterline{\input{\GraphicsFolder/jacobi.tex}}, \\[1ex] 
0&=\quad \vcenterline{\input{\GraphicsFolder/cojacobi.tex}}+
\vcenterline{\input{\GraphicsFolder/cojacobi2.tex}}
+\quad\vcenterline{\input{\GraphicsFolder/cojacobi3.tex}}, \\[1ex] 
0&=\quad \vcenterline{\input{\GraphicsFolder/drinfeld.tex}}+\quad\vcenterline{\input{\GraphicsFolder/drinfeld2.tex}},\\[1ex]
0& =\quad \vcenterline{\input{\GraphicsFolder/involutivity.tex}}\quad+\quad\vcenterline{\input{\GraphicsFolder/involutivity2.tex}}\quad+\quad\vcenterline{\input{\GraphicsFolder/involutivity3.tex}}.
\end{align*}
\endgroup}
\end{Proposition}
\begin{proof}
The proof is clear by specializing \eqref{Eq:IBLInfRel}, \eqref{Eq:MaurerCartanEquation} and \eqref{Eq:TwistedOperations}.
\end{proof}

\begin{Remark}[Higher operations]\phantomsection\label{Rem:Higher}
\begin{RemarkList}
\item We see from Proposition~\ref{Prop:dIBL} that if $\OPQ_{120}^\PMC \circ_1 \OPQ_{120}^\PMC = 0$ and $\OPQ_{210}^\PMC \circ_{2} \OPQ_{120}^\PMC = 0$, then $[\Bdd^\PMC, \OPQ_{130}^\PMC] = 0$ and $[\Bdd^\PMC, \OPQ_{111}^\PMC] = 0$, respectively, and hence the operations $\OPQ_{130}^\PMC: \hat{\Ext}_1\HIBL^\PMC\rightarrow \hat{\Ext}_3\HIBL^\PMC$ and $\OPQ_{111}^\PMC: \hat{\Ext}_1\HIBL^\PMC\rightarrow \hat{\Ext}_1\HIBL^\PMC$ are well-defined (provided that the assumption of Definition \ref{Def:HomIBL} holds). Likewise, the higher operation~$\OPQ_{1lg}^\PMC$ defines a map $\hat{\Ext}_1\HIBL^\PMC \rightarrow \hat{\Ext}_l\HIBL^\PMC$, provided that the following equation holds:
\[ \OPQ^\PMC_{210}\circ_2 \OPQ_{1,l+1,g-1}^\PMC + \sum_{\substack{l_1, l_2 \ge 1 \\ g_1, g_2 \ge 0 \\ l_1 + l_2 = l+1 \\ g_1 + g_2 = g \\ (l_i,g_i)\neq (1,0)}} \OPQ^\PMC_{1l_1 g_1}\circ_1 \OPQ^\PMC_{1 l_2 g_2} = 0. \]
This expression is just the left-over after subtracting the commutator $[\OPQ_{1lg}^\PMC,\OPQ_{110}^\PMC] = \OPQ_{110}^\PMC \circ_1 \OPQ_{1lg}^\PMC + \OPQ_{1lg}^\PMC \circ_1 \OPQ_{110}^\PMC$ from \eqref{Eq:IBLInftydIBL}.
\item In the genus-$0$ case, i.e., $\OPQ_{1lg}^\PMC = 0$ whenever $g\ge 1$, relations \eqref{Eq:IBLInftydIBL} reduce to
\begin{align*}
 0 & = \OPQ_{210}\circ_1\OPQ_{210}, \\
 0 & = \OPQ_{1lg}^\PMC\circ_1\OPQ_{210} + \OPQ_{210}\circ_1\OPQ_{1lg}^\PMC, \\
 0 & = \sum_{\substack{l_1,l_2\ge 1\\ g_1, g_2 \ge 0 \\ l_1 + l_2 = l + 1 \\ g_1 + g_2 = g}} \OPQ_{1l_1 g_1}^\PMC \circ_1 \OPQ_{1l_2 g_2}^\PMC.
\end{align*}
The first relation is the Jacobi identity for $\OPQ_{210}$, the second relation is a generalization of the Drinfeld identity to higher coproducts $\OPQ_{1l0}^\PMC$, and the third relations are \emph{$\CoLInfty$-relations} for $\OPQ_{1lg}^\PMC$.
Therefore, allowing $g\ge 0$, we see that the twisted $\dIBL$-algebra $\OPQ_{110}^\PMC$, $\OPQ_{210}$, $(\OPQ_{1lg}^\PMC)$ is, in fact, a \emph{quantum $\CoLInfty$-algebra.}\qedhere
\end{RemarkList}
\end{Remark}

\section{Dual cyclic bar complex and cyclic (co)homology
}
\allowdisplaybreaks
\label{Sec:Alg2}
\Correct[caption={DONE Notation for cyclic Hochschild}]{Notation for cyclic Hochschild}
\Correct[caption={DONE Reduced is defined wrt. normalized!},inline]{Reduced is defined as coker or ker in normalized and not in the full complex! It is OK because we have the cyclic symmetry!}
\Modify[caption={DONE Convention for duals},inline]{The convention for cochain complexes will be as follows. Chain complex will be denoted by $C$ and cochain complex by $C^*$. The components will be denoted by $C_q$ and $C^q$. }
\begin{Definition}[Bar complexes] \label{Def:BarComplex}
Let $V$ be a graded vector space. The \emph{bar- and dual bar-complex of $V$} are the weight-graded vector spaces defined by 
$$ \B V\coloneqq \RTen(V[1])\quad\text{and}\quad\DB V \coloneqq (\B V)^{\WGD}, $$
respectively, where $\bar{T}V \coloneqq \bigoplus_{k=1}^\infty V^{\otimes k}$ is the weight-reduced tensor algebra. For every $k\in \N$, let $t_k \in \Perm_k$ be the cyclic permutation $t_k : (1,\dotsc,k) \mapsto (2,\dotsc,k,1)$,
so that for all $v_1$, $\dotsc$, $v_k \in V[1]$ we have
\begin{equation*}
t_k(v_1 \otimes \dotsb \otimes v_k) = (-1)^{\Abs{v_k}(\Abs{v_1} + \dotsb + \Abs{v_{k-1}})} v_k \otimes v_1 \otimes \dotsb \otimes v_{k-1}.
\end{equation*}
We set
$$ t\coloneqq \sum_{k=1}^\infty t_k : \B V \longrightarrow \B V. $$
The \emph{cyclic bar-complex} is defined by 
$$ \BCyc V \coloneqq \B V / \Im(1-t). $$
We denote the image of $v_1 \otimes \dotsb \otimes v_k \in \B V$ under the canonical projection $\pi: \B V \rightarrow \BCyc V$ by $v_1\dots v_k$. If $v_i\in V[1]$ are homogenous, then $v_1\dots v_k$ is called a \emph{generating word}; we have
\begin{equation*}
v_1 \dots v_k = (-1)^{\Abs{v_k}(\Abs{v_1}+\dotsb + \Abs{v_{k-1}})} v_k v_1 \dots v_{k-1}.
\end{equation*}
We define the section $\iota: \BCyc V \rightarrow \B V$ of $\pi$ by
$$ \iota(v_1\dots v_k) \coloneqq \frac{1}{k} \sum_{i=0}^{k-1} \underbrace{t_k^i}_{\mathrlap{\displaystyle \eqqcolon t_k \circ \dotsb \circ t_k\ i\text{-times}}}(v_1\otimes\dotsb \otimes v_k) $$
and use it to identify $\BCyc V$ with the subspace $\Im \iota = \Ker(1-t) \subset \B V$ consisting of cyclic symmetric tensors.

We define the \emph{dual cyclic bar-complex} by 
$$ \DBCyc V \coloneqq \{ \psi\in \DB V \mid \psi \circ t = \psi \}. $$
\end{Definition}

\begin{Remark}[Non-weight-reduced bar complex]\label{Rem:NWG}
In fact, our $\DBCyc V$ is weight-reduced. The non-weight-reduced version would be $\DBCyc V \oplus \R$ with $\R$ of degree~$0$. This might play a role in the theory of  weak $\AInfty$-algebras ($\coloneqq$\,operation $\mu_0$ added; c.f., Definition~\ref{Def:CyclicAinfty}), and it might also be possible to consider $\IBLInfty$-algebras on non-weight-reduced cyclic cochains (c.f., Section~\ref{Sec:Alg3}).
\end{Remark}

Notice that $\psi \in \DB V$ is homogenous of degree $\Abs{\psi}\in \Z$ if and only if for all homogenous $v_1$,~$\dotsc$, $v_k \in V[1]$ the following implication holds:
\begin{equation*}
\Abs{v_1} + \dotsb + \Abs{v_k} \neq \Abs{\psi}\quad\Implies\quad \psi(v_1\otimes \dotsb \otimes v_k) = 0.
\end{equation*}
This is the cohomological grading convention.

\begin{Notation}[Degree shifts of bar complexes] \label{Def:Notation}
Let $A\in \Z$. In the following, we write $\DBCyc V$, but the convention applies to all complexes from Definition\,\ref{Def:BarComplex}. We denote by $\Susp_A$ and $\SuspU$ the formal symbols of degrees 
$$ \Abs{\Susp_A} = -A \quad \text{and}\quad\Abs{\SuspU} = -1, $$
respectively. The degree shift $V \mapsto V[1]$ will be realized as  multiplication with~$\SuspU$ and the degree shift $\DBCyc V\mapsto \DBCyc V[A]$ as multiplication with~$\Susp_A$. In addition, the following notation will be used consistently:
\begin{itemize}
 \item $\tilde{v}\in V \longleftrightarrow v = \SuspU \tilde{v} \in V[1]$
 
  To clarify this, given $\tilde{v} \in V$, then~$v$ automatically means $v = \SuspU \tilde{v} \in V[1]$, and the other way round. Recall that the degree of $\tilde{v}\in V$ is denoted by~$\Deg(\tilde{v})$ or simply by $\tilde{v}$ in the exponent, e.g., $(-1)^{\tilde{v}}$.
 \item  $\psi\in\DBCyc V\longleftrightarrow \Psi = \Susp_A \psi\in \DBCyc V[A]$.
 \item A generating word of $\BCyc V$ of weight $k$ will be denoted by the symbol $w$ and written as $w= v_1 \dots v_k$, where $v_i = \SuspU \tilde{v}_i \in V[1]$. A generating word of $\Ext_k \BCyc V$ is an element $w_1 \dotsb w_k \in \Ext_k \BCyc V$ such that each~$w_i$ is a generating word of $\BCyc V$.
 \item $w\in \BCyc V\longleftrightarrow \text{\footnotesize W} = \Susp_A w \in \BCyc V[A]$.
\end{itemize}
We abbreviate
$$ \DBCyc V[A]\coloneqq (\DBCyc V)[A]. $$
In contrast to this, we would write $\DBCyc(V[A])$ for the dual cyclic bar-complex of~$V[A]$. We also identify $(\DBCyc V[A])[1] = \DBCyc V[A+1]$ in $\Ext \DBCyc V[A]$.
\end{Notation}

\begin{Definition}[Pairing of tensor powers of bar complexes]\label{Def:Pairings}
For every $A\in \Z$ and  $k\in \N$, we define the pairing as follows:
\begin{equation} \label{Eq:Pairing}
\begin{aligned}
(\DB V[A])^{\otimes k} \otimes (\B V[A])^{\otimes k} & \longrightarrow \R \\ 
(\Psi_1 \otimes \dotsb \otimes \Psi_k, \W_1 \otimes \dotsb \otimes \W_k) & \longmapsto \underbrace{\psi_1(w_1) \dots \psi(w_k)}_{\mathllap{\textstyle{(\Psi_1 \otimes \dotsb \otimes \Psi_k)(\W_1\otimes \dotsb\otimes \W_k)\coloneqq}}}.
\end{aligned}
\end{equation}
This means that we evaluate elements from the left-hand side on the elements from the right-hand side in this way without any signs (see the discussion in Remark~\ref{Rem:BadConvention}). We extend the pairing by $0$ if the number of $\Psi_i$'s and the number of $\W_i$'s differ.
\end{Definition}

\begin{Remark}[Dual bar complex and dual of the bar complex] \label{Rem:Identifications}
Because the pairing~\eqref{Eq:Pairing} is non-degenerate, we can embed the space on the left into the the linear dual of the space on the right.
From Definition~\ref{Def:BarComplex} we have $\DBCyc V \subset \DB V$, and $\BCyc V$ is identified with $\Im \iota \subset \B V$. Therefore, we can restrict \eqref{Eq:Pairing} to obtain the pairing of $\DBCyc V$ and $\BCyc V$. It is easy to see that for any $\psi\in \DBCyc V$ and any generating word $v_1\dots v_k \in \BCyc V$, we have
\begin{equation*}
\psi(v_1\dots v_k) = \psi(v_1 \otimes \dotsb \otimes v_k).
\end{equation*}
The subspace of $(\BCyc V)^{\LD}$ corresponding to $\DBCyc V$ is then precisely $(\BCyc V)^{\WGD}$.

More generally, for every $k\in \N$, the spaces $\Ext_k \DBCyc V$ and $\Ext_k \BCyc V$ are embedded into $(\DBCyc V[1])^{\otimes k}$ and $(\BCyc V[1])^{\otimes k}$, respectively, using $\iota$ and $\pi$ from Definition~\ref{Def:SymAlgebra}. Therefore, the restriction of~\eqref{Eq:Pairing} gives the pairing of $\Ext_k \DBCyc V$ and $\Ext_k \BCyc V$. It is easy to see that for any generating word $w_1\dotsb w_k \in \Ext_k \BCyc V$ and any $\psi_1\dotsb \psi_k\in \Ext_k \DBCyc V$, we have
$$ (\psi_1\dotsb \psi_k)(w_1\dotsb w_k) = \frac{1}{k!}\sum_{\sigma\in \Perm_k} \varepsilon(\sigma,w) \psi_1(w_{\sigma_1^{-1}})\dotsc \psi_k(w_{\sigma_k^{-1}}). $$
The subspace of $(\Ext_k \BCyc V)^{\LD}$ corresponding to $\Ext_k \DBCyc V$  lies in $(\Ext_k \BCyc V)^{\WGD}$; it is equal to $(\Ext_k \BCyc V)^{\WGD}$, provided that $V$ is finite-dimensional.\footnote{The problem is that if $\dim(V) = \infty$, then $(V\otimes V)^* \neq V^* \otimes V^*$.}
\end{Remark}

The weight-graded vector spaces $\B V$ and $\BCyc V$ are canonically filtered by the filtration by weights \eqref{Eq:FiltrWeights}. Their weight-graded duals $\DB V$ and $\DBCyc V$ are filtered by the dual filtrations and the exterior powers $\Ext_k \DB V$ and $\Ext_k \DBCyc V$ by the induced filtration from Definition~\ref{Def:Filtrations}. 

\begin{Proposition}[Completed dual cyclic bar complex] \label{Prop:Compl}
Let $V$ be a graded vector space and $A\in \Z$. The filtration of $\DBCyc V$ dual to the weight-filtration of $\BCyc V$ is $\Z$-gapped, Hausdorff,  decreasing and bounded from above. Moreover, the following holds:
$$ \dim(V)<\infty\quad\Implies\quad (WG1)\ \&\ (WG2)\text{ are satisfied.} $$
The same holds for the induced filtration of $\Ext_k \DBCyc V[A]$.

In the sense of Remark~\ref{Rem:Identifications}, we have\Correct[caption={E},noline]{Here should be $A$ instead of $A+1$}
$$ \nCDBCyc V \simeq (\BCyc V)^{\GD}\quad\text{and}\quad \hat{\Ext}_k \DBCyc V[A] \subset (\Ext_k \BCyc V[A+1])^{\GD}, $$
where ``='' holds if $V$ is finite-dimensional.

The \emph{filtration degree} of $\Psi\in \hat{\Ext}_m \DBCyc V[A]$ satisfies
$$ \Norm{\Psi} = \min\{k\in \N_0 \mid \exists \W\in (\Ext_m \BCyc V[A])_k: \Psi(\!\W)\neq 0 \}.$$
\end{Proposition}

\begin{proof}
The proof is clear.
\end{proof}

\begin{Def}[Cyclic $\AInfty$-algebra] \label{Def:CyclicAinfty}
A graded vector space $V$ together with a pairing\Correct[noline,caption={Degree of pairing}]{This should be standardized with the degree of Poincare duality algebra, canonical dIBL algebra, .... The degree should be probably minus the degree of the pairing on $V$ (not $V[1]$)} 
$$ \Pair: V[1]\otimes V[1] \rightarrow \R $$
of degree $d\in \Z$ and a collection of homogenous linear maps 
$$\mu_k: V[1]^{\otimes k} \rightarrow V[1]\quad\text{for }k\ge 1$$
is called a \emph{cyclic $\AInfty$-algebra of degree~$d$} if the following conditions are satisfied:
\begin{PlainList}
 \item The pairing $\Pair$ is non-degenerate and graded antisymmetric; i.e., we have
  $$ \Pair(v_1,v_2) = (-1)^{1+\Abs{v_1}\Abs{v_2}} \Pair(v_2,v_1) \quad\text{for all }v_1, v_2 \in V[1]. $$
 \item The degrees satisfy $\Abs{\mu_k}=1$ for all $k\ge 1$.
 \item The \emph{$\AInfty$-relations} are satisfied: for all $k\ge 1$, we have
\begin{equation} \label{Eq:AInftyDef}
 \sum_{\substack{k_1, k_2 \ge 1 \\ k_1+k_2 = k+1}} \sum_{p=1}^{k_1} \mu_{k_1} \circ_1^p \mu_{k_2} = 0,
 \end{equation}
where for all $p=1$, $\dotsc$, $k$ and $v_1$, $\dotsc$, $v_{k}\in V[1]$ we define
$$(\mu_{k_1} \circ_1^p \mu_{k_2})(v_1, \dotsc, v_{k}) \coloneqq \begin{multlined}[t] (-1)^{\Abs{v_1} + \dotsb + \Abs{v_{p-1}}} \mu_{k_1}(v_1, \dotsc, v_{p-1},\\ \mu_{k_2}(v_p,\dotsc,v_{p+k_2-1}),v_{p+k_2}\dotsc,v_{k}). \end{multlined}$$

 \item The operations $\mu_k^+: V[1]^{\otimes k+1} \rightarrow \R$ defined by 
 $$ \mu_k^+\coloneqq \Pair\circ (\mu_k \otimes \Id) $$
 for all $k\ge 1$ are cyclic symmetric; i.e., we have
 $$ \mu_k^+ \circ t_{k+1} = \mu_k^+. $$
\end{PlainList}
We denote by $\tilde{\Pair}: V\otimes V \rightarrow \R$ and $\tilde{\mu}_k: V^{\otimes k} \rightarrow \R$ the operations before the degree shift; i.e., for all $k\ge 1$ and $\tilde{v}_1$, $\dotsc$, $\tilde{v}_k \in V$ with $v_i = \SuspU \tilde{v}_i$, we have
\begin{align*}
\tilde{\Pair}(\tilde{v}_1, \tilde{v}_2) &\coloneqq (-1)^{\tilde{v}_1} \Pair(v_1, v_2)\quad\text{and} \\[\jot]
\tilde{\mu}_k(\tilde{v}_1, \dotsc, \tilde{v}_k) &\coloneqq \varepsilon(\SuspU,\tilde{v}) \mu_k(v_1,\dotsc, v_k).
\end{align*}
We define $\tilde{\mu}_k^+: V^{\otimes k+1}\rightarrow \R$ similarly.

If $\mu_k \equiv 0$ for all $k\ge 2$, then $(V,\Pair, \mu_1)$ is called a \emph{cyclic cochain complex}. If $\mu_k \equiv 0$ for all $k\ge 3$, then $(V,\Pair,\mu_1,\mu_2)$ is called a \emph{cyclic dga}. We use the same terminology but omit ``cyclic'' if there is no pairing $\Pair$ and 1) and 4) are thus irrelevant.
\end{Def}

\begin{Remark}[A difference in sign conventions]\label{Rem:mukplus}
Our definition of $\mu_k^+$ differs from the definition of $\mathrm{m}_k^+$ in~\cite[Definition 12.1]{Cieliebak2015} by a sign. To compensate this, we have to add this artificial sign in the definitions of Maurer-Cartan elements later; e.g., in Definition~\ref{Def:CanonMC} or in the formula~\eqref{Eq:PushforwardMC}.
\end{Remark}
\Correct[caption={DONE Change sub to sup},noline]{Change $\Hd^k$ to $\Hd_k$ and so on. Why are the indices upstairs?}
\begin{Definition}[Cyclic (co)homology of an $\AInfty$-algebra]\label{Def:CycHom}
Let $\mathcal{A}=(V,(\mu_k))$ be an $\AInfty$-algebra. For every $k\ge 1$, we consider the maps $\Hd'_k$, $R_k: V[1]^{\otimes k} \rightarrow \B V$ given by 
\begin{equation}\label{Eq:bRH} \begin{aligned}\Hd'_k & \coloneqq \sum_{j=1}^k \sum_{i=0}^{k-j} t^i_{k-j+1}\circ(\mu_j \otimes \Id^{k-j})\circ t_k^{-i}\quad\text{and}\\
R_k &\coloneqq \sum_{j=2}^k \sum_{i=1}^{j-1} (\mu_j\otimes \Id^{k-j})\circ t_k^{i}, \end{aligned}
\end{equation}
respectively, and define the following maps $\B V \rightarrow \B V$:
\begin{equation*}
\Hd'\coloneqq \sum_{k=1}^{\infty} \Hd'_k, \quad R\coloneqq \sum_{k=2}^\infty R_k\quad\text{and}\quad \Hd\coloneqq \Hd' + R.
\end{equation*}
We denote by $\Hd^*: \CDB V = (\B V)^{\GD} \rightarrow \CDB V$ the dual map to $\Hd: \B V \rightarrow \B V$.  The following holds:\footnote{The facts~\eqref{Eq:HH}, in some form, are generally known; see \cite{Mescher2016} or \cite{Lazarev2003}. We prove them in our setting in Appendix~\ref{App:AInfty}.}
\begin{equation} \label{Eq:HH}
\Abs{\Hd} = 1\ (\Abs{\Hd^*}=-1), \quad \Hd\circ \Hd = 0 \quad\text{and}\quad \Hd(1-t) = (1-t)\Hd'.
\end{equation} 
From the last equation we see that $\Hd$ restricts to $\BCyc V = \B V / \Im(1-t)$.
We define the following graded vector spaces:
\begin{align*}
D(V) &\coloneqq r(\B V)[1], & D^*(V) &\coloneqq r(\CDB V)[1], \\ D^\lambda(V) &\coloneqq r(\BCyc V)[1], & D_\lambda^*(V) &\coloneqq r(\CDBCyc V)[1], \end{align*}
where $r$ denotes the grading reversal.
For instance, we have
$$ D_\lambda^q(V) = r(\CDBCyc V)^{q+1} = (\CDBCyc V)^{-q-1}\quad \text{for all } q\in \Z.$$
Then $(D(V),\Hd)$ and $(D^\lambda(V),\Hd)$ are chain complexes and $(D^*(V),\Hd^*)$ and $(D_\lambda^*(V),\Hd^*)$ the dual cochain complexes, respectively.
We define the following (co)homologies:
\begin{align*}
\H\H(\mathcal{A};\R)& \coloneqq \H(D(V), \Hd), & \H\H^*(\mathcal{A};\R) &\coloneqq \H(D^*(V),\Hd^*),\\  
\H^\lambda(\mathcal{A};\R)& \coloneqq \H(D^\lambda(V), \Hd), & \H^*_\lambda(\mathcal{A};\R), &\coloneqq \H(D^*_\lambda(V),\Hd^*).
\end{align*}
We call $\H\H$ the \emph{Hochschild homology} and $\H^\lambda$ the \emph{cyclic homology} of the $\AInfty$-algebra $\mathcal{A}$. We call $\H\H^*$ the \emph{Hochschild cohomology} and $\H_\lambda^*$ the \emph{cyclic cohomology} of~$\mathcal{A}$.
\end{Definition}

For a dga $\mathcal{A} = (V,\mu_1,\mu_2)$, we have for all $v_1$, $\dotsc$, $v_k\in V[1]$ the formula
\begin{align*}
 \Hd(v_1 \dots v_k) &= \sum_{i=1}^k (-1)^{\Abs{v_1} + \dotsb + \Abs{v_{i-1}}} v_1 \dots \mu_1(v_i) \dots v_k  \\ 
   &+ \sum_{i=1}^{k-1} (-1)^{\Abs{v_1} + \dotsb + \Abs{v_{i-1}}} v_1 \dots \mu_2(v_i,v_{i+1}) \dots v_k \\
   &+ (-1)^{\Abs{v_k}(\Abs{v_1} + \dotsb + \Abs{v_{k-1}})} \mu_2(v_k,v_1)v_2\dots v_{k-1}.
\end{align*}

\begin{Definition}[Strict units and strict augmentations]\label{Def:AugUnit}
Let $\mathcal{A}= (V, (\mu_k))$ be an $\AInfty$-algebra. A non-zero homogenous element $\NOne \in V[1]$ with $\Abs{\NOne} = -1$ is called a \emph{strict unit} for $\mathcal{A}$ if the following holds:
\begin{align*} \mu_2(\NOne, v) = (-1)^{\Abs{v} + 1}\mu_2(v,\NOne) &= v\qquad\forall v\in V[1], \\[\jot]
\mu_k(v_1, \dotsc, v_{i-1}, \NOne, v_{i+1}, \dotsc, v_k) &= 0\qquad\forall\ k\neq 2,\ 1\le i \le k,\ v_j \in V[1]. \end{align*}
The pair $(\mathcal{A},\NOne)$ is called a \emph{strictly unital  $\AInfty$-algebra.}

A strictly unital $\AInfty$-algebra $(\mathcal{A},\NOne)$ is called \emph{strictly augmented} if it is equipped with a linear map $\varepsilon: V[1] \rightarrow \R[1]$ which satisfies
$$ \varepsilon(\NOne_V) = \NOne_\R, \quad \varepsilon \circ \mu_1 = 0\quad\text{and}\quad \varepsilon \circ \mu_2 = \mu_2\circ(\varepsilon \otimes \varepsilon), $$
where $\NOne_\R$ is the strict unit for~$\R$ endowed with the standard multiplication. The map $\varepsilon$ is called a \emph{strict augmentation.}.

If the \emph{homological dga} $\H(\mathcal{A})\coloneqq (\H(V,\tilde{\mu}_1), \mu_1 \equiv 0, \mu_2)$ of $\mathcal{A}$ is strictly unital and strictly augmented, then~$\mathcal{A}$ is called \emph{homologically unital} and \emph{homologically augmented}, respectively. A strictly unital and strictly augmented cochain complex $(V,\mu_1,\NOne,\varepsilon)$ is called just augmented. 
\end{Definition}

We denote by $u: \R[1] \rightarrow V[1]$ the injective linear map defined by $u(\NOne_\R)\coloneqq \NOne_V$, and by $u^*: \DBCyc V \rightarrow \DBCyc \R$ and $\varepsilon^*: \DBCyc \R \rightarrow \DBCyc V$ the precompositions with $u^{\otimes k}$ and $\varepsilon^{\otimes k}$ in every weight-$k$ component, respectively. 

\begin{Remark}[On units and augmentations]\phantomsection\label{Rem:AugUnit}
\begin{RemarkList}
\item A strict unit $\NOne_V$ for $\mathcal{A}$ induces an $\AInfty$-morphism $(u_k): \R \rightarrow V$ given by $u_1(\NOne_\R)\coloneqq \NOne_V$ and $u_k \equiv 0$ for all $k\ge 2$. A (general) augmentation of $(\mathcal{A},\NOne_V)$ is by definition any $\AInfty$-morphism $(\varepsilon_k): V \rightarrow \R$ such that $(\varepsilon_k) \circ (u_k) = \Id$ as $\AInfty$-morphisms (see~\cite{Keller1999}). Strict augmentations are precisely the maps $\varepsilon_1$ coming from augmentations $(\varepsilon_k)$ with $\varepsilon_k \equiv 0$ for all~$k\ge 2$.

\item As for $(V,\mu_1,\NOne,\varepsilon)$, we need the chain map $\varepsilon$ to provide the splitting of the short exact sequence of chain complexes
$$\begin{tikzcd}
0 \arrow{r} & \R[1] \arrow[hook]{r}{u} & \arrow[bend left=50]{l}{\varepsilon} V[1] \arrow[two heads]{r} & \coker(u) \arrow{r} & 0,
\end{tikzcd}$$
so that we get $\H(V) \simeq \H_{\RedMRM}(V)\oplus \R$, where $\H_{\RedMRM}(V)\coloneqq \H(\coker(u))$. If $(V,\mu_1)$ is non-negatively graded and we are given an injective chain map $u: \R[1] \rightarrow V[1]$ ($\eqqcolon$\,the classical augmentation), then one can show that such $\varepsilon$ always exists. \qedhere
\end{RemarkList}
\end{Remark}

\begin{Definition}[Reduced dual cyclic bar complex]\label{Def:ReducedDual}
Let $(\mathcal{A}, \NOne)$ be a strictly unital $\AInfty$-algebra. Consider the injection $\iota_{\NOne}: \B V \rightarrow \B V$, $v_1 \otimes \dotsb \otimes v_k \mapsto \NOne \otimes v_1 \otimes \dotsb \otimes v_k$. We define the \emph{reduced dual cyclic bar-complex} by
$$ \RedDBCyc V \coloneqq \{\psi \in \DBCyc V \mid \psi\circ \iota_{\NOne} = 0\}. $$
Under the assumption of strict unitality, $\Hd^*$ preserves $\RedDBCyc V$, and hence we can consider the reduced cyclic cochain complex\Correct[caption={DONE Missing red},noline]{Correct missing red in the definition of $D_r$} 
$$ D_{\lambda,\RedMRM}^*(V) \coloneqq r(\CRedDBCyc V)[1]$$
and define the \emph{reduced cyclic cohomology of $\mathcal{A}$} by
$$ \H_{\lambda, \RedMRM}^*(\mathcal{A};\R)\coloneqq \H(D_{\lambda, \RedMRM}^*(V), \Hd^*). $$ 
\end{Definition}

\begin{Proposition}[Reduction to the reduced cyclic cohomology]\label{Prop:Reduced}
Let $\mathcal{A}= (V,(\mu_k))$ be an $\AInfty$-algebra with a strict unit $\NOne$ and a strict augmentation $\varepsilon$. Then the inclusions $\RedDBCyc V$, $\varepsilon^*(\DBCyc \R) \subset \DBCyc V$ induce the decomposition
\begin{align*}
\H_\lambda^*(\mathcal{A};\R) &\simeq \H_{\lambda, \RedMRM}^*(\mathcal{A};\R) \oplus \H_\lambda^*(\R;\R).
\end{align*}
Here we have
\begin{equation*}
 \H_\lambda^{q}(\R; \R) = \begin{cases} \langle \NOne^{q+1*} \rangle & \text{for }q\ge 0 \text{ even}, \\
0 & \text{for }q> 0 \text{ odd and }q<0, \\
\end{cases}
\end{equation*}
where $\NOne^{i*}: \R[1]^{\otimes i} \rightarrow \R$ is defined by $\NOne^{i*}(\NOne^{i}) \coloneqq 1$.
\end{Proposition}
\begin{proof}[Sketch of the proof]
The maps $\varepsilon^*: D_\lambda(\R) \rightarrow D_\lambda(V)$ and $u^*: D_\lambda(V) \rightarrow D_\lambda(\R)$ are chain maps with $u^*\circ \varepsilon^* = \Id$. Therefore, we have the sequence of cochain complexes
\begin{equation}\label{Eq:UnitAugSS}
\begin{tikzcd}
 0 \arrow{r} &D_{\lambda,\RedMRM}(V) \arrow[hook]{r} & D_\lambda(V) \arrow[two heads]{r}{u^*} & \arrow[bend left=50]{l}{\varepsilon^*} D_\lambda(\R) \arrow{r} & 0, 
\end{tikzcd}
\end{equation}
which is exact everywhere except for the middle, and where $\varepsilon^*$ is a splitting map. The idea of \cite{LodayCyclic} is to replace these cochain complexes with quasi-isomorphic bicomplexes consisting of normalized Hochschild cochains $\bar{D}(V)$ such that the sequence becomes exact. The work then reduces to proving that $\bar{D}(V)$ computes $\H\H(\mathcal{A};\R)$; a variant of this result for $\AInfty$-algebras was proven in~\cite{Lazarev2003}.
See Appendix~\ref{App:AInfty} for the full proof.
\end{proof}

We will now compare our version of the cyclic cohomology of a dga $(V,\mu_1, \mu_2)$ to the version from~\cite[Section 5]{LodayCyclic}.
In order to do this, we have to undo the degree shift~$V[1]$ first since it is not considered in \cite{LodayCyclic}.

Let $\tilde{\Hd}$, $\tilde{\delta}: \bar{T}V \rightarrow \bar{T}V$ be the linear maps defined for all $\tilde{v}_1$, $\dotsc$, $\tilde{v}_k \in V$ by
\allowdisplaybreaks
\begin{align*}
   \tilde{\Hd}(\tilde{v}_1\otimes \dotsb \otimes \tilde{v}_k) & \coloneqq \begin{multlined}[t] \sum_{i=1}^{k-1} (-1)^{i-1} \tilde{v}_1 \otimes \dotsb \otimes \tilde{\mu}_2(\tilde{v}_i, \tilde{v}_{i+1}) \otimes \dotsb \otimes \tilde{v}_k  \\ {}+ (-1)^{k-1+ \tilde{v}_k(\tilde{v}_1 + \dotsb + \tilde{v}_{k-1})}\tilde{\mu}_2(\tilde{v}_k, \tilde{v}_1)\otimes\tilde{v}_2\otimes\dotsb\otimes\tilde{v}_{k-1}, 
\end{multlined} \\ 
\tilde{\delta}(\tilde{v}_1\otimes \dotsb \otimes \tilde{v}_k) & \coloneqq  \sum_{i=1}^k (-1)^{\tilde{v}_1 + \dotsb + \tilde{v}_{i-1}} \tilde{v}_1\otimes\dotsb \otimes \tilde{\mu}_1(\tilde{v}_i)\otimes \dotsb \otimes \tilde{v}_k.
\end{align*}
For all $q\ge 0$, we define
\begin{equation}\label{Eq:NDSComplex}
\tilde{D}_q(V) \coloneqq \bigoplus_{\substack{k\ge 1 \\ d\in \Z \\k-d= q + 1}} (V^{\otimes k})^d
\end{equation}
and $\tilde{\Bdd}: \tilde{D}_{q+1}(V) \rightarrow \tilde{D}_{q}(V)$ by   
$$ \tilde{\Bdd}(\tilde{v}_1\dotsb \tilde{v}_k) = \tilde{\Hd}(\tilde{v}_1\dotsb \tilde{v}_k) + (-1)^{k+1} \tilde{\delta}(\tilde{v}_1\dotsb \tilde{v}_k). $$
It can be checked that $\tilde{\Bdd}\circ\tilde{\Bdd}=0$ and $\tilde{\Bdd}(\Im(1-\tilde{t}))\subset \Im(1-\tilde{t})$, so that $\tilde{\Bdd}$ induces a boundary operator on
the chain complexes \Correct[caption={DONE Wrong cyclic permutation}]{Here the $t$ is modified i.e. $\tilde{t}(v_1\dotsc v_k) = (-1)^{k-1} t(v_1 \dotsc v_k)$}
$$ \tilde{D}(V)\coloneqq \bigoplus_{q\in \Z} \tilde{D}_q(V)\quad\text{and}\quad\tilde{D}^\lambda(V) \coloneqq \tilde{D}(V)/\Im(1-\tilde{t}). $$
Here, we have $\tilde{t}(\tilde{v}_1 \dotsb \tilde{v}_k) \coloneqq (-1)^{k + \Abs{\tilde{v}_k}(\Abs{\tilde{v}_1} + \dotsb + \Abs{\tilde{v}_{k-1}})} \tilde{v}_k \tilde{v}_1 \dotsb \tilde{v}_{k-1}$.

We call $(\tilde{D}(V),\tilde{\Bdd})$ the \emph{non-degree-shifted Hochschild complex} and $(\tilde{D}^\lambda(V), \tilde{\Bdd})$ the \emph{non-degree-shifted cyclic complex} of the dga $(V,\mu_1,\mu_2)$.
We denote their homologies by $\ClasHH(V)$ and $\ClasCycH(V)$, respectively.

Looking at \eqref{Eq:NDSComplex}, the chain complex $(\tilde{D}(V),\tilde{\Bdd})$ is the total complex of the bicomplex
\begin{equation}\label{Eq:TotComplNDS}
\begin{tikzcd}
{} & \arrow{d} &\arrow{d} & \arrow{d} \\
{} &\arrow{l} \arrow{d}{\tilde{\Hd}} (V^{\otimes 3})^2 & \arrow{l}{\tilde{\delta}} \arrow{d}{\tilde{\Hd}} (V^{\otimes 3})^1 & \arrow{l}{\tilde{\delta}} \arrow{d}{\tilde{\Hd}} (V^{\otimes 3})^{0}  \\
{} &\arrow{l} \arrow{d}{\tilde{\Hd}} (V^{\otimes 2})^2 & \arrow{l}{-\tilde{\delta}} \arrow{d}{\tilde{\Hd}} (V^{\otimes 2})^1 & \arrow{l}{-\tilde{\delta}} \arrow{d}{\tilde{\Hd}} (V^{\otimes 2})^{0} \\
{} &\arrow{l} V^2 & \arrow{l}{\tilde{\delta}} V^1 & \arrow{l}{\tilde{\delta}} V^{0}
\end{tikzcd}
\end{equation}
with chain groups being the direct sums of the top-left/right-bottom diagonals.
This differs from the bicomplex \cite[Equation (5.3.2.1)]{LodayCyclic}, whose total complex is used to define the Hochschild homology of $V$ in \cite{LodayCyclic}, by the reversed grading in degree.
The convention of \cite{LodayCyclic} is namely $\Abs{\tilde{\mu}_1} = -1$, whereas ours is $\Abs{\tilde{\mu}_1}=1$.
The total complex of \cite{LodayCyclic} corresponds to the bottom-left/right-top diagonal in \eqref{Eq:TotComplNDS}.
Therefore, the homologies might differ!

We also warn the careful reader that the degree is called ``weight'' in \cite{LodayCyclic}.

The next proposition shows that our $\CycH(V)$ indeed computes $\ClasCycH(V)$.

\begin{Proposition}[Non-degree-shifted case] \label{Prop:DGA}
Let $\mathcal{A} = (V,\mu_1,\mu_2)$ be a dga. Then the degree shift map
\begin{align*} 
 U: \tilde{D}_q (V) & \longrightarrow D_q(V), \\
        \tilde{v}_1 \otimes \dotsb \otimes \tilde{v}_k & \longmapsto \varepsilon(\SuspU, \tilde{v}) v_1 \otimes \dotsb \otimes v_k,  \end{align*}
where we denote $v_i = \SuspU \tilde{v}_i$ for a formal symbol $\theta$ with $\Abs{\theta}=-1$, is an isomorphism of the chain complexes $(\tilde{D}(V),\tilde{\Bdd})$ and $(D(V), \Hd)$, resp.~$(\tilde{D}^\lambda(V),\tilde{\Bdd})$ and $(D^\lambda(V),\Hd)$.
\end{Proposition}
   
\begin{proof} 
First of all, it holds $\Abs{\tilde{\mu}_j} = 2 - j$ for every $j\ge 1$. For every $j$, $k$, $l\ge 1$ such that $j+l \le k+1$ and for every $\tilde{v}_1$, $\dotsc$, $\tilde{v}_k \in V$, we compute
\begin{align*}
&\bigl[U^{-1}(\Id^{l-1}\otimes \mu_j \otimes \Id^{k-j-l+1})U\bigr](\tilde{v}_1\dotsb \tilde{v}_k) \\[\jot] &\quad = (-1)^{l-1 + (j-2)(\tilde{v}_1 + \dotsb + \tilde{v}_{l-1} + k - l - j +1)} \tilde{v}_1\dotsb\tilde{v}_{l-1}\tilde{\mu}_j(\tilde{v}_l\dotsb \tilde{v}_{l+j-1})\tilde{v}_{l+j}\dotsb \tilde{v}_k, \\[\jot]
& [U^{-1} t_k U](\tilde{v}_1\dotsb \tilde{v}_k) = (-1)^{k-1} \tilde{v}_1 \dotsb \tilde{v}_k,
\end{align*}
where we use the Koszul convention $(f_1\otimes f_2)(v_1\otimes v_2) = (-1)^{\Abs{f_2}\Abs{v_1}} f_1(v_1)\otimes f_2(v_2)$. Using this, we obtain
\begin{align*}
U^{-1} \Hd'_k U &= \sum_{j=1}^k \sum_{i=0}^{k-1} (-1)^{i+j(i+k+1)} t^i_{k-j+1}(\tilde{\mu}_j \otimes \Id^{k-j})t_k^{-i}\quad\text{and} \\
U^{-1} R_k U &= \sum_{j=1}^k \sum_{i=1}^{j-1} (-1)^{(i+j)(k+1)} (\tilde{\mu}_j\otimes \Id^{k-j})t_k^i.
\end{align*}
It is now easy to check that $U^{-1}\circ \Hd\circ U = \tilde{\Bdd}$.

If $k\in \N$ is a weight and $d\in\Z$ a degree such that $k-d-1 = q$ for some $q\in \Z$, we have schematically $U: (k,d)\mapsto (k, d - k) = (k,-q-1)$. Therefore, $U$ preserves the grading of chain complexes. This finishes the proof.
\end{proof}

\begin{Proposition}[Reduced cochains are complete in $0$,\,$1$-connected case]\label{Prop:SimplCon}
Suppose that $V = \bigoplus_{d\ge 0} V^d$ is a non-negatively graded vector space with $V^0=\langle 1 \rangle$ for some $1\in V$ ($\eqqcolon$\,$V$ is \emph{connected}) and $V^1 = 0$ ($\eqqcolon$\,$V$ is \emph{simply-connected}). Then for all $m\ge 1$, we have
$$ \hat{\Ext}_m \RedDBCyc V = \Ext_m \RedDBCyc V. $$
\end{Proposition}
\begin{proof}
Let $\bar{V}\coloneqq \bigoplus_{d\ge 2} V^d$. We clearly have $\RedDBCyc V \simeq \DBCyc \bar{V}$. Since $\bar{V}[1]$ is positively graded, we have $(\B \bar{V})_{k}^d = 0$ whenever $k>d$. Therefore, a map $\Psi\in \hat{\Ext}_m \bar{V}$, which is non-zero only on finitely many homogenous components of $\BCyc V[1]^{\otimes m}$, will be non-zero only on finitely many weights. This implies that $\Psi\in \Ext_m \bar{V}$.
\end{proof}

\begin{Remark}[Universal coefficient theorem]\label{Rem:UCT}
We have 
$$ \B V = \bigoplus_{d\in\Z} \bigoplus_{k=1}^\infty (V[1]^{\otimes k})^d\quad\text{and}\quad \DB V = \bigoplus_{d\in \Z} \bigoplus_{k=1}^\infty (V[1]^{\otimes k})^{d*}, $$
and hence
$$ (\B V)^{\GD} = \bigoplus_{d\in\Z} \prod_{k=1}^\infty (V[1]^{\otimes k})^{d*} = \bigoplus_{d\in \Z} \reallywidehat{(\DB V)^d} = \CDB V. $$
Therefore, $(D^*_\lambda(V), \Hd^*)$ is dual to $(D^\lambda(V), \Hd)$ as a chain complex. Now, because we work over $\R$, the universal coefficient theorem gives
\begin{equation*}
 \H^q_\lambda(\mathcal{A},\Hd^*) \simeq [\H_q^\lambda(\mathcal{A},\Hd)]^*\quad\text{for all } q\in \Z. 
\end{equation*}
Suppose that we have found closed homogenous elements $(w_i)_{i\in I}\subset D^\lambda(V)$ for some index set~$I$ which induce a basis of $\H^\lambda(\mathcal{A}; \R)$. For every $i\in I$, we define the linear map $w_i^*: D^\lambda(V) \rightarrow \R$ by prescribing
$$ w_i^*(w_j) = \delta_{ij}\qquad\text{for all }j\in I $$
and $w_i^* \equiv 0$ on $\Im \Hd$ and on a complement \Correct[caption={DONE Universal coefficient theorem}]{Here is enough an arbitrary complement of $\Ker)(b)$. That means that for every $i$, we can have a different complement $Z_i$} of $\Ker(\Hd)$ in $D^\lambda(V)$. Then $(w_i^*)_{i\in I} \subset D_\lambda^*(V)$ are closed homogenous elements which generate linearly independent cohomology classes in $\H_\lambda^*(\mathcal{A}; \R)$; if we denote $I_q \coloneqq \{i\in I \mid w_i \in C^\lambda_q(V)\}$, then we can write
\begin{equation*}
\H_\lambda^q(\mathcal{A}; \R) = \Bigl\{ \sum_{i\in I_q} \alpha_i w_i^* \bigMid \alpha_i\in \R \Bigr\}\quad\text{for all }q\in\Z.\qedhere
\end{equation*}
\end{Remark}

\section{Canonical dIBL-structure on cyclic cochains
}

\label{Sec:Alg3}
 
In this section, we will consider a cyclic dga $(V,\Pair,m_1,m_2)$ of degree $2-n$ for some $n\in \N$ which is finite-dimensional. %

For all $v_1$, $v_2$, $v_3 \in V[1]$, the following relations holds:
\begin{equation}\label{Eq:CycDGA}{\hbadness=10000 \text{cyc. dga}\left\{ \begin{aligned} \Pair(v_1,v_2) &= (-1)^{1+ \Abs{v_1} \Abs{v_2}} \Pair(v_2, v_1), &
\mathclap{
\smash{
\raisebox{-0.58cm}{$
\hspace{-0.8cm}
\left.\begin{aligned}\mathstrut\\ \mathstrut\\ \mathstrut \end{aligned}\right\}
\text{\parbox{5em}{cyc. cochain complex}}
$}}}
\\
m_1(m_1(v_1)) &= 0, &\\
m_1^+(v_1, v_2) &= (-1)^{\Abs{v_1}\Abs{v_2}} m_1^+(v_2, v_1), &\\
 m_1(m_2(v_1,v_2)) &= \begin{multlined}[t]- m_2(m_1(v_1),v_2)  \\ - (-1)^{\Abs{v_1}} m_2(v_1,m_1(v_2)), \end{multlined} &\\
m_2(m_2(v_1, v_2), v_3) &= (-1)^{\Abs{v_1} + 1} m_2(v_1,m_2(v_2,v_3)), &\\
m_2^+(v_1, v_2, v_3) &= (-1)^{\Abs{v_3}(\Abs{v_1}+\Abs{v_2})}m_2^+(v_3, v_1, v_2).
\end{aligned}\right.}
\end{equation}
The facts (A) and (C) from the Overview apply, and we get the canonical $\dIBL$-algebra $\dIBL(\DBCyc V[2-n])$ of bidegree $(n-3,2)$ and the canonical Maurer-Cartan element $\MC = (\MC_{10})$. We will denote\Correct[caption={Definition of cyclic cochains},noline]{Define cyclic cochains as completion, and then distinguish long and short cyclic cochains.}
$$ \CycC(V)\coloneqq \DBCyc V[2-n] $$
and call it the space of \emph{cyclic cochains on $V$}. If $V$ is fixed, we will write just $\CycC$.

\begin{Def}[Canonical $\dIBL$-algebra] \label{Def:CanonicaldIBL}
Let $(V,\Pair,m_1)$ be a cyclic cochain complex of degree $2-n$ which is finite-dimensional. Let $(e_0, \dotsc, e_m)\subset V[1]$ be a basis of $V[1]$, and let $(e^0,\dotsc, e^m)$ be the dual basis with respect to $\Pair$; this means that
$$ \Pair(e_i,e^j) = \delta_{ij}\quad\text{for all }i, j =0, \dotsc,  m. $$
We define the tensor $T = \sum_{i,j=0}^m T^{ij} e_i \otimes e_j \in V[1]^{\otimes 2}$ by\footnote{See Appendix~\ref{Section:Appendix} for the invariant meaning of $T$ as the Schwartz kernel of $\pm \Id$.}
\begin{equation} \label{Eq:PropagatorT}
 T^{ij} = (-1)^{\Abs{e_i}} \Pair(e^i,e^j) \quad\text{for all }i,j = 0,\dotsc,m.
\end{equation}
The \emph{canonical $\dIBL$-algebra} on $\CycC(V)$ is the quadruple
$$ \dIBL(\CycC(V)) \coloneqq (\CycC(V),\OPQ_{110}, \OPQ_{210}, \OPQ_{120}), $$
where the operations $\OPQ_{110}$, $\OPQ_{210}$, $\OPQ_{120}$ are defined for all $\psi$, $\psi_1$, $\psi_2 \in \CDBCyc V$ and generating words $w = v_1 \dots v_k$, $w_1 = v_{11}\dots v_{1k_1}$, $w_2 = v_{21}\dots v_{2k_2}\in \BCyc V$ with $k$, $k_1$, $k_2\ge 1$ as follows:
\begin{itemize}
\item The \emph{$\dIBL$-boundary operator} $\OPQ_{110}: \hat{\Ext}_1 \CycC \rightarrow \hat{\Ext}_1 \CycC$ of degree $\Abs{\OPQ_{110}} = -1$ is defined by\Correct[noline,caption={Should here be $\Susp$?}]{There should not be $\Susp$, it is a number.}
\begin{equation}\label{Eq:Diff}
\OPQ_{110}(\Susp \psi)(\Susp w) \coloneqq \Susp\sum_{i=1}^k (-1)^{\Abs{v_1}+ \dotsb + \Abs{v_{i-1}}} \psi(v_1 \dots v_{i-1}m_1(v_i)v_{i+1} \dots v_k).
\end{equation}
\item The \emph{product} $\OPQ_{210}: \hat{\Ext}_2 \CycC \longrightarrow \hat{\Ext}_1 \CycC$ of degree $\Abs{\OPQ_{210}}= -2(n-3)-1$ is written schematically as
$$ \OPQ_{210}(\Susp^2 \psi_1 \otimes \psi_2)(\Susp w) \coloneqq \sum \varepsilon(w\mapsto w^1 w^2)(-1)^{\Abs{e_j}\Abs{w^1}}T^{ij}\psi_{1}(e_i w^1) \psi_2(e_j w^2) $$
and defined ``algorithmically'' as follows:

For every cyclic permutation $\sigma\in\Perm_k$, consider the tensor 
$$\sigma(w) \coloneqq \varepsilon(\sigma,w) v_{\sigma_1^{-1}}\otimes \dotsb \otimes v_{\sigma_k^{-1}}, $$
and split it into two parts $w^1$ and $w^2$ of possibly zero length such that $v_{\sigma_1^{-1}}\otimes \dotsb \otimes v_{\sigma_k^{-1}} = w^1 \otimes w^2$. Feed $w^1$ and $w^2$ into $\psi_1$ and $\psi_2$ preceded by~$e_i$ and~$e_j$, respectively, and multiply the result with the sign $(-1)^{\Abs{e_j}\Abs{w^1}}$, which is the Koszul sign to order 
$$ e_i e_j w^1 w^2 \longmapsto e_i w^1 e_j w^2. $$
Finally, sum over all $\sigma \in \Perm_k$, all splittings of $\sigma(w)$ and all indices $i,j = 0$,~$\dotsc$, $m$. The sign $\varepsilon(\sigma,w)$ is denoted by $\varepsilon(w\mapsto w^1w^2)$ to indicate the splitting.

\item The \emph{coproduct} $\OPQ_{120}: \hat{\Ext}_1 \CycC \longrightarrow \hat{\Ext}_2 \CycC$ of degree $\Abs{\OPQ_{120}} = -1$ is written schematically~as
$$ \begin{aligned} &\OPQ_{120}(\Susp \psi)(\Susp^2 w_1 \otimes w_2) \\ & \qquad = \frac{1}{2} \sum \varepsilon(w_1\mapsto w_1^1)\varepsilon(w_2\mapsto w_2^1) (-1)^{\Abs{e_j}\Abs{w_1^1}} T^{ij} \psi(e_i w_1^1 e_j w_2^1) \end{aligned}$$
and defined ``algorithmically'' as follows:

For all cyclic permutations $\sigma\in \Perm_{k_1}$ and $\mu\in \Perm_{k_2}$, denote $w_1^1\coloneqq \sigma(w_1)$ and $w_2^1\coloneqq \mu(w_2)$ and let $\varepsilon(w_1\mapsto w_1^1)$ and $\varepsilon(w_2\mapsto w_2^1)$ be the corresponding Koszul signs, respectively. Feed~$w_1^1$ and~$w_2^1$ into~$\psi$ in the indicated order interleaved by~$e_i$ and~$e_j$ and multiply the result with the sign $(-1)^{\Abs{e_j}\Abs{w_1^1}}$, which is the Koszul sign to order
$$ e_i e_j w_1^1 w_2^1 \mapsto e_i w_1^1 e_j w_2^1. $$
Finally, sum over all $\sigma\in \Perm_{k_1}$, $\mu\in\Perm_{k_2}$ and all indices $i$, $j = 0$, $\dotsc$, $m$.
\end{itemize}
The operations are extended continuously to the completion.
\end{Def}

\begin{Definition}[Canonical Maurer-Cartan element] \label{Def:CanonMC}
Let $(V,\Pair,m_1,m_2)$ be a finite-dimensional cyclic dga of degree $2-n$. The \emph{canonical Maurer-Cartan element} $\MC$ for $\dIBL(\CycC(V))$  consists of only one element $\MC_{10}\in \hat{\Ext}_1 \CycC$ of degree $\Abs{\MC_{10}} = 2(n-3)$ which is defined by
\begin{equation*}
\MC_{10}(\Susp v_1 v_2 v_3) \coloneqq (-1)^{n-2}  \mu_2^+(v_1,v_2,v_3)\quad\text{for all } v_1, v_2, v_3\in V[1]
\end{equation*}
on the weight-three component of $\BCyc V[3-n]$ and extended by $0$ to other weight-$k$ components.
\end{Definition}

\begin{Remark}[On canonical $\dIBL$-structure]\phantomsection
\begin{RemarkList}
\item Elements of the completion $\hat{\CycC}(V)$ which are not in $\CycC(V)$ will be called \emph{long cyclic cochains}. Because there are no infinite sums in Definition~\ref{Def:CanonicaldIBL}, $\dIBL(\CycC)$ is completion-free. Clearly, the twist $\dIBL^{\PMC}(\CycC)$ remains completion-free as long as $\PMC_{lg}\in \Ext_l \CycC$ for all $l$, $g$.
\item The constructions of $\OPQ_{210}$ and $\OPQ_{120}$ do not depend on the choice of a basis and can be rephrased in terms of summation over ribbon graphs (see Example~\ref{Ex:Canon}).
\item According to Proposition~\ref{Prop:Compl}, the filtration on $\CycC(V)$ satisfies (WG1) \& (WG2). Thus, the $\IBL$-structures $\IBL(\HIBL(\CycC))$ and $\IBL(\HIBL^{\MC}(\CycC))$ are well-defined (see Definition~\ref{Def:HomIBL}).\qedhere
\end{RemarkList}
\end{Remark}

\begin{Proposition}[Formulas for twisted operations]\label{Prop:Formulafortwisted}
Let $\dIBL(\CycC(V))$ be the canonical $\dIBL$-algebra for a finite-dimensional cyclic cochain complex $(V,\Pair,m_1)$ of degree $2-n$, and let $\PMC=(\PMC_{lg})$ be a Maurer-Cartan element. Then for all $l\ge 1$, $g\ge 0$, $\Psi\in \CDBCyc V[3-n]$ and generating words $\W_1$, $\dotsc$, $\W_l \in \BCyc V [3-n]$, we have 
\begin{equation}\label{Eq:TwisteddIBL}
\begin{aligned}
& [(\OPQ_{210}\circ_1 \PMC_{lg})(\Psi)](\W_1 \otimes \dotsb \otimes \W_l) \\ 
& \quad = \begin{multlined}[t]\sum_{j=1}^l \sum \varepsilon' \varepsilon(w_j \mapsto w_j^1 w_j^2) T^{ab} \Psi(\Susp e_a w_j^1) \PMC_{lg}(\W_1 \otimes \dotsb \W_{j-1} \otimes (\Susp e_b w_j^2)  \\ \otimes \W_{j+1} \otimes \dotsb \otimes \W_l), \end{multlined}
\end{aligned}
\end{equation}
where the sum without limits is the sum in Definition \ref{Def:CanonicaldIBL} for $\OPQ_{210}$ and $\varepsilon'$ is the Koszul sign of the following operation:
$$\begin{multlined}(\Susp e_a e_b) \W_1 \dots \W_{j-1}(\Susp w_j^1 w_j^2) \W_{j+1} \dots \W_l \\ 
\longmapsto (\Susp e_a w_j^1) \W_1 \dots \W_{j-1} (\Susp e_b w_j^2)\W_{j+1} \dots \W_l.\end{multlined}$$
In particular, for $l=1$, $g\ge 0$ and $\W\in \BCyc V[3-n]$, we have 
\begin{equation}\label{Eq:TwistDif}
(\OPQ_{210} \circ_1 \PMC_{1g})(\W)= (-1)^{n-3} \sum T^{ab} \varepsilon(w \mapsto w^1 w^2) \PMC_{1g}(\Susp e_a w^1)\psi(e_b w^2),
\end{equation}
and for $l=2$, $g\ge 0$ and $\W_1$, $\W_2\in \BCyc V[3-n]$, we have
\begin{equation}\label{Eq:Twistn2}
\begin{aligned}
 & [(\OPQ_{210}\circ_1 \PMC_{2g})(\Psi)](\W_1 \otimes \W_2) \\
 &\quad = \begin{multlined}[t](-1)^{(n-3)(\Abs{\Psi} + 1)} \Bigl[ \sum T^{ab} \varepsilon(w_1 \mapsto w_1^1w_{1}^{2}) (-1)^{\Abs{e_b}\Abs{w_1^1}}\Psi(\Susp e_a w_1^1)  \\ \PMC_{20}(\Susp e_b w_1^2 \otimes \W_2) + (-1)^{\Abs{\W_1}\Abs{\W_2}}\sum T^{ab}  \varepsilon(w_2\mapsto w_2^1w_2^2) \\ (-1)^{\Abs{e_b}\Abs{w_2^1}} \Psi(\Susp e_a w_2^1)\PMC_{20}(\Susp e_b w_2^2 \otimes \W_1)\Bigr].\end{multlined} \end{aligned}
\end{equation}
\end{Proposition}

\begin{proof}
Let us first discuss the completions. Given $\PMC_{lg}\in \hat{\Ext}_l \CycC$, we can write it as $\PMC_{lg} = \sum_{i=1}^\infty \Phi_1^i \dotsb \Phi_l^i$ with generating words $\Phi_1^i\dotsb \Phi_l^i\in \Ext_l \CycC$ of weights approaching $\infty$. The canonical extension of $\circ_h$ to maps with finite filtration degree commutes with convergent infinite sums, and hence we have $\OPQ_{klg} \circ_h \PMC_{lg} = \sum_{i=1}^\infty \OPQ_{klg}\circ_h (\Phi_1^i \dotsb \Phi_l^i)$. Therefore, it suffices to prove the formulas for generating words $\Phi_1^i \dotsb \Phi_l^i \in \Ext_l \CycC$.%

From \eqref{Eq:CompositionSimple}, we get for every $\Psi$, $\Phi_1$, $\dotsc$, $\Phi_l \in \CycC$ the equation\Correct[noline,caption={This can't be correct!}]{There must be a sign problem, set $l=1$!}
$$ [\OPQ_{210}\circ_1(\Phi_1\dotsb \Phi_l)](\Psi) = \sum_{i=1}^l (-1)^{\Abs{\Phi_i}(\Abs{\Phi_1}+\dotsb+\Abs{\Phi_{i-1}})}\OPQ_{210}(\Psi,\Phi_i) \Phi_1 \dotsb \hat{\Phi}_i \dotsb \Phi_l, $$
where $\Phi_1\dotsb \Phi_l$ on the left-hand-side is considered as a map $\Ext_0 \CycC = \R \rightarrow \Ext_l \CycC$ mapping $1$ to $\Phi_1\dotsb \Phi_l$.
For $\W_1$, $\dotsc$, $\W_l\in \BCyc V[3-n]$ and $\sigma\in \Perm_l$, we use
$$ [\sigma(\Phi_1\otimes \dotsb \otimes \Phi_l)](\W_1 \otimes \dotsb \otimes \W_l) = (\Phi_1 \otimes \dotsb \otimes \Phi_l)[\sigma^{-1}(\W_1\otimes \dotsb \otimes \W_l)] $$
and Definition~\ref{Def:CanonicaldIBL} to get
\allowdisplaybreaks
\begin{align*}
&\bigl([\OPQ_{210}\circ_1(\Phi_1\dotsb \Phi_l)](\Psi)\bigr)(\W_1 \otimes \dotsb \otimes \W_l) = \\
&\quad= \begin{multlined}[t] \sum_{i=1}^l (-1)^{\Abs{\Phi_i}(\Abs{\Phi_1}+\dotsb+\Abs{\Phi_{i-1}})} \frac{1}{l!}\sum_{\sigma\in \Perm_l} \varepsilon(\sigma^{-1},\W) [\OPQ_{210}(\Psi,\Phi_i)](\W_{\sigma_1}) \\ \Phi_1(\W_{\sigma_2}) \dotsb \hat{\Phi}_i(\emptyset) \dotsb \Phi_l(\W_{\sigma_l})\end{multlined} \\
&\quad=\begin{multlined}[t] \sum_{i=1}^l (-1)^{\Abs{\Phi_i}(\Abs{\Phi_1}+\dotsb+\Abs{\Phi_{i-1}})} \frac{1}{l!}\sum_{\sigma\in \Perm_l} \varepsilon(\sigma^{-1},\W) (-1)^{\Abs{\Susp}\Psi}\\ \sum \varepsilon(w_{\sigma_1}\mapsto w_{\sigma_1}^1 w_{\sigma_1}^2) 
(-1)^{\Abs{e_b}\Abs{w_{\sigma_1}^1}}T^{ab} \Psi(e_a w_{\sigma_1}^1) \Phi_i(e_b w_{\sigma_1}^2) \\ \Phi_1(\W_{\sigma_2}) \dots \hat{\Phi}_i(\emptyset) \dots \Phi_l(\W_{\sigma_l})\end{multlined} \\
&\; \eqqcolon (*),
\end{align*}
where $\hat{\Phi}_i(\emptyset)$ means omission of the corresponding term. Consider the bijection 
$$\begin{aligned}
I: \{1,\dotsc, l\} \times \Perm_l &\longrightarrow \{1,\dotsc, l\} \times \Perm_l \\
(i,\sigma) &\longmapsto \Biggl(j\coloneqq \sigma_1,\mu\coloneqq \begin{pmatrix} 1 & \dots & i-1 & i & i+1 & \dots & l \\ \sigma_2 & \dots & \sigma_{i} & \sigma_1 & \sigma_{i+1} & \dots & \sigma_l \end{pmatrix}\Biggr).
\end{aligned}$$
Given $(i,\sigma) \in \{1,\dotsc,l\}\times \Perm_l$ and $b\in \{1,\dotsc,m\}$, let $(j,\mu)\coloneqq I(i,\sigma)$ and
$$ \W'\coloneqq \W_1 \otimes \dotsb \otimes \W_{j-1} \otimes (\Susp e_b w_{j}^2)\otimes \W_{j+1}  \otimes \dotsb \otimes \W_l. $$
Suppose that $(\Phi_1 \otimes \dotsb \otimes \Phi_l)(\W')\neq 0$. We compute the Koszul sign $\varepsilon(\mu^{-1},\W')$ in the following way:
$$ \begin{aligned}
\W' &\mapsto (-1)^{(\Abs{w_{j}^1} + \Abs{e_b} + \Abs{\W_j})(\Abs{\W_1} + \dotsb + \Abs{\W_{j-1}})} (\Susp e_b w_j^2)\W_1 \dots \hat{\W}_j \dots \W_l \\
 & \mapsto \underbrace{(-1)^{(\Abs{w_{j}^1} + \Abs{e_b})(\Abs{\W_1} + \dotsb + \Abs{\W_{j-1}})}}_{\eqqcolon\varepsilon_1}\varepsilon(\sigma^{-1},\W) (\Susp e_b w_j^2) \W_{\sigma_2} \dots \W_{\sigma_l} \\
& \mapsto \underbrace{\varepsilon_1 \varepsilon(\sigma^{-1},\W) (-1)^{\Abs{\Phi_i}(\Abs{\Phi_1} + \dotsb + \Abs{\Phi_{i-1}})}}_{=\varepsilon(\mu^{-1},\W')}  \underbrace{\W_{\sigma_2} \dots \W_{\sigma_{i}} (\Susp e_b w_j^2) \W_{\sigma_{i+1}}\dots\W_{\sigma_l}}_{=\W'_{\mu_1}\dots \W'_{\mu_l}}.
\end{aligned}$$
Using this, we can rewrite $(*)$ as
$$\begin{aligned}
(*) &= \begin{multlined}[t](-1)^{\Abs{s}\Abs{\Psi}}\sum_{j=1}^l \sum\varepsilon(w_j \mapsto w_j^1 w_j^2)(-1)^{\Abs{e_b}\Abs{w_j^1}}T^{ab} \Psi(e_a w_j^1)  \\ 
\varepsilon_1 \frac{1}{l!}\sum_{\mu \in \Perm_l} \varepsilon(\mu^{-1}, \W')\Phi_1(\W'_{\mu_1}) \dots \Phi_l(\W'_{\mu_l}) \end{multlined} \\
 & = \begin{multlined}[t]\sum_{j=1}^l \sum \varepsilon(w_j \mapsto w_j^1 w_j^2) (-1)^{\Abs{s}\Abs{\Psi} + \Abs{e_b}\Abs{w_j^1} + (\Abs{w_j^1} + \Abs{e_b})(\Abs{\W_1}+ \dotsb + \Abs{\W_{j-1}}) }T^{ab} \\ 
 \Psi(\Susp e_a w_j^1) (\Phi_1 \dotsb \Phi_l)(\W_1 \otimes \dotsb \otimes \W_{j-1} \otimes (\Susp e_b w_j^2) \otimes \W_{j+1} \otimes \dotsb \otimes \W_l). \end{multlined}
\end{aligned}$$
Finally, we use
$$ T^{ab} \neq 0\; \Implies\; \Abs{e_a} + \Abs{e_b} = n-2 $$
to write
$$\begin{aligned}
\Abs{s}\Abs{\Psi} &= \Abs{s}(\Abs{w_{j}^1} + \Abs{e_a}) = (n-3)(\Abs{w_{j}^1} + n-2 - \Abs{e_b}) \\ 
&= \Abs{s}(\Abs{w_{j}^1} + \Abs{e_b}) \mod 2, \end{aligned}$$
and the formula \eqref{Eq:TwisteddIBL} follows.

As for \eqref{Eq:TwistDif}, we first compute $\varepsilon'$ for $l=1$ as follows:
$$ \begin{aligned}\ln_{-1}(\varepsilon')  &= \Abs{w_1} \Abs{e_b} + (\Abs{e_b} + \Abs{w_1})\Abs{s}  \underset{\mathclap{\substack{\uparrow\rule{0pt}{1.5ex} \\2(n-3) = \Abs{\PMC_{10}} = \Abs{s} + \Abs{e_b} + \Abs{w^2}}}}{=} \Abs{w^1}\Abs{w^2} + \Abs{s}\Abs{e_b} \\ &\underset{\mathclap{\substack{\uparrow\rule{0pt}{1.5ex} \\ \Abs{e_a} + \Abs{e_b} = \Abs{s} + 1}}}{=} \Abs{w^1}\Abs{w^2} + \Abs{e_a}\Abs{e_b} \mod 2.  \end{aligned}$$
Using this, we obtain
$$ \begin{aligned}
[(\OPQ_{210}\circ_1 \PMC_{1g})(\Psi)](W) &= \sum \varepsilon' \varepsilon(w\mapsto w^1 w^2) T^{ab} \Psi(\Susp e_a w^1) \PMC_{1g}(\Susp e_b w^2) \\
&\underset{\mathclap{\substack{\uparrow\rule{0pt}{1.5ex} \\ T^{ab} = (-1)^{\Abs{s} + \Abs{e_a}\Abs{e_b}} T^{ba} \\
\varepsilon(w\mapsto w^1 w^2) =  (-1)^{\Abs{w^1} \Abs{w^2}} \varepsilon(w\mapsto w^2 w^1)}}}{=} (-1)^{\Abs{s}} \sum \varepsilon(w\mapsto w^2 w^1) T^{ba} \PMC_{1g}(\Susp e_b w^2) \Psi(\Susp e_a w^1),
\end{aligned}$$
which implies \eqref{Eq:TwistDif}.

The proof of \eqref{Eq:Twistn2} is a combination of the same arguments.
\end{proof}

We will now relate homology of the twisted boundary operator $\OPQ_{110}^\PMC$ to cohomology of an $\AInfty$-algebra on $V$ induced by $\PMC_{10}$. 

\begin{Definition}[$\AInfty$-operations and compatible Maurer-Cartan element]\label{Def:MukDef}
Suppose that $(V,\Pair,m_1)$ is a finite-dimensional cyclic cochain complex of degree $2-n$, and let $\PMC = (\PMC_{lg})$ be a Maurer-Cartan element for $\dIBL(\CycC(V))$. We define the operations $\mu_k: V[1]^{\otimes k} \rightarrow V[1]$ for all $k\ge 1$ by
\begin{equation*}
\mu_k(v_1,\dotsc, v_k) \coloneqq (-1)^{n-3}\sum_{i, j} T^{ij} \PMC_{10}(\Susp e_i v_1 \dots v_k) e_j
\end{equation*}
for all $v_1$,~$\dotsc$, $v_k \in V[1]$, where $T^{ij}$ is the matrix from Definition~\ref{Def:CanonicaldIBL}.

If $(V,\Pair,m_1,m_2)$ is in addition a cyclic dga and $\MC$ the canonical Maurer-Cartan element for $\dIBL(\CycC(V))$, then we say that $\PMC$ is \emph{compatible with $\MC$} if
$$ \PMC_{10}(\Susp v_1 v_2 v_3) = \MC_{10}(\Susp v_1 v_2 v_3)\quad\text{for all }v_1, v_2, v_3 \in V[1]. $$
\end{Definition}

\begin{Proposition}[Twisted boundary operator $\OPQ^\PMC_{110}$ and $\AInfty$-cyclic cohomology]\label{Prop:CyclicHom}
In the setting of Definition~\ref{Def:MukDef}, the triple $\mathcal{A}_\PMC(V) \coloneqq (V,\Pair, (\mu_k))$ is a cyclic $\AInfty$-algebra. We always have $\mu_1 = m_1$, and if $\PMC$ is compatible with $\MC$ for a cyclic dga $(V,\Pair,m_1,m_2)$, then also $\mu_2 = m_2$. 

The following holds for the homologies:
\begin{equation*}
 \HIBL^\PMC(\CycC(V))=  r(\H^*_\lambda(\mathcal{A}_\PMC(V);\R))[3-n].
\end{equation*}
\end{Proposition}

\begin{proof}
First of all, according to Definition~\ref{Def:MaurerCartan} we must have $\Norm{\PMC_{10}} > 2$, and hence 
$$ \PMC_{10}(\Susp v_1 v_2) = \PMC_{10}(\Susp v_1) = 0 \quad\text{for all }v_1, v_2 \in V[1].  $$
This implies $\mu_1 = m_1$.

Now, let $e_0$, $\dotsc$, $e_m$ be a basis of $V[1]$ and let $e^0$, $\dotsc$, $e^m$ be the dual basis with respect to $\Pair$. For all $k\ge 2$ and $v_1$, $\dotsc$, $v_k \in V[1]$, we compute the following:
\allowdisplaybreaks
\begin{align*}
& \Pair(\mu_k(v_1,\dotsc,v_k),v_{k+1}) \\
&\qquad = (-1)^{n-3}\sum_{ij} (-1)^{\Abs{e_i}}\Pair(e^i, e^j) \PMC_{10}(\Susp e_i v_1\dots v_k) \Pair(e_j, v_{k+1}) \\
&\qquad  \underset{\mathclap{\substack{\uparrow\rule{0pt}{1.5ex}\\\forall v\in V[1]: \ \sum_j \Pair(v,e^j)e_j = v}}}{=} (-1)^{n-3}\sum_i (-1)^{\Abs{e_i}} \PMC_{10}(\Susp e_i v_1\dotsc v_k) \Pair(e^i,v_{k+1}) \\
&\qquad \underset{\mathclap{\substack{\uparrow\rule{0pt}{1.5ex}\\(\Abs{v}_1 + \dotsb + \Abs{v_k}) \Abs{e_i} = \\ (\Abs{\PMC_{10}} + \Abs{s}+\Abs{e_i})\Abs{e_i} = (\Abs{s}+1)\Abs{e_i}}}}{=} (-1)^{n-3}\sum_i (-1)^{\Abs{e_i}(n-3)} \PMC_{10}(\Susp v_1 \dotsc v_k e_i) \Pair(e^i,v_{k+1}) \\
&\qquad \underset{\mathclap{\substack{\big\uparrow\rule{0pt}{2.5ex}\\ 1+\Abs{v_{k+1}}\Abs{e^i} = 1+(\Abs{e}^i + 2 - n)\Abs{e^i} \\= 1+(3-n)\Abs{e^i} = 1+(3-n)\Abs{e_i}}}}{=}  (-1)^{n-2}\sum_i \PMC_{10}(\Susp v_1 \dotsc v_k e_i) \Pair(v_{k+1},e^i) \\
&\qquad = (-1)^{n-2}\PMC_{10}(\Susp v_1 \dotsc v_{k+1}).
\end{align*}
Therefore, we have
\begin{equation*}
 \PMC_{10} = (-1)^{n-2}\sum_{k\ge 2} \mu_k^+.
\end{equation*}
In this case,~\cite[Proposition 12.3]{Cieliebak2015} asserts that the $\AInfty$-relations~\eqref{Eq:AInftyDef} for $(\mu_k)_{k\ge 1}$ are equivalent to the ``lowest'' Maurer-Cartan equation~\eqref{Eq:MCEq} for $\PMC_{10}$. 
The degree condition $\Abs{\mu_k}=1$ and the cyclic symmetry of $\mu_k^+$ are easy to check. Therefore, $\mathcal{A}_\PMC(V)$ is a cyclic $\AInfty$-algebra.

As for the compatibility with $\MC$, we have for all $v_1$, $v_2\in V[1]$ the following:
\begin{equation*}
 \begin{aligned}
  m_2(v_1, v_2) & = \sum_i \Pair(e_i, m_2(v_1,v_2)) e^i \\
 &\underset{\mathclap{\substack{\big\uparrow\rule{0pt}{2.5ex} \\ T^{ij} = (-1)^{\Abs{e_i}} \Pair(e^i,e^j)}}}{=}  \sum_{i,j} (-1)^{\Abs{e_i}} T^{ij} \Pair(e_i,m_2(v_1,v_2))e_j \\
 &\underset{\mathclap{\substack{\big\uparrow\rule{0pt}{2.5ex}\\ \Pair(v_1,v_2) = (-1)^{1+(n-3)\Abs{v_1}} \Pair(v_2,v_1)}}}{=} \sum_{i,j} (-1)^{1 + (n-2)\Abs{e_i}}T^{ij} \underbrace{\Pair(m_2(v_1,v_2),e_i)}_{\displaystyle{\mathclap{=(-1)^{n-2}\MC_{10}(\Susp v_1 v_2 e_i)}}} e_j\\
 & \underset{\mathclap{\substack{\big\uparrow\rule{0pt}{2.5ex} \\  (\Abs{v_1} + \Abs{v_2})\Abs{e_i} =(\Abs{\MC_{10}} - \Abs{s} - \Abs{e_i})\Abs{e_i}\\ = (n-2)\Abs{e_i} }}}{=} (-1)^{n-3} \sum_{i,j} T^{ij} \MC_{10}(\Susp e_i v_1 v_2)e_j \\ 
 &= \mu_2(v_1,v_2).
\end{aligned}
\end{equation*}

We will now clarify the relation to the cyclic cohomology of $\mathcal{A}_\PMC(V)$. Recall from Proposition~\ref{Prop:dIBL} that $\OPQ_{110}^\PMC(\Psi) = \OPQ_{110}(\Psi) + \OPQ_{210}(\PMC_{10}, \Psi)$ for $\Psi\in\CDBCyc V[3-n]$, where the first term is given by~\eqref{Eq:Diff} and the second by~\eqref{Eq:TwistDif}. Consider now ${\Hd'}^k$ and $R^k$ from~\eqref{Eq:bRH}, whose sum gives the Hochschild boundary operator $\Hd$. Using the cyclic symmetry, we can rewrite a summand of ${\Hd'}^k$ for $j=1$, $\dotsc$, $k$ and $i=0$, $\dotsc$, $k-j$ applied to a generating word $v_1\dots v_k \in \BCyc V$ as follows: 
\begin{equation}\label{Eq:Tmp1}
\begin{aligned}
& [t^i_{k-j+1} \circ (\mu_j \otimes \Id^{k-j}) \circ t_k^{-i}](\underbrace{v_1 \dots v_k}_{\eqqcolon w}) = \\
&\quad =(-1)^{\Abs{v_1} + \dotsb + \Abs{v_{i}}} v_1 \dots v_i \mu_j(v_{i+1} \dots v_{i+j})v_{i+j+1} \dots v_k \\[\jot]
&\quad = \varepsilon(w\mapsto w^1 w^2) \mu_j(\underbrace{v_{i+1}\dots v_{i+j}}_{\eqqcolon w^1}) \underbrace{v_{i+j+1} \dots v_k v_1 \dots v_i}_{\eqqcolon w^2}
\end{aligned}
\end{equation}
Clearly, summing \eqref{Eq:Tmp1} over $j=1$ and $i=0$, $\dotsc$, $k-1$ gives the dual to $\OPQ_{110}$. For $j=2$, $\dotsc$, $k$, we can write \eqref{Eq:Tmp1} as
$$ 
(-1)^{n-3} \sum_{i,j} \varepsilon(w\mapsto w^1 w^2) T^{ij}\PMC_{10}(\Susp e_i w^1) e_j w^2.
$$
Therefore, the sum over $j=2$, $\dotsc$, $k$ and $i=0$, $\dotsc$, $k-j$ gives the part of the dual to $\OPQ_{210}(\PMC_{10},\Psi)$ corresponding to the cyclic permutations $\sigma\in \Perm_{k}$ with $\sigma_1 = 1$, $j+1$, $\dotsc$, $k$. The rest, i.e., the cyclic permutations with $\sigma_1 = 2$, $\dotsc$, $j$, is obtained analogously from the summands $(\mu_j \otimes \Id^{k-j})\circ t_k^i$ of $R^k$ for $j=2$,~$\dotsc$,\,$k$ and $i=1$, $\dotsc$, $j-1$. We conclude that $\OPQ_{110}^\PMC: \CDBCyc V[3-n] \rightarrow \CDBCyc V[3-n]$ is a degree shift of $\Hd^*: \CDBCyc V \rightarrow \CDBCyc V$. As for the gradings, we have:
$$ \begin{aligned}
r(D_\lambda(V))[3-n]^i &= r(D_\lambda(V))^{i+3-n} = (D_\lambda(V))^{-i-3+n} = (\CDBCyc V)^{i+3-n-1} \\
 &= \CDBCyc V[2-n]^i.
\end{aligned}$$
This finishes the proof.
\end{proof}

We will now turn to units and augmentations.

\begin{Definition}[Reduced canonical $\dIBL$-algebra]\label{Def:ReduceddIBL}
Let $(V, \Pair, m_1, \NOne, \varepsilon)$ be an augmented cyclic cochain complex of degree $2-n$ from Definition~\ref{Def:AugUnit}. We define the space of \emph{reduced cyclic cochains} on $V$ by
$$ \CycC_{\RedMRM}(V)\coloneqq \RedDBCyc V[2-n]. $$
We define the \emph{reduced canonical $\dIBL$-algebra} by
$$ \dIBL(\RedCycC(V))\coloneqq (\RedCycC(V), \OPQ_{110}, \OPQ_{210}, \OPQ_{120}), $$
where $\OPQ_{110}$, $\OPQ_{210}$, $\OPQ_{120}$ are restrictions of the operations of $\dIBL(\CycC(V))$.
\end{Definition}

\begin{Definition}[Strictly reduced Maurer-Cartan element] \label{Def:StrictlyReduced}
In the setting of Definition~\ref{Def:ReduceddIBL}, we call a Maurer-Cartan element $\PMC=(\PMC_{lg})$ for $\dIBL(\CycC(V))$ \emph{strictly reduced} if $\PMC_{lg}\in \hat{\Ext}_l \RedCycC(V)$ for all $(l,g)\neq (1,0)$ and if the $\AInfty$-algebra $(\mathcal{A}_\PMC(V),\NOne,\varepsilon)$ induced by $\PMC_{10}$ is strictly unital and strictly augmented.
Given a strictly reduced Maurer-Cartan element $\PMC$, we can define the twisted $\IBLInfty$-algebra
$$ \dIBL^\PMC(\CycC_{\mathrm{red}}(V)) \coloneqq (\CycC_{\mathrm{red}}(V), (\OPQ_{klg}^\PMC)), $$
where $\OPQ_{klg}^\PMC$ are the restrictions of the operations of $\dIBL^\PMC(\CycC(V))$. We denote the homology of $\dIBL^\PMC(\RedCycC)$ by $\HIBL^\PMC(\RedCycC)$ or $\HIBL^{\PMC,\mathrm{red}}(\CycC)$.\footnote{The latter option suggests that it might be possible to define the reduced homology with the induced $\IBL$-algebra even if $\PMC$ is not strictly reducible, e.g., if $(\mathcal{A}_\PMC(V),\NOne,\varepsilon)$ is only homologically unital and augmented.}
\end{Definition}

\Modify[caption={DONE Put remark to text}]{Include this remark to text because it is in fact a definition.}

\begin{Remark}[On strictly reduced Maurer-Cartan element]\phantomsection
\begin{RemarkList}
\item We can say that the $\IBLInfty$-algebra $\dIBL^\PMC(\CycC_{\mathrm{red}})$ is a \emph{subalgebra} of $\dIBL^\PMC(\CycC)$, which means that the inclusion $\CycC_{\mathrm{red}}\xhookrightarrow{}\CycC$ induces the following commutative diagram for all $k,l\ge 1$, $g\ge 0$:
$$\begin{tikzcd}
\hat{\Ext}_k \CycC \arrow{r}{\OPQ_{klg}^\PMC} & \hat{\Ext}_l \CycC \\
\hat{\Ext}_k \CycC_{\mathrm{red}} \arrow{r}{\OPQ_{klg}^\PMC}\arrow[hook]{u} & \hat{\Ext}_l \CycC_{\mathrm{red}}.\arrow[hook]{u}
\end{tikzcd}$$
We denote this fact by $\dIBL^\PMC(\CycC_{\mathrm{red}})\subset \dIBL^\PMC(\CycC)$.

\item The canonical Maurer-Cartan element $\MC$ of a strictly augmented strictly unital dga $(V,m_1,m_2,\NOne,\varepsilon)$ is strictly reduced (this follows from Proposition~\ref{Prop:CyclicHom}).

\item In the situation of Definition~\ref{Def:StrictlyReduced}, we denote 
$$ \bar{V}[1]\coloneqq \Ker(\varepsilon),$$
so that $V=\bar{V}\oplus \langle 1 \rangle$. We use  the canonical projection $\pi: V \rightarrow \bar{V}$ to identify $\CDBCyc \bar{V} \xrightarrow{\simeq} \CRedDBCyc V$ via the componentwise pullback $\pi^*$. In this way, we obtain the $\IBLInfty$-algebras $\dIBL(\CycC(\bar{V}))$ and $\dIBL^\PMC(\CycC(\bar{V}))$, which are isomorphic to $\dIBL(\CycC_{\mathrm{red}}(V))$ and $\dIBL^\PMC(\CycC_{\mathrm{red}}(V))$, respectively.
\qedhere
\end{RemarkList}
\end{Remark}

In the following list, we sum up our main reasons for considering units, augmentations and reduced Maurer-Cartan elements. Suppose that we are in the situation of Definition~\ref{Def:StrictlyReduced}, then:
\begin{itemize}
 \item  Proposition~\ref{Prop:Reduced} implies the splitting
\begin{equation}\label{Eq:IBLSPlit}
 \HIBL^\PMC(\CycC)[1] = \HIBL^\PMC(\CycC_{\mathrm{red}})[1] \oplus \langle \Susp \NOne^{2q-1 *} \mid q\in \N \rangle.
\end{equation}
Here $\NOne^{i*} \in \DBCyc V$ is the componentwise pullback $\varepsilon^*$ of $\NOne^{i*}\in \DBCyc(\R)$. To get this, we used
\begin{equation*} 
\HIBL^\PMC(\CycC_{\mathrm{red}}) =  r(\RedCycCoH^*(\mathcal{A}_\PMC))[3-n],
\end{equation*}
which can be seen by redoing the proof of Proposition~\ref{Prop:CyclicHom} with reduced cochains.
\item The subalgebra $\dIBL^\PMC(\CycC_{\mathrm{red}})\subset \dIBL^\PMC(\CycC)$ induces the subalgebra 
$$\IBL(\HIBL^\PMC(\CycC_{\mathrm{red}}))\subset \IBL(\HIBL^\PMC(\CycC)), $$
and any higher operation $\OPQ_{1lg}^\PMC$ which induces a map $\hat{\Ext}_1 \HIBL(\CycC) \rightarrow \hat{\Ext}_l \HIBL(\CycC)$ induces a map $\hat{\Ext}_1\HIBL(\CycC_{\mathrm{red}}) \rightarrow \hat{\Ext}_l\HIBL(\CycC_{\mathrm{red}})$ as well.
\item If $V$ is non-negatively graded, connected and simply-connected, then we have $\hat{\Ext}_k \CycC_{\mathrm{red}} \simeq \Ext_k \RedCycC$ for all $k\in \N_0$ by Proposition~\ref{Prop:SimplCon}, and hence $\dIBL^\PMC(\RedCycC)$ is completion-free. 
\end{itemize}

\begin{Proposition}[Operations on units]\label{Prop:Ones}
Suppose that $(V,\Pair,m_1,\NOne,\varepsilon)$ is a finite-dimensional augmented cyclic cochain complex of degree $2-n$ such that $n\ge 1$, and let~$\PMC$ be a strictly reduced Maurer-Cartan element for $\dIBL(\CycC(V))$. The following relations are the only relations containing $\NOne^{i*}$ which may be non-zero on the homology $\HIBL^\PMC(\CycC)$: 

For all $\Psi \in \RedCycC(V)$ and $l\ge 1$, $g\ge 0$, we have
\begin{align*}
\OPQ_{210}(\Susp\NOne^* \otimes \Psi) &= (-1)^{(n-2)\Abs{\Psi}} \OPQ_{210}(\Psi \otimes \Susp\NOne^*)  = (-1)^{n-2}\Psi \circ \iota_\NVol\quad\text{and} \\[\jot]
\OPQ_{1lg}^\PMC(\Susp\NOne^*) & = - \PMC_{lg} \circ \iota_\NVol,
\end{align*}
where $\iota_\NVol$ is defined as follows:
\begin{itemize}
\item The element $\NVol \in V[1]$ is the unique vector such that $\Pair(\NOne,\NVol) = 1$ and $\NVol \perp \bar{V}[1]$ with respect to $\Pair$. Note that $\Abs{\NVol} = n-1$ and that such $\NVol$ always exists due to non-degeneracy.
\item We start by defining $\iota_\NVol^0 : \BCyc V \rightarrow \BCyc V$ by
$$ \iota_\NVol^0(v_1\dots v_k) \coloneqq \sum_{i=1}^k (-1)^{\Abs{\NVol}(\Abs{v_1} + \dotsb + \Abs{v_{i-1}})} v_1 \dots v_{i-1} \NVol v_i \dots v_k $$
for all generating words $v_1 \dots v_k \in \BCyc V$. Next, for all $k\ge 1$, we define $\iota_\NVol : (\BCyc V)^{\otimes k} \rightarrow (\BCyc V)^{\otimes k}$ by
$$ \iota_\NVol(w_1 \otimes \dotsb \otimes w_k) \coloneqq  \begin{multlined}[t](-1)^{\Abs{\NVol}k}\sum_{j=1}^k (-1)^{\Abs{\NVol}(\Abs{w_1} + \dotsb + \Abs{w_{j-1}})} w_1 \otimes \dotsb \otimes w_{j-1} \\ \otimes  \iota_\NVol^0(w_j) \otimes w_{j+1} \otimes \dotsb \otimes w_k \end{multlined}$$
for all generating words $w_1$, $\dotsc$, $w_k \in \BCyc V$. Finally, we take the degree shift $\iota_\NVol: (\BCyc V[3-n])^{\otimes k} \rightarrow (\BCyc V[3-n])^{\otimes k}$ according to the degree shift convention~\eqref{Eq:DegreeShiftConv}. 
\end{itemize}
\end{Proposition}
\begin{proof}
Pick a basis $(e_0, \dotsc, e_m)$ of $V[1]$ such that $e_0 = \NOne$ and $\bar{V}[1] = \langle e_1, \dotsc, e_m \rangle$. If $(e^0,\dotsc,e^m)$ is the dual basis, then we have $\NVol = e^0$. We will often use the following relation:
\begin{equation}\label{Eq:NiceFormula}
 \sum_{j=0}^m T^{1j} e_j = \sum_{j=0}^m (-1)^{\Abs{\NOne}} \Pair(\NVol,e^j) e_j =  - \NVol.
\end{equation}
We consider only those generating words $w = v_1 \dotsc v_k$ of $\BCyc V$ with either $v_i\in \bar{V}$ for each $i$ (shortly $w\in \BCyc \bar{V}$) or $v_i = \NOne$ for each $i$ with $k$ odd (i.e., $w=\NOne^{2j-1}$ for some $j$). Let $w_1$, $\dotsc$, $w_l$ with $w_j = v_{j 1} \dots v_{j k_j}$ denote such generating words. Clearly, if $\Phi \in \hat{\Ext}_l \CycC(V)$ is a $\OPQ_{110}^\PMC$-closed element which vanishes on all $w_1 \otimes \dotsb \otimes w_l$, then~\eqref{Eq:IBLSPlit} implies that $[\Phi] = 0$ in $\hat{\Ext}_l\HIBL(\CycC)$.

For $\Psi\in \RedCycC(V)$ and $q\ge 1$ odd, we compute using \eqref{Eq:NiceFormula} the following:
\allowdisplaybreaks
\begin{align*}
\OPQ_{210}(\Susp^2 \NOne^{q*}\otimes \psi)(\Susp w)  &= \sum \varepsilon( w\mapsto w^1 w^2)(-1)^{(n-1)\Abs{w^1}}T^{1j}\NOne^{q*}(\NOne w^1) \psi(e_j w^2) \\[\jot]
&= -\sum \varepsilon( w\mapsto w^1 w^2)(-1)^{(n-1)\Abs{w^1}}\NOne^{q*}(\NOne w^1) \psi(\NVol w^2)\\[\jot]
&\eqqcolon(*). 
\end{align*}
Now, in order to get $(*)\neq 0$, we need either $q=1$ and $w\in \BCyc \bar{V}$, in which case
\allowdisplaybreaks
\begin{align*}
 (*)&=-\sum \varepsilon(w\mapsto \underbrace{w^1}_{=\emptyset} w^2)\psi(\NVol w^2) \\ &= -\sum_{j=1}^k (-1)^{\Abs{v}(\Abs{v_1} + \dotsb + \Abs{v_{j-1}})} \psi(v_1 \dotsc v_{j-1} \NVol v_{j+1}\dotsc v_k) \\[\jot]
 &=  - (\psi \circ \iota^0_\NVol)(w) \\
 &= (-1)^{n-2} (\Psi \circ \iota_\NVol)(\W),
\end{align*}
or $q>1$ odd and $w = \NOne^{q-1}$, in which case
\allowdisplaybreaks
\begin{align*}
 (*) &= \sum \varepsilon(w\mapsto w^1 \underbrace{w^2}_{=\emptyset}) \NOne^{q*}(\NOne^q) \psi(\NVol) \\
 &= \psi(\NVol) \sum_{j=1}^{q-1} (-1)^j \\
 & = 0.
\end{align*}

Next, because $n\ge 1$, we get $T^{11} = 0$, and hence 
$$ \OPQ_{120}(\NOne^{q*}) = 0\quad\text{for all }q\in \N$$
on the chain level. Therefore, we have $\OPQ_{1lg}^\PMC = \OPQ_{210}\circ_1 \PMC_{lg}$ for all $l\ge 1$, $g\ge 0$, and  using Proposition~\ref{Prop:Formulafortwisted} and \eqref{Eq:NiceFormula}, we obtain 
\allowdisplaybreaks
\begin{align*}
&[(\OPQ_{210}\circ_1 \PMC_{lg})(\NOne^{q*})](\W_1 \otimes \dotsb \otimes \W_l) \\[\jot]
&\quad= \begin{multlined}[t]-\smash{\sum_{j=1}^l} \sum \varepsilon' \varepsilon(w_j \mapsto w_j^1 w_j^2) \NOne^{q*}(\NOne w_j^1) \PMC_{lg}(\W_1 \otimes \dotsb \otimes \W_{j-1} \otimes (\Susp \NVol w_j^2) \\[\jot] \otimes \W_{j+1}\otimes \dotsb \otimes \W_l)\end{multlined}\\
&\quad\eqqcolon(**).
\end{align*}
In order to get $(**)\neq 0$, we need either $q=1$ and $w_j \in \BCyc \bar{V}$ for all $j$, in which case
\allowdisplaybreaks
\begin{align*}
(**) &= \begin{multlined}[t]-\smash{\sum_{j=1}^l} \smash{\sum_{i=1}^{k_j}} (-1)^{\Abs{\NVol}(\Abs{\W_1}+ \dotsb + \Abs{\W_{j-1}} + \Abs{\Susp})}(-1)^{\Abs{\NVol}(\Abs{v_1} + \dotsb + \Abs{v_{i-1}})}\PMC_{lg}(\W_1 \otimes \dotsb \\[\jot] \otimes \W_{j-1} \otimes (\Susp v_1 \dots v_{i-1} \NVol v_{i} \dots v_{k_j}) \otimes \W_{j+1}\otimes \dotsb\otimes \W_l)\end{multlined} \\[\jot]
 &= -(\PMC_{lg} \circ \iota_\NVol)(\W_1 \otimes \dotsb \otimes \W_l),
\end{align*}
or $q>1$ odd and $w_j = \NOne^{q-1}$ for some $j$, in which case
\allowdisplaybreaks
\begin{align*}
(**) &=\begin{multlined}[t]-\smash{\sum_{\substack{1\le j \le l \\ w_j = \NOne^{q-1}}}} \varepsilon' \Bigl(\sum \varepsilon(w_j \mapsto w_j^1 \underbrace{w_{j}^2}_{=\emptyset} \NOne^{q*}(\NOne w_j^1)) \Bigr) \PMC_{lg}(\W_1\otimes \dotsb \\ \otimes\W_{j-1}\otimes (\Susp\NVol)\otimes \W_{j+1} \otimes \dotsb \otimes \W_l )\end{multlined}\\[\jot]
&=\begin{multlined}[t]-\smash{\sum_{\substack{1\le j \le l \\ w_j = \NOne^{q-1}}}} \varepsilon' \smash{\underbrace{\Bigl( \sum_{i=1}^{q-1} (-1)^i \Bigr)}_{=0}} \PMC_{lg}(\W_1 \otimes \dotsb \otimes \W_{j-1} \otimes (\Susp \NVol) \otimes \W_{j+1} \otimes \\[\jot] \dotsb \otimes \W_l)\end{multlined}\\[\jot]
&=0.
\end{align*}

The only relation left to check is
$$ \OPQ_{210}(\Susp \NOne^{q_1 *},\Susp \NOne^{q_1 *}) = 0 \quad\text{for all } q_1, q_2 \in \N.$$
However, this is easy to see, and the proof is done. 
\end{proof}

\begin{Remark}[Finite type]
Everything in this section works if $V$ is just of \emph{finite type}, i.e., if $\dim V^d < \infty$ for all $d\in \Z$. The only difference is that $T$ is not a tensor in $V[1]^{\otimes 2}$ anymore.
\end{Remark}

\chapter{Chern-Simons Maurer-Cartan element and string topology}

In Section~\ref{Sec:Manifold1}, we consider the cyclic dga's $\DR(M)$, $\H_{\mathrm{dR}}(M)$ and $\Harm(M)$ for a closed oriented $n$-manifold $M$ (Proposition~\ref{Prop:DGAs}) and apply the theory from Section~\ref{Sec:Alg3} to the last two, which are finite-dimensional.

In Section~\ref{Section:Proof1}, we define the admissible Hodge propagator $\Prpg$ (Definition~\ref{Def:GreenKernel}).
It is a primitive to the Schwartz kernel $\HKer$ of the harmonic projection~$\pi_\Harm$ (see Proposition~\ref{Lemma:HKer}) outside the diagonal and extends smoothly to the spherical blow-up of the diagonal. 
These ideas come from an early version of~\cite{Cieliebak2018}.
We consider conditions (P1)--(P5) on a linear operator $\Htp$ and its Schwartz kernel $\Prpg$ (see p.~\pageref{ConditionsG}) and show that $\Prpg$ satisfying all these conditions always exists (Proposition~\ref{Prop:ExistenceG}).
We also mention the standard Hodge propagator $\PrpgStd$ (see \eqref{Eq:GStdStd}), which might be a canonical Hodge propagator satisfying (P1)--(P5). 

In Section~\ref{Section:Proof2}, we review ribbon graphs, labelings, compatibility of the order and orientation of internal edges, and the edge and vertex order from~\cite{Cieliebak2015} (Definitions~\ref{Def:Ribbon}, \ref{Def:Labeling}, \ref{Def:CompatLabeling} and~\ref{Def:EdgeVertex}).
We then define $\PMC$ as a signed sum of integrals of products of Hodge propagators and harmonic forms which are associated to labeled trivalent ribbon graphs (Definition~\ref{Def:PushforwardMCdeRham}).
We do not show that $\PMC$ satisfies the Maurer-Cartan equation, but we do show all other properties of a Maurer-Cartan element (Lemma~\ref{Lem:MCCond} and Proposition~\ref{Prop:FormalPushforwardProp}).
We define the $Y$-graph, trees, circular graphs, vertices of types $A$, $B$, $C$ and their contributions $A_{\alpha_1, \alpha_2}$, $B_\alpha$, $C$, respectively (Definitions~\ref{Def:Graphs} and~\ref{Def:Contributions}). 

In Section~\ref{Sec:Vanishing}, we observe that vanishing of some subintegrals associated to special configurations of vertices in a graph implies $\PMC_{lg} = \MC_{lg}$.
For example, if all configurations with $\NOne$ at an external vertex vanish, which holds if $\Prpg$ satisfies (P4) and (P5) (Proposition~\ref{Prop:COne}), then all higher operations $\OPQ_{1lg}^\PMC$ vanish on the chain level in dimensions $n>3$ (Proposition~\ref{Prop:PMCEqualsMC}).
Next, if all configurations with an $A$-vertex vanish, then $\PMC_{10}=\MC_{10}$, and hence $\OPQ_{110}^\PMC = \OPQ_{110}^\MC$ (Proposition~\ref{Prop:Avertexvanish}).
We show that $\PMC = \MC$ for simply-connected geometrically formal manifolds with $n\neq 2$ (Proposition~\ref{Prop:GeomForm}).
Using the results of \cite{Cieliebak2018}, we argue that the chain complexes of $\OPQ_{110}^\MC$ and~$\OPQ_{110}^\PMC$ are quasi-isomorphic provided~$M$ is simply-connected and formal (Proposition~\ref{Prop:Formal}).

In Section~\ref{Sec:StringTopology}, we recall basic facts about the Chas-Sullivan operations $\StringOp_2$ and $\StringCoOp_2$ on the $\Sph{1}$-equivariant homology of the free loop space and formulate a version of the string topology conjecture for simply-connected manifolds (Conjecture~\ref{Conj:StringTopology}).
\Correct[caption={DONE Mistake in reduced notation}]{Correct the notation here, if it is not already corrected.
Anyway, define something like RedNTwistHIBL and replace the construction.}

\section{Canonical dIBL-structures on cyclic cochains of de Rham cohomology} \label{Sec:Manifold1}
\allowdisplaybreaks
Let $M$ be an oriented closed Riemannian manifold of dimension $n$.
We consider the following graded vector spaces:
$$\begin{aligned}
\DR^*(M) &\;\,\dots && \text{smooth de Rham forms}, \\
\Harm^*(M) &\;\,\dots &&\text{harmonic forms}, \\
\H_{\mathrm{dR}}^*(M) &\;\,\dots && \text{de Rham cohomology}.
\end{aligned}$$
Since $M$ is fixed, we often write just $\DR$, $\Harm$ and~$\HDR$.
We consider the Hodge decomposition $\DR=\Harm\oplus\Im\Dd\oplus\Im\CoDd$, where~$\Dd$ is the de Rham differential and~$\CoDd$ the codifferential.
We call the corresponding projection
\begin{equation}\label{Eq:HarmProj}
\pi_\Harm: \DR^*(M) \longrightarrow \Harm^*(M) 
\end{equation}
the \emph{harmonic projection} and the induced isomorphism $\pi_\Harm: \H_{\mathrm{dR}} \rightarrow \Harm$ mapping a cohomology class into its unique harmonic representative the \emph{Hodge isomorphism}.

\begin{Notation}[Updated notation for bar complexes] \label{Def:DeRham}
We use Notation~\ref{Def:Notation} for $V=\DR{}$, $\Harm{}$, $\H_{\mathrm{dR}}{}$ and $A=n-3$ with the following changes:
$$\tilde{v}\sim \eta \in V,\quad v \sim \alpha\in V[1],\quad w\sim \omega \in \BCyc V, \quad \W\sim \Omega\in \BCyc V[n-3]. $$
We use the formal symbols $\Susp$ and $\SuspU$ with $\Abs{\Susp} = n - 3$ and $\Abs{\SuspU} = -1$, so that $\alpha = \SuspU \eta$ and $\Omega = \Susp \omega$.
\end{Notation}

\begin{Proposition}[De Rham cyclic dga's]\label{Prop:DGAs}
Let $M$ be an oriented closed Riemannian manifold of dimension $n$. The quadruple $(\DR(M), \Pair, m_1, m_2)$ with the operations from~\eqref{Eq:DeRhamDGA} is a cyclic dga of degree $2-n$.
For the operations before the degree shift, we have
\begin{align*}
\tilde{m}_1(\eta_1) & = \Dd \eta_1, \\[\jot]
\tilde{m}_2(\eta_1, \eta_2) &= \eta_1 \wedge \eta_2, \\ 
\tilde{\Pair}(\eta_1, \eta_2) &=  \int_M \eta_1 \wedge \eta_2 \eqqcolon (\eta_1,\eta_2),
\end{align*}
where $\Dd$ is the de Rham differential, $\wedge$ the wedge product and $\tilde{\mathcal{P}}$ the \emph{intersection pairing}. 
The operations restrict to $\H_{\mathrm{dR}}(M)$ and make $(\H_{\mathrm{dR}}(M),\Pair,m_1\equiv 0,m_2)$ into a cyclic dga.
If we define $\mu_1 \equiv 0$ and
\begin{equation}\label{Eq:HarmProd}
\mu_2(\alpha_1, \alpha_2) \coloneqq \pi_{\Harm}(m_2(\alpha_1, \alpha_2))\quad \text{for all }\alpha_1, \alpha_2 \in \Harm(M)[1],
\end{equation}
then $(\Harm(M), \Pair, \mu_1, \mu_2)$ is a cyclic dga as well, and $\pi_\Harm: \H_{\mathrm{dR}} \rightarrow \Harm$ is an isomorphism of cyclic dga's.
All three dga's $\DR$, $\H_{\mathrm{dR}}$ and $\Harm$ are strictly unital and strictly augmented with the unit $\NOne\coloneqq \SuspU 1 \in \DR[1]$, where $1$ is the constant one.
\end{Proposition}

\begin{proof}
The relations \eqref{Eq:CycDGA} follow from the classical properties of $\Dd$ and $\wedge$ and from the Stokes' theorem for oriented closed manifolds.
The Poincar\'e duality asserts that $(\cdot,\cdot)$ is non-degenerate on $\H_{\mathrm{dR}}{}$ and~$\Harm{}$, and thus they are cyclic dga's as well.
The fact that $\pi_\Harm : \H_{\mathrm{dR}}{}\rightarrow \Harm{}$ is an isomorphism of vector spaces follows from the Hodge theory.
As for compatibility with the product, given $\eta_1$, $\eta_2 \in \Harm{}$, then $\eta_1 \wedge \eta_2$ is closed, and since $\Ker \Dd = \Harm \oplus \Im \Dd$, we see that $\pi_{\Harm}(\eta_1 \wedge \eta_2) = \eta_1 \wedge \eta_2 + \Dd \eta$ for some $\eta\in \DR$ is a harmonic representative of the cohomology class $[\eta_1 \wedge \eta_2] = [\eta_1] \wedge [\eta_2]$.
Unitality is obvious, and the construction of an augmentation map clear.
Note that a strict augmentation for $\DR(M)$ is the evaluation at a point, for instance. 
\end{proof}

The facts (A) and (C) from the Overview apply to the cyclic dga's $\Harm{}$ and~$\H_{\mathrm{dR}}{}$ (not to $\DR{}$ because it is infinite-dimensional!), and we get the canonical $\dIBL$-algebras $\dIBL(\CycC(\Harm))$ and $\dIBL(\CycC(\H_{\mathrm{dR}}))$ of bidegrees $(n-3,2)$ with the canonical Maurer-Cartan element $\MC = (\MC_{10})$.
The Hodge isomorphism induces an isomorphism of these $\dIBL$-algebras, and hence we can use $\Harm$ and~$\HDR$ interchangeably.
The reduced versions (see Definition~\ref{Def:ReduceddIBL}) also satisfy $\dIBL(C_{\mathrm{red}}(\Harm)) \simeq \dIBL(C_{\mathrm{red}}(\HDR))$, $\dIBL^\MC(C_{\mathrm{red}}(\Harm)) \simeq \dIBL^{\MC_{\mathrm{dR}}}(C_{\mathrm{red}}(\HDR))$.
We have $\OPQ_{110} \equiv 0$, and hence $\dIBL(\CycC(\Harm))$ is, in fact, an $\IBL$-algebra.
However, we will denote it by $\dIBL$ and call it a $\dIBL$ algebra as a reminder of the canonical $\dIBL$-structure.
The canonical Maurer-Cartan element~$\MC$ satisfies
\begin{equation}\label{Eq:CanonMC}
\MC_{10}(\Susp \alpha_1 \alpha_2 \alpha_3) = (-1)^{n-2 + \eta_2} \int_M \eta_1 \wedge \eta_2 \wedge \eta_3 \quad \text{for all } \alpha_1, \alpha_2, \alpha_3 \in \Harm[1].
\end{equation}
We get the canonical twisted $\dIBL$-algebra $\dIBL^\MC(\CycC(\Harm))$ from~\eqref{Eq:CanonMCTwist} with, in general, non-trivial boundary operator $\OPQ_{110}^\MC$ whose homology is the cyclic homology of $\HDR$ up to degree shifts.

\section{Hodge propagator} \label{Section:Proof1}
\allowdisplaybreaks
\Correct[inline,caption={DONE Notation for piH}]{We should define the special deformation retract by requiring only $\Htp \iota_\Harm = 0$!! The notation $\Htp \pi_\Harm = 0$ does not make much sense! It makes.}

We will use fiberwise integration and spherical blow-ups, which we now recall.

\begin{Definition}[Fiberwise integration] \label{Def:FibInt}
Let $\Pr: E \rightarrow B$ be a smooth oriented fiber bundle with an oriented fiber $F$ over an oriented manifold~$B$ with $\Bdd B = \emptyset$.
We orient $E$ as $F\times B$.
Let 
$\DR_{\mathrm{c}} (B)$ denote the space of forms with compact support and $\DR_{\mathrm{cv}}(E)$ the space of forms with compact vertical support.
For any $\kappa\in \DR_{\mathrm{cv}}(E)$, let $\FInt{F} \kappa \in \DR(B)$ be the unique smooth form such that
\begin{equation*} \label{Eq:FiberInt}
\int_E \kappa \wedge \Pr^*\eta = \int_B \Bigl(\FInt{F}\kappa \Bigr) \wedge \eta\quad\text{for all }\eta\in \DR_{\mathrm{c}} (B).
\end{equation*}
\end{Definition}
 
\begin{Definition}[Spherical blow-up]\label{Def:SphBlow}Let $X$ be a smooth $n$-dimensional manifold and $Y\subset X$ a smooth $k$-dimensional submanifold.
The \emph{blow-up} of $X$ at $Y$ is as a set defined by
\[ \Bl_Y X \coloneqq X\backslash Y \sqcup P^+N Y, \]
where $P^+N Y$ is the real oriented projectivization of the normal bundle $NY$ of~$Y$ in $X$.
This means that $P^+NY$ is the quotient of $\{ v\in NY\,
\vert\, v\neq 0\}$ by the relation $v\sim a v$ for all $a\in (0,\infty)$.
The \emph{blow-down map} is defined by
\begin{align*}
 \pi : \Bl_Y X &\longrightarrow X \\
     p\in X\backslash Y &\longmapsto p, \\
     [v]_p\in P^+N Y &\longmapsto p.
\end{align*}
\end{Definition}

In the following, we will equip the blow-up with the structure of a smooth manifold with boundary such that its interior becomes diffeomorphic to $X\backslash Y$ via the blow-down map and the boundary becomes $P^+ NY$.
Consider an adapted chart $(U,\psi)$ for $Y$ in $X$ with $\psi(U)=\R^{n}$ and $\psi(U\cap Y)=\{(0,y)\mid y\in \R^k \}$.
It induces the bijection\Correct[caption={DONE Blow up charts}]{It is not a bijection if $\psi(U\cap Y) \neq \emptyset$! We need it centered at $(0,0)$! No we need actually only the intersection with $Y$!}
\begin{align*}
 \tilde{\psi}: \Bl_{U\cap Y}U &\longrightarrow [0,\infty) \times \Sph{n-k-1} \times \R^k \\
  p\in U\backslash Y &\longmapsto \Bigl (\Abs{\pi_1 \psi(p)}, \frac{\pi_1\psi(p)}{\Abs{\pi_1\psi(p)}},\pi_2 \psi(p) \Bigr), \\
  [v]\in P^+N_p Y & \longmapsto \Bigl(0,\frac{\pi_1 \Diff{\psi}(v)}{\Abs{\pi_1 \Diff{\psi} (v)}},\pi_2\psi(p)\Bigr),
\end{align*}
where $\pi_1$ and $\pi_2$ are the canonical projections to the factors of $\R^{n-k} \times \R^k$.
Notice that we have the canonical inclusion $\Bl_{U\cap Y} U \subset \Bl_Y X$.
It can be checked that for any two overlapping adapted charts $(U_1,\psi_1)$ and $(U_2,\psi_2)$, the transition function $\tilde{\psi}_1\circ \tilde{\psi}_2^{-1}$ is a diffeomorphism of manifolds with boundary.
Therefore, we can use the charts $(\Bl_{U\cap Y}U, \tilde{\psi})$ to define a smooth atlas on~$\Bl_Y X$.
If $X$ is oriented, we orient $\Bl_Y X$ so that $\pi$ restricts to an orientation preserving diffeomorphism of the interior.

An important obvious fact is that if $X$ is compact, then \emph{$\Bl_Y X$ is compact}.

We are interested in the case when $X = M\times M$ for an oriented closed manifold $M$ and $Y = \Delta \coloneqq \{(m,m) \mid m\in M\}$ is the \emph{diagonal}.
Given a chart $\varphi: U \rightarrow \R^n$ on~$M$, the following is a smooth chart on $\Bl_\Diag(M\times M)$:
\begin{equation} \label{Eq:BlowUpChart}
\begin{aligned}
\tilde{\varphi}: \Bl_{\Diag}(U\times U) & \longrightarrow [0,\infty) \times \Sph{n-1} \times \R^n \\
(x,y)\in (U\times U)\backslash \Diag & \longmapsto  (r,w,u)\coloneqq \begin{multlined}[t] \Bigl(\frac{1}{2}\Abs{\varphi(x)-\varphi(y)}, \frac{\varphi(x)-\varphi(y)}{\Abs{\varphi(x)-\varphi(y)}},\\\frac{1}{2}(\varphi(x)+\varphi(y))\Bigr),\end{multlined} \\
[(v,-v)]_{(x,x)} &\longmapsto \Bigl(0,\frac{\Dd \varphi_x(v)}{\Abs{\Dd \varphi_x(v)}},\varphi(x)\Bigr).
\end{aligned}
\end{equation}
The inverse relations for $r>0$ read
\begin{equation*} \label{Eq:BlowupRelations}
\varphi(x) = u+w r\quad\text{and}\quad\varphi(y)=u-w r.
\end{equation*}
We will denote by $M_i$ the $i$-th factor of $M\times M$; i.e., we will write $M \times M = M_1 \times M_2$.
We denote the corresponding projection by $\Pr_i$.
We define $\widetilde{\Pr}_i \coloneqq \Pr_i\circ\pi$, where $\pi: \Bl_\Diag(M\times M) \rightarrow M \times M$ is the blow-down map.
We also identify $(M\times M)\backslash \Diag$ with the interior of $\Bl_\Diag(M\times M)$ via $\pi$.

\begin{Example}[Product minus open thickening of diagonal]
Have a look at Figure~\ref{Fig:SimpleBlow} for the simplest examples of what kind of manifold $\Bl_\Diag(M\times M)$ is.
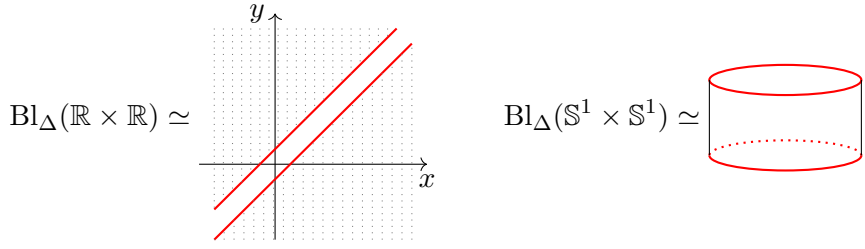
\begin{figure}
\[\Bl_\Diag(\R\times \R) \simeq \vcenterline{\begin{tikzpicture}
\draw[->] (0,-1) -- (0,2) node[left] {$y$};
\draw[->] (-1,0) -- (2,0) node[below] {$x$};
\draw[thick,red, domain=-.8:1.6] plot (\x, {\x + .2});
\foreach \i in {0,...,20}
{\draw[dotted,gray] ($({2.4/20*\i - .8},1.8)$) -- ($({2.4/20*\i - .8},{2.4/20*\i - .8+ .2})$);
}
\draw[thick,red, domain=-.8:1.8] plot (\x, {\x - .2});
\foreach \i in {0,...,20}
{\draw[dotted,gray] ($({2.6/20*\i - .8},-1)$) -- ($({2.6/20*\i - .8},{2.6/20*\i - .8 - .2})$);
}
\end{tikzpicture}}
\qquad
\Bl_\Diag(\Sph{1}\times \Sph{1}) \simeq 
\vcenterline{\begin{tikzpicture}
\draw[thick,red] (0,1) arc (0:360:{1} and {.2});
\draw[thick,red,dotted] (0,0) arc (0:180:{1} and {.2});
\draw[thick,red] (0,0) arc (360:180:{1} and {.2});
\draw (0,0) -- (0,1);
\draw (-2,0) -- (-2,1);
\end{tikzpicture}}\]
\caption{The simplest blow-up examples.}
\label{Fig:SimpleBlow}
\end{figure}
In fact, it is always diffeomorphic to $M\times M$ minus an open thickening of the diagonal.
The blow-down map is therefore an essential part of the blow-up construction.
\end{Example}

The map $\widetilde{\Pr}_2: \Bl_\Diag(M\times M) \rightarrow M_2$ is an oriented fiber bundle with fiber $\Bl_{*}(M_1)$, which is the blow-up of $M_1$ at a point (we shall assume that $M$ is connected).
The \emph{fiberwise integration along~$\widetilde{\Pr}_2$} will be denoted by~$\FInt{\Bl_* M_1}$. 

\begin{Def}[Hodge propagator] \label{Def:GreenKernel}
Let $M$ be an oriented closed $n$-dimensional Riemannian manifold.
Consider the harmonic projection $\pi_{\Harm}$ from \eqref{Eq:HarmProj}, and let $\iota_{\Harm}: \Harm(M) \xhookrightarrow{} \DR(M)$ be the inclusion.
A smooth $(n-1)$-form $\Prpg$ on $(M\times M)\backslash \Diag$ is called an \emph{admissible Hodge propagator} if the following conditions are satisfied:
\begin{PlainList}
\item The form $\Prpg$ admits a smooth extension to $\Bl_\Diag(M\times M)$.
More precisely, the pullback $(\Restr{\pi}{\mathrm{int}})^* \Prpg$ along the blow-down map restricted to the interior is a restriction of a smooth form on $\Bl_\Diag(M\times M)$.
We denote this form by $\Prpg$ again by uniqueness.
\item The operator $\Htp: \DR^*(M) \rightarrow \DR^{*-1}(M)$ defined by
\begin{equation} \label{Eq:SchwKer}
\Htp(\eta) \coloneqq \FInt{\Bl_* M_1} \Prpg \wedge \widetilde{\Pr}_1^* \eta \quad \text{for all }\eta\in \DR(M)
\end{equation}
satisfies
\begin{equation} \label{Eq:CochainHomotopy}
\Dd\circ\Htp + \Htp\circ\,\Dd = \iota_{\Harm}\circ \pi_{\Harm} - \Id.
\end{equation}
Any homogenous linear operator $\Htp: \DR^*(M) \rightarrow \DR^{*-1}(M)$ satisfying~\eqref{Eq:CochainHomotopy} will be called a \emph{Hodge homotopy}.%
\item For the twist map $\tau: M \times M \rightarrow M \times M$ defined by $(x,y)\mapsto (y,x)$, the following symmetry property holds:
\begin{equation}\label{Eq:SymProp}
\tau^* \Prpg = (-1)^n \Prpg. 
\end{equation}
\end{PlainList}
\end{Def}

\begin{Remark}[On Hodge propagator]\phantomsection\label{Rem:GKer}
\begin{RemarkList}
\item Given a homogenous linear operator $\Htp: \DR^*(M) \rightarrow \DR^{*-1}(M)$, if there is a $\Prpg \in \DR^{n-1}(\Bl_{\Diag}(M\times M))$ such that \eqref{Eq:SchwKer} holds, then it is unique.
\item Because $\tau: M\times M \rightarrow M\times M$ preserves $\Diag$, it extends to a diffeomorphism $\tilde{\tau}$ of $\Bl_\Diag(M\times M)$.
The condition \eqref{Eq:SymProp} is then equivalent to $\tilde{\tau}^*\tilde{\Prpg} = (-1)^n \tilde{\Prpg}$ for the extension $\tilde{\Prpg}$ of $\Prpg$ to $\Bl_{\Diag}(M\times M)$.
We denote both extensions by $\tau$ and $\Prpg$, respectively.

\item Using the intersection pairing $(\eta_1,\eta_2) = \int_M \eta_1 \wedge \eta_2$, we have 
\[ \begin{aligned}(\Htp(\eta_1),\eta_2) &= \int_{\Bl_{\Diag}(M\times M)} \Prpg \wedge \widetilde{\Pr}_1^*\eta_1 \wedge \widetilde{\Pr}_2^*\eta_2 \\ 
&= (-1)^n \int_{\Bl_{\Diag}(M\times M)} \tau^*\Prpg \wedge \widetilde{\Pr}_2^*\eta_1 \wedge \widetilde{\Pr}_1^*\eta_2 \end{aligned}\]
and 
\[ \begin{aligned}(\eta_1, \Htp(\eta_2)) &= (-1)^{\eta_1(\eta_2 - 1)} (\Htp(\eta_2),\eta_1) \\ &= (-1)^{\eta_1} \int_{\Bl_\Diag(M\times M)} \Prpg \wedge \widetilde{\Pr}_2^* \eta_1 \wedge \widetilde{\Pr}_1^* \eta_2 \end{aligned}\]
for all $\eta_1$, $\eta_2\in \DR(M)$.
This implies the following:
\begin{equation}\label{Eq:GSA}
\tau^* \Prpg = (-1)^n \Prpg \quad \Longleftrightarrow \quad \begin{multlined}[t](\Htp(\eta_1),\eta_2) = (-1)^{\eta_1}(\eta_1, \Htp(\eta_2)) \\ \forall \eta_1, \eta_2 \in\DR(M). \end{multlined}
\end{equation}

\item Because $\Bl_\Diag(M\times M)$ is compact, $\Prpg\in \DR(\Bl_\Diag(M\times M))$ induces an $L^1$-integrable form on $M\times M$.

\item In the original article \cite{Cieliebak2015}, Hodge propagator was called ``Green kernel''.
However, this name is standardly used for the Schwartz kernel of the generalized inverse of an elliptic pseudo-differential operator, e.g., of the Laplacian $\Delta$.
Because this was a source of confusion in various discussions, we decided to change the name to Hodge propagator.
Doing a random search on the internet, we found that this name appeared already in~\cite{Cattaneo2015}.
This is the story of how we discovered \cite{Mnev2009}, where Hodge propagators were also treated, and where we took the trick in the proof of Proposition~\ref{Prop:ExistenceG} from.
\qedhere
\end{RemarkList}
\end{Remark}

We will now prove three propositions which will allow us to rewrite~\eqref{Eq:CochainHomotopy} equivalently as a differential equation for $\Prpg$ on $M\times M \backslash \Diag$.

\begin{Proposition}[Identities for fiberwise integration]\label{Prop:StokesForm}
In the situation of Definition~\ref{Def:FibInt},
assume that $F$ has a boundary $\Bdd F$.
We orient $\Bdd F$ using $T_p F = N(p)\oplus T_p \Bdd F$ for $p\in \Bdd F$, where $N$ is an outward pointing normal vector field.
The following formulas hold for all $\kappa\in \DR_{\mathrm{cv}}(E)$ and $\eta\in \DR_{\mathrm{c}}(B)$:
\begin{itemize}
\item The \emph{projection formula}
\[\FInt{F}(\kappa\wedge \pi^* \eta) = \Bigl(\FInt{F}\kappa\Bigr)\wedge \eta, \]
\item \emph{Stokes' formula}
\[(-1)^F \Dd \FInt{F}\kappa  = \FInt{F} \Dd \kappa -  \FInt{\Bdd F} \kappa, \]
where $F$ in the exponent denotes the dimension of $F$.
\end{itemize}
\end{Proposition} 
\begin{proof} %
The projection formula is proven by a straightforward calculation from the definition.

As for Stokes' formula, we get the oriented fiber bundle $\Bdd E \rightarrow B$ with fiber~$\Bdd F$ locally by restricting an oriented local trivialization of $E$.
There are two orientations on $\Bdd E$ --- as the total space of $\Bdd E \rightarrow B$ and as the boundary of $E$.
They agree due to our orientation convention.
Using standard Stokes' theorem, we get
\allowdisplaybreaks
\begin{align*}
 (-1)^F \int_B \Dd \Bigl(\FInt{F}\kappa\Bigr) \wedge \eta & = (-1)^{\kappa + 1} \int_B \Bigl(\FInt{F}\kappa\Bigr)\wedge \Dd \eta \\ &=  (-1)^{\kappa + 1} \int_{E} \kappa \wedge \Dd \pi^*\eta \\
  & = \int_E \bigl(\Dd \kappa \wedge \pi^*\eta - \Dd(\kappa \wedge\pi^*\eta)\bigr) \\ 
  &= \int_B \Bigl(\FInt{F} \Dd \kappa\Bigr) \wedge \eta - \int_{\Bdd E} \kappa \wedge \pi^*\eta \\ 
  &= \int_B \Bigl(\FInt{F} \Dd \kappa - \FInt{\Bdd F}\kappa \Bigr) \wedge \eta.
\end{align*}
This proves the proposition.
\end{proof}

In what comes next, we will view the canonical projection $\Pr_2 : M_1 \times M_2 \rightarrow M_2$ as an oriented fiber bundle such that the orientation of the total space agrees with the product orientation.
The fiberwise integration for this bundle will be denoted by $\FInt{M_1}$. 

\begin{Proposition}[Schwartz kernel of the harmonic projection] \label{Lemma:HKer}
Let $M$ be an oriented closed $n$-dimensional Riemannian manifold.
Let $\Le_1$,~$\dotsc$, $\Le_m$ be a homogenous basis of $\Harm(M)$ which is orthonormal with respect to the $L^2$-inner product 
\[ (\eta_1, \eta_2)_{L^2} \coloneqq \int_M \eta_1 \wedge \Star\eta_2\quad\text{for }\eta_1, \eta_2 \in \DR(M), \]
where $\Star$ denotes the Hodge star.
The smooth form $\HKer \in \DR^n(M\times M)$ defined by
\begin{equation}\label{Eq:HKerHKerHKer}
\HKer \coloneqq \sum_{i=1}^m (-1)^{n \Le_i}\Pr_1^*(\Star \Le_i) \wedge \Pr_2^*(\Le_i)
\end{equation}
satisfies the following properties:
\begin{ClaimList}
\item For all $\eta\in \DR(M)$, we have
\begin{equation*}
 \HPr(\eta) = \FInt{M_1} \HKer\wedge \Pr_1^*\eta.
\end{equation*} 
\item The form $\HKer$ is closed and Poincar\'e dual to $\Diag \subset M\times M$.
\item The following symmetry condition is satisfied:
\begin{equation} \label{Eq:SymHKer}
\tau^* \HKer = (-1)^n \HKer.
\end{equation}
\end{ClaimList}
\end{Proposition}
\begin{proof} %
\begin{ProofList}
\item For the purpose of the proof, we denote $\Harm(\eta) \coloneqq \FInt{M_1} \HKer \wedge \Pr_1^* \eta$.
For every $k=1$,~$\dotsc$, $m$, we use the projection formula to compute
\allowdisplaybreaks
\begin{align*}
\Harm(\Le_k) &= \sum_{i=1}^m (-1)^{\Le_i n+\Le_i \Le_k}\FInt{M_1} \Pr_1^*(\Star \Le_i \wedge \Le_k)\wedge \Pr_2^*(\Le_i) \\ &=\sum_{i=1}^m (-1)^{\Le_i(n+\Le_k) + \Le_k(n+\Le_i)}\Bigl(\int_M \Le_k \wedge \Star \Le_i\Bigr) \Le_i \\
 & = \Le_k.
\end{align*}
It is easy to see that $\Harm(\eta) \in \Harm(M)$ for all $\eta\in \DR(M)$.
Therefore, $\Harm$ is a projection to $\Harm(M)$.
Relations $\Harm(\Dd \eta) = \Harm(\CoDd \eta) = 0$ for all $\eta\in \DR(M)$ follow from the second line of the computation above with $\Le_k$ replaced by $\Dd\eta$ and $\CoDd\eta$ using that $\Im \CoDd \oplus \Im \Dd$ is $L^2$-orthogonal to $\Harm(M)$.
We see that $\Harm = \pi_{\Harm}$. 

\item Using $\Dd \circ \Harm = \Harm \circ \Dd = 0$ and Stokes' theorem, we get 
\[  \FInt{M_1} \Dd \HKer \wedge \Pr_1^* \eta = (-1)^n \Dd \Harm(\eta) - \Harm(\Dd \eta) = 0\quad \text{for all }\eta\in \DR(M). \]
It follows that $\Dd \HKer = 0$.
Using the K\"unneth formula, we can write a given $\kappa\in \DR(M\times M)$ with $\Dd \kappa = 0$ as $\kappa = \Pr_1^* \eta_1 \wedge \Pr_2^* \eta_2 + \Dd \eta$ for some $\eta_1$, $\eta_2 \in \Harm(M)$ and $\eta\in \DR(M)$.
Then 
\[ \begin{aligned} \int_{M\times M} \HKer \wedge \kappa & = \int_{M\times M} \HKer \wedge \Pr_1^* \eta_1 \wedge \Pr_2^* \eta_2 \\ &= \int_M \Harm(\eta_1)\wedge \eta_2 \\ &= \int_M \eta_1 \wedge \eta_2 = \int_{\Diag} \kappa. \end{aligned} \]
This shows that $\HKer$ is Poincar\'e dual to $\Diag$.
\item It follows from the Hodge decomposition that
\begin{equation}\label{Eq:HKerHKer}
(\pi_\Harm(\eta_1), \eta_2) = (\pi_{\Harm}(\eta_1), \pi_{\Harm}(\eta_2)) = (\eta_1,\pi_\Harm(\eta_2))\quad\text{for all }\eta_1, \eta_2 \in \DR(M),
\end{equation}
where $(\cdot,\cdot)$ is the intersection pairing.
As in (iii) of Remark \ref{Rem:GKer}, one shows that this is equivalent to \eqref{Eq:SymHKer}.\qedhere
\end{ProofList}
\end{proof}

\begin{Proposition}[Differential condition] \label{Prop:GKer}
Let $M$ be an oriented closed $n$-dimensional Riemannian manifold.
For $\Prpg \in \DR^{n-1} (\Bl_\Delta(M\times M))$,  the following claims are equivalent:
\begin{PlainList} 
\item The operator $\Htp: \DR^*(M) \rightarrow \DR^{*-1}(M)$ defined by $\Htp(\eta) \coloneqq \FInt{\Bl_* M_1} \Prpg \wedge \widetilde{\Pr}_1^*\eta$ for $\eta\in \DR(M)$ is a Hodge homotopy.
\item It holds \begin{equation} \label{Eq:GKer}
 \Dd \Prpg = (-1)^{n} \HKer\quad\text{on }(M\times M)\backslash \Delta.
\end{equation} 
\end{PlainList}
\end{Proposition}
\begin{proof} %
Before we begin, note that~\eqref{Eq:GKer} is equivalent to the equation $\Dd \tilde{\Prpg} = (-1)^n \pi^* \HKer$ on $\Bl_\Diag(M\times M)$ for the extension $\tilde{\Prpg}$ of $\Prpg$; we denote~$\tilde{\Prpg}$ by~$\Prpg$ and $\pi^*\HKer$ by $\HKer$ by uniqueness.

We will first prove $2) \Longrightarrow 1)$.
Using Stokes' formula, we get for every $\eta\in \DR(M)$ the following:
\begin{align*}
  \Dd \Htp (\eta) &= \Dd \FInt{\Bl_* M_1} \Prpg\wedge \widetilde{\Pr}_1^* \eta \\ 
  &= (-1)^n \Bigl( \FInt{\Bl_* M_1} \Dd(\Prpg\wedge\widetilde{\Pr}_1^* \eta) - \FInt{\Bdd \Bl_* M_1} \Prpg \wedge \widetilde{\Pr}_1^* \eta \Bigr) \\ 
  &= \HPr(\eta) - \Htp(\Dd\eta) +  \FInt{\Bdd \Bl_* M_1} (-1)^{n+1}\Prpg\wedge \widetilde{\Pr}_1^*\eta.
\end{align*}
Since $\widetilde{\Pr}_1 = \widetilde{\Pr}_2$ on $\Bdd \Bl_\Diag (M\times M)$, we get with the help of the projection formula the following:
\begin{equation*}
 \FInt{\Bdd \Bl_* M_1} \Prpg\wedge \widetilde{\Pr}_1^*\eta  = \FInt{\Bdd \Bl_* M_1} \Prpg\wedge \widetilde{\Pr}_2^*\eta = \Bigl(\FInt{\Bdd \Bl_* M_1} \Prpg \Bigr)\wedge\eta.
\end{equation*}
We will show that the $0$-form $\FInt{\Bdd \Bl_* M_1} \Prpg$ is constant $(-1)^n$.
Stokes' formula implies
\begin{equation*}
 \FInt{\Bdd \Bl_* M_1} \Prpg = \FInt{\Bl_* M_1} \Dd \Prpg = (-1)^n \FInt{\Bl_* M_1} \HKer. 
\end{equation*}
Using that $\HKer$ is Poincar\'e dual to $\Diag$, we get for every $\eta\in \DR^n(M)$ the following:
\begin{align*}
 \int_M \Bigl(\FInt{\Bl_* M_1} \HKer \Bigr) \wedge \eta &= \int_{\Bl_\Diag(M\times M)} \HKer\wedge \widetilde{\Pr}_2^*\eta 
 \\ &= \int_{M\times M} H\wedge \Pr_2^*\eta  \\ &= \int_\Diag \Pr_2^*\eta  = \int_M 1\wedge \eta.
\end{align*}
The implication follows.

We will now prove $1) \Longrightarrow 2)$.
Assume that~\eqref{Eq:CochainHomotopy} holds and that~$\Prpg$ extends smoothly to the blow-up.
Denote
\[K\coloneqq (-1)^n \Dd \Prpg - \HKer\quad\text{and}\quad L\coloneqq -1 +  \int^{\Bdd \Bl_*(M_1)} (-1)^n \Prpg. \]
Notice that $L$ is a function on $M$.
From the previous computations, we deduce that 
\[ \int^{\Bl_*(M_1)} K\wedge \widetilde{\Pr}_1^*\eta = L\eta\quad\text{for all }\eta\in \DR(M), \]
and hence
\[ \int_{\Bl_{\Diag}(M\times M)} K \wedge \widetilde{\Pr}_1^*(\eta_1) \wedge \widetilde{\Pr}_2^*(\eta_2)  = \int_M L \eta_1 \wedge \eta_2 \quad\text{for all }\eta_1, \eta_2 \in \DR(M). \]
If $K(x,y) \neq 0$ for some $(x,y)\in (M\times M)\backslash\Delta$, we can choose $\eta_1$, $\eta_2$ with disjoint supports such that the left-hand side is non-zero.
This is a contradiction.
Consequently, we have $K\equiv 0$.
\end{proof}

In general, the \emph{Schwartz kernel} of a linear operator $\Htp: \DR(M) \rightarrow \DR(M)$ is a distributional form $\Prpg$ on $M\times M$ which satisfies\footnote{We may consider such class of $\Htp$'s, e.g., pseudo-differential operators, such that $\Prpg$ exists and is unique (c.f., the well-known Schwartz kernel theorem).} 
\begin{equation*} 
\Htp(\eta)(x) = \int_{y\in M_1} \Prpg(y,x)\eta(y)\quad\text{for all }\eta\in \DR(M)\text{ and }x\in M_2.
\end{equation*}
We consider the following conditions on $\Htp$ and $\Prpg$:\label{ConditionsG}\Correct[caption={Typo},disable]{There is ''of'' missing downstairs.} 
\begin{center}
\begin{minipage}{.9\textwidth}
\begin{enumerate}[label=(P\arabic*)]
\item The Schwartz kernel $\Prpg$ of $\Htp$ is a restriction of a smooth form on $\Bl_\Diag(M\times M)$.
\item $\Dd\circ \Htp + \Htp\circ \Dd = \iota_{\Harm} \circ \pi_\Harm - \Id$.
\item $(\Htp(\eta_1),\eta_2) = (-1)^{\eta_1}(\eta_1,\Htp(\eta_2))$ for all $\eta_1$, $\eta_2\in \DR(M)$.
\item $\Htp \circ \iota_\Harm = 0$, $\pi_\Harm \circ \Htp = 0$.
\item $\Htp \circ \Htp = 0$.
\end{enumerate}
\end{minipage}
\end{center}
Clearly, (P1)--(P3) are equivalent to $\Prpg$ being a Hodge propagator from Definition~\ref{Def:GreenKernel}.
Conditions (P4) and (P5) play a crucial role in the vanishing results for the Chern-Simons Maurer-Cartan element $\PMC$ in Section~\ref{Sec:Vanishing} --- the more conditions are satisfied, the more vanishing we get.

\begin{Definition}[Special Hodge propagator]
	We call a Hodge propagator satisfying (P1)--(P5) \emph{special.}
\end{Definition}

The following lemma will be used in the proof of the upcoming proposition.
\begin{Lemma}\label{Lem:Smoothing}
Let $\Htp_1$, $\Htp_2$ be two linear operators $\DR(M) \rightarrow \DR(M)$ with Schwartz kernels $\Prpg_1$, $\Prpg_2 \in \DR(\Bl_\Diag(M\times M))$.
Then $\Htp\coloneqq \Htp_1 \circ \Htp_2$ is a smoothing operator, i.e., its Schwartz kernel $\Prpg$ is a smooth form on $M\times M$. 
\end{Lemma}
\begin{proof}
It holds $\Prpg(x_1,x_2) = \pm \int_x \Prpg_2(x_1,x) \Prpg_1(x,x_2)$.
The lemma follows from properties of convolution.\ToDo[caption={Add more detailed proof here!},noline]{Add more detailed proof here.}
\end{proof}

A version of the following proposition can be found in \cite{Mnev2009}.
\begin{Proposition}[Existence of special Hodge propagator] \label{Prop:ExistenceG}
Every oriented closed Riemannian manifold $M$ admits a special Hodge propagator.
\end{Proposition}

\begin{proof}
Because $\HKer$ is Poincar\'e dual to $\Diag$, we have for any closed $\kappa \in \DR_{\mathrm{c}}((M\times M)\backslash \Delta)$ the following:
\[ \int_{(M\times M)\backslash \Diag} \HKer \wedge \kappa = \int_{M\times M} \HKer \wedge \kappa = \int_{\Diag} \kappa = 0. \]
Poincar\'e duality for non-compact oriented manifolds (see \cite{BottTu1982}) implies that $\HKer$ is exact on $(M\times M)\backslash \Diag$.
Because a manifold with boundary is homotopy equivalent to its interior,  the restriction of the blow-down map induces an isomorphism $\pi^* : \HDR((M\times M)\backslash \Diag) \rightarrow \HDR(\Bl_{\Diag}(M\times M))$.
Poincar\'e duality for non-compact oriented manifolds (see \cite{BottTu1982}) implies that $\HKer$ is exact on $(M\times M)\backslash \Diag$. Because a manifold with boundary is homotopy equivalent to its interior,  the restriction of the blow-down map induces an isomorphism $\pi^* : \HDR((M\times M)\backslash \Diag) \rightarrow \HDR(\Bl_{\Diag}(M\times M))$. It follows that $(-1)^n \pi^* \HKer$ admits a primitive $\Prpg \in \DR(\Bl_{\Diag}(M\times M))$. According to Proposition~\ref{Prop:GKer}, the corresponding $\Htp$ satisfies (P1) and (P2).

If we define
\[ \tilde{\Prpg} \coloneqq \frac{1}{2}(\Prpg + (-1)^n \tau^* \Prpg)\in \DR^{n-1}(\Bl_{\Diag}(M\times M)), \]
then $\tilde{\Prpg}$ satisfies $\tau^* \tilde{\Prpg} = (-1)^n \tilde{\Prpg}$ and is still a primitive to $(-1)^n \pi^* \HKer$. Proposition~\ref{Prop:GKer} and \eqref{Eq:GSA} imply that the corresponding $\Htp$ satisfies (P1)--(P3).

Given $\Htp$ satisfying (P1)--(P3), we will now show that we can arrange (P4). Let us define
\[ \tilde{\Htp}\coloneqq (\Id - \pi_{\Harm}) \circ \Htp \circ (\Id - \pi_{\Harm}), \]
where we view $\pi_{\Harm}$ as a map $\pi_{\Harm}: \DR \rightarrow \Harm \subset \DR$. Then $\tilde{\Htp}$ is a Hodge homotopy because
\[ \Dd\circ\tilde{\Htp} + \tilde{\Htp} \circ \Dd = (\Id-\pi_{\Harm})\circ (\Dd\circ \Htp + \Htp \circ \Dd) \circ (\Id-\pi_{\Harm}) = \Id-\pi_{\Harm}. \]
Using \eqref{Eq:HKerHKer} and \eqref{Eq:GSA}, we see that $\tilde{\Htp}$ satisfies (P3). Using the intersection pairing and Proposition~\ref{Lemma:HKer}, we can write 
\[ \pi_{\Harm}(\eta) = \sum_{i=1}^m (-1)^{(n+\eta)\Le_i} (\Star \Le_i, \eta) \Le_i \quad\text{for all }\eta\in\DR(M),\]
and hence we have for all $\eta_1$, $\eta_2\in \DR(M)$ the following:
\begin{align*}
&\bigl(\Htp(\pi_{\Harm}(\eta_1)),\eta_2\bigr) \\
&\qquad = \sum_{i=1}^m (-1)^{(n+\eta_1)\Le_i} (\Star \Le_i, \eta_1) \bigl(\Htp(\Le_i), \eta_2\bigr) \\ 
& \qquad = \sum_{i=1}^m (-1)^{(n+1)\Le_i}  \int_{M\times M}\Pr_1^*(\Star \Le_i)\wedge \Pr_2^*(\Htp(\Le_i))\wedge \Pr_1^*(\eta_1)\wedge \Pr_2^*(\eta_2).
\end{align*}
It follows that the Schwartz kernel of $\Htp \circ \pi_{\Harm}$ is the smooth form
\[ \Kern_{\Htp\circ \pi_{\Harm}} \coloneqq \sum_{i=1}^m (-1)^{(n+1)\Le_i}\Pr_1^*(\Star \Le_i)\wedge \Pr_2^*(\Htp(\Le_i)). \]
Moreover, if we replace $\Htp$ with 
$\pi_{\Harm}\circ \Htp$, we get the smooth Schwartz kernel $\Kern_{\pi_{\Harm}\circ\Htp\circ \pi_{\Harm}}$ of $(\pi_{\Harm}\circ \Htp) \circ \pi_{\Harm}$. In the same way, but now using in addition \eqref{Eq:GSA}, we can write
\begin{align*}
&\bigl(\pi_{\Harm}(\Htp(\eta_1)),\eta_2 \bigr) \\[1ex]
& \qquad = (-1)^{\eta_1}\bigl( \eta_1, \Htp(\pi_{\Harm}(\eta_2))\bigr) \\[1ex]
& \qquad = (-1)^{\eta_1 \eta_2} \bigl(\Htp(\pi_{\Harm}(\eta_2)),\eta_1\bigr) \\[.5ex]
 &\qquad = \sum_{i=1}^m (-1)^{\eta_1 \eta_2 + (n+1)\Le_i} \int_{M\times M}\Pr_1^*(\Star \Le_i)\wedge \Pr_2^*(\Htp(\Le_i))\wedge \Pr_1^*(\eta_2)\wedge \Pr_2^*(\eta_1) \\
 &\qquad = \sum_{i=1}^m (-1)^{(n+1) \Le_i + n} \int_{M \times M} \Pr_2^*(\Star \Le_i) \wedge \Pr_1^*(\Htp(\Le_i))\wedge \Pr_1^*(\eta_1)\wedge\Pr_2^*(\eta_2),
\end{align*}
where in the last equality we pulled back the integral along the twist map. It follows that the Schwartz kernel of $\pi_{\Harm}\circ \Htp$ is the smooth form
\[ \Kern_{\pi_{\Harm}\circ\Htp} \coloneqq \sum_{i=1}^m (-1)^{\Le_i}\Pr_1^*(\Htp(\Le_i))\wedge \Pr_2^*(\Star \Le_i). \]
The Schwartz kernel of $\tilde{\Htp} = \Htp - \pi_{\Harm}\circ \Htp - \Htp \circ \pi_{\Harm} + \pi_{\Harm}\circ\Htp\circ\pi_{\Harm}$ is then
\[ \tilde{\Prpg} = \Prpg - \pi^* \Kern_{\Htp \circ \pi_{\Harm}} - \pi^* \Kern_{\pi_{\Harm}\circ\Htp} + \pi^* \Kern_{\pi_{\Harm}\circ\Htp\circ\pi_{\Harm}}, \]
which is a smooth form on $\Bl_{\Diag}(M\times M)$. Therefore, $\tilde{\Prpg}$ satisfies (P1)--(P4).

Given $\Htp$ satisfying (P1)--(P4), we will show that we can arrange (P5). The trick from \cite{Mnev2009} is to define
\[ \tilde{\Htp} = \Htp \Dd \Htp. \]
Applying (P1) and (P2) repeatedly, we compute
\begin{equation}\label{Eq:ExprGreen}
\begin{aligned}
\Dd \Htp \Htp \Htp \Dd &= \Dd \Htp \Htp - \Dd \Htp \Htp \Dd \Htp \\
&= \Dd \Htp \Htp - \Dd \Htp \Htp + \Dd \Htp \Dd \Htp \Htp \\ &= \Dd \Htp \Htp - \Dd \Dd \Htp \Htp \Htp \\
&= \Dd \Htp \Htp \\
& = \Htp - \Htp \Dd \Htp, 
\end{aligned}
\end{equation}
and hence
\[ \tilde{\Htp} = \Htp - \Dd \Htp \Htp \Htp \Dd. \]
Clearly, $\tilde{\Htp}$ satisfies (P1) and (P2). As for (P3), we compute
\[ (\eta_1, \tilde{\Htp} \eta_2) = (-1)^{\eta_1}(\Htp \eta_1, \Dd \Htp \eta_2) = (\Dd \Htp \eta_1, \Htp \eta_2) = (-1)^{\eta_1}(\tilde{\Htp}\eta_1,\eta_2). \]
As for (P5), we have
\begin{equation}\label{Eq:GSquared}
\begin{aligned}
\tilde{\Htp} \tilde{\Htp} &= \Htp \Dd \Htp (\Htp\Dd) \Htp \\
&= \Htp \Dd \Htp \Htp - \Htp \Dd (\Htp \Dd) \Htp \Htp \\
&= \Htp \Dd \Htp \Htp - \Htp \Dd \Htp \Htp + \Htp \Dd \Dd \Htp \Htp \\
&= 0.
\end{aligned}
\end{equation}
In order to show (P4), we have to compute the Schwartz kernel of $\Dd\Htp\Htp\Htp\Dd$. By Lemma \ref{Lem:Smoothing}, the Schwartz kernel $\TKer$ of $\TOp\coloneqq \Htp \Htp \Htp$ is a smooth form on $M \times M$. Therefore, Stokes' formula without the boundary term applies, and we get 
\[ (\Dd \TOp \Dd)\eta = \Dd \int^{M_1} \TKer \wedge \Dd \pi_1^*(\eta) = \int^{M_1} \Dd \TKer \wedge \Dd \pi_1^*(\eta) = (-1)^{\TKer} \int^{M_1} \Dd_1 \Dd \TKer \wedge \pi_1^*(\eta). \]
Here $\Dd_1: \DR(M\times M) \rightarrow \DR(M\times M)$ is the operator defined in local coordinates by
\[ \Dd_1\bigl(f(x,y)\Diff{x}^I\Diff{y}^J\bigr) = \sum_{i=1}^n \frac{\partial f}{\partial x^i}(x,y)\Diff{x^i}\Diff{x}^I \Diff{y}^J. \]
It follows that the Schwartz kernel $\tilde{\Prpg}$ of $\tilde{\Htp}$ satisfies 
\[ \tilde{\Prpg} = \Prpg + (-1)^{n} \Dd_1 \Dd \TKer \]
and is a smooth $(n-1)$-form on $\Bl_\Diag(M\times M)$. Conditions (P1)--(P5) are satisfied. \qedhere
\end{proof}

\begin{Remark}[Property (P5) in dimensions $1$ and $2$]
In dimension $1$, every operator of degree $-1$ satisfies (P5) from degree reasons. In dimension $2$, every operator satisfying (P1) and (P2) satisfies (P5) as well, which follows from \eqref{Eq:ExprGreen} and \eqref{Eq:GSquared}.
\end{Remark}

\begin{Remark}[The standard Hodge propagator]
Consider the Hodge-de Rham Laplacian $\Delta = \Dd \circ \CoDd + \CoDd \circ \Dd : \DR(M) \rightarrow \DR(M)$ and its ``Green operator'' $\GOp$ of degree~$0$ (see (v) of Remark \ref{Rem:GKer} for the collision of terminology) which was defined in~\cite[Definition~6.9]{Warner1983} by
\[\GOp \coloneqq (\Restr{\Delta}{\Harm(M)^{\perp}})^{-1} \circ \pi_{\Harm(M)^\perp}, \]
where $\perp$ denotes the $L^2$-orthogonal complement. We introduce the \emph{standard Hodge homotopy} by
\begin{equation}\label{Eq:GStdStd}
\HtpStd \coloneqq -\CoDd\circ \Htp_{\Delta}.
\end{equation}
Using the properties of $\GOp$, $\Dd$ and $\CoDd$, one can show that $\HtpStd$ satisfies (P2)--(P5).

As for (P1), there is the following formula inspired by~\cite{Harris2004}:\Correct[noline,caption={Sign problems with std propagator}]{Inconsitency of this and a sign in chapter 7}
\begin{equation}\label{Eq:PrpgUsingHeatKernel}
\PrpgStd = (-1)^{n+1}\lim_{t\to 0} \int_t^\infty \frac{1}{2}\CoDd K_\tau \Diff{\tau},
\end{equation}
where $\KKer_t(x,y) = \sum_i (-1)^{n e_i} e^{-\lambda_i t} (\Star e_i)(x) \wedge e_i(y)$ is the heat kernel of $\Delta$ and~$e_i$ the $L^2$-orthonormal eigenbasis of $\Delta$ with eigenvalues $\lambda_i$ (the signs come from our convention for fiberwise integration, c.f., \eqref{Eq:HKerHKerHKer}).
In order to show (P1), a plan is to consider \eqref{Eq:PrpgUsingHeatKernel} in local coordinates and transform it to blow-up coordinates.
For $\R^n$ with the standard Euclidean metric $g_0$, the integral and limit can be computed explicitly, see Section~\ref{Sec:HeatRN}, and~ $\PrpgStd$ indeed extends smoothly to the blow-up.
See Chapter~\ref{Chap:Prpg} for some more theory on Hodge propagator.
There is a ``rather condensed proof'' of a version of (P1) in \cite[Section~4.3]{Axelrod1993} using the Hadamard parametrix construction.
\end{Remark}

\section{Chern-Simons Maurer-Cartan element} \label{Section:Proof2} 

We first recall ribbon graphs and their labelings based on~\cite{Cieliebak2015}. 

\begin{Definition}[Ribbon graph]\label{Def:Ribbon}
A \emph{graph} $\Gamma$ is a quadruple $(V,H,\mathcal{V},\mathcal{E})$, where~$V$ is a finite set of vertices, $H$ a finite set of half-edges, $\mathcal{V}: H \rightarrow V$ the ``vertex map'' and $\mathcal{E}: H \rightarrow H$ with $\mathcal{E} \circ \mathcal{E} = \Id$ and without fixed points the ``edge map''. The preimage $\mathcal{E}^{-1}(h_1) = \{h_1, h_2\}$ for some $h_1$, $h_2\in H$ is called an \emph{edge}; the set of edges is denoted by $E$. We assume that the graphs are \emph{connected}, i.e., that for any $\Vert_1$, $\Vert_2 \in V$ there exists a path in $E$ connecting $\Vert_1$ to $\Vert_2$.

A \emph{ribbon graph} is a graph $\Gamma$ together with a free transitive action $\Z_{d(\Vert)}\ \rotatebox[origin=c]{-90}{$\circlearrowright$}\  \mathcal{V}^{-1}(\Vert)$ for every $\Vert\in V$, where $$d(\Vert) \coloneqq \Abs{\mathcal{V}^{-1}(\Vert)} $$
is the \emph{valency of $\Vert$}. We denote by $\Succ: H \rightarrow H$ the bijection induced by $1\in \Z_{d(\Vert)}$ for every $\Vert\in V$.

For a ribbon graph $\Gamma$, consider the set of sequences $(h_n)_{n\in \Z} \subset H$ such that the following conditions holds:
$$ \forall n\in \Z: \quad  h_{n+1}= \begin{cases} 
\mathcal{E}(h_n) & n\text{ even,}\\                                
\Succ(h_n) & n\text{ odd.}
\end{cases}$$
Two such sequences $(h_n)_{n\in \Z}$ and $(h'_n)_{n\in \Z}$ are equivalent if and only if there exist $n_0$, $n'_0 \in \Z$ both even or both odd such that $h_{n_0} = h'_{n_0'}$. An equivalence class $[(h_n)_{n\in \Z}]$ is called a \emph{boundary (or a boundary component) of $\Gamma$.} The set of boundaries of $\Gamma$ is denoted by $\Bdd \Gamma$.

An \emph{IE ribbon graph} is a ribbon graph $\Gamma$ together with the decomposition $V = V_{\mathrm{int}} \sqcup V_{\mathrm{ext}}$ into \emph{internal} and \emph{external vertices} $V_{\mathrm{int}}$ and $V_{\mathrm{ext}}$, respectively, such that $d(\Vert) = 1$ for all $\Vert\in V_{\mathrm{ext}}$. This decomposition induces the decomposition $E = E_{\mathrm{int}} \sqcup E_{\mathrm{ext}}$, where an edge $\Edge$ is internal if it connects two internal vertices and is external otherwise. We allow only graphs with \emph{at least one internal vertex}. We often identify an external vertex with its unique adjacent half-edge or the unique adjacent external edge; we call either of these an \emph{external leg}. For any $\Boundary \in \Bdd \Gamma$, we define the \emph{valency of $\Boundary$} by
$$ s(\Boundary) \coloneqq \Abs{\mathcal{V}(\Boundary) \cap V_{\mathrm{ext}}}, $$
where $\mathcal{V}(\Boundary) = \{\mathcal{V}(h_n) \mid n\in \Z\}$. We also have the free transitive $\Z_{s(\Boundary)}$-action on $\mathcal{V}(\Boundary)\cap V_{\mathrm{ext}}$ mapping $\Vert\in \mathcal{V}(\Boundary) \cap V_{\mathrm{ext}}$ to the next external vertex in the sequence $(\mathcal{V}(h_n))_{n\in\Z}$. We will denote this action by $\mathcal{N}$ again.

We say that an IE ribbon graph $\Gamma$ is \emph{reduced} if $s(\Boundary) \ge 1$ for all $\Boundary\in \Bdd \Gamma$.

The following letters will be used to denote the numerical invariants of a graph:
$$\begin{aligned}
k &\;\,\dots && \text{the number of internal vertices}, \\
s &\;\,\dots && \makebox[\widthof{the number of}]{---\ditto---} \text{ external vertices.}\\
l &\;\,\dots && \makebox[\widthof{the number of}]{---\ditto---}\text{ boundary components,} \\
e &\;\,\dots && \makebox[\widthof{the number of}]{---\ditto---} \text{ internal edges.}
\end{aligned}$$
Moreover, we define the \emph{genus $g\in \N_0$} so that the following \emph{Euler formula} holds:
\begin{equation} \label{Eq:EulerFormula}
 k - e + l = 2 - 2g.
\end{equation}\par
We denote by $\RG_{klg}$ the set of isomorphism classes of connected IE ribbon graphs with fixed $k$, $l$, $g$. We let $\RRG_{klg} \subset \RG_{klg}$ be the subset of reduced graphs. For $m\in \N_0$, we denote by $\RG_{klg}^{(m)} \subset \RG_{klg}$ the set of isomorphism classes of connected IE ribbon graphs with all internal vertices \emph{m-valent}, i.e., with 
$$ d(\Vert) = m\quad\text{for all }\Vert\in V_{\mathrm{int}}. $$
The notation $\Gamma \in \RG_{klg}$ means that $\Gamma$ is a representative of an equivalence class $[\Gamma]\in \RG_{klg}$.
\end{Definition}

\begin{Remark}[On ribbon graphs]\phantomsection
\begin{RemarkList}
\item An $m$-valent ribbon graph with $m\ge 2$ has a unique decomposition $V = V_{\mathrm{int}}\sqcup V_{\mathrm{ext}}$, and hence we can omit writing IE.
\item In this text, we will use only reduced ribbon graphs. Non-reduced ribbon graphs may play a role in the extension of the theory of $\dIBL^\PMC(\CycC(\Harm))$ to non-reduced cyclic cochains  or in the weak $\IBLInfty$-theory (see Remarks \ref{Rem:BVForm} and~\ref{Rem:NWG}).\qedhere
\end{RemarkList}
\end{Remark}

\begin{Def}[Labeling] \label{Def:Labeling}
A \emph{labeling} of an IE ribbon graph $\Gamma$ is the triple $L = (L_1,L_2,L_3)$, where $L_i$ have the following meanings: 
\begin{itemize}
 \item The symbol $L_1$ represents an ordering of internal vertices ($\eqqcolon L_1^v$), and of boundary components ($\eqqcolon L_1^b$). Given $L_1$, we write $V_{\mathrm{ext}} = \{\Vert_1, \dotsc, \Vert_k\}$, $\Bdd \Gamma = \{\Boundary_1, \dotsc, \Boundary_l\}$ and denote
$$ d_i \coloneqq d(\Vert_i)\quad\text{and}\quad s_j\coloneqq s(\Boundary_j). $$
 \item The symbol $L_2$ represents an ordering and orientation of internal edges. Given $L_2$, we write $E_{\mathrm{int}} = \{\Edge_1, \dotsc, \Edge_e\}$ and $\Edge_i =\{h_{i,1}, h_{i,2}\}$ for $h_{i,1}$, $h_{i,2}\in H$.
 \item The symbol $L_3$ represents an ordering of half-edges at every internal vertex ($\eqqcolon L_3^v$) and of external vertices at every boundary component ($\eqqcolon L_3^b$), both compatible with the \emph{ribbon structure} ($\coloneqq$\,the $\Z_m$-actions). Given $L_3$, we write $\mathcal{V}^{-1}(\Vert) = \{h_{\Vert,1}, \dotsc,h_{\Vert,d(\Vert)} \}$ and $\mathcal{V}(\Boundary) \cap V_{\mathrm{ext}}= \{\Vert_{\Boundary,1},\dotsc,\Vert_{\Boundary,s(\Boundary)}\}$ with $\Succ(h_{\Vert,i}) = h_{\Vert,i+1}$ and $\Succ(\Vert_{\Boundary,j}) = \Vert_{\Boundary,j+1}$ for all $i$, $j$, respectively. 
\end{itemize}
We sometimes call $L_i$ \emph{partial labelings} and $L$ a \emph{full labeling}. A ribbon graph $\Gamma$ together with a labeling $L$ is called a \emph{labeled ribbon graph}.
\end{Def}

Given a ribbon graph $\Gamma$, one can construct an oriented surface with boundary~$\Sigma_\Gamma$ --- the thickening of $\Gamma$ --- in the obvious way and a closed oriented surface~$\hat{\Sigma}_\Gamma$ by gluing oriented disks to the oriented boundaries of $\Sigma_\Gamma$. If partial labelings~$L_1$ and~$L_2$ are given, we obtain the following chain complex with oriented chain groups (vector spaces over $\R$):
\begin{equation} \label{Eq:OrientationComplex}
\begin{tikzcd}
C_2\coloneqq \langle \Boundary_1,\dotsc,\Boundary_l \rangle \arrow{r}{\Bdd_2} & C_1\coloneqq \langle \Edge_1,\dotsc, \Edge_e \rangle \arrow{r}{\Bdd_1} &  C_0\coloneqq \langle \Vert_k,\dotsc, \Vert_1 \rangle.
\end{tikzcd}
\end{equation}
Here $\Boundary_i$ stands for the oriented disc glued to the $i$-th boundary component of~$\Sigma_\Gamma$ and now being mapped into $\hat{\Sigma}_\Gamma$, $\Edge_i$ stands for the $1$-simplex in $\hat{\Sigma}_\Gamma$ corresponding to the $i$-th internal edge, $\Vert_i$ stands for the $0$-simplex in $\hat{\Sigma}_\Gamma$ corresponding to the $i$-th internal vertex, and the boundary map $\Bdd$ is the ``geometric'' boundary operator. The homology of this chain complex is isomorphic to the singular homology $\H(\hat{\Sigma})\coloneqq\H(\hat{\Sigma}_\Gamma;\R)$.

The orientation of $C_i$ ($\coloneqq$\,the order of generators in \eqref{Eq:OrientationComplex}) induces naturally an orientation of $\H(\hat{\Sigma}_\Gamma)$. The construction from \cite[Appendix A]{Cieliebak2015} is as follows. We pick complements $H_i$ of $\Im(\Bdd_{i+1})$ in $\Ker(\Bdd_{i})$ and complements $V_i$ of $\Ker(\Bdd_i)$ in $C_i$ and write 
$$\begin{tikzcd}
C_2 = V_2 \oplus H_2 \arrow{r}{\Bdd_2} & C_1 = V_1 \oplus H_1 \oplus \Im(\Bdd_2) \arrow{r}{\Bdd_1} & C_0 = \Im(\Bdd_1) \oplus H_0.
\end{tikzcd}$$
We orient $V_i$ arbitrarily and transfer the orientation to $\Im(\Bdd_{i})$ via  $\Bdd_{i} : V_i \overset{\simeq}{\rightarrow} \Im(\Bdd_{i})$. Then, assuming the direct sum orientation, orienting $H_i$ is equivalent to orienting $C_i$, and we obtain the orientation of $\H_i(\hat{\Sigma}_\Gamma)$ via the canonical projection $\pi: H_i \overset{\simeq}{\rightarrow} \H_i(\hat{\Sigma}_\Gamma) = \Ker(\Bdd_i)/\Im(\Bdd_{i+1})$. This construction does not depend on the choices of complements and orientations of $V_i$.

\begin{Definition}[Compatibility of $L_1$ and $L_2$]\label{Def:CompatLabeling}
Given a ribbon graph $\Gamma$ with partial labelings $L_1$ and $L_2$, we say that \emph{$L_2$ is compatible with $L_1$} if the orientation on $\H(\widehat{\Sigma}_\Gamma)$ induced by \eqref{Eq:OrientationComplex} agrees with the canonical orientation $$\H(\widehat{\Sigma}_\Gamma) =  \langle \Vert_1 + \dotsb + \Vert_k \rangle  \oplus \H_1(\widehat{\Sigma}_\Gamma) \oplus \langle \Boundary_1 + \dotsb + \Boundary_l\rangle, $$
where $\H_1(\widehat{\Sigma}_\Gamma)$ is oriented using the canonical symplectic intersection form.
\end{Definition}

Given a labeled IE ribbon graph $\Gamma$, the set of half-edges adjacent to internal vertices $\mathcal{V}^{-1}(V_{\mathrm{int}})$ can be ordered in two ways corresponding to writing 
$$ 2e + (s_1+ \dotsb +s_l) = d_1+ \dotsb +d_k. $$
This leads to the following definition.
\begin{Definition}[Edge order and vertex order]\label{Def:EdgeVertex}
For a labeled IE ribbon graph~$\Gamma$, we define the following two orders on the set of half-edges $H$:
\begin{itemize}
\item \emph{Edge order:} The first $2e$ half-edges $h_{i,j}^{\mathrm{e}}$ are the ones from internal edges; they are ordered according to $L_2$. They are followed by blocks of $s_1$,~$\dotsc$, $s_l$ half-edges $h_{i,j}^{\Boundary}$ which come from the boundary components $i=1$,~$\dotsc$, $l$, respectively, and which are ordered according to $L_3^b$ inside the blocks. Schematically, we have
$$ (h_{1,1}^{\Edge} h_{1,2}^{\Edge})\dots(h_{e,1}^{\Edge} h_{e,2}^{\Edge})(h^{\Boundary}_{1,1}\dots h^{\Boundary}_{1,s_1})\dots(h^{\Boundary}_{l,1}\dots h^{\Boundary}_{l,s_l}). $$
\item \emph{Vertex order:} It consists of blocks of $d_1$,~$\dotsc$, $d_k$ half-edges $h_{i,j}^{\Vert}$ which come from internal vertices $1$,~$\dotsc$, $k$, and which are ordered according to $L_3^v$ inside the blocks. Schematically, we have
$$ (h_{1,1}^{\Vert}\dots h_{1,d_1}^{\Vert})\dots (h_{k,1}^{\Vert}\dots h_{k,d_k}^{\Vert}). $$
\end{itemize}
	
	We denote by $\sigma_L \in \Perm_{\Abs{H}}$ the \emph{permutation from the edge to the vertex order} which is constructed such that the $i$-th half-edge in the edge order is the same as the $\sigma_L(i)$-th half-edge in the vertex order.
\end{Definition}

From now on, we will consider only reduced trivalent ribbon graphs $\TRRG_{klg}$ with $k$, $l \ge 1$, $g\ge 0$. We will often use the equation
\begin{equation} \label{Eq:TrivalentFormula}
2e + s = 3k.
\end{equation}

\begin{Def}[Chern-Simons Maurer-Cartan element]
\label{Def:PushforwardMCdeRham}
Let $M$ be an oriented closed Riemannian manifold, and let $\Prpg \in \DR^{n-1}(\Bl_\Diag(M\times M))$ be an admissible Hodge propagator from Definition~\ref{Def:GreenKernel}. The \emph{Chern-Simons Maurer-Cartan element $\PMC$}, or \emph{formal pushforward Maurer-Cartan element}, is the collection of 
$$ \PMC_{lg}\in \hat{\Ext}_l \CycC(\Harm(M))\quad\text{for all }l\ge 1, g\ge 0 $$
defined on generating words $\omega_i = \alpha_{i1} \dots \alpha_{is_i}\in \BCyc \Harm(M)$, where $\alpha_{ij} = \SuspU \eta_{ij}$ with $\eta_{ij}\in \Harm(M)$ for $s_i\ge 1$ and $i=1, \ldots, l$, 
by the formula
\begin{equation}
\label{Eq:PushforwardMCdeRham}
\begin{aligned} 
& \PMC_{lg}(\Susp^{l} \omega_1 \otimes \dotsb \otimes \omega_l) \\ 
& \qquad \coloneqq \frac{1}{l!}\sum_{[\Gamma]\in \TRRG_{klg}} \frac{1}{\Abs{\Aut(\Gamma)}} (-1)^{s(k,l) + P(\omega)} \sum_{L_1,\,L_3^b} (-1)^{\sigma_L} I(\sigma_L),
\end{aligned}
\end{equation}
which we explain as follows:
\begin{itemize}
\item %
The second sum is over all partial labelings $L_1$ and $L_3^b$ of a representative~$\Gamma$ of~$[\Gamma]$. In every summand, we complete $L_1$ and $L_3^b$  to a full labeling $L = (L_1, L_2, L_3)$ by picking an arbitrary $L_3^v$ and an arbitrary $L_2$ compatible with $L_1$.

\item Suppose that $\Gamma$ and $L_1$ are \emph{admissible} with respect to the input $\omega_1$, $\dotsc$, $\omega_l$; this means that $\Gamma$ has $l$ boundary components and that the $i$-th boundary component has valency $s_i$ for every $i=1$, $\dotsc$, $l$. In this case, denoting $\sigma = \sigma_L$, we define
\begin{equation} \label{Eq:ISigma}
I(\sigma_L) \coloneqq \begin{multlined}[t]\int_{x_1,\dotsc,x_k} \Prpg(x_{\xi(\sigma_1)},x_{\xi(\sigma_2)}) \dotsm \Prpg(x_{\xi(\sigma_{2e-1})},x_{\xi(\sigma_{2e})}) \\ \eta_{11}(x_{\xi(\sigma_{2e+1})}) \dotsm \eta_{ls_{l}}(x_{\xi(\sigma_{2e+s})}),\end{multlined}
\end{equation}
where $\xi: \{1,\dotsc, 3k\} \rightarrow \{1,\dotsc,k\}$ is the function defined by 
$$ \xi(3j-2) = \xi(3j-1) = \xi(3j) \coloneqq j $$ for all $j=1, \dotsc, k$, $s = s_1 + \dotsb + s_l$, $\eta_{}(x_i)$ denotes the pullback of $\eta$ along the canonical projection $\pi_i: M^{\times k} \rightarrow M$ to the $i$-th component $M_i$, $\Prpg(x_i, x_j)$ denotes the pullback of $\Prpg$ along $\pi_{i} \times \pi_j : M^{\times k} \rightarrow M_i \times M_j$, and $\int_{x_1,\dotsc,x_k}$ denotes the integral of an $nk$-form over $k$ copies of $M$.

If $\Gamma$ and $L_1$ are not admissible, then we set $I(\sigma_L)\coloneqq 0$.
\item $s(k,l) \coloneqq k + kl(n-1) + \frac{1}{2} k(k-1) n \mod 2$.
\item $P(\omega) \coloneqq  \sum_{i=1}^l \sum_{j=1}^{s_i} (s - s_1 - \dotsb - s_{i-1} - j) \eta_{ij} \mod 2$.
\end{itemize}
\end{Def}
\noindent %
In order to show that $\PMC_{lg}$ is well-defined and that the collection $(\PMC_{lg})$ satisfies Definition~\ref{Def:MaurerCartan} for $\dIBL(\CycC(\Harm(M)))$, there are several things to check:
\begin{PlainList}
 \item The integral $I(\sigma_L)$ converges.
 \item The sums are finite.
 \item The product $(-1)^{\sigma_L} I(\sigma_L)$ is independent of the choice of $L_3^v$ and $L_2$ compatible with $L_1$.
 \item The sum over labelings is independent of the chosen representative $\Gamma$ in an isomorphism class from $\TRRG_{klg}$.
  \item The map $\PMC_{lg}: \BCyc\Harm(M)[3-n]^{\otimes l} \rightarrow \R$ is graded symmetric on permutations of its inputs $\Susp\omega_i$.
 \item The map $\PMC_{lg}$ is graded symmetric on cyclic permutations of the components $\alpha_{ij}$ of each $\omega_{i}$.
 \item The degree condition 1) from Definition \ref{Def:MaurerCartan} holds with $d = n-3$.
 \item The filtration-degree condition 2) from Definition \ref{Def:MaurerCartan} holds with $\gamma = 2$.
 \item The Maurer-Cartan equation \eqref{Eq:MaurerCartanEquation} holds.
\end{PlainList}

Condition 9) will be proven in~\cite{Cieliebak2018} using the theory of iterated blow-ups.

Condition 1) follows from the property (P1) of a Hodge propagator~ $\Prpg$.
The idea is to associate one integration variable to each internal half-edge, obtaining the space $\{x_i^{(1)},x_i^{(2)},x_i^{(3)}\mid i=1, \dotsc, k\}$, and blowing-up pairs of variables connected by internal edges.
By (P1), the integrand of \eqref{Eq:ISigma} lifts to a smooth form on such a space.
The integral $\int_{x_1 \dotsc x_k}$ is then realized as an integral over the closure of a lift of the subspace $\{x_i^{(1)} = x_i^{(2)} = x_i^{(3)}\}$.
By compactness, the integral converges.
A detailed proof is under preparation in~\cite{Cieliebak2018}.

In this text, \underline{we will take 1) and 9) for granted.} 

\begin{Lemma} \label{Lem:MCCond}
Conditions 2) -- 8) hold.
\end{Lemma}
\begin{proof}

As for 2), the fixed input $\omega_1$, $\dotsc$, $\omega_l$ fixes the number $s$ of external vertices of $\Gamma$ by admissibility. Expressing $e$ from~\eqref{Eq:EulerFormula} and plugging it  in~\eqref{Eq:TrivalentFormula} gives
\begin{equation} \label{Eq:MCk}
k = s + 2 l + 4g - 4.
\end{equation}
We see that all parameters are fixed. Now, there is only finitely many elements with fixed $s$ in $\TRRG_{klg}$, and each of them has only finitely many labelings. Therefore, the sums are finite.

As for 3), we have to consider the orientation of the complex~\eqref{Eq:OrientationComplex}. Clearly, if two $L_2$'s are compatible with $L_1$, then they differ by an even number of the following operations: a~transposition of two internal edges or a change of the orientation of an internal edge. The former operation introduces no sign in $(-1)^{\sigma_L}$ but generates the sign $(-1)^{n-1}$ in $I(\sigma_L)$ from swapping the corresponding~$\Prpg$'s. The latter operation induces the sign $-1$ in $(-1)^{\sigma_L}$ and the sign $(-1)^n$ in $I(\sigma_L)$ from the symmetry $\Prpg(x,y) = (-1)^n \Prpg(y,x)$. Because the overall signs in $(-1)^{\sigma_L} I(\sigma_L)$ are the same, an even number of these operations preserves $(-1)^{\sigma_L} I(\sigma_L)$. This implies the independence of an $L_2$ compatible with~$L_1$. A~change in $L_3^v$ produces no sign in $(-1)^{\sigma_L}$ because every internal vertex is trivalent and a cyclic permutation of an odd number of elements is even. The integral $I(\sigma_L)$ remains unchanged because the change in $\sigma_L$ is compensated by the composition with $\xi$. Independence of the choice of $L_3^v$ follows.

As for 4), every isomorphism of ribbon graphs $\Gamma \rightarrow \Gamma'$ induces the bijection $L\mapsto L'$ of compatible labelings such that $\sigma_L = \sigma_{L'}$ ($L'$ is the ``pushforward'' labeling). The independence of the choice of a representative of $[\Gamma]$ follows.

As for 5), let $\mu\in \Perm_l$ be a permutation of the inputs $\Susp \omega_1$, $\dotsc$, $\Susp \omega_l$. The set of graphs which admit an admissible labeling is the same for both $\PMC_{lg}(\Susp^l \omega_1 \otimes \dotsb \otimes \omega_l)$ and $\PMC_{lg}(\Susp^l \omega_{\sigma_{1}^{-1}} \otimes \dotsb \otimes \omega_{\sigma_l^{-1}})$; we will pick one such $\Gamma$ and study the admissible labelings $L$ and $L'$, respectively. We write $\eta_i = \eta_{i1}\dots \eta_{i s_i}$ and $\Omega_i = \Susp \omega_i$ for all $i$, $j$, and denote by $I'(\sigma_{L'})$ the integral in the definition of $\PMC_{lg}(\Susp^l \omega_{\mu_1^{-1}} \otimes \dotsb \otimes \omega_{\mu_l^{-1}})$. Let $\tilde{\mu}\in \Perm_{3k}$ be the permutation which acts as the identity on $1$, $\dotsc$, $2e$ and as the block permutation determined by $\mu$ on $2e + 1$,~$\dotsc$, $2e+s$ divided into $l$ blocks of lengths $s_1$, $\dotsc$, $s_l$. For any $\sigma\in \Perm_{3k}$, we have
$$ \begin{aligned}
I'(\sigma) & = \begin{multlined}[t]\int_{x_1,\dotsc,x_k}\!\! \Prpg(x_{\xi(\sigma_1)},x_{\xi(\sigma_2)}) \dotsm \Prpg(x_{\xi(\sigma_{2e-1})},x_{\xi(\sigma_{2e})}) \\ \eta_{\mu_1^{-1} 1}(x_{\xi(\sigma_{2e+1})}) \dots \eta_{\mu_l^{-1} s_{\mu_l^{-1}}}(x_{\xi(\sigma_{2e+s})})\end{multlined} \\
 & = \begin{multlined}[t]\varepsilon(\mu,\eta) \int_{x_1,\dotsc,x_k} \!\! \Prpg(x_{\xi((\sigma\circ\tilde{\mu})_1)},x_{\xi((\sigma\circ\tilde{\mu})_2)}) \dotsm \Prpg(x_{\xi((\sigma\circ \tilde{\mu})_{2e-1})},x_{\xi((\sigma\circ\tilde{\mu})_{2e})}) \\ \eta_{11}(x_{\xi((\sigma\circ \tilde{\mu})_{2e+1})})\dots  \eta_{ls_l}(x_{\xi((\sigma\circ \tilde{\mu})_{2e+s})})\end{multlined} \\
 & = \varepsilon(\mu,\eta) I(\sigma \circ \tilde{\mu}).
\end{aligned} $$
The precomposition with $\tilde{\mu}$ corresponds to a bijection $(L_1,L_3^b)\mapsto (L_1', {L_3^{b}}')$ of partial labelings for $\PMC_{lg}(\Susp^l \omega_1 \dots \omega_l)$ and $\PMC_{lg}(\Susp^l \omega_{\mu_1^{-1}} \otimes \dotsb \otimes \omega_{\mu_l^{-1}})$, respectively. However, if $L_2$ is compatible with $L_1$, then in order to get an $L_2'$ compatible with $L_1'$, the labeling~$L_2$ has to be altered by as many operations of switching two internal edges or changing the orientation of an internal edge as there are transpositions in $\mu$. We explained in the proof of 3) that this produces the sign $(-1)^{(n-1)\mu}$ in $(-1)^{\sigma_{L'}}I(\sigma_{L'})$. Therefore, after the choice of compatible $L_2$ and~$L_2'$, we have
$$ (-1)^{\sigma_{L'}}I'(\sigma_{L'}) = (-1)^{(n-1)\mu} (-1)^{\tilde{\mu}} \varepsilon(\mu, \eta) (-1)^{\sigma_L} I(\sigma_L). $$ 
If we view $\eta$ as $\eta_{11} \dots \eta_{l s_l}$, we can understand $(-1)^{P(\omega)}$ as the Koszul sign $\varepsilon(\SuspU, \eta)$. Similarly, we write $(-1)^{P(\mu(\omega))}=\varepsilon(\SuspU,\mu(\eta))$, where we first view $\eta$ as $\eta_1 \otimes \dotsb \otimes \eta_l$ to apply $\mu$ and then as the list of components $\eta_{ij}$ to compute the Koszul sign (this is a little ambiguity in our notation). If we denote by $\widebar{\mu}$ the permutation of $1$, $\dotsc$, $s$ permuting the $l$ blocks of lengths $s_1$, $\dotsc$, $s_l$ according to $\mu$, then $\widebar{\mu}$ has the same sign as $\tilde{\mu}$, and the decomposition of $\varepsilon(\SuspU,\mu(\eta))$ into the moves
$$ \begin{aligned}
\SuspU_{1} \dots \SuspU_{s} \eta_{\mu_1^{-1}1} \dots \eta_{\mu_l^{-1} s_{\mu_l^{-1}}} &\xrightarrow{(1)} \SuspU_{\widebar{\mu}_{1}} \dots \SuspU_{\widebar{\mu}_{s}} \eta_{11} \dots \eta_{l s_l} \xrightarrow{(2)} \SuspU_{\widebar{\mu}_1} \eta_{11} \dots \SuspU_{\widebar{\mu}_s} \eta_{l s_l} \\ &\xrightarrow{(3)} \SuspU_{1} \eta_{\mu_1^{-1}1} \dots \SuspU_{s} \eta_{\mu_l^{-1} s_l} \end{aligned} $$
shows that
$$ (-1)^{P(\mu(\omega))} = \underbrace{(-1)^{\tilde{\mu}}\varepsilon(\mu,\eta)}_{(1)} \underbrace{(-1)^{P(\omega)}}_{(2)}\underbrace{\varepsilon(\mu,\omega)}_{(3)}. $$
Using this, we write
$$ (-1)^{P(\mu(\omega))} (-1)^{\sigma_{L'}} I'(\sigma_{L'}) = \varepsilon(\mu,\omega) (-1)^{(n-1)\mu} (-1)^{P(\omega)}(-1)^{\sigma_L} I(\sigma_L), $$
and compute
$$ \begin{aligned}
& \PMC_{lg}(\Omega_{\mu_1^{-1}} \otimes \dotsb \otimes \Omega_{\mu_l^{-1}}) \\ & \qquad =\varepsilon(\mu(\Susp), \mu(\omega)) \PMC_{lg}(\Susp^l \omega_{\mu_1^{-1}} \otimes \dotsb \otimes \omega_{\mu_l^{-1}}) \\ 
& \qquad= \varepsilon(\mu(\Susp), \mu(\omega)) (-1)^{\Abs{s}\mu} \varepsilon(\mu,\omega) \PMC_{lg}(\Susp^l \omega_{1} \otimes \dotsb \otimes \omega_{l}) \\
& \qquad= \underbrace{\varepsilon(\mu(\Susp), \mu(\omega))}_{(1)} \underbrace{(-1)^{\Abs{s}\mu} \varepsilon(\mu,\omega)}_{(2)} \underbrace{\varepsilon(\Susp,\omega)}_{(3)} \PMC_{lg}(\Susp\omega_{1} \otimes \dotsb \otimes \Susp \omega_{l}) \\
& \qquad= \varepsilon(\mu,\Omega) \PMC_{lg}(\Omega_{1} \otimes \dotsb \otimes \Omega_{l}).
\end{aligned} $$
We used $\Abs{s} = n-1\text{ mod }2$, and the last equality follows from the decomposition of $\varepsilon(\mu,\Omega)$ into the moves
$$  \begin{multlined} \Susp_1 \omega_1 \dots \Susp_l \omega_l \xrightarrow{(3)} \Susp_1 \dots \Susp_l \omega_1 \dots \omega_l \xrightarrow{(2)} \Susp_{\mu_1^{-1}}\dots \Susp_{\mu_l^{-1}} \omega_{\mu_1^{-1}} \dots \omega_{\mu_l^{-1}} \\ \xrightarrow{(1)} \Susp_{\mu_1^{-1}} \omega_{\mu_1^{-1}} \dots \Susp_{\mu_l^{-1}} \omega_{\mu_l^{-1}}. \end{multlined}$$
This proves the symmetry of $\PMC_{lg}$.

As for 6), fix an $i=1$, $\dotsc$, $l$ and let $\mu\in \Perm_{s_i}$ be a cyclic permutation permuting the components of $\omega_i = \alpha_{i1}\dots \alpha_{is_i}$. Similarly to the previous case, we denote by~$\tilde{\mu}$ the corresponding permutation of $1$, $\dotsc$, $3k$ and get a bijection $(L_1,L_3^b) \mapsto (L_1'=L_1,{L_3^b}')$ of admissible labelings of a given graph $\Gamma$ for $\PMC_{lg}(\Susp^l \omega_1 \otimes \dotsb \otimes \alpha_{i1}\dots \alpha_{is_i} \otimes \dotsb \otimes \omega_l)$ and $\PMC_{lg}(\Susp^l \omega_1 \otimes \dotsb \otimes \alpha_{i\mu_1^{-1}} \dots \alpha_{i\mu_{s_i}^{-1}}\otimes \dotsb \otimes \omega_l)$, respectively. This time, there is no change in $L_1$, and thus we can take $L_2'=L_2$, producing no sign. Therefore, we have
$$ (-1)^{\sigma_{L'}}I'(\sigma_{L'}) = (-1)^{\tilde{\mu}}\varepsilon(\mu,\eta_i)(-1)^{\sigma_{L}}I(\sigma_{L}), $$
where $\varepsilon(\mu,\eta_i)$ comes from permuting the forms in $I'(\sigma_{L'})$. Further, we deduce
$$ (-1)^{P(\mu(\omega))} = (-1)^{\tilde{\mu}} \varepsilon(\mu,\eta_i) (-1)^{P(\omega)} \varepsilon(\mu,\omega_i), $$
and hence
$$ \begin{aligned}
&\PMC_{lg}(\Susp^l \omega_1 \otimes \dotsb \otimes \alpha_{i\mu_1^{-1}} \dots \alpha_{i\mu_{s_i}^{-1}}\otimes \dotsb \otimes \omega_l) \\
&\qquad = \varepsilon(\mu,\omega_i) \PMC_{lg}(\Susp^l \omega_1 \otimes \dotsb \otimes \alpha_{i1}\dots \alpha_{is_i}\otimes\dotsb\otimes \omega_l ). \end{aligned}$$
This shows the symmetry of $\PMC_{lg}$ on cyclic permutations of the components of~$\omega_i$.

As for 7), suppose that $\PMC_{lg}(\Susp^l \omega_1 \otimes \dotsb \otimes \omega_l) \neq 0$, and let $D$ denote the total form-degree of the input $\eta_{11}$, $\dotsc$, $\eta_{l s_l}\in \Harm(M)$; i.e., we define
\begin{equation*}
 D \coloneqq \deg(\eta_{11}) + \dotsb + \deg(\eta_{1s_1}) + \dotsb + \deg(\eta_{l 1}) + \dotsb + \deg(\eta_{l s_l}).
\end{equation*}
Clearly, we must have
\begin{equation}  \label{Eq:TotDeg}
nk = (n-1) e  + D, 
\end{equation}
where the left-hand side is the dimension of $M^{\times k}$ and the right-hand side the form-degree of the integrand of $I(\sigma_L)$. If we plug in $e$ from~\eqref{Eq:EulerFormula} and $k$ from~\eqref{Eq:MCk}, we get
$$ \begin{aligned} D &= nk - (n-1) e \\ 
   &= nk - (n-1)(k+l+2g-2) \\
   & = k - (n-1)(l+2g-2) \\
   & = s + 2l +4g - 4 - (n-1)(l+2g-2) \\
   & = s - (n-3)(l+2g-2). \end{aligned}$$
It follows that
$$ \Abs{\PMC_{lg}} = \Abs{\Susp^l} + \Abs{\omega_1} +  \dotsb + \Abs{\omega_l} =  l(n-3) + D - s = - 2(n-3)(g-1). $$
This is exactly the degree from Definition~\ref{Def:MaurerCartan}. 

As for 8), if $\PMC_{lg}(\Susp^l \omega_1 \otimes \dotsb\otimes \omega_l)\neq 0$, then
$$ s = k - 2l -4g -4 \ge 1 + 2(2-2g-l) = 1 + 2 \chi_{0lg}, $$
and hence $\PMC_{lg}\in \Filtr^{1+2\chi_{0lg}}\hat{\Ext}_l \CycC$ for the filtration induced from the dual of the filtration of $\BCyc \Harm$ by weights. Therefore, we get 
$$ \|\PMC_{lg}\| \ge 1 + 2\chi_{0lg} > 2\chi_{0lg} \quad\text{for all }l\ge 1, g\ge 0. $$
This finishes the proof.
\end{proof}

\Modify[inline,caption={Superfluous signs}]{Add here a remark that we do not have to sum over $L_1^v$!.}

\begin{Definition}[Vertices of types A, B, C and some special graphs]\label{Def:Graphs}
Let $\Gamma \in \TRG_{klg}$ be a trivalent ribbon graph and $\Vert$ its internal vertex. We say that~$\Vert$ is of \emph{type $A$, $B$ or~$C$} if it is connected to precisely $1$, $2$ or~$3$ internal vertices, respectively (see Figure~\ref{Fig:Vertices}). The graph $\Gamma$ is called (see Figures~\ref{Fig:Types} and~\ref{Fig:YOk}):
\begin{itemize}
\item a \emph{tree} if $[\Gamma] \in \RG_{k10}$ for some $k\ge 1$;
\item \emph{circular} if $[\Gamma]\in \RG_{k20}$ for some $k\ge 1$;
\item the \emph{$Y$-graph} is the unique tree with $k=1$;
\item an \emph{$O_k$-graph} if $\Gamma$ is circular with $k$ internal vertices and no $A$-vertex.
\end{itemize}
We denote the $Y$-graph simply by $Y$.
\end{Definition}

{ \begingroup
\begin{figure}[t]
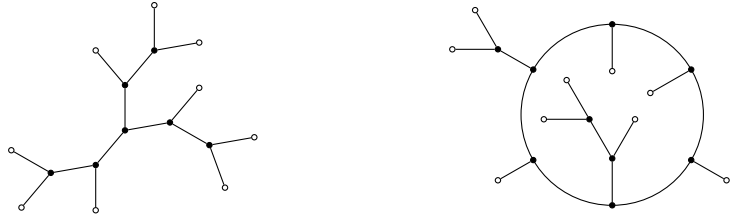

\centering
\begin{subfigure}{0.4\textwidth}
\centering
\input{\GraphicsFolder/tree.tex}
\end{subfigure}
\begin{subfigure}{0.4\textwidth}
\centering
\input{\GraphicsFolder/circular.tex}
\end{subfigure}
\caption[A tree and a circular graph.]{A tree and a circular graph. Internal vertices are denoted with a full dot and external vertices with an empty dot.}\label{Fig:Types}
\end{figure}
\endgroup }
{ \begingroup
\begin{figure}[t]
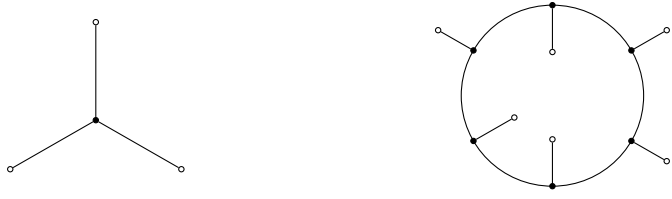

\centering
\begin{subfigure}{0.4\textwidth}
\centering
\input{\GraphicsFolder/ygraphgen.tex}
\end{subfigure}
\begin{subfigure}{0.4\textwidth}
\centering
\input{\GraphicsFolder/ok.tex}
\end{subfigure}
\caption{The $Y$-graph and an $O_6$-graph.}\label{Fig:YOk}
\end{figure}
\endgroup }

\begin{Remark}[On $A$, $B$, $C$ vertices and special graphs]
We observe the following:
\begin{RemarkList}
\item A trivalent graph $\Gamma \neq Y$ has each internal vertex of type $A$, $B$ or $C$.
\item The term $\PMC_{10}$ is a sum over trees, and the term $\MC_{10}$ is the contribution of the $Y$-graph to $\PMC_{10}$ (see Proposition~\ref{Prop:FormalPushforwardProp} below). The term $\PMC_{20}$ is a sum over circular graphs.\qedhere
\end{RemarkList}
\end{Remark}

Wee will also denote by $A$, $B$, $C$ the numbers of internal vertices of the corresponding type. Under the change of variables
\begin{equation} \label{Eq:ChangeOfVariables}
\begin{aligned}
  s &= 2 A + B, \\
  e &= B + \frac{1}{2} A + \frac{3}{2} C, \\
  k &= A + B + C,
\end{aligned}
\end{equation}
the trivalent formula~\eqref{Eq:TrivalentFormula} becomes trivial and the Euler formula~\eqref{Eq:EulerFormula} becomes
\begin{equation} \label{Eq:GenusFormulaa}
 C - A = 2l - 4 + 4g.
\end{equation}

\begin{Proposition}[Chern-Simons Maurer-Cartan element]\label{Prop:FormalPushforwardProp}
The collection $\PMC = (\PMC_{lg})$ from Definition~\ref{Def:PushforwardMCdeRham} is a Maurer-Cartan element for $\dIBL(\Harm(M))$ which is compatible with $\MC$. In particular, the $\AInfty$-algebra $\Harm(M)_\PMC$ is homologically unital and augmented.
\end{Proposition}
\begin{proof}
The fact that $\PMC$ is a Maurer-Cartan element follows from Lemma~\ref{Lem:MCCond} assuming 1) and 9) from \cite{Cieliebak2018}. \Correct[caption={DONE Redundant check of vanishing for $w\le 2$}]{Change it here because the definition of compatible changed!! In particular, the vanishing on words of length $1$ and $2$ follows from the filtration degree condition. Do not have to show it here.}

As for the compatibility with $\MC$,
the only graph contributing to $\PMC_{10}(\Susp \alpha_1 \alpha_2 \alpha_3)$  is the $Y$-graph with $k=1$. The group $\Aut(Y)$ consists of three rotations, and there is only one possible~$L_1$, no~$L_2$ and three~$L_3^b$. In Definition~\ref{Def:PushforwardMCdeRham}, we get $s(1,1) = n-2$, $(-1)^{\sigma_L} = 1$ because a cyclic permutation of an odd number of elements is even, and also $P(\alpha_1\alpha_2\alpha_3) =\eta_2$. Finally, we compute
\begin{align*}%
 \PMC_{10}(\Susp \alpha_1 \alpha_2 \alpha_3) &= \frac{1}{3} (-1)^{n-2 + \eta_2}\sum_{L_3^b} \int_x \alpha_1(x_{\xi(\sigma_1)})\alpha_2(x_{\xi(\sigma_2)})\alpha_3(x_{\xi(\sigma_3)}) \\ 
 & = (-1)^{n-2 + \eta_2}\int_M \eta_1 \wedge \eta_2 \wedge \eta_3\\[\jot]
 & = \MC_{10}(\Susp \alpha_1 \alpha_2 \alpha_3).\qedhere
\end{align*}
\end{proof}
\begin{figure}[t]
\centering
\input{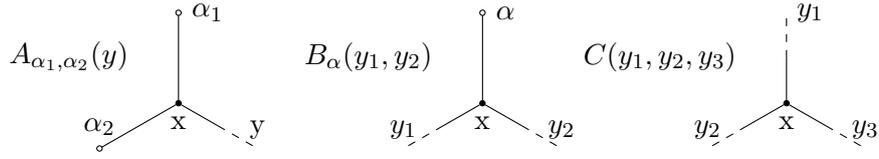}
\caption[Trivalent vertices of types $A$, $B$ and $C$.]{Trivalent vertices of types $A$, $B$ and $C$ with the corresponding forms $A_{\alpha_1,\alpha_2}$, $B_\alpha$ and $C$, respectively.}\label{Fig:Vertices}
\end{figure}
\begin{Definition}[Contributions of A, B, C vertices]\label{Def:Contributions}
Consider an internal vertex of type A, B or C as in Figure~\ref{Fig:Vertices}. We define the following smooth forms on $M$, $M^{\times 2}$ and $M^{\times 3}$, respectively:\footnote{The definitions can be made precise in local coordinates. Smoothness of $A_{\alpha_1, \alpha_2}$ is clear, smoothness of $B_\alpha$ follows from Lemma~\ref{Lem:Smoothing}, and smoothness of $C$ can be shown by a similar argument.}
$$\begin{aligned} 
 A_{\alpha_1, \alpha_2}(y)&\coloneqq \int_x \Prpg(y,x)\eta_1(x)\eta_2(x),\\ 
 B_\alpha(y_1,y_2) &\coloneqq \int_x \Prpg(y_1,x)\Prpg(x,y_2)\eta(x), \\
 C(y_1,y_2,y_3) &\coloneqq \int_x \Prpg(x,y_1)\Prpg(x,y_2)\Prpg(x,y_3).
\end{aligned}$$
\end{Definition}

\section{Results about vanishing of Chern-Simons Maurer-Cartan element}
\label{Sec:Vanishing}

\Modify[inline]{Do I really need that $\pi_\Harm \Htp = 0$? What is precisely the defininition?}

In the situation of Definition~\ref{Def:PushforwardMCdeRham}, let $\Gamma \in \TRRG_{klg}$ be a reduced trivalent ribbon graph, $L=(L_1,L_2,L_3)$ its labeling, $x_i$ the integration variable associated to the $i$-th internal vertex, $\Prpg(x_i,x_j)$ an admissible Hodge propagator on the oriented internal edge between $x_i$ and~$x_j$, and $\alpha_{ij}\in \Harm(M)[1]$ the harmonic form on the $j$-th external vertex on the $i$-th boundary component. Recall that we denote by $\omega_i = \Susp \alpha_{i1}\dotsc\alpha_{is_i}$ the $i$-th input of $\PMC_{lg}$ and by $D$ the total form-degree of all inputs. 

By saying ``\emph{a graph vanishes}'' we mean that $I(\sigma_L) = 0$ in the given context.

\begin{Proposition}[Vanishing of graphs with $\NOne$] \label{Prop:PMCEqualsMC}
In the setting of Definition~\ref{Def:PushforwardMCdeRham}, suppose that the following condition is satisfied: 
\begin{description}
\item[($V_{\NOne}$)] Every graph $\Gamma \in \TRRG_{klg}$, $\Gamma \neq Y$ which has $\NOne = \SuspU 1\in \Harm(M)[1]$ at an external vertex vanishes. 
\end{description}
Then $\PMC$ is strictly reduced, and the following holds depending on the dimension $n$:
\begin{enumerate}[label=(\alph*)]
 \item For $n>3$: All graphs which are not trees or circular vanish. Therefore, $\PMC_{lg} = 0$ for all $(l,g)\neq (1,0)$, $(2,0)$, and it follows that all higher operations~$\OPQ_{1lg}^\PMC$ vanish on the chain level.
  \item For $n=3$: A tree vanishes unless all $\eta_{1}$,~$\dotsc$, $\eta_{s}$ are one-forms. Therefore, $\PMC_{10}(\Susp \alpha_1 \dots \alpha_s) \neq 0$ implies $\deg(\eta_i)=1$ for all $i$.
 \item For $n<3$: All trees except for $Y$ vanish. Therefore, we have $\PMC_{10} = \MC_{10}$, and consequently $\OPQ_{110}^\PMC = \OPQ_{110}^\MC$.
\end{enumerate}
Moreover, we have 
\begin{enumerate}[resume,label=(\alph*)]
 \item A circular graph vanishes unless all $\eta_{11}$, $\dotsc$, $\eta_{2s_2}$ are one-forms. Therefore, $\PMC_{20}(\Susp^2 \alpha_{11}\dots \alpha_{1s_1} \otimes \alpha_{21}\dots \alpha_{2s_2})\neq 0$ implies $\deg(\eta_{ij})=1$ for all $i$, $j$.
\end{enumerate}
In addition to $(V_{\NOne})$, suppose that $\HDR^1(M) = 0$. Then:
\begin{enumerate}[resume,label=(\alph*)]
 \item All circular graphs vanish. Therefore, we have $\PMC_{20} = 0$, and consequently $\OPQ_{120}^\PMC = \OPQ_{120}$.
\item For $n\le 6$: All trees except for $Y$ vanish. Therefore, we have $\PMC_{10} = \MC_{10}$, and consequently $\OPQ_{110}^\PMC = \OPQ_{110}^\MC$. 
\end{enumerate} 
\end{Proposition}

\begin{proof}
The proof is just combinatorics with $D$. Suppose that a trivalent ribbon graph $\Gamma\neq Y$ does not vanish on the input $\omega_1$, $\dotsc$, $\omega_l$. Because all external vertices of $\Gamma$ are adjacent to an $A$-vertex or a $B$-vertex, the assumption $(V_{\NOne})$ implies $D\ge s$, where $s$ is the total number of external vertices. A combination of~\eqref{Eq:TotDeg} and~\eqref{Eq:TrivalentFormula} yields
$$ nk - (n-1)e = D \ge s = 3k - 2e\quad\Equiv\quad(n-3)k \ge (n-3)e. $$
\begin{ProofList}[label=(\alph*)]
\item For $n>3$, we get $k \ge e$, which implies that $\Gamma$ is either a tree or a circular graph.
\item If $\Gamma$ is a tree, then $s = k + 2$ and $e = k-1$. From~\eqref{Eq:TotDeg} we get
\begin{equation} \label{Eq:TreeEq}
D = nk - (n-1)(k-1) = k+n-1.  
\end{equation}
Now $D$ is the sum of $s=k+2$ form-degrees $\deg(\eta_{ij})>0$, and hence~\eqref{Eq:TreeEq} for $n=3$ implies that $\deg(\eta_{ij}) = 1$ for all $i$, $j$.
\item For $n<3$, we get $e \ge k$, which implies that $\Gamma$ is not a tree.
\item If $\Gamma$ is a circular graph, then $e=k=s$, and we get using~\eqref{Eq:TotDeg} that
$$ D = nk - (n-1)k = k. $$
Here $D$ is the sum of $s=k$ form-degrees $\deg(\eta_{ij})>0$, and hence $\deg(\eta_{ij})=1$ for all $i$, $j$.
\end{ProofList}
We will now assume, in addition, that $\Harm^1(M) \simeq \HDR^1(M) = 0$.\Add[caption={DONE Add in addition}]{This is now assumed in addition to (1)!}
\begin{ProofList}[resume, label=(\alph*)]
\item We must have $D\ge 2 s$, which is in contradiction with $D = s$ for a circular graph. Therefore, $\PMC_{20} = 0$.
\item Finally, for a tree $\Gamma \neq Y$, we have
\begin{equation*}
 k+n-1 = D \ge 2 s = 2(k + 2)\quad\Equiv\quad  n-5 \ge k. \end{equation*}
This finishes the proof of the proposition.\qedhere
\end{ProofList}
\end{proof}

\begin{Proposition}[Special Hodge propagator]\label{Prop:COne}
In the setting of Definition~\ref{Def:PushforwardMCdeRham}, suppose that the Hodge propagator $\Prpg$ is special. Then the condition ($V_{\NOne}$), and hence Proposition~\ref{Prop:PMCEqualsMC} holds.
\end{Proposition}
\begin{proof}
It is easy to see that $A_{\alpha_1,\alpha_2} = \Htp(\eta_1 \wedge \eta_2)$ for all $\alpha_1$, $ \alpha_2\in \Harm(M)[1]$, and that $-B_{\NOne}$ is the Schwartz kernel of $\Htp \circ \Htp$. Therefore, (P4) and (P5) imply $A_{\alpha_1,\NOne}=0$ and $B_{\NOne} = 0$, respectively.

As for the integral $I(\sigma_L)$,  one has to apply the Fubini theorem in order to integrate out single vertices $A_{\alpha_1, \NOne}$ and $B_{\NOne}$. This step relies on $L^1$-integrability of the integrand which follows from \cite{Cieliebak2018} (the integrand comes from a smooth form on a compact manifold with corners).
\end{proof}
\begin{Proposition}[Vanishing of $A$-vertices]\label{Prop:Avertexvanish}
In the setting of Definition~\ref{Def:PushforwardMCdeRham}, suppose that the following condition is satisfied:
\begin{description}
\item[($V_A$)] Every graph with an $A$-vertex vanishes.
\end{description}
Then we have $\PMC_{10} = \MC_{10}$, and the only contribution to $\PMC_{20}(\Susp^2 \alpha_{11}\dots \alpha_{1s_1} \otimes \alpha_{21}\dots \alpha_{2s_2})$ comes from $O_k$-graphs with $k = s_1 + s_2 = D$.
\end{Proposition}

\begin{proof}
The only trees and circular graphs which are not excluded by the assumption are the $Y$-graph and $O_k$-graphs, respectively (the external branches contract). The condition on form-degrees is obtained as in the proof of Proposition~\ref{Prop:PMCEqualsMC}.

To argue that $I(\sigma_L)=0$, we again need $L^1$-integrability as in the proof of Proposition \ref{Prop:COne}.
\end{proof}

\begin{Proposition}[$1$-connected geometrically formal manifolds] \label{Prop:GeomForm}
Let $M$ be a geometrically formal $n$-manifold and $\Prpg$ a special Hodge propagator (it exists by Proposition~\ref{Prop:ExistenceG}). If $\HDR^1(M) = 0$, then the following holds:
\begin{description}
\item[$(n\neq 2)$]  All $Y \neq \Gamma \in \RRG_{klg}$ with $k$, $l\ge 1$, $g\ge 0$ vanish, and hence $\PMC = \MC$.
\item[$(n=2)$] All $Y\neq \Gamma \in \RRG_{kl0}$ with $k$, $l\ge 1$ vanish, and hence $\PMC_{l0} = \MC_{l0}$ for all~$l\ge 1$.
\end{description}
\end{Proposition}
\begin{proof}
Given $\eta_1$, $\eta_2 \in \Harm$, geometric formality implies $\eta_1 \wedge \eta_2 \in \Harm$, and hence $A_{\alpha_1,\alpha_2} = \Htp(\eta_1\wedge\eta_2) = 0$. We see that $(V_{\NOne})$ and $(V_{A})$ are satisfied, and hence the implications of Propositions~\ref{Prop:PMCEqualsMC} and~\ref{Prop:Avertexvanish} hold. The claim for $n>3$ follows.

As for $n=3$, Poincar\'e duality implies $\HDR^2(M;\R)=0$. \Correct[caption={DONE Wrong reference}]{Here is not Eq:GenusFormula but the relationf of A B C vertices to graph variables}Therefore, the total form-degree $D$ satisfies $D= n B$, where $B$ is the number of $B$-vertices. We see using \eqref{Eq:ChangeOfVariables} that \eqref{Eq:TotDeg} is equivalent to
\begin{equation}\label{Eq:VerticesEq}
B+\frac{1}{2}(3-n) C = D = nB\quad\Equiv\quad (n-1)B = \frac{1}{2}(3-n) C.
\end{equation}
It follows that $B=0$, and hence all reduced graphs vanish.

As for $n=2$, we get from \eqref{Eq:VerticesEq} and \eqref{Eq:GenusFormulaa} that $B\ge l$ is equivalent to $g\ge 1$.
\qedhere
\end{proof}

\begin{Remark}[$\AInfty$-homotopy transfer]  \label{Rem:RemMu}
In~\cite{Cieliebak2018}, it will be shown that the $\AInfty$-algebra $\Harm(M)_\PMC = (\Harm(M),(\mu_k))$ induced by $\PMC_{10}$ agrees with the $\AInfty$-algebra obtained by the $\AInfty$-homotopy transfer
$$\begin{tikzcd}
\biggl(\ \begin{gathered}\DR(M) \\ m_1,\  m_2\end{gathered}\ \biggr)\arrow[rightsquigarrow]{r} & 
\biggl(\ \begin{gathered}
\Harm(M) \\
\mu_1\equiv 0,\ \mu_2 = \pi_\Harm m_2 (\iota_\Harm, \iota_\Harm),\ \mu_3,\ \dotsc
\end{gathered}\ \biggr)
\end{tikzcd}$$
 using the homotopy retract (see~\cite{Vallette2012})
$$\begin{tikzcd}[column sep=large]
(\DR(M),m_1)  \arrow[loop left]{l}{\Htp}  \arrow[shift left]{r}{\pi_\Harm}  & \arrow[shift left]{l}{\iota_\Harm} (\Harm(M),m_1 \equiv 0).
\end{tikzcd}$$
The operation $\mu_k$ of the transferred $\AInfty$-structure is computed as a sum over planar trees with a root and $k$ leaves decorated by $\iota_\Harm$ at the leaves, $\pi_\Harm$ at the root and~$\Htp$ at the internal edges (see \cite{Akaho2007}). The result of \cite{Cieliebak2018} is plausible because the part of $\PMC_{10}$ contributing to $\mu_k$ is a sum over trivalent ribbon trees with $k+1$ leaves.

In~\cite{Cieliebak2018}, they will also show that $\iota_1\coloneqq \iota_\Harm: \Harm \rightarrow \DR$ extends to an $\AInfty$-quasi-isomorphism $(\iota_k)_{k\ge 1}$ from $(\Harm,(\mu_k))$ to $(\DR,m_1,m_2)$. The induced chain map on the dual cyclic bar complexes is then the map $\HTP_{110}^\MC$ coming from the $\IBLInfty$-theory in the Overview.
\end{Remark}

\begin{Proposition}[Twisted boundary operator for formal manifolds]\label{Prop:Formal}
In the setting of Definition~\ref{Def:PushforwardMCdeRham}, suppose that $M$ is formal in the sense of rational homotopy theory. Then there is a quasi-isomorphism
$$\begin{tikzcd}
\HHTP_{110}: (\CDBCyc \HDR(M)[3-n], \OPQ_{110}^\MC)\arrow{r}{} & (\CDBCyc \Harm(M)[3-n],\OPQ_{110}^\PMC). \end{tikzcd}$$
\end{Proposition}

\begin{proof}
Formality of $M$ is equivalent to the existence of a zig-zag of quasi-isomorphisms of dga's (see \cite{Vallette2012}) 
$$\begin{tikzcd}[column sep=normal] (\H_{\mathrm{dR}}(M),m_1\equiv 0, m_2) \arrow[rightsquigarrow]{r} &\bullet\quad\dotsb\quad\bullet &\arrow[rightsquigarrow]{l} (\DR(M),m_1,m_2). \end{tikzcd}$$
Because a dga-quasi-isomorphism has a homotopy inverse in the category of $\AInfty$-algebras, we get a direct $\AInfty$-quasi-isomorphism 
$$\begin{tikzcd}
(g_k):\quad (\DR(M),m_1,m_2) \arrow[rightsquigarrow]{r} & (\H_{\mathrm{dR}}(M),m_1\equiv 0, m_2).
\end{tikzcd}$$
Precomposing with $(\iota_k)$ from Remark~\ref{Rem:RemMu}, we get the $\AInfty$-isomorphism 
$$\begin{tikzcd}
(h_k):\quad (\Harm(M),(\mu_k)) \arrow[rightsquigarrow]{r} & (\HDR(M),m_1\equiv 0,m_2). \end{tikzcd}$$
This induces the quasi-isomorphism $\HHTP_{110}$ of the cyclic cochain complexes.
\end{proof}

\begin{Remark}[On formality]\phantomsection
Geometrically formal manifolds include $\Sph{n}$, $\C P^n$ and Lie groups (see~\cite{Kotschick2000}). Any geometrically formal manifold is formal. Every simply-connected manifold of dimension at most $6$ is formal (see \cite{Miller1979}).
\end{Remark}

\section{Conjectured relation to string topology}\label{Sec:StringTopology}
\Add[inline,caption={DONE Simplify string top.
and cyc. hom. comparison}]{Add somewhere that the map from $(\CDBCyc \HDR(M),b^*)$ to $(C(\StringSpace M),\Bdd)$ with no degree shifts does the job (after removing all degree shifts from definitions).}

Given a smooth connected oriented $n$-dimensional manifold $M$, we consider the equivariant homology of the free loop space $\Loop M \coloneqq \{\gamma: \Sph{1} \rightarrow M \text{ continuous}\}$ with respect to the reparametrization action of $\Sph{1}$. It is defined as the singular homology of the Borel construction  
$$ \LoopBorel M \coloneqq \EG\Sph{1} \times_{\Sph{1}} \Loop M \coloneqq (\EG\Sph{1}\times \Loop M)/\Sph{1}, $$
where $\EG\Sph{1} = \Sph{\infty} \rightarrow \BG \Sph{1} = \CP^\infty$ is a model for the universal bundle for $\Sph{1}$, and we quotient out the diagonal action. We denote this homology by
$$ \StringH(\Loop M) \coloneqq \H(\LoopBorel M). $$
Recall that $(\EG\Sph{1}\times \Loop M)/\Sph{1}$ is the homotopically correct quotient replacing $\Loop M/\Sph{1}$, which we prefer to use because the diagonal action of $\Sph{1}$ on $\EG\Sph{1}$ is free, and hence $\EG\Sph{1}\times \Loop M \rightarrow (\EG\Sph{1}\times \Loop M)/\Sph{1}$ is a principal $\Sph{1}$-bundle (in contrast to the pathological map $\Loop M \rightarrow \Loop M /\Sph{1}$). Recall also that $\EG \Sph{1}$ is contractible.

The ``geometric versions'' of the homologies were defined in \cite{Sullivan1999} as the degree shifts
$$ \GeomH(\Loop M) \coloneqq \H(\Loop M)[n]\quad\text{and}\quad \GeomStringH(\Loop M)\coloneqq \StringH(\Loop M)[n]. $$

There is the \emph{loop product} $\LoopPr: \GeomH(\Loop M)^{\otimes 2} \rightarrow \GeomH(\Loop M)$ of degree $0$ which makes $\GeomH(\Loop M)$ into a graded commutative associative algebra. There is also the \emph{loop coproduct} $\LoopCoPr: \ConstRedGeomH(\Loop M) \rightarrow \ConstRedGeomH(\Loop M)^{\otimes 2}$ of degree $1-2n$ which is graded cocommutative and coassociative and is a derivation of $\LoopPr$. The geometric construction of~$\LoopPr$ and~$\LoopCoPr$ on transverse smooth chains in $\Loop M$ was described in~\cite{Sullivan1999} and~\cite{Basu2011}, respectively. Here, the symbol $\ConstRedGeomH(\Loop M)$ stands for the degree shifted relative homology
$$  \ConstRedGeomH(\Loop M)\coloneqq\H(\Loop M, M)[n] $$
with respect to constant loops $M \hookrightarrow \Loop M$. The geometric construction of~$\LoopCoPr$ does not work on the whole $\GeomH(\Loop M)$ because of the phenomenon of ``vanishing of small loops'' depicted in \cite[Figure 4, p.\,13]{Cieliebak2007}.
\Add[caption={DONE Reduced loop product}]{Is it really true? What about the example of torus.}

The projection $\EG\Sph{1} \times \Loop M \rightarrow \LoopBorel M$ is an $\Sph{1}$-principal bundle and thus induces a Gysin sequence. This sequence written using the geometric versions reads
\begin{equation}\label{Eq:Gysin}
\begin{tikzcd}
\dots\arrow{r}& \GeomH_i \arrow{r}{\Erase} & \GeomStringH_i \arrow{r}{\cap c} & \GeomStringH_{i-2} \arrow{r}{\Mark} & \GeomH_{i-1} \arrow{r} & \dots,
\end{tikzcd}
\end{equation}
where the map $\Mark$ adds a marked point in each string in a family in all possible positions, the map $\Erase$ erases the marked point of each string in a family, $c\in \StringCoH^{2}(\Loop M)$ is the Euler class of the circle bundle and $\cap$ the cap product.

The \emph{string bracket} $\tilde{\StringOp}_2: \GeomStringH(\Loop M)^{\otimes 2}\rightarrow \GeomStringH(\Loop M)$ and the \emph{string cobracket} $\tilde{\StringCoOp}_2: \ConstRedGeomStringH(\Loop M) \rightarrow \ConstRedGeomStringH(\Loop M)^{\otimes 2}$ are defined by
$$ \tilde{\StringOp}_2 \coloneqq \Erase \circ \LoopPr \circ \Mark^{\otimes 2}\quad\text{and}\quad\tilde{\StringCoOp}_2 \coloneqq \Erase^{\otimes 2} \circ \nu \circ \Mark. $$
Here, the symbol $\ConstRedGeomStringH(\Loop M)$ stands for the degree shifted relative $\Sph{1}$-equivariant homology
$$ \ConstRedGeomStringH(\Loop M) \coloneqq \underbrace{\StringH(\EG \Sph{1} \times_{\Sph{1}} \Loop M, \EG \Sph{1} \times_{\Sph{1}} M)}_{\displaystyle \eqqcolon \ConstRedStringH(\Loop M)}[n]. $$
Because $\Abs{\Mark} = 1$ and $\Abs{\Erase} = 0$, we have for all $\xi \in \ConstRedGeomStringH(\Loop M)$ and $\xi_1$, $\xi_2 \in \GeomStringH$ the relations\Correct[caption={DONE Typo}]{$\xi$ in the exponent should be $\Abs{\xi_1}$}
\begin{equation}\label{Eq:StringOpCoOp}
\begin{aligned}
\tilde{\StringOp}_2(\xi_1,\xi_2) &= (-1)^{\Abs{\xi_1}} \Erase(\Mark(\xi_1)\LoopPr\Mark(\xi_2)), \\
\tilde{\StringCoOp}_2(\xi) & = \sum \Erase(\nu^{1}) \otimes \Erase(\nu^2),
\end{aligned}
\end{equation}
where we write $\nu(\Mark(\xi)) = \sum \nu^1 \otimes \nu^2$. The operations $\tilde{\StringOp}_2$ and $\tilde{\StringCoOp}_2$ have degrees~$2$ and $2-2n$ with respect to the grading on $\GeomStringH(\Loop M)$, respectively. In fact, we will consider $\tilde{\StringOp}_2$ and $\tilde{\StringCoOp}_2$ given by \eqref{Eq:StringOpCoOp} as operations on the even degree shift $\StringH(\Loop M)[2-n] = \GeomStringH(\Loop M)[2-2n]$, which have degrees $2(2-n)$ and $0$, respectively. The symbols $\StringOp_2$ and $\StringCoOp_2$ will denote their degree shifts to $\StringH(\Loop M)[3-n]$, which have degrees of an $\IBL$-algebra from Definition~\ref{Def:IBLInfty}.

In work in progress \cite{Cieliebak2018b}, they consider the map 
$$I_{\lambda,*}: \CycH_{-\bullet- 1}(\DR(M)) \longrightarrow \StringCoH^\bullet(\Loop M; \R)$$ defined on the chain level as a cyclic version of Chen's iterated integrals $I_\lambda$. Recall that $\CycH_{-\bullet-1}(\DR) = \H_\bullet(\BCyc \DR, \Hd)$ \Correct[caption={DONE Homology exchangend with cohomology}]{DONE The homology of $\OPQ_{110}$ should be the cyclic cohomology},
where $\Hd: \B \DR = \bigoplus_{k\ge 1} \DR[1]^{\otimes k} \rightarrow \B \DR $ is the Hochschild differential of the de Rham dga $(\DR,m_1,m_2)$, see Section~\ref{Sec:Alg2}. They prove in \cite{Cieliebak2018b} that if $M$ is simply-connected, then the map~$I_{\lambda,*}$ induces an isomorphism $\RedCycH_{-\bullet-1}(\DR(M))) \simeq \RedStringCoH^\bullet(\Loop M)$, where 
$$ \RedStringCoH(\Loop M) \coloneqq \StringCoH(\EG \Sph{1} \times_{\Sph{1}} \Loop M, \EG \Sph{1} \times_{\Sph{1}} \{x_0\}) $$
is the \emph{reduced $\Sph{1}$-equivariant cohomology} with respect to a base point $x_0 \in M$ (the constant loop at $x_0$). Dualizing their map, we obtain the isomorphism 
\begin{equation}\label{Eq:StringIsom}
\RedCycCoH^{-\bullet-1}(\DR(M))\simeq \RedStringH_\bullet(\Loop M; \R).
\end{equation}

Suppose from now on that $M$ is closed. Pick a Riemannian metric and an admissible Hodge propagator $\Prpg \in \DR^{n-1}(\Bl_\Diag(M\times M))$. We will assume that $\Prpg$ is special, i.e., that it satisfes (P1)--(P5) from Section~\ref{Section:Proof1}, so that the Chern-Simons Maurer-Cartan element~$\PMC$ is strictly reduced, and hence the twisted reduced $\IBLInfty$-algebra $\dIBL^\PMC\bigl(\RedCycC(\Harm)\bigr)$ and the induced $\IBL$-algebra $\IBL(\HIBL^{\PMC,\mathrm{red}}(\CycC(\Harm)))$ are well-defined.
Recall that $\HIBL^\PMC_\bullet(\CycC(\Harm)) = \CycH_{n-3-\bullet}(\Harm_\PMC)$, where $\Harm_\PMC$ is the $\AInfty$-algebra on $\Harm$ twisted by $\PMC_{10}$. From~\cite{Cieliebak2018}, we have
\begin{equation}\label{Eq:IBLIsom}
\CycCoH(\Harm(M)_\PMC) \simeq \CycCoH(\DR(M)).
\end{equation}
Now, \eqref{Eq:StringIsom} and \eqref{Eq:IBLIsom} give the following version of the string topology conjecture.

\begin{Conjecture}[String topology conjecture for simply-connected manifold]\label{Conj:StringTopology}
Let $M$ be an oriented closed manifold of dimension $n$. There is a chain map 
$$(C^{\mathrm{sing}}(\Loop_{\Sph{1}}M; \R),\Bdd)\longrightarrow (\CDBCyc\Harm(M),\OPQ_{110}^\PMC), $$
where $C^{\mathrm{sing}}$ denotes the (smooth) singular chain complex and $\Bdd$ the standard boundary operator, which, if $M$ is simply-connected, satisfies the following:
\begin{itemize}
\item It induces an isomorphism $\RedStringH(\Loop M; \R)[2-n] \simeq \HIBL^{\PMC,\mathrm{red}}\bigl(\CycC(\Harm(M))\bigr)$.
\item It intertwines $\StringOp_2$ on $\StringH(\Loop M; \R)$ and $\OPQ_{210}$.
\item The pullback of $\OPQ_{120}^\PMC$ to $\RedStringH(\Loop M; \R)$ is compatible with $\StringCoOp_2$ on $\ConstRedStringH(\Loop M; \R)$ under the morphism induced by the inclusion $(\Loop M, x_0) \rightarrow (\Loop M,M)$.
\end{itemize}
\end{Conjecture}

\begin{Remark}[On string topology conjecture]\phantomsection
\begin{RemarkList}
\item The conjecture can be interpreted as follows. There is an $\IBL$-structure on $\RedStringH(\Loop M;\R)$ compatible with Chas-Sullivan operations, and the $\IBLInfty$-algebra $\dIBL^\PMC(\RedCycC(\Harm(M)))$ is its chain model.
\item The loop coproduct $\tau$ is geometrically defined only on $\ConstRedStringH(\Loop M)$; the conjecture thus provides an extension of $\StringCoOp_2$ to $\RedStringH(\Loop M)$. In~\cite{Basu2011}, it is shown that the geometric definition of $\tau$ can be extended to $\H(\Loop M)$ for manifolds with zero Euler characteristic, i.e., $\chi(M) = 0$. This extension depends on the choice of a non-vanishing vector field on~$M$. By homotopy invariance (see (v) below), our extension of $\StringCoOp_2$ should not depend on the admissible Hodge propagator $\Prpg$.
\item The loop product $\LoopPr$ is geometrically defined on $\H(\Loop M)$; however, it does not always induce an associative product on $\RedH(\Loop M) = \H(\Loop M, x_0)$. Indeed, the examples of $\T^2$ (see \cite{Basu2011}) and $\Sph{3}$ (see \cite{Sullivan1999}) show that $\H(x_0;\R) \subset \H(\Loop M;\R)$ is not an ideal with respect to $\LoopPr$. By \cite{Tamanoi2010}, this does not happen when $\chi(M) \neq 0$, and hence, in this case,~$\LoopPr$ restricts to $\H(\Loop M, x_0; \R)$.
\item The computation for $\Sph{n}$ with $n\ge 2$ and the computation for $\CP^n$ in Section~\ref{Section:Computation} support the conjecture. The computation for $\Sph{1}$ in Section~\ref{Section:HomSphere} provides a counterexample for non-simply-connected $M$. Surfaces of genus~$g\ge 1$ should be considered.

\item  We expect that if $M_1$ and $M_2$ are homotopy equivalent, then the $\IBLInfty$-algebras $\dIBL^{\PMC_1}(\CycC(\HDR(M_1)))$ and $\dIBL^{\PMC_2}(\CycC(\HDR(M_2)))$ are $\IBLInfty$-homotopy equivalent.
\qedhere
\end{RemarkList}
\end{Remark}

\chapter{Explicit computations}

\label{Section:Computation}
In Section~\ref{Sec:GreenSphere}, we solve the differential equation for the Hodge propagator~$\Prpg$ for~$\Sph{n}$ (Proposition~\ref{Prop:GKerSph}) using the Relative Poincar\'e Lemma (Lemma~\ref{Lem:ChainHtpy}). In the rest of the section, we will be showing that $\Prpg$ is admissible (Proposition~\ref{Proposition:GreenKernel}); the most work is to show that $\Prpg$ extends smoothly to the blow-up (Proposition~\ref{Prop:GKerBdd}). Another Hodge propagator for~$\Sph{1}$ can be obtained in an alternative simple way by writing $\Sph{1} = \R / \Z$, and there are nice geometric formulas for $\Prpg$ for $\Sph{2}$ (Example~\ref{Example:Circle}).

In Section \ref{Section:MCSphere}, we use $\Prpg$ from Section~\ref{Sec:GreenSphere} to compute the Chern-Simons Maurer-Cartan element $\PMC$ for $\Sph{n}$ (Proposition~\ref{Proposition:MCSphere}). We first prove that the condition~$(V_{\NOne})$ from Proposition~\ref{Prop:PMCEqualsMC} is satisfied (Lemma~\ref{Lemma:ABVanishing}) and then perform combinatorics with degrees to show vanishing of some more integrals (Proposition~\ref{Prop:TotalVanishing}). In fact, all the integrals vanish for $\Sph{n}$ with $n\ge 3$, and the only non-vanishing integrals for $\Sph{1}$ are the $O_k$-graphs with even $k$. We compute these integrals explicitly together with all signs and combinatorial coefficients required to obtain $\PMC_{20}$ (Lemmas~\ref{Lemma:IntegralFor1}, \ref{Lemma:Independence}, \ref{Lemma:SignForMCOnCircle} and  \ref{Lemma:CombinatorialCoefficientForMCOnCircle}). There might be some non-vanishing integrals associated to reduced graphs for $\Sph{2}$ as well as some non-vanishing integrals associated to graphs without external vertices for $\Sph{3}$; however, the simplest examples vanish (Remarks~\ref{Rem:GraphsTwoSphere} and~\ref{Rem:GraphsThreeSphere}).  

In the remaining Sections~\ref{Section:HomSphere} and~\ref{Section:CPn1}, we compute $\IBL(\HIBL^\PMC(\CycC(\Harm(M))))$ and the higher operations~$\OPQ_{1lg}^\PMC$ on $\HIBL^\PMC$ for $M = \Sph{n}$, $\CP^n$. As soon as we argue that $\PMC_{10} = \MC_{10}$ due to geometric formality, the computation of $\HIBL^\MC(\CycC(\Harm(\Sph{n})))$ and $\HIBL^\MC(\CycC(\Harm(\CP^n)))$ is an easy exercise in cyclic homology. The operations for $\Sph{2m}$ and $\CP^n$ vanish for degree reasons (Remark~\ref{Rem:DegRes}). Therefore, the integrals from Section \ref{Section:MCSphere} help only in the case of $\Sph{2m-1}$. We compare our results to Chas-Sullivan string topology from~\cite{Basu2011} and confirm Conjecture~\ref{Conj:StringTopology} for~$\Sph{n}$ with~$n\ge 2$ and for~$\CP^n$.

\section{Construction of a Hodge propagator for spheres}
\allowdisplaybreaks
\label{Sec:GreenSphere}

The standard Riemannian volume form on the round sphere $\Sph{n}\subset \R^{n+1}$ is the restriction of the following closed form on $\R^{n+1}\backslash \{0\}$:  
\[ \Vol(x) \coloneqq \frac{1}{|x|^{n+1}}\sum_{i=1}^{n+1} (-1)^{i+1} x^i \Diff x_1 \dotsm \widehat{\Diff x_i} \dotsm \Diff x_{n+1}. \]
Here $\widehat{\Diff x_i}$ means that $\Diff{x_i}$ is omitted.
We denote the Riemannian volume of $\Sph{n}$ by
\[ V\coloneqq \int_{\Sph{n}} \Vol. \]
The $n$-form $\HKer$ from Proposition~\ref{Lemma:HKer} reads
\begin{equation*}
\HKer =  \frac{1}{V}\bigl(\Pr_1^*\Vol +  (-1)^{n} \Pr_2^*\Vol\bigr).
\end{equation*}
According to Proposition~\ref{Prop:GKer}, the equation which we want to solve reads
\begin{equation} \label{Eq:GreenKernel}
\Dd \Prpg = \frac{1}{V}\bigl((-1)^{n}\Pr_1^*\Vol +  \Pr_2^*\Vol\bigr).
\end{equation}
We denote 
\[\tilde{\Prpg}\coloneqq V\Prpg\quad\text{and}\quad \tilde{\HKer}\coloneqq V\HKer.\]

The following lemma will be used to construct a solution to~\eqref{Eq:GreenKernel}. 
\begin{Lem}[Relative Poincar\'e Lemma]\label{Lem:ChainHtpy}
Let $M$ be a smooth oriented manifold and $\psi: [0,1]\times M \rightarrow M$ a smooth map. Consider the operator $T: \DR^*(M) \rightarrow \DR^{*-1}(M)$ defined by
\[ T(\eta) \coloneqq \FInt{[0,1]} \psi^*\eta\quad\text{for all }\eta\in \DR(M), \]
where we integrate along the fiber of the oriented fiber bundle $\Pr_2: [0,1]\times M \rightarrow M$. Then we have
\[ \Dd\circ T + T\circ \Dd = \psi_1^* - \psi_0^*. \]
\end{Lem}
\begin{proof}
Stokes' formula from Proposition~\ref{Prop:StokesForm} gives
\[ \Dd \FInt{[0,1]}\psi^*\eta = -\Bigl( \FInt{[0,1]} \Dd \psi^* \eta - \FInt{\Bdd [0,1]} \psi^* \eta \Bigr) = - \FInt{[0,1]} \psi^* \Dd \eta + \psi^*_1 \eta - \psi^*_0 \eta\]
for all $\eta \in \DR(M)$.
\end{proof}

\begin{Proposition}[Solution to \eqref{Eq:GreenKernel}] \label{Prop:GKerSph}
For all $(x,y)\in (\Sph{n}\times \Sph{n}) \backslash \Diag$, let
\begin{equation} \label{Eq:GreenKernelMC1}
\Prpg (x,y) \coloneqq (-1)^{n} \sum_{k=0}^{n-1} g_k(x,y)\omega_k(x,y),
\end{equation}
where
\begin{equation}\label{Eq:FunctionsGk1}
 g_k(x,y) \coloneqq \int_{0}^1 \frac{t^k(t-1)^{n-1-k}}{(2t(t-1)(1+x\cdot y) + 1)^{\frac{n+1}{2}}} \Diff{t}
\end{equation}
and 
\begin{equation} \label{Eq:FormOmega1}
\omega_k(x,y) \coloneqq \begin{multlined}[t]\frac{1}{k!} \frac{1}{(n-1-k)!} \sum_{\sigma \in \Perm_{n+1}} (-1)^\sigma x^{\sigma_1} y^{\sigma_2} \Diff{x^{\sigma_3}} \dotsm \Diff{x^{\sigma_{2+k}}} \\ \Diff{y^{\sigma_{3+k}}} \dotsm \Diff{y^{\sigma_{n+1}}}.\end{multlined}
\end{equation}
The form~\eqref{Eq:GreenKernelMC1} is a smoooth solution to~\eqref{Eq:GreenKernel} on $(\Sph{n}\times \Sph{n}) \backslash \Diag$.
\end{Proposition}
\begin{proof} %
Define the set
\[ N \coloneqq (\R^{n+1}_{\neq 0}\times\R^{n+1}_{\neq 0})\backslash \{(x,a x) \mid x\in \R^{n+1}, a>0\}. \]
It is an open thickening of $(\Sph{n}\times \Sph{n})\backslash \Diag$ in $\R^{n+1} \times \R^{n+1}\backslash \Diag $. Consider the smooth deformation retraction
\begin{align*}
 \psi : [0,1] \times N & \longrightarrow N \\
 (t,x,y) &\longmapsto \psi_t(x,y) \coloneqq (x,(1-t)y-tx)
\end{align*}
with 
\[ \psi_0(x,y)=(x,y) \quad \text{and}\quad \psi_1(x,y)= (x,-x)\quad\text{for all }(x,y)\in N. \]
The retraction is depicted in Figure~\ref{Fig:Retraction}.
\begin{figure}
\centering
\input{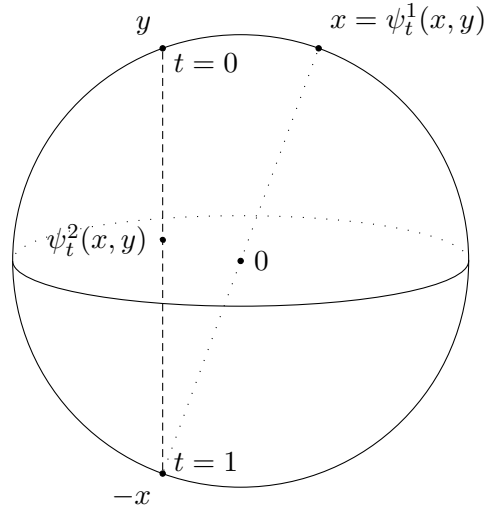}
\caption[Retraction of the configuration space $C_2(\Sph{n})$ to $\Sph{n}$.]{Retraction $\psi_t = (\psi_t^1, \psi_t^2)$. A point of $\Sph{n}\times \Sph{n}$ is visualized as a pair of points on $\Sph{n}$.}\label{Fig:Retraction}
\end{figure}
Denote by $A: \R^{n+1} \rightarrow \R^{n+1}$, $x\mapsto -x$ the antipodal map. It is easy to see that 
\[ A^* \Vol = (-1)^{n+1} \Vol, \]
and hence
\begin{equation*}
\psi_1^* \tilde{\HKer} = \psi_1^* \Pr_1^*\Vol + (-1)^n \psi_1^*\Pr_2^* \Vol  = \Pr_1^*\Vol + (-1)^n \Pr_1^*A^* \Vol = 0. 
\end{equation*}
Define 
\begin{equation}  \label{Eq:GExpr}
\Prpg \coloneqq (-1)^{n+1} \FInt{[0,1]} \psi^*\HKer.
\end{equation}
Let $T: \DR^*(N)\rightarrow \DR^{*-1}(N)$ be the cochain homotopy from Lemma~\ref{Lem:ChainHtpy} associated to $\psi$. Because $\Dd \HKer=0$, we get
\[ \Dd \Prpg = (-1)^{n+1} \Dd T(\HKer) = (-1)^{n+1}(\Dd T + T \Dd)\HKer = (-1)^{n+1}(\psi_1^*-\psi_0^*)\HKer = (-1)^{n}\HKer. \]
For every $i=1$,~$\dotsc$, $n+1$, we have
\[ \psi^*(\Diff{x^i}) = \Diff{x^i}\quad\text{and}\quad\psi^*(\Diff{y^i}) = (1-t)\Diff{y^i} - t\Diff{x^i} - (y^i+x^i)\Diff{t}. \]
We compute
\begin{align*}
 &  \FInt{[0,1]} \psi^* \tilde{\HKer}\\
 & \quad = (-1)^{n}\FInt{[0,1]} \psi^* \Pr_2^* \Vol \\
 & \quad = (-1)^{n+1}\FInt{[0,1]} \sum_{i=1}^{n+1} (-1)^{i}\frac{((1-t)y^i - t x^i)}{{\Abs{(1-t)y-tx}^{n+1}}} \psi^*(\Diff{y^1} \dotsm \widehat{\Diff{y^i}} \dotsm \Diff{y^{n+1}}) \\
 & \quad = (-1)^{n+1} \sum_{1\le i<j \le n+1} (-1)^{i+j}(x^i y^j - y^i x^j) \FInt{[0,1]}  \frac{\Diff{t} \psi^*(\Diff{y^1} \dotsm \widehat{\Diff{y^i}} \dotsm \widehat{\Diff{y^j}} \dotsm \Diff{y^{n+1}})}{\Abs{(1-t)y-tx}^{n+1}} \\ 
 & \quad = \begin{multlined}[t] - \sum_{k=0}^{n-1} \Bigl(\int_0^1 \frac{t^k(t-1)^{n-1-k}}{\Abs{(1-t)y-tx}^{n+1}} \Diff{t}\Bigr) \sum_{1\le i<j\le n+1} (-1)^{i+j+1} (x^i y^j - y^i x^j) \\
 \sum_{\mathclap{\substack{\sigma: \{1,\dotsc,n-1\}\rightarrow \{1,\dotsc,\hat{i},\dotsc,\hat{j},\dotsc,n+1\} \\ \sigma_1<\dots < \sigma_k \\ \sigma_{k+1}<\dots < \sigma_{n-1}}}} (-1)^\sigma \Diff{x^{\sigma_1}} \dotsm \Diff{x^{\sigma_k}}\Diff{y^{\sigma_{k+1}}} \dotsm \Diff{y^{\sigma_{n-1}}}.\end{multlined}
\end{align*}
The formulas~\eqref{Eq:FunctionsGk1} and~\eqref{Eq:FormOmega1} are obtained from this by writing 
\[ \Abs{(1-t)y - tx}^2 = 2t(t-1)(1+x \cdot y) + 1 \]
in the denominator of the integrand and by simple combinatorics in the form part, respectively. Smoothness of~$\Prpg$ on $(\Sph{n}\times \Sph{n})\backslash \Diag$ follows from the expression~\eqref{Eq:GExpr}.
\end{proof}
Note that $g_k$ are smooth functions on $(\Sph{n}\times \Sph{n})\backslash \Diag$.
\begin{Example}[$\Prpg$ for $\Sph{1}$ and $\Sph{2}$]\phantomsection\label{Example:Circle}%
\begin{ExampleList}
\item Let 
\[\alpha: (\Sph{1}\times\Sph{1})\backslash \Diag \rightarrow (0,2\pi)\]
be the smooth function assigning to a pair $(x,y)\in  (\Sph{1}\times\Sph{1})\backslash \Diag$ the counterclockwise angle from $x$ to $y$. Let $\alpha_1$, $\alpha_2 \in [0,2\pi)$ be such that $x=\cos(\alpha_1)\StdBasis_1 + \sin(\alpha_1)\StdBasis_2$ and $y=\cos(\alpha_2)\StdBasis_1+\sin(\alpha_2)\StdBasis_2$ for the standard Euclidean basis $\StdBasis_1$, $\StdBasis_2$ of $\R^2$. It is easy to see that
\[ \alpha(x,y) = \begin{cases} \alpha_2 - \alpha_1 & \text{if }\alpha_1<\alpha_2, \\
\alpha_2-\alpha_1+2\pi & \text{if }\alpha_1>\alpha_2. \end{cases} \]
Therefore, we get
\[ \Diff{\alpha} = \Diff{\alpha_2} -\Diff{\alpha_1} = -2\pi H \quad\text{on } (\Sph{1}\times\Sph{1})\backslash \Diag. \]
On the other hand, we can compute $\Prpg$ from~\eqref{Eq:GreenKernelMC1} as follows. Using the substitution $u=2t-1$, we get for all $x$, $y\in \Sph{1}$ with $x\neq \pm y$ the following:
\begin{align*}
 g_{0}(x,y) &= \int_0^1 \frac{\Diff{t}}{2t(t-1)(1+x\cdot y) +1} \\
 & =  \frac{1}{1-x\cdot y}\int_{-1}^1 \frac{\Diff{u}}{\frac{1+x\cdot y}{1- x\cdot y}u^2 + 1} \\
 & = \frac{2}{\sqrt{1-(x\cdot y)^2}}\arctan\Bigl(\sqrt{\frac{1+x\cdot y}{1-x\cdot y}}\Bigr) \\
 & = \frac{\pi - \arccos(x\cdot y)}{\sqrt{1-(x\cdot y)^2}} \\ 
 & = \frac{\pi - \arccos(x\cdot y)}{ \Abs{x^1 y^2 - x^2 y^1}} \\
 & = \frac{\pi - \alpha(x,y)}{x^1 y^2 - x^2 y^1}.
\end{align*}
The third from last equality can be obtained by trigonometric considerations and the second from last equality by an algebraic manipulation with the denominator. We will explain the last equality. Consider the matrix 
\[ R=\begin{pmatrix}
0 & -1 \\ 1 & 0
\end{pmatrix} \]
representing the counterclockwise rotation by $\frac{\pi}{2}$. The function $\arccos: (-1,1) \rightarrow (0,\pi)$ satisfies
\[ \arccos(x\cdot y) = \begin{cases}
                        \alpha(x,y) & \text{if }y\cdot Rx > 0, \\
                        2\pi - \alpha(x,y) & \text{if }y\cdot Rx<0. 
                       \end{cases}\]
The last equality becomes clear when we notice that $x^1 y^2 - x^2 y^1 = y\cdot Rx$. 

Finally, we have $\omega_{0}(x,y) = x^1 y^2 - x^2 y^1$, and hence
\begin{align*}
2\pi \Prpg(x,y) &=  - g_{0}(x,y) \omega_{0}(x,y) \\
& = \alpha(x,y) - \pi \\
& = \pi - \alpha(y,x).
\end{align*}
\item For $n=2$, we get the formulas
\allowdisplaybreaks
\begin{align*}
g_0(x,y) & = - g_1(x,y) = \frac{1}{x\cdot y - 1}\quad\text{and} \\
\omega_0(x,y) &=(x^2 y^3 - x^3 y^2) \Diff{y^1} +(x^3 y^1 - x^1 y^3) \Diff{y^2} + (x^1 y^2 - x^2 y^1) \Diff{y^3} \\ &=\sum_{i=1}^3 (x\times y)^i \Diff{y^i}.
\end{align*}
The formula for $\omega_1(x,y)$ is obtained from the formula for $\omega_0(x,y)$ by replacing~$\Diff{y}$ with~$\Diff{x}$.\qedhere
\end{ExampleList}
\end{Example}

Consider the diagonal action of the orthogonal group $O(n+1)$ on $\R^{n+1}\times\R^{n+1}$ by matrix multiplication.
\begin{Proposition}[Symmetries of $\Prpg$]\label{Prop:SymmetryOfG}
Consider $\Prpg$ from Proposition~\ref{Prop:GKerSph}. For all $R\in O(n+1)$, we have
\[ R^* \Prpg = (-1)^R \Prpg, \]
where $(-1)^R = \det(R)$. Moreover, if $\tau$ denotes the twist map, then
\[\tau^*\Prpg = (-1)^{n}\Prpg. \] 
\end{Proposition}
\begin{proof} %
We will use the thickening $N$, the antipodal map $A$ and the expression~\eqref{Eq:GExpr} for~$\Prpg$ from the proof of Proposition~\ref{Prop:GKerSph}.

It is easy to check that both $\tau$ and $R$ preserve $N$. Let~$\tilde{\tau}$ and $\tilde{R}$ be the isomorphisms of the fiber bundle $\Pr_2: [0,1]\times N \rightarrow N$ given by 
\[ \tilde{\tau}(t,x,y) \coloneqq (1-t,y,x)\quad \text{and}\quad \tilde{R}(t,x,y) \coloneqq (t,Rx,Ry) \]
for all $(t,x,y)\in [0,1]\times N$. Then $\tilde{\tau}$ covers $\tau$ and $\tilde{R}$ covers $R$. A simple computation directly from Definition~\ref{Def:FibInt} shows that the fiberwise integration commutes with the pullback along a bundle morphism if the bundle map and the base map are both either orientation preserving or reversing. In our case, we have 
\[ (-1)^{\tau + \tilde{\tau}} = -1\quad \text{and}\quad (-1)^{R+\tilde{R}} = 1. \]
Using this and the equation
\[ \Pr_2 \circ \psi \circ \tilde{\tau} = A\circ \Pr_2 \circ \psi, \]
we get firstly
\allowdisplaybreaks
\begin{align*}
\tau^* \FInt{[0,1]} \psi^*\tilde{\HKer} &=  - \FInt{[0,1]} \tilde{\tau}^* \psi^*\Pr_2^*\Vol \\ &= - \FInt{[0,1]} \psi^*\Pr_2^*A^* \Vol \\ &= (-1)^n \FInt{[0,1]} \psi^*\Pr_2^*\Vol \\ & = (-1)^n \FInt{[0,1]} \psi^*\tilde{\HKer}
\end{align*}
and secondly
\[ R^* \FInt{[0,1]} \psi^* \HKer = \FInt{[0,1]} \tilde{R}^* \psi^* \HKer = \FInt{[0,1]} \psi^* R^*\HKer = (-1)^{n+1} \FInt{[0,1]} \psi^* \HKer. \]
This proves the proposition.
\end{proof}
Both diffeomorphisms $R$ and $\tau$ preserve $\Delta$, and hence they extend to diffeomorphisms of $\Bl_\Diag(\Sph{n}\times\Sph{n})$. If also $\Prpg$ extends, then the statement of Proposition~\ref{Prop:SymmetryOfG} holds for $\Prpg$ on $\Bl_\Diag(\Sph{n}\times\Sph{n})$.

In the rest of the section, we will be proving that $\Prpg$ extends smoothly to $\Bl_{\Diag}(\Sph{n}\times \Sph{n})$. This is a local problem at the boundary, where we introduce the following radial coordinates. Define the set
\[ X \coloneqq \{(r,\eta,x)\in [0,\infty)\times \Sph{n}\times\Sph{n} \mid \eta\cdot x = 0\}, \]
and let $\kappa: X \longrightarrow \Bl_\Diag(\Sph{n}\times \Sph{n})$ be the map defined by
\begin{align*}
\kappa(r,\eta,x) &\coloneqq \begin{cases} 
\Bigl(x,\dfrac{x+r\eta}{\Abs{x+r\eta}}\Bigr)\in (\Sph{n}\times \Sph{n})\backslash \Diag & \text{for }r>0, \\[2ex]
[(-\eta,\eta)]\in P^+ N_{(x,x)}\Diag & \text{for }r=0.
\end{cases}
\end{align*}
Recall that the oriented projectivization $P^+$ was defined in Definition~\ref{Def:SphBlow}.
For the upcoming computations, it is convention to define the map $\gamma: \R \rightarrow (-1,1)$ by
\[ \gamma(r) \coloneqq \frac{r}{\sqrt{1+r^2}+1} \quad\text{for all }r\in \R. \]
It is a diffeomorphism with inverse $r = \frac{2 \gamma}{1-\gamma^2}$.

\begin{figure}[t]\centering
\input{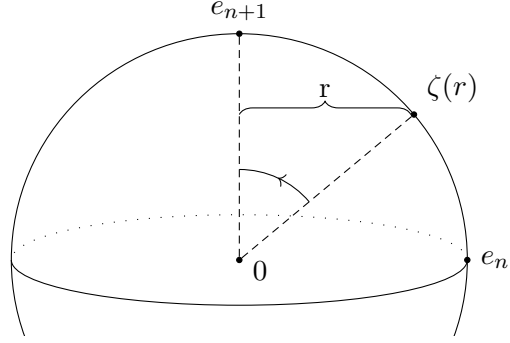}
\caption[A curve approaching the diagonal in the configuration space $C_2(\Sph{n})$.]{The curve $\zeta\coloneqq \kappa \circ \zeta'$ is given by $\zeta(r)=\bigl(e_{n+1},\frac{e_{n+1}+r e_n}{\Abs{e_{n+1}+r e_n}}\bigr)$ for $r>0$.}\label{Fig:CurveOnSphere}
\end{figure}

\begin{Lem}[Parametrization of collar neighborhood] \label{Lem:NewBlowupParam}
The subset $X\subset \R\times \R^{n+1}\times\R^{n+1}$ is a submanifold with boundary, and the map $\kappa: X \longrightarrow \Bl_\Diag(\Sph{n}\times \Sph{n})$ is an embedding onto a neighborhood of $\Bdd \Bl_\Diag(\Sph{n}\times \Sph{n})$.
\end{Lem}
\begin{proof}
The set $X$ is a Cartesian product of $[0,\infty)$ and a regular level set; therefore, it is a submanifold with boundary. The inclusion $\Sph{n}\times \Sph{n} \subset \R^{n+1}\times \R^{n+1}$ induces an embedding of manifolds with boundary $\Bl_\Diag(\Sph{n}\times\Sph{n})\subset \Bl_\Diag(\R^{n+1}\times \R^{n+1})$.  Consider the global chart $\tilde{\Id}: \Bl_\Diag(\R^{n+1}\times \R^{n+1}) \rightarrow [0,\infty) \times \Sph{n} \times \R^{n+1}$ from~\eqref{Eq:BlowUpChart} induced by the identity. We have
\[ \begin{aligned}Y &\coloneqq \tilde{\Id}(\Bl_\Diag(\Sph{n}\times \Sph{n})) \\ &= \{(\tilde{r},w,u)\in [0,\infty) \times \Sph{n} \times \R^{n+1} \mid \Abs{u}^2+\tilde{r}^2=1,\ w\cdot u = 0\}, \end{aligned}\] 
where we denote $r$ on $Y$ by $\tilde{r}$ in order to distinguish it from $r$ on $X$. It suffices to prove the claim for the map $\mu\coloneqq \tilde{\Id}\circ\kappa: X \rightarrow Y$. For $(r,\eta,x)\in X$, we compute
\[ \mu(r,\eta,x) = \biggl( \frac{\gamma}{\sqrt{1+\gamma^2}}, \frac{1}{\sqrt{1+\gamma^2}}(\gamma x - \eta), \frac{1}{1+\gamma^2}(x+\gamma \eta) \biggr). \]
This formula defines a smooth map of $\R\times \R^{n+1}\times\R^{n+1}$.
It is a local diffeomorphism because its Jacobian is non-vanishing:
\[ |\Jac{\mu}| = \frac{\partial \tilde{r}}{\partial r} \Bigl(\frac{\partial w}{\partial \eta}\frac{\partial u}{\partial x}  - \frac{\partial w}{\partial x}\frac{\partial u}{\partial \eta}\Bigr)^{n+1} = (-1)^{n+1}(1+\gamma^2)^{-\frac{n+4}{2}} \frac{\partial \gamma}{\partial r}. \]
Moreover, the map $\mu$ is injective, maps $X$ into $Y$ and $\Bdd X$ onto $\Bdd Y$. The claim follows.
\end{proof}
Consider the action of $O(n+1)$ on $X$ defined by
\[ R\cdot(r,\eta,x) \coloneqq (r,R\eta,Rx)\quad\text{for all }(r,\eta,x)\in X\text{ and }R\in O(n+1). \]
Via $\kappa$, this agrees with the diagonal action of $O(n+1)$ on $\Bl_\Diag(\Sph{n}\times\Sph{n})$. Denote
\[ \Prpg'\coloneqq \kappa^* \Prpg \in \Omega^{n-1}(\Int(X)). \]
From Proposition~\ref{Prop:SymmetryOfG} we get
\begin{equation} \label{Eq:SymmetryOfGPrime}
R^* \Prpg' = (-1)^R \Prpg'\quad \text{for all }R\in O(n+1).
\end{equation}
Consider the smooth curve (see Figure~\ref{Fig:CurveOnSphere})
\begin{equation*}\label{Eq:CurveZetaDef}
 \begin{aligned} \zeta': [0,\infty) &\longrightarrow  X \\
                    r &\longmapsto (r,e_n,e_{n+1}). \end{aligned}
\end{equation*}
We have the following lemma.
\begin{Lem}[Smooth extension along curve] \label{Lem:ExtAlongCurve}
The form $\Prpg'$ extends smoothly to $X$ if and only if the map $\Prpg'\circ \zeta' : (0,\infty)\rightarrow \Lambda^{n-1}T^* X$ extends smoothly to the interval $[0,\infty)$. 
\end{Lem}
\begin{proof}
As for the non-trivial implication, let $(0,\eta_0,x_0) \in X$ be a boundary point. Pick vectors $v_1$,~$\dotsc$, $v_{n-1}\in \R^{n+1}$ so that the vectors $v_1$,~$\dotsc$, $v_{n-1}$, $\eta_0$, $x_0$ are linearly independent, and define the set 
\[ U\coloneqq\{(r,\eta,x)\in X \mid v_1,\,\dotsc,\,v_{n-1},\,\eta,\,x \text{ are linearly independent}\}. \]
It is an open neighborhood of $(0,\eta_0,x_0)$ in $X$. Applying the Gram-Schmidt orthogonalization to $v_1$,~$\dotsc$, $v_{n-1}$, $\eta$, $x$, we find a smooth map $R: U \rightarrow O(n+1)$ such that 
\[ R(r,\eta,x)\cdot (r,\eta,x) = (r,e_n,e_{n+1}) \quad \text{for all }(r,\eta,x)\in U. \]
The equation~\eqref{Eq:SymmetryOfGPrime} implies
\[ \Prpg'(r,\eta,x) =(-1)^R R(r,\eta,x)^*\bigl(\Prpg'(r,e_n,e_{n+1})\bigr)\quad\text{for all }(r,\eta,x)\in \Int(U), \]
where $R(r,\eta,x)^*: \Lambda^* T^* X \rightarrow \Lambda^* T^* X$ is the smooth cotangential map which is induced by the diffeomorphism $R(r,\eta,x): X\rightarrow X$, and which maps the fiber over $z\in X$ to the fiber over $R(r,\eta,x)^{-1} z$. By the assumption, all maps in the composition are smooth in their arguments. The lemma follows.
\end{proof}

\begin{Lem}[Local expression at boundary] \label{Lem:FormulaAlongCurve}
On the interval $(0,\infty)$, we have
\begin{equation*}\label{Eq:FormulaAlongCurve}
\tilde{\Prpg}'\circ\zeta' = (-1)^{n+1}(1+\gamma^2)^{-\frac{n-1}{2}} \sum_{k=0}^{n-1} \gamma^{n-k} (h_{k}\circ\gamma)(\nu_{k}\circ\zeta'), 
\end{equation*}
where the functions $h_{k}: (0,1)\rightarrow \R$ are defined by
\begin{equation*}
  h_{k}(\gamma)\coloneqq \int_{-1}^1 \frac{(u+\gamma^2)^{k}(u-1)^{n-1-k}}{(u^2+\gamma^2)^{\frac{n+1}{2}}} \Diff{u} \quad \text{for all }\gamma\in (0,1)
\end{equation*}
and the forms $\nu_{k}\in \Omega(X)$ are defined by
\begin{align*}
 \nu_{k}(r,x,\eta) &\coloneqq \frac{1}{k!(n-1-k)!}\sum_{\sigma\in \Perm_{n-1}} (-1)^\sigma \Diff{x^{\sigma_1}}\dotsm
\Diff{x^{\sigma_k}} \Diff{\eta^{\sigma_{k+1}}} \dotsm \Diff{\eta^{\sigma_{n-1}}}.
\end{align*}
\end{Lem}

\begin{proof}
We start with the following formula from the proof of Proposition~\ref{Prop:GKerSph}:
\[ \tilde{\Prpg} =  \sum_{1\le i<j \le n+1} (-1)^{i+j}(x^i y^j - y^i x^j) \FInt{[0,1]}  \frac{\Diff{t} \psi^*(\Diff{y^1} \dotsm \widehat{\Diff{y^i}} \dotsm \widehat{\Diff{y^j}} \dotsm \Diff{y^{n+1}})}{\Abs{(1-t)y-tx}^{n+1}}. \]
We restrict to the points $(x,y)=\kappa(r,e_n,e_{n+1})$ with $r>0$. There, we have
\begin{align*}
&x^1= \dotsb =x^n=0,\ x^{n+1}=1, \\
&y^1 = \dotsb = y^{n-1} =0,\ y^n= \frac{2\gamma}{1+\gamma^2},\ y^{n+1} = \frac{1-\gamma^2}{1+\gamma^2}.
\end{align*}
Under the substitution $u = 2t-1$, we get
\[\Abs{(1-t)y-t x}^2 = \frac{4 t (t-1)}{1+\gamma^2}+1=\frac{u^2+\gamma^2}{1+\gamma^2}. \]
We make the following preliminary computations:
\[ \begin{aligned}
x^i y^j - y^i x^j & = 0\quad\text{for }1\le i \le n-1\text{ and }i<j\le n+1, \\
x^n y^{n+1} - y^n x^{n+1} & = -\frac{2\gamma}{1+\gamma^2}, \\
\kappa^*(\Diff{y^i}) & = \frac{1}{1+\gamma^2}\bigl((1-\gamma^2)\Diff{x^i}+ 2\gamma \Diff{\eta^i}\bigr)\quad\text{for } 1\le i \le n-1. \end{aligned} \]
We plug these in the formula for $\tilde{\Prpg}$ and get
\begin{align*}
 \tilde{\Prpg}'(\zeta'(r)) &= 2\gamma(1+\gamma^2)^{\frac{n-1}{2}} \FInt{[0,1]}\Diff{t}\frac{\prod_{i=1}^{n-1} \bigl((1-t)\kappa^*(\Diff{y^i}) - t\Diff{x^i}\bigr)}{(u^2+\gamma^2)^{\frac{n+1}{2}}}  \\ 
&=(-1)^{n+1}\gamma(1+\gamma^2)^{-\frac{n-1}{2}}\FInt{[-1,1]} \Diff{u} \frac{\prod_{i=1}^{n-1}\bigl((u+\gamma^2)\Diff{x^i} + \gamma(u-1)\Diff{\eta^i}\bigr)}{(u^2+\gamma^2)^{\frac{n+1}{2}}}  \\
& =(-1)^{n+1}(1+\gamma^2)^{-\frac{n-1}{2}}\sum_{k=0}^{n-1} \gamma^{n-k}\Bigl( \int_{-1}^1 \frac{(u+\gamma^2)^{k}(u-1)^{n-1-k}}{(u^2+\gamma^2)^{\frac{n+1}{2}}} \Diff{u}\Bigr) \nu_{k}.
\end{align*}
The lemma follows.
\end{proof}
\begin{Lem}[Integrals depending on parameter] \label{Lem:GeneralIntegralExtension}
Let $n\in \N$, and let $l\in\{ 0, 1,~\dotsc, n-1\}$. The function $F_{n,l}: (0,\infty)\rightarrow \R$ defined by
\begin{equation}\label{Eq:GeneralIntegral}
F_{n,l}(t) \coloneqq \int_{-1}^1 \frac{t^{n-l} u^l}{(u^2+t^2)^{\frac{n+1}{2}}} \Diff{u}\quad\text{for all }t\in(0,\infty)
\end{equation}
extends smoothly to $[0,\infty)$.
\end{Lem} 
\begin{proof}
We have
\[ F_{1,0}(t) = 2 \arctan\Bigl(\frac{1}{t}\Bigr)=\pi - 2 \arctan(t) \quad\text{for all }t\in (0,\infty). \]
The right-hand side is a smooth function on $\R$.

For $n\ge 2$, we deduce the recursive formula
\[ F_{n,0}(t) = \frac{1}{n-1}\Bigl((n-2)F_{n-2,0}(t)+\frac{2 t^{n-2}}{(1+t^2)^{\frac{n-1}{2}}}\Bigr).\]
If $l$ is odd, then $F_{n,l}\equiv 0$ for all $n$ because the integrand of \eqref{Eq:GeneralIntegral} is odd as a function of $u$.

For $n\ge 3$ and even $2\le l \le n-1$, we deduce yet another recursive formula
\[  F_{n,l}(t) = \frac{1}{n-l}\Bigl((l-1) F_{n,l-2}(t) -  \frac{2 t^{n-l}}{(1+t^2)^{\frac{n-1}{2}}} \Bigr).
\]
The claim for all $F_{n,l}$ follows by induction.
\end{proof}

\begin{Proposition}[Smooth extension to boundary]\label{Prop:GKerBdd}
The form $\Prpg$ from~\eqref{Eq:GreenKernelMC1} extends smoothly to $\Bl_\Diag(\Sph{n}\times\Sph{n})$.
\end{Proposition}
\begin{proof}
According to Lemmas~\ref{Lem:NewBlowupParam} and~\ref{Lem:ExtAlongCurve}, it suffices to show that the curve $\Prpg'\circ \zeta': (0,\infty)\rightarrow \Lambda^{n-1}T^* X$ extends smoothly to $[0,\infty)$. Lemma~\ref{Lem:FormulaAlongCurve} gives an expression for $\Prpg'\circ \zeta'$ as a linear combination of smooth forms $\nu_{k}\in \Omega^{n-1}(X)$ with coefficients $\gamma^{n-k}(h_{k}\circ \gamma)$ for $k=0$,~$\dotsc$, $n-1$ multiplied by the overall coefficient $(-1)^n (1+\gamma^2)^{-\frac{n-1}{2}}$. We expand
\begin{equation*}
\gamma^{n-k}(h_{k}\circ\gamma) = \sum_{a=0}^k\sum_{b=0}^{n-1-k}(-1)^{n-1-k-b} \binom{k}{a} \binom{n-1-k}{b} \int_{-1}^1 \frac{\gamma^{n+k-2a}u^{a+b}}{(u^2+\gamma^2)^{\frac{n+1}{2}}} \Diff{u}
\end{equation*}
and notice that we can write
\begin{equation*}
 \int_{-1}^1 \frac{\gamma^{n+k-2a}u^{a+b}}{(u^2+\gamma^2)^{\frac{n+1}{2}}} \Diff{u} = \gamma^{k-a+b}(F_{n,a+b}\circ \gamma)
\end{equation*}
for the function $F_{n,l}$ from~\eqref{Eq:GeneralIntegral} with $l\coloneqq a+b$. Because $0\le l \le n-1$, Lemma~\ref{Lem:GeneralIntegralExtension} asserts that $F_{n,l}$ extends smoothly to $[0,\infty)$. Because $k-a+b\ge 0$, the entire coefficient at $\nu_{k}$ extends smoothly to $[0,\infty)$ for every $k=0$,~$\dotsc$, $n-1$. The lemma follows.
\end{proof}

We summarize our results in the following proposition:

\begin{Proposition}[Hodge propagator for $\Sph{n}$]\label{Proposition:GreenKernel}
The form $\Prpg$ from~\eqref{Eq:GreenKernelMC1} defines a Hodge propagator for $\Sph{n}$ satisfying Definition~\ref{Def:GreenKernel}. Moreover, we have the symmetries
\begin{align*}
R^* \Prpg &= (-1)^R \Prpg\quad \text{for all }R\in O(n+1)\text{ and} \\
\tau^* \Prpg & = (-1)^n \Prpg.
\end{align*}
\end{Proposition}
\begin{proof}
The proposition is a summary of Propositions~\ref{Prop:GKerSph},~\ref{Prop:SymmetryOfG} and~\ref{Prop:GKerBdd}.
\end{proof}

\begin{Remark}[Better notation due to R. Bryant, see  \cite{MO291535}] \label{Remark:Bryant}
Pick an oriented basis $e_1$,~$ \dotsc$, $e_{n+1}$ of $\R^{n+1}$ as generators of the exterior algebra $\Lambda^*(\R^{n+1})$, and view $x$, $y$, $\Diff{x}$, $\Diff{y}$ as $\Lambda^*(\R^{n+1})$-valued forms on~$\R^{n+1}$. For example, we view $x$ as the map $x\in \R^{n+1} \mapsto \sum_{i=1}^{n+1} x^i e_i \in \Lambda^1(\R^{n+1})$ and $\Diff{x}$ as the map $x\in \R^{n+1} \mapsto \sum_{i=1}^{n+1} (\Diff{x}_i)_x e_i \in \Lambda^1(\R^{n+1})$. There is a natural wedge product on the space of $\Lambda^*(\R^{n+1})$-valued forms. If $\omega$ is a top-form, we denote by $[\omega]$ the coefficient of $\omega$ at $e_1 \wedge \dotsm \wedge e_{n+1}$. Then it holds
\[ \omega_k(x,y) = \frac{1}{k!}\frac{1}{(n-1-k)!}[x\wedge y \wedge (\Diff{x})^{k} \wedge (\Diff{y})^{n-1-k}]. \]

Note that if we view $e_i$ as odd variables, then $[\cdot]$ corresponds to the odd integration~$\int \mathrm{D}e(\cdot)$. It would be interesting to know whether this notation simplifies some proofs, especially if Lemma~\ref{Lemma:ABVanishing} can be deduced from abstract algebraic facts or rules valid for odd integration.
\end{Remark}

\section{Computation of Chern-Simons Maurer-Cartan element for spheres}
\allowdisplaybreaks
\label{Section:MCSphere}

We recall from Definition~\ref{Def:PushforwardMCdeRham} that the Chern-Simons Maurer-Cartan element~$\PMC$ is computed as a sum over trivalent ribbon graphs decorated with the admissible Hodge Propagator~$\Prpg$ at internal edges, integration variables $x_i$ at internal vertices and, in the case of $\Sph{n}$, with $\NOne$ or $\NVol$ at external vertices.
\begin{figure}[t]
\centering
\input{\GraphicsFolder/ygraph.tex}
\caption{The $Y$-graph for $\Sph{n}$.}\label{Fig:YGraph}
\end{figure}

The canonical Maurer-Cartan element $\MC$ is the contribution of the Y-graph (see Figure~\ref{Fig:YGraph}), and it is easy to see that
\begin{equation*}
 \MC_{10}(\Susp \NVol \NOne \NOne ) = (-1)^n \MC_{10}(\Susp\NOne \NVol \NOne) = \MC_{10}(\Susp \NOne \NOne \NVol) = (-1)^{n-2}.
\end{equation*}
Throughout this section, we will be in the setting of Definition~\ref{Def:PushforwardMCdeRham}.
In particular, $\Gamma \in \TRG_{klg}$ is a ribbon graph, $L$ its compatible labeling admissible with respect to an input $\omega_1$, $\dotsc$, $\omega_l$ and $I(\sigma_L)$ the corresponding integral.
\begin{Lemma}[Condition $(V_{\NOne})$ holds] \label{Lemma:ABVanishing}
For the Hodge Propagator $\Prpg$ from~\eqref{Eq:GreenKernelMC1} for $\Sph{n}$ for $n\ge 1$, every graph $\Gamma \neq Y$ with~$\NOne$ at an external vertex vanishes.
\end{Lemma}
\begin{proof}
The only contribution of an $A$-vertex which does not vanish for degree reasons is
\[ A_{\NVol,\NOne}(y) = \int_x \Prpg(x,y) \Vol(x). \]
From the symmetry of $\Prpg$ and $\Vol$ under the action of $O(n+1)$, we get
\[ R^* A_{\NVol, \NOne} = (-1)^R A_{\NVol, \NOne}\quad\text{for all }R\in O(n+1). \]
Therefore, it suffices to check that $A_{\NVol, \NOne}(\StdBasis_1) = 0$, where $\StdBasis_1$, $\dotsc$, $\StdBasis_{n+1}$ denotes the standard basis of $\R^{n+1}$.
Evaluation of~\eqref{Eq:FormOmega1} at $(x,\StdBasis_1)$ gives
\[ \omega_0(x,\StdBasis_1) = \frac{1}{(n-1)!} \sum_{\substack{\sigma\in\Perm_{n+1}\\ \sigma_2 = 1}} (-1)^\sigma x^{\sigma_1} \Diff{y^{\sigma_3}} \dotsm \Diff{y^{\sigma_{n+1}}}. \]
Therefore, we get
\begin{align*}
A_{\NVol, \NOne}(\StdBasis_1) &= (-1)^n \int_x g_0(x\cdot \StdBasis_1) \omega_0(x,\StdBasis_1) \Vol(x)  \\ 
&= (-1)^{n+1} \int_x g_0(x^1) \Bigl( \sum_{j=2}^{n+1} (-1)^{j} x^j\Bigr) \Vol(x) \Diff{y^2} \dotsm  \widehat{\Diff{y^j}} \dotsm \Diff{y^{n+1}},
\end{align*}
where we view $g_0$ as a function of $x\cdot y$.
Consider the cyclic permutation of $x^2$, $\dotsc$, $x^{n+1}$:
\begin{align*} 
I:\Sph{n} &\longrightarrow \Sph{n} \\
   (x^1, x^2, \dotsc, x^n, x^{n+1}) &\longmapsto (x^1, x^{n+1}, x^2 \dotsc, x^{n}). 
\end{align*}
Then we have
\begin{align*}
\int_x g_0(x^1) \sum_{j=2}^{n+1} (-1)^{j} x^j \Vol(x)  &= (-1)^{n-1}\int_x I^*\Bigl(g_0(x^1) \sum_{j=2}^{n+1} (-1)^{j} x^j \Vol(x)\Bigr) \\
& = (-1)^{n-1} \int_x g_0(x^1)\Bigl(-\sum_{j=2}^{n+1} (-1)^j x^j\Bigr)\bigl((-1)^{n-1}\Vol(x)\bigr) \\
& = - \int_x g_0(x^1) \sum_{j=2}^{n+1} (-1)^{j} x^j \Vol(x).
\end{align*}
It follows that $A_{\NVol,\NOne}(\StdBasis_1)=0$.\footnote{Notice that we avoided using $\sum \int = \int \sum$ because the convergence of single summands is not guaranteed.}

Let us now consider the contribution of a $B$-vertex with $\NOne$:
\[ B_{\NOne}(y,z) = \int_{x} \Prpg(y,x) \Prpg(x,z) = (-1)^n \int_x \Prpg(y,x) \Prpg(x,z). \]
For $n=1$, the degree of $\Prpg(y,x)\Prpg(x,z)$ is $0$, and hence $B_{\NOne}= 0$ trivially.
Suppose that $n\ge 2$.
As in the case of $A_{\NVol,\NOne}$, we get that 
\[ R^* B_{\NOne} = (-1)^R B_{\NOne} \quad\text{for all }R\in O(n+1). \]
Therefore, it suffices to check that $B_{\NOne}(\StdBasis_1, c_1 \StdBasis_1 + c_2 \StdBasis_2) = 0$ for all $(c_1,c_2)\in \Sph{1}$.
We have
\[  B_{\NOne}(\StdBasis_1, c_1 \StdBasis_1 + c_2 \StdBasis_2) = \begin{multlined}[t] (-1)^n \int_x \sum_{a=1}^{n-1} g_a(x^1) g_{n-a}(c_1 x^1 + c_2 x^2)\omega_a(x,\StdBasis_1) \\ \omega_{n-a}(x,c_1 \StdBasis_1 + c_2 \StdBasis_2).\end{multlined}\]
We will show that for every $a=1$,~$\dotsc$, $n-1$ we can write
\begin{equation} \label{Eq:Mu}
\mu_a(x) \coloneqq \omega_a(x,\StdBasis_1) \omega_{n-a}(x,c_1 \StdBasis_1 + c_1 \StdBasis_2) = \Bigl(\sum_{i=3}^{n+1} \pm  x^i \Vol(x)\Bigr) \eta_a(y,z)
\end{equation}
with alternating signs for some form $\eta_a(y,z)$.
Then we can use the same argument as for~$A_{\NVol,\NOne}$ with the cyclic permutation of $x^3$, $\dotsc$, $x^{n+1}$ given by
\begin{align*} 
I':\Sph{n} &\longrightarrow \Sph{n} \\
   (x^1, x^2, x^3,\dotsc, x^n, x^{n+1}) &\longmapsto (x^1, x^2, x^{n+1}, x^3 \dotsc, x^{n}) 
\end{align*}
to conclude that $B_{\NOne}(\StdBasis_1, c_1 \StdBasis_1 + c_2 \StdBasis_2) = 0$.

In order to show \eqref{Eq:Mu}, we have to study the product of $\omega_i$'s.
From~\eqref{Eq:FormOmega1} we get
\begin{equation}\label{Eq:GenProd}
\begin{aligned}
&\omega_a(x,y)\omega_{n-a}(x,z)\\
&\quad =\begin{multlined}[t] \frac{1}{a! (n-1-a)! (n-a)! (a-1)!} \sum_{\substack{\sigma,\,\mu\in \Perm_{n+1}}} (-1)^{\sigma + \mu} x^{\sigma_1} x^{\mu_1} y^{\sigma_2} z^{\mu_2}  \\ \Diff{x}^{\sigma_3} \dotsm \Diff{x}^{\sigma_{2+a}} \Diff{x}^{\mu_3}\dotsm\Diff{x}^{\mu_{2+n-a}} 
\Diff{y}^{\sigma_{3+a}}\dotsm \Diff{y}^{\sigma_{n+1}}\\\Diff{z}^{\mu_{3+n-a}}\dotsm \Diff{z}^{\mu_{n+1}}.
\end{multlined}
\end{aligned}
\end{equation}
In order to simplify this expression, we decompose $\sigma\in \Perm_{n+1}$ as
\begin{equation} \label{Eq:Decomp}
 \sigma = \sigma^5 \circ \sigma^4 \circ \sigma^3 \circ \sigma^2 \circ \sigma^1,
 \end{equation}
where $\sigma^1$, $\dotsc$, $\sigma^5\in \Perm_{n+1}$ are permutations defined as follows:
\begin{itemize}
 \item The permutation $\sigma^1$ is a shuffle permutation $\sigma^1 \in \Perm_{2+a,n-a-1}$ such that its first block denoted by $\sigma^1(1) = (\sigma^1_1,\dotsc,\sigma^1_{2+a})$ is equal to the ordered set $\{\sigma_1, \dotsc, \sigma_{2+a}\}$.
 The second block $\sigma^1(2)$ is then the ordered set $\{\sigma_{3+a},\dotsc,\sigma_{n+1}\}$, which will be denoted by $J_\sigma$.
 \item The permutation $\sigma^2$ acts on the block $\sigma^1(1)$ by moving $\sigma_2$ in front.
 We denote the new block $\sigma^1(1)\backslash\{\sigma_2\}$ by $I_\sigma$, so that we can write $\sigma^2 : \sigma^1(1) \mapsto (\sigma_2, I_\sigma)$. 
 \item The permutation $\sigma^3$ acts on the block $I_\sigma$ by moving $\sigma_1$ in front.
 Together with the previous step we get $\sigma^1(1)\mapsto (\sigma_2, \sigma_1,I_\sigma\backslash\{\sigma_1\})$.
 \item The permutation $\sigma^4$ is a transposition of $\sigma_1$ and $\sigma_2$.
 \item The permutation $\sigma^5$ is determined by the pair $(\sigma^{51},\sigma^{52}) \in \Perm_{a} \times \Perm_{n-1-a}$ of permutations $\sigma^{51}$ and $\sigma^{52}$ acting on blocks $I_\sigma\backslash\{\sigma_1\}$ and $J_\sigma$ to get $(\sigma_3,\dotsc,\sigma_{2+a})$ and $(\sigma_{3+a},\dotsc,\sigma_{n+1})$, respectively.
\end{itemize}

We define the decomposition $\mu^1$,~$\dotsc$, $\mu^5$ for $\mu\in \Perm_{n+1}$ from \eqref{Eq:GenProd} analogously with $a$ replaced by $n-a$.
Using~\eqref{Eq:Decomp}, the product~\eqref{Eq:GenProd} can be written as
\begin{align*}
  &\begin{multlined}\frac{1}{a! (n-1-a)! (n-a)! (a-1)!} \sum_{\substack{\sigma^1,  \dotsc,\, \sigma^5 \\ \mu^1, \dotsc,\, \mu^5}} (-1)^{\sigma^1 + \dotsb + \sigma^5 + \mu^1 + \dotsb + \mu^5} x^{\sigma_1} x^{\mu_1} y^{\sigma_2} z^{\mu_2} \\ \Diff{x}^{\sigma^{51}(I_\sigma\backslash \{\sigma_1\})} \Diff{x}^{\mu^{51}(I_\mu\backslash\{\mu_1\})} \Diff{y}^{\sigma^{52}(J_\sigma)} \Diff{z}^{\mu^{52}(J_\mu)} \end{multlined} \\[5pt]
 &\qquad =\begin{multlined}[t]- \sum_{\sigma^1,\, \mu^1} (-1)^{\sigma^1 + \mu^1} \Bigl(\sum_{\sigma^2,\, \mu^2} (-1)^{\sigma^2 + \mu^2} \sum_{\sigma^3,\, \mu^3} (-1)^{\sigma^3 + \mu^3} x^{\sigma_1} x^{\mu_1} y^{\sigma_2} z^{\mu_2} \\ \Diff{x}^{I_\sigma \backslash \{\sigma_1\}} \Diff{x}^{I_\mu\backslash \{\mu_1\}}\Bigr) \Diff{y}^{J_\sigma} \Diff{z}^{J_\mu},\end{multlined} \end{align*}
where $-1$ comes from $(-1)^{\sigma^4}$ and $\sigma^5$ is compensated by permutations of forms.
For fixed $\sigma^1$ and $\mu^1$, consider the coefficient at $\Diff{y}^{J_\sigma}\Diff{z}^{J_\mu}$ in the brackets.
If we evaluate it at $y=\StdBasis_1$, $z=c_1 \StdBasis_1 + c_2 \StdBasis_2$, we get
\begin{align*}
&  c_1 \overbrace{\sum_{\substack{\sigma^3,\, \mu^3 \\ \sigma_2 = 1 \\ \mu_2 = 1}} (-1)^{\sigma^3 + \mu^3} x^{\sigma_1} x^{\mu_1}  \Diff{x}^{I_\sigma\backslash\{\sigma_1\}} \Diff{x}^{I_\mu \backslash\{\mu_1\}}}^{=:\displaystyle\mathrm{I}} \\ 
& {} + (-1)^{\mu^2} c_2  \overbrace{\sum_{\substack{\sigma^3,\, \mu^3 \\ \sigma_2 = 1 \\ \mu_2 = 2}} (-1)^{\sigma^3 + \mu^3} x^{\sigma_1} x^{\mu_1}  \Diff{x}^{I_\sigma\backslash\{\sigma_1\}} \Diff{x}^{I_\mu \backslash\{\mu_1\}}}^{=:\displaystyle\mathrm{II}},
 \end{align*}
where $(-1)^{\mu^2} = -1$ if and only if $1\in I_{\mu}$.

More generally, for multiindices $I_1$, $I_2 \subset \{1,\dotsc, n+1\}$ of lengths $a+1$ and $n-a+1$,  respectively, consider the sum
\begin{equation} \label{Eq:Sum}
S(I_1,I_2) \coloneqq \sum_{\substack{i_1 \in I_1 \\ i_2 \in I_2}} (-1)^{(i_1,I_1) + (i_2,I_2)} x^{i_1} x^{i_2} \Diff{x}^{I_1\backslash\{i_1\}} \Diff{x}^{I_2\backslash\{i_2\}},
\end{equation}
where $(i_j, I_j)$ is the number of transpositions required to move $i_j$ in front of $I_j$.
The following implication holds: 
\begin{equation*}
 S(I_1, I_2)\neq 0\; \Implies \; 1 \le \Abs{I_1 \cap I_2} \le 2.
\end{equation*}
We distinguish the two cases left: 
\begin{description}[font=\normalfont\itshape]
\item[Case $I_1 \cap I_2 =\{i,j\}$ with $i < j$: ] We get
\begin{align*} 
S(I_1, I_2) &= \begin{multlined}[t] (-1)^{(i,I_1) + (j,I_2)} x^{i} x^j \Diff{x}^{I_1 \backslash\{i\}} \Diff{x}^{I_2 \backslash \{j\}} \\ {}+ (-1)^{(j,I_1) + (i,I_2)} x^{j} x^i \Diff{x}^{I_1 \backslash\{j\}} \Diff{x}^{I_2 \backslash \{i\}} \end{multlined} \\
&=\begin{multlined}[t] (-1)^{(i,I_1) + (j,I_2) + (j,I_1)+1 + (i,I_2)} x^i x^j \Diff{x}^j \Diff{x}^{I_1\backslash\{i,j\}}\\\Diff{x}^i \Diff{x}^{I_2\backslash\{i,j\}} + (-1)^{(j,I_1) + (i,I_2)+(i,I_1)+ (j,I_2)+1} x^i x^j \\ \Diff{x}^i \Diff{x}^{I_1\backslash\{i,j\}}\Diff{x}^j \Diff{x}^{I_2\backslash\{i,j\}} \end{multlined} \\
&= \pm(-1 +  1) x^i x^j \Diff{x}^i \Diff{x}^{I_1\backslash\{i,j\}} \Diff{x}^j\Diff{x}^{I_2\backslash\{i,j\}},
\end{align*}
where in the last step we switched $\Diff{x}^i \leftrightarrow \Diff{x}^j$ in the first summand.
Therefore, it holds $S(I_1, I_2) = 0$.
\item[Case $I_1 \cap I_2 = \{i\}$:] We must have $I_1 \cup I_2 = \{1,\dotsc,n+1\}$.
A non-zero summand in~\eqref{Eq:Sum} has either $i_1 = i$ and $i_2\in I_2$, in which case \[I_1\backslash \{i_1\} \cup I_2\backslash\{i_2\}= \{1, \dotsc, \widehat{i_2}, \dotsc, n+1\}, \] or $i_2 = i$ and $i_1\in I_1$ with $i_1 \neq i$, in which case \[I_1\backslash \{i_1\} \cup I_2\backslash\{i_2\}= \{1, \dotsc, \widehat{i_1}, \dotsc, n+1\}.\] Indices $i_2$ from the first case and $i_1$ from the second case constitute $\{1,\dotsc,n+1\}$.
Therefore, for some signs $\pm$, we can write
\begin{equation*} \label{Eq:AlternatingSum}
S(I_1,I_2) = x^i \sum_{j=1}^{n+1}  \pm  x^j \Diff{x^1} \dotsm \widehat{\Diff{x^j}} \dotsm \Diff{x^{n+1}}.
\end{equation*}
We will prove that the signs alternate, and hence $S(I_1,I_2)= \pm x^i \Vol(x)$.
Suppose that $j$, $j+1 \in I_{1}$ for some $j\in \{1,\ldots,n\}$.
The two summands in~\eqref{Eq:Sum} with $(i_1, i_2)=(j, i)$ and $(i_1, i_2)= (j+1,i)$, respectively, give
\begin{align*}
& \begin{multlined}[t](-1)^{(j,I_1) + (i,I_2)} x^{j} x^{i} \Diff{x}^{I_1\backslash\{j\}} \Diff{x}^{I_2\backslash\{i\}} + (-1)^{(j+1,I_1) + (i,I_2)} x^{j+1} x^{i} \\ \Diff{x}^{I_1\backslash\{j+1\}}\Diff{x}^{I_2\backslash\{i\}} \end{multlined}
\\ &\quad= \begin{multlined}[t] (-1)^{(i,I_2)} x^{j} x^{i} \Diff{x}^{j+1}\Diff{x}^{I_1\backslash\{j, j+1\}} \Diff{x}^{I_2\backslash\{i\}}  \\ {}+ (-1)^{1 + (i,I_2)} x^{j+1} x^{i} \Diff{x}^j \Diff{x}^{I_1\backslash\{j, j+1\}} \Diff{x}^{I_2\backslash\{i\}} \end{multlined} \\
&\quad = (-1)^{(i,I_2)}x^i (x^j \Diff{x}^{j+1} - x^{j+1} \Diff{x}^j) \Diff{x}^{I_1\backslash\{j, j+1\}} \Diff{x}^{I_2\backslash\{i\}}.
\end{align*}
The signs clearly alternate.
A symmetric argument holds when $j$, $j+1\in I_2$.
Now assume that $j \in I_1$ and $j+1\in I_2$.
The two summands in~\eqref{Eq:Sum} which have $(i_1, i_2)=(j,i)$ and $(i_1, i_2) = (i,j+1)$, respectively, give
\begin{align*}
& \begin{multlined}[t] (-1)^{(j,I_1)+(i,I_2)} x^j x^i \Diff{x}^{I_1\backslash\{j\}}\Diff{x}^{I_2\backslash\{i\}} + (-1)^{(i,I_1) + (j+1,I_2)} x^i x^{j+1} \\ \Diff{x}^{I_1\backslash\{i\}}\Diff{x}^{I_2\backslash\{j+1\}} \end{multlined}\\
&\quad=\begin{multlined}[t]
(-1)^{(j,I_1)+(i,I_1\backslash\{j\}) + (a + 1)} x^j x^i \Diff{x^{I_1\backslash\{i, j\}}}\Diff{x}^{I_2} \\ {}+ (-1)^{(i,I_1) + (j+1,I_2) + (j,I_1\backslash\{i\})} x^i x^{j+1} \Diff{x}^j \Diff{x}^{I_1\backslash\{i, j\}} \Diff{x}^{I_2\backslash\{j+1\}}
\end{multlined}\\&\quad=
\begin{multlined}[t]
(-1)^{(j,I_1) + (i,I_1\backslash\{j\})+(j+1,I_2)} x^i x^j \Diff{x}^{j+1} \Diff{x}^{I_1\backslash\{i,j\}} \Diff{x}^{I_2\backslash\{j+1\}} \\ {}+ (-1)^{1 + (j,I_1) + (i,I_1\backslash\{j\})+(j+1,I_2)} x^i x^{j+1} \Diff{x}^j \Diff{x}^{I_1\backslash\{i,j\}} \Diff{x}^{I_2\backslash\{j+1\}} 
\end{multlined}\\
&\quad= \begin{multlined}[t] (-1)^{(j,I_1) + (i,I_1\backslash\{j\})+(j+1,I_2)} x^i (x^j \Diff{x}^{j+1} - x^{j+1} \Diff{x}^j) \\ \Diff{x}^{I_1\backslash\{i,j\}} \Diff{x}^{I_2\backslash\{j+1\}}. \end{multlined}
\end{align*}
The signs alternate again.
A symmetric argument holds for $j\in I_2$ and $j+1\in I_1$.
\end{description}	 

 Back to the original problem, we have $\mathrm{I} = S(I_\sigma, I_\mu)$ with $I_\sigma, I_\mu \subset \{2,\dotsc, n+1\}$.
 It follows that the first case applies, and hence $\mathrm{I} = 0$. 
 We have $\mathrm{II} = S(I_\sigma, I_\mu)$ with $I_\sigma \subset \{2,\dotsc, n+1\}$ and $I_\mu \subset \{1, \widehat{2}, \dotsc, n+1\}$.
 It follows that either the first case or the second case with $i\ge 3$ applies.
 This proves~\eqref{Eq:Mu} up to signs.
 Tracing back the signs, one can convince himself that the signs in \eqref{Eq:Mu} indeed alternate.

The last paragraph of the proof of Proposition~\ref{Prop:COne} finishes the proof.
\end{proof}

We summarize the consequences in the following proposition.
The main argument is the same as in the proof of Proposition~\eqref{Prop:GeomForm}.

\begin{Proposition}[Vanishing of graphs for $\Sph{n}$] \label{Prop:TotalVanishing}
Consider $\Sph{n}$ with the Hodge Propagator~\eqref{Eq:GreenKernelMC1}.
Only the following trivalent ribbon graphs $\Gamma \neq Y$ do not necessarily vanish:
\begin{description}[font=\normalfont\itshape]
 \item[($n=1$):] The $O_{k}$-graph with $k\in 2\N$ internal vertices of type $B$ with $\NVol$ at the external vertex (see Figure~\ref{Fig:Gamma0}).
 \item[($n=2$):] It must hold $A=0$, $C=2B$ and all $B$ vertices must have $\NVol$ at the external vertex.
 Moreover, if $\Gamma$ is reduced, it must have $g\ge 1$.
 \item[($n=3$):] There is no external vertex and $4 \mid C$ holds.
 \item[($n>3$):] All graphs vanish.
\end{description}
\end{Proposition}
\begin{proof}
Lemma~\ref{Lemma:ABVanishing} implies that $A=0$ and that the total form-degree $D$ satisfies $D= n B$.
Therefore, we get from \eqref{Eq:VerticesEq} the following: for $n>3$ there is neither a $B$-vertex nor a $C$-vertex; for $n=3$, there is no $B$-vertex; for $n=2$, we have $C=2B$; and for $n=1$, there is no $C$-vertex.

Consider the pullback of $I(\sigma_L)$ along the (multi)diagonal action of an $R\in O(n+1)$ with $\Det(R)=-1$ on $(\Sph{n})^{\times k}$.
We get schematically
\[ \int_{(\Sph{n})^{\times k}} \Prpg^{e}\Vol^{s} = (-1)^{k +e + s} \int_{(\Sph{n})^{\times k}} G^{e}\Vol^{s}. \]
Therefore, $k+ e + s$ has to be even.
If we plug-in from~\eqref{Eq:ChangeOfVariables}, we get
\[ k+ e + s =  \begin{cases}
3 B & \text{for }n=1, \\
8 B & \text{for }n=2, \\
\frac{5}{2} C & \text{for }n=3.
\end{cases} \]

A non-vanishing reduced graph must have $B\ge l$.
For $n=2$, so that $C=2B$, the formula~\eqref{Eq:GenusFormulaa} gives $g\ge 1$.
\end{proof}

\begin{Remark}[Graphs for $\Sph{2}$]\phantomsection\label{Rem:GraphsTwoSphere}

{ \begingroup
\begin{figure}
\centering
\begin{subfigure}{0.45\textwidth}
\centering
\input{\GraphicsFolder/P1.tex}
\caption{$P_1$}
\end{subfigure}
\begin{subfigure}{0.45\textwidth}
\centering
\input{\GraphicsFolder/P2.tex}
\caption{$P_2$}
\end{subfigure}
\caption{Vanishing graphs $P_1$ and $P_2$ for $\Sph{2}$.}\label{Fig:P1P2}
\end{figure}
\endgroup }
The simplest possibly non-vanishing graph for $\Sph{2}$ has $A= 0$, $B=1$, $C=2$. If it is reduced, we must have $l = g = 1$, and hence it will contribute to $\PMC_{11}$. Up to an isomorphism, there is only one such graph, which we denote by $P_1$ (see Figure~\ref{Fig:P1P2}). However, we see that the pair of internal vertices $x_1$ and $x_2$ is connected by two edges, which implies that $P_1 =0$. Indeed, $\Prpg(x,y)$ has odd degree, and hence we have\footnote{We recall from Section~\ref{Section:Proof2} that the notation $\Prpg(x_i,x_j)$ means $(\pi_{i} \times \pi_j)^* \Prpg$ and not just the evaluation at $(x_i,x_j)$.}
\[ \Prpg(x,y)\Prpg(y,x) = \Prpg(x,y)^2 = 0 \]
by the symmetry on the pullback along the twist map. It follows that $\PMC_{11} = 0$.

The second simplest possibly non-vanishing reduced graph is the graph~$P_2$ from Figure~\ref{Fig:P1P2}. Let 
\begin{equation*}
\eta(x_1,x_2,x_3,x_4,x_5) \coloneqq\begin{multlined}[t]\Prpg(x_1,x_2)\Prpg(x_1,x_3)\Prpg(x_4,x_2)\Prpg(x_4,x_3)\Prpg(x_3,x_5)\\ \Prpg(x_2,x_5)\Vol(x_5) \end{multlined}
\end{equation*}
denote the form in the integrand coming from the part of the graph on the right-hand side of the vertical axis going through $x_1$, $x_4$. If $\tau_{1,4}$ denotes the exchange of $x_1$ and $x_4$, then clearly $\tau_{1,4}^* \eta = \eta$ because the graph is symmetric with respect to the horizontal axis going through $x_5$, $x_6$. Using this, we compute
\begin{align*}
& \int_{x_1,x_2,x_3,x_4,x_5,x_6} \NVol(x_6)\Prpg(x_1,x_6)\Prpg(x_4,x_6) \eta(x_1,x_2,x_3,x_4,x_5) \\
& = \int_{\tau_{1,4}(x_1,x_2,x_3,x_4,x_5,x_6)} \tau_{1,4}^*\bigl(\NVol(x_6)\Prpg(x_1,x_6)\Prpg(x_4,x_6) \eta(x_1,x_2,x_3,x_4,x_5)\bigr) \\
& = \int_{x_4,x_2,x_3,x_1,x_5,x_6} \NVol(x_6)\Prpg(x_4,x_6)\Prpg(x_1,x_6) \eta(x_4,x_2,x_3,x_1,x_5)\\
& = 
 -\int_{x_1,x_2,x_3,x_4,x_5,x_6} \NVol(x_6)\Prpg(x_1,x_6)\Prpg(x_4,x_6)  \eta(x_1,x_2,x_3,x_4,x_5),
\end{align*}
where the minus sign comes from switching the first two $\Prpg$'s. We see that $P_2$ vanishes. The other variants with $x_5$ moved on the edge $x_3$, $x_4$ and~$x_2$, $x_4$ vanish by a similar argument using the compositions $\tau_{1,3}\circ\tau_{5,6}$ and $\tau_{1,2}\circ\tau_{5,6}$, respectively. We conclude that $\PMC_{21} = 0$, and hence $\OPQ_{121}^\PMC = 0$. 

We sum up some general observations about the integrals for $\Sph{2}$:

\begin{itemize}
\item We have $B_{\NOne} \neq 0$ and $C \neq 0$ for the corresponding forms.
\item We have the multiplication formula (c.f., Example~\ref{Example:Circle})
\[ \omega_1(x,y) \omega_1(x,z) = x\cdot(y\times z) \Vol(x). \]
\item The number $(-1)^{\sigma_L} I(\sigma_L)$ does not depend on the choice of $L_1$ provided a compatible $L_2$ is chosen.
\item It holds $\sum_{L_3^b} (-1)^{\sigma_L} I(\sigma_L) = 0$ whenever there is a boundary component with even number of $\NVol$'s.
\item If there is a $B$-vertex $x$ such that the underlying graph (after forgetting the ribbon structure) is symmetric on the reflection along an axis going through $x$, then $I(\sigma_L) = 0$. \qedhere
\end{itemize}

\Correct[caption={DONE Nonsense about lower bound}]{$g\ge 1$ is equivalent to $B \ge l$ and moreover we must have $C=2B$. What else was I thinking here?}
\end{Remark}

\begin{Remark}[Graphs for $\Sph{3}$]\label{Rem:GraphsThreeSphere}
{ \begingroup
\begin{figure}
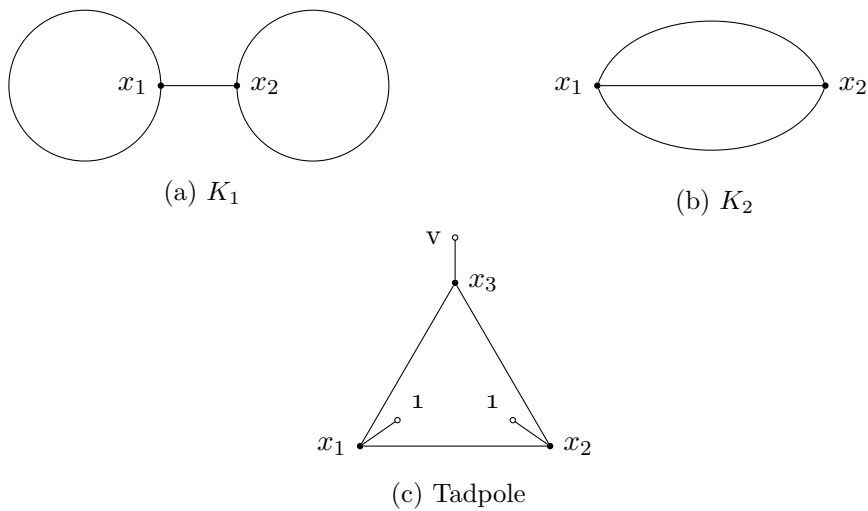

\centering
\begin{subfigure}{0.45\textwidth}
\centering
\input{\GraphicsFolder/K1.tex}
\caption{$K_1$}
\end{subfigure}
\begin{subfigure}{0.45\textwidth}
\centering
\input{\GraphicsFolder/K2.tex}
\caption{$K_2$}
\end{subfigure}
\begin{subfigure}{0.45\textwidth}
\centering
\input{\GraphicsFolder/Tadpole.tex}
\caption{Tadpole}
\end{subfigure}
\caption[Graphs $K_1$, $K_2$ and the tadpole graph from Chern-Simons theory.]{Graphs $K_1$ and $K_2$ from the Chern-Simons theory and the tadpole graph with $(l,g)=(2,0)$ for $n=3$.}\label{Fig:K1K2}
\end{figure}
\endgroup }

For $\Sph{3}$, we consider the non-reduced graphs $K_1$ and $K_2$ and the tadpole graph from Figure~\ref{Fig:K1K2}. The graphs $K_1$ and $K_2$ appear in the definition of the Chern-Simons topological invariant in~\cite{Kohno2002} (with a gauge group). The corresponding integrals from our theory vanish ``algebraically'', i.e., at the level of wedge products of $\omega_i$. Indeed, every summand in $K_1$ contains 
\[ \omega_a(x_1,x_1) = 0\quad \text{for some }a=0, 1, 2, \]
and, for degree reasons, the form part of $K_2$ can contain   only
\[ \omega_1(x_1,x_2)^3=0\quad\text{or}\quad\omega_0(x_1,x_2)\omega_{1}(x_1,x_2)\omega_{2}(x_1,x_2)=0.\] 
The tadpole graph contains only
\begin{equation*}
 \omega_2(x_1,x_3)\omega_1(x_1,x_2)\omega_2(x_2,x_3) = 0.\qedhere \end{equation*}
\qedhere
\end{Remark}

Equations in Remark~\ref{Rem:GraphsThreeSphere} were checked by the computer---this is possible as the vanishing is purely algebraic.
Any author's attempts to show numerically that some integrals are non-zero or get a hint for ``analytic vanishing'' failed so far.
The program for Wolfram Mathematica~10.4 will be made available at \cite{sourcecode} and possibly updated.

We will now compute $\PMC_{20}$ for $\Sph{1}$, which according to Proposition~\ref{Prop:TotalVanishing} consists only of contributions from the $O_k$-graphs with $k$ even. 
By analogy with the finite dimensional case (see~Appendix~\ref{Section:Appendix}), we expect that the number $(-1)^{\sigma_L}I(\sigma_L)$ does not depend on $L$. All inputs are namely the same and the degrees even, i.e., $|m_2^+| = -2$, $|\SuspU^2\Prpg| = -2$ and $|\NVol| = 0$. 

We fix $s_1$, $s_2\ge 1$ such that $k=s_1+s_2$ is even and make the ansatz
\begin{equation*}
n_{20}(\Susp \NVol^{s_1}\otimes  \Susp \NVol^{s_2}) \coloneqq \varepsilon(s_1,s_2) C(s_1,s_2) I(k),
\end{equation*}
where $I(k)$ is the integral
\begin{equation} \label{Eq:Ik}
\frac{1}{V^k}\int_{x_1, \dotsc, x_{k}} G(x_1,x_2) \dotsm G(x_{k-1},x_{k})G(x_{k},x_1)\Vol(x_1) \dotsm \Vol(x_{k}), \end{equation}
$\varepsilon(s_1,s_2)$ a sign and $C(s_1,s_2)$ a combinatorial coefficient to be determined.

We fix a circle in the plane with $k$ points (=\,internal vertices) and denote by $O(s_1,s_2)$ the set of ribbon graphs constructed by attaching external legs from which $s_1$ points in the interior and $s_2$ in the exterior, or the other way round, so that $O(s_1,s_2) = O(s_2, s_1)$ (see Figure~\ref{Fig:Gamma0}). Recall that the ribbon structure is induced from the counterclockwise orientation of the plane. It is easy to see that all graphs in $O(s_1,s_2)$ admit a labeling which is admissible with respect to $\Susp \NVol^{s_1} \otimes \Susp \NVol^{s_2}$, and that $O(s_1,s_2)$ contains a representative of every such $O_k$-graph. 

\Correct[inline,caption={DONE Sign not necessary}]{There is no need to write $(-1)^{k+1}$ when we know that $k$ is even!!! Who cares? Let's keep it so that it is clear where the sign comes from.}
\begin{Lemma}[Integral for the $O_k$-graph for $\Sph{1}$] \label{Lemma:IntegralFor1}
For every even $k\ge 2$, the integral $I(k)$ is equal to
 \begin{equation}\label{Eq:TheFormulaForIk}
 (-1)^{k+1} \frac{1}{2^k}\sum_{i=2, 4, \dotsc ,k} \frac{i}{(i+1)!} \sum_{\substack{i_1+ \dotsb +i_r = k-i \\ i_1, \dotsc, i_r \in 2\N,\, r\in \N}} (-1)^r \frac{1}{(i_1+1)! \dotsm (i_r+1)!}.
 \end{equation}
\end{Lemma}

\begin{proof}
Denote $\bar{\Prpg} \coloneqq -2\pi \Prpg$. For all $k$, $l \ge 1$, we consider the more general integral
\[ I(k,l) \coloneqq \int_{x_1, \dotsc, x_k} \bar{\Prpg}(x_1,x_2) \dotsm \bar{\Prpg}(x_{k-1},x_k) \bar{\Prpg}(x_k,x_1)^l\Vol(x_1) \dotsm \Vol(x_k). \]
Taking the pullback along $(x_1, x_2, \dotsc, x_{k-1}, x_k) \mapsto (x_k, x_{k-1}, \dotsc, x_2, x_1)$ and using the antisymmetry of $\bar{\Prpg}(x,y)$, we get $I(k,l) = 0$ whenever $k+l$ is even. We will compute $I(k,1)$ for $k\in 2\N$ from a recursive relation which arises from successive integration.

For the recursion step, we need to evaluate the integral 
\[\int_{y} \bar{\Prpg}(x,y)\bar{\Prpg}(y,z)^l \Vol(y)\]
for fixed $(x,z)\in (\Sph{1}\times \Sph{1})\backslash\Diag$. Pick the chart $g: \Sph{1}\backslash\{z\} \rightarrow (-\pi,\pi)$ defined by 
\[ g(y)= \bar{\Prpg}(y,z) =  \pi - \alpha(y,z) \quad\text{for } y\in \Sph{1}\backslash\{z\}, \]
where the angle $\alpha$ was defined in Example~\ref{Example:Circle}. It holds $\Diff{g}(y) = \Vol(y)$ and
\[ \bar{\Prpg}(x,y) = \begin{cases} \bar{\Prpg}(x,z) - g(y) - \pi & \text{for }-\pi< g(y)< \bar{\Prpg}(x,z), \\
\bar{\Prpg}(x,z) - g(y) + \pi & \text{for }\bar{\Prpg}(x,z)<g(y)<\pi.
 \end{cases}\]
We compute
\begin{align*}
\int_{y} \bar{\Prpg}(x,y)\bar{\Prpg}(y,z)^l \Vol(y) &= \begin{multlined}[t] \int_{-\pi}^{\pi} (\bar{\Prpg}(x,z) - g)g^l \Diff{g} - \pi \int_{-\pi}^{\bar{\Prpg}(x,z)} g^l \Diff{g} \\ {}+ \pi \int_{\bar{\Prpg}(x,z)}^\pi g^l \Diff{g} \end{multlined} \\
 &=  \frac{2\pi}{l+1} \begin{cases}
  \pi^l \bar{\Prpg}(x,z) - \bar{\Prpg}(x,z)^{l+1} & \text{for }l\text{ even},\\[2ex]
  \dfrac{\pi^{l+1}}{l+2} - \bar{\Prpg}(x,z)^{l+1} & \text{for }l \text{ odd}.
 \end{cases}
\end{align*}

From now on, $\int$ will stand for the Riemannian integral, i.e., $\int f \coloneqq \int f\Vol$ for a function~$f$. We compute
\begin{align*}
I(2,l) &= \int_{x_1, x_2} \bar{\Prpg}(x_1,x_2)\bar{\Prpg}(x_2,x_1)^{l} \\
 &= - \int_{y z} \bar{\Prpg}(y,z)^{l+1} \\
 &=- 2\pi \int_{-\pi}^\pi g^{l+1} \Diff{g} \\
 &= \begin{cases} 0 & \text{for }l \text{ even}, \\[2ex] 
 - \dfrac{4 \pi^{l+3}}{l+2} &\text{for }l\text{ odd}. \end{cases}
\end{align*}
For $k\ge 4$ even and $l$ odd, we compute %
\begin{align*}
 I(k,l) &= \begin{multlined}[t]\frac{2\pi}{l+1}\int_{x_1, \dotsc, x_{k-1}}\bar{\Prpg}(x_1,x_2)\dotsm \bar{\Prpg}(x_{k-2},x_{k-1}) \\ \Bigl(\frac{\pi^{l+1}}{l+2} - \bar{\Prpg}(x_{k-1},x_1)^{l+1}\Bigr) \end{multlined} \\ 
&\begin{multlined}
=\frac{-4\pi^2}{(l+1)(l+2)} \int_{x_1, \dotsc, x_{k-2}}\ \bar{\Prpg}(x_1,x_2) \dotsm \bar{\Prpg}(x_{k-3},x_{k-2})\\ \Bigl(\pi^{l+1} \bar{\Prpg}(x_{k-2},x_1)
 -\bar{\Prpg}(x_{k-2},x_1)^{l+2}\Bigr)
\end{multlined} \\
&= \frac{4\pi^2}{(l+1)(l+2)}\bigl(-\pi^{l+1} I(k-2,1)+I(k-2,l+2)\bigr).
\end{align*}
For the second equality, we used $\int_{x_1} \bar{\Prpg}(x_1,x_2) = 0$ to show that the term multiplied by $\frac{\pi^{l+1}}{l+2}$ vanishes. It follows that
\begin{align*}
I(k,1) &= \frac{(2\pi)^{k-2}}{(k-1)!} I(2,k-1) - \sum_{l=2, 4, \dotsc, k-2} \frac{(2\pi^2)^{k-l}}{(k-l+1)!} I(l,1) \\ 
&=-\frac{k(2\pi^2)^k}{(k+1)!}-\sum_{l=2, 4, \dotsc, k-2} \frac{(2\pi^2)^{k-l}}{(k-l+1)!} I(l,1)\qquad\text{for all }k=2,\,4,\,\dotsc
\end{align*}
This is a recursive equation of the form $a_k = c_k + \sum_{l=1}^{k-1} d_{k-l} a_l$. Its solution is $a_k = \sum_{i=1}^k c_i D_{k-i}$ with $D_0\coloneqq 1$ and $D_i = \sum d_{i_1} \dotsm d_{i_r}$, where we sum over all $r=1$,~$\dotsc$, $i$ and $i_1$,~$\dotsc$, $i_r\in \N$ such that $i_1+ \dotsb + i_r = i$. Therefore, we get %
\[ I(k,1) = -(2\pi^2)^k \sum_{i=2, 4, \dotsc, k} \frac{i}{(i+1)!} \sum_{\substack{i_1 + \dotsb + i_r = k-i \\ i_1, \dotsc, i_r \in 2\N, r\in \N}} (-1)^r \frac{1}{(i_1+1)! \dotsm (i_r+1)!}.\]
The result has to be multiplied by $(-1)^k(2\pi)^{-2k}$ in order to get $I(k)$. 
\end{proof} %

\begin{Lemma}[Independence of labeling]\label{Lemma:Independence}
The summand $(-1)^{\sigma_L} I(\sigma_L)$ in the definition of $\PMC_{20}(\Susp\NVol^{s_1}\otimes \Susp\NVol^{s_2})$ for $\Sph{1}$ is independent of the choice of $\Gamma\in O(s_1,s_2)$ and its labeling $L$ which is compatible and admissible with respect to the input.
\end{Lemma}
\begin{proof}
Pick $\Gamma\in O(s_1,s_2)$ and its admissible labeling $L$. Let $L'$ be an other admissible labeling of $\Gamma$. We distinguish the following situations:
\begin{itemize}
 \item Suppose that $L$ and $L'$ differ by a permutation $\mu$ in $L_3^b$. A similar argument as in the proof of Lemma~\ref{Lem:MCCond} shows that $(-1)^{\sigma_{L'}} = (-1)^\mu (-1)^{\sigma_L}$ and $I(\sigma_L') = (-1)^\mu I(\sigma_L)$, where the sign in the integral comes from the permutation of $\Vol$'s, which have form-degree $1$. Hence $(-1)^{\sigma_{L'}}I(\sigma_{L'}) = (-1)^{\sigma_{L}}I(\sigma_{L})$.
\item Suppose that the boundaries are permuted, i.e., that~$L$ and~$L'$ differ in~$L_1^b$. Notice that~$s_1=s_2$ because otherwise one of~$L$ or~$L'$ would not be admissible. The sign from changing~$L_1^b$ cancels as in the previous case. 
\item Suppose that $L$ and $L'$ differ in $L_2$. It was explained in the proof of Lemma~\ref{Lem:MCCond} that a single change of $L_2$ induces the sign $(-1)^{n-1} = 1$ in $(-1)^{\sigma_L} I(\sigma_L)$.  
\item A cyclic permutation in $L_3^v$ induces a sign neither in $(-1)^{\sigma_L}$ nor in $I(\sigma_L)$.
 \item A permutation $\mu$ in $L_1^v$ induces $(-1)^\mu$ in $(-1)^{\sigma_L}$ and a change in $I(\sigma_L)$, which can be realized by taking the pullback along $\mu: (x_1,\ldots,x_k)\mapsto (x_{\mu_1},\ldots,x_{\mu_k})$. However, the sign of the Jacobian is $(-1)^\mu$, which cancels the sign from $(-1)^{\sigma_L}$.
\end{itemize}

\begin{figure}[t]
\centering
\input{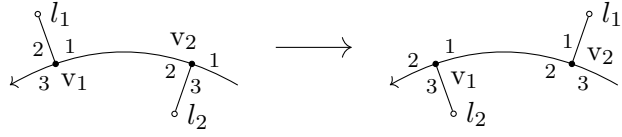}
\caption[Swap of adjacent vertices on a circular graph.]{Swapping adjacent legs.}\label{Fig:TwoLegs}
\end{figure}

Next, we prove the independence of $\Gamma\in O(s_1,s_2)$. Let $L$ be an admissible and compatible labeling of $\Gamma$. Pick two adjacent internal vertices with external legs pointing to different regions, i.e., one to the interior of the circle and the other to the exterior. Suppose that the vertices are labeled by $\Vert_1$ and $\Vert_2$ and the legs by $l_1$ and $l_2$, respectively. Let $\Gamma'\in O(s_1,s_2)$ be the graph with the two legs turned inside out (see Figure~\ref{Fig:TwoLegs}). We can construct an admissible and compatible labeling $L'$ of $\Gamma'$ by making the following changes to $L$: The new leg at $\Vert_1$ will be labeled by $l_2$ and the new leg at $\Vert_2$ by $l_1$.
The cyclic orderings at $\Vert_1$ and $\Vert_2$, respectively, have to be modified by a transposition in order to get compatibility with the new ribbon structure. All other labelings can be copied from $L$.
In total, we get
\[ (-1)^{\sigma_L - \sigma_{L'}} = -1. \]
This sign is compensated by swapping the one-forms in $I(\sigma_L)$:
\[ \Vol (x_{\Vert_1}) \dots \Vol(x_{\Vert_2}) \;\longleftrightarrow\;\Vol(x_{\Vert_2}) \dots \Vol(x_{\Vert_1}). \]
The independence of $\Gamma\in O(s_1,s_2)$ follows from the fact that we can span the entire $O(s_1,s_2)$ by repeating the swap-of-legs operation.
\end{proof}

\begin{Lemma}[Sign]\label{Lemma:SignForMCOnCircle}
We have
\[\varepsilon(s_1,s_2) = (-1)^{s_1+1}. \]
\end{Lemma}
\begin{proof} 

\begin{figure}
\centering
\input{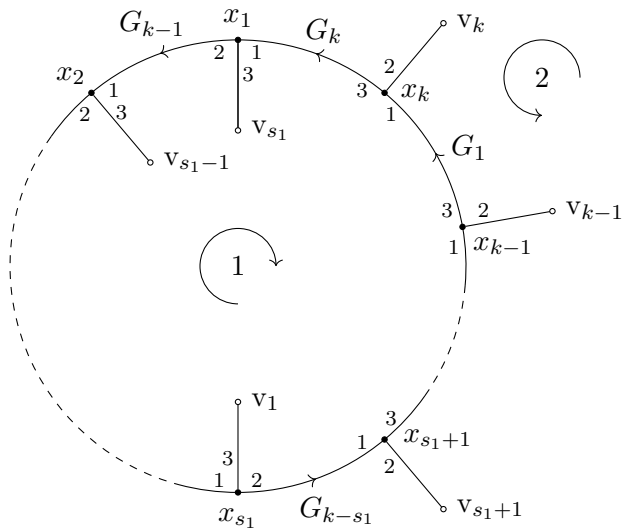}
\caption[A fully labeled circular ribbon graph for $\Sph{1}$.]{The graph $\Gamma^*$ with the labeling $L^*$. It can be checked that~$L_1$ and~$L_2$ are compatible.}\label{Fig:Gamma0}
\end{figure}

By Lemma~\ref{Lemma:Independence}, in order to compute $(-1)^{\sigma_L}I(\sigma_L)$, we can pick $\Gamma^*\in O(s_1,s_2)$ and its admissible and compatible labeling $L^*$ from Figure~\ref{Fig:Gamma0}. We abbreviate $\sigma_0 = \sigma_{L^*}$. The corresponding integral \eqref{Eq:ISigma} reads
\[ I(\sigma_0) = \begin{multlined}[t] \frac{1}{V^k}\int_{x_1, \dotsc, x_k} G(x_{k-1},x_k) \dotsm G(x_{1},x_2)G(x_k,x_1) \Vol(x_{s_1}) \dotsm \Vol(x_1)\\ \Vol(x_{s_1+1}) \dotsm \Vol(x_k). \end{multlined} \]
It differs from $I(k)$ from~\eqref{Eq:Ik} in the order of $\Prpg$'s and $\Vol$'s. A reordering produces the sign
\[(-1)^{\frac{1}{2}s_1(s_1-1)}.\] 
We will compute $(-1)^{\sigma_0}$ by ordering half-edges from the edge order back to the vertex order while looking at Figure~\ref{Fig:Gamma0}. The steps are as follows: 
\begin{itemize}
 \item Transpose half-edges at internal vertices so that the first half-edge goes inside the vertex and the third outside with respect to the counterclockwise orientation. This gives $(-1)^a$.
 \item Permute external legs so that $\NVol_i$ is at $x_i$ for all $i=1$, $\dotsc$, $k$. This gives 
 \[(-1)^{\frac{1}{2}s_1(s_1-1)}. \]
 \item Permute internal edges so that $\Prpg_i$ starts at the third half-edge of $x_i$ and ends at the first half-edge of $x_{i+1}$ for all $i=1$, $\dotsc$, $k-1$. This does not produce any sign as swapping of two $\Prpg$'s requires two transpositions.
 \item  At this point, we have the permutation
 \[\begingroup\setlength\arraycolsep{4pt}      \begin{pmatrix}
 1 & 2 &\dots & 2(e-1)-1 & 2(e-1) & 2e-1 & 2e & 2e + 1 & \dots & 3k \\
 3 & 4 &\dots & 3k-3 & 3k-2 &  3k &  1 & 2 & \dots & 3k-1
 \end{pmatrix}.\endgroup\]
We interpret the last line as $\Prpg_1\dots \Prpg_k \NVol_1\dots \NVol_k$ and permute it to the sequence $\NVol_1 \Prpg_1 \NVol_2 \Prpg_2\dots \NVol_k \Prpg_k$, which does not produce any sign. We end up with 
\[ \sigma_0' = \begin{pmatrix}
 1 & 2 & 3 & \dots & 3k-1 & 3k \\
 2 & 3 & 4 & \dots & 3k & 1
\end{pmatrix}. \]
It is now easy to see that
\[ (-1)^{\sigma_0'} = (-1)^{3k-1}. \]
\end{itemize}
In total, we get
\[ (-1)^{\sigma_0} = (-1)^{s_1 + \frac{1}{2}s_1(s_1-1) + k + 1}. \]
As for the other signs in Definition~\ref{Def:PushforwardMCdeRham}, we have $s(k,l) = k + \frac{1}{2}k(k-1)$ and $P(\NVol^k) = \frac{1}{2}k(k-1)$. There is no sign from $\Susp^2 \NVol^{s_1}\otimes \NVol^{s_2} = \Susp \NVol^{s_1} \otimes \Susp \NVol^{s_2}$ since $\Abs{\Susp} = -2$. Multiplying everything together, we get $\varepsilon(s_1,s_2)$.
\end{proof}
\begin{figure}
\centering
\input{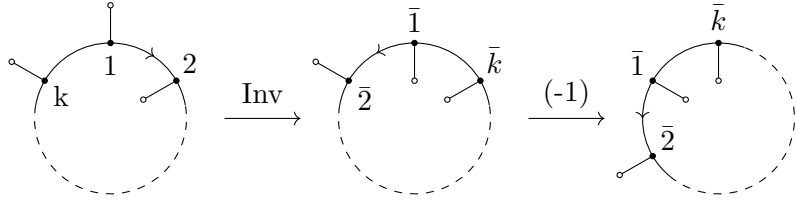}
\caption[The mirror isomorphism for the $O_k$-graph.]{The mirror isomorphism $M: 1 \dots k \mapsto \bar{k} \dots\bar{1}$ is a composition of the inversion and the counterclockwise rotation by one place.}\label{Fig:Mirror}
\end{figure}
\begin{Lemma}[Combinatorial coefficient]
\label{Lemma:CombinatorialCoefficientForMCOnCircle}
It holds \[C(s_1,s_2) = \frac{1}{2} a k! \binom{k-1}{s_1}.\]
\end{Lemma}
\begin{proof}
Every isomorphism of ribbon graphs $\Gamma$ and $\Gamma'$ from $O(s_1,s_2)$ is a composition of the clockwise rotation $(r)$ for $r\in \Z$ and the mirror operation $M$ defined as follows: If $1$,~$\dotsc$, $k$ label internal vertices in the clockwise direction starting from the north-pole, then the result of $M$ is $\bar{k}$,~$\dotsc$, $\bar{1}$, where $\bar{i}$ means that the external leg is reversed (see Figure~\ref{Fig:Mirror}). These operations satisfy
\[ (r+k) = (r),\; (r)(-r)=\Id,\; M^2 = \Id,\; (r)M = M(-r), \]
and hence generate a group $\mathrm{G}$ which is isomorphic to the dihedral group $\Z_k \rtimes \Z_2$. The orbit space $O(s_1,s_2)/\mathrm{G}$ is in $1:1$ correspondence with isomorphism classes of admissible $O_k$-graphs and $\Aut(\Gamma)$ is in $1:1$ correspondence with $\Stab{\Gamma}$. From the orbit-stabilizer formula, we get
\begin{align*}
\sum_{\substack{[\Gamma] \text{ admiss. }\\O_k-\text{graph}}} \frac{1}{\Abs{\Aut(\Gamma)}} &= \sum_{[\Gamma]\in O(s_1,s_2)/\mathrm{G}} \frac{1}{\Abs{\Stab{\Gamma}}} \\
& = \sum_{\Gamma\in O(s_1,s_2)} \frac{1}{\Abs{\Orb{\Gamma}}\Abs{\Stab{\Gamma}}} \\[2ex]
&= \frac{\Abs{O(s_1,s_2)}}{\Abs{\mathrm{G}}}\\[1ex] 
&= \frac{1}{2k}{\binom{k}{s_1}}\times\begin{cases} 1 & \text{for }s_1=s_2, \\ 2 & \text{for }s_1\neq s_2.\end{cases}
\end{align*}
The two cases are compensated in the sum over labelings: For $s_1=s_2$, both labelings $L_1^b$ are admissible, and hence we get the factor $2$.

Next, we multiply by $k! s_1(k-s_1)$, which is the number of $L_1^v$ and $L_3^b$. There is also the factor $\frac{1}{l!} = \frac{1}{2}$. Multiplying everything together, we get $C(s_1,s_2)$.
\end{proof}

Before we summarize the results of our computations (see Proposition~\ref{Proposition:MCSphere} below), we show directly that $\PMC$ is a Maurer-Cartan element; i.e., in this special case, we do not need general results from \cite{Cieliebak2018} at all.

\begin{Lemma}[Maurer-Cartan equation for $\Sph{n}$] \label{Lem:MCEquation}
For $n\ge 1$, consider $\Sph{n}$ with the Hodge Propagator from~\eqref{Eq:GreenKernelMC1}. The collection $(\PMC_{lg})$ satisfies the Maurer-Cartan equation~\eqref{Eq:MaurerCartanEquation} for $\dIBL(\CycC(\Harm(\Sph{n})))$.
\end{Lemma}

\begin{proof}
We will show that for every $l\ge 1$, $g\ge 0$ all summands in the relation corresponding to $(l,g)$ vanish. The summands for $(l,g)= (1,0)$ are $\OPQ_{110}(\PMC_{10})$ and $\frac{1}{2}\OPQ_{210}(\PMC_{10},\PMC_{10})$, and the summand for $(l,g)=(2,0)$ is $\OPQ_{120}(\PMC_{10})$.
The first term vanishes trivially as $\OPQ_{110}=0$, while the other two terms vanish by~\cite[Proposition 12.5]{Cieliebak2015} because $\PMC_{10} = \MC_{10}$
is the canonical Maurer-Cartan element. For $(l,g)\neq (1,0)$, we have the following four situations:
\begin{description}[font=\normalfont\itshape]
\item[$\OPQ_{210}\circ_2 \PMC_{lg}$, $l\ge 2$:] 
Let $\Psi = \Psi_1 \cdots \Psi_l\in \Ext_l \CycC$ be a summand of $\PMC_{lg}$. 
 From Proposition~\ref{Prop:TotalVanishing} it follows that the summands can be chosen such that $\Psi_1$,~$\dotsc$, $\Psi_l \in \DBCycRed\Harm(\Sph{n})[3-n]$, i.e., such that $\Psi_i$ evaluates to $0$ whenever $\NOne$ is a part of its argument. From Definition~\ref{Def:CircS}, we compute
\[\OPQ_{210}\circ_2 (\Psi_1\cdots\Psi_l) = \sum_{\sigma\in \Perm_{2,l-2}} \varepsilon(\sigma,\Psi)\OPQ_{210}(\Psi_{\sigma_1^{-1}}\cdot\Psi_{\sigma_2^{-1}})\cdot\Psi_{\sigma_3^{-1}}\cdots \Psi_{\sigma_l^{-l}}. \]
We clearly have $\OPQ_{210}(\Psi_{\sigma_1^{-1}}\cdot\Psi_{\sigma_2^{-1}})=0$ because $\OPQ_{210}$ feeds $\NOne$ into one of its inputs. It follows that $\OPQ_{210}\circ_2 \PMC_{lg} = 0$.

\item[$\OPQ_{210}\circ_{1,1}(\PMC_{l_1 g_1} \odot \PMC_{l_2 g_2})$, $(l_i,g_i) \neq (1,0)$:] A similar argument as above.
\item[$\OPQ_{120}\circ_1 \PMC_{lg}$, $(l,g)\neq(1,0)$:] A similar argument as above using that $\OPQ_{120}$ also feeds~$\NOne$ into its input. 
\item[$\OPQ_{210}\circ_{1,1} (\PMC_{10}\odot \PMC_{lg})$, $(l,g)\neq (1,0)$:] 
As in the case of $\OPQ_{210}\circ_2 \PMC_{lg}$, suppose that $\Psi_1$,~$\dotsc$, $\Psi_l \in \DBCycRed\Harm(\Sph{n})[3-n]$. Recall that we write $\Omega_i = \Susp \omega_i \in \BCyc \Harm(\Sph{n})[3-n]$ and $\Omega = \Omega_1\otimes \dotsb \otimes \Omega_l$. From Definition~\ref{Def:CircS}, we compute
\begin{align*}
&[\OPQ_{210} \circ_{1,1}(\PMC_{10}\odot \Psi)](\Omega_1 \otimes \dotsb \otimes \Omega_l) 
\\ &\quad = \begin{multlined}[t] \Bigl[ \sum_{i=1}^l (-1)^{\Abs{\Psi_i}(\Abs{\Psi_1} + \dotsb + \Abs{\Psi_{i-1}})} \OPQ_{210}(\PMC_{10}\Psi_i)\Psi_1\cdots \widehat{\Psi_i}\cdots \Psi_l\Bigr]\\ (\Omega_1 \otimes \dotsb \otimes \Omega_l) \end{multlined} \\ 
 &\quad = \begin{multlined}[t] \vphantom{\sum_{i}}\smash{\sum_{\substack{\mu \in \Perm_l \\ i=1, \dotsc, l}}} \frac{1}{l!} (-1)^{\Abs{\Psi_i}(\Abs{\Psi_1} + \dotsb + \Abs{\Psi_{i-1}})} \varepsilon(\mu,\Omega) (\OPQ_{210}(\PMC_{10}\cdot\Psi_i)\otimes \Psi_1\otimes \dotsb \\ \widehat{\Psi_i}\dotsb\otimes \Psi_l)(\Omega_{\mu_1^{-1}} \otimes \dotsb \otimes \Omega_{\mu_l^{-1}}). \end{multlined}
\end{align*}
For every $i=1$, $\dotsc$, $l$, we have
 \begin{align*} 
 & \OPQ_{210}(\PMC_{10}\cdot \Psi_i)(\Omega) = \OPQ_{210}(\PMC_{10}\otimes \Psi_i)(\Omega) \\
 & \qquad = \begin{multlined}[t]- \sum \varepsilon(\omega\mapsto \omega^1 \omega^2)[(-1)^{(n-1)\Abs{\omega^1}} \PMC_{10}(\Susp \NOne \omega^1) \Psi_i(\Susp \NVol \omega^2) \\ {}+ (-1)^{\Abs{\omega^1}} \PMC_{10}(\Susp \NVol \omega^1) \Psi_i(\Susp \NOne \omega^2)] \end{multlined} \\ 
 & \qquad = - \sum \varepsilon(\omega\mapsto \omega^1 \omega^2)(-1)^{(n-1)\Abs{\omega^1}} \PMC_{10}(\Susp \NOne \omega^1) \Psi_i(\Susp \NVol \omega^2).
\end{align*}
This can be non-zero only if $\omega = \NOne \NVol^{s-1}$ for some $s\ge 2$ (up to a cyclic permutation). For this input, we get
\begin{align*} &\OPQ_{210}(\PMC_{10}\cdot \Psi_i)(\Susp \NOne \NVol^{s-1}) \\ &\quad = \begin{aligned}[t]
&-[\varepsilon(\NOne \NVol^{s-1} \mapsto \NOne\NVol^{s-1})\PMC_{10}(\Susp \NOne\NOne\NVol)\Psi_i(\Susp\NVol^{s-1}) \\
&{}+ \varepsilon(\NOne \NVol^{s-1} \mapsto \NVol\NOne\NVol^{s-2})\PMC_{10}(\Susp \NOne\NVol\NOne)  \Psi_i(\Susp\NVol^{s-1})] \end{aligned} \\
 &\quad= (-1)^{n-3}[1+(-1)^{ns+s-1}] \Psi_i(\Susp\NVol^{s-1}).
\end{align*}
The prefactor in brackets is $0$ for $n$ odd or $s$ even, whereas $\NVol^{s-1} = 0$ for $n$ even and~$s$ odd. Therefore,  we have $\OPQ_{210}\circ_{1,1} (\PMC_{10}\odot \PMC_{lg}) = 0$. \qedhere
\end{description}
\end{proof}

\begin{Proposition}[Chern-Simons Maurer-Cartan element for $\Sph{n}$] \label{Proposition:MCSphere}
For $n\ge 1$, consider the round sphere $\Sph{n}$ with the Hodge Propagator~\eqref{Eq:GreenKernelMC1}. The Chern-Simons Maurer-Cartan element $\PMC$ is a strictly reduced Maurer-Cartan element for $\dIBL(\CycC(\Harm(\Sph{n})))$ which satisfies
\[ \PMC_{10} = \MC_{10}\quad\text{for all }n\in \N\]
plus the following properties depending on the dimension:
\begin{description}[font=\normalfont\itshape]
 \item[($n = 1$):] It holds $\PMC_{lg}=0$ for all $l\ge 1$, $g\ge 0$ such that $(l,g) \neq (1,0)$, $(2,0)$; the only non-trivial relation for $\PMC_{20}$ is
  \begin{equation}\label{Eq:MC20}
  \PMC_{20}(\Susp \NVol^{s_1}\otimes  \Susp \NVol^{s_2}) = (-1)^{s_1+1} \frac{1}{2} s_1 k! \binom{k-1}{s_1} I(k),
  \end{equation}
  where $s_1$, $s_2\ge 1$ are such that $k = s_1 + s_2$ is even, and $I(k)$ is given by~\eqref{Eq:TheFormulaForIk}.
 \item[($n=2$):] It holds $\PMC_{l0}=0$ for all $l\ge 2$. We also have $\PMC_{11}=0$.
 \item[($n\ge 3$):] It holds $\PMC_{lg} = 0$ for all $l\ge 1$, $g\ge 0$ such that $(l,g) \neq (1,0)$.
\end{description}
\end{Proposition}

Notice that $\PMC_{20}\not\in \Ext_2 \CycC(\Harm(\Sph{1}))$, i.e.~$\PMC_{20}$ is a ``long cochain'' because it is non-zero in infinitely many weights.

\section{Twisted IBL-infinity-structure for spheres
}
\label{Section:HomSphere}
Let $e_0$, $e_1$ be the basis of $\Harm(\Sph{n})[1]$ defined by
\begin{equation*} \label{Eq:BasisOfHarm}
 e_0 \coloneqq \NOne \coloneqq \SuspU 1, \quad e_1\coloneqq\NVol \coloneqq \frac{1}{V}\SuspU\Vol.
\end{equation*}
The degrees satisfy
$$ \Abs{\NOne} = -1, \quad \Abs{\NVol} = n-1. $$
The matrix of the pairing $\Pair$ with respect to the basis $e_0$, $e_1$ reads
$$ \Pair=\begin{pmatrix}
 0 & 1 \\
 (-1)^{n} & 0
\end{pmatrix}. $$
The dual basis $e^0$, $e^1$ to $e_0$, $e_1$ with respect to $\Pair$ is thus
$$ e^0= \NVol,\quad e^1 = (-1)^{n} \NOne. $$
It follows that the matrix $(T^{ij})$ from~\eqref{Eq:PropagatorT} satisfies
\begin{equation*} 
 (T^{ij}) = - \begin{pmatrix}
0 & 1 \\
1 & 0
\end{pmatrix}.
\end{equation*}

We clearly have
$$ \CRedDBCyc \Harm(\Sph{1}) = \Bigl\{ \sum_{k=1}^\infty c_k \NVol^{k*} \BigMid c_k \in \R \Bigr\}, $$
where $\NVol^{k*}$ is the dual to the cyclic word $\NVol^k = \NVol \dots \NVol$ of length $k$. Observe that the cyclic symmetry gives
\begin{equation*}
\NVol^i = (-1)^{(n-1)(i-1)} \NVol^i\quad\text{for all }i\ge 1.
\end{equation*}
Therefore, $\NVol^{i*} = 0$ holds if both $n$ and $i$ are even.

For $n\ge 2$, the vector space $\Harm(\Sph{n})$ is connected and simply-connected, and Proposition~\ref{Prop:SimplCon} implies that there are no long reduced cyclic cochains (i.e., we have only finite sums of $\NVol^{k*}$'s).

The product $\mu_2: \Harm[1]^{\otimes 2} \rightarrow \Harm[1]$ from \eqref{Eq:HarmProd} has the following matrix with respect to the basis $\NOne$, $\NVol$:
\begin{equation*} 
 \mu_2 = \begin{pmatrix} \NOne &  \NVol \\ (-1)^n \NVol & 0 \end{pmatrix}.
\end{equation*}
Because $\mu_2(\NVol, \NVol) = 0$, we get
\begin{equation*}
\HIBL^\MC(\RedCycC(\Harm(\Sph{n})))[1] = \begin{cases}
                        \langle \Susp \NVol^{i*} \mid i \ge 1 \rangle & \text{for }n\ge 3\text{ odd}, \\
                        \langle  \Susp \NVol^{2i-1*} \mid i\ge 1 \rangle & \text{for }n\text{ even},\\
 \bigl\{ \Susp\sum_{k=1}^\infty c_k \NVol^{k*} \mid c_k\in \R \bigr\} & \text{for }n=1. 
\end{cases}
\end{equation*}
Because we are in the strictly unital and strictly augmented case, we obtain
\begin{equation} \label{Eq:HIBLSn}
\HIBL^\MC(\CycC)[1] = \begin{cases}
\langle \Susp \NVol^{i*}, \Susp \NOne^{2j-1*} \mid i, j \ge 1\rangle & \text{for }n\ge 3 \text{ odd}, \\
\langle \Susp \NVol^{2 i-1*}, \Susp \NOne^{2j-1*} \mid i, j \ge 1\rangle &\text{for }n\text{ even}, \\
 \bigl\langle \Susp\sum_{k=1}^\infty c_k \NVol^{k*}, \Susp \NOne^{2j-1*}\mid c_k\in \R, j \ge 1\bigr\rangle & \text{for }n=1. 
\end{cases}
\end{equation}
The canonical $\IBL$-operations can be written as
\begin{align*}
\OPQ_{210}(\Susp^2 \psi_1 \otimes \psi_2)(\Susp \omega) &= \begin{multlined}[t]-\sum \varepsilon(\omega\mapsto \omega^1 \omega^2)[(-1)^{(n-1)\Abs{\omega^1}} \psi_1(e_0 \omega^1) \\ \psi_2(e_1 \omega^2) + (-1)^{\Abs{\omega_1}}\psi_1(e_1 \omega^1) \psi_2(e_0 \omega^2)], \end{multlined} \\
\OPQ_{120}(\Susp \psi)(\Susp^2 \omega_1\otimes \omega_2) & = \begin{multlined}[t] - \frac{1}{2} \sum \varepsilon(\omega_1\mapsto \omega_{1}^{1}) \varepsilon(\omega_2\mapsto \omega_{2}^{1}) [(-1)^{(n-1)\Abs{\omega_{1}^{1}}} \\ \psi(e_0 \omega_{1}^{1} e_1 \omega_{2}^{1})  + (-1)^{\Abs{\omega_{1}^{1}}}\psi(e_1 \omega_{1}^{1} e_0 \omega_{2}^{1})] \end{multlined}
\end{align*}
for all $\psi$, $\psi_1$, $\psi_2 \in \CDBCyc\Harm$ and generating words $\omega$, $\omega_1$, $\omega_2\in \BCyc\Harm$. For all $k$, $k_1$, $k_2 \ge 1$, we have
$$ \OPQ_{210}((\Susp \NVol^{k_1*}) \cdot (\Susp \NVol^{k_2*})) = 0\quad\text{and}\quad \OPQ_{120}(\Susp \NVol^{k*}) = 0$$
because both $\OPQ_{210}$ and $\OPQ_{120}$ feed $\NOne$ into their inputs. For the \emph{canonically twisted reduced $\IBL$-algebra}, this implies the following:
$$ \IBL\bigl(\HIBL^\MC(\RedCycC)\bigr) = \bigl(\HIBL^\MC(\RedCycC), \OPQ_{210} \equiv 0, \OPQ_{120} \equiv 0 \bigr)\quad \text{for all }n\in \N.  $$
By Proposition~\ref{Prop:Ones}, the only possibly non-zero relation  of $\IBL(\HIBL^\MC(\CycC))$ is   
$$\begin{aligned}
& \OPQ_{210}(\Susp \NOne^* \otimes \Susp \NVol^{k*}) \\[\jot]
&\qquad = (-1)^{n-2} \Susp (\NVol^{k*} \circ \iota_\NVol) \\ 
&\qquad = (-1)^{n-2}\bigl(\sum_{i=1}^{k-1} (-1)^{i \Abs{\NVol}}\bigr)\Susp\NVol^{k-1 *}
= \begin{cases}
   -(k-1) \Susp \NVol^{k-1*} & \text{for }n\text{ odd},\\
    0 & \text{for }n\text{ even}.
  \end{cases}\end{aligned}$$
The reason for $0$ for even $n$ is that either $k$ is odd, in which case $\sum_{i=1}^{k-1} (-1)^i = 0$, or $k$ is even, in which case $\NVol^{k*} = 0$. Therefore, for the \emph{canonically twisted $\IBL$-algebra}, we have
\begin{equation*}
\IBL\bigl(\HIBL^\MC(\CycC)\bigr) = \bigl(\HIBL^\MC(\CycC), \OPQ_{210}, \OPQ_{120} \equiv 0 \bigr)\quad \text{for all }n\in \N,
\end{equation*}
where $\HIBL^\MC(\CycC)$ is given by \eqref{Eq:HIBLSn} and $\OPQ_{210}$ satisfies the following:
\begin{description}[font=\normalfont\itshape]
\item[($n$ even):] $\OPQ_{210} \equiv 0$.
\item[($n\ge 3$ odd):] The non-trivial relations are
$$ \OPQ_{210}(\Susp \NOne^* \otimes \Susp \NVol^{k*}) = \OPQ_{210}(\Susp \NVol^{k*} \otimes \Susp \NOne^*) = -(k-1) \NVol^{k-1*}\quad\text{for }k\ge 2.  $$
\item[($n=1$):]  The non-trivial relations are
$$ \OPQ_{210}\Bigl(\Susp \NOne^* \otimes \Susp\sum_{k=1}^\infty c_k \NVol^{k*}\Bigr) = - \Susp \sum_{k=1}^\infty k c_{k+1} \NVol^{k*} \quad\text{for }c_k\in \R. $$
\end{description}
Recall that the twist by $\MC$ does not produce any higher operation $\OPQ_{1lg}^\MC$.

We will now consider $\dIBL^\PMC(\CycC(\Harm(\Sph{n})))$. Recall that $\OPQ_{110}^\PMC = \OPQ_{210}\circ_1 \PMC_{10}$, $\OPQ_{210}^\PMC = \OPQ_{210}$ and $\OPQ_{120}^\PMC = \OPQ_{120} + \OPQ_{210}\circ_1 \PMC_{20}$. By Proposition~\ref{Proposition:MCSphere}, we have $\PMC_{10} = \MC_{10}$ for all $n\in \N$ and $\PMC_{20} = 0$ for all $n\ge 2$. It follows that $\OPQ_{110}^\PMC = \OPQ_{110}^\MC$ for all $n\in \N$ and that the only non-trivial twist may occur in $\OPQ_{120}^\PMC$ for $\Sph{1}$. Using~\eqref{Eq:Twistn2}, we get for all $\psi\in \CDBCyc \Harm(\Sph{n})$ and generating words $\omega_1$, $\omega_2 \in \BCyc \Harm(\Sph{n})$ the following:
\begin{equation}
\begin{aligned}
& (\OPQ_{210}\circ_1 \PMC_{20})(\Susp \psi)(\Susp \omega_1 \otimes \Susp \omega_2) \label{Eq:CoprodTwist}\\
& \quad = \begin{multlined}[t] (-1)^{n-2}\Bigl[ \sum \varepsilon(\omega_1 \mapsto \omega_1^1 \omega_1^2)\psi(\NOne \omega_1^1)\PMC_{20}(\Susp \NVol \omega_1^2 \otimes \Susp \omega_2) \\ {}+ (-1)^{(n-3+\Abs{\omega_1})(n-3+\Abs{\omega_2})} \sum \varepsilon(\omega_2 \mapsto \omega_2^1 \omega_2^2) \psi(\NOne \omega_2^1)  \\ \PMC_{20}(\Susp \NVol \omega_2^2 \otimes \Susp \omega_1)\Bigr].  \end{multlined}\end{aligned}
\end{equation}

In this paragraph, we suppose that $n=1$ and compute $\OPQ_{120}^\PMC$. Clearly, $(\OPQ_{210}\circ_1\PMC_{20})(\Susp \NVol^{k*}) = 0$ for all $k\ge 1$ since~$\NOne$ is fed into $\NVol^{k*}$. A non-zero evaluation of $(\OPQ_{210}\circ_1\PMC_{20})(\Susp\NOne^{k*})$ for some $k\ge 1$ odd is possible only on $\Susp \NOne^{k-1}\NVol^{k_1}\otimes \Susp\NVol^{k_2}$ for $k_1$, $k_2\ge 0$ (up to a transposition of arguments and their cyclic permutation). If $k>1$, only the first summand of~\eqref{Eq:CoprodTwist} contributes, and we get
\begin{equation*}
\begin{aligned}
&(\OPQ_{210}\circ_1\PMC_{20})(\Susp\NOne^{k*})(\Susp \NOne^{k-1}\NVol^{k_1}\otimes \Susp\NVol^{k_2})  \\
& \qquad= \begin{multlined}[t] (-1)^{n-2} \sum \varepsilon(\NOne^{k-1}\NVol^{k_{1}} \mapsto \omega_1 \omega_2) \NOne^{k*}(\NOne \omega_1) \PMC_{20}(\Susp \NVol \omega_2\otimes \Susp \NVol^{k_2}) \end{multlined} \\ & \qquad = (-1)^{n-2} \NOne^{k*}(\NOne\NOne^{k-1}) \PMC_{20}(\Susp \NVol\NVol^{k_{1}}\otimes \Susp \NVol^{k_2})  \\ & \qquad = - \PMC_{20}(\Susp \NVol^{k_{1}+1}\otimes \Susp\NVol^{k_2}).
\end{aligned}
\end{equation*}
According to Proposition~\ref{Proposition:MCSphere}, this is non-zero if and only if $k_{1}+k_2$ is odd. It follows that 
$$ \OPQ_{120}^\PMC \neq \OPQ_{120}^\MC = \OPQ_{120}\quad\text{on the chain level for }\Sph{1}. $$
However, the chains $\Susp \NOne^{k-1}\NVol^{k_1}\otimes \Susp \NVol^{k_2}$ for $k>1$ do not survive to the homology (c.f., \eqref{Eq:HIBLSn}). The only possibility is thus $k=1$. In this case, both summands of~\eqref{Eq:CoprodTwist} contribute, and  using~\eqref{Eq:MC20}, we get for all $k_1$, $k_2 \ge 1$ the following:
\allowdisplaybreaks
\begin{align*}
&(\OPQ_{210}\circ_1\PMC_{20})(\Susp\NOne^*)(\Susp\NVol^{k_1} \otimes \Susp\NVol^{k_2}) \\ &\qquad = \begin{multlined}[t](-1)^{n-2}\Bigl[\sum \varepsilon(\NVol^{k_1} \mapsto \NVol^0 \NVol^{k_1}) \NOne^*(\NOne) \PMC_{20}(\Susp\NVol^{k_1+1}\otimes \Susp \NVol^{k_2}) \\ {}+ (-1)^{(n-3 + k_1(n-1))(n-3 + k_2(n-1))} \sum \varepsilon(\NVol^{k_2} \mapsto \NVol^0 \NVol^{k_2})  \NOne^*(\NOne)\\ \PMC_{20}(\Susp \NVol^{k_2 + 1} \otimes\Susp\NVol^{k_1})\Bigr] \end{multlined}
\\&\qquad = -  k_1 \PMC_{20}(\Susp\NVol^{k_1+1}\otimes\Susp\NVol^{k_2}) -  k_2 \PMC_{20}(\Susp\NVol^{k_2+1}\otimes\Susp\NVol^{k_1}) \\ 
&\qquad = \begin{multlined}[t] -\frac{1}{2}(k_1+k_2+1)!I(k_1+k_2+1)\Bigl[(-1)^{k_1} k_1 (k_1+1) \binom{k_1+k_2}{k_1+1} \\ {}+ (-1)^{k_2} k_2  (k_2+1) \binom{k_1+k_2}{k_2+1}\Bigr] \end{multlined} \\ 
&\qquad =  -\frac{1}{2}(k_1+k_2+1)! k_1 k_2 \binom{k_1+k_2}{k_1} \underbrace{I(k_1 + k_2 + 1) [(-1)^{k_1} + (-1)^{k_2}]}_{=:(*)}.
\end{align*}
Denoting $k\coloneqq k_1 + k_2 + 1$, we have that $(-1)^{k_1} + (-1)^{k_2} = 0$ for $k$ even and $I(k) = 0$ for~$k$ odd. Therefore, $(*) = 0$ for any $k_1$, $k_2\ge 1$.
This implies that 
$$ \OPQ_{120}^\PMC = \OPQ_{120}^\MC = \OPQ_{120}\quad\text{on the homology for }\Sph{1}. $$
We conclude that the \emph{twisted $\IBL$-algebra} satisfies
$$ \IBL\bigl(\HIBL^\PMC(\CycC(\Harm(\Sph{n})))\bigr) = \IBL\bigl(\HIBL^\MC(\CycC(\Harm(\Sph{n})))\bigr) \quad\text{for all }n\in \N. $$

As for the \emph{higher twisted operations}, combining Propositions~\ref{Prop:dIBL} and~\ref{Proposition:MCSphere}, we see that for~$\Sph{n}$ with $n\in \N\backslash\{2\}$ all higher operations~$\OPQ_{1lg}^\PMC$ vanish already on the chain level. For $n=2$, we have that $\OPQ_{1l0}^\PMC = 0$ for all $l\ge 3$ and $\OPQ_{111}^\PMC = 0$ on the chain level. However, we did not prove that all higher operations vanish on the chain level. As for the operations induced on the homology, the graded vector space~$\HIBL^\PMC(\CycC(\Harm(\Sph{2})))$ is concentrated in even degrees and $\OPQ_{1lg}^\PMC$ are odd (see Definition~\ref{Def:IBLInfty}). Therefore, all higher operations vanish also on $\HIBL^\PMC(\CycC(\Harm(\Sph{2})))$.

The string topology $\StringH(\Sph{n})$ and the string operations $\StringOp_2$ and $\StringCoOp_2$ were computed in \cite{Basu2011} for all $n\in \N$.
We review their results and basic ideas below:

We will consider \emph{even spheres} first. The minimal model for the Borel construction $\LoopBorel \Sph{2m}$ for $m\in \N$ is denoted by $\Lambda^{\Sph{1}}(2,m)$ --- it is the free graded commutative dga (=:cdga) over~$\R$ generated by homogenous vectors $x_1$, $y_1$, $x_2$, $y_2$, $u$ of degrees
$$ \Abs{x_1} = 2m,\quad \Abs{y_1} = 2m - 1, \quad\Abs{x_2} = 4m-1,\quad \Abs{y_2} = 2(2m - 1),\quad\Abs{u} = 2, $$
whose differential $\Dd$ satisfies
$$ \Dd y_1 = 0, \quad \Dd x_1 = y_1 u, \quad \Dd y_2 = - 2 x_1 y_1, \quad\Dd x_2 = x_1^2 + y_2 u. $$
The minimal model for the loop space $\Loop \Sph{2m}$ is the dga $\Lambda(2,m)$ which is obtained from $\Lambda^{\Sph{1}}(2,m)$ by setting $u = 0$. A computation (see \cite[Theorem 3.6]{Basu2011}) gives the following for all $m\in \N$:
\begin{equation}\label{Eq:EvenSphereString}
\begin{aligned}
\H^*(\Loop \Sph{2m}; \R) &\simeq \H(\Lambda(2,m), \Dd) =\langle y_2^i x_1 - 2 i y_1 x_2 y_2^{i-1}, y_1 y_2^j, 1 \mid i, j\in \N_0 \rangle,  \\
\StringCoH^*(\Loop \Sph{2m}; \R) &\simeq \H(\Lambda^{\Sph{1}}(2,m),\Dd) = \langle y_1 y_2^i, u^j \mid i, j\in \N_0\rangle,
\end{aligned}
\end{equation}
where $y_2^0 \coloneqq u^0 \coloneqq 1$ is the unit in $\Lambda^{\mathrlap{\Sph{1}}\hphantom{S}}(2,m)$ and $\langle \cdot \rangle$ denotes the linear span over~$\R$. Clearly, the cohomology groups are degree-wise finite-dimensional, and hence, using the universal coefficient theorem, they are isomorphic to the corresponding homology groups. We can thus identify $\H(\Loop \Sph{2m}; \R)$ and $\StringH(\Loop\Sph{2m}; \R)$ with the vector spaces on the right hand side of \eqref{Eq:EvenSphereString}. We have $\StringH_{2k}= \langle u^k \rangle$ for all $k\in \N_0$, and hence the multiplication with $u$ induces an isomorphism $\StringH_{2k} \simeq \StringH_{2k+2}$. This corresponds to the cap product with the Euler class in \eqref{Eq:Gysin}, and exactness of the sequence implies $\Mark(\StringH_{2k}) = \Erase(\StringH_{2k}) = 0$. Using this and degree considerations, we get $\StringOp_2=\StringCoOp_2 = 0$.

We will now consider \emph{odd spheres} with $n\ge 3$. The minimal model for $\LoopBorel \Sph{2m+1}$ 
for $m\in \N$ is denoted simply by $\Lambda(x,y,u)$ --- it is the free cdga on homogenous vectors $x$, $y$, $u$ of degrees
$$ \Abs{x} = 2m+1, \quad \Abs{y} = 2m,\quad \Abs{u} = 2, $$
such that
$$ \Dd x = y u, \quad \Dd y = \Dd u = 0. $$
We get immediately
$$\begin{aligned}
\H^*(\Loop \Sph{2m+1}; \R) & \simeq \langle x^i, y^j \mid i, j \in \N_0 \rangle,  \\
\StringCoH^*(\Loop \Sph{2m+1}; \R) &\simeq \langle y^i, u^j \mid i,j \in \N_0 \rangle, \end{aligned}$$
and we can again identify $\H$ and $\StringH$ with the vector spaces on the right hand side. Clearly, $\StringH_{2k-1} = 0$ for all $k\in \N$, and hence $\StringOp_2 = \StringCoOp_2 = 0$ for degree reasons (the operations are odd).

We will now consider \emph{the circle} $\Sph{1}$. For every $i\in \Z$, let $\alpha_i : \Sph{1} \rightarrow \Sph{1}$ and $\theta_i : \Sph{1} \rightarrow \Loop \Sph{1}$ be the maps defined by
$$ \alpha_i(z) \coloneqq z^i\quad\text{and}\quad\theta_i(w) \coloneqq w \alpha_i \quad \text{for all }w,z\in \Sph{1}\subset \C. $$
By examining the equivariant homology of connected components of $\Loop \Sph{1}$ containing~$\alpha_i$ separately as in \cite[Section 2.1.4]{Basu2011}, we get 
$$\begin{aligned}
\H(\Loop \Sph{1}; \R) &=\langle \alpha_i, \theta_j \mid i,j\in \Z\rangle, \\
\StringH(\Loop \Sph{1}; \R) &= \langle u^i, \theta_0 u^j, \alpha_k \mid i, j \in \N_0, k\in \Z\backslash\{0\} \rangle,
\end{aligned}$$
where $u$ corresponds to the Euler class and
$$ \Abs{u} = 2, \quad \Abs{\theta_i} =1, \quad \Abs{\alpha_i} = 0 $$
are the degrees in the singular chain complex. On \cite[p. 21]{Basu2011} they show that the string cobracket $\StringCoOp_2$ is $0$ and that all non-trivial relations for the string bracket $\StringOp_2: \StringH(\Loop \Sph{1})[2]^{\otimes 2}\rightarrow \StringH(\Loop \Sph{1})[2]$ are the following:
\begin{equation*}
\StringOp_2( \Susp \alpha_{k}, \Susp \alpha_{-k}) = k^2 \Susp\theta_0 \quad\forall k \in \N.
\end{equation*}

We will now compare the reduced $\IBL$-structures motivated by Conjecture~\ref{Conj:StringTopology}. The point-reduced versions $\RedStringH(\Loop \Sph{n})$ for $n\ge 2$ are obtained from $\StringH(\Loop \Sph{n})$ by deleting $u^i$. We have the following isomorphisms of graded vector spaces:
$$ \begin{aligned}
  \HIBL^{\PMC}(\RedCycC(\Harm(\Sph{n})))[1] &\longrightarrow  \RedStringH(\Loop \Sph{n})[3-n] && \\ 
       \Susp \NVol^i &\longmapsto \Susp y^i  &&\text{for }n> 1\text{ odd}, \\  
    \Susp \NVol^{2i+1} &\longmapsto \Susp y_1 y_2^i  &&\text{for }n\text{ even}.
  \end{aligned}$$
Because all operations are trivial, it induces the isomorphism
$$ \IBL\bigl(\HIBL^{\PMC}(\RedCycC(\Harm(\Sph{n})))\bigr) \simeq \IBL\bigl(\RedStringH(\Loop \Sph{n})[2-n]\bigr) \quad\text{for }n\ge 2. $$
For $n = 1$, the reduced homology is seemingly different.

\begin{Remark}[Triviality for degree reasons]\label{Rem:DegRes}\Modify[caption={DONE Too dense text}]{Add paragraphs here --- too dense. Ans also add $\Sph{1}$!!}
The graded vector spaces 
$$ \StringH(\Loop \Sph{2m-1})[3-n]\quad\text{and}\quad\HIBL^\PMC(\CycC(\Harm(\Sph{2m})))[1] $$
are concentrated in even degrees, and so any $\IBLInfty$-structure must be trivial for degree reasons. On the other hand, the graded vector spaces 
$$ \StringH(\Loop \Sph{2m})\quad\text{and}\quad\HIBL^\PMC(\CycC (\Harm(\Sph{2m-1})))[1] $$
have both even and odd degrees, and hence an additional argument is needed to prove vanishing of the $\IBL$-structure. This is not the case of the reduced homology, which is again concentrated in even degree.\qedhere

\Add[caption={DONE Non-trivial degree}]{Add here the computation of which relations on homology are not implied automatically by degree reasons.}
\end{Remark}

\section{Homology of twisted IBL-infinity structure for complex projective space}
\label{Section:CPn1}
Let $\Kaehler\in \DR^2(\CP^n)$ be the Fubini--Study K\"ahler form on $\CP^n$ (see \cite[Examples 3.1.9]{Huybrechts2004}). The powers of $\Kaehler$ are harmonic,\footnote{This follows by induction on the power of $\Kaehler$ using the fact that, on a general K\"ahler manifold $M$, the Lefschetz operator $\Lef: \DR(M)\rightarrow \DR(M)$ defined by $\Lef(\eta)\coloneqq \eta \wedge \Kaehler$ for all $\eta\in \DR(M)$ commutes with the Hodge--de Rham Laplacian~$\Delta$ (see \cite[Chapter 3]{Huybrechts2004}). } and we get easily
\[ \Harm(\CP^n) = \langle 1,\Kaehler, \dotsc, \Kaehler^n \rangle. \]
We denote the Riemannian volume of $\CP^n$ by
\[ V\coloneqq \int_{\CP^n} \frac{1}{n!}\Kaehler^n. \]
Consider the basis $e_0$, $\dotsc$, $e_n$ of $\Harm(\CP^n)[1]$ defined for all $i=0$, $\dotsc$, $n$ by
\[ e_i\coloneqq \frac{\NK^i}{(n! V)^{\frac{i}{n}}}, \quad \text{where}\quad\NK^i\coloneqq \SuspU \Kaehler^i. \]
The matrix of the pairing $\Pair$ from~\eqref{Eq:DeRhamDGA} with respect to the basis $e_0$,~$\dotsc$,~$e_{n}$ reads:
\[ (\Pair^{ij}) = \begin{pmatrix}
0 & \dotsb & 1 \\
\vdots & {\displaystyle\, .^{{\displaystyle \, .^{\displaystyle\,.}}}} & \vdots \\
1 & \dotsb & 0
\end{pmatrix}. \]
The basis $e^0$, $\dotsc$, $e^n$ dual to $e_0$, $\dotsc$, $e_n$ with respect to $\Pair$ thus satisfies
\[ e^i = e_{n-i}\quad\text{for all }i=0,\dotsc,n. \]
Therefore, the following holds for the matrix $(T^{ij})$ from~\eqref{Eq:PropagatorT}:
\begin{equation*}
(T^{ij}) = - (\Pair^{ij}).
\end{equation*}
For all $1\le i, j, k \le n$, we have
\[ \mu_2(e_i, e_j) = e_{i+j}\quad \text{and}\quad
\MC_{10}(\Susp e_{i}e_j e_k) = \delta_{i+j+k,n}. \]
For $\psi$, $\psi_1$, $\psi_2 \in \CDBCyc \Harm$ and generating words $\omega$, $\omega_1$, $\omega_2 \in \BCyc\Harm$, we chave
\begin{equation*}
\begin{aligned}
\OPQ_{210}(\Susp^2 \psi_1 \otimes \psi_2)(\Susp \omega) &= -\sum_{i=0}^n \sum \varepsilon(\omega\mapsto \omega^1\omega^2)(-1)^{\Abs{\omega^1}} \psi_1(e_i \omega^1)\psi_2(e_{n-i}\omega^2), \\
\OPQ_{120}(\Susp \psi)(\Susp^2 \omega_1 \otimes \omega_2) & = - \sum_{i=0}^n \sum \varepsilon(\omega_1 \mapsto \omega_1^1)\varepsilon(\omega_2\mapsto \omega_2^1) (-1)^{\Abs{\omega_1}} \psi(e_i \omega_1^1 e_{n-i} \omega_2^1).
\end{aligned}
\end{equation*}

The cyclic homology of $\Harm(\CP^n)$ is that of the truncated polynomial algebra
\[ A \coloneqq\R[x]/(x^{n+1})\quad\text{with }\Deg(x)=2. \]
The following lemma computes its cyclic homology.
\begin{Lemma}[Cyclic homology of truncated graded polynomial algebra]
Consider $A\coloneqq\R[x]/(x^{n+1})$ with $\Deg(x)=d\in \Z$. For all $i=1$, $\dotsc$, $n$ and $k\in \N_0$, there are cycles $\tilde{t}_{2k+1,i}\in  \tilde{D}_q(A)$ of weights $w(\tilde{t}_{2k+1,i}) = 2k+1$ and degrees $\Abs{\tilde{t}_{2k+1,i}} = d(i+(n+1)k)$, where $q = w(\tilde{t}_{2k+1,i})-\Abs{\tilde{t}_{2k+1,i}}-1$, which form a basis of $\ClasCycH(A)$ (the non-degree shifted cyclic homology defined on page~\pageref{Eq:NDSComplex}).
\end{Lemma}
\begin{proof}
A computation of the cyclic homology of $A$ for $\Abs{x}=0$ is the goal of \cite[Exercise 4.1.8.]{LodayCyclic} or \cite[Exercise 9.1.1]{Weibel1994}.
The hint is to compute the Hochschild homology $\H\H_n(A) = \Tor_n^{A_e}(A,A)$, where $A_e$ is the enveloping algebra of $A$, using a non-canonical (i.e., not the bar complex) projective resolution of the $A_e$-module $A$ given by
\begin{equation}\label{Eq:ProjRes}
\begin{tikzcd}
\dotsb \arrow[r] & A_e \arrow{r}{\cdot v} & A_e \arrow{r}{\cdot u} & A_e \arrow{r}{\mu} & A \arrow{r} & 0,
\end{tikzcd}
\end{equation}
where $u = x \otimes 1 - 1 \otimes x$ and $v = \sum_{i=0}^{n} x^i \otimes x^{n-i}\in A_e$.
The resolution continues to the left with $\cdot u$ und $\cdot v$ periodically.
Clearly, $\mu$ composed with $\cdot u$ vanishes and $u\cdot v = v\cdot u = x^{n+1}\otimes 1 - 1 \otimes x^{n+1} = 0$; one can check that \eqref{Eq:ProjRes} is indeed a resolution.
If $\Deg(x)=d$, then $\Deg(u) = d$ and $\Deg(v) = n d$.

We lift the resolution \eqref{Eq:ProjRes} to the graded category (i.e., we require that the maps are homogenous) by taking the degree shifts
\[
\begin{tikzcd}
\dotsb \arrow{r} &
A_e[-(n+1)di] \arrow{r}{d_{2i}} \ar[draw=none]{d}[name=X, anchor=center]{}& 
A_e[-(n+1)di + nd] \arrow{r}{d_{2i-1}} &
A_e[-d(n+1)(i-1)]  \ar[rounded corners,
            to path={ -- ([xshift=2ex]\tikztostart.east)
                      |- (X.center) \tikztonodes
                      -| ([xshift=-2ex]\tikztotarget.west)
                      -- (\tikztotarget)}]{dlll}[at end]{}\\
\dotsb\arrow{r} &
A_e[-(n+1)d] \arrow{r}{d_{2}} &
A_e[-d] \arrow{r}{d_1}\ar[draw=none]{d}[name=Y, anchor=center]{} &
A_e  \ar[rounded corners,
            to path={ -- ([xshift=2ex]\tikztostart.east)
                      |- (Y.center) \tikztonodes
                      -| ([xshift=-2ex]\tikztotarget.west)
                      -- (\tikztotarget)}]{dl}{\mu} \\ 
&  & A \arrow{r} & 0,
\end{tikzcd}
\]
where we denoted by $d_i$ the maps from \eqref{Eq:ProjRes}.
Tensoring with $A$, we get the graded vector spaces $A \otimes_{A_e} A_e[\cdot] \simeq A[\cdot]$, and the maps $d_j$ become multiplications with $u$, $v \in A_e$ in $A$ as a right $A_e$-module.
For all polynomials $p\in A$, we have
$$ \begin{aligned}
    d_{2i-1}(p) = p\cdot (x\otimes 1 - 1 \otimes x) &= px - (-1)^{\Abs{x} \Abs{p}}xp = 0, \\
    d_{2i}(p) = p\cdot (\sum_{i=0}^n x^i \otimes x^{n-i}) & = (-1)^{n \Abs{x}\Abs{p}}(n+1) x^n p.
\end{aligned}$$
The homology of this chain complex consists of graded vector spaces $\H\H_{(l)}$ for $l\ge 0$ which correspond to the homology of the bar complex graded by weights, i.e., $\H\H_{(l)}$ would be represented by cycles in $A^{\otimes l+1}$ if the bar resolution was taken. We compute
\begin{equation}\label{Eq:HochCPn}
\H\H_{(l)} = \begin{cases} 
 (x A)[-(n+1)di] & \text{for }l=2i, \\
 \R[x]/(x^n)[-(n+1)di + nd] & \text{for }l = 2i - 1, \\
 A & \text{for }l = 0.
\end{cases}
\end{equation}
Now, because the differential $\tilde{\delta}$ is zero and $\tilde{\Hd}$ is degree preserving, it holds (c.f., \eqref{Eq:TotComplNDS})
\[\ClasHH(A) = \bigoplus_{l\in \N_0} \H\H_{(l)}.\] 
Clearly, the same will hold for $\ClasCycH(A)$, and hence we can ignore the gradation by degree and just use the gradation by weights in the bar complex, i.e., the non-graded theory.

In order to compute $\CycH_{(l)}(A)$, we consider the Connes' exact sequence in homology, or $\mathrm{ISB}$-sequence, see \cite[Theorem~2.2.1]{LodayCyclic}.
It arises from the exact sequence of bicomplexes $0 \rightarrow \LodCycBi^{\{2\}} \hookrightarrow \LodCycBi \twoheadrightarrow \LodCycBi[2,0] \rightarrow 0$, where~$\LodCycBi$ is the Loday's cyclic bicomplex from~\cite[Paragraph~2.1.2]{LodayCyclic} (bar complexes in columns), $\LodCycBi^{\{2\}}$ is the sub-bicomplex consisting of the first two columns of $\LodCycBi$ and $\LodCycBi[2,0]$ is the part of $\LodCycBi$ starting with the third column. 
Because $A$ is augmented and unital, \cite[Theorem~4.1.13]{LodayCyclic} guarantees a splitting of the ISB-sequence into
\begin{equation}\label{Eq:ConnexIBS}
0 \longrightarrow \widebar{\H}^\lambda_{(l-1)} \longrightarrow \widebar{\H\H}_{(l)} \longrightarrow \widebar{\H}^\lambda_{(l)} \longrightarrow 0\quad\text{for }l\ge 1,
\end{equation}
where the bar $\bar{\cdot}$ denotes the reduced homology. It holds $\CycH_{(0)}= \H\H_{(0)}= A$, and hence $\widebar{\H}^\lambda_{(0)} = \langle x, \dotsc, x^n \rangle$.
Using \eqref{Eq:HochCPn}, the first map for $l=1$ in \eqref{Eq:ConnexIBS} reads $\langle x,\dotsc, x^n \rangle \hookrightarrow \langle 1, x, \dotsc, x^{n-1}\rangle[-d]$, and hence it is an isomorphism.
It follows that $\widebar{\H}^\lambda_{(1)} = 0$.
For $k\ge 1$, we obtain inductively $\widebar{\H}^\lambda_{(2k)} \simeq \widebar{\H\H}_{(2k)} = \langle x,\dotsc,x^n\rangle[-(n+1)dk] \hookrightarrow \widebar{\H\H}_{(2k+1)} = \langle 1, x, \dotsc, x^{n-1} \rangle[-(n+1)dk - d]$. This again has to be an isomorphism, and hence $\widebar{\H}^\lambda_{(2k+1)} = 0$.
\end{proof}
We apply the degree shift $U: \tilde{D}(A) \rightarrow D(A)$ from Proposition~\ref{Prop:DGA} to get the generators
\[ t_{w,i} \coloneqq U(\tilde{t}_{w,i}) \in D^\lambda( \Harm(\CP^n)) \]
of weights $w$ and degrees $2i+ (w-1)n -1$, so that
\[ \H^\lambda(\Harm(\CP^n)) = \langle t_{w,i}, \NOne^{w} \mid w\in \N \text{ odd}, i=1,\dotsc, n\rangle. \]
By the universal coefficient theorem, we have $\H_\lambda^* = (\H^\lambda)^{\GD}$ with respect to the grading by the degree. Given $d\in \Z$, the equation $d= 2i + (w-1)n - 1$ has only finitely many solution $(w,i) \in \N \times \{1,\dotsc,n\}$, and hence we get
\begin{equation}\label{Eq:CPnHom}
\HIBL^\MC(\CycC(\Harm(\CP^n))) = \langle \Susp t_{w,i}^*, \Susp\NOne^{w*} \mid w\in \N \text{ odd}, i=1,\dotsc, n \rangle,
\end{equation}
where $t_{w,i}^*$ and $\NOne^{w*} \in \DBCyc \Harm$ are the duals to $t_{w,i}$ and $\NOne^{w}$, respectively (see Remark~\ref{Rem:UCT}). Notice that both $\Abs{\Susp t_{w,i}^*}$ and $\Abs{\Susp \NOne^{w*}}$ are even since $\Abs{\Susp} = 2n-3$.

Because $\CP^n$ is geometrically formal, Proposition~\ref{Prop:GeomForm} implies that $\PMC_{10} = \MC_{10}$. Because $\HIBL^\MC(\CycC)$ is concentrated in even degrees and because a general $\IBLInfty$-operation $\OPQ_{klg}$ is odd, all operations vanish on the homology. Therefore, for the \emph{twisted $\IBL$-algebras} we have
\begin{equation*}
\IBL(\HIBL^\PMC(\CycC)) = \IBL(\HIBL^\MC(\CycC)) = (\HIBL^\MC(\CycC), \OPQ_{210} \equiv 0, \OPQ_{120}\equiv 0),
\end{equation*}
where $\HIBL^\MC(\CycC)$ is given by \eqref{Eq:CPnHom}.

According to \cite[Section 3.1.2]{Basu2011}, the minimal model for the Borel construction $\LoopBorel \CP^n$ is the cdga $\Lambda^{\mathrlap{\Sph{1}}\hphantom{S}}(n+1,1)$, which is freely generated (over $\R$) by the homogenous vectors $x_1$, $x_2$, $y_1$, $y_2$, $u$ of degrees
\[ \Abs{x_1} = 2,\quad \Abs{x_2} = 2n+1,\quad \Abs{y_1} = 1,\quad \Abs{y_2} = 2n, \quad \Abs{u} = 2, \]
whose differential $\Dd$ satisfies
\[ \Dd y_1 = 0,\quad \Dd x_1 = y_1 u,\quad \Dd y_2 = -(n+1) x_1^n y_1,\quad \Dd x_2 = x_1^{n+1} + y_2 u. \]
By \cite[Theorem 3.6]{Basu2011}, the string cohomology $\StringCoH^*(\Loop \CP^n; \R)\simeq \H(\Lambda^{\mathrlap{\Sph{1}}\hphantom{S}}(n+1,1),\Dd)$ satisfies for all $m\in \N_0$ the following:
\begin{align*}
\StringCoH^m(\Loop \CP^n; \R) = \begin{cases} 
\langle u^j \rangle & \text{if }m=2j, \\
\langle y_1 y_2^p x_1^q \mid 0\le q \le n-1, p\ge 0; q + n p = j\rangle & \text{if }m=2j+1.
\end{cases}
\end{align*}
The right-hand side can be identified with $\StringH(\Loop \CP^n; \R)$ by the universal coefficient theorem. According to \cite[Proposition 3.7]{Basu2011}, we have $\StringOp_2 = 0$ and $\StringCoOp_2 = 0$. We conclude that the map
\[\begin{aligned}
 \HIBL^\PMC(\RedCycC(\Harm(\CP^n)))[1] & \longrightarrow \RedStringH(\Loop \CP^n; \R)[3-n] \\
\Susp t^*_{2k+1,l} & \longmapsto \Susp y_1 y_2^k x_1^{l-1}\qquad\text{for }k\ge 0\text{ and }l=1,\dotsc, n
\end{aligned} \]
induces an isomorphism of $\IBL$-algebras
\[ \IBL(\HIBL^\PMC(\RedCycC(\Harm(\CP^n)))) \simeq \IBL(\RedStringH(\Loop \CP^n; \R)[3-n]). \]

\part{Follow-up topics}

\chapter{IBL-infinity formality and Poincar\'e duality models}\label{Chap:5}

In this chapter, we study the following question:
\begin{Question}\label{Q:Form}
Let $M$ be a connected oriented closed $n$-manifold which is formal in the sense of rational homotopy theory.
Let $\MC$ be the canonical and $\PMC$ the formal pushforward (or Chern-Simons) Maurer-Cartan element for $\dIBL(\CycC(\HDR(M)))$.
Are $\dIBL^\MC(\CycC(\HDR(M)))$ and $\dIBL^\PMC(\CycC(\HDR(M)))$ homotopy equivalent as $\IBLInfty$-algebras?
\end{Question}
We remind that Theorem~\ref{IntroThm:B} states that for a geometrically formal manifold~$M$ with $\HDR^1(M) = 0$, one can pick a special Hodge propagator $\PrpgStd$ such that $\PMC = \MC$ at least for $n\neq 2$.\footnote{We did not prove that the homotopy class of $\dIBL^\PMC(\CycC(\HDR(M)))$ does not depend on $\PrpgStd$, but we expect so.} Question~\ref{Q:Form} asks for a generalization of Theorem~\ref{IntroThm:B}.
In the upcoming sections, we will propose a strategy to answer Question~\ref{Q:Form} in the case of $\HDR^1(M) = 0$, when the expected answer is ``yes''.
Our main tool is the theory of Poincar\'e duality models from~\cite{Lambrechts2007}.
Nevertheless, we also study $\DGA$'s of Hodge type and their small subalgebras from~\cite{Van2019}, which, as we think, naturally fit in the picture.

In Section~\ref{SubSec:CycStr}, we consider $\DGA$'s and define the notions of an orientation and cyclic structure (Definition~\ref{Def:CycStr}), non-degeneracy and Poincar\'e duality (Definition~\ref{Def:PoincDual}), degenerate subspace and non-degenerate quotient (Definition~\ref{Def:NonDegQ}), Hodge decomposition and Hodge pair (Definition~\ref{Def:HodgeDecomp}) and Hodge and small subalgebra from~\cite{Van2019} (Definition~\ref{Def:SmallSubalg}).
We prove that in some cases orientations and cyclic structures are in one-to-one correspondence (Proposition~\ref{Prop:OrAndCyc}) and that being of Hodge type is equivalent to acyclicity of the degenerate subspace (Proposition~\ref{Prop:HodgeAcyc}).
We describe the small subalgebra of~\cite{Van2019} in terms of Kontsevich-Soibelman--like evaluations of rooted binary trees with internal edges labeled with the identity or the standard Hodge homotopy (Proposition~\ref{Prop:SmallDescription}).
We study what happens when we take small subalgebras and non-degenerate quotients iteratively (Proposition~\ref{Prop:PropPropertiessd}).
We give examples when small subalgebras and their non-degenerate quotients differ for different Hodge decompositions of the same $\DGA$ (Examples~\ref{Ex:NonUniqueSmall} and~\ref{Ex:SUsix}).
We prove an important lemma allowing to extend an oriented $\DGA$ to a $\DGA$ of Hodge type later (Lemma~\ref{Lemma:Exte}).
We also prove several little lemmas (Lemmas~\ref{Lem:Pom}, \ref{Lem:PomLemma}, \ref{Lem:AutomaticInjectivity} and~\ref{Eq:LemSmallSub}) and make several remarks (Remarks~\ref{Rem:Eq}, \ref{Rem:VolForms}, \ref{Rem:NonDegPD}, \ref{Rem:RemarkHarm} and~\ref{Rem:OnHodgeSubalg}) which might be useful for future investigations.
We finish with some open questions (Questions~\ref{Q:QuestOnPoinc})

In Section~\ref{SubSec:PoincModel}, we define differential Poincar\'e duality algebras\footnote{The same as cyclic $\DGA$'s for unital commutative $\DGA$'s up to a degree shift and the correspondence of orientations and cyclic structures} and Poincar\'e $\DGA$'s (Definition~\ref{Def:PDGA}), i.e., $\DGA$'s whose homology is a Poincar\'e duality algebra.
We consider Poincar\'e duality models (Definition~\ref{Def:PDModel}).
We argue that a Poincar\'e $\DGA$ is formal if and only if it is formal as a $\DGA$ (Proposition~\ref{Prop:PoincModelOfFormal}).
We prove that an oriented $\DGA$ extends under some conditions to a $\DGA$ of Hodge type (Proposition~\ref{Prop:ExtensionOfHodgeType}) and use it to prove the existence and uniqueness of Poincar\'e duality models in some cases (Propositions~\ref{Prop:ExOfLambrStan} and~\ref{Prop:LambrechtUnique}).
We discuss minimal models (Remark~\ref{Rem:Models}).
We give examples of manifolds whose de Rham algebras do not admit a Poincar\'e duality model with just one arrow (Example~\ref{Ex:NoOneArrow}).
We finish with some open questions (Questions~\ref{Q:QuestionsPonc}).

In Section~\ref{Section:FuncIBL}, we prove functoriality of the canonical $\dIBL^\MC$ construction on differential Poincar\'e duality algebras up to $\IBLInfty$-homotopy (Proposition~\ref{Prop:Functorialityo}) and the uniqueness up to $\IBLInfty$-homotopy in a weak homotopy equivalence class of $\PDGA$'s in some cases (Proposition~\ref{Prop:ExteofDFS}).
We define $\IBLInfty$-formality of a $\PDGA$ (Definition~\ref{Def:IBLFormality}) and conjecture that $\IBLInfty$-formality is implied by $\DGA$-formality (Conjecture~\ref{Con:DGAIBLForm}); in fact, this conjecture holds when Proposition~\ref{Prop:ExteofDFS} holds.
We discuss another relevant notions of formality (Remark~\ref{Rem:Intfor}).
For the de Rham complex of a smooth manifold, we conjecture that the $\IBLInfty$-algebra on cyclic cochains of de Rham cohomology twisted by the Chern-Simons Maurer-Cartan element is homotopy equivalent to the $\dIBL$-algebra on cyclic cochains of the non-degenerate quotient of the small subalgebra of the de Rham complex twisted by the canonical Maurer-Cartan element (Conjecture~\ref{Conj:SmallSimplCon}).
Assuming that the conjectures hold, we obtain a positive answer to Question~\ref{Q:Form} when $\HDR^1(M)=0$ (Conjecture~\ref{Cor:FormCorollary}).

\section{Orientation, Poincar\'e duality, Hodge decomposition}
\label{SubSec:CycStr}

In this section, a $\DGA$ $V$ is a triple $(V,\Dd,\wedge)$, where 
\begin{itemize}
\item $V = \bigoplus_{n\in \Z} V^n$ is a $\Z$-graded vector space,
\item $\Dd: V\rightarrow V$ a differential of degree $1$, and
\item $\wedge: V\otimes V \rightarrow V$ an associative product of degree $0$ such that $\Dd$ and $\wedge$ satisfy the Leibnitz identity $\Dd(v_1 \wedge v_2) = \Dd v_1 \wedge v_2 + (-1)^{v_1} v_1 \wedge\Dd v_2$ for all homogenous $v_1$, $v_2\in V$.
\end{itemize}
In other words, we consider general, possibly non-unital and non-commutative, $\Z$-graded $\DGA$'s. We denote by $\deg(v)$ the degree of a homogenous element $v\in V$ and write $(-1)^v$ in the exponent.

\begin{Definition}[Orientations and cyclic structures]\label{Def:CycStr} 
Let $(V,\Dd)$ be a $\Z$-graded cochain complex. We define the following:
\begin{itemize}
\item An \emph{orientation in degree $n$} is a linear function $\Or : V^n \rightarrow \R$ such that
\[\Or \neq 0\quad\text{and}\quad\Or \circ \Dd = 0. \]
In other words, it is a surjective chain map $\Or : (V,\Dd) \rightarrow (\R[-n],0)$.
\item A~\emph{cyclic structure of degree $n$} is a homogenous bilinear form $\langle \cdot, \cdot \rangle : V \otimes V \rightarrow \R$ of degree $-n$ as a map such that for all homogenous $v_1$, $v_2\in V$, we have
\begin{equation}\label{Eq:CycStr}
\langle \Dd v_1, v_2\rangle = (-1)^{1+v_1 v_2} \langle \Dd v_2, v_1 \rangle.
\end{equation}
\end{itemize}
A cyclic structure on a $\DGA$ $(V,\Dd,\wedge)$ is additionally required to satisfy  
\begin{equation}\label{Eq:CycStrII}
\langle v_1 \wedge v_2, v_3 \rangle = (-1)^{v_3(v_1 + v_2)}\langle v_3\wedge v_1, v_2 \rangle
\end{equation}
for all homogenous $v_1$, $v_2$, $v_3\in V$.
\end{Definition}

Analogously, one can define a cyclic structure on an $\AInfty$-algebra.

\begin{Remark}[Cyclic $\DGA$]\label{Rem:Eq}
A cyclic $\DGA$ $(V,\Pair,\mu_1,\mu_2)$ of degree $n$ from Definition~\ref{Def:CyclicAinfty} is the same as a $\DGA$ equipped with a cyclic structure~$\langle\cdot,\cdot\rangle$ of degree $n$ which is symmetric, i.e., 
\[ \langle v_1, v_2 \rangle = (-1)^{v_1 v_2} \langle v_2, v_1\rangle \]
for all homogenous $v_1$, $v_2\in V$, and non-degenerate (see Definition~\ref{Def:PoincDual}). The correspondence is via the degree shift
\begin{align*}
\mu_1(\SuspU v) &= \SuspU \Dd(v),\\
\mu_2(\SuspU v_1, \SuspU v_2) &= (-1)^{v_1}\SuspU(v_1\wedge v_2),\\
\Pair(\SuspU v_1, \SuspU v_2) & = (-1)^{v_1} \langle v_1,v_2\rangle,
\end{align*}
where $\SuspU$ is a formal symbol of degree $-1$. From this reason, we sometimes call a $\DGA$ equipped with a non-degenerate symmetric cyclic structure a cyclic $\DGA$, although there are degree shifts involved.
\end{Remark}

The homology $\H(V)\coloneqq \H(V,\Dd)$ of a $\DGA$ $(V,\Dd,\wedge)$ is also a $\DGA$ with the induced product $\wedge$ and with zero differential. Given an orientation $\Or$ or a cyclic structure $\langle\cdot,\cdot\rangle$ on~$V$, we define the maps $\Or^\H: \H(V) \rightarrow \R$ and $\langle\cdot,\cdot\rangle^\H: \H(V)\otimes\H(V)\rightarrow\R$ for all closed $h_1$, $h_2\in V$ by
\begin{equation}\label{Eq:IndOnHom}
\begin{aligned}
\Or^\H([h_1]) &\coloneqq \Or(h_1)\text{ and}\\
\langle[h_1],[h_2]\rangle^\H &\coloneqq \langle h_1, h_2\rangle,
\end{aligned}
\end{equation}
respectively, where $[\cdot]$ denotes the cohomology class. It is easy to see that $\langle\cdot,\cdot\rangle^\H$ is a cyclic structure on $\H(V)$ and that $\Or^\H$ is an orientation on $\H(V)$ provided that $\Restr{\Or}{\Ker \Dd} \neq 0$. 

\begin{Proposition}[Orientation on homology]\label{Prop:OrOnHomG}
We have the following:
\begin{ClaimList}
\item Let $(V,\Dd)$ be a $\Z$-graded cochain complex, and let $n\in \Z$. Given an orientation $\tilde{\Or}: \H^n(V) \rightarrow \R$, there is an orientation $\Or^V: V^n \rightarrow \R$ such that~$\tilde{\Or} = \Or^\H$. If $\Dd V^n = 0$, then we have the correspondence
\[\text{Orientations on }V\text{ in degree }n\ \overset{1:1}{\simeq}\ \text{Orientations on }\H(V)\text{ in degree }n. \]  
\item If $(V_1,\Dd_1,\Or_1)$ and $(V_2,\Dd_2,\Or_2)$ are $\Z$-graded cochain complexes oriented in degree $n$ with $\Dd V_1^n = 0 = \Dd V_2^n$, then a chain map $f: V_1 \rightarrow V_2$ preserves the orientation if and only if the induced map $f_*: \H(V_1) \rightarrow \H(V_2)$ preserves the induced orientation.
\end{ClaimList}
\end{Proposition}
\begin{proof}
\begin{ProofList}
\item We define $\Or^V(v) \coloneqq \tilde{\Or}([v])$ for all closed $v\in V$ and extend it by $0$ to a complement of $\Ker \Dd$ in~$V$. It is obvious that $\Or^\H = \tilde{\Or}$. If $\Dd V^n = 0$, any complement is trivial and $\Or^V$ is uniquely determined by $\Or^\H$.

\item Let $v\in V_1^n$. Because $\Dd v = 0$, we have
\[ \Or_2(f(v)) = \Or^{\H}_2([f(v)]) = \Or^{\H}_2(f_*[v]) = \Or^{\H}_1([v]) = \Or_1(v). \]
This finishes the proof.\qedhere
\end{ProofList}
\end{proof}

Suppose that $1$ is a unit for $(V,\Dd,\wedge)$, i.e., $1\in V^0$, $\Dd 1 = 0$ and $1\wedge v = v \wedge 1 = v$ for all $v\in V$, and let $\langle \cdot,\cdot\rangle$ by a cyclic structure on $V$. For all homogenous $v_1$, $v_2\in V$, we have 
\begin{align*}
\langle v_1, v_2 \rangle &= \langle v_1\wedge 1,v_2\rangle \\
& = (-1)^{v_1 v_2}\langle 1\wedge v_2, v_1 \rangle \\
& = (-1)^{v_1 v_2}\langle v_2, v_1 \rangle,  
\end{align*}
and hence $\langle \cdot,\cdot\rangle$ is automatically symmetric.

Recall that a $\DGA$ is called commutative if $v_1 \wedge v_2 = (-1)^{v_1 v_2} v_2\wedge v_1$ for all homogenous $v_1$, $v_2\in V$. Commutativity of a $\DGA$ and symmetry of a general cyclic structure on it seem to be unrelated.

\begin{Proposition}[Correspondence of orientations and cyclic structures on $\DGA$'s]\label{Prop:OrAndCyc}
Let $(V,\Dd,\wedge)$ be a $\DGA$. Then the following holds:
\begin{ClaimList}
\item If $\wedge$ is commutative, then any orientation $\Or$ in degree $n$ induces a cyclic structure~$\langle\cdot,\cdot\rangle$ of degree $n$ which is given for all homogenous $v_1$, $v_2\in V$ by
\begin{equation}\label{Eq:PairForm}
\langle v_1, v_2 \rangle \coloneqq \begin{cases}
    \Or(v_1 \wedge v_2) & \text{if }\deg(v_1) + \deg(v_2) = n, \\
    0 & \text{otherwise}. \end{cases}
\end{equation}
\item If $1$ is a unit, then any non-zero cyclic structure $\langle\cdot,\cdot\rangle$ of degree $n$ induces an orientation~$\Or$ in degree $n$ by defining 
\begin{equation}\label{Eq:OrForm}
\Or(v) \coloneqq \begin{cases}
 \langle v, 1 \rangle & \text{for }v\in V^n,\\
 0 & \text{otherwise.}
 \end{cases}
\end{equation}
\item For a unital commutative $\DGA$ $(V,\Dd,\wedge,1)$, formulas \eqref{Eq:PairForm} and \eqref{Eq:OrForm} define the correspondence
\[ \text{Orientations in degree }n\ \overset{1:1}{\simeq}\ \text{Non-zero cyclic structures of degree }n.\]
\end{ClaimList}
\end{Proposition}
\begin{proof}
\begin{ProofList}
\item  Using the Leibnitz identity, properties of an orientation and commutativity, we check that
\begin{align*}
\langle \Dd v_1, v_2 \rangle &= \Or(\Dd v_1 \wedge v_2) \\
&= \Or\bigl(\Dd(v_1\wedge v_2) - (-1)^{v_1} v_1 \wedge \Dd v_2\bigr) \\
&= (-1)^{1+v_1}\Or(v_1\wedge\Dd v_2)\\
&= (-1)^{1+v_1 v_2} \Or(\Dd v_2 \wedge v_1) \\
&= (-1)^{1+v_1 v_2} \langle \Dd v_2, v_1\rangle
\end{align*}
and 
\begin{align*}
\langle v_1\wedge v_2, v_3\rangle &= \Or(v_1\wedge v_2\wedge v_3)\\
&= (-1)^{v_3(v_1+v_2)} \Or(v_3 \wedge v_1 \wedge v_2)\\
&= (-1)^{v_3(v_1+v_2)} \langle v_3\wedge v_1, v_2\rangle
\end{align*}
for all homogenous $v_1$, $v_2$, $v_3\in V$.
\item For all $v\in V$, we have
\begin{align*}
\Or(\Dd v) = \langle \Dd v, 1 \rangle = - \langle \Dd 1, v \rangle = 0.
\end{align*}
From $\langle\cdot,\cdot\rangle \neq 0$ it follows that there are homogenous $v_1$, $v_2\in V$ with $\deg(v_1) + \deg(v_2) = n$ such that $\langle v_1, v_2 \rangle \neq 0$. Then $v_1\wedge v_2\in V^n$ and
\begin{align*}
\Or(v_1 \wedge v_2) &= \langle v_1\wedge v_2,1\rangle\\
&= \langle 1 \wedge v_1, v_2\rangle\\
&=\langle v_1,v_2\rangle \neq 0.
\end{align*}
Therefore, $\Or$ is an orientation on $V$.
\item This is a combination of (a) and (c) plus the uniqueness, which is easy to check.\qedhere
\end{ProofList}
\end{proof}

\begin{Remark}[Volume forms]\label{Rem:VolForms}
If $\H^n(V) \simeq \R$, then orientations $\Or: \H^n(V)\rightarrow\R$  and elements $0\neq [\Vol] \in \H^n(V)$ called \emph{volume forms} are in one-to-one correspondence via
\[ \Or([\Vol]) = 1. \]
A consequence is the following:
Suppose that $(V_1,\Dd_1,\Or_1)$ and $(V_2,\Dd_2,\Or_2)$ are cochain complexes oriented in degree $n$ which satisfy 
\begin{equation}\label{Eq:SDDFG}
\H^n(V_1)\simeq\H^n(V_2)\simeq\R\quad\text{and}\quad\Dd_1 V_1^n \simeq \Dd_2 V_2^n = 0,
\end{equation}
so that Proposition~\ref{Prop:OrOnHomG} applies. Then a chain map $f: V_1 \rightarrow V_2$ preserves orientation if and only if the induced map $f_*: \H^n(V_1) \rightarrow \H^n(V_2)$ maps $[\Vol_1]$ to $[\Vol_2]$. In the category of unital commutative $\DGA$'s satisfying \eqref{Eq:SDDFG}, so that also Proposition~\ref{Prop:OrAndCyc} holds, if the orientations come from cyclic structures, then $f$ preserves cyclic structure if and only if~$f_*$ preserves volume form.
\todo[noline,caption={DONE Augmented case}]{In the augmented unital case, if the pairing is non-deg, one defines $\Vol$ uniquely by requiring that $\Vol \perp \bar{V}$ and $\Or(\Vol) = 1$. Let us not write it.}
\end{Remark}

\begin{Definition}[Non-degeneracy and Poincar\'e duality]\label{Def:PoincDual}
Given a graded vector space~$V$, a homogenous bilinear form ($\eqqcolon$\,pairing) $\langle\cdot,\cdot\rangle: V\otimes V \rightarrow \R$ of degree $-n$ as a map which is graded symmetric is called \emph{non-degenerate} if for every $v\in V$, the following implication holds:
\[ \langle v,w \rangle= 0\quad\text{for all }w\in V\quad\Implies\quad v=0. \]
We say that $\langle\cdot,\cdot\rangle$ satisfies \emph{Poincar\'e duality} if the map $\flat: V \rightarrow V^{\GD}$ (graded dual) defined by 
\[\flat(v)(w) \coloneqq \langle v,w\rangle\quad \text{for all }v,w\in V\]
is a graded isomorphism (it has degree $-n$ as a map) of graded vector spaces.
\end{Definition}

\begin{Remark}[On non-degeneracy and Poincar\'e duality]\phantomsection\label{Rem:NonDegPD}
\begin{RemarkList}
\item Clearly, Poincar\'e duality implies non-degeneracy. If the degree $k$ component $V^k$ of $V$ is finite-dimensional for every~$k\in \Z$ --- we say that $V$ is of \emph{finite type} --- then the opposite is true as well. If $n=0$, then Poincar\'e duality implies that $V$ is of finite type.
\item If $V$ is non-negatively graded, then non-degeneracy of $\langle\cdot,\cdot\rangle: V\otimes V \rightarrow \R$ implies $V = V^0 \oplus \dotsb \oplus V^n$. Therefore, non-negatively graded vector spaces of finite type which admit a non-degenerate homogenous bilinear form are finite-dimensional.
\item The de Rham complex $(\DR(M),\Dd,\wedge)$ of an oriented closed $n$-manifold~$M$ with the orientation $\int: \DR^n(M) \rightarrow \R$ is an oriented $\DGA$ whose cyclic structure is non-degenerate but does not satisfy Poincar\'e duality. On the other hand, the induced structure on homology $\H(\DR(M))$ satisfies Poincar\'e duality.
\qedhere
\end{RemarkList}
\end{Remark}

An analog of the following definition is used in \cite{Van2019} and also in \cite{Lambrechts2007} (under the name ``set of orphans''). 

\begin{Definition}[Degenerate subspace and non-degenerate quotient]\label{Def:NonDegQ}
Given a symmetric pairing $\langle\cdot,\cdot\rangle: V\otimes V\rightarrow\R$ on a graded vector space~$V$, we define the \emph{degenerate subspace}~$V^\perp\subset V$ by 
\begin{align*}
V^\perp\coloneqq \{ v\in V \mid \langle w,v \rangle = 0\text{ for all }w\in V\}.
\end{align*}
If $(V,\Dd,\wedge)$ is a $\DGA$ and $\langle\cdot,\cdot\rangle$ a cyclic structure, then \eqref{Eq:CycStr} implies that $V^\perp$ is a differential graded ideal in $V$, and thus we obtain the short exact sequence of $\DGA$'s
\begin{equation}\label{Eq:ImportantSES}
\begin{tikzcd}
0 \arrow{r} & V^\perp \arrow[hook]{r}{\iota} & V \arrow[two heads]{r}{\pi^\VansQuotient} & \VansQuotient(V) \coloneqq V / V^\perp \arrow{r} & 0,
\end{tikzcd}
\end{equation}
where $\iota$ is the inclusion and $\pi^\VansQuotient$ the canonical projection. We call the $\DGA$ $\VansQuotient(V)$ together with the induced non-degenerate cyclic structure $\langle\cdot,\cdot\rangle^\VansQuotient$ such that $\langle\pi^\VansQuotient(\cdot),\pi^\VansQuotient(\cdot)\rangle^\VansQuotient = \langle\cdot,\cdot\rangle$ the \emph{non-degenerate quotient.}
\end{Definition}

It was observed in~\cite{Van2019} that the question whether $(V^\perp,\Dd)$ is acyclic, and hence~$\pi^\VansQuotient$ is a quasi-isomorphism, turns out to be related to the existence of Hodge decomposition.

\begin{Definition}[Hodge decomposition]\label{Def:HodgeDecomp}
A cochain complex $(V,\Dd)$ with a symmetric cyclic structure $\langle\cdot,\cdot\rangle: V\otimes V \rightarrow \R$ is \emph{of Hodge type} if there exist subspaces $\Harm \subset \Ker \Dd$ and $C\subset V$ such that 
\begin{equation}\label{Eq:HodgeDecomp}
V = \Ker \Dd \oplus C, \quad \Ker \Dd = \Im \Dd \oplus \Harm\quad\text{and}\quad C \perp \Harm \oplus C,
\end{equation}
where $\perp$ denotes the relation of being perpendicular with respect to $\langle\cdot,\cdot\rangle$. Such decomposition is called a \emph{Hodge decomposition.} We call $\Harm$ the \emph{harmonic subspace} and $C$ the \emph{coexact part}.

Given a Hodge decomposition, we define the \emph{standard Hodge homotopy} $\HtpStd: V \rightarrow V$~by 
\[\HtpStd \coloneqq \begin{cases}
    -(\Restr{\Dd}{C})^{-1} & \text{on }\Im \Dd, \\
    0 & \text{on } \Harm \oplus C.
   \end{cases}\]
Then we have $\Dd \HtpStd = - \pi_{\Im \Dd}$, $\HtpStd \Dd = -\pi_{C}$, and hence 
\[ [\Dd,\HtpStd] = \Dd \HtpStd + \HtpStd \Dd = \pi_\Harm - \Id.\]
 We call $(\Harm,\HtpStd)$ the \emph{Hodge pair} associated to the Hodge decomposition \eqref{Eq:HodgeDecomp}.
\end{Definition}

\begin{Proposition}[Non-deg., fin.~type implies Hodge type]\label{Prop:NDegFin}
Any cochain complex of finite type with a non-degenerate symmetric cyclic structure is of Hodge type.
\end{Proposition}
\begin{proof}
This is \cite[Lemma~11.1]{Cieliebak2015}, and the proof uses formal Hodge theory.
\end{proof}

\begin{Remark}[Harmonic subspaces]\label{Rem:RemarkHarm}
In the situation of Proposition~\ref{Prop:NDegFin}, it was shown in \cite[Remark~2.6]{Van2019} that for any complement $\Harm$ of $\im\Dd$ in $\ker\Dd$ (in other words, the image of a section $\H(V)\rightarrow \Ker\Dd$) there is a coexact part $C$ such that $V=\Harm \oplus \Im \Dd \oplus C$ is a Hodge decomposition. From this reason, we call any complement of $\im \Dd$ in $\ker \Dd$ a \emph{harmonic subspace.}\footnote{Given a Hodge decomposition $V=\Harm\oplus\im\Dd\oplus C$ and a harmonic subspace $\Harm'$, then it holds $\Harm' = \Graph(\alpha: \Harm \rightarrow \Dd V)$ because $\Harm \oplus \Dd V = \Harm' \oplus \Dd V$, and one can take $C'=\Graph(- \alpha^\dagger - \frac{1}{2}\alpha\alpha^\dagger: C \rightarrow \Harm\oplus\Dd V)$.}
\end{Remark}

The following lemma will be used in the proof of Proposition~\ref{Prop:HodgeAcyc}.

\begin{Lemma}[Complement of acyclic subcomplex over $\R$]\label{Lem:Pom}
Let $f: V_1 \rightarrow V_2$ be an injective chain map of cochain complexes $(V_1,\Dd_1)$ and $(V_2,\Dd_2)$ over $\R$ such that $(V_1,\Dd_1)$ is acyclic. Then there is a chain map $g: V_2 \rightarrow V_1$ such that $g\circ f=\Id$.\footnote{This lemma can be used to prove that over $\R$, every surjective quasi-isomorphism is a deformation retraction and every injective quasi-isomorphism is a section of a deformation retraction.}
\end{Lemma}

\begin{proof}
For every $i\in \Z$, consider the diagram
\[\begin{tikzcd}
\Ker \Dd_1^i \oplus C^i_1 \arrow{r}{f^i} \arrow{d}{\Dd_1^i} & \Ker \Dd_2^i \oplus C^i_2 \arrow{d}{\Dd_2^i} \\
\Ker \Dd^{i+1}_1 \oplus C^{i+1}_1 \arrow{r}{f^{i+1}} & \Ker \Dd_2^{i+1} \oplus C^{i+1}_2,
\end{tikzcd}\]
where $C^i_1$ is a complement of $\Ker \Dd_1^i$ in $V_1^i$ and $C^i_2$ is a complement of $\Ker \Dd_2^i$ in~$V_2^i$. With respect to this decomposition, we write
\begin{align*}
f^i & = \begin{pmatrix}
f^{i}_{11} & f^i_{12} \\
f^{i}_{21} & f^i_{22}
\end{pmatrix},  & g^i &= \begin{pmatrix}
g^{i}_{11} & g^i_{12} \\
g^{i}_{21} & g^i_{22}
\end{pmatrix}, \\
\Dd^i_1 &= \begin{pmatrix}
0 & d^i_1 \\
0 & 0
\end{pmatrix},& \Dd^i_2 &= \begin{pmatrix}
0 & d^i_2 \\
0 & 0
\end{pmatrix}.
\end{align*}

The assumption $\H(V_1)=0$ implies that $d^i_1$ is an isomorphism. The fact that~$f$ is a chain map translates to
\begin{equation}\label{Eq:EqEqEq}
d_2^i f^i_{21} = 0,\quad f^{i+1}_{21} d_1^i = 0,\quad f^{i+1}_{11} d^i_1 = d^i_2 f^i_{22}.
\end{equation}
From the second relation and surjectivity of $d_1^i$, we get that $f_{21}^{i+1} = 0$. Now,~$f^{i+1}_{11}$ has to be injective because it is the only possibly non-zero part of $f$ on $\Ker d_1^{i+1}$. From the third relation of \eqref{Eq:EqEqEq} and the fact that $d_1^i$ is injective, we get that $f_{22}^i$ is injective as well.
Relations \eqref{Eq:EqEqEq} hold also for $g$ with $d_1$ and $d_2$ switched. In particular, we have $g_{21}^{i}=0$. The relation $g \circ f = \Id$ translates using $f^i_{21} = g^i_{21} = 0$ to
\begin{equation}\label{Eq:EqEq}
g^i_{11} f^i_{11} = \Id, \quad g^i_{11} f^i_{12} + g^i_{12}f^i_{22}=0, \quad g^i_{22} f^i_{22} = \Id.
\end{equation}
Because $d_1^i$ is an isomorphism, the last equation is equivalent to 
\[ \Id = d_1^i g^i_{22}f^i_{22} (d_1^i)^{-1} = g^{i+1}_{11} d_2^i f^i_{22} (d_1^i)^{-1} = g^{i+1}_{11} f^{i+1}_{11} d_1^i (d_1^i)^{-1} = g^{i+1}_{11} f^{i+1}_{11}. \]

We see that $g$ can be constructed as follows. For all $i\in \Z$, let $g_i^{11}$ be an arbitrary left inverse of $f_i^{11}$. Set $g_{22}^i \coloneqq (d_1^i)^{-1}g_{11}^{i+1} d_2^i$ and $g_i^{21}\coloneqq0$. Finally, $g_{12}^i$ has to be chosen such that the second equation of \eqref{Eq:EqEq} is satisfied. This is possible since we can first define~$g_{12}^i$ on $\Im f^i_{22}$ because~$f^i_{22}$ injective and then extend it by $0$ to a complement.
\end{proof}

Claim (b) of the following proposition corresponds to \cite[Lemma~2.8]{Van2019}. Claim (c) was suggested by Prof.~Hông Vân Lê via e-mail correspondence.

\begin{Proposition}[Hodge decomposition and acyclicity of $V^\perp$]\label{Prop:HodgeAcyc}
Let $(V,\Dd)$ be a cochain complex with a symmetric cyclic structure $\langle\cdot,\cdot\rangle: V\otimes V\rightarrow\R$. If $V$ is of Hodge type, then the following implications hold:
\begin{ClaimList}
\item If $\langle \cdot,\cdot \rangle$ is non-degenerate, then $\langle \cdot,\cdot \rangle^\H$ is non-degenerate.
\item If $\langle \cdot,\cdot\rangle^\H$ is non-degenerate, then $(V^\perp,\Dd)$ is acyclic.
\end{ClaimList}
Moreover, the following reverse implication holds:
\begin{ClaimList}[resume]
\item If $V$ is of finite type and $(V^\perp,\Dd)$ is acyclic, then $V$ is of Hodge type.
\end{ClaimList}
\end{Proposition}
\begin{proof}
\begin{ProofList}
\item Let $V = \Im \Dd \oplus \Harm \oplus C$ be a Hodge decomposition. Then $\Im \Dd \oplus C \subset \Harm^\perp$, and hence
\[ \langle \Dd \eta + b + c, b' \rangle = \langle b,b'\rangle\quad\text{for all }\eta\in V, c\in C\text{ and }b, b'\in \Harm. \]
The claim follows easily. Notice that having a Hodge decomposition, it holds
\[ \Restr{\langle \cdot,\cdot\rangle}{\Harm\otimes \Harm}\text{ non-degenerate}\quad\Equiv\quad \Im\Dd \oplus C=\Harm^\perp. \] 
\item Consider a Hodge decomposition as above, and let $v\in V^\perp \cap \Ker \Dd$ be a non-zero vector. Suppose that $v\not\in \Im\Dd$. Then $[v] \neq 0$ in $\H(V)$, and hence there is a $b\in \Harm$ such that $\langle v,b \rangle \neq 0$ by non-degeneracy of $\langle\cdot,\cdot\rangle^\H$. This is a contradiction with $v\in V^\perp$. Therefore, it holds $V^\perp \cap \Ker \Dd = V^\perp \cap \Im \Dd$. In particular, there is an $\eta \in C$ such that $v = \Dd \eta$. Now, for any $\eta'\in C$, $b\in \Harm$ and $c\in C$, we have using $C\perp \Harm\oplus C$ and $v\in V^\perp$ the following:
\begin{align*}
\langle \Dd \eta' + b + c, \eta \rangle &= \langle \Dd \eta',\eta \rangle \\
&=(-1)^{1+\eta\eta'}\langle\Dd\eta,\eta'\rangle\\
&= (-1)^{1+\eta\eta'}\langle v,\eta'\rangle\\
&= 0.
\end{align*}
Therefore, it holds $\eta\in V^\perp$, and we have shown that $V^\perp \cap \Im\Dd = \Dd V^\perp$. The claim follows.
\item Because $V^\perp \subset V$ is an acyclic subcomplex and we work over $\R$, there is a complementary subcomplex $Z\subset V$; i.e., $\Dd Z\subset Z$ and $V = V^\perp\oplus Z$. This follows from Lemma~\ref{Lem:Pom} by setting $V^\perp = \Im f$ and $Z=\Ker g$. \Add[caption={Deformation retract over $\R$ add},noline]{Add here deformation retract somewhere.}Now, the restriction of $\langle\cdot,\cdot\rangle$ to $Z$ is non-degenerate, and Proposition~\ref{Prop:NDegFin} provides its Hodge decomposition $Z=\Dd Z \oplus \Harm \oplus D$. Let $E\subset V^\perp$ be a graded vector space which is complementary to $\Dd V^\perp$ in $V^\perp$; i.e., $V^\perp = \Dd V^\perp \oplus E$. It is easy to check that $V=\Im\Dd\oplus\Harm\oplus C$ with $C\coloneqq D \oplus E$ is a Hodge decomposition.
\qedhere
\end{ProofList}
\end{proof}
\todo[noline,caption={PD implies Hodge without fin type}]{Does the following hold without assuming finite type? I.e. does Poincar\'e duality imply Hodge type?}
The following notions were taken from \cite{Van2019}.

\begin{Definition}[Hodge subalgebra and small subalgebra]\label{Def:SmallSubalg}
Consider a $\DGA$ $(V,\Dd,\wedge)$ with a symmetric cyclic structure $\langle\cdot,\cdot\rangle$. Suppose that it admits a Hodge decomposition with the Hodge pair $(\Harm,\HtpStd)$. A \emph{Hodge subalgebra} is a differential graded subalgebra $W\subset V$ which satisfies
\[ \Harm\subset W \quad\text{and}\quad \HtpStd W \subset W. \]
We denote the smallest Hodge subalgebra of $V$ by $\VansSmall(V)$ and call it the \emph{small subalgebra}. We stress that the definition of $\VansSmall(V)$ depends on $(\Harm,\HtpStd)$!
\end{Definition}

The following is a version of \cite[Proposition~3.3]{Van2019} which generalizes to the non-simply-connected case (see (iii) of Remark~\ref{Rem:OnHodgeSubalg} below for the comparison).

\begin{Proposition}[Description of small subalgebra]\label{Prop:SmallDescription}
Consider the situation of Definition~\ref{Def:SmallSubalg}. The small subalgebra $\VansSmall(V)$ is generated as a graded vector space by Kontsevich-Soibelman--like evaluations of rooted binary trees with $k\ge 1$ leaves labeled with homogenous elements of~$\Harm$, interior vertices labeled with $\wedge$ and interior edges labeled either with~$\StdHtp$ or with~$\Id$.
\end{Proposition}
\begin{proof}
We abbreviate $\VansSmall \coloneqq \VansSmall(V)$, denote the set of labeled trees by $\Trees$ and denote the vector space generated by evaluations of elements of $\Trees$ by $\langle \Trees\rangle$.

Clearly, we have $\langle \Trees\rangle \subset \VansSmall$. For ``$=$'', it suffices to check that for any $T$, $T_1$, $T_2\in\Trees$, it holds $\Dd T$, $\HtpStd T$, $T_1 \wedge T_2 \in \langle\Trees\rangle$, i.e., that $\langle \Trees\rangle$ is a Hodge subalgebra.

As for $\Dd T$, we imagine $\Dd$ propagating from the root to the leaves. When it encounters $L\wedge R$, where $L$ stands for the left and $R$ for the right sub-branch, it duplicates the tree and continues propagating in $L$ and $R$, respectively (we take the sum of the two copies in the end). This is justified by the Leibnitz identity $\Dd(L\wedge R) = \Dd L\wedge R + (-1)^L L \wedge \Dd R$. When it encouners $\HtpStd$, it triples the tree and either goes past $\HtpStd$ and continues propagating, or exchanges $\HtpStd$ for $\Id$ and stops, or exchanges $\HtpStd$ for $\pi_\Harm$ and stops. This is justified by $\Dd \HtpStd = -\HtpStd \Dd - \Id + \pi_\Harm$. If it encounters $\Id$, nothing happens and it keeps propagating. If it reaches a leaf with $h\in \Harm$, then the corresponding tree evaluates to $0$. We see that we alway obtain an element of $\langle\Trees\rangle$.

As for $\HtpStd T$, we have either $\HtpStd T = 0$ if the interior edge adjacent to the root ($\eqqcolon$ the root edge) is labeled with $\HtpStd$, or $\HtpStd T$ is a new tree in $\Trees$ which arises from $T$ by replacing $\Id$ by $\HtpStd$ on the root edge.

As for $T_1 \wedge T_2$, using $\Id = \pi_\Harm - \Dd \HtpStd - \HtpStd \Dd$ and the Leibnitz identity, we get
\begin{align*}
T_1 \wedge T_2 &= \pi_\Harm(T_1 \wedge T_2) - (\Dd \HtpStd)(T_1 \wedge T_2) - (\HtpStd \Dd)(T_1 \wedge T_2) \\
&= \pi_\Harm(T_1 \wedge T_2) - \Dd\bigl(\HtpStd(T_1 \wedge T_2)\bigr) - \HtpStd(\Dd T_1 \wedge T_2)
- (-1)^{T_1} \HtpStd(T_1 \wedge \Dd T_2).
\end{align*}
We see that $T_1\wedge T_2 \in \langle \Trees\rangle$.
\end{proof}

\begin{Remark}[On Hodge subalgebra and small subalgebra]\phantomsection\label{Rem:OnHodgeSubalg}
\begin{RemarkList}
\item Any Hodge subalgebra~$W$ inherits the Hodge decomposition
\begin{equation}\label{Eq:InducedHodgeDecomp}
W = \Harm \oplus \Dd W \oplus \HtpStd W
\end{equation}
with the Hodge pair $(\Harm,\Restr{\HtpStd}{W})$.
\item We will be in the situation of Definition~\ref{Def:SmallSubalg} and use the notation of the proof of Proposition~\ref{Prop:SmallDescription}. In addition, we suppose that $V$ is non-negatively graded. Having established $\VansSmall = \langle \Trees \rangle$, consider the Hodge decomposition \eqref{Eq:InducedHodgeDecomp} for $W = \VansSmall$. Let $\widebar{\VansSmall} = \bigoplus_{k\ge 1} \VansSmall^k$ denote the reduced part of $\VansSmall$, and let $\langle\widebar{\VansSmall} \wedge \widebar{\VansSmall}\rangle$ denote the graded vector space generated by products $v_1 \wedge v_2$ for $v_1$, $v_2\in \widebar{\VansSmall}$. We have
\begin{align*}
\HtpStd\langle\Trees\rangle &= \bigl\langle\{ T\in \Trees \text{ with }\HtpStd\text{ on the root edge}\}\bigr\rangle \\
&= \HtpStd\langle\widebar{\VansSmall} \wedge \widebar{\VansSmall}\rangle
\end{align*}
and
\begin{align*}
 \Dd\langle\Trees\rangle &= \Dd (\pi_\Harm - \Dd \HtpStd - \HtpStd\Dd)\langle\Trees\rangle \\
 &= \Dd \HtpStd \Dd \langle\Trees\rangle \\
 &\subset \Dd \HtpStd \langle\Trees\rangle \\
 &\subset \Dd \HtpStd\langle\widebar{\VansSmall}\wedge\widebar{\VansSmall}\rangle.
\end{align*}
It holds even $\Dd\langle\Trees\rangle = \Dd\HtpStd\langle\widebar{\VansSmall}\wedge\widebar{\VansSmall}\rangle$ due to \eqref{Eq:InducedHodgeDecomp}. We can now write \eqref{Eq:InducedHodgeDecomp} as 
\begin{equation}\label{Eq:VansFormula}
\VansSmall^k = \Harm^k \oplus \Dd \HtpStd \langle\widebar{\VansSmall}\wedge\widebar{\VansSmall}\rangle^k \oplus \HtpStd \langle\widebar{\VansSmall}\wedge\widebar{\VansSmall}\rangle^{k+1}.
\end{equation}
If $\H^1 = 0$, this agrees with the formula from \cite[Proposition~3.3]{Van2019}. In this case, it holds $\VansSmall^1 = 0$, and hence $\langle\widebar{\VansSmall}\wedge\widebar{\VansSmall}\rangle^{k+1}$ depends only on $\VansSmall^i$ for $i<k$. Therefore, we can compute~$\VansSmall^k$ from \eqref{Eq:VansFormula} inductively starting with $\VansSmall^0 = \langle 1\rangle$.
\item The previous remark implies the following: Let $(V,\Dd,\wedge)$ be a non-negatively graded $\DGA$ with cyclic structure $\langle\cdot,\cdot\rangle$ of Hodge type. Suppose that $\H^1(V)=0$. If $\H(V)$ is of finite type, then so is $\VansSmall(V)$.
\item The advantage of $\VansSmall(V)$ is that we have the diagram of pairing preserving $\DGA$-quasi-isomorphisms
\[\begin{tikzcd}
 V & \VansSmall(V)\arrow[hook']{l} \arrow[two heads]{r} & \VansQuotient(\VansSmall(V)).
\end{tikzcd}\]
Moreover, in the case of (iii), the non-degenerate quotient is finite-dimensional, and hence a Poincar\'e duality model of $V$ (see the next section).
\item Notice that if there is a quasi-isomorphism $f:\H(V)\rightarrow V$, then $\VansSmall(V) = V$ for any Hodge decomposition with harmonic subspace $\im f$.  \qedhere
\end{RemarkList}
\end{Remark}

\begin{Proposition}[Properties of $\VansSmall$ and $\VansQuotient$]\label{Prop:PropPropertiessd}
Let $(V,\Dd,\wedge)$ be a $\DGA$ with a symmetric cyclic structure $\langle\cdot,\cdot\rangle$. Suppose that it admits a Hodge decomposition with Hodge pair $(\Harm,\HtpStd)$. Then $\VansSmall(V)$ and $\VansQuotient(V)$ admit Hodge decompositions with the Hodge pairs $(\Harm,\HtpStd^\VansSmall\coloneqq\Restr{\HtpStd}{\VansSmall(V)})$ and $(\pi^\VansQuotient(\Harm),\HtpStd^\VansQuotient)$, respectively, where $\pi^\VansQuotient: V \rightarrow \VansQuotient(V)$ is the canonical projection and $\HtpStd^\VansQuotient$ the unique map on $\VansQuotient(V)$ satisfying $\HtpStd^\VansQuotient \circ \pi^\VansQuotient=\pi^\VansQuotient\circ \HtpStd$. Furthermore, with respect to the induced Hodge decompositions, we have
\begin{equation}\label{Eq:QSRelations}
\VansSmall(\VansSmall(V)) = \VansSmall(V), \quad \VansQuotient(\VansQuotient(V)) = \VansQuotient(V)\quad\text{and}\quad \VansSmall(\VansQuotient(\VansSmall(V))) = \VansQuotient(\VansSmall(V)).\end{equation}
\end{Proposition}
\begin{proof}
The fact that $\VansSmall\coloneqq\VansSmall(V)$ admits a Hodge decomposition with Hodge pair $(\Harm,\HtpStd^\VansSmall)$ was stated in (i) of Remark~\ref{Rem:OnHodgeSubalg} and is easy to check.

We prove that $\VansQuotient\coloneqq \VansQuotient(V)$ has a Hodge decomposition with Hodge pair $(\pi^\VansQuotient(\Harm),\HtpStd^\VansQuotient)$. Because $\pi^\VansQuotient$ is a quasi-isomorphism, $\pi^\VansQuotient(\Harm)$ is a harmonic subspace of $\VansQuotient$, and we have $\Ker \Dd^\VansQuotient = \pi^\VansQuotient(\Harm)\oplus \Im \Dd^\VansQuotient$. Let $c\in C$ with $\Dd c \in V^\perp$ such that $\pi^\VansQuotient(c)\in\Ker\Dd^\VansQuotient \cap \pi^\VansQuotient(C)$. From the cyclicity of $\langle\cdot,\cdot\rangle$ with respect to $\Dd$ and from $C\perp\Harm\oplus C$, it follows that $c\in V^\perp$, and thus $\pi^\VansQuotient(c)=0$. Together with surjectivity of $\pi^\VansQuotient$ this implies that $\VansQuotient = \ker \Dd^\VansQuotient \oplus \pi^\VansQuotient(C)$. From $\langle \pi^\VansQuotient(\cdot),\pi^\VansQuotient(\cdot)\rangle^\VansQuotient$, we see that $\pi^\VansQuotient(C) \perp \pi^\VansQuotient(\Harm) \oplus \pi^\VansQuotient(C)$. Therefore, $\VansQuotient = \pi^\VansQuotient(\Harm)\oplus\Im\Dd^\VansQuotient\oplus\pi^\VansQuotient(C)$ is a Hodge decomposition, and it is easy to see that its standard Hodge homotopy $\HtpStd^\VansQuotient$ satisfies $\HtpStd^\VansQuotient \circ \pi^\VansQuotient=\pi^\VansQuotient\circ \HtpStd$. This defines $\HtpStd^\VansQuotient$ uniquely because $\pi^\VansQuotient$ is surjective.

As for the relations \eqref{Eq:QSRelations}, the first two are clear. The third can be seen as follows. Since $\pi^\VansQuotient: V \rightarrow \VansQuotient$ is a $\DGA$-morphism mapping the harmonic subspaces isomorphically onto each other and commuting with $\Dd$ and $\HtpStd$, the assignments $Y\subset V \mapsto \pi^\VansQuotient(Y)\subset \VansQuotient$ and $Z\subset \VansQuotient \mapsto (\pi^{\VansQuotient})^{-1}(Z)\subset V$ preserve Hodge subalgebras. Therefore, if $Z\subset \VansQuotient(\VansSmall)$ is a Hodge subalgebra, then $(\pi^\VansQuotient)^{-1}(Z)\subset\VansSmall$ is a Hodge subalgebra. It holds even $(\pi^\VansQuotient)^{-1}(Z)=\VansSmall$ by minimality of $\VansSmall$, and hence $Z=\VansQuotient(\VansSmall)$ by surjectivity of~$\pi^\VansQuotient$.  
\end{proof}

A natural question is, how does $\VansQuotient(\VansSmall(V))$ depend on the chosen Hodge pair and how does it behave under quasi-isomorphisms? The following lemmas might be useful.

\begin{Lemma}[Kernel of pairing-preserving morphism]\label{Lem:PomLemma}
Let $V_1$ and $V_2$ be vector spaces with symmetric bilinear forms $\langle \cdot,\cdot \rangle_1: V_1 \otimes V_1 \rightarrow \R$ and $\langle \cdot,\cdot\rangle_2: V_2\otimes V_2 \rightarrow \R$, respectively. Let $f: V_1 \rightarrow V_2$ be a linear map such that
\begin{equation}\label{Eq:Isometryyy}
\langle v_1, v_2 \rangle_1 = \langle f(v_1), f(v_2) \rangle_2 \quad \text{for all }v_1, v_2 \in V_1.
\end{equation}
Then it holds 
\[ \Ker f \subset V_1^\perp\quad\text{and}\quad f(V_1^\perp)\subset f(V_1)^\perp. \]
In particular, the following statements are true:
\begin{ClaimList}
\item If $\langle\cdot,\cdot\rangle_1$ is non-degenerate, then $f$ is injective.
\item If $\langle\cdot,\cdot\rangle_2$ is non-degenerate and $f$ is surjective, then $\Ker f = V_1^\perp$.
\end{ClaimList}
\end{Lemma}
\begin{proof}
Clear.
\end{proof}

\begin{Lemma}[Injectivity on domain with non-degenerate orientation]\label{Lem:AutomaticInjectivity}
Let $(V_1,\Dd_1,\wedge_1)$ be a non-negatively graded commutative $\DGA$ with an orientation $\Or_1$ in degree $n$ such that the induced cyclic structure is non-degenerate. Let $(V_2,\Dd_2,\wedge_2)$ be any $\DGA$, and let $f: V_1 \rightarrow V_2$ be a morphism of $\DGA$'s. Then injectivity of $f_*: \H^n(V_1) \rightarrow \H^n(V_2)$ implies injectivity of $f: V_1 \rightarrow V_2$.
\end{Lemma}
\begin{proof}
Firstly, injectivity of a homogenous map is equivalent to degree-wise injectivity. Secondly, because $V_1$ is non-negatively graded and the cyclic structure of degree $n$ is non-degenerate, we have $V_1 = V_1^0\oplus\dotsb\oplus V_1^n$. Now, suppose that $v\in V_1^k$ for some $k=0$,~$\dotsc$, $n$ satisfies $f(v) = 0$. For any $w\in V^{n-k}$, the product $v\wedge w$ lies in $V^n$, and thus $\Dd(v\wedge w)=0$. We compute
\begin{align*}
f_* [v\wedge w] = [f(v\wedge w)] = [f(v)\wedge f(w)] = 0,
\end{align*}
and hence $v\wedge w = \Dd \eta$ for some $\eta \in V_1^{n-1}$ by injectivity of $f_*$. Consequently, we have
\begin{align*}
\Or_1(v\wedge w) &= \Or_1(\Dd \eta) =0,
\end{align*}
and hence $v = 0$ by non-degeneracy of $\Or_1$.
\end{proof}

\begin{Lemma}[Small subalgebra of cyclic $\DGA$ and quasi-iso.]\label{Eq:LemSmallSub}
Let $(V_1,\Dd_1,\wedge_1,\langle\cdot,\cdot\rangle_1)$ and~$(V_2,\Dd_2,\wedge_2,\langle\cdot,\cdot\rangle_2)$ be non-negatively graded unital commutative $\DGA$'s of finite type with non-degenerate cyclic structures of degree $n$ (hence finite-dimensional). Let $f: V_1 \rightarrow V_2$ be a $\DGA$-morphism such that $f_*: (\H(V_1),\Or_1^\H)\rightarrow (\H(V_2),\Or_2^\H)$ is an isomorphism. Then a Hodge decomposition of $V_1$ with Hodge pair $(\Harm_1,\HtpStd^1)$ induces a Hodge decomposition of~$V_2$ with Hodge pair $(f(\Harm_1),\HtpStd^2)$, where~$\HtpStd^2$ satisfies $\HtpStd^2 \circ f = f\circ \HtpStd^1$. Consequently,~$f$ induces an isomorphism $\VansSmall(V_1) \simeq \VansSmall(V_2)$.
\end{Lemma}
\begin{proof}
Because $f_*$ preserves orientation and it holds $V_1^{n+1} = 0 = V_2^{n+1}$, Proposition~\ref{Prop:OrOnHomG} implies that $f$ preserves cyclic structure and hence is injective by Lemma~\ref{Lem:PomLemma}. Let $V_1 = \Harm_1 \oplus \Im \Dd_1 \oplus C_1$ be a Hodge decomposition. From the injectivity of $f$, it follows that
\[ f(V_1) = f(\Harm_1) \oplus \Dd f(V_1) \oplus f(C_1) \]
and that the restriction of $\langle\cdot,\cdot\rangle_2$ to $f(V_1)\otimes f(V_1)$ is non-degenerate. Because $V_1$ is of finite type, $f(V_1)$ is of finite type, and so non-degeneracy implies Poincar\'e duality $f(V_1)^{\GD}\simeq f(V_1)$. It follows that
\[ V_2 = f(V_1) \oplus f(V_1)^\perp.\]
Cyclicity of $\Dd_2$ with respect to $\langle\cdot,\cdot\rangle_2$ implies that $f(V_1)^\perp\subset V_2$ is a subcomplex. Because~$f$ is a quasi-isomorphism, we have $\H(V_1)\simeq \H(f(V_1))\simeq \H(V_2)$; because the homology is additive, we have $\H(V_2)\simeq\H(f(V_1))\oplus\H(f(V_1)^\perp)\simeq \H(V_2)\oplus\H(f(V_1)^\perp)$; finally, because $\H(V_2)$ is of finite type, we have $\H(f(V_1)^\perp) = 0$. Because $V_2 = f(V_1) \oplus f(V_1)^\perp$ and $\langle\cdot,\cdot\rangle_2$ is non-degenerate, its restriction to $f(V_1)^\perp$ is non-degenerate too. As $V_2$ and hence $f(V_1)^\perp$ is of finite type, Proposition~\ref{Prop:NDegFin} gives a Hodge decomposition $f(V_1)^\perp = \Dd f(V_1)^\perp \oplus C_2'$. It is easy to check that 
\[ V_2 = f(\Harm_1)\oplus\underbrace{\bigl(\Dd f(V_1) \oplus \Dd f(V_1)^\perp)}_{\displaystyle=\Im\Dd_2}\oplus(\underbrace{f(C_1)\oplus C_2'}_{\displaystyle\eqqcolon C_2})\]
is a Hodge decomposition of $V_2$. The corresponding standard Hodge homotopy clearly satisfies $\HtpStd^2 \circ f = f\circ \HtpStd^1$, and it holds $\HtpStd^2(f(V_1)) \subset f(V_1)$, so that $f(V_1)$ is a Hodge subalgebra. By injectivity, it follows that $\VansSmall(V_1) \simeq \VansSmall(V_2)$.
\end{proof}

Since $\VansSmall=\VansSmall(V)$ is ``small'', a question whether it can be fit inside the image of a Sullivan's minimal model arose. This would imply, under some additional assumptions, that for any two Hodge decompositions of $V$, the non-degenerate quotients $\VansQuotient_1$, resp.~$\VansQuotient_2$ of the corresponding small subalgebras $\VansSmall_1$, resp.~$\VansSmall_2$ would be isomorphic as Poincar\'e duality algebras. Let us sketch the idea of this construction assuming that $f_1: \Lambda U_1 \twoheadrightarrow \VansSmall_1$ and $f_2: \Lambda U_2 \twoheadrightarrow \VansSmall_2$ are surjective Sullivan's minimal models. Uniqueness from \cite[Theorem~2.24]{Felix2008} gives an isomorphism $\Lambda U \coloneqq \Lambda U_1 \simeq \Lambda U_2$ such that the following diagram commutes up to homotopy of $\DGA$'s:
\begin{equation}\label{Eq:HpyCommutDiag}
\begin{tikzcd}
& & \VansSmall_1 \arrow[hook]{ld}\arrow{r} & \VansQuotient_1 \\
\Lambda U \arrow[bend left,two heads]{rru}{f_1}\arrow[bend right,two heads]{rrd}[below]{f_2}  &  V & &  \\
& & \VansSmall_2 \arrow[hook]{lu}\arrow{r}& \VansQuotient_2.
\end{tikzcd}
\end{equation}
Now, $\Lambda U \xrightarrow{f_1} \VansSmall_1 \xhookrightarrow{} V$ and $\Lambda U \xrightarrow{f_2} \VansSmall_2 \xhookrightarrow{} V$ induce the same isomorphism on homology; this can be used to pullback the orientation~$\Or^\H$ on $\H(V)$ to an orientation $\tilde{\Or}$ on $\H(\Lambda U)$, so that all maps in \eqref{Eq:HpyCommutDiag} will preserve the orientation on homology. Under the assumptions $\Dd V^n = 0$ and $\Dd (\Lambda U)^n = 0$, Proposition~\ref{Prop:OrOnHomG} applies, and we obtain a cyclic structure $\langle\cdot,\cdot\rangle^{\Lambda U}$ on $\Lambda U$ which is preserved by both $f_1$ and~$f_2$ on the chain level. We denote $f_i^\VansQuotient\coloneqq \pi^{\VansQuotient}_i \circ f_i$ and write down the following diagram with pairing-preserving $\DGA$-quasi-isomorphisms:
\begin{equation}\label{Eq:Diagram}
\begin{tikzcd}
& \arrow[two heads,swap]{ld}{f_1^\VansQuotient} \bigl(\Lambda U,\langle\cdot,\cdot\rangle^{\Lambda U}\bigr)\arrow[two heads]{rd}{f_2^\VansQuotient}& \\
\bigl(\VansQuotient_1,\langle\cdot,\cdot\rangle^{\VansQuotient}_1\bigr) & & \bigl(\VansQuotient_2,\langle\cdot,\cdot\rangle^{\VansQuotient}_2 \bigr).
\end{tikzcd}
\end{equation}
Claim (b) of Lemma~\ref{Lem:PomLemma} implies that $\ker f_1^\VansQuotient = \ker f_2^\VansQuotient = (\Lambda U)^\perp$, and hence
\[ \VansQuotient_1 \simeq  \Lambda U / (\Lambda U)^\perp\simeq \VansQuotient_2. \]
Unfortunately, the next two examples show that one can not, in general, expect~$\VansSmall_1$ and~$\VansSmall_2$, or even $\VansQuotient_1$ and $\VansQuotient_2$ to be isomorphic and $\VansSmall$ to fit inside the image of the minimal model.\footnote{Note that it is always possible to construct a surjective (non-minimal) Sullivan model $f: \Lambda U \twoheadrightarrow V$ by taking the minimal Sullivan model and inductively adding generators $\xi_i$ and $\mu_i$ with $\Dd \mu_i = 0$ and $\Dd \xi_i = \mu_i$ (the minimality condition on a Sullivan's algebra would require $\Dd \xi_i$ to be decomposable), and mapping them to $v\in V$ and $\Dd v$, respectively. Nevertheless, the uniqueness property of non-minimal Sullivan models is much weaker, see \cite[Lemma~2.20]{Felix2008}.}

\begin{Example}[Small algebras are, in general, not unique and not contained in images of Sullivan minimal models]\label{Ex:NonUniqueSmall}
Consider $M=\CP^2$, and let $\Kaehler$ be the Fubini--Study form on~$M$.
For $\alpha\in\DR^1(M)$, set $\Kaehler_\alpha \coloneqq \Kaehler + \Dd\alpha$.
Then $K_\alpha \wedge K_\alpha = K \wedge K + \Dd(2\alpha\wedge K + \alpha \wedge \Dd\alpha)$.
We can choose $\alpha$ such that $\Dd(2\alpha\wedge K + \alpha\wedge\Dd\alpha)\neq 0$.
Consider the Riemannian Hodge decomposition of $\DR(M)$ with $\Harm = \langle 1, \Kaehler, \Kaehler\wedge \Kaehler\rangle$.
According to Remark~\ref{Rem:RemarkHarm}, there is also a ``twisted'' Hodge decomposition with the harmonic subspace $\Harm_\alpha \coloneqq \langle 1, \Kaehler_\alpha, \Kaehler \wedge \Kaehler\rangle$.

The small subalgebra $\VansSmall$ for the Riemann Hodge decomposition is $\VansSmall=\langle 1, \Kaehler, \Kaehler\wedge\Kaehler\rangle$.
The small subalgebra $\VansSmall_\alpha$ for the twisted Hodge decomposition must contain $1$, $\Kaehler_\alpha$ and both $\Kaehler_\alpha\wedge\Kaehler_\alpha$ and $\Kaehler\wedge\Kaehler$ (we require $\Harm_\alpha\subset\VansSmall_\alpha$ by definition).
Therefore, it contains 
\[
\Kaehler_\alpha \wedge \Kaehler_\alpha - \Kaehler\wedge\Kaehler = \Dd(2\alpha\wedge K + \alpha\wedge\Dd\alpha)
\]
and also
\[
\StdPrpg^\alpha\Dd(2\alpha\wedge K + \alpha\wedge\Dd\alpha)=\pi_{C_\alpha}(2\alpha\wedge K + \alpha\wedge\Dd\alpha).
\]
Proposition~\ref{Prop:SmallDescription} asserts that these vectors, together with $1$ and $\Kaehler_\alpha$, generate $\VansSmall_\alpha$ as a graded vector space.
Clearly, $\VansSmall_\alpha$ is not isomorphic to $\VansSmall$ for generic $\alpha$, but 
\[
\VansQuotient(\VansSmall_\alpha)=\langle1,\Kaehler_\alpha,\Kaehler_\alpha\wedge\Kaehler_\alpha\rangle\simeq \langle 1, \Kaehler, \Kaehler\wedge\Kaehler\rangle= \VansQuotient(\VansSmall).
\]
In fact, the previous argument works in general and implies that for $\CP^2$, the non-degenerate quotients of two small subalgebras are isomorphic as Poincar\'e duality algebras (this does not hold for any $M$, see Example~\ref{Ex:SUsix}).

The Sullivan minimal model of $M$ is the free $\DGA$ $\Model \coloneqq \Lambda(\eta,\mu)$ with $\Abs{\eta}=2$, $\Abs{\mu}=5$, $\Dd\eta = 0$ and $\Dd\mu=\eta\wedge\eta$.
A $\DGA$-quasi-isomorphism $f: \Model\rightarrow\DR(M)$ is specified by its values on~$\eta$ and~$\mu$; for example, $f(\eta)\coloneqq\Kaehler$ and $f(\mu)\coloneqq 0$.
We see that neither $\VansSmall_\alpha$ nor $\Harm_\alpha$ can lie in $\im f$ because $\dim (\im f)^{4}=1$ for any $f$.
\end{Example}

\begin{Example}[Non-degenerate quotients of small subalgebras for different Hodge decompositions are, in general, not isomorphic]\label{Ex:SUsix}
Consider $M=\mathrm{SU}(6)$.
It is a compact simply-connected Lie group of dimension~$35$ ($=n^2 - 1$ for $n=6$) whose cohomology ring is freely generated by single elements in degrees  $3$, $5$,~$\dotsc$,~$11$; see \cite[Corollary~3.11]{Mimura1991}.
There is a biinvariant Riemannian metric and there are biinvariant differential forms $x_3$, $x_5$,~$\dotsc$, $x_{11}$ in the corresponding degrees such that 
\[
\Harm \coloneqq \Lambda(x_3, \dotsc, x_{11})\subset \DR(M)
\]
is the algebra of harmonic forms, see \cite[Chapter~1]{Felix2008}.

For $\eta_6\in \DR^6(M)$ and $\eta_8\in\DR^8(M)$, which are going to be specified later, set
\[
\tilde{x}_7 \coloneqq x_7 + \Dd \eta_6\quad\text{and}\quad\tilde{x}_9 \coloneqq x_9 + \Dd \eta_8,
\]
and let $\Harm'$ denote the graded vector space obtained from $\Harm$ by replacing the vectors $x_{7}$ and $x_{9}$ with $\tilde{x}_7$ and $\tilde{x}_9$, respectively.
We emphasize that we are replacing just the vectors, not the products; e.g., $x_7\wedge x_9$ is an element of $\Harm'$ but $\tilde{x}_7\wedge \tilde{x}_9$ might not be.
Let $\VansSmall = \Harm$ be the small subalgebra corresponding to the Riemannian Hodge decomposition, and let~$\VansSmall'$ be the small subalgebra corresponding to a Hodge decomposition based on~$\Harm'$ (such always exists by Remark~\ref{Rem:RemarkHarm}).
The following elements in degrees $15$, resp.~$20$ must be contained in $\VansSmall'$:
\begin{align*}
y&\coloneqq\StdPrpg'(\tilde{x}_7 \wedge x_9 -x_7 \wedge x_9)\\
 &=\StdPrpg'\Dd(\eta_6 \wedge x_9),\\
z&\coloneqq\tilde{x}_9 \wedge x_{11} - x_9 \wedge x_{11}\\
 &=\Dd(\eta_8 \wedge x_{11}).
\end{align*}
Using Stokes theorem and $\Dd\circ\StdPrpg' = \pi_{\Im \Dd}$, we get
\begin{align*}
\langle y,z\rangle & = \pm \int_{M} \StdPrpg'\Dd(\eta_6 \wedge x_9) \wedge\Dd(\eta_8 \wedge x_{11})\\
&= \pm \int_{M} \Dd\eta_6\wedge \eta_8 \wedge x_9 \wedge x_{11}.
\end{align*}
We claim that the integral can be made non-zero by a choice of $\eta_6$ and $\eta_8$.
Indeed, because $x_9\wedge x_{11}$ generates non-zero homology, there is an $m\in M$ such that $x_9(m)\wedge x_{11}(m)\neq 0$.
Pick local coordinates $(x^i)$ centered at $m$, and let $\alpha^I\Diff{x}^I$ be a non-zero coefficient in $x_9(m)\wedge x_{11}(m)$. Consider the complement $J = I^C$ and decompose $J= J_1 \cup J_2$ into two parts with $\Abs{J}_1 = 7$ and $\Abs{J}_2 = 8$.
For some $0\neq c\in\R$, set locally
\[
\eta_6 \coloneqq ((x^{J_{11}}+c)\Diff{x}^{J_1\backslash\{J_{11}\}})\quad\text{and}\quad\eta_8=\Diff{x}^{J_2},
\]
and extend them to the whole of $M$ by multiplying with a bump function which is constant non-zero in a neighborhood of $m$.
We have achieved that the integrand $\omega \coloneqq \Dd\eta_6\wedge \eta_8 \wedge x_9 \wedge x_{11}$ is non-zero around $m$.
Because the intersection pairing is non-degenerate and because $\omega \neq 0$, there is a function $f\in C^\infty(M)$ such that $\int_M f\omega \neq 0$.
We can now just rescale $\eta_8$ by $f$.

We have shown that $z$ induces a non-zero element $\pi_{\VansQuotient}'(z)\in \VansQuotient(\VansSmall')^{20}$, where $\pi_{\VansQuotient}': \VansSmall' \rightarrow \VansSmall'/{\VansSmall'}^\perp$ is the canonical projection.
Now, $\pi_{\VansQuotient}'$ is a $\DGA$-morphism, and hence $\pi_{\VansQuotient}'(z)$ is exact.
Because $\VansSmall'$ is of Hodge type, $\pi_{\VansQuotient}'$ is also a quasi-isomorphism, and hence $\pi_{\VansQuotient}'(x_{9}\wedge x_{11})$ generates non-trivial homology.
It follows that $\pi_{\VansQuotient}'(z)$ is not a multiple of $\pi_{\VansQuotient}'(x_9\wedge x_{11})$, and thus $\dim \VansQuotient(\VansSmall')^{20} \ge 2$.
However, we have $\VansQuotient(\VansSmall)^{20} = \Harm^{20} = \langle x_9\wedge x_{11} \rangle$.
This shows that $\VansQuotient(\VansSmall)$ and $\VansQuotient(\VansSmall')$ can not be isomorphic as vector spaces.
\end{Example}

It is not hard to come up with artificial examples of oriented $\DGA$'s which are not of Hodge type.\Add[noline,caption={Not of Hodge type}]{Example of an oriented dga without a Hodge decomposition.}

The following lemma and proposition show that in some cases it is possible to extend a $\DGA$ to a $\DGA$ of Hodge type.

\begin{Lemma}[Giving partners to non-degenerates]\label{Lemma:Exte}
 Let $(V,\Dd,\wedge,\Or)$ be a unital commutative $\DGA$ which is non-negatively graded and oriented in degree $n$.
 For $k=\lceil\frac{n}{2}\rceil$, $\dotsc$, $n$, consider the following property $(P_k)$ of a direct sum decomposition%
\begin{equation}\label{Eq:DecompOfV}
V=\Harm\oplus \Dd V \oplus C,
\end{equation}
where $\Harm$ is a harmonic subspace and $C$ a complement of $\ker \Dd$ in $V$ perpendicular to~$\Harm$ with respect to the induced cyclic structure $\langle\cdot,\cdot\rangle$:
\begin{description}
\item[$(P_{k})$] There is a complement $E$ of
\[
C^\perp\coloneqq \{ c^\perp \in C \mid \langle c^\perp,c\rangle=0\text{ for all }c\in C\}
\]
in $C$ and a homogenous linear map 
\[
\rho:  E^{\lceil n/2\rceil}\oplus \dotsb \oplus E^{k} \longrightarrow \Dd V
\]
such that for all $e'\in E$ and $c^\perp\in C^\perp$, the following holds:
\begin{align}
\langle e', \rho(e) \rangle &=\langle e', e \rangle, \label{Eq:ConditionTemp} \\
\langle c^\perp, \rho(e) \rangle &=  0.\label{Eq:ConditionTempII}
\end{align}
\end{description}
Suppose that $V$ is non-negatively graded, of finite type and satisfies $V^0=\Span\{1\}$ and $V^1 = 0$.
Suppose that $n \ge 5$ and that $(\H(V), \wedge, \Or^\H)$ is a Poincar\'e duality algebra.
Given $\lceil\frac{n}{2}\rceil\le l\le n$, suppose that $V$ admits a decomposition of type \eqref{Eq:DecompOfV} such that either $l=\lceil\frac{n}{2}\rceil$ or $l>\lceil\frac{n}{2}\rceil$ and $(P_{l-1})$ holds.
Then there is an $m\in \N_0$ and a Sullivan $\DGA$ 
 \begin{equation}\label{Eq:SullAlg}
 \Lambda \coloneqq \Lambda(w_1,\dotsc,w_m,z_1,\dotsc,z_m)
 \end{equation}
 specified by $\deg w_i = l-1$, $\deg z_i = l$, $\Dd z_i = 0$ and $\Dd w_i = z_i$ for all $i=1$, $\dotsc$, $m$ such that the tensor product $\DGA$ 
\[
 \hat{V}\coloneqq V\otimes \Lambda
 \]
 admits an orientation $\hat{\Or}: \hat{V}\to\R$ which extends $\Or: V \rightarrow \R$ on the canonical inclusion $V \hookrightarrow\hat{V}$, and there is a decomposition
\begin{equation}\label{Eq:DevompOfVHat}
\hat{V} = \hat{\Harm}\oplus \Dd \hat{V}\oplus \hat{C}
\end{equation}
of type \eqref{Eq:DecompOfV} for which $(P_{l})$ holds.
\end{Lemma}

\begin{proof}
The proof consists of a construction of $\hat{\Or}: \hat{V}\rightarrow\R$, a construction of a harmonic subspace $\hat{\Harm}$ in $\hat{V}$, a construction of a complement $\hat{C}$ of $\ker \Dd$ in $\hat{V}$, a degreewise description of $\hat{C}$ and $\hat{C}^\perp$, a degreewise construction of a complement $\hat{E}$ of $\hat{C}^\perp$ in $\hat{C}$ and a proof of the property $(P_l)$ for the constructed decomposition.
\begin{description}[leftmargin=0pt,font=\normalfont\itshape]
\item[Construction of $\hat{\Or}$:]
Consider the decomposition \eqref{Eq:DecompOfV}.
Because $\Dd : C \rightarrow \Dd V$ is an isomorphism, we can write
\begin{equation}\label{Eq:VDecomp}
V = \Harm \oplus \underbrace{\Dd E \oplus \Dd C^\perp}_{\displaystyle \Dd V}\oplus \underbrace{E\oplus C^\perp}_{\displaystyle C}.
\end{equation}
The restriction of $\langle\cdot,\cdot\rangle$ to $E$ is non-degenerate, and $V$ is of finite type by assumption; hence, $E$ is finite-dimensional.
Set
\[
m \coloneqq \dim E^l.
\]
Let $\xi_1$, $\dotsc$, $\xi_m$ be a basis of $E^{l}$ and $\xi^1$, $\dotsc$, $\xi^m$ its dual basis in $E^{n-l}$.
The Sullivan algebra~$\Lambda$ can be written as a direct sum 
\begin{equation}\label{Eq:LambdaDecomp}
\Lambda = \bigoplus_{k=0}^\infty \Lambda_k\quad\text{with}\quad
\Lambda_k = \bigoplus_{\substack{r, m \ge 0 \\ r + m = k}}\Lambda_r(w)\otimes \Lambda_m(z),
\end{equation}
where $\Lambda_r(w)$ and $\Lambda_m(z)$ are the graded vector spaces generated by monomials $w_I = w_{i_1}\dotsc w_{i_r}$ and $z_J = z_{j_1}\dotsc z_{j_m}$ for all multiindices $I=\{i_1, \dotsc, i_r\}$ and $J=\{j_1,\dotsc,j_m\}$, respectively.
The direct sum decompositions~\eqref{Eq:VDecomp} and~\eqref{Eq:LambdaDecomp} induce a direct sum decomposition of $\hat{V} = V \otimes \Lambda$ via the distributivity of $\otimes$ and $\oplus$.
We denote 
\[
\hat{V}_k \coloneqq V \otimes \Lambda_k\quad\text{for }k\ge 0.
\]
Let $\hat{\Or}: \hat{V} \to \R$ be the linear map satisfying
\begin{align}
	\hat{\Or}(v) &\coloneqq \Or(v) && \text{for all }v\in V,\\
	\hat{\Or}(\xi^i \wedge z_j) &\coloneqq \Or(\xi^i\wedge \xi_j) && \text{and} \\
	\hat{\Or}(\Dd \xi^i \wedge w_j) &\coloneqq (-1)^{\deg \xi^i + 1}\Or(\xi^i \wedge \xi_j) && \text{for all }i, j = 1, \dotsc, m,
\end{align}
and which is zero on $(\Harm\oplus\Dd V\oplus C^\perp \oplus \bigoplus_{i\ge 0, i \neq n-l}E^i)\otimes \Lambda_1(z)$, on $(\Harm\oplus C \oplus \Dd C^\perp \oplus \bigoplus_{i\ge 0, i\neq n-l} \Dd E^i)\otimes\Lambda_1(w)$ and on $\hat{V}_k$ for $k \ge 2$.

In order to show that $\hat{\Or}$ is an orientation, we must check that $\hat{\Or}\neq 0$ and $\hat{\Or}\circ \Dd = 0$.
The first condition is clear from $\Restr{\hat{\Or}}{V} = \Or \neq 0$. 
As for the second condition, $\hat{V}$ is generated by elements $v \wedge w_I \wedge z_J$ for $v\in V$ and multiindices $I$, $J$.
It holds $\Dd \hat{V}_k \subset \hat{V}_k$ for all~$k\ge 0$, and hence $\Dd \hat{V} = \bigoplus_{k=0}^\infty \Dd \hat{V}_k$.
From the definition of $\hat{\Or}$, we have immediately $\Dd \hat{V}_0 = \Dd V \subset \ker \Or \subset \ker \hat{\Or}$ and $\bigoplus_{k=2}^\infty \Dd \hat{V}_k \subset \ker\hat{\Or}$.
As for $\Dd\hat{V}_1$, we write $\hat{V}_{1} = \Span \{v\wedge w_j, v\wedge z_j \mid v\in V, j=1,\dotsc, m\}$ as a graded vector space and compute
\begin{equation}\label{Eq:DVI}
\begin{aligned}
	\Dd \hat{V}_1 &= \Span\{\Dd(v\wedge w_j), \Dd(v\wedge z_j) \mid v\in V, j=1, \dotsc, m \}\\
	&=\Span\{\Dd v \wedge w_j + (-1)^{\Deg v} v \wedge z_j \mid v\in V, j=1, \dotsc, m\}.
\end{aligned}
\end{equation}
Write $v\in V^{n-l}$ as $v = h  + \Dd c + c^\perp + \sum_{i=1}^m \alpha_i \xi^i$ for $h\in \Harm^{n-l}$, $c\in C^{n-l-1}$, $c^\perp\in C^{\perp n-l}$ and $\alpha_i\in \R$, and compute for every $j=1$, $\dotsc$, $m$ the following:
\begin{align*}
\hat{\Or}(\Dd v\wedge w_j) & = \hat{\Or}\Bigl(\Dd c^\perp \wedge w_j + \sum_{i=1}^m \alpha_i \Dd \xi^i \wedge w_j\Bigr)\\
&= \sum_{i=1}^m \alpha_i \hat{\Or}(\Dd \xi^i \wedge w_j)\\
&= \sum_{i=1}^m (-1)^{\deg \xi^i + 1}\alpha_i \hat{\Or}(\xi^i \wedge \xi_j) \\
&= (-1)^{n-l+1}\sum_{i=1}^m \alpha_i \hat{\Or}(\xi^i \wedge z_j)\\
&= (-1)^{n-l+1}\hat{\Or}\Bigl((h + \Dd c + c^\perp)\wedge z_j +\sum_{i=1}^m \alpha_i \xi^i \wedge z_j \Bigr) \\
&= (-1)^{\deg v + 1}\hat{\Or}(v\wedge z_j).
\end{align*}
Consequently, $\Dd \hat{V}_{1}\subset \ker \hat{\Or}$.
This shows $\hat{\Or}\circ\Dd = 0$.

The inclusion $V\hookrightarrow\hat{V}$ is clearly orientation preserving. 

\item[Construction of $\hat{\Harm}$:]
It holds $\bar{\H}(\Lambda) = 0$ for the reduced homology, and hence $\H(\hat{V}) \simeq \H(V)\otimes\H(\Lambda) = \H(V)$ by K\"unneth's formula. Because $\Harm\subset \ker \Dd$, $\Harm\cap\im\Dd = 0$ and $\dim(\Harm) = \dim \H(\hat{V})$, $\Harm$ is a complement of $\Dd\hat{V}$ in $\ker \Dd$. Therefore, 
\[
\hat{\Harm} \coloneqq \Harm
\]
is a harmonic subspace of $\hat{V}$.
Also note that $\H(\hat{V})=\bigoplus_{k=0}^\infty \H(\hat{V}_k)$ and $\dim \H(\hat{V}) = \dim \H(\hat{V}_0)$, and hence $\H(\hat{V}_k) = 0$ for all $k\ge 1$.

\item[Construction of $\hat{C}$:]
We construct $\hat{C}$ as a direct sum $\hat{C} = \bigoplus_{k=0}^\infty \hat{C}_k$. 
Set 
\[
\hat{C}_0 \coloneqq C.
\]
For $k=1$, define
\begin{equation}\label{Eq:CIDef}
\begin{aligned}
 \tilde{C}_1 &\coloneqq \Span\{v \wedge w_i \mid v\in V, i=1, \dotsc,m\} \subset \hat{V}_1, \\
  \hat{C}_1 &\coloneqq \{\tilde{c} - \pi(\tilde{c}) \mid \tilde{c}\in \tilde{C}_1\}\subset\hat{V}_0\oplus\hat{V}_1,
\end{aligned}
\end{equation}
where $\pi: \hat{V}\rightarrow\hat{\Harm}$ is the orthogonal projection.
For all $k\ge 2$, let $\hat{C}_k \subset \hat{V}_k$ be an arbitrary complement of $\ker\Dd$ in $\hat{V}_k$ as a graded vector space.

We show first that $\hat{C}_i$ for $i\ge 0$ are disjoint.
Clearly, $\hat{C}_j \cap \sum_{i=0, i\neq j}^\infty \hat{C}_i = 0$ for all $j\ge 2$.
Because $(\Harm + C)\cap\tilde{C}_1 = 0$ and $\Harm \cap C = 0$, it holds $\hat{C}_1 \cap \hat{C}_0 = 0$, and $(\hat{C}_0 + \hat{C}_1) \cap \sum_{i\ge 2} \hat{C}_i = 0$ implies that $\hat{C}_j \cap \sum_{i=0, i\neq j}^\infty \hat{C}_i = 0$ holds also for $j=0$, $1$.

We show that $\hat{C}=\bigoplus_{k\ge 0}\hat{C}_k$ is a complement of $\ker \Dd$ in~$\hat{V}$.
For $k\ge 2$, $\hat{C}_k$ are complements of $\ker \Dd$ in $\hat{V}_k$ by construction.
For $k=0$, $\hat{C}_0 \oplus \ker \Dd \cap \hat{V}_0 = \hat{V}_0$ follows from \eqref{Eq:DecompOfV}. 
For $k=1$, we compare \eqref{Eq:DVI} and \eqref{Eq:CIDef} to see that $\tilde{C}_1 \oplus \Dd \hat{V}_1 = \hat{V}_1$.
Because $\Harm\subset \ker \Dd$, it is easy to see that $\hat{C}_0\oplus\hat{C}_1$ is a complement of $\ker \Dd$ in $\hat{V}_0\oplus\hat{V}_1$.

Finally, $\hat{C}_0 \perp \hat{\Harm}$ holds by \eqref{Eq:DecompOfV}, $\hat{C}_1\perp\Harm$ holds by the construction of $\hat{C}_1$ from $\tilde{C}_1$ and $\hat{C}_k \perp \Harm$ follows from the definition of $\hat{\Or}$.

\item[Degreewise description of $\hat{C}$ and $\hat{C}^\perp$:]
We are interested in $\hat{C}^i$ for $0\le i\le l$.
For all $k\ge 2$, the graded vector space $\hat{C}_k$ is concentrated in degrees $i\ge 2(l-1)$.
But $2(l-1)>l$ due to $n\ge 5$.
Therefore, $\hat{C}_k^i = 0$ for $k\ge 2$ and $0\le i \le l$.
We denote $\bar{V} \coloneqq \bigoplus_{i=1}^\infty V^i$ and write $\tilde{C}_1 = \Lambda_1(w) \oplus (\bar{V}\otimes \Lambda_1(w))$.
Because $V^1=0$, the graded vector space $\bar{V}\otimes\Lambda_1(w)$ is concentrated in degrees $i\ge 2 + (l-1) = l + 1$. We obtain  
\begin{equation}\label{Eq:CDegreewise}
\hat{C}^i = (\hat{C}_0 \oplus \hat{C}_1)^i =
 \begin{cases}
 	C^i & \text{for }0 \le i \le l-1, \\
	C^i + \Span\{ w_j \mid j = 1, \dotsc, m\} & \text{for }i = l - 1, \\
	C^i & \text{for }i = l.
 \end{cases}
\end{equation}
Here we used that $\hat{C}_1^{l-1} = \tilde{C}_1^{l-1} = \Span\{ w_i \mid i = 1, \dotsc, m\}$, which is true because $\Lambda_1(w)\perp\Harm$ from the definition of $\hat{\Or}$.

We are now interested in $\hat{C}^{\perp i}$ for $n-l\le i \le l$.
Note that $n-l \le i \le l$ is equivalent to $n-l \le n-i \le l$.
By definition, $\hat{\Or}$ vanishes on $C\wedge\Lambda_1(w)=C\otimes\Lambda_1(w)$ and $\Lambda_1(w)\wedge\Lambda_1(w) \subset \Lambda_2(w)$.
Looking at \eqref{Eq:CDegreewise}, we see the following: 
\begin{equation}\label{Eq:CPerpDegreewise}
	\hat{C}^{\perp i} =
		\begin{cases}
			C^{\perp i} &  \text{for }n-l \le i \le l - 2, \\
			C^{\perp i} + \Span\{w_i\mid i=1,\dotsc,m\} & \text{for }i = l-1, \\
			C^{\perp i} & \text{for }i = l.
		\end{cases}		
\end{equation}
Because $\hat{V}$ is non-negatively graded, it holds $\hat{C}^{\perp i} = \hat{C}^i$ for $i>n$.
Notice that $\hat{C}^{\perp i}$ might be smaller than $C^i$ for $0\le i \le n-l-1$.
The reason for this is a possible existence of $v_1$, $v_2\in V$ such that $v_1 \wedge v_2$ has a non-trivial $\Dd E$-component and $\langle v_1, v_2\wedge w_i \rangle = \hat{\Or}((v_1\wedge v_2)\wedge w_i) \neq 0$.

\item[Construction of $\hat{E}$:]
Because $\hat{V}_i \wedge \hat{V}_j \subset \hat{V}_{i+j}$ and $\hat{V}_k \subset \ker \hat{\Or}$ for $k\ge 2$, we have $\hat{C}_k \subset \hat{C}^\perp$ for all $k\ge 2$.
It follows that 
\[
\hat{C}^\perp = (\hat{C}_0\oplus\hat{C}_1)\cap\hat{C}^\perp \oplus \bigoplus_{k\ge 2}\hat{C}_k,
\]
and hence it is enough to construct $\hat{E} \subset \hat{C}_0\oplus\hat{C}_1$ such that $\hat{E}\oplus (\hat{C}_0\oplus\hat{C}_1)\cap\hat{C}^\perp = \hat{C}_0\oplus\hat{C}_1$.
Because $E\cap \hat{C}^\perp \subset E \cap C^\perp = 0$, we can get $\hat{E}$ by extending $E$.
For $0\le i \le n$, we define
\begin{equation}\label{Eq:EDegreewise}
\hat{E}^i \coloneqq
	\begin{cases}
		E^i\oplus \Span\{c_1^i,\dotsc, c_{k_i}^i\}\text{ for }c_1^i, \dotsc, c_{k_i}^i\in C^i & \text{for }0\le i \le n-l-1,\\
		E^i & \text{for }n-l\le i \le l,\\
		E^i\oplus\Span\{\hat{v}_1,\dotsc,\hat{v}_k\}\text{ for }\hat{v}_1^i,\dotsc,\hat{v}_{k_i}^i\in(\hat{C}_0\oplus\hat{C}_1)^i & \text{for }l+1\le i \le n,
	\end{cases}
\end{equation}
where the existence of $c^i_j$ and $\hat{v}^i_j$ and the fact that $\hat{C} = \hat{E}\oplus\hat{C}^\perp$ are justified by \eqref{Eq:CDegreewise}, \eqref{Eq:CPerpDegreewise} and \eqref{Eq:ConditionTempII}.

\item[Property $(P_l)$:] We define $\hat{\rho}: \hat{E}^{\lceil \frac{n}{2}\rceil}\oplus\dotsb\oplus\hat{E}^l \rightarrow \Dd\hat{V}$ by $\hat{\rho} \coloneqq \rho$ on $\hat{E}^{\lceil \frac{n}{2}\rceil}\oplus\dotsb\oplus\hat{E}^{l-1}=E^{\lceil \frac{n}{2}\rceil}\oplus\dotsb\oplus E^{l-1}$ and by
\[
\hat{\rho}(\xi_i) \coloneqq z_i\quad\text{for all }i=1, \dotsc, m.
\]
Conditions \eqref{Eq:ConditionTemp} and \eqref{Eq:ConditionTempII} are checked easily using $(P_{l-1})$, \eqref{Eq:CDegreewise}, \eqref{Eq:CPerpDegreewise}, \eqref{Eq:EDegreewise} and the definition of $\hat{\Or}$.\qedhere
\end{description}
\end{proof}

\begin{Questions}\phantomsection\label{Q:QuestOnPoinc}
\begin{RemarkList}
\item Given a cochain complex $(V,\Dd)$ with a symmetric pairing $\langle\cdot,\cdot\rangle$, which conditions have to be satisfied by the maps $\pi_\Harm$, $\StdHtp: V\rightarrow V$ so that $V=\Im \pi_\Harm \oplus \Im \Dd \oplus \Im \StdHtp$ is a Hodge decomposition with Hodge pair $(\Im \pi_\Harm, \StdHtp)$? This would characterize Hodge decompositions in terms of Hodge pairs.
\item It should be possible to prove Lemma~\ref{Lemma:Exte} also for $n\le 4$ by hand. Check that! 
\qedhere
\end{RemarkList}
\end{Questions}

\section{Poincar\'e DGA's and Poincar\'e duality models}\label{SubSec:PoincModel}
\allowdisplaybreaks
\Correct[noline,caption={DONE Hom to weak Hom}]{Homotopy to weak homotopy}
\Correct[noline,caption={DONE Add orientation to the data of PDGA}]{Add orientation to the data of PDGA}
\Correct[noline,caption={DONE Change integral to or}]{Change integral to or}

In this section, we restrict to non-negatively graded unital commutative $\DGA$'s, which we denote by $\nnuCDGA$. In this case, the notions of orientation and of cyclic structure agree by Proposition~\ref{Prop:OrAndCyc}.

We modify and combine definitions from~\cite{Van2019} and~\cite{Lambrechts2007} as follows.

\begin{Definition}[Dif.~Poincar\'e duality algebra, $\PDGA$ and formality]\label{Def:PDGA}
A \emph{differential Poincar\'e duality algebra of degree $n$} is a $\nnuCDGA$ $(V,\Dd,\wedge)$ of finite type with orientation~$\Or$ in degree~$n$ such that the induced pairing on $V$ satisfies Poincar\'e duality. If $\Dd = 0$, we call it just \emph{Poincar\'e duality algebra.}

A \emph{Poincar\'e~$\DGA$ (shortly $\PDGA$) of degree $n$} is a $\nnuCDGA$ $(V,\Dd,\wedge)$ together with an orientation $\Or^\H: \H(V) \rightarrow \R$  of degree $n$ which makes $\H(V)$ into a Poincar\'e duality algebra.

A \emph{morphism of $\PDGA$'s} $(V_1,\Dd_1,\wedge_1,\Or^\H_1)$ and $(V_2,\Dd_2,\wedge_2,\Or^\H_2)$ is a $\DGA$-morphism $f: V_1 \rightarrow V_2$ such that the induced map $f_*: \H(V_1) \rightarrow \H(V_2)$ preserves orientation, i.e., it holds $\Or_2^\H \circ f_* = \Or_1^\H$.

A \emph{quasi-isomorphism (or weak equivalence)} of $\PDGA$'s is a morphism of $\PDGA$'s $f: V_1 \rightarrow V_2$ such that $f_*: \H(V_1) \rightarrow \H(V_2)$ is an isomorphism of oriented $\DGA$'s.

Two $\PDGA$'s are \emph{weakly homotopy equivalent (or isomorphic in the homotopy category)} if they are connected by a zig-zag of $\PDGA$-quasi-isomorphisms.

A $\PDGA$ is \emph{formal} if it is weakly homotopy equivalent (as a $\PDGA$) to its homology.
\end{Definition}

It follows from Proposition~\ref{Prop:OrAndCyc} that a differential Poincar\'e duality algebra according to Definition~\ref{Def:PDGA} is precisely a cyclic $\DGA$ from Part~I. In particular, it is finite dimensional. However, if we relax finite type, unitality or commutativity, we obtain a different notion.

\begin{Remark}[Frobenius algebra]
A differential Poincar\'e duality algebra, resp.~a cyclic $\DGA$, is precisely a finite-dimensional symmetric dg-Frobenius algebra from \cite[p.~13]{Vallette2012} or \cite[Theorem~1.1]{Cohen2006}. 
\end{Remark}

\begin{Definition}[Poincar\'e duality model]\label{Def:PDModel}
A \emph{Poincar\'e duality model} of a $\PDGA$ $(V,\Dd,\wedge,\Or^\H)$ is a differential Poincar\'e duality algebra $(\Model,\Dd^\Model,\wedge^\Model,\Or^\Model)$ which is weakly homotopy equivalent to $V$ as a $\PDGA$.

We call a Poincar\'e duality model \emph{small} if $\VansQuotient(\VansSmall(\Model)) \simeq \Model$ for every Hodge decomposition.
\end{Definition}

\begin{Remark}[On Poincar\'e duality models]
\begin{RemarkList}
\item The definition of a Poincar\'e duality model in~\cite{Lambrechts2007} requires only weak homotopy equivalence of $\DGA$'s, i.e., it does not require quasi-isomorphisms to preserve orientation on homology.

\item In general, a Sullivan minimal model $\Lambda U$ fails easily to be a Poincar\'e duality model because it often has non-zero elements in degree $>n$, e.g., powers of even generators; see Example~\ref{Ex:SphereModel} for $\Sph{2}$. 

For a compact connected Lie group $G$, the subalgebra of harmonic forms $\Harm$ for any biinvariant Riemannian metric is isomorphic to a free algebra on odd generators, see \cite[Chapter~1]{Felix2008}. Therefore, $\Harm$ with zero differential and the induced cyclic structure is at the same time the Sullivan minimal model and a Poincar\'e duality model for $\DR(G)$.

\item Poincar\'e duality models are not ``strongly unique'' in the sense that two Poincar\'e duality models of the same algebra must not be isomorphic; see Example~\ref{Ex:SUsix} for $\mathrm{SU}(6)$. However, there is a ``weak uniqueness'' statement in Proposition~\ref{Prop:LambrechtUnique} below. The situation is similar to the situation with Sullivan and minimal Sullivan models; see \cite{Felix2008}. We introduced ``smallness'' in Definition~\ref{Def:PDModel} as a candidate for a minimality condition on a Poincar\'e duality model which might imply its ``strong uniqueness''; see Question~\ref{Q:QuestionsPonc}.
\qedhere
\end{RemarkList}
\end{Remark}

Consider the functor from $\PDGA$'s to $\DGA$'s which forgets the orientation on homology. We have the following trivial yet somewhat surprising observation.

\begin{Proposition}[$\PDGA$-formality is the same as $\DGA$-formality]\label{Prop:PoincModelOfFormal}
A Poincar\'e $\DGA$ $(V,\Dd,\wedge,\Or^\H)$ is formal (as a $\PDGA$) if and only if it is formal as a $\DGA$.
\end{Proposition}
\begin{proof}
The ``only if'' part is clear.

As for the ``if'' part, let 
\begin{equation}\label{Eq:ZZ}
V \longleftarrow \bullet \dotsb \bullet \longrightarrow \H(V)
\end{equation}
be a weak homotopy equivalence of $\DGA$'s. Denote by $f: \H(V) \rightarrow \H(V)$ the isomorphism on homology induced by \eqref{Eq:ZZ} from the left to the right. We adjoin $\H(V)$ to the right of~\eqref{Eq:ZZ} to obtain the homotopy
\begin{equation}\label{Eq:ZZII}
V \longleftarrow \bullet \dotsb \bullet \longrightarrow \H(V) \xrightarrow{f^{-1}} \H(V)
\end{equation}
whose induced map on homology from the left to the right is the identity. Therefore, we can orient homologies of the inner nodes of~\eqref{Eq:ZZII} so that all maps preserve orientation on homology.
\end{proof}

The next proposition will be used to show the existence of Poincar\'e duality models.

\begin{Proposition}[Extension of Hodge type]\label{Prop:ExtensionOfHodgeType}
Let $V$ be a $\PDGA$ of degree $n\ge 5$ which is of finite type and satisfies $V^0=\Span\{1\}$ and $V^1 = 0$. Then it is a retract of an oriented $\DGA$ $\LambrechtsExtension(V)$ of Hodge type in the category of $\PDGA$'s.
\end{Proposition}

\begin{proof}
Pick an arbitrary harmonic subspace $\Harm$ and an arbitrary complement $C$ of $\ker \Dd$ in $V$.
If $C$ is not perpendicular to $\Harm$, replace it with $\{c - \pi(c)\mid c\in C\}$, where $\pi: V \rightarrow \Harm$ is the orthogonal projection.
We start with $l=\lceil \frac{n}{2} \rceil$ and apply Lemma~\ref{Lemma:Exte} inductively to get an extension $\hat{V} = V \otimes \Lambda$ which admits a decomposition 
\begin{equation}\label{Eq:HatDecomp}
\hat{V} = \hat{\Harm}\oplus\Dd\hat{V}\oplus\hat{C}
\end{equation}
of type \eqref{Eq:DecompOfV} such that there is a complement $\hat{E}$ of $\hat{C}^\perp$ in $\hat{C}$ and a linear map $\hat{\rho}: \hat{E}^{\lceil n/2\rceil}\oplus\dotsb\oplus\hat{E}^n\rightarrow\Dd\hat{V}$ such that \eqref{Eq:ConditionTemp} and \eqref{Eq:ConditionTempII} hold.
We consider a linear map
\[
\kappa: \hat{C} \longrightarrow \Dd \hat{V}
\]
such that
\[
\kappa(e)=
\begin{cases}
	0 & \text{for }e\in \hat{E}^i\text{ with }i<\lceil\frac{n}{2}\rceil,\\
	\hat{\rho}(e) & \text{for }e\in \hat{E}^i\text{ with } i>\lceil\frac{n}{2}\rceil, \\
\end{cases}\quad\text{and}\quad \kappa(c^\perp) = 0\quad\text{for }c^\perp\in\hat{C}^\perp.
\]
The case $n = 2k$ and $e\in \hat{E}^k$ is specified as follows.
If $k$ is even, then $\langle \cdot,\cdot \rangle: \hat{E}^{k}\otimes \hat{E}^{k} \rightarrow \R$ is an inner product, and there is an orthonormal basis $\eta_1$, $\dotsc$, $\eta_m$ for some $m\in\N$.
We require
\begin{equation}\label{Eq:InnerProdCase}
 \kappa(\eta_i) = \frac{1}{2}\hat{\rho}(\eta_i)\quad\text{for all }i=1, \dotsc, m.
\end{equation}
If $k$ is odd, then $\langle \cdot,\cdot \rangle: \hat{E}^{k}\otimes \hat{E}^{k} \rightarrow \R$ is a symplectic form, and there is a symplectic basis $\eta_1$, $\theta_1$, $\dotsc$, $\eta_m$, $\theta_m$ for some $m\in\N$.
We use the convention $\langle \theta_i,\eta_j\rangle = \delta_{ij}$ for $i$, $j=1$,~$\dotsc$, $m$.
We require 
\begin{equation}\label{Eq:SymplCase}
 \kappa(\eta_i) = \hat{\rho}(\eta_i)\quad\text{and}\quad\kappa(\theta_i)= 0\quad\text{for }i=1,\dotsc,m.
\end{equation}
Let 
\[
\hat{C}' \coloneqq \{c - \kappa(c) \mid c\in \hat{C}\}.
\]
This is a complement of $\ker \Dd$ in $\hat{V}$ perpendicular to $\hat{\Harm}$ because $\hat{C}$ is and $\im \kappa \subset\Dd\hat{V}$.
Given homogenous $c_1$, $c_2\in \hat{C}$ with $\deg c_1 + \deg c_2 = n$ and $\deg c_1 \le \deg c_2$, write $c_1 = c^\perp_1 + e_1$ and $c_2 = c^\perp_2 + e_2$ for $c_1^\perp$, $c_2^\perp \in \hat{C}^\perp$ and $e_1$, $e_2\in\hat{E}$, and compute
\begin{align*}
\langle c_1 - \kappa(c_1), c_2 - \kappa(c_2) \rangle &= \langle c_1, c_2 \rangle - \langle \kappa(c_1), c_2 \rangle - \langle c_1, \kappa(c_2) \rangle\\
&=\begin{aligned}[t]
&\underbrace{\langle e_1, e_2 \rangle - \langle \kappa(e_1), e_2 \rangle - \langle e_1, \kappa(e_2) \rangle}_{\eqqcolon(*)} \\ &{}-\underbrace{\langle\kappa(e_1),c_2^\perp\rangle - \langle c_1^\perp, \kappa(e_2)\rangle}_{\eqqcolon(**)}.
\end{aligned}
\end{align*}
Now, $(**)=0$ because of \eqref{Eq:ConditionTempII}.
As for $(*)$, if $\deg c_1 < \deg c_2$, then
\begin{align*}
(**) &= \langle e_1, e_2 \rangle - \langle e_1,\kappa(e_2)\rangle \\
     &= \langle e_1, e_2 \rangle - \langle e_1,\hat{\rho}(e_2)\rangle \\ 
     &= 0
\end{align*}
because of \eqref{Eq:ConditionTemp}.
If $\deg c_1 = \deg c_2 = k$ and $k$ is even, we plug in the orthonormal basis and get using \eqref{Eq:InnerProdCase} that
\begin{align*}
e_1 = \eta_i,\ e_2 = \eta_j:  && (**) &= \langle \eta_i, \eta_j \rangle - \langle \kappa(\eta_i),\eta_j\rangle - \langle \eta_i,\kappa(\eta_j)\rangle \\
&& & = \langle \eta_i, \eta_j\rangle - \langle \eta_j, \kappa(\eta_i)\rangle - \langle \eta_i, \kappa(\eta_j)\rangle \\
&& & = \langle \eta_i, \eta_j \rangle - \frac{1}{2}\langle \eta_j, \hat{\rho}(\eta_i)\rangle - \frac{1}{2}\langle\eta_i,\hat{\rho}(\eta_j)\rangle\\
&& & = \langle \eta_i, \eta_j \rangle - \frac{1}{2}\langle \eta_j, \eta_i \rangle - \frac{1}{2}\langle \eta_i, \eta_j \rangle \\
&& & = 0.
\end{align*}
If $k$ is odd, we plug in the symplectic basis and get using \eqref{Eq:SymplCase} that
\begin{align*}
e_1 = \eta_i,\ e_2 = \eta_j: && (**) &= \langle \eta_j, \kappa(\eta_i) \rangle - \langle \eta_i, \kappa(\eta_j) \rangle \\
&& &= \langle \eta_j, \hat{\rho}(\eta_j) \rangle - \langle \eta_i, \hat{\rho}(\eta_j) \rangle \\
&& &= 0, \\
e_1 = \theta_i,\ e_2 = \eta_j: && (**) &= \langle \theta_i, \eta_j\rangle - \langle \theta_i, \kappa(\eta_j) \rangle \\
&& &= \langle \theta_i, \eta_j\rangle - \langle \theta_i, \hat{\rho}(\eta_j) \rangle \\
&& &= 0, \\
e_1 = \theta_i,\ e_2 = \theta_j: && (**) &= 0.
\end{align*}
This shows that $\hat{C}\perp\hat{C}$, and hence \eqref{Eq:HatDecomp} is a Hodge decomposition.

Finally, because $\hat{V} = V \otimes \Lambda$ as a $\DGA$, both the inclusion $\iota: V \rightarrow \hat{V}$ of $V$ into $\hat{V}_0$ and the projection $\pi: \hat{V} \rightarrow V$ from $\hat{V}_0$ onto $V$ are $\DGA$ morphisms.
Because $\pi \circ \iota = \Id$ and because $\iota_*$ is an orientation preserving isomorphism, $\pi_*$ is an orientation preserving isomorphism as well.
Therefore, $V$ is a retract of $\LambrechtsExtension(V)\coloneqq\hat{V}$ in the category of $\PDGA$'s.
\end{proof}

The next example shows that the Sullivan minimal model is sometimes of Hodge type.

\begin{Example}[Sullivan minimal model of $\Sph{2}$ is of Hodge type]\label{Ex:SphereModel}
The Sullivan minimal model of $\Sph{2}$ is the free graded commutative algebra $\Model\coloneqq\Lambda(\eta_2,\eta_3)$ with $\Abs{\eta_2}=2$, $\Abs{\eta_3}=3$, $\Dd \eta_2 = 0$ and $\Dd\eta_3=\eta_2\wedge\eta_2$.
From degree reasons, it holds $\Model = \Span\{\eta_2^k \eta_3^l \mid k\ge 0, l \in \{0,1\}\}$ as a graded vector space.
We have a canonical decomposition $\Model = \Harm \oplus \Dd \Model \oplus C$, where $\Harm = \Span\{\eta_2\}$, $\Dd \Model = \Span\{\eta_2^k \mid k \ge 2 \}$ and $C = \Span\{\eta_2^{k}\eta_3 \mid k \ge 0\}$.
We define an orientation $\Or : \Model \rightarrow \R$ in degree $2$ by $\Or(\eta_2) \coloneqq 1$ on $\Harm$ and by $0$ on $\Dd \Model$ and $C$.
It is easy to see that $C\perp \Harm$ and $C\perp C$ with respect to the induced cyclic structure $\langle \cdot,\cdot\rangle$. 
Consider the $\DGA$-quasi-isomorphism $f: \Model \rightarrow \DR(\Sph{2})$ defined by $f(\eta_2)\coloneqq \Vol$ and $f(\eta_3)\coloneqq 0$.
Clearly, it is orientation preserving.
We have $\Model/\Model^\perp\simeq \Lambda(\eta_2)$
\end{Example}

The following proposition about the existence of a Poincar\'e dualiy model of a Poincar\'e $\DGA$ $V$ with $\H^1(V)=0$ was originally proven in \cite[Theorem~1.1]{Lambrechts2007}.
It was formulated in the category of $\DGA$'s, i.e., not checking whether the arrows are orientation preserving on homology.
The idea was to construct an extension $\LambrechtsExtension(\Lambda U )$ of the Sullivan minimal model $\Lambda U$ of~$V$ and an orientation on it such that the degenerate subspace is acyclic; the Poincar\'e duality model is then obtained by taking the quotient.
The extension is constructed by adding elements which kill the so called orphans.

By Proposition~\ref{Prop:HodgeAcyc}, we know that $V^\perp$ is acyclic if and only if $V$ is of Hodge type ($V$ needs to be of finite type for the direct implication).
Based on this, we give a new construction of $\LambrechtsExtension(\Lambda U)$ using Lemma~\ref{Lemma:Exte}, i.e., by adding exact partners to non-degenerates.
Our construction works for $n\ge 5$, whereas the assumption of \cite{Lambrechts2007} is $n\ge 7$.
We also do not need $\Dd (\Lambda U)^2 = 0$, although it follows from $\H^1(V) = 0$.
It is also clear from our construction that the arrows preserve orientation on homology.
However, this can be checked for the construction of \cite{Lambrechts2007} as well.

\begin{Proposition}[Existence of Poincar\'e duality model for $\H^1 = 0$]\label{Prop:ExOfLambrStan}
A~Poincar\'e $\DGA$ $V$ with $\H^0(V) = \Span\{1\}$ and $\H^1(V)=0$ admits a Poincar\'e duality model $\Model$.
\end{Proposition}
\begin{proof}
If $\Or^\H$ comes from a pairing on $V$ which is of Hodge type, then we can take $\Model=\VansQuotient(\VansSmall(V))$. The weak homotopy equivalence of $\PDGA$'s looks like
\begin{equation}\label{Eq:ModOne}
\begin{tikzcd}
 &  \VansSmall(V) \arrow[two heads]{dl}\arrow[hook]{dr} & \\
 \VansQuotient(\VansSmall(V)) & & V.
\end{tikzcd}
\end{equation}
If $V$ is not of Hodge type, we proceed as follows. Let $n$ be the degree of the orientation on $\H(V)$.
If $n\le 6$, then $V$ is formal as a $\DGA$ by~\cite{Miller1979}, and we can take $\Model=\H(V)$ by Proposition~\ref{Prop:PoincModelOfFormal}.
The weak homotopy equivalence of $\PDGA$'s looks like
\begin{equation}\label{Eq:ModTwo}
\begin{tikzcd}
 & \Lambda U \arrow{dl}\arrow{dr} & \\
 \H(V) & & V,
\end{tikzcd}
\end{equation}
where $\Lambda U$ is the Sullivan minimal model of $V$.
The Sullivan minimal model $W\coloneqq \Lambda U$ is of finite type and satisfies $W^0 = \R$, $W^1 = 0$ and $\Dd W^2 = 0$.
Suppose that $n\ge 7$.
Let $\LambrechtsExtension(W)$ be the extension of $W$ of Hodge type either from Proposition~\ref{Prop:ExtensionOfHodgeType} or from \cite[Section~4]{Lambrechts2007}.
This extension is of finite type, the inclusion $W\hookrightarrow\LambrechtsExtension(W)$ is a quasi-isomorphism of $\DGA$'s and $\LambrechtsExtension(W)^\perp$ is acyclic.
Moreover, $W\hookrightarrow\LambrechtsExtension(W)$ preserves orientation on homology.
We take $\Model = \VansQuotient(\LambrechtsExtension(\Lambda U))$ and obtain the following weak homotopy equivalence of $\PDGA$'s:
\begin{equation}\label{Eq:ModThree}\begin{tikzcd}
 & \Lambda U \arrow{dl}\arrow{dr} & \\
\VansQuotient(\LambrechtsExtension(\Lambda V))& & V.
\end{tikzcd}\end{equation}
This proves the proposition.
\end{proof}

The following is mostly \cite[Theorem~7.1]{Lambrechts2007}.
In addition, we check that the orientation on homology is preserved.
Also, by using our extension of Hodge type, we can improve from $n\ge 7$ to $n\ge 5$.

\begin{Proposition}[``Weak uniqueness'' of Poincar\'e duality model]\label{Prop:LambrechtUnique}
Let $V_1$ and $V_2$ be differential Poincar\'e duality algebras of degree $n$ which are weakly homotopy equivalent as $\PDGA$'s.
Suppose that $\H^0(V_1)=\H^0(V_2)=\Span\{1\}$ and $\H^1(V_1) = \H^1(V_2) = 0$.
In addition, suppose that $\H^2(V_1) = \H^2(V_2) = 0$, $V_1^1 = V_2^1 = 0$ and $n\ge 5$.
Then there is a differential Poincar\'e duality algebra $V_3$ and $\PDGA$-quasi-isomorphisms\footnote{These are automatically injective and orientation preserving.}
\begin{equation}\label{Eq:LambrechtsZigZag}
\begin{tikzcd}
& V_3 & \\
V_1\arrow{ur}& & \arrow{ul}V_2.
\end{tikzcd}
\end{equation}
\end{Proposition}
\begin{proof}
By the assumption, there is $k\ge 1$ and a zig-zag of $\PDGA$-quasi-isomorphisms
\begin{equation}\label{Eq:ZigZag}
V_1 \longleftarrow Z_1 \longrightarrow Z_2 \longleftarrow Z_3 \longrightarrow Z_4 \longleftarrow \dotsb \longleftarrow Z_k \longrightarrow V_2.
\end{equation}
Consider the Sullivan minimal model $\Lambda U \rightarrow Z_2$ and use the Lifting Lemma \cite[Lemma~2.15]{Felix2008} to construct $\DGA$-quasi-isomorphisms $\Lambda U \rightarrow Z_1$ and $\Lambda U \rightarrow Z_3$ such that the diagram
$$\begin{tikzcd}
& \Lambda U \arrow{d} \arrow{ld} \arrow{rd} & \\
Z_1 \arrow{r} & Z_2 & \arrow{l} Z_3
\end{tikzcd}$$
commutes up to homotopy of $\DGA$'s. It is easy to see that there is an orientation on $\H(\Lambda U)$ such that all morphisms preserve orientation on homology. Therefore, we can replace the segment $V_1 \longleftarrow Z_1 \longrightarrow Z_2 \longleftarrow Z_3 \longrightarrow Z_4$ in \eqref{Eq:ZigZag} by $V_1 \longleftarrow \Lambda U \longrightarrow Z_4$. Repeating this process, we can shorten \eqref{Eq:ZigZag} to 
\begin{equation}\label{Eq:DiagDiag}
\begin{tikzcd}
& \Lambda U \arrow{dr}{f_2} \arrow[swap]{dl}{f_1} & \\
V_1 & & V_2,
\end{tikzcd}
\end{equation}
where $f_1$ and $f_2$ are $\PDGA$-quasi-isomorphisms. In order to take $\VansQuotient(\LambrechtsExtension(\Lambda U))$ and obtain a zig-zag with three terms, we have to revert the arrows in \eqref{Eq:DiagDiag}. The trick from~\cite{Lambrechts2007} is the following:

Consider the relative minimal model of the multiplication $\mu: \Lambda U \otimes \Lambda U \rightarrow \Lambda U$; from \cite[Example~2.48]{Felix2008}, it is given by
\begin{equation}\label{Eq:RelMinMod}
\begin{tikzcd}
\Lambda U \otimes \Lambda U\arrow{r}{\mu} \arrow{rd}{i} & \Lambda U \\
& M(\mu)\coloneqq \Lambda U \otimes \Lambda U \otimes \Lambda(U[1]), \arrow{u}{p}
\end{tikzcd}
\end{equation}
where $i$ is the inclusion into the first two factors, which is a cofibration, and $p$ is a surjective quasi-isomorphism. Let $\iota_i : \Lambda U \rightarrow \Lambda U \otimes \Lambda U$ for $i=1$, $2$ be the inclusions to the first and the second factor, respectively. Because $\mu \circ \iota_i = \Id$ and $p$ is a quasi-isomorphism, the maps $i \circ \iota_i : \Lambda U \rightarrow M(\mu)$ for $i=1$, $2$ are quasi-isomorphisms. Moreover, it is easy to see that $\H(M(\mu))$ inherits an orientation such that $p_*$ and $(i\circ \iota_i)_*$ are orientation preserving. To transfer this situation to $V_i$, we use the diagram
\begin{equation}\label{Eq:Pushout}
\begin{tikzcd}
\Lambda U \arrow{r}{f_i}\arrow{d}{\iota_i}& V_1 \arrow{d}{\iota_i^V} \\
\Lambda U \otimes \Lambda U \arrow{r}{f_1\otimes f_2} \arrow{d}{i} & V_1 \otimes V_2 \arrow{d}{g_2} \\
M(\mu) \arrow{r}{g_1} & \tilde{V}_3 \coloneqq  M(\mu) \otimes_{\Lambda U \otimes \Lambda U} (V_1\otimes V_2),
\end{tikzcd}
\end{equation}
where $\iota_i^V : V_i \rightarrow V_1 \otimes V_2$ for $i=1$, $2$ are inclusions. The lower square, i.e., the maps $g_1$, $g_2$ and the $\DGA$ $\tilde{V}_3$, is a pushout diagram (see \cite[Example~1.4]{LoopSpaces}). According to~\cite{MO204414}, the model category of $\nnuCDGA$ is proper, and hence pushouts along cofibrations preserve quasi-isomorphisms. Therefore, $f_1\otimes f_2$ being a quasi-isomorphism implies that $g_1$ is a quasi-isomorphism. We push the orientation to~$\H(\tilde{V}_3)$ via $g_{1*}$. Since $i\circ \iota_i$, $g_1$ and $f_i$ are quasi-isomorphisms preserving orientation on homology, it follows that $h_i\coloneqq g_2\circ \iota_i^V:  V_i \rightarrow V_3$ are quasi-isomorphisms preserving orientation of homology as well. It holds
\[
\tilde{V}_3 \simeq \Lambda(U[1]) \otimes V_1 \otimes V_2.
\]
Clearly, $\tilde{V}_3$ is of finite type.
Using $\H^1(V_i) = \H^2(V_i) = 0$, we have $U^1 = U^2 = 0$, and hence $(\Lambda(U[1]))^1 = 0$.
This together with $V_i^1 = 0$ implies that $\tilde{V}_3^1 = 0$.
Therefore, all conditions for an application of the Hodge extension $\LambrechtsExtension$ from Proposition~\ref{Prop:ExtensionOfHodgeType} are satisfied, and we can set
$$ V_3\coloneqq \VansQuotient(\LambrechtsExtension(\tilde{V}_3)). $$
This finishes the proof.
\end{proof}

\begin{Conjecture}\label{Conj:PDGALST}
The additional assumptions of Proposition~\ref{Prop:LambrechtUnique} can be dropped.
\end{Conjecture}

\begin{Remark}[Weak uniqueness in the case of $\H^1(V)=0$ and $n\le 3$]
 For $n=1$, there is no differential Poincar\'e duality algebra $V$ with $\H^1(V) = 0$.
 
 For $n=2$, a general differential Poincar\'e duality algebra can be written in terms of its Hodge decomposition as
 \begin{align*}
 	V^2 & = \Span\{\Vol\}\oplus \Dd C^1\\
	V^1 & = \Dd C^0 \oplus C^1\\
	V^0 & = \Span\{1\} \oplus C^0.
 \end{align*}
 Now, $\Harm = \Span\{1\}\oplus \Span\{\Vol\}$ is a dg-subalgebra which is itself a Poincar\'e duality algebra. Therefore, two differential Poincar\'e duality algebras with $\H(V_1)\simeq \H(V_2)$ and $\H^1(V_i) = 0$ are connected via the zig-zag
\[
\begin{tikzcd}
& \Harm \arrow{dr}{} \arrow[swap]{dl}{} & \\
V_1 & & V_2.
\end{tikzcd}
\]
For $n=3$, we have 
\begin{align*}
	V^3 & = \Span\{\Vol\} \oplus \Dd C^2\\
 	V^2 & =\Dd C^1 \oplus C^2 \\
	V^1 & = \Dd C^0 \oplus C^1\\
	V^0 & = \Span\{1\} \oplus C^0,
\end{align*}
and the same situation as for $n=2$ occurs.

For $n=4$ and $V^1 = 0$, we have $V \simeq \Harm$.
\end{Remark}

We would like to define a ``minimal Poincar\'e duality model''. We motivate this notion in the following remark.

\begin{Remark}[Model and minimal model]\label{Rem:Models}
We shall understand models and minimal models in terms of model categories and their homotopy categories.

Let us illustrate this on Sullivan models. A Sullivan $\DGA$ is a free graded commutative algebra $\Lambda U$ over a positively graded vector space $U$ which admits a well-ordered homogenous basis $(v_\alpha)$ such that $\Dd v_\alpha \in \Lambda(v_\beta \mid \beta < \alpha)$ ($\coloneqq$\,the subalgebra of $\Lambda U$ generated by the $v_\beta$'s) for all $\alpha$. A Sullivan $\DGA$ is called minimal if $\im \Dd \subset \Lambda_{\ge 2} U$ ($\coloneqq$\,the set of decomposable elements).

According to \cite[Theorem~4.3]{Bousfield1976}, the category $\nnuCDGA$ is a model category with weak equivalences being $\DGA$-quasi-isomorphisms, fibrations being degree-wise surjective $\DGA$-morphisms and cofibrations being retracts of relative Sullivan algebras (see \cite[Proposition~2.22 and Proposition~2.28]{Felix2008}). Cofibrant objects are then precisely Sullivan algebras. 

So, the homotopy extension property holds already for Sullivan $\DGA$'s. To see the role of minimality, we shall descent to the homotopy category. The homotopy category is constructed from a model category by localizing morphisms at weak equivalences. An isomorphism in the homotopy category, called weak homotopy equivalence, corresponds to a zig-zag of weak equivalences. If $V_1$ is weakly homotopy equivalent to $V_2$, we say that $V_2$ is a model of $V_1$. If there is a weak equivalence $V_2 \rightarrow V_1$, we say that $V_2$ is a resolution of~$V_1$. We understand minimality as a condition which is in each weak homotopy equivalence class satisfied by at most one object up to isomorphism in the model category. If minimal models exist, they form a skeleton of the homotopy category. This is precisely the case of $\nnuCDGA$ and minimal Sullivan algebras. Indeed, by \cite[Theorem~2.24]{Felix2008}, every connected $\nnuCDGA$ is resolved by a minimal Sullivan algebra. Next, by \cite[Proposition~2.26]{Felix2008}, a $\DGA$-morphism lifts to resolutions by Sullivan algebras, and by \cite[Corollary~2.13]{Felix2008}, quasi-isomorphic minimal Sullivan $\DGA$'s are isomorphic. Finally, this implies that weakly homotopy equivalent minimal Sullivan algebras are isomorphic, and hence are minimal in the sense above.

As another example, for an operad (or properad) $\Operad$, one wants to construct a dg-operad~$\Operad_\infty$ which is a quasi-free resolution of $\Operad$ (see \cite{Vallette2012}). Quasi-free means that after forgetting the differential, the operad $\Operad_\infty$ is free over $\Perm$-bimodules. This is similar to Sullivan models which are free over vector spaces. For quadratic operads, $\Operad_\infty$ is often constructed as the cobar construction $\Omega$ of the Koszul dual cooperad $\Operad^{\mbox{!`}}$. This is the case of $\AInfty$, $\LInfty$ or of the properad $\IBLInfty$. The differential on $\Omega \Operad^{\mbox{!`}}$ is the extension of the decomposition on~$\Operad^{\mbox{!`}}$ to a derivative, and hence it has decomposable image (c.f., the explicit formula~\cite[Formula~(2)]{Peksova2018}). This is similar to minimal Sullivan models. It would be interesting to know whether $\Operad_\infty$ can be constructed using the same inductive method of ``killing'' and ``adding'' generators of homology as Sullivan minimal models. 

Finally, let us note that in \cite{Cirici2019}, they use the inductive method to construct (minimal) models of $\Operad$-algebras for a wide class of operads $\Operad$. Note that cyclic $\AInfty$-algebras can be formulated in the language of cyclic operads. However, in the case of $\PDGA$'s, we have the non-degenerate pairing on homology and an operadic description is not clear.
\end{Remark}

The following example shows that the zig-zag of a Poincar\'e duality model can not always be shortened to one arrow.

\begin{Example}[No Poincar\'e dualiy model with one arrow]\phantomsection\label{Ex:NoOneArrow}
\begin{ExampleList}
\item The closed genus $2$ surface $\Sigma_2$ does not admit a Poincar\'e duality model $A$ with just one arrow $A \rightarrow \DR(\Sigma_2)$. Suppose the contrary. We consider the quasi-isomorphism $f:A\rightarrow \DR(\Sigma_2)$ and compute for homogenous $v_1$, $v_2\in A$ the following:
\begin{align*}
\langle f(v_1),f(v_2)\rangle &= \pm \langle f(v_1)\wedge f(v_2),1 \rangle\\
&= \pm \langle f(v_1\wedge v_2),1 \rangle\\
&= \pm \langle [f(v_1\wedge v_2)],[1]\rangle\qquad\text{(*)}\\
&= \pm \langle f_*[v_1\wedge v_2], f^*[1] \rangle\\
&= \pm \langle [v_1\wedge v_2],[1]\rangle\\
&= \pm \langle v_1\wedge v_2, 1 \rangle\qquad(*) \\
&= \langle v_1, v_2 \rangle.
\end{align*}
Stars hold because the only non-zero case is when $\deg(v_1) + \deg(v_2) = \deg(\langle\cdot,\cdot\rangle)$, and hence $\Dd (v_1\wedge v_2) = 0$ because non-degeneracy implies vanishing of higher degrees. We see that $f$ preserves the pairing on the chain-level and is injective by non-degeneracy. We can thus assume that $A\subset\DR(\Sigma_2)$ is a dg-subalgebra equipped with the restriction of the intersection pairing. Since $\dim(A)<\infty$, there is a Hodge decomposition
\begin{align*}
	A^2 &= \Harm^2 \oplus \Dd C^1\\
	A^1 &= \Harm^1 \oplus \Dd C^0 \oplus C^1\\
	A^0 &= \Harm^0 \oplus C^0.
\end{align*}
If $1\neq f\in C^{\infty}(\Sigma_2)$, then $f^k$, $k\in \N_0$ are linearly independent over $\R$. Therefore, it must hold $\Harm^0 = \Span\{1\}$ and $C^0 = 0$. Duality implies $\Dd C^1 = 0$, and hence $\Harm^2$ is spanned by $\omega \in \DR^2(\Sigma_2)$ with $\int_{\Sigma_2}\omega = 1$. It holds even $C^1=0$ as $\ker \Dd \cap C = 0$ in a Hodge decomposition. Since $A\simeq \H(\Sigma_2)$ as a $\DGA$, there are closed $\alpha_1$, $\beta_1$, $\alpha_2$, $\beta_2\in\DR^1(\Sigma_2)$ such that $\Harm^1 = \Span\{\alpha_1, \beta_1, \alpha_2, \beta_2\}$ and such that for all $x\in\Sigma_2$ the following holds:
\begin{align*}
	\alpha_1(x) \wedge \alpha_2(x) &=  \alpha_1(x) \wedge \beta_2(x) = 0\\
	\quad\alpha_1(x)\wedge\beta_1(x) &= \alpha_2(x)\wedge\beta_2(x) = \omega(x).
\end{align*}
Taking an $x\in \Sigma_2$ with $\omega(x) \neq 0$ gives a contradiction.

\item  The simply-connected $4$-manifold $\CP^{2\# 7}$, where $\#$ denotes the connected sum, does not admit a Poincar\'e duality model $A$ with just one arrow $A \rightarrow \DR(M)$. Similarly as in the proof for $\Sigma_2$, we can restrict to the case $A\subset\DR(\CP^{2 \# 7})$. We obtain $A^4 = \Harm^4 = \langle \omega \rangle$ for a somewhere non-vanishing $4$-form $\omega$ and $\Harm^2 = \langle K_0, \dotsc, K_6 \rangle$ such that for all $x\in \CP^{2\#7}$ the following holds:
\begin{align*}
	K_i(x)\wedge K_i(x) &= \pm \omega(x) \\
	K_i(x) \wedge K_j(x) &= 0.
\end{align*}
We now view $K_i(x)$ as vectors in $\R^6$ so that $K_0(x)\wedge K_j(x) = 0$ corresponds to taking the scalar product. If $\sum_{i=0}^6 \lambda_i K_i(x) = 0$ for some $\lambda_i\in \R$ with $\lambda_{i_0} \neq 0$, then
\[
0 = K_{i_0}(x) \wedge \Bigl(\sum_{i=0}^6 \lambda_i K_i(x)\Bigr) = \pm \lambda_{i_0} \omega(x).
\]
Therefore, $\omega(x) \neq 0$ implies that $K_i(x)$ are linearly independent. Hence, in this case, $K_0(x) \wedge K_i(x) = 0$ for all $i=1$, $\dotsc$, $6$ implies $K_0(x) = 0$, which is a contradiction with $K_0(x) \wedge K_0(x) = \omega(x)$.\qedhere
\end{ExampleList}
\end{Example}

\begin{Questions}\phantomsection\label{Q:QuestionsPonc}
\begin{RemarkList}
\item Give an example of a Sullivan minimal model $f: \Lambda U \rightarrow V$ of a $\PDGA$ $V$ which is not of Hodge type for any orientation such that $f_*: \H(\Lambda U) \rightarrow \H(V)$ is orientation preserving.
\item Can the additional assumptions in Lemma~\ref{Lemma:Exte} and Proposition~\ref{Prop:ExtensionOfHodgeType} be dropped?
\item Is smallness or minimal dimension the correct notion of ``minimality'' of a Poincar\'e duality model?  \qedhere
\end{RemarkList}
\end{Questions}

\section{Consequences for IBL-infinity theory}\label{Section:FuncIBL}

Consider the construction $\dIBL^\MC(\CycC(\cdot))$ on cyclic $\DGA$'s ($=$\,differential Poincar\'e duality algebras) from Section~\ref{Sec:Alg3}. We first need the following technical result.

\begin{Proposition}[Functoriality of $\dIBL^\MC$-construction up to homotopy]\label{Prop:Functorialityo}
Suppose that $V_1$ and~$V_2$ are differential Poincar\'e duality algebras, and let $f: V_1 \rightarrow V_2$ be a $\PDGA$-quasi-isomorphism.
Then there is an $\IBLInfty$-quasi-isomorphism $\HTP: \dIBL(\CycC(V_2)) \rightarrow \dIBL(\CycC(V_1))$ which is also an $\IBLInfty$-quasi-isomorphism $\HTP: \dIBL^\MC(\CycC(V_2)) \rightarrow \dIBL^\MC(\CycC(V_1))$ and satisfies $\HTP_{110} = f^*$. 
\end{Proposition}

\begin{proof}
Lemma~\ref{Lem:PomLemma} implies that $f$ is injective, and hence we can consider the case when $V_1\subset V_2$ is a dg-subalgebra and $f=\iota: V_1 \hookrightarrow V_2$ is the inclusion.
In the proof of Lemma~\ref{Eq:LemSmallSub} it was argued that there are Hodge decompositions $V_1 = \Harm \oplus \Dd C_1 \oplus C_1$ and $V_1^\perp = \Dd C_2 \oplus C_2$ such that $V_2 = \Harm \oplus (\Dd C_1 \oplus \Dd C_2) \oplus (C_1 \oplus C_2)$ is a Hodge decomposition of $V$.
Define $\Prpg: V_2 \rightarrow V_2$ by
\[
 \Prpg(\Dd c_2) \coloneqq - c_2\quad\text{for all }c_2\in C_2
\]
and by $0$ on other direct summands.
Let $\pi: V_2 \rightarrow V_1$ be the canonical projection induced by $V_2 = V_1 \oplus V_1^\perp$.
For $v$, $v'\in V_2$, write $v = h + \Dd c_{11} + c_{12} + \Dd c_{21} + c_{22}$, $v'=h' + \Dd c_{11}' + c_{12}' + \Dd c_{21}' + c_{22}'$ for $h$, $h'\in \Harm$, $c_{11}$, $c_{11}'$, $c_{12}$, $c_{12}'\in C_1$, $c_{21}$, $c_{21}'$, $c_{22}$, $c_{22}'\in C_2$ and compute
\begin{align*}
 (\Prpg \circ \Dd + \Dd\circ\Prpg)(v) &= \Prpg(\Dd c_{21}) + \Dd\Prpg(\Dd c_{22}) \\
 & = - c_{21} - \Dd c_{22} \\
 & = (\pi - \Id)(v)
\end{align*}
and
\begin{align*}
	\langle \Prpg v, v' \rangle &= \langle - c_{21},\Dd c_{22}'\rangle \\
	& = (-1)^{\deg c_{21}} \langle \Dd c_{21}, c_{22}' \rangle \\
	& = (-1)^{\deg c_{21} + 1} \langle \Dd c_{21}, \Prpg v' \rangle \\
	& = (-1)^{\deg v} \langle v, \Prpg v' \rangle.
\end{align*}
These are conditions (11.2), resp.~(11.3) from the beginning of \cite[Section~11]{Cieliebak2015}.
We can now apply \cite[Theorem~11.3]{Cieliebak2015} to obtain an $\IBLInfty$-quasi-isomorphism $\HTP: \dIBL(\CycC(V_1)) \rightarrow \dIBL(\CycC(V_2))$ with $\HTP_{110} = \iota^*$.
Because $\iota$ is a morphism of algebras, $\iota^*$ commutes also with the Hochschild differential, i.e., with $\OPQ_{110}^\MC$.
The claim follows because $\dIBL = (\OPQ_{110},\OPQ_{210},\OPQ_{120})$ and $\dIBL^\MC = (\OPQ_{110}^\MC,\OPQ_{210},\OPQ_{120})$. 
\end{proof}

For any $\PDGA$ $(V,\dd,\wedge,\Or^\H)$ with $\H^0(V)=\R$ and $\H^1(V)=0$, we can pick its Poincar\'e duality model $\Model(V)$ and associate to it the $\dIBL$-algebra $\dIBL^\MC(\CycC(\Model(V)))$. This extends the domain of $\dIBL^\MC(\CycC(\cdot))$ to $\PDGA$'s. However, removing choices requires us to relax the codomain to $\IBLInfty$-homotopy equivalence classes.

\begin{Proposition}[Uniqueness of $\dIBL^\MC$ for $\PDGA$ up to homotopy]\label{Prop:ExteofDFS}
Let $V$ be a $\PDGA$ of degree $n$ with $\H^0(V)=\Span\{1\}$ and $\H^1(V)=0$.
Let $\Model_1$ and $\Model_2$ be two Poincar\'e duality models of $V$.
In addition, suppose that $\Model_1^1 = \Model_2^1 = 0$ if $n\ge 4$ and that $\H^2(V)=0$ if $n\ge 5$.
Then $\dIBL^\MC(\CycC(\Model_1))$ and $\dIBL^\MC(\CycC(\Model_2))$ are $\IBLInfty$-quasi-isomorphic.
\end{Proposition}
\begin{proof}
By Proposition~\ref{Prop:LambrechtUnique}, there is a differential Poincar\'e duality algebra~$V_3$ and $\PDGA$-quasi-isomorphisms $f_1: \Model_1\rightarrow V_3$ and $f_2: \Model_2\rightarrow V_3$.
Proposition~\ref{Prop:Functorialityo} gives $\IBLInfty$-quasi-isomorphisms $\HTP_1: \dIBL^\MC(\CycC(\Model_1)) \rightarrow \dIBL^\MC(\CycC(V_3))$ and $\HTP_2: \dIBL^\MC(\CycC(\Model_2)) \rightarrow \dIBL^\MC(\CycC(V_3))$ extending $f_1^*$ and $f_2^*$, respectively. Application of \cite[Theorem 1.2]{Cieliebak2015} (existence of $\IBLInfty$-homotopy inverses for $\IBLInfty$-quasi-isomorphisms) finishes the proof.
\end{proof}

\begin{Conjecture}\label{Conj:UniqModelFollow}
	The additional assumptions of Proposition~\ref{Prop:ExteofDFS} can be removed.
\end{Conjecture}

If we had the notion of a minimal Poincar\'e duality model $\Model_0(V)$ which is unique in every weak homotopy class of $\PDGA$'s (c.f., Questions~\ref{Q:QuestionsPonc}), we could associate to a weak homotopy equivalence class of a $\PDGA$ $V$ a canonical $\dIBL$-algebra $\dIBL^\MC(\CycC(\Model_0(V)))$ up to an $\IBLInfty$-isomorphism.\footnote{In order to think about functors, we would first have to find a canonical construction of an $\IBLInfty$-morphism induced by a $\PDGA$-morphism of cyclic $\DGA$'s. The construction of \cite[Section~11]{Cieliebak2015} applies only to quasi-isomorphisms, and even if we pick the standard Hodge propagator $\StdPrpg$, it still depends on the choice of the Hodge decomposition. Therefore, the best we can hope for is an assignment of $\PDGA$-morphisms to $\IBLInfty$-homotopy classes of $\IBLInfty$-morphisms.}

\begin{Definition}[$\IBLInfty$-formality]\label{Def:IBLFormality}
We say that a differential Poincar\'e duality algebra $V$ is \emph{$\IBLInfty$-formal} if $\dIBL^\MC(\CycC(V))$ and $\dIBL^\MC(\CycC(\H(V)))$ are weakly $\IBLInfty$-homotopy equivalent.
\end{Definition}

Due to the existence of homotopy inverses of quasi-isomorphisms in the $\IBLInfty$-category, being weakly $\IBLInfty$-homotopy equivalent (the existence of zig-zag of quasi-isomorphisms) is equivalent to being $\IBLInfty$-quasi-isomorphic (the existence of a~direct quasi-isomorphism), which is equivalent to being $\IBLInfty$-homotopy equivalent (the existence of a quasi-isomorphism with a homotopy inverse).

We formulate the following statement as a conjecture although a special case holds by Proposition~\ref{Prop:ExteofDFS}.
The proof might be easier than the proof of Conjecture~\ref{Conj:UniqModelFollow}.

\begin{Conjecture}[Algebraic formality conjecture]\label{Con:DGAIBLForm}
Let $V$ be a $\PDGA$ with $\H^0(V)=\Span\{1\}$ and $\H^1(V)=0$ which is formal as a $\DGA$.
Then any its Poincar\'e duality model $\Model(V)$ is $\IBLInfty$-formal.
\end{Conjecture}
\begin{proof}
By Proposition~\ref{Prop:PoincModelOfFormal}, $\DGA$-formality is equivalent to $\PDGA$-formality. Therefore,~$\H(V)$ is a Poincar\'e duality model of $V$. The rest follows from Conjecture~\ref{Conj:UniqModelFollow}.
\end{proof}

\begin{Remark}[$\IBLInfty$-formality of $V$ and $\CycC(V)$]\label{Rem:Intfor}
Let $\Operad$ be a dg-operad. If $V$ is an $\Operad$-algebra, then $\H(V)$ is an $\Operad$-algebra with zero differential. We say that $V$ is $\Operad$-formal if $V$ and $\H(V)$ are weakly homotopy equivalent as $\Operad$-algebras; i.e., if there is a zig-zag of $\Operad$-quasi-isomorphisms between $V$ and $\H(V)$.

Consider $\DGA$ and $\AInfty$. If a $\DGA$ $V$ is formal, then it is also $\AInfty$-formal because $\DGA$ is a subcategory of $\AInfty$.\footnote{Note that $\AInfty$, similarly as $\IBLInfty$, admits homotopy inverses of quasi-isomorphisms, and hence $\AInfty$-formality is equivalent to the existence of a direct $\AInfty$-quasi-isomorphism $V \rightarrow \H(V)$.} The converse is also true by the rectification procedure (see~\cite{MSE2719961}). Therefore, $\DGA$-formality and $\AInfty$-formality of a $\DGA$ $V$ are equivalent. This should be true for any suitable (pro)perad~$\Operad$ and its quasi-free resolution~$\Operad_\infty$.

In the spirit above, we can speak about $\dIBL$-, resp.~$\IBLInfty$-formality of the $\dIBL$-algebra $\dIBL^\MC(\CycC(V))$ on $\CycC(V)$, which is about the existence of zig-zags of $\dIBL$-, resp.~$\IBLInfty$-quasi-isomorphisms between $\CycC(V)$ and $\H(\CycC(V),\OPQ_{110}^\MC)$. These notions are most likely equivalent as in the case of $\DGA$- and $\AInfty$-formality.

On the other hand, $\IBLInfty$-formality of a differential Poincar\'e duality algebra $V$ from Definition~\ref{Def:IBLFormality} is a different notion because $V$ is not an $\IBLInfty$-algebra.

We conclude that the following three types of ``formalities'' might be interesting: 
\begin{enumerate}[label=\arabic*)]
 \item $\DGA$-formality of a $\PDGA$ $V$, 
 \item $\IBLInfty$-formality of a $\PDGA$ $V$ (resp.~of its Poincar\'e duality model),
 \item $\IBLInfty$-formality of $\CycC(V)$ as the $\dIBL$-algebra $\dIBL^\MC(\CycC(V))$.
\end{enumerate}
Except for the fact that (1) implies (2), we do not have any other guesses. Notions (1)--(3) should also be interpreted geometrically in the context of the String Topology Conjecture.
\end{Remark}

Let $M$ be a connected oriented closed $n$-manifold with $\HDR^1(M)=0$. We know that $\DR(M)$ is of Hodge type and that a Riemannian metric on $M$ induces a canonical Hodge decomposition $\DR(M) = \Harm \oplus \Im\Dd \oplus \Im\CoDd$, where $\Harm$ is the subspace of harmonic forms and~$\CoDd$ the codifferential. So, in this case, we have a canonical Poincar\'e duality model 
$$ \Model(\DR(M)) = \VansQuotient(\VansSmall(\DR(M))). $$
Due to $\HDR^1(M)=0$, it is of finite type and hence finite-dimensional. Therefore, $\dIBL^\MC(\CycC(\Model(\DR(M))))$ is well-defined. Recall that $(\DR,\Dd,\wedge,\int)$ is not of finite type and $\dIBL^\MC(\CycC(\cdot))$ is not well-defined;
that was the reason to define the Chern-Simons, aka formal pushforward, Maurer-Cartan element $\PMC$ for $\dIBL(\CycC(\HDR(M)))$ directly as a summation over Feynman integrals. The idea was that $\dIBL^\PMC(\CycC(\HDR(M)))$ is ``homotopy equivalent'' to ``$\dIBL^\MC(\CycC(\DR(M)))$''.

We conjecture the following:

\begin{Conjecture}[Equivalence of algebraic and geometric approach]\label{Conj:SmallSimplCon}
Let $M$ be a connected oriented closed Riemannian $n$-manifold with $\HDR^1(M) = 0$. Then there is an admissible Hodge propagator $\Prpg$ such that if we consider the Chern-Simons Maurer-Cartan element $\PMC$, then the $\IBLInfty$-algebras 
$$\dIBL^\PMC(\CycC(\HDR(M)))\quad\text{and}\quad\dIBL^\MC(\CycC(\VansQuotient(\VansSmall(\DR(M))))) $$
are $\IBLInfty$-homotopy equivalent.
\end{Conjecture}
\begin{proof}[Comment on proof]
By the vanishing result from Theorem~\ref{IntroThm:B}, we can find $\Prpg$ such that the only non-zero component of $(\PMC_{lg})$ is $\PMC_{10}$.
It satisfies $\PMC_{10} = \sum \pm m_k^+$, where $m_k: \Harm^{\otimes k} \rightarrow \Harm$ are the $\AInfty$-operations obtained by homotopy transfer from $\DR$ (we use the canonical Hodge isomorphism $\HDR\simeq\Harm$).
By lifting the zig-zag of $\PDGA$'s to $\AInfty$, we obtain an $\IBLInfty$-quasi-isomorphism $\GTP: (\VansQuotient(\VansSmall(\DR)),\Dd,\wedge) \rightarrow (\HDR,(m_k))$.
The idea is to show that $\GTP_*\MC$ and $\PMC$, where $\MC$ is the canonical Maurer-Cartan element, are gauge equivalent Maurer-Cartan elements (see \cite[Definition~9.7]{Cieliebak2015}).
For this, it might be useful to describe a path object of the $\IBL$-algebra $\dIBL(\CycC(\HDR))$ (it is a $\dIBL$-algebra), how it arises from the path object of a $\DGA$, and how an overlying Maurer-Cartan element can be constructed from the $\AInfty$-morphism.
\end{proof}

The following is a consequence of Conjecture~\ref{Conj:SmallSimplCon}.
However, we formulate it separately as it might be easier to prove.

\begin{Conjecture}[Formality conjecture for geometric construction]\label{Cor:FormCorollary}
In the situation of Conjecture~\ref{Conj:SmallSimplCon}, suppose that~$M$ is formal. Then $\dIBL^\PMC(\CycC(\HDR(M)))$ is $\IBLInfty$-homotopy equivalent to $\dIBL^\MC(\CycC(\HDR(M)))$.
\end{Conjecture}
\begin{proof}[Comment on proof]
One has to prove the following. If there is an $\AInfty$-isomorphism $(f_i)$ of $(\HDR,\wedge)$ and $(\HDR,\wedge,m_3,\dotsc)$ with $f_1 = \Id$, then $\MC_{10}$ and $\PMC_{10} = \sum \pm \MC_k^+$ are gauge equivalent.
\end{proof}

Together with the String Topology Conjecture~\ref{Conj:StringTopology}, this gives a canonical chain model for the equivariant Chas-Sullivan string topology for formal $M$ with $\HDR^1(M)=0$.

\chapter{Standard Hodge propagator}
\label{Chap:Prpg}

In this chapter, we tackle the question whether the standard Hodge propagator $\StdPrpg$ extends smoothly to the blow-up.

In Section~\ref{Sec:SchwFrom}, we recall the Schwartz kernel theorem (Proposition~\ref{Prop:SchwKer}) and define the Schwartz form of an operator between sections of the exterior bundle of the cotangent bundle (Proposition~\ref{Prop:SchwForm} and Definition~\ref{Def:SDFDF}). We mention that the Schwartz form of a pseudo-differential operator is smooth outside of the diagonal (Proposition~\ref{Prop:ASD}). On the example of the identity (Example~\ref{Ex:IdFE}) we illustrate that the smooth part does not always determine the pseudo-differential operator. We recall basic facts about regularity of the Laplace Green kernel and standard Hodge propagator (Proposition~\ref{Prop:BasicFactsGP}). We formulate the problem of smooth extension to the blow-up (Definition~\ref{Def:Esdas}) and ask whether it is satisfied by $\GKer$ and $\StdPrpg$.

In Section \ref{Sec:TZ}, we study uniqueness of the Hodge homotopy (Proposition~\ref{Prop:UniHO}). We would like to characterize the standard Hodge propagator as a unique primitive to the harmonic kernel satisfying certain properties. We give some ideas how to do it. We also illustrate that one has to be careful with blow-ups (Proposition~\ref{Prop:PatEm}).

In Section \ref{Sec:Hwe}, we consider the heat form approximation. We define the heat form and sum up its computational properties (Proposition~\ref{Prop:Heasd}). We write down formulas for the heat form approximations (Proposition~\ref{Sec:Hwe}). We postulate that the standard Hodge propagator is the codifferential of the Laplacian Green kernel with respect to one variable (Proposition~\ref{Prop:StdCodifInt}).

In Section \ref{Sec:HeatRN}, we study the standard Hodge propagator on $\R^n$. We recall the well-known formulas for the heat form and the Green form on $\R^n$ (Proposition~\ref{Prop:GreenKernelRn}). We compute the standard Hodge propagator in two ways --- as a coderivative of the Green form and as an integral using heat form approximation (Proposition~\ref{Prop:StdHodgePropRn} for $n\ge 2$ and Example~\ref{Ex:SDFSDF} for $n=1$).

In Section \ref{Sec:GrSpgh}, we study the standard Hodge propagator on $\Sph{n}$. We compute it explicitly for $\Sph{1}$ (Example~\ref{Ex:SADQQ}). Next, we study the structure of the space of Hodge propagators on $\Sph{n}$ (Proposition~\ref{Prop:SpaceOfSolnSn}). We show that the Hodge propagator for $\Sph{n}$ constructed in Part~I is coexact (Proposition~\ref{Prop:ArtProsCoexact}). Finally, using the previous results, we prove that the standard Hodge propagator for $\Sph{2}$ extends smoothly to the blow-up and give an explicit formula with an unknown constant (Proposition~\ref{Prop:StdS2}).

\section{Schwartz form and smooth extension to blow-up}\label{Sec:SchwFrom}
\allowdisplaybreaks

For a smooth vector bundle $\pi: E\rightarrow M$, let $\TestF(E)$ be the space of ``test sections of $E$'', and let $\TestF'(E^*)$ be the space of distributions.  Let $\langle\cdot,\cdot\rangle: \TestF'(E^*)\otimes\TestF(E)\rightarrow \R$ be the natural pairing. Proper definitions can be formulated easily based on \cite[Section~6]{HormanderI}.
\begin{Proposition}[Schwartz kernel theorem, see {\cite[Theorem~5.2.1]{HormanderI}} for the local version]\label{Prop:SchwKer}
Let $E_1$ and $E_2$ be smooth vector bundles over smooth manifolds $M_1$ and $M_2$, respectively. Then there is a one-to-one correspondence between continuous operators $\TOp: \TestF(E_1)\rightarrow\TestF'(E_2)$ and elements $\KKer_\TOp\in\TestF'(E_1\boxtimes E_2^*)$ which is given by the equation
$$ \langle \TOp s_1, \psi_2 \rangle = \langle \KKer_\TOp, s_1 \boxtimes \psi_2 \rangle\quad\text{for all }s_1\in \TestF(E_1)\text{ and }\psi_2\in \TestF(E_2^*). $$
The distributional section $\KKer_{\TOp}$ is called the \emph{Schwartz kernel} of $\TOp$. 
\end{Proposition}

An $L^1_{\text{loc}}$-integrable function $k: M\times M \rightarrow \Hom(E_1,E_2)=E_1^*\boxtimes E_2$ on an oriented Riemannian manifold $M$ defines the distribution $\KKer\in \TestF'(E_1\boxtimes E_2^*)$ by
$$ \langle\KKer,s_1\boxtimes \psi_2\rangle = \int_{x_1, x_2} \langle k(x_1,x_2)s_1(x_1),\psi_2(x_2)\rangle \Vol(x_1)\Vol(x_2) $$
In the case of exterior bundles, we introduce an equivalent notion of a Schwartz form. This name was proposed by Dr.~A.~Hermann in a discussion in Potsdam.

\begin{Proposition}[Schwartz kernel and Schwartz form]\label{Prop:SchwForm}
In the setting of Proposition~\ref{Prop:SchwKer}, suppose that $M_1 = M_2 = M$ is a smooth oriented Riemannian manifold and $E_1 = E_2 = \Lambda T^* M$. We consider the isomorphism of vector bundles $\Psi:\Lambda T^*M\boxtimes(\Lambda T^*M)^*\rightarrow\Lambda T^*(M\times M)$ which is for every $x_1$, $x_2\in M$ given by
\begin{align*}
 \Psi: \Lambda T^*_{x_1}M\otimes\bigl(\Lambda T_{x_2}^*M\bigr)^* &\longrightarrow \Lambda T^*_{(x_1,x_2)}(M\times M) \\
 \omega_1\otimes\xi_2&\longmapsto\omega_1\wedge\sharp\xi_2.
\end{align*}
Here, $\sharp: \Lambda T^*M \rightarrow \bigl(\Lambda T^*M \bigr)^* $ denotes the musical isomorphism with respect to the natural pointwise inner product on $\Lambda T^*M$. We obtain the isomorphism
\begin{align*}
\Psi_*:\DR'(M\times M)&\longrightarrow\TestF'(E_1^*\boxtimes E_2)\\
\DR_\TOp&\longmapsto\KKer_\TOp,
\end{align*}
where $\DR'$ denotes the space of de Rham currents.

If $\TOp$ is homogenous of degree~$\Abs{\TOp}$, then the degree of $\DR_\TOp$ satisfies 
$$ \Deg(\DR_\TOp) = \dim(M) + \Abs{\TOp}. $$

If $\KKer_T$ is represented by an $L^1_{\mathrm{loc}}$-integrable section of $\Hom(E_1,E_2)\simeq E_1^*\boxtimes E_2$, then $\DR_T$ is represented by an $L^1_{\mathrm{loc}}$-integrable form, and the other way round. In this case, for all $\omega_1 \in \DR(M)$, $\omega_2\in \DR_c(M)$, we have
\begin{equation}\label{Eq:SchwarzKerForm}
\begin{aligned}
\int_{x_2} \bigl((\TOp \omega_1)(x_2)\bigr)\bigl[\omega_2(x_2)\bigr] &= \int_{x_1, x_2} \KKer_T(x_1,x_2)[\omega_1(x_1)]\wedge \Vol(x_1)\wedge \omega_2(x_2)\\
&= \int_{x_1, x_2} \DR_T(x_1,x_2)\wedge\omega_1(x_1)\wedge\omega_2(x_2).
\end{aligned}
\end{equation}
\end{Proposition}
\begin{proof}
Straightforward computations similar to Example~\ref{Ex:IdFE} below.
\end{proof}

\begin{Definition}[Schwartz form]\label{Def:SDFDF}
The current $\DR_\TOp\in \DR'(M\times M)$ from Proposition~\ref{Prop:SchwForm} is called the \emph{Schwartz form} of $\TOp$.
\end{Definition}

We will consider pseudo-differential operators $\TOp:\DR(M)\rightarrow\DR(M)$ on a Riemannian manifold $M$; this class of operators generalizes differential operators and contains generalized inverses of elliptic operators (see \cite{Hormander} for thorough treatment).

\begin{Proposition}[Schwartz form of pseudo-differential operators] \label{Prop:ASD}
Let $\TOp: \DR(M) \rightarrow \DR(M)$ be a pseudo-differential operator on a smooth oriented Riemannian manifold $M$. Then the Schwartz form $\DR_{\TOp}$ restricts to a smooth form on $M\times M\backslash\Diag$.
\end{Proposition}
\begin{proof}
Well known fact, proof based on \cite{Hormander}.
\end{proof}
 
\begin{Example}[Schwartz form of $\Id$]\label{Ex:IdFE}The Schwartz form of the identity $\Id: \DR(\R^n)\rightarrow \DR(\R^n)$ reads
\begin{equation}\label{Eq:SchwartzFormOfId}
\DR_{\Id}(x,y) = \delta(x-y)(\Diff{x}^1 - \Diff{y}^1)\dotsb(\Diff{x}^n-\Diff{y}^n),
\end{equation}
where $\delta$ denotes the Dirac delta function on $\R^n$ centered at $0$. In order to prove this, we start by rewriting
\begin{align*}
(\Diff{x}^1 - \Diff{y}^1) \dotsb (\Diff{x}^n - \Diff{y}^n) &=\sum_{I}(-1)^{\Abs{I}} \varepsilon(I^c I \mapsto[n])\Diff{x}^{I^c} \wedge \Diff{y}^I \\
&=\sum_{I} (-1)^{n \Abs{I}} (\Star \Diff{x}^I) \wedge \Diff{y}^I.
\end{align*}
Here, the sum is over all multiindices $I\subset \{1,\dotsc,n\}$, and we use that $\Star(\Diff{x}^I) = \varepsilon(I, I^c) \Diff{x}^{I^c}$, where $\varepsilon(I,I^c)$ denotes the sign to order $I I^c$ to $\{1,\dotsc,n\}$. Now, for any $\omega\in \DR_c(\R^n)$, which we write as $\omega(x) = \sum_K \omega_K(x)\Diff{x}^K$, we compute using $\Diff{x}^K\wedge\Star(\Diff{x}^I)=(\Diff{x}^K,\Diff{x}^I)\Vol(x) = \delta^{KI}\Vol(x)$ the following:
\begin{align*}
&\int_x \delta(x-y)  (\Diff{x}^1-\Diff{y}^1)\dotsb(\Diff{x}^n-\Diff{y}^n) \omega(x)  \\
&\qquad=\sum_{I,K} \int_x \delta(x-y)(-1)^{n\Abs{I}}  \omega_K(x)(\Star \Diff{x}^I)\Diff{y}^I\Diff{x}^K \\
&\qquad=\sum_{I,K}\int_x \delta(x-y) (-1)^{n\Abs{I} + n\Abs{K}} \delta(x-y)  \omega_K(x) \Diff{x}^K \Star(\Diff{x}^I) \Diff{y}^I \\
&\qquad=\sum_{I} \int_x \delta(x-y) \omega_I(x)\Vol(x) \Diff{y}^I \\
&\qquad=\omega(y).
\end{align*}
This shows \eqref{Eq:SchwarzKerForm}, and \eqref{Eq:SchwartzFormOfId} follows. Notice that
$$ \Restr{\DR_\Id}{\R^n\times \R^n\backslash \Diag} = 0, $$
and thus the smooth part of $\DR_\Id$ does not recover the data of the operator $\Id$.
\end{Example}

We consider the Green operator $\GOp$ for the Laplacian $\Laplace$ (see \cite{Warner1983}) and the standard Hodge propagator $\HtpStd$. They are both pseudo-differential operators, and it holds $\HtpStd = - \CoDd \GOp$. We will study their Schwartz forms $\GKer$ and $\PrpgStd$, which are called the \emph{Green kernel} and the \emph{standard Hodge propagator,} respectively.

\begin{Proposition}[Basic facts about $\GKer$ and $\StdPrpg$]\label{Prop:BasicFactsGP}
The Green kernel $\GKer$ represents an $L^2$-integrable form.
The standard Hodge propagator $\StdPrpg$ represent an $L^1$-integrable form.
\end{Proposition}
\begin{proof}
The fact that $\GKer$ is $L^2$ is an exercise in \cite{Warner1983}. The fact that~$\StdPrpg$ defines an $L^1$-integrable form on $M\times M$ for a compact manifold $M$ was proved in \cite{Harris2004} using the heat kernel approximation (see the next section).
\end{proof}

A consequence of Proposition~\ref{Prop:BasicFactsGP} is that the Schwartz forms $\GKer$ and $\StdPrpg$ are determined by their smooth restrictions to $M\times M\backslash \Diag$. This follows from \eqref{Eq:SchwarzKerForm} because the integral does not depend on sets of zero measure. Therefore, we will write $\GKer$, $\StdPrpg \in \DR^{n-1}(M\times M\backslash\Diag)$.

\begin{Definition}[Smooth extension to the blow-up]\label{Def:Esdas}
Let $M$ be a smooth manifold, and let $\pi: \Bl_\Diag(M\times M)\rightarrow M\times M$ be the spherical blow-up of $M\times M$ at the diagonal $\Diag$.\footnote{Another name of this construction suggested to me by Dr.~Oliver Lindblad Petersen after explaining him our setting should be ``Melrose blow-up''.} Consider the blow-up diagram
\[\begin{tikzcd}
 & \Bl_\Diag(M\times M)\arrow[two heads]{d}{\pi} \\
 M\times M\backslash \Diag \arrow[hook]{r}{\iota}\arrow[hook]{ru}{\tilde{\iota}} & M\times M,
\end{tikzcd}\]
where $\iota$ is the inclusion and $\tilde{\iota}$ its unique smooth lift --- the embedding of the interior. We say that a smooth form $\omega\in\DR(M\times M\backslash\Diag)$ \emph{extends smoothly to the blow-up} if there is a smooth form $\tilde{\omega}\in \DR(\Bl_\Diag(M\times M))$ such that 
$$ \tilde{\iota}^*\tilde{\omega}=\omega. $$
\end{Definition}

Note that the extension, if it exists, is necessarily unique.

\begin{Question}
Do $\GKer$ and $\StdPrpg$ extend smoothly to the blow-up?
We expect that $\GKer$ does not and $\StdPrpg$ does.
\end{Question}

\section{Uniqueness of Hodge propagator}\label{Sec:TZ}
\allowdisplaybreaks
\begin{Proposition}[Uniqueness of Hodge homotopy]\label{Prop:UniHO}
Let $M$ be a closed oriented manifold, and let $\DR(M)=\Im\Dd\oplus\Im\CoDd\oplus\Harm$ be its Hodge decomposition. For any linear map $\Htp:\DR(M)\rightarrow\DR(M)$ satisfying
\begin{equation}\label{Eq:GreenOp}
\Dd\circ\Htp + \Htp\circ\Dd = \iota_\Harm\circ\pi_\Harm - \Id,
\end{equation}
there exist linear maps $R_1: \Im \Dd \rightarrow \Im \Dd$, $R_2: \Im \CoDd \rightarrow \Ker \Dd$ and $R_3: \Harm \rightarrow \Ker \Dd$ such that
\begin{equation} \label{Eq:GreenOpForm}
\Htp = \begin{cases}
 -(\Restr{\Dd}{\Im\CoDd})^{-1} + R_1 & \text{on }\Im \Dd, \\
 (\Restr{\Dd}{\Im\CoDd})^{-1} R_1 \Dd + R_2 & \text{on } \Im \CoDd, \\
 R_3 & \text{on }\Harm.
\end{cases}
\end{equation}
Note that $\Restr{\Dd}{\Im \CoDd} : \Im \CoDd \rightarrow \Im \Dd$ is an isomorphism.
Moreover, the following facts are equivalent for an operator $\Htp$ satisfying \eqref{Eq:GreenOp}:
\begin{enumerate}[label=(\arabic*)]
\item $\Htp = \StdHtp$ is the standard Hodge homotopy,
\item $R_1 = R_2 = R_3 = 0$,
\item $\Im\Htp\subset\Im\CoDd$, 
\end{enumerate}
\end{Proposition}
\begin{proof}
Suppose first that $\omega \in \Im\Dd$. Plugging $\omega$ in \eqref{Eq:GreenOp}, we get that $\Htp \omega$ has to satisfy
$$ \Dd \Htp \omega = \omega. $$
We see that 
$$ \Htp\omega \in \eta + \Ker \Dd, $$
where $\eta$ is the unique coexact form with $\Dd \eta = \omega$. In other words, $\eta$ is the preimage of $\omega$ under the isomorphism $\Dd: \Im \CoDd \rightarrow \Im \Dd$, and we can write 
\begin{equation}\label{Eq:GOpOnClosed}
\Htp \omega = (\Restr{\Dd}{\Im \CoDd})^{-1} \omega + R_1 \omega,
\end{equation}
where $R_1: \Im \Dd \rightarrow \Ker \Dd$.

Suppose now that $\omega \in \Im\CoDd$. Using \eqref{Eq:GreenOp} and \eqref{Eq:GOpOnClosed}, we obtain
$$ \Dd \Htp \omega = \omega - \Htp \Dd \omega = \underbrace{\omega - (\Restr{\Dd}{\Im \CoDd})^{-1} \Dd \omega}_{=0}- R_1 \Dd \omega, $$
where the two terms cancel because $\omega$ is the unique coexact primitive to $\Dd \omega$. Notice that this equation restricts $R_1$ to $R_1: \Im \Dd \rightarrow \Im \Dd$. Similarly as in the first case, we obtain
$$ \Htp \omega = - (\Restr{\Dd}{\Im\CoDd})^{-1} R_1 \Dd \omega + R_2 \omega, $$
where $R_2: \Im\CoDd \rightarrow \Ker \Dd$.

If $\omega\in \Harm$, then \eqref{Eq:GreenOp} gives
$$ \Dd \Htp \omega = 0, $$
and hence $\Htp \omega = R_3 \omega$ for some $R_3: \Harm \rightarrow \Ker \Dd$.

Therefore, \eqref{Eq:GreenOp} implies \eqref{Eq:GreenOpForm}. The other direction clearly holds as well.

As for the equivalent facts, because $\Ker \Dd \perp \Im \CoDd$ (with respect to the $L^2$-inner product), it is clear from \eqref{Eq:GreenOpForm} that $\Im(\Htp)\subset \Im(\CoDd)$ is equivalent to $R_1 = R_2 = R_3 = 0$. Therefore, if there is a $\Htp$ satisfying \eqref{Eq:GreenOp} and (3), then it is unique; it must be $\HtpStd = -\CoDd\GOp$.  
\end{proof}

We see that the standard Hodge homotopy $\HtpStd$ can be characterized as the unique Hodge homotopy with coexact image.

We would like to use the equation
\begin{equation}\label{Eq:DFGDFG}
\Dd \StdPrpg = \pm \HKer \quad\text{on }(M\times M)\backslash\Diag 
\end{equation}
and characterize $\StdPrpg$ as its unique coexact solution. This is probably not enough and additional assumptions on the asymptotic behavior near $\Diag$ are needed.

Another idea is to lift the equation \eqref{Eq:DFGDFG} to the blow-up and study primitives to $\pi^*\HKer$ on $\Bl_\Diag(M\times M)$. The advantage is that $\Bl_\Diag(M\times M)$ is a compact manifold with boundary, and hence for a given Riemannian metric, we have the Hodge decomposition with boundary conditions.  However, the pull-back metric $g$ along $\pi : \Bl_\Diag(M\times M)\rightarrow M\times M$ is singular at the boundary. We would have to approximate it with a family of metrics~$g_t$ in a correct way, use Hodge theory to find a $\CoDd$-primitive $\eta_t$ to a unique coexact $\Dd$-primitive~$\Prpg_t$ to $\pi^*\HKer$ with certain boundary conditions, and finally prove that $\Prpg_t$ converges to a solution of \eqref{Eq:DFGDFG} on the interior and $\eta_t$ to its $\CoDd$-primitive.

Natural differential operators on $\DR(M\times M)$ do not always pull-back to differential operators on $\DR(\Bl_\Diag(M\times M))$ via the blow-down map. The total differential $\Dd$ does, but the operators $\Id \otimes \Dd_y$ and $\Dd_x \otimes \Id$ do not. We suppose that none of $\CoDd$, $\Id \otimes \CoDd_y$ and $\CoDd_x \otimes \Id$ does. We illustrate the consequences on the following seemingly pathological example.
\begin{Proposition}[Pathological example]\label{Prop:PatEm}
For any $n\in \N$, there exists a smooth form $\eta \in \DR(\Bl_\Diag(\R^n\times \R^n))$ with compact vertical support with respect to the fiber bundle $\tilde{\pi}_2 = \Pr_2 \circ \pi : \Bl_\Diag(\R^n\times \R^n) \rightarrow \R^n$ such that $\Dd_y \eta = 0$ on $(\R^n\times \R^n)\backslash \Diag$ but $\Dd \tilde{\pi}_{2*}\eta \neq 0$.

If $\eta \in \DR(\Bl_\Diag(\R\times \R))$ is as above and satisfies in addition $\tau^* \eta = \pm \eta$, then $\Dd_y \eta = 0$ on $\R\times \R \backslash \Diag$ implies $\Dd \tilde{\pi}_{2*}\eta = 0$.
\end{Proposition}
\begin{proof}
Let us start with $n=1$. Let $f: \R\times \R\backslash\Diag\rightarrow \R$ be a function such that 
\begin{enumerate}
 \item $f: \R\times \R\backslash\Diag\rightarrow \R$ is smooth.
 \item It holds $\frac{\partial f}{\partial y}(x,y) = 0$ for all $(x,y)\in \R\times \R\backslash \Diag$.
 \item For every $y\in \R$, the function $f(\cdot,y)$ defined on $\R\backslash\{y\}$ has compact support in $\R\backslash\{y\}$.
\end{enumerate} 
Clearly, (1) and (2) implies
\begin{equation}\label{Eq:PartialConst}
 f(x,y) = \begin{cases} f^-(x) & \text{for }x<y, \\ f^+(x) & \text{for }x>y, \end{cases}\quad\text{for all }(x,y)\in \R\times \R\backslash \Diag,
\end{equation}
where $f^+$, $f^-: \R \rightarrow \R$ are smooth functions. It is easy to see that (3) implies the existence of $x^+$, $x^-\in \R$ such that 
$$ f^+(x) = 0\quad\text{for all }x>x^+ \quad\text{and}\quad f^-(x) = 0\quad\text{for all }x<x^-. $$
We consider the form
\begin{equation}
\eta(x,y) = f(x,y) \Diff{x}\quad\text{on }\R\times\R\backslash\Diag.
\end{equation}
In general, a form $\eta(x,y)$ on $\R^n\times\R^n\backslash\Diag$ is a restriction of a smooth form on $\Bl_\Diag(\R^n\times \R^n)$ if and only if the form $(\Phi^*\eta)(r,\omega,u)$ for the diffeomorphism $\Phi: (0,\infty) \times \Sph{n-1} \times \R^n \rightarrow \R^n \times \R^n \backslash \Diag$ given by $\Phi(r,\omega,u) = (u+r\omega,u)$ is a restriction of a smooth form on $[0,\infty) \times \Sph{n-1}\times \R^n$. In our case, we have
$$ (\Phi^*\eta)(r,u) = \begin{cases} 
 f^+(u+r)\bigl(\Diff{u}+\Diff{r}\bigr) & \text{on } D^+:= [0,\infty) \times \{1\}\times \R, \\
 f^-(u-r)\bigl(\Diff{u}-\Diff{r}\bigr) & \text{on }D^-:=[0,\infty) \times \{-1\} \times \R, 
\end{cases} $$
where we used the fact that $\Sph{0} = \{\pm 1\}$ and splitted the domain in two connected components. We see that $\Phi^* \eta$ extends smoothly to $[0,\infty) \times \Sph{0}\times \R$ in the obvious way. We denote the extension by $\widetilde{\Phi^* \eta}$ and the induced extension of $\eta$ to $\Bl_\Diag(\R\times \R)$ by $\tilde{\eta}$. It holds $\tilde{\Phi}^* \tilde{\eta} = \widetilde{\Phi^* \eta}$ under the extended diffeomorphism $\tilde{\Phi}: [0,\infty) \times \Sph{0} \times \R \rightarrow \Bl_\Diag(\R\times\R)$. The fiberwise integral along the smooth oriented fiber bundle $\tilde{\pi}_2: \Bl_\Diag(\R\times \R) \rightarrow \R$ transforms under $\tilde{\Phi}$ to the fiberwise integral along $\tilde{p}_3: [0,\infty) \times \Sph{0} \times \R \rightarrow \R$, and we get for all $y=u \in \R$ the following:
$$ (\tilde{\pi}_{2*}\tilde{\eta})(y) = \tilde{p}_{3*}(\widetilde{\Phi^*\eta})(u) = \int_0^\infty \bigl(f^+(u+r) + f^-(u-r) \bigr)\Diff{r}. $$
The algebraic sign of the second term was canceled by the geometric sign coming from different orientations of $[0,\infty)$ in $[0,\infty)\times\{0\}$ and in $[0,\infty)\times \{1\}$. We think of $\tilde{\Phi}$ as of an isomorphism of fiber bundles $\tilde{p}_3$ and $\tilde{\pi}_2$ covering the identity, which explains the notation $y=u$. The fiberwise integration along $\tilde{p}_3: [0,\infty) \times \Sph{0} \times
 \R \rightarrow \R$ reduces to the Lebesgue integration of a smooth function $g$ on $[0,\infty) \times \Sph{\omega} \times \R$ with respect to $(r,\omega)$. Clearly, we can permute any differential operator acting on $u$ with the integral.
 In our case, we use $\Dd$ and obtain
\begin{align*}
(\Dd \tilde{\pi}_{2*}\tilde{\eta})(y) &= (\Dd \tilde{p}_{3*}\widetilde{\Phi^*\eta})(u) \\
 &= \frac{\partial }{\partial u}\Bigl(\int_{0}^\infty \bigl(f^+(u+r)+f^-(u-r)\bigr) \Diff{r}\Bigr) \Diff{u} \\
&= \Bigl(\int_0^\infty (f^+)'(u+r) + (f^-)'(u-r) \Diff{r}\Bigr)\Diff{u} \\
& = \Bigl(\int_u^\infty (f^+)'(z) \Diff{z} + \int_{-\infty}^u (f^-)'(z) \Diff{z}\Bigr) \Diff{u} \\
& = \bigl(f^-(u)- f^+(u)\bigr) \Diff{u}
\end{align*}
This is zero precisely when $f$ extends continuously to $\R \times \R$.

A general form on $\R\times \R\backslash \Diag$ is 
$$ \eta(x,y) = f_0(x,y) + f_1(x,y) \Diff{x} + f_2(x,y) \Diff{y} + f_3(x,y) \Diff{x}\Diff{y}. $$
Imposing the condition $\Dd_y\eta= 0$, we get \eqref{Eq:PartialConst} for both $f_0$ and $f_1$.
Imposing $\tau^* \eta = - \eta$, we get $f_3(x,y) = f_3(y,x)$, $f_2(x,y) = - f_1(y,x)$ and $f_0(x,y) = - f_0(y,x)$.
Now, requiring that $\eta$ has compact vertical support in $x$ implies $f_0 = f_1 \equiv 0$ because of \eqref{Eq:PartialConst}, and hence also $f_2 \equiv 0$.
However, the form $\eta = f_3(x,y) \Diff{x} \Diff{y}$ satisfies $\Dd_y \eta = 0$ and $\Dd\tilde{\pi}_{2*}\eta = 0$ trivially.

As for the higher dimensions, let $n\ge 1$, and consider
\begin{equation*}
\eta(x,y) := \begin{multlined}[t]\frac{\lambda(x)}{\Abs{x-y}^n} \sum_{i=1}^n (-1)^{i+1}(x^i - y^i) \Diff{x}^1\dotsb\Diff{x}^n\Diff{y}^1\dotsb \widehat{\Diff{y}^i}\dotsb\Diff{y}^n\end{multlined}
\end{equation*}
for $(x,y)\in \R^n\times\R^n\backslash\Diag$, where $\lambda: \R^n \rightarrow \R$ is a smooth bump function. Then $\eta$ extends smoothly to $\Bl_\Diag(\R^n\times \R^n)$, it holds $\Dd_y \eta(x,y) = 0$, but 
$$ \Dd \int_x \eta(x,y) = \Vol(\Sph{n}) \lambda(y) \Diff{y}^1 \dotsb \Diff{y}^n. $$
This finishes the proof.
\end{proof}

\section{Approximation using heat form}\label{Sec:Hwe}
\allowdisplaybreaks

Given an oriented Riemannian manifold $(M,g)$, consider the \emph{heat form}
\begin{equation}\label{Eq:HK}
\KKer_t(x,y) = \sum (-1)^{kn}e^{-\lambda_i t} (\star e_i)(x) \wedge e_i(y),
\end{equation}
where $(e_i)$ are eigenvectors of $\Delta$ and $\lambda_i$ the corresponding eigenvalues. It is equivalently the Schwartz form of the operator $\exp(-\Delta t)$ (see \cite[Chapter~3]{Harris2004}) or a unique solution of the equation $\Laplace \KKer_t(x,y) = - \frac{\partial}{\partial t}\KKer_t(x,y)$ with $\lim_{t\to 0} \KKer_t = \DR_\Id$, where $\DR_\Id$ is the Schwartz form of the identity (see \cite{Hein2006}).
\begin{Proposition}[Properties of the heat kernel]\label{Prop:Heasd}
Let $M$ be an oriented Riemannian manifold. The heat form $\KKer_t(x,y)$ is smooth on $M\times M \times (0,\infty)$ and satisfies
$$ \Dd \KKer_t = 0, \quad \tau^*\KKer_t = (-1)^n\KKer_t\quad\text{and}\quad \frac{1}{2}\Delta \KKer_t = \Diag_x \KKer_t = - \frac{\partial }{\partial t} \KKer_t. $$
\end{Proposition}
\begin{proof}
Straightforward computation and a nice combinatorial argument.
\end{proof}
\begin{Proposition}[Approximation using heat form]\label{Prop:HeatKerFormulas}
Let $M$ be an oriented Riemannian manifold and $\KKer_t(x,y)$ the heat form.
For all $(t,x,y)\in \bigl([0,\infty)\times M \times M\bigr)\backslash \{0\}\times \Diag =: D(\KKer)$, define
\begin{equation}\label{Eq:HeatKerApprox}
\begin{aligned} 
\GKer_t(x,y) &\coloneqq \int_t^\infty \KKer_\tau(x,y)\Diff{\tau}\quad\text{and}\\
\Prpg_t(x,y) &\coloneqq (-1)^{n+1}\int_t^\infty (\Id\COtimes\CoDd_y) \KKer_\tau(x,y)\Diff{\tau}.
\end{aligned}
\end{equation}
Then:
\begin{ClaimList}
\item The forms $\GKer_t$ and $\Prpg_t$ are smooth on $D(\KKer)$, the point-wise limits $\GKer'$ and $\Prpg'$ as $t\to 0$ exist, and it holds $\GKer_t \darrow[t]\GKer'$ and $\Prpg_t\darrow[t]\Prpg'$ as $t\to 0$ (uniform convergence) in $C^\infty_{\text{loc}}(M\times M\backslash\Diag)$.
\item On $D(\KKer)$, the following relations hold:
\begin{align*}
\Dd \GKer_t &= 0 & \Prpg_t &= (-1)^{n+1}\frac{1}{2} \CoDd\GKer_t \\
\Laplace \GKer_t &= \KKer_t - \HKer & \Dd \Prpg_t &= (-1)^n(\HKer - \KKer_t) \\
\tau^*\GKer_t &=(-1)^n\GKer_t  & \tau^* \Prpg_t &= (-1)^n \Prpg_t.
\end{align*}
It follows that $\GKer' = \GKer$ is the (Laplace) Green form and $\Prpg' = \StdPrpg$ the standard Hodge propagator. 
\end{ClaimList}
\end{Proposition}

\begin{proof}
The formal computation is clear. An honest proof uses the standard heat kernel estimates.
\ToDo[caption={Say more},noline]{Say more about the proofs!}
\end{proof}

\begin{Proposition}[$\StdPrpg$ is codifferential of $\GKer$]\label{Prop:StdCodifInt}
Let $M$ be an oriented Riemannian manifold, and let $\GKer\in \DR^n(M\times M\backslash\Diag)$ be the Green form. Then the standard Hodge propagator $\StdPrpg$ satisfies
\begin{equation}\label{Eq:FormForPUsingG}
\StdPrpg(x,y)= (-1)^{n+1}(\Id\otimes \CoDd_y)\GKer(x,y),
\end{equation}
where $\Id\otimes \CoDd_y: \DR^\bullet(M\times M) \rightarrow \DR^{\bullet-1}(M \times M)$ is the differential operator defined in local coordinates by commuting $\CoDd$ over the first factor with the Koszul sign and applying it to the second factor.
\end{Proposition}
\begin{proof}
As for the signs, $(-1)^n$ comes from $\TOp \GOp \omega(y) = (-1)^{nT}\int_x(\Id \otimes \TOp_y)\GKer(x,y)\omega(x)$ with $T=\CoDd$ and $-1$ from $\StdHtp = - \CoDd\GOp$.
The rest can be proven using the heat kernel approximation and standard heat kernel estimates.
There is an other method using the asymptotic expansion of $\GKer$, which was shown to the author by Prof.~Dr.~Christian~Bär.
\ToDo[caption={Say more},noline]{Say more about the proofs!}
\end{proof}

\section{Standard Hodge propagator for Euclidean space}\label{Sec:HeatRN}
\allowdisplaybreaks

We will use Propositions~\ref{Prop:StdCodifInt} and~\ref{Prop:HeatKerFormulas} to prove in two ways that $\StdPrpg$ for $\R^n$ extends smoothly to the blow-up. We will start with the following well known formulas.
\begin{Proposition}[Green form and heat kernel for $\R^n$]\label{Prop:GreenKernelRn}
The Green form for $\R^n$, which can be equivalently characterized as the unique solution $\GKer\in \DR^n(\R^n\times\R^n\backslash\Diag)$ of
\begin{equation}\label{Eq:GrRn}
\Laplace_y \GKer(x,y) = \delta(x-y)(\Diff{x}^1-\Diff{y}^1)\dotsb(\Diff{x}^n-\Diff{y}^n),
\end{equation}
where $\delta(x-y)$ is the Dirac delta function, satisfies
$$ \GKer(x,y) = \begin{cases}
\displaystyle\frac{1}{(n-2) \Vol(\Sph{n-1})} \frac{1}{\Abs{x-y}^{n-2}}(\Diff{x}^1 - \Diff{y}^1) \dotsb (\Diff{x}^n - \Diff{y}^n) & \text{for }n\ge 3, \\[1em]
-\dfrac{1}{2\pi}\ln \Abs{x-y}(\Diff{x}^1 - \Diff{y}^1)(\Diff{x}^2 - \Diff{y}^2)& \text{for }n=2.
\end{cases} $$
The heat kernel for $\R^n$, i.e., the solution of $\Laplace_y \QKer_t(x,y) = -\frac{\partial}{\partial t}\QKer_t(x,y)$,  is given by
\begin{equation}\label{Eq:EuclidHeatKer}
\QKer_t(x,y) = (4\pi t)^{-\frac{n}{2}}\exp\Bigl(-\frac{\Abs{x-y}^2}{4t}\Bigr) (\Diff{x}^1 - \Diff{y}^1)\ldots (\Diff{x}^n-\Diff{y}^n).
\end{equation}
\end{Proposition}
\begin{proof}
See \cite{BGV}.
\end{proof}

\begin{Proposition}[Standard Hodge propagator for $\R^n$]\label{Prop:StdHodgePropRn}
The standard Hodge propagator for $\R^n$ with $n\ge 2$ satisfies
$$\StdPrpg(x,y) = \frac{(-1)^{n+1}}{\Vol(\Sph{n-1})} \sum_{i=1}^n (-1)^{i-1} \frac{x^i-y^i}{\Abs{x-y}^n}(\Diff{x}^1 - \Diff{y}^1)\dotsb\reallywidehat{(\Diff{x}^i - \Diff{y}^i)}\dotsb(\Diff{x}^n-\Diff{y}^n) $$
and extends smoothly to $\Bl_\Diag(\R^n\times \R^n)$. 
\end{Proposition}
\begin{proof}[Proof using the Green form]
Consider the formulas from Proposition~\ref{Prop:GreenKernelRn}. Recall the formula for the codifferential
\begin{equation}\label{Eq:CodiffRn}
\CoDd f(x)\Diff{x}^I = - \sum_{i\in I} \varepsilon(i,I) \frac{\partial f(x)}{\partial x^i} \Diff{x}^{I\backslash\{i\}}.
\end{equation}
For $n\ge 3$, we compute
\begin{align*}
&(\Id\otimes\CoDd_y)\GKer(x,y) \\
&\quad=\frac{1}{(n-2)\Vol(\Sph{n-1})} \sum_I (-1)^{n\Abs{I}} (\Id \otimes \CoDd_y)\Bigl(\frac{1}{\Abs{x-y}^{n-2}}(\Star\Diff{x}^I)\wedge\Diff{y}^I\Bigr) \\
&\quad=\frac{1}{\Vol(\Sph{n-1})} \sum_{I} (-1)^{n\Abs{I} + \Abs{I} + n+1} \sum_{i\in I}\frac{x^i-y^i}{\Abs{x-y}^n}\varepsilon(i,I)(\Star\Diff{x}^{I})\wedge\Diff{y}^{I\backslash\{i\}}\\
&\quad=\frac{1}{\Vol(\Sph{n-1})} \sum_{I} (-1)^{n\Abs{I} + \Abs{I} + n+1} \sum_{i\in I}\frac{x^i-y^i}{\Abs{x-y}^n}\varepsilon(i,I)\varepsilon(I,I^c) \Diff{x}^{I^c}\wedge\Diff{y}^{I\backslash\{i\}}\\
&\quad=\begin{multlined}[t]\frac{1}{\Vol(\Sph{n-1})} \sum_{i=1}^n \frac{x^i-y^i}{\Abs{x-y}^n} \sum_{\substack{J\subset \{1,\dotsc,\hat{i},\dotsc,n\}}}(-1)^{n\Abs{J} + \Abs{J}}\varepsilon(i,J)\varepsilon(J\cup\{i\},J^c\backslash\{i\})\\ \Diff{x}^{J^c\backslash\{i\}} \wedge \Diff{y}^{J}\end{multlined}\\
&\quad=\frac{1}{\Vol(\Sph{n-1})} \sum_{i=1}^n (-1)^{i-1} \frac{x^i-y^i}{\Abs{x-y}^n}(\Diff{x}^1 - \Diff{y}^1)\dotsb\reallywidehat{(\Diff{x}^i - \Diff{y}^i)}\dotsb(\Diff{x}^n-\Diff{y}^n).
\end{align*}
In the last step, we used that
$$ \varepsilon(i,J)\varepsilon(J\cup\{i\},J^c\backslash\{i\}) = (-1)^{i-1}\varepsilon(J,J^c\backslash\{i\}) $$
and
\begin{align*}
&\Diff{x}^{J^c\backslash\{i\}} \wedge \Diff{y}^{J}\\
& = (-1)^{\Abs{J}} \varepsilon(J^c\backslash\{i\},J)[(\Diff{x}^1-\Diff{y}^1)\dotsb\reallywidehat{(\Diff{x}^i-\Diff{y}^i)}\dotsb (\Diff{x}^n-\Diff{y}^n)]_{J^{c}\backslash\{i\}, J}\\
& = (-1)^{n \Abs{J} + \Abs{J}}[(\Diff{x}^1-\Diff{y}^1)\dotsb\reallywidehat{(\Diff{x}^i-\Diff{y}^i)}\dotsb (\Diff{x}^n-\Diff{y}^n)]_{J^{c}\backslash\{i\}, J},
\end{align*}
where $[\cdot]_{I_1,I_2}$ denotes the part of the product which picks the first variable at positions $I_1$ and the second at positions $I_2$. The computation for $n=2$ gives the same result, and the formula for $\StdPrpg$ is justified by Proposition~\ref{Prop:StdCodifInt}.

We will now study whether $\StdPrpg$ extends smoothly to the blow-up. Consider the polar coordinates in one variable
\begin{align*}
\phi: [0,\infty) \times \Sph{n} \times \R^n  &\longrightarrow \R^n \times \R^n \\
(r,\omega,u) &\longmapsto (u+r\omega, u).
\end{align*}
There is a unique smooth map $\tilde{\phi}$ which fits in the blow-up diagram
$$\begin{tikzcd}
 & \Bl_\Diag(\R^n\times \R^n)\arrow{d}{\pi}\\{}
 [0,\infty)\times\Sph{n}\times\R^n\arrow{r}{\phi}\arrow{ru}{\tilde{\phi}}& \R^n\times \R^n,
\end{tikzcd}$$
and it is a diffeomorphism of manifolds with boundary. We denote by
$$ \phi_0 \coloneqq \Restr{\phi}{(0,\infty) \times \Sph{n}\times\R^n} : (0,\infty)\times\Sph{n}\times\R^n\longrightarrow \R^n \times \R^n\backslash\Diag $$
the restriction of $\phi$ to the interior. The commutative diagram
$$\begin{tikzcd}{}
[0,\infty)\times\Sph{n}\times\R^n\arrow{r}{\tilde{\phi}}& \Bl_\Diag(\R^n\times\R^n) \\
(0,\infty)\times\Sph{n}\times\R^n\arrow{r}{\phi_0}\arrow[hook]{u} & \R^n\times\R^n\backslash\Diag \arrow[hook]{u}{\tilde{\iota}}
\end{tikzcd}$$
shows that we can equivalently study whether the form 
$$ \StdPrpgPol\coloneqq \phi_0^* \StdPrpg $$
admits a smooth extension $\widetilde{\StdPrpgPol}$ to $[0,\infty)\times\Sph{n-1}\times\R^n$. We have
\begin{align*}
\StdPrpgPol(r,\omega,u) &= \phi_0^*\bigl(\CoDd_y \GKer)(r,\omega,u) \\
& = \begin{multlined}[t] 
\frac{(-1)^{n+1}}{r^{n-1}\Vol(\Sph{n-1})}\sum_{i=1}^n (-1)^{i-1} \omega^i (r\Diff{\omega}^1 + \omega^1 \Diff{r}) \dotsb \reallywidehat{(r\Diff{\omega}^i + \omega^i \Diff{r})} \dotsb \\ (r\Diff{\omega}^n + \omega^n \Diff{r})
\end{multlined}\\
 & = \begin{aligned}[t]
 \frac{(-1)^{n+1}}{r^{n-1}\Vol(\Sph{n-1})} \Bigl(r^{n-1}&\sum_{1\le i \le n}(-1)^{i-1}\omega^i\Diff{\omega}^1\dotsb\widehat{\Diff{\omega}^i}\dotsb\Diff{\omega}^n    \\
  - r^{n-2}&\sum_{1\le i < j \le n}(-1)^{i+j} \omega^i \omega^j \Diff{r}\Diff{\omega}^1 \dotsb \widehat{\Diff{\omega}^i} \dotsb \widehat{\Diff{\omega}^j} \dotsb\Diff{\omega}^n \\
  + r^{n-2}&\sum_{1 \le j < i \le n}(-1)^{i+j} \omega^j\omega^i \Diff{r} \Diff{\omega}^1 \dotsb \widehat{\Diff{\omega}^j} \dotsb \widehat{\Diff{\omega}^i} \dotsb\Diff{\omega}^n\Bigr)
  \end{aligned} \\
 & = (-1)^{n+1}\frac{\Vol_{\Sph{n-1}}(\omega)}{\Vol(\Sph{n-1})},
\end{align*} 
where the last two sums canceled. The result extends smoothly beyond $r=0$ because it does not depend on $r$ at all.
\end{proof}
\begin{proof}[Proof using the heat kernel]
Consider the polar coordinates with respect to the diagonal
$$ \begin{aligned}
   \varphi: [0,\infty)\times \Sph{n-1} \times \R^n &\longrightarrow \R^n \times \R^n \backslash \Diag \\ 
 (r,\omega,u) &\longmapsto (x,y) = (u+r\omega, u-r\omega).
   \end{aligned} $$
There is a unique smooth map $\tilde{\varphi}$ which fits in the blow-up diagram
$$\begin{tikzcd}
 & \Bl_\Diag(\R^n\times \R^n)\arrow{d}{\pi}\\{}
 [0,\infty)\times\Sph{n}\times\R^n\arrow{r}{\varphi}\arrow{ru}{\tilde{\varphi}}& \R^n\times \R^n,
\end{tikzcd}$$
and it is a diffeomorphism of manifolds with boundary. The heat kernel \eqref{Eq:EuclidHeatKer} transforms under $\varphi$ as
\begin{align*}
\QKerPol_t(r,\omega,u) &\coloneqq (\varphi^* \QKer_t)(r,\omega,u) \\
 & = (4 \pi t)^{-\frac{n}{2}} \exp(-\frac{r^2}{t}) 2^n r^{n-1} \Diff{r} \Vol(\omega) \\
 &= (\pi t)^{-\frac{n}{2}} \exp(-\frac{r^2}{t}) r^{n-1} \Diff{r} \Vol(\omega).
\end{align*}
It follows from Proposition~\ref{Prop:HeatKerFormulas} that for $(r,\omega,u)\in (0,\infty) \times \Sph{n-1}\times \R^n$ we have
\begin{equation}\label{Eq:StdPrpgPol}
\StdPrpgPol(r,\omega,u) = (-1)^{n+1}\frac{1}{2}\lim_{t\to 0} \int_t^\infty \CoDdPol \QKerPol_\tau(r,\omega,u) \Diff{\tau},
\end{equation}
where $\CoDdPol$ is the codifferential on $\R^n\times \R^n$ computed with respect to the pullback of the product Riemannian metric. If $g: \R^n \otimes \R^n\rightarrow \R$ is an inner product, then the pullback Riemannian metric satisfies
\begin{equation}\label{Eq:PullbackMetric}
g^{\mathrm{pol}}\coloneqq \varphi^*(g\oplus g)(r,\omega,u) = 2\begin{pmatrix}
g(\omega,\omega) & 0 & 0 \\
0 & r^2 g & 0 \\
0 & 0 & g
\end{pmatrix};
\end{equation}
it is degenerate at the boundary $r=0$. Next, it is easy to check that the conformal transformation $g\mapsto\lambda g$ on vectors induces the conformal transformation $g\mapsto\frac{1}{\lambda^k}g$ on $k$-forms. Thus, if $\Norm{\cdot}^\mathrm{pol}$ denotes the point-wise norm with respect to $g^{\mathrm{pol}}$ for the standard Euclidean metric, then
$$ \Norm{\Diff{r}}^{\mathrm{pol}} = 2^{-\frac{1}{2}},\quad\Norm{\Vol(u)}^{\mathrm{pol}} = 2^{-\frac{n}{2}},\quad\Norm{\Vol(\omega)}^{\mathrm{pol}} = 2^{-\frac{n-1}{2}} r^{-(n-1)}, $$
and hence
$$ \VolPol(r,\omega,u) = 2^{n}r^{n-1}\Diff{r}\Vol(\omega)\Vol(u). $$
In order to get the Hodge star $\StarPol$, we compute
\begin{align*}
\Diff{r}\Vol(\omega)\wedge \StarPol \bigl(\Diff{r}\Vol(\omega)\bigr) & \overset{!}{=} {\Norm{\Diff{r}\Vol(\omega)}^{\mathrm{pol}}}^2 \Vol^{\mathrm{pol}}(r,\omega,u) \\
&=r^{-(n-1)} \Diff{r}\Vol(\omega)\Vol(u), \\
\Diff{r}\Vol(u) \wedge \StarPol(\Diff{r}\Vol(u)) &\overset{!}{=} {\Norm{\Diff{r} \Vol(u)}^{\mathrm{pol}}}^2\Vol^{\mathrm{pol}}(r,\omega,u) \\ 
&=\frac{1}{2} r^{n-1} \Diff{r} \Vol(\omega) \Vol(u).
\end{align*}
Using the product structure \eqref{Eq:PullbackMetric}, it follows that
\begin{align*}
\StarPol\bigl(\Diff{r}\Vol(\omega)\bigr)&=r^{-(n-1)} \Vol(u)\quad\text{and}\\
\StarPol(\Diff{r}\Vol(u))&= \frac{1}{2} r^{n-1} \Vol(\omega).
\end{align*}
Recalling the definition of the codifferential $\CoDd \alpha = (-1)^{n(k-1)+1} \Star\Dd\Star \alpha$, we compute
\begin{align*}
(\StarPol \Dd \StarPol)\bigl(\QKerPol_t(r,\omega,u)\bigr) &=  \StarPol \Dd \Bigl( (\pi t)^{-\frac{n}{2}} \exp(-\frac{r^2}{t}) \Vol(u) \Bigr)  \\ 
& = \StarPol\Bigl(\frac{-2r}{t}(\pi t)^{-\frac{n}{2}}\exp(-\frac{r^2}{t}) \Diff{r} \Vol(u)\Bigr)\\
&= -\frac{1}{t}(\pi t)^{-\frac{n}{2}} \exp(-\frac{r^2}{t}) r^n \Vol(\omega),
\end{align*}
and because the product dimension is even, we obtain
$$ \CoDdPol \QKerPol_t(r,\omega,u) = \frac{1}{t}(\pi t)^{-\frac{n}{2}} \exp(-\frac{r^2}{t}) r^n \Vol(\omega). $$
In order to integrate this according to \eqref{Eq:StdPrpgPol}, we will make use of the $\Gamma$-function $\Gamma(z) = \int_0^\infty x^{z-1} e^{-x} \Diff{x}$. Then, we have
\begin{align*}
(-1)^{n+1}\StdPrpgPol(r,\omega,u)&=\frac{1}{2}\lim_{t\to 0} \int_t^\infty \CoDdPol \QKerPol_\tau(r,\omega,u)\Diff{\tau}\\
&=\frac{1}{2}\pi^{-\frac{n}{2}}r^n \Bigl(\lim_{t\to 0}\int_{t}^\infty \tau^{-\frac{n}{2} - 1} \exp(-\frac{r^2}{\tau})\Diff{\tau}\Bigr)\Vol(\omega)\\
&=\frac{1}{2}\pi^{-\frac{n}{2}}r^n \Bigl(\lim_{t\to 0}\int_{\frac{r^2}{t}}^0 \Bigl(\frac{r^2}{z}\Bigr)^{-\frac{n}{2} - 1}\exp(-z)\bigl(-z^{-2}r^2\Diff{z}\bigr)\Bigr)\Vol(\omega)\\
&= \frac{1}{2}\pi^{-\frac{n}{2}}\Bigl(\lim_{t\to 0} \int_0^{\frac{r^2}{t}} z^{\frac{n}{2}-1} \exp(-z) \Diff{z}\Bigr)\Vol(\omega) \\
&= \frac{\Gamma(\frac{n}{2})}{2\pi^{\frac{n}{2}}}\Vol(\omega) \\
& = \frac{\Vol(\omega)}{\Vol(\Sph{n-1})},
\end{align*}
which recovers the formula in the other proof of Proposition~\ref{Prop:StdHodgePropRn}.
\end{proof}

\begin{Example}[The case $n=1$]\label{Ex:SDFSDF}
It is easy to check directly that
$$ \Prpg(x,y) \coloneqq \theta(x-y), $$
where $\theta$ is the Heavyside step function, is the Schwartz form of a Hodge homotopy $\Htp$. Indeed, for any smooth $f: \R\rightarrow\R$ with compact support, we have 
$$ \Htp(f \Diff{x})(y) = \int_x \Prpg(x,y)f(x)\Diff{x} = \int_{y}^{\infty} f(x)\Diff{x}, $$
and hence $\Dd\circ\Htp = -\Id$. It is also easy to check that the following is the Laplace Green form:
$$ \GKer(x,y) = - \frac{1}{2}\Abs{x-y}(\Diff{x}-\Diff{y}). $$
Indeed, using $\Sgn(x) = 2\theta(x) - 1$, we get
$$ \frac{\partial^2}{\partial y^2} \frac{1}{2}\Abs{x-y} = \frac{\partial}{\partial y}\frac{1}{2}\Sgn(y-x) = \delta(y-x) = \delta(x-y), $$
which shows \eqref{Eq:GrRn}. We carefully compute
\begin{align*}
(\Id\otimes \CoDd_y)\GKer(x,y) & = \CoDd_y \Bigl(\frac{1}{2}\Abs{x-y}\Diff{y}\Bigr) \\
& = -\frac{1}{2} \frac{\partial}{\partial y}\Abs{x-y} \\
& = -\frac{1}{2}(2\theta(y-x)-1) \\ 
& = - \theta(y-x) + \frac{1}{2} \\
& = \theta(x-y) - \frac{1}{2}
\end{align*}
and apply Proposition~\ref{Prop:StdCodifInt} to get
$$ \StdPrpg(x,y) = \theta(x-y) - \frac{1}{2}. $$
In particular, we see that the sign agrees with the direct computation of the Hodge propagator above. Smoothness on the blow-up is for $n=1$ equivalent to smoothness on the closures of connected components of $\R\times\R\backslash\Diag$; this is readily satisfied. 
\end{Example}

\begin{Question}
	Is it possible to use the asymptotic heat kernel expansion and formula~\eqref{Eq:PrpgUsingHeatKernel} or the asymptotic Green kernel expansion and formula \eqref{Eq:FormForPUsingG} to infer from the case of $\R^n$ with the standard Euclidean metric $g_0$ that $\StdPrpg$ extends smoothly to the blow-up for flat manifolds, i.e., locally isometric to $(\R^n,g_0)$?  
\end{Question}

\section{Standard Hodge propagator for 1- and 2-sphere}\label{Sec:GrSpgh}
\allowdisplaybreaks

In this section, we denote the Hodge propagator for $\Sph{n}$ constructed in Part~I by $\ArtPrpg$. We would like to use $\ArtPrpg$ to study the standard Hodge propagator $\StdPrpg$ for $\Sph{n}$.

For $\Sph{1}$, we can compute $\KKer_t$ and hence $\StdPrpg$ explicitly. 

\begin{Example}[$\StdPrpg$ for $\Sph{1}$]\label{Ex:SADQQ}
We write $\Sph{1} = \R/2\pi\Z$ and use the coordinate $x\in [0,2\pi)$. Because $\Sph{1}$ is flat, we have
$$ \Laplace = - \frac{\partial^2}{\partial x^2}. $$
Solving the eigenvalue problem $\Laplace \omega = \lambda \omega$ for $\omega\in \DR(\Sph{1})$ and $\lambda\in \R$, we get $\lambda\in \{0, n^2 \mid n\in\N\}$ and the corresponding eigenvectors
\begin{align*}
\Bigl\{\frac{1}{\sqrt{2 \pi}}, \frac{1}{\sqrt{\pi}} \cos(nx), \frac{1}{\sqrt{\pi}} \cos(nx) \Diff{x}, \frac{1}{\sqrt{\pi}} \sin(nx), \frac{1}{\sqrt{\pi}} \sin(nx) \Diff{x} \mid n\in \N\Bigr\},
\end{align*}
which we normalized in the $L^2$-norm. Plugging in \eqref{Eq:HK}, we get
\begin{align*}
\KKer_t(x_1,x_2) & = \frac{1}{2\pi} + \frac{1}{\pi}\sum_{n=1}^\infty e^{-n^2 t}\bigl(\cos(nx_1) \cos(nx_2) + \sin(nx_1)\sin(nx_2) \bigr) (\Diff{x_1}-\Diff{x_2}) \\
& = \frac{1}{2\pi} + \frac{1}{\pi}\sum_{n=1}^\infty e^{-n^2 t}\cos(nx_1 - nx_2)(\Diff{x_1}-\Diff{x_2}).
\end{align*}
Applying the product codifferential, we get
\begin{align*}
\CoDd \KKer_t(x_1,x_2) &= \frac{1}{\pi}\sum_{n=1}^\infty e^{-n^2 t}\Bigl[- \frac{\partial}{\partial x_1}\cos(nx_1-nx_2)+\frac{\partial}{\partial x_2}\cos(nx_1-nx_2)\Bigr] \\
& = \frac{2}{\pi} \sum_{n=1}^\infty e^{-n^2 t} n \sin(n x_1 - n x_2). 
\end{align*}
Finally, the integration gives
\begin{align*}
\StdPrpg(x_1,x_2) &= \frac{1}{2}\int_0^\infty \CoDd\KKer_t(x_1,x_2) \Diff{t} \\
&=\frac{1}{\pi}\sum_{n=1}^\infty \Bigl(\int_0^\infty e^{-n^2 t} \Diff{t}\Bigr)n\sin(nx_1 - nx_2) \\
&= \frac{1}{\pi}\sum_{n=1}^\infty\frac{\sin\bigl(n(x_1-x_2)\bigr)}{n} \\
& = \frac{1}{2\pi} \begin{cases}
\pi - (x_1 - x_2) & x_1 > x_2, \\
-\pi - (x_1-x_2) & x_1< x_2.
\end{cases} \\
& = \frac{1}{2\pi}\bigl(\alpha(x_1,x_2) - \pi\bigr).
\end{align*}
This is precisely $\ArtPrpg$ from Part~I.
\end{Example}

For $n\ge 2$, an explicit formula for any of $\StdPrpg$, $\LapGKer$ or $\KKer_t$ seems to be unknown. For $\Sph{2}$, the formula for $\StdPrpg$ on functions was derived by Dr.~A.~Hermann.

Our idea to study $\StdPrpg$ via $\ArtPrpg$ is to examine the uniqueness of Hodge propagators. Consider the Schwartz form $\HKer(x,y) = \frac{1}{V}(\Vol(x) + (-1)^n\Vol(y))$ of the harmonic projection for $\Sph{n}$. Let $C_2(\Sph{n})\coloneqq\Sph{n}\times\Sph{n}\backslash\Diag$ denote the configuration space, and let
\begin{equation}\label{Eq:DifEq}
\Soln_n\coloneqq\{\Prpg\in\DR^{n-1}(C_2(\Sph{n}))\mid\Dd\Prpg=(-1)^n\HKer\}
\end{equation}
be the space of primitives to $(-1)^n\HKer$. We know that $\StdPrpg\in \Soln_n$. The following holds.

\begin{Proposition}[The space of primitives to $\HKer$ for $\Sph{n}$]\label{Prop:SpaceOfSolnSn}
Let
$$ V_n\coloneqq \begin{cases}
\DR^{n-2}(C_2(\Sph{n}))/\Dd \DR^{n-3}(C_2(\Sph{n})) & \text{for } n\ge 3, \\
\DR^{0}(C_2(\Sph{n}))/\R & \text{for }n=2,
\end{cases}$$
where $\R\subset\DR^0(C_2(\Sph{n}))$ denotes the constants. The action $\rho: V_n \times \Soln_n \rightarrow \Soln_n$, $(\lambda,\Prpg)\mapsto \Prpg + \Dd\lambda$ of the additive group $V_n$ on $\Soln_n$ defines the structure of an affine space on $\Soln_n$ for $n\ge 2$. If we require $SO(n+1)$ or $(-1)^n\tau^*$-invariance, then the same holds with $\DR$ replaced by the correspondingly invariant forms.
\end{Proposition}
\begin{proof}
We have to check that the action $\rho$ is free and transitive. For $\Prpg_1$, $\Prpg_2\in \Soln_n$, the difference $\eta \coloneqq \Prpg_1 - \Prpg_2$ is a closed $(n-1)$-form; it is exact because $C_2(\Sph{n})$ is homotopy equivalent to $\Sph{n}$. A primitive $\lambda_1$ is an $n-2$ form. If $\lambda_2$ is another primitive, then $\lambda_1 - \lambda_2$ is closed, and hence it is a constant for $n=2$ and an exact form for $n \ge 3$. Therefore,~$\Soln_n$ is an affine space over $V_n$. As for the invariance, we can average a primitive of an invariant form over $SO(n+1)$ or take $\frac{1}{2}(\Id + (-1)^n \tau^*)$.
\end{proof}

Note that $\Soln_1 \simeq \R$ by adding the constant and that all functions on $C_2(\Sph{1})$ are coexact. Therefore, $\StdPrpg$ for $\Sph{1}$ can not be characterized as a unique coexact solution of the differential equation for the Hodge propagator.

\begin{Proposition}[Coexactness of artificial Hodge propagator]\label{Prop:ArtProsCoexact}
The Hodge propagator $\ArtPrpg\in\DR^{n-1}(\Sph{n}\times\Sph{n}\backslash\Diag)$ constructed in Part~I is coexact for every $n\in \N$.
\end{Proposition}
\begin{proof}
First of all, we rewrite
\begin{align*}
\omega_k(x,y) &= \frac{1}{k!(n-1-k)!}\sum_{\sigma\in \Perm_{n+1}} x^{\sigma_1} y^{\sigma_1} \Diff{x}^{\sigma_3} \dotsb\Diff{x}^{\sigma_{2+k}} \Diff{y}^{\sigma_{3+k}}\dotsb \Diff{y}^{\sigma_{n+1}} \\
& = (-1)^k \sum_{\substack{I\subset \{1,\dotsc,n+1\} \\ \Abs{I} = k + 1}} \iota_x(\Diff{x}^I) \wedge \underbrace{\iota_y \Star^{\R^{n+1}}}_{\Star^{\Sph{n}}}(\Diff{y}^I).
\end{align*}
Recall the formulas $\CoDd \alpha = (-1)^{d(k-1)+1}\Star \Dd \Star \alpha$ and $\Star \Star \alpha= (-1)^{k(n-k)}\alpha$ for $\alpha\in \DR^k(M)$, where $d=\dim(M)$. For all $(x,y)\in \Sph{n}\times\Sph{n}\backslash\Diag$, we compute 
\begin{align*}
 \CoDd_y \ArtPrpg(x,y) &= \CoDd_y\Bigl(\sum_{k=0}^{n-1} (-1)^k g_k(x\cdot y) \sum_{\substack{I\subset\{1,\dotsc,n+1\}\\\Abs{I}=k+1}} (\iota_x \Diff{x}^I)\wedge \Star^{\Sph{n}}(\Diff{y}^I) \Bigr) \\
 & =\sum_{k=0}^{n-1}\sum_{\substack{I\subset\{1,\dotsc,n+1\}\\\Abs{I}=k+1}} (\iota_x \Diff{x}^I)\wedge\CoDd_y\bigl(g_k(x\cdot y) \Star^{\Sph{n}}(\Diff{y}^I)\bigr) \\
 & \underset{\mathclap{\qquad\ \; \qquad\subalign{& \Big\uparrow\rule{0pt}{5.5ex} \\ \CoDd_y &= (-1)^{n(n-k)+1}\Star^{\Sph{n}}\Dd \Star^{\Sph{n}}\\
\Star^{\Sph{n}} \Star^{\Sph{n}} &= (-1)^{(k+1)(n-k-1)} \Id\\
\text{tot.~sign} &= (-1)^k}}}{=} \sum_{k=0}^{n-1} (-1)^k \sum_{\substack{I\subset\{1,\dotsc,n+1\}\\\Abs{I}=k+1}} (\iota_x \Diff{x}^I) \wedge \Star^{\Sph{n}} \Dd_y\bigl(g_k(x\cdot y) \Diff{y}^I\bigr) \\
& = \sum_{k=0}^{n-1}(-1)^k g_k'(x\cdot y) \sum_{\substack{I \subset \{1,\dotsc,n+1\}\\\Abs{I}=k+1\\}}\underbrace{\begin{multlined}[t] \sum_{i\in I} \sum_{j\in \{1,\dotsc,n+1\}\backslash I} \varepsilon(i,I)\varepsilon(j,I) x^i x^j \\ \Diff{x}^{I\backslash\{i\}}\wedge\Star^{\Sph{n}}(\Diff{y}^{I\cup \{j\}}) \end{multlined}}_{=0} \\
& = 0.
\end{align*}
The cancellation occurs because the summand $(I, i, j)$ contains the same terms as the summand $(I'=I\backslash\{i\}\cup j, j, i)$, and the signs satisfy
$$ \varepsilon(j,I')\varepsilon(i,I') = - \varepsilon(i,I)\varepsilon(j,I). $$
We have
$$ \H_{n-1}(\DR(\Sph{n}\times\Sph{n}\backslash\Diag),\CoDd) \simeq \HDR^{n+1}(\Sph{n}\times\Sph{n}\backslash\Diag) = 0, $$
and hence any coclosed $(n-1)$-form is coexact.
\end{proof}

\begin{Proposition}[Smooth extension to the blow-up for $\Sph{2}$]\label{Prop:StdS2}
The standard Hodge propagator for $\Sph{2}$ extends smoothly to the blow-up.
\end{Proposition}
\begin{proof}
Let $\Prpg_1$, $\Prpg_2 \in \Soln_2$ be two $\SO(3)$-symmetric solutions. Proposition~\ref{Prop:SpaceOfSolnSn} asserts that there is a smooth $\SO(3)$-symmetric function $\lambda: C_2(\Sph{2})\rightarrow \R$ such that $\Prpg_1 - \Prpg_2 = \Dd \lambda$. Because $\SO(3)$ acts on $\Sph{2}$ transitively, there is a smooth function $f: [-1,1)\rightarrow \R$ such that
$$ \lambda(x,y) = f(x\cdot y)\quad\text{for all }(x,y)\in C_2(\Sph{2}). $$
Note that one can let $f$ explode at $1$ and obtain Hodge propagators which do not extend smoothly to the blow-up. Let us assume, in addition, that $\Prpg_1 - \Prpg_2$ is coexact. We obtain 
$$ 0 = \CoDd(\Prpg_1 - \Prpg_2) = \CoDd \Dd\lambda = \Laplace \lambda. $$
Therefore, $\lambda$ is a harmonic function on $C_2(\Sph{2})$. Denoting 
\[
B(x,y) \coloneqq x\cdot y,
\]
we can write $\lambda = f\circ B$,  which implies
$$ \Laplace(f \circ B) = f'' \Norm{\Grad B}^2 + f' \Laplace B. $$
The computation of $\Norm{\Grad B}$ and $\Laplace B$ is straightforward and we will do it for any $n\in \N$. If $\tilde{f}: \Sph{n} \rightarrow \R$ is a smooth function and $f: \R^{n+1} \rightarrow \R$ is defined by 
$$ f(x)\coloneqq \tilde{f}\Bigl(\frac{x}{\Abs{x}}\Bigr)\quad\text{for all }x\in \R^{n+1}\backslash\{0\}, $$
then
$$ \Laplace^{\Sph{n}} \tilde{f} = \Restr{\bigl(\Laplace^{\R^{n+1}}f\bigr)}{\Sph{n}}\quad\text{and}\quad \Grad^{\Sph{n}} \tilde{f} = \Restr{\bigl(\Grad^{\R^{n+1}}f\bigr)}{\Sph{n}}. $$
Here $\Laplace^{\Sph{n}}$, resp.~$\Grad^{\Sph{n}}$ are the Laplacian, resp.~the gradient on $\Sph{n}$ expressed in terms of the corresponding operators $\Laplace^{\R^{n+1}}$ and $\Grad^{\R^{n+1}}$ on $\R^{n+1}$, where $\Sph{n}$ is embedded into. 
We compute
\begin{align*}
\Laplace_x^{\R^{n+1}} \Bigl( \frac{x}{\Abs{x}}\cdot y \Bigr) &= \sum_{i=1}^{n+1} - \frac{\partial}{\partial x^i}\Bigl(\frac{y^i}{\Abs{x}} - \frac{x^i}{\Abs{x}^3}x\cdot y\Bigr) \\
 &= \sum_{i=1}^{n+1} \frac{x^i y^i}{\Abs{x}^3} - 3 \frac{x^i x^i}{\Abs{x}^5} x \cdot y + \frac{x\cdot y}{\Abs{x}^3} + \frac{x^i y^i}{\Abs{x}^3} \\
 & = 0,\\
\Grad^{\R^{n+1}}_x\Bigl(\frac{x}{\Abs{x}}\cdot y\Bigr) & = \sum_{i=1}^{n+1} \Bigl(\frac{y^i}{\Abs{x}} - \frac{x^i}{\Abs{x}^3}x\cdot y \Bigr)\frac{\partial}{\partial x^i}\quad\text{and}\\
\Norm{\Grad (x\cdot y)}^2 &= \Bigl\|\sum_{i=1}^{n+1}(y^i - (x\cdot y) x^i) \frac{\partial}{\partial x^i} + \sum_{i=1}^{n+1}(x^i - (x\cdot y) y^i) \frac{\partial}{\partial y^i}\Bigr\|^2 \\
& = \sum_{i=1}^{n+1} (y^i - (x\cdot y) x^i)^2 + (x^i - (x\cdot y)y^i)^2 \\
& = 1 - 2(x\cdot y)^2 + (x\cdot y)^2 + 1 - 2(x\cdot y)^2 + (x\cdot y)^2 \\
& = 2(1- (x\cdot y)^2).
\end{align*}
Therefore, $\Laplace B = 0$, $\Norm{\Grad B(x_1,x_2)}^2=2(1-(x_1\cdot x_2)^2)$, and we arrive to the equation
$$ 2(1-u^2) f''(u) = 0\quad \text{for all }u\in[-1,1) $$
and a smooth function $f: [-1,1) \rightarrow \R$. The only solution is a linear function, and it must hold
$$ \lambda(x_1,x_2) = a B(x_1, x_2) + b \quad\text{for some }a, b\in \R. $$
We see that $\lambda$ extends smoothly to the blow-up.
\end{proof}

If we determine the constant $a$ in the proof of Proposition~\ref{Prop:StdS2}, then we get a formula relating $\HtpStd$ to $\ArtPrpg$ for $\Sph{2}$.
\ToDo[caption={Do more here!}]{This needs to be computed.}

\chapter{BV-formalism for IBL-infinity theory on cyclic cochains}

In this chapter, we use the $\BV$-formalism to formulate the theory of $\dIBL$-algebras on cyclic cochains of an odd symplectic vector space and its twisting with a Maurer-Cartan element.
The ``fields'' are cyclic Hochschild chains and the $\BV$-operator comes from the algebraic string bracket and cobracket.
We hope that this point of view will help to understand relations between string topology, Chern-Simons theory, symplectic field theory and string field theory (let's say ``S(F)T'').

In Section~\ref{BV:Summary}, we summarize relevant $\IBLInfty$-theory from \cite{Cieliebak2015}.

In Section~\ref{Sec:BVAction}, we define the canonical string $\BV$-operator on the space of ``observables'' and the total action; it consists of the free part and of the interaction part which corresponds to the Maurer-Cartan element.
We argue that the quantum master equation is satisfied for the canonical and the Chern-Simons Maurer-Cartan element and that the string $\BV$-operator twisted by the action corresponds to the twisted $\dIBL$-algebra (Proposition~\ref{Prop:BVActForCanMC}).
In general, we show that the action satisfies the quantum master equation if and only if its interaction part encodes a Maurer-Cartan element and that the twistings in both formalisms agree (Proposition~\ref{Prop:BVActForAnyMCElement}).

In Section~\ref{Sec:HPL}, we sketch how to apply ideas from \cite{Doubek2018} for quantum $\LInfty$-algebras to $\IBLInfty$-algebras.
We ask whether their formulas for the effective action and the path integral which come from the Homological Perturbation Lemma (and Wick's Theorem) agree with the formulas from \cite{Cieliebak2015} with summations over ribbon graphs (Question~\ref{Q:EqForm}).

\section{Summary of IBL-infinity theory for cyclic DGA}\label{BV:Summary}
Let $(V,\Pair,m_1,m_2)$ be a cyclic $\DGA$ of degree $n$ of finite type.
We have:
\begin{itemize}
    \item Cyclic cochains $\CycC(V) = \CDBCyc V[2-n]$,
    \item Canonical dIBL-algebra $\OPQ_{110}$, $\OPQ_{210}$, $\OPQ_{120}$ on $\CycC(V)$ denoted by $\dIBL(\CycC(V))$,
    \item Canonical Maurer-Cartan element $\MC = (\MC_{10})$,
    \item Twisted dIBL-algebra $\OPQ_{110}^\MC$, $\OPQ_{210}$, $\OPQ_{120}$ on $\CycC(V)$ denoted by $\dIBL^\MC(\CycC(V))$.
\end{itemize}
Recall from Appendix~\ref{Section:AppEqDefPrCoPr} that, up to signs and permutation of variables, $\OPQ_{210}$ is the cyclization ${}^+$ of the Gerstenhaber bracket $[\cdot,\cdot]$ and that it holds $\BVOp_{\CycMRM} = \ProdCyc\circ\OPQ_{120}$, where~$\BVOp_{\CycMRM}$ is an extension of the Schwarz's $\BV$-operator on polynomial functions on $V$ to $\CycC(V)$ and~$\ProdCyc$ is the cyclic shuffle coproduct. 

Let $(V',\Pair',m_1',m_2')$ be another cyclic $\DGA$ of degree $n$ of finite type, and let 
\begin{equation}\label{Eq:DefRetr}
\begin{tikzcd}[column sep=large]
(V,m_1)\arrow[loop left]{l}{\Htp}\arrow[shift left]{r}{\pi}&\arrow[shift left]{l}{\iota} (V',m_1')
\end{tikzcd}
\end{equation}
be a deformation retraction.
We have the following:
\begin{itemize}
    \item $\IBLInfty$-morphism $\HTP: \dIBL(\CycC(V))\rightarrow \dIBL(\CycC(V'))$ with components $\HTP_{klg}: \hat{\Ext}_k \CycC(V) \rightarrow \hat{\Ext}_l \CycC(V')$ such that 
 $$ \HTP_{110}=\iota^*: (\CycC(V),\hat{\OPQ}_{110})\longrightarrow(\CycC(V'),\hat{\OPQ}'_{110}) $$
 is a quasi-isomorphism.
 The number $\HTP_{klg}(\Susp^k \psi_1\otimes\dotsb\otimes\psi_k)(\Susp^{l}\omega_1\otimes\dotsb\otimes\omega_l)$
for $\psi_1$,~$\dotsc$, $\psi_k \in \CDBCyc V$ and $\omega_1 = \omega_{11}\dotsb\omega_{1 s_1}$,~$\dotsc$, $\omega_l = \omega_{l1}\dotsb\omega_{ls_l}\in \BCyc V'$, where $\Susp$ is the formal symbol of degree $n-3$, is computed via summation over ribbon graphs with interior vertices decorated with $\psi_1$, $\dotsc$, $\psi_k$, interior edges decorated with the propagator --- the Schwartz kernel $\Prpg$ of $\Htp$ --- and exterior vertices decorated with~$\omega_{ij}$ at the $i$-th boundary component and with $j$ respecting the orientation.
We will refer to such graphs shortly as \emph{Feynman graphs.}
    \item Pushforward Maurer-Cartan element $\PMC \coloneqq\HTP_* \MC$ with components $\PMC_{lg} \in \hat{\Ext}_l\CycC(V)$.
    The number $\PMC_{lg}(\Susp^l\omega_1\otimes \dotsc \otimes \omega_l)$ is computed via summation over Feynman graphs as above with trivalent internal vertices $\MC_{10}$.
    \item Twisted $\IBLInfty$-morphism $\HTP^\MC: \dIBL^\MC(\CycC(V)) \rightarrow \dIBL^\PMC(\CycC(V'))$ with components $\HTP^\MC_{klg}: \hat{\Ext}_k \CycC(V) \rightarrow \hat{\Ext}_l \CycC(V')$ such that
 $$\begin{multlined}[t]\HTP_{110}^\MC: (\CycC(V),\hat{\OPQ}_{110}^\MC)\longrightarrow(\CycC(V'),{\hat{\OPQ}'^\PMC}_{110}) \\ =\iota^* + \HTP_{210}\circ_1 \MC_{10} + \frac{1}{2!} \HTP_{310} \circ_{1,1}(\MC_{10}, \MC_{10}) + \dotsb\end{multlined} $$
is a quasi-isomorphism.
The number $\HTP_{klg}(\Susp^k \psi_1\otimes\dotsb\otimes\psi_k)(\Susp^{l}\omega_1\otimes\dotsb\otimes\omega_l)$ is computed via summation over Feynmann graphs with interior vertices $\psi_1$, $\dotsc$, $\psi_k$ and $\MC_{10}$.
\end{itemize}
Recall that an $\IBLInfty$-quasi-isomorphism is automatically an $\IBLInfty$-homotopy equivalence; this has been proven in \cite[Theorem~1.2]{Cieliebak2015} via obstruction theory.
Obstruction theory also gives the existence of a minimal model of any $\IBLInfty$-algebra, which is the content of~\cite[Theorem~1.3]{Cieliebak2015}.
What is not proven yet but what we suspect is true is the following:
\begin{enumerate}[label=(\arabic*)]
 \item For a different choice of the deformation retraction \eqref{Eq:DefRetr}, the corresponding $\IBLInfty$-morphisms $\HTP$ and $\HTP'$ are $\IBLInfty$-homotopic.
 \item If $\iota$ is in addition a morphism of Poincar\'e $\DGA$'s (perhaps simply-connected), then $\dIBL^\MC(\CycC(V))$ and $\dIBL^{\MC'}(\CycC(V'))$ are $\IBLInfty$-homotopy equivalent (perhaps even the Maurer-Cartan elements $\MC'$ and $\HTP_{*}\MC$ are gauge equivalent).
\end{enumerate}

\section{BV-formulation for dIBL-algebra on cyclic cochains}\label{Sec:BVAction}

In the setting of Section~\ref{BV:Summary}, we define 
$$ \CycB(V) := \BCyc V [3-n]\quad\text{and}\quad \Fun(\CycB(V)) := \hat{\Ext} \CycC(V) ((\hbar)) = \Ext \CycC(V) \COtimes \K((\hbar)). $$
We call $\Fun(\CycB(V))$ the \emph{space of functions on $\CycB(V)$}. This makes sense because $\CycC(V) \subset (\BCyc V [2-n])^{\GD}$, $\Ext \CycC(V) = \Sym(\CycC(V)[1])$ and the symmetric algebra of the dual can be viewed as polynomial functions.
In contrast to \cite{Doubek2018}, we do not have any canonical odd symplectic\Correct[noline,caption={Graded symplectic form}]{Define and use graded symplectic form and $dg$-symplectic vector space.} form on $B(V)$, and hence there is no Schwarz's $\BV$-operator on $\Fun(B(V))$.
However, we have the following $\BV$-operators from~\cite{Cieliebak2015}:
\begin{align*}
 \BVOp_0 & \coloneqq \hat{\OPQ}_{120} + \hbar \hat{\OPQ}_{210}, \\
 \BVOp &\coloneqq \hat{\OPQ}_{110} + \BVOp_0, \\
 \BVOp^\MC & \coloneqq \reallywidehat{\OPQ_{210}\circ_1 \MC_{10}} + \BVOp.
\end{align*}
The first $\BV$-operator is canonical for an odd symplectic vector space $(V,\Pair)$, and we call it the \emph{string $\BV$-operator.}
It is a $\BV$-operator because $\hat{\OPQ}_{120}$ is a derivative of order $1$, $\hat{\OPQ}_{210}$ a derivative of order $2$, and it holds $\BVOp_0\circ\BVOp_0 = 0$.
The second $\BV$-operator is canonical for a cyclic cochain complex $(V,\Pair,m_1)$ and the third for a cyclic $\DGA$ $(V,\Pair,m_1,m_2)$ (in both cases, $(V,\Pair)$ is an odd symplectic vector space).
Note that the perturbations $\hat{\OPQ}_{110}$, resp.~$\reallywidehat{\OPQ_{210}\circ_1\MC_{10}}$ are first order differential operators.

For $\BVOp$, we consider the associated Gerstenhaber bracket $\{\cdot,\cdot\}: \Fun(\CycB(V))^{\otimes 2} \rightarrow \Fun(\CycB(V))$, which is for all $\Psi_1$, $\Psi_2\in \Fun(\CycB(V))$ given by
$$ \{\Psi_1, \Psi_2\} := (-1)^{\Abs{\Psi_1}}\bigl(\BVOp(\Psi_1 \Psi_2)- \BVOp(\Psi_1)\Psi_2 - (-1)^{\Abs{\Psi_1}}\Psi_1\BVOp(\Psi_2)\bigr). $$
It is easy to see that
$$ \{\Psi_1,\Psi_2\} = (-1)^{1 + (\Abs{\Psi_1} + 1)(\Abs{\Psi_2}+1)}\{\Psi_2,\Psi_1\}. $$
If $\{\cdot,\cdot\}_0$ and $\{\cdot,\cdot\}^\MC$ are the Gerstenhaber brackets for $\BVOp_0$ and $\BVOp^\MC$, respectively, then it holds
$$ \{\cdot,\cdot\}_0 = \{\cdot,\cdot\}^\MC = \{\cdot,\cdot\} $$
because $\BVOp$, $\BVOp_0$ and $\BVOp^\MC$ differ only by differential operators of order $\le 1$.
 
Consider the cyclizations $m^+_1$, $m^+_2: \BCyc V \rightarrow \R$ defined for all $v_1$, $v_2$, $v_3 \in V[1]$ by
$$ m_1^+(v_1,v_2)\coloneqq \Pair(m_1(v_1),v_2)\quad\text{and}\quad m^+_2(v_1,v_2,v_3)\coloneqq \Pair(m_2(v_1,v_2),v_3). $$
They have degree $3-n$ as maps, thus degree $n-3$ as elements of $\DBCyc V$ with the cohomological grading, and thus $\Susp m^+_1$, $\Susp m^+_2\in \Fun(\CycB(V))$ have degree $2(n-3)$.
They are ``linear functions'' in the sense that $\Susp m^+_1$, $\Susp m^+_2 \in \hat{\Ext}_1\CycC(V)$.
We define the \emph{total action}
\begin{equation}\label{Eq:Action}
 \Action \coloneqq \underbrace{(-1)^{n-2}(\Susp m_1^+)\hbar^{-1}}_{\displaystyle\eqqcolon \FreeAction} + \underbrace{(-1)^{n-2}(\Susp m_2^+)\hbar^{-1}}_{\displaystyle \eqqcolon \IntAction}\in \Fun(\CycB(V))
\end{equation}
and call $\FreeAction$ the \emph{free action} and $\IntAction$ the \emph{interaction.} The total action is linear and has degree $0$.
As a function, we can write 
$$ S(\Susp b) = (-1)^{n-2}m_1^+(b) \hbar^{-1} + (-1)^{n-2} m_2^+(b) \hbar^{-1}\quad\text{for }b\in \BCyc V.$$
Recall that $\MC_{10} = (-1)^{n-2}\Susp m_2^+$ is the canonical Maurer-Cartan element.

\begin{Proposition}[$\BV$-action for canonical Maurer-Cartan element]\label{Prop:BVActForCanMC}
Let $(V,\Pair,m_1,m_2)$ be a cyclic $\DGA$ of degree $n$ of finite type, and let $\BVOp_0 = \hat{\OPQ}_{120} + \hbar \hat{\OPQ}_{210}$ be the string $\BV$-operator on $\Fun(\CycB(V))$.
The total action $S\in \Fun(\CycB(V))$ from \eqref{Eq:Action} satisfies the \emph{quantum master equation} 
\begin{equation}\label{Eq:QME}
 \BVOp_0 S + \frac{1}{2}\{S,S\}_0 = 0,
\end{equation}
and it holds
\begin{equation}\label{Eq:Twist}
 \BVOp^\MC(f)= \BVOp_0(f) + \{S,f\}_0\quad\text{for all }f\in \Fun(\CycC(V)),
\end{equation}
where $\BVOp^\MC = \hat{\OPQ}_{110}^\MC + \hat{\OPQ}_{120} + \hbar \hat{\OPQ}_{210}$ is the $\BV$-operator associated to the twisted $\dIBL$-algebra $\dIBL^\MC(\CycC(V))$.
\end{Proposition}

\begin{proof}
Let $f\in \Fun(\CycB(V))$, and consider the left multiplication $L_f: \Fun(\CycB(V)) \rightarrow \Fun(\CycB(V))$ given by  $g\mapsto fg$.
If we view $f$ as a linear map $\Fun(\CycB(V)) \rightarrow \Fun(\CycB(V))$ which maps the constant $1$ to $f$ and vanishes on the reduced part, then we can write $L_f  = f\odot \Id$, where~$\odot$ is the convolution product on the symmetric algebra (see Appendix~\ref{App:IBLMV}
).
Using formulas from Proposition~\ref{Prop:PartCompositions}, we compute
\begin{align*}
\hat{\OPQ}_{210} \circ L_f &= \hat{\OPQ}_{210}(f\odot \Id) \\
&= \OPQ_{210}\circ_{0,2}(f,\Id) + \OPQ_{210}\circ_{1,1}(f,\Id) + \OPQ_{210}\circ_{2,0}(f,\Id) \\
&=(-1)^{\Abs{f}} L_f \circ \hat{\OPQ}_{210} + \reallywidehat{\OPQ_{210}\circ_1 f} + L_{\hat{\OPQ}_{210}(f)}.
\end{align*}
Therefore, it holds
\begin{align*}
 \{f,\cdot\}_0 & = \hbar\bigl(\hat{\OPQ}_{210} \circ L_f - L_{\hat{\OPQ}_{210}(f)} - (-1)^{\Abs{f}} L_f \circ \hat{\OPQ}_{210}\bigr)  \\
 & =  \hbar\reallywidehat{\OPQ_{210}\circ_1 f}.
\end{align*}
We compute 
\begin{align}\label{Eq:TerribleComputationProblem}
&\bigl(\OPQ_{210}\circ_1 (\Susp m_1^+)\bigr)(\Susp\psi)(\omega) \nonumber\\ 
&\qquad= \OPQ_{210}(\Susp m_1^+, \Susp \psi)(\omega) \nonumber\\
&\qquad= (-1)^{\Abs{\Susp}} \OPQ_{210}(\Susp^2 m_1^+, \psi) \nonumber\\
&\qquad \underset{\mathclap{\substack{\uparrow\rule{0pt}{1.5ex} \\ T^{ij} = (-1)^{\Abs{s} + \Abs{e_i}\Abs{e_j}} T^{ji} \\
T^{ji} = (-1)^{\Abs{e_j}}\Pair(e^j,e^i) \\
\Pair(e^j,e^i)\neq 0\ \Implies\ \Abs{e_j} + \Abs{e_i} = \Abs{s} + 1}}}{=}
(-1)^{\Abs{\Susp}} \sum \varepsilon(\omega\to\omega^1\omega^2)(-1)^{\Abs{e_j}\Abs{\omega^1}}T^{ij} m_1^+(e_i \omega^1)\psi(e_j\omega^2)\nonumber\\
&\qquad \underset{\mathclap{\substack{\uparrow\rule{0pt}{1.5ex} \\ \sum_i \Pair(v,e^i)e_i = v\\ 
m_1^+(e^j \omega^1) = (-1)^{\Abs{e^j}\Abs{\omega^1}}m_1^+(\omega^1 e^j) \\
}}}{=} \sum \varepsilon(\omega\to\omega^1\omega^2)(-1)^{\Abs{e_j}(\Abs{\Susp}+1 + \Abs{\omega^1}) + \Abs{e^j}\Abs{\omega^1}} \Pair(m_1(\omega^1),e^j)\psi(e_j\omega^2) \nonumber\\
&\qquad\underset{\mathclap{\substack{\uparrow\rule{0pt}{1.5ex} \\ \hspace{1cm}\Pair(m_1(\omega^1),e^j)\neq 0\ \Implies\ 1 + \Abs{\omega^1} + \Abs{e^j} = \Abs{\Susp} + 1 \\ 
\Abs{e^j} + \Abs{e_j} = \Abs{\Susp} + 1
}}}{=} (-1)^{\Abs{\Susp} + 1} \sum \varepsilon(\omega^1 \omega^2) \psi(m_1(\omega^1)\omega^2) \nonumber\\
&\qquad = (-1)^{\Abs{\Susp} + 1} \sum_{i=1}^k (-1)^{\Abs{\omega_1} + \dotsb + \Abs{\omega_{i-1}}} \psi(\omega_1 \dotsb m_1(\omega_i) \dotsb \omega_k) \nonumber\\
&\qquad = (-1)^{\Abs{\Susp} + 1}\OPQ_{110}(\Susp \psi)(\Susp \omega).
\end{align}
Therefore, we have 
\begin{equation}\label{Eq:FreeActionBracket}
 \{\FreeAction,\cdot\}_0 = \hat{\OPQ}_{110}.
\end{equation}
Since $m_1 \circ m_1 = 0$ and since $m_1^+$ is cyclically symmetric, it holds
\begin{equation*}
\{\FreeAction,\FreeAction\}_0 = (-1)^{n-2}\hbar^{-1}\OPQ_{110}(\Susp m_1^+) = 0.
\end{equation*}
Because $m_1^+ \in \DBCyc V$ has weight $2$, i.e., it vanishes on words of length $3$ and more, $\OPQ_{120}$ decreases the weight by $2$, and $\DBCyc V$ is reduced, i.e., its weight-$0$ part is $0$, we have
\begin{equation*}
\BVOp_0 \FreeAction = (-1)^{n-2}\OPQ_{120}(\Susp m_1^+) = 0.
\end{equation*}
We now compute
\begin{equation}\label{Eq:SSBracket}\begin{aligned}
\{\Action,\Action\}_0 &= \{\FreeAction,\FreeAction\}_0 + 2 \{\FreeAction,\IntAction\}_0 + \{\IntAction,\IntAction\}_0 \\
& = 2 \{\FreeAction,\IntAction\}_0 + \{\IntAction,\IntAction\}_0  \\
& = 2\hbar^{-1}\Bigl(\OPQ_{110}(\MC_{10}) + \frac{1}{2}\OPQ_{210}(\MC_{10},\MC_{10})\Bigr)
\end{aligned}\end{equation}
and
\begin{equation}\label{Eq:SBVOp}
 \BVOp_0 \Action = \hbar^{-1}\OPQ_{120}(\MC_{10}).
\end{equation}
The right-hand side of \eqref{Eq:SSBracket} vanishes due to \cite[Proposition~12.3]{Cieliebak2015}, and the right-hand side of \eqref{Eq:SBVOp} vanishes by the discussion preceding \cite[Proposition 12.5]{Cieliebak2015} (for degree reasons). Therefore, $\Action$ satisfies the quantum master equation \eqref{Eq:QME}.
The twisting equation~\eqref{Eq:Twist} is also clear.
\end{proof}

The inclusion of $\hbar^{-1}$ into $S$ is the convention from \cite{Cieliebak2015}.
After the transformation $\BVOp_0 \to \BVOp_0' = \BVOp_0$, $\Action\to\Action' = \hbar \Action$, we get the well-known equations
\begin{equation*}
 \hbar \BVOp_0' \Action' + \frac{1}{2} \{\Action',\Action'\}_0' = 0\quad\text{and}\quad \hbar {\BVOp'}^\MC= \hbar \BVOp_0' + \{\Action',\cdot\}_0'.
\end{equation*}

\begin{Proposition}[$\BV$-action for general Maurer-Cartan element]\label{Prop:BVActForAnyMCElement}
Let $(V,\Pair,m_1,m_2)$ be a cyclic $\DGA$ of degree $n$ of finite type, and let $\BVOp_0 = \hat{\OPQ}_{120} + \hbar \hat{\OPQ}_{210}$ be the string $\BV$-operator on $\Fun(\CycB(V))$.
Let $\MC_{lg}\in \hat{\Ext}_l \CycC(V)$ for all $l$, $g\ge 0$ satisfy the degree and filtration degree conditions of a Maurer-Cartan element for an $\IBLInfty$-algebra of bidegree $(n-3,2)$.%
Then $\MC = (\MC_{lg})$ satisfies the Maurer-Cartan equation for $\dIBL(\CycC(V))$ if and only if the total action
\begin{equation}\label{Eq:MyAction}
 \Action := \FreeAction + \sum_{l,g\ge 0} \MC_{lg} \hbar^{g-1}  \in \Fun(\CycB(V))
\end{equation}
satisfies the quantum master equation \eqref{Eq:QME}.
The $\BVInfty$-operator $\BVOp^\MC$ of the twisted $\IBLInfty$-algebra $\dIBL^\MC(\CycC(V))$ is given by \eqref{Eq:Twist}.
The elements $\MC_{0g}$ appear neither in \eqref{Eq:QME} nor in~\eqref{Eq:Twist}.
\end{Proposition}
\begin{proof}
Using $\BVOp_0 \FreeAction = \{\FreeAction,\FreeAction\}_0 = 0$, we compute
\begin{align*}
\BVOp_0 \Action + \frac{1}{2}\{\Action,\Action\}_0 &= \BVOp_0 \IntAction + \{\FreeAction,\IntAction\}_0 + \frac{1}{2}\{\IntAction,\IntAction\}_0 \\
& = \begin{multlined}[t]\sum_{l,g\ge 0}\bigl(\hat{\OPQ}_{120}(\MC_{lg})\hbar^{g-1} + \hat{\OPQ}_{210}(\MC_{lg})\hbar^g + \hat{\OPQ}_{110}(\MC_{lg})\hbar^{g-1}\bigr) \\ 
{}+ \frac{1}{2} \sum_{l_1,l_2,g_1,g_2\ge 0} \reallywidehat{\OPQ_{210}\circ_1\MC_{l_1 g_1}}(\MC_{l_2 g_2})\hbar^{g_1 + g_2 - 1}\end{multlined}\\
& = \begin{multlined}[t]\sum_{l,g\ge 0} (\iota_l \pi_l)\Bigl(\OPQ_{120}\circ_1 \MC_{l-1, g} + \OPQ_{210}\circ_2\MC_{l+1,g-1} + \OPQ_{110}\circ_1\MC_{lg} \\ 
{}+ \frac{1}{2} \sum_{\substack{l_1,l_2,g_1,g_2\ge 0 \\ 
l_1 + l_2 - 1 = l\\
g_1 + g_2 = g}} \OPQ_{210}\circ_{1,1}(\MC_{l_1 g_1},\MC_{l_2 g_2})\Bigr)\hbar^{g-1}.\end{multlined}
\end{align*} 
Comparing to Proposition~\ref{Prop:dIBL}, we see that the $(l,g)$-components for $l\ge 1$, $g\ge 0$ give precisely the Maurer-Cartan equation in $\dIBL(\CycC(V))$.
We compute
\begin{align*}
\BVOp_0 + \{S,\cdot\}_0 &= \hat{\OPQ}_{120} + \hbar\hat{\OPQ}_{210} + \hat{\OPQ}_{110} + \sum_{l,g\ge 0}\reallywidehat{\OPQ_{210}\circ_1 \MC_{lg}} \hbar^g \\
&=\begin{aligned}[t]
&(\hat{\OPQ}_{110} + \reallywidehat{\OPQ_{210}\circ_1 \MC_{10}}) + (\hat{\OPQ}_{120} + \reallywidehat{\OPQ_{210}\circ_1 \MC_{20}}) + (\reallywidehat{\OPQ_{210}\circ_1 \MC_{30}}) + \dotsb \\
&{}+\bigl[(\hat{\OPQ}_{210}) + (\reallywidehat{\OPQ_{210}\circ_1 \MC_{11}}) + (\reallywidehat{\OPQ_{210}\circ_1 \MC_{21}})+\dotsb\bigr]\hbar\\
&{}+\sum_{l\ge 0, g\ge 2} (\reallywidehat{\OPQ_{210}\circ_1 \MC_{lg}})\hbar^g
\end{aligned} \\
&=\begin{aligned}[t]&(\hat{\OPQ}_{110}^\MC) + (\hat{\OPQ}_{120}^\MC) + (\hat{\OPQ}_{130}^\MC) + \dotsb \\
&{}+\bigl[(\hat{\OPQ}_{210}^\MC) + (\hat{\OPQ}_{111}^\MC) + (\hat{\OPQ}_{121}^\MC) + \dotsb\bigr]\hbar \\
&{}+\sum_{\substack{l\ge 0, g\ge 2}}(\hat{\OPQ}_{1lg}^\MC) \hbar^g
\end{aligned}
\end{align*}
and see that this is indeed the $\BVInfty$-operator $\BVOp^\MC$ of $\dIBL^\MC(\CycC(V))$.
\end{proof}

\section{Homotopy transfer and effective action}\label{Sec:HPL}

In \cite{Doubek2018}, they consider multilinear operations $l_{lg}: V^{\otimes l} \rightarrow \R$ for $l\ge 1$, $g\ge 0$ on an odd symplectic vector space~$V$ and write down an action $\Action\in \Fun(V)$ in the form of~\eqref{Eq:MyAction} with~$\MC_{lg}$ replaced by~$l_{lg}^+$. We remark that their ``vertices'' are $l^+_{lg}(v,\dotsc,v)$ for a ``field'' $v\in V$, whereas ours are $\MC_{lg}(v_1,\dotsc,v_l)$ for a ``string of fields'' $v_1\dotsb v_l\in \CycB(V)$. They consider the Schwarz's canonical $\BV$-operator on $\Fun(V)$ from~\cite{Schwarz1992} and show that $\Action$ satisfies the quantum master equation if and only if $(l_{lg})$ satisfy the relations of a quantum $\LInfty$-algebra.\footnote{This is equivalent to the notion of a loop homotopy algebra from \cite{Markl1997} and to string brackets in closed string field theory from \cite{Zwiebach1992}.} On the other hand, we have the string $\BV$-operator on $\Fun(\CycB(V))$ and our action \eqref{Eq:MyAction} satisfies the quantum master equation if and only if $\MC=(\MC_{lg})$ is a Maurer-Cartan element for $\dIBL(\CycC(V))$. Next, they consider a deformation retract ($\eqqcolon\mathrm{DR}$) as in \eqref{Eq:DefRetr} and obtain explicit formulas for the homotopy transfered quantum $\LInfty$-algebra on $V'$ together with all maps and homotopies via the Homological Perturbation Lemma ($\eqqcolon\mathrm{HPL}$). In what follows, we will sketch how to apply their construction to $\IBLInfty$-algebras. We stress that the details have NOT been done yet!

A DR \eqref{Eq:DefRetr} induces a DR
\begin{equation*}
\begin{tikzcd}
\bigl(\CycC(V),\OPQ_{110}\bigr) \arrow[loop left]{l}{K_\CycC}\arrow[shift left]{r}{P_\CycC} & \arrow[shift left]{l}{I_\CycC} \bigl(\CycC(V'),\OPQ_{110}'\bigr),
\end{tikzcd}
\end{equation*}
which further induces a DR
\begin{equation}\label{Eq:SDRSDR}
\begin{tikzcd}
\bigl(\Fun(B(V)),\hat{\OPQ}_{110}\bigr) \arrow[loop left]{l}{K}\arrow[shift left]{r}{P} & \arrow[shift left]{l}{I} \bigl(\Fun(B(V')),\hat{\OPQ}_{110}'\bigr).
\end{tikzcd}
\end{equation}
In~\cite[Remark~3]{Doubek2018}, they write down a formula for $K$ on $\Fun(V)$ given $\Htp$ using a ``tensor trick'' due to Eilenberg Mac-Lane; the same method may apply to get~$K_C$ from~$\Htp$ on $\CycC(V)$ and $K$ on $\Fun(\CycB(V))$ from $K_C$ in our case. With such $K$'s, one can take $P_{\CycC}$ and $P$, resp.~$I_{\CycC}$ and~$I$ to be the natural extensions of $\iota^*$, resp.~$\pi^*$. Note that any surjective quasi-isomorphism over $\R$ is a deformation retraction, but their formula is explicit and preserves special DR's, i.e., DR's satisfying $\Htp^2 = \Htp \iota = \pi \Htp = 0$ ($\eqqcolon\mathrm{SDR}$). \ToDo[caption={injectiv qi},noline]{Are injective quasi-isomorphisms sections of deformatino retractions?} The crucial idea of~\cite{Doubek2018} translated to our situation is to view $\BVOp$ and $\BVOp^\MC$ as perturbations of $\{\FreeAction,\cdot\} = \hat{\OPQ}_{110}$, which we denote by~$\delta^{(1)}$ and~$\delta^{(2)}$, respectively, and apply the HPL from~\cite{Crainic2004}:

For $i=1$, $2$, suppose that $\delta^{(i)}$ is ``small'', i.e., that $(\Id - \delta^{(i)} K)$ is invertible, and consider the maps
\begin{align*} 
 \BVOp^{(i)} &\coloneqq \hat{\OPQ}_{110}' + P(\Id - \delta^{(i)}K)^{-1}\delta^{(i)} I = \hat{\OPQ}_{110}' + P \delta^{(i)} I + P \delta^{(i)} K \delta^{(i)} I + \dotsb ,\\
 I^{(i)} &\coloneqq I + K(\Id - \delta^{(i)}K)^{-1}\delta^{(i)} I = I + K \delta^{(i)} I + K \delta^{(i)} K \delta^{(i)} I + \dotsb,\\
 P^{(i)} &\coloneqq P + P(\Id-\delta^{(i)}K)^{-1}\delta^{(i)} K = P + P \delta^{(i)} K + P \delta^{(i)} K \delta^{(i)} K + \dotsb,\\
 K^{(i)} &\coloneqq K + K(\Id-\delta^{(i)}K)^{-1}\delta^{(i)}K = K + K\delta^{(i)} K + K \delta^{(i)} K \delta^{(i)} K + \dotsb.
\end{align*}
The HPL asserts that if \eqref{Eq:SDRSDR} is an SDR, then the following are SDR's as well:
\begin{equation*}\begin{tikzcd}[execute at end picture={
\draw[->,dashed] (3.5,1.5) to[out=0,in=90] node[midway,right,xshift=.5cm]{$\delta^{(1)} = \BVOp_0$} (5,.75) to[out=-90,in=0] (3.5,0);
\draw[->,dashed] (3.5,1.5) to[out=0,in=0] node[pos=0.8,right,xshift=.3cm]{$\delta^{(2)} = \BVOp_0 + \{\IntAction,\cdot\}$} (3.5,-1.5);
}]
\bigl(\Fun(B(V)),\hat{\OPQ}_{110}\bigr) \arrow[loop left]{l}{K}\arrow[shift left]{r}{P} & \arrow[shift left]{l}{I} \bigl(\Fun(B(V')),\hat{\OPQ}_{110}'\bigr) \\
\bigl(\Fun(B(V)),\BVOp \bigr) \arrow[loop left]{l}{K^{(1)}}\arrow[shift left]{r}{P^{(1)}} & \arrow[shift left]{l}{I^{(1)}} \bigl(\Fun(B(V')),\BVOp^{(1)}\bigr)\\
\bigl(\Fun(B(V)),\BVOp^\MC \bigr) \arrow[loop left]{l}{K^{(2)}}\arrow[shift left]{r}{P^{(2)}} & \arrow[shift left]{l}{I^{(2)}} \bigl(\Fun(B(V')),\BVOp^{(2)}\bigr)
\end{tikzcd}\end{equation*}
One defines the \emph{effective action} 
$$W \coloneqq \log\bigl(P^{(1)}(e^{\IntAction})\bigr) \in \Fun(\CycB(V'))$$
and the \emph{path integral}
$$ Z \coloneqq L_{e^{-W}} \circ P^{(1)} \circ L_{e^{\IntAction}}: \Fun(\CycB(V)) \rightarrow \Fun(\CycB(V')). $$
Under certain circumstances, the following formulas hold:
\begin{equation}\label{Eq:NiceEqns}
\BVOp^{(1)} = P\BVOp_0 I,\quad \BVOp^{(2)}=\BVOp^{(1)} + \{W,\cdot\}^{(1)}\quad\text{and}\quad P^{(2)} = Z.
\end{equation}
This is proven in \cite{Doubek2018} on $\Fun(V)$ when $V'=\Harm$ is a ``harmonic'' subspace in a Hodge decomposition $V = \Harm \oplus C$ into odd symplectic subspaces, $\pi$ and $\iota$ are the canonical projection and inclusion, respectively, the homotopy $\Htp$ is such that $(\pi,\iota,\Htp)$ is an SDR, and the homotopy $K$ was constructed from $\Htp$ via the tensor trick. Since we deal with the string $\BV$-operator on $\Fun(\CycB(V))$, we can not talk about this ``symplectic compatibility'' and the proof of \eqref{Eq:NiceEqns} might be based on another arguments.

\begin{Question}[HPL and $\IBLInfty$]\label{Q:EqForm}
If one picks an SDR of $V$ onto a harmonic subspace~$\Harm$ as above (basically equivalent to the setting of \cite[Section~11]{Cieliebak2015}) and constructs $K_\CycC$ and $K$ in a particular way, can one achieve that \eqref{Eq:NiceEqns} and the following identities hold?
$$ \BVOp^{(1)} = \BVOp_0',\quad \BVOp^{(2)} = \BVOp^\MC, \quad W = \Action_{\HTP_* \MC},\quad P^{(1)} = e^\HTP,\quad P^{(2)}= e^{\HTP^\MC}. $$
Here, $\Action_{\HTP_*\MC}$ denotes the action \eqref{Eq:MyAction} for the Maurer-Cartan element $\HTP_*\MC$.
\end{Question}

\begin{Remark}[On $\BV$-formalism for $\IBLInfty$]
\begin{RemarkList}
\item As summarized in \cite[Section~5]{Doubek2018}, given a $\BV$-action $\Action$, there are various approaches to obtain $W$ and $Z$ as summations over Feynman graphs (see \cite{Mnev2017} for the stationary phase formula approach).
\item The appearance of Feynman graphs can be explained from the proof of \cite[Theorem~2]{Doubek2018}, where they show that in the special setting above, it holds
$$ P^{(1)} = P e^{D_{\Prpg}} $$
for an order $\le 2$ differential operator $D_{\Prpg}$ which ``connects'' two legs with the propagator~$\Prpg$ (obtained by ``rising one index'' of $\Htp$ using the odd symplectic form). This is reminiscent of the Wick's Theorem for the (formal) perturbative expansion of the path integral (the classical approach to quantum field theories).
\item Having a $\BV$-formulation of the $\IBLInfty$-theory, it is intriguing to compare it to~\cite{Muenster2011}, where certain $\IBLInfty$-structures are considered in the context of open-closed string field theory.\qedhere
\end{RemarkList}
\end{Remark}

\part{Appendices}
\appendix

\chapter{Evaluation of labeled ribbon graphs}
\label{Section:Appendix}
\Correct[noline,caption={DONE Propagator}]{Change propagator to admissible propagator or symmetric propagator. Let's leave it as a propagator. It is the propagator in our theory, hence, we can have some definitions.}
In this appendix, we consider an algebraic propagator $\Prpg$ and the graph pairing $\langle \cdot, \cdot \rangle^{\Prpg}_\Gamma$ (Definition~\ref{Def:EvalRibGraph}), which encapsulates the contribution of a ribbon graph~$\Gamma$ to the map $f_{klg}: (\DBCyc V)^{\otimes k} \rightarrow (\DBCyc V)^{\otimes l}$ defined as a sum of contributions of ribbon graphs (Proposition~\ref{Prop:GraphPairing}). Such maps were already defined in \cite[Section 11]{Cieliebak2015} using coordinates; here we use an invariant framework inspired by~\cite{Mnev2017}. As an example, we work out in details expressions for the canonical $\dIBL$-operations $\OPQ_{210}$ and~$\OPQ_{120}$ (Example~\ref{Ex:Canon}). We also explain the technicality of identifying symmetric maps with maps on symmetric powers (Remark~\ref{Rem:SymMaps}).

Next, we show that the matrix $(T^{ij})$ from Definition~\ref{Def:CanonicaldIBL} corresponds to the algebraic Schwartz kernel (Definition~\ref{Def:LinSchw}) of the identity $\Id$ up to a sign. Assuming that the Hodge propagator $\Prpg$ from Definition~\ref{Def:GreenKernel} is algebraic, we deduce natural candidates for the signs in Definition~\ref{Def:PushforwardMCdeRham} based on the formula from \cite[Remark 12.10]{Cieliebak2015} for the genuine pushforward Maurer-Cartan element~$\PMC$ in the finite-dimensional case. Establishing the formal analogy between the de Rham case and the finite-dimensional case is our main application of the invariant framework. Finally, we sketch how to obtain signs for the Fr\'echet $\dIBL$-structure on $\DR(M)$ (Remark~\ref{Rem:Frechet}).

Throughout this appendix, we will use Notation~\ref{Def:Notation} without further remarks.

\section{Finite dimensional case}

\begin{Def}[Propagator \& graph pairing]\label{Def:EvalRibGraph}
Let $V$ be a graded vector space. We will call a tensor $\Prpg\in V[1]^{\otimes 2}$ an \emph{algebraic propagator}. We call it \emph{admissible} if the symmetry condition
\begin{equation}\label{Eq:SymmetryCondition}
 \tau(\Prpg) = (-1)^{\Abs{\Prpg}} \Prpg
\end{equation}
is satisfied, where $\tau: V[1]^{\otimes 2} \rightarrow V[1]^{\otimes 2}$ is the twist map defined by $\tau(v_1 \otimes v_2) = (-1)^{\Abs{v_1}\Abs{v_2}} v_2 \otimes v_1$ for all $v_1$, $v_2\in V[1]$.

For a ribbon graph $\Gamma \in \RRG_{klg}$ and its labeling $L$, consider the permutation~$\sigma_L$ from Definition~\ref{Def:EdgeVertex}. It acts on tensor powers according to Definition~\ref{Def:Permutations} and thus defines the map
\begin{equation*} \sigma_L: (V[1]^{\otimes 2})^{\otimes e} \otimes V[1]^{\otimes s_1} \otimes \dotsb \otimes V[1]^{\otimes s_l} \longrightarrow V[1]^{\otimes d_1} \otimes \dotsb \otimes V[1]^{\otimes d_k},
\end{equation*}
where $d_i$ and $s_i$ are the valencies of internal vertices $1$, $\dotsc$, $k$ and boundary components $1$, $\dotsc$, $l$, respectively, and $e$ is the number of internal edges. We extend $\sigma_L$ by $0$ to other combinations of tensor powers.

The \emph{graph pairing}
\[ \langle \cdot, \cdot \rangle_{\Gamma}^{\Prpg}\ :\ (\DBCyc V)^{\otimes k} \otimes (\BCyc V)^{\otimes l} \longrightarrow \R  \]
is defined for all $\psi_1$, $\dotsc$, $\psi_{k} \in \DBCyc V$ and generating words $w_i = v_{i1} \dots v_{i m_i}$ with $v_{ij} \in V[1]$ for $m_{i} \in \N$ and $i=1$, $\dotsc$, $l$ by the following formula:
\begin{align*}
&\langle \psi_1\otimes \dotsb \otimes \psi_{k}, w_1 \otimes \dotsb \otimes w_l \rangle_{\Gamma}^{\Prpg} \\
 &\quad \coloneqq \begin{multlined}[t]\smash{\sum_{L_1,\,L_3^b}}\vphantom{\sum_{L}} (\psi_1 \otimes \dotsb \otimes \psi_k)\bigl(\sigma_L(\Prpg^{\otimes e}\otimes  (v_{11}\otimes \dotsb \otimes v_{1 m_1}) \otimes \dotsb \\ \otimes (v_{l1}\otimes \dotsb \otimes v_{l m_{l}})\bigr), \end{multlined}
\end{align*}
where we use the pairing from Definition~\ref{Def:Pairings} and in every summand an $L_2$ compatible with $L_1$ and an $L_{3}^v$ are chosen arbitrarily to get a full labeling $L$ of~$\Gamma$. The graph pairing extends to $\langle \cdot, \cdot\rangle_\Gamma^{\Prpg}\ :\ \RTen \DBCyc V \otimes \RTen \BCyc V \rightarrow \R$, where $\RTen W = \bigoplus_{k=1}^\infty W^{\otimes k}$ is the reduced tensor product.
\end{Def}

\begin{Proposition}\label{Prop:GraphPairing}
In the setting of Definition~\ref{Def:EvalRibGraph}, we denote $w = w_1 \otimes \dotsb \otimes w_l$ and $\psi = \psi_1 \otimes \dotsb\otimes \psi_k$. If $\Prpg$ is admissible, then the following holds:
\begin{ClaimList}
\item The number $\psi(\sigma_L(\Prpg^{\otimes e}\otimes w))$ does not depend on the choice of $L_{3}^v$ and an $L_2$ compatible with $L_1$. Moreover, $\langle\cdot, \cdot \rangle_\Gamma^{\Prpg}$ does not depend on the representative of $[\Gamma]\in \RRG_{klg}$.
\item If $V$ is finite-dimensional, then for every $k$, $l \ge 1$, $g \ge 0$ there is a unique linear map
\[ f_{klg}: (\DBCyc V)^{\otimes k} \longrightarrow (\DBCyc V)^{\otimes l} \] such that
\begin{equation*}
\begin{aligned}
&f_{klg}(\psi_1\otimes \dotsb \otimes \psi_k)(w_1 \otimes \dotsb \otimes w_l) \\
 &\qquad\qquad =  \frac{1}{l!} \sum_{[\Gamma]\in \RRG_{klg}} \frac{1}{\Abs{\Aut(\Gamma)}} \langle \psi_1\otimes \dotsb \otimes \psi_k, w_1 \otimes \dotsb \otimes w_l\rangle^{\Prpg}_{\Gamma}. \end{aligned}
\end{equation*}
\item The following holds for the map $f_{klg}$ from (b):
\begin{itemize}
\item It is homogenous of degree
\begin{equation} \label{Eq:DegreeForm} \Abs{f_{klg}} = -\Abs{\Prpg}(k+l-2+2g). \end{equation}
\item The filtration degree satisfies
\begin{equation} \label{Eq:FiltrDegreeForm} 
 \Norm{f_{klg}} \ge - 2 (k+l-2+2g).
\end{equation}
\item For all $\eta\in\Perm_l$ and $\mu\in\Perm_k$, we have
\begin{equation} \label{Eq:SymmetryForm} \eta \circ f_{klg} \circ \mu = (-1)^{\Abs{\Prpg}(\eta + \mu)}f_{klg}.
\end{equation}
\end{itemize}
\end{ClaimList}
\end{Proposition}

\begin{proof}
\begin{ProofList}
\item Let us denote by $\bar{i}$ and $ij$ the operations on $L_2$ given by $\mathrm{e}_i \mapsto -\mathrm{e}_i$ and $\mathrm{e}_i \leftrightarrow \mathrm{e}_j$, respectively. An even number of these operations does not change the orientation of the complex~\eqref{Eq:OrientationComplex}. Their effect in $\sigma_L$ acting on $\Prpg^{\otimes e}\otimes w$ is
\begin{equation*}
\bar{i}: \Prpg_i\mapsto \tau(\Prpg_i) = (-1)^{\Abs{\Prpg}} \Prpg_i\quad\text{and}\quad ij: \Prpg_i \dots \Prpg_j\mapsto (-1)^{\Abs{\Prpg}} \Prpg_j \dots \Prpg_i.
\end{equation*}
Therefore, an even number of them does not change $\sigma_L(\Prpg^{\otimes e}\otimes w)$. This proves the independence of the choice of a compatible $L_2$. The independence of the choice of $L_{3}^v$ is clear since $\psi_i$ are cyclic symmetric.

An isomorphism of ribbon graphs $\eta: \Gamma \rightarrow \Gamma'$ induces the map of compatible labelings $L \mapsto L' = \eta_* L$ such that $\sigma_{L} = \sigma_{L'}$. The independence of the choice of a representative of $[\Gamma]$ follows.

\item Suppose that $\psi = \psi_1 \otimes \dotsb \otimes \psi_k$ with $\psi_i \in (\DBCyc V)_{r_i}^{c_i}$, where $r_i\in \N$ and $c_i \in \Z$ for $i=1$, $\dotsc$, $k$. A~general element of $(\DBCyc V)^{\otimes k}$ is then a finite linear combination of such~$\psi$'s. 

First of all, let us argue that the sum $\sum_{\RRG_{klg}}$ is finite. The number of internal edges~$e$ is fixed from the Euler formula \eqref{Eq:EulerFormula}. Therefore, the number of contributing graphs $(V_{\mathrm{int}}, E_{\mathrm{int}})$ is finite. In order to bound the number of external vertices, we notice that $d_1 = r_1$, $\dotsc$, $d_k = r_k$ must hold for $\psi(\sigma_L(\Prpg^{\otimes e}\otimes w))$ to be non-zero. Therefore, the sum is finite.

We now have the linear functional 
\[ f_{klg}(\psi) \coloneqq \frac{1}{l!}\sum_{[\Gamma]\in \RRG_{klg} } \frac{1}{\Abs{\Aut(\Gamma)}} \langle \psi \mid \cdot \rangle_\Gamma^{\Prpg}: (\BCyc V)^{\otimes l} \longrightarrow \R \] 
and need to show that $f_{klg}(\psi)\in (\DBCyc V)^{\otimes l} \subset (\BCyc V)^{\otimes l*}$. Because $V$ has finite dimension, the weight-filtration of $\BCyc V$ satisfies (WG1) \& (WG2) (see \eqref{Eq:WGs} and Proposition~\ref{Prop:Compl}), and hence we have
\[ (\DBCyc V)^{\otimes l} = (\BCyc V)^{\WGD \otimes l} =\bigl((\BCyc V)^{\otimes l}\bigr)^{\WGD} \]
for the weight-graded duals. Therefore, it suffices to show that $f_{klg}(\psi)$ vanishes on all but finitely many degrees and weights of $(\BCyc V)^{\otimes k}$. However, the relation $f_{klg}(\psi)(w) \neq 0$ for a generating word $w\in (\BCyc V)^{\otimes k}$ implies
\begin{equation}\label{Eq:UUUU}
\begin{aligned}
\Abs{w} &= \Abs{\psi} - e \Abs{\Prpg}\quad\text{and} \\
k(w) &= k(\psi) - 2 e,
\end{aligned}
\end{equation}
where $k$ denotes the weight, and hence $f_{klg}(\psi)\in (\DBCyc V)^{\otimes l}$ indeed holds.

\item The formulas~\eqref{Eq:DegreeForm} and~\eqref{Eq:FiltrDegreeForm} follow from~\eqref{Eq:UUUU} and from~\eqref{Eq:EulerFormula}.

As for the symmetry \eqref{Eq:SymmetryForm}, suppose that $L$ and $L'$ are compatible labelings of the same graph $\Gamma$ such that $L_1'$ differs from $L_1$ by a permutation $\mu\in \Perm_{k}$ of internal vertices and a permutation $\eta\in \Perm_l$ of boundary components. Viewing $\mu$ and $\eta$ as block permutations in the vertex and edge order, respectively, we get
\[ \sigma_{L'}(\Prpg^{\otimes e}\otimes w)=(-1)^{\Abs{\Prpg}(\eta+\mu)}\mu(\sigma_L(\Prpg^{\otimes e}\otimes \eta( w))). \]
The sign comes from the difference of $L_2$ and $L_2'$ which compensates the change of the orientation of~\eqref{Eq:OrientationComplex} caused by $\mu$ and $\eta$.\qedhere
\end{ProofList}
\end{proof}

Given $\mu\in \Perm_k$ and $\psi = \psi_1 \otimes \dotsb \otimes \psi_k \in (\DBCyc V)^{\otimes k}$, it is easy to see that
\[ \varepsilon(\mu, \Psi) = \varepsilon(\mu(\Susp),\mu(\psi)) \varepsilon(\mu,\Susp)\varepsilon(\Susp,\psi)\varepsilon(\mu,\psi), \]
where $\Psi = (\Susp \psi_1) \otimes \dotsb \otimes (\Susp \psi_k) \in (\DBCyc V[A])^{\otimes k}$ and $\varepsilon(\mu,\Susp) = (-1)^{\Abs{s}\mu}$. If $A= - \Abs{\Prpg}$, then we get from \eqref{Eq:SymmetryForm} that the degree shift $\HTP_{klg}: (\DBCyc V [A])^{\otimes k} \rightarrow (\DBCyc V[A])^{\otimes l}$ has the following symmetries:
\begin{equation}\label{Eq:SymMap}
\forall \mu \in \Perm_k, \eta\in \Perm_l: \quad \eta \circ \HTP_{klg} \circ \mu = \HTP_{klg}.
\end{equation}
Note that the degrees satisfy
\begin{equation}\label{Eq:DegDegShift}
\Abs{\HTP_{klg}} = \Abs{f_{klg}} + (k-l) A.
\end{equation}

\begin{Remark}[Symmetric maps versus maps on symmetric powers]\label{Rem:SymMaps}
In the situation above, we define~$\tilde{\HTP}_{klg}$ as the unique map such that the solid lines of the following diagram commute:    
\[\begin{tikzcd}
(\DBCyc V[A])^{\otimes k} \arrow{r}{\HTP_{klg}} \arrow[two heads]{d}{\pi} & (\DBCyc V[A])^{\otimes l} \arrow[two heads,swap]{d}{\pi}\\
\arrow[bend left,dotted]{u}{\iota}\Sym_k \DBCyc V[A] \arrow{r}{\tilde{\HTP}_{klg}} & \Sym_l \DBCyc V[A].\arrow[bend right,swap,dotted]{u}{\iota}
\end{tikzcd}\]
The symmetry condition \eqref{Eq:SymMap} provides the existence of $\tilde{\HTP}_{klg}$ and implies commutativity of the dotted diagram as well. Moreover, for all $\psi_1$, $\dotsc$, $\psi_k \in \DBCyc V$ and $w_1$, $\dotsc$, $w_l\in \BCyc V$, we have 
\[ \tilde{\HTP}_{klg}(\Susp^k \psi_1\cdots \psi_k)(\Susp^l w_1\cdots w_l) = \HTP_{klg}(\Susp^k \psi_1 \otimes \dotsb \otimes \psi_k)(\Susp^l w_1 \otimes \dotsb\otimes w_l), \]
where we use the pairing from Definition~\ref{Def:Pairings}. We denote $\tilde{\HTP}_{klg}$ again by $\HTP_{klg}$.
\end{Remark}

\begin{Definition}[Algebraic Schwartz kernel]\label{Def:LinSchw}
Let $V$ be a graded vector space and $\Pair: V\otimes V \rightarrow \R$ a non-degenerate pairing on $V$. We extend $\Pair$ to a non-degenerate pairing $\Pair: V^{\otimes k}\otimes V^{\otimes k}\rightarrow \R$ for $k\ge 1$ by setting
\begin{equation*} 
\Pair(v_{11} \otimes \dotsb \otimes v_{1k}, v_{21}\otimes \dotsb \otimes v_{2k}) \coloneqq \varepsilon(v_1, v_2)\Pair(v_{11},v_{21}) \dots \Pair(v_{1k},v_{2k})
\end{equation*}
for all $v_{11}$, $\dotsc$, $v_{1k}$, $v_{21}$, $\dotsc$,  $v_{2k}\in V$, where $\varepsilon$ is the Koszul sign (see Definition~\ref{Def:Koszul}). For $k=0$, we let $\Pair: \R\otimes \R \rightarrow \R$ be the multiplication on $\R$.

For $k$, $l \ge 0$, we say that $\Kern_L\in V^{\otimes k + l}$ is the \emph{algebraic Schwartz kernel} of a linear operator $L: V^{\otimes k} \rightarrow V^{\otimes l}$ if the following is satisfied:
\begin{equation} \label{Eq:KernelEquation}
\forall w_1 \in V^{\otimes k},  w_2 \in V^{\otimes l}:\quad \Pair(L(w_1),w_2) = \Pair(\Kern_L, w_1\otimes w_2).
\end{equation}
We usually omit writing ``algebraic'' if it is clear from the context (i.e., if we do not consider any ``extensions'' of $V^{\otimes k}$). 
\end{Definition}

In the situation of Definition~\ref{Def:LinSchw}, let $(e_i)\subset V$ be a basis and $(e^i)$ its dual basis such that $\Pair(e_i,e^j)=\delta_{ij}$. We define the coordinates $K_L^{ij} \in \R$ and $L^{ij}\in \R$ by 
\[ \Kern_L = \sum_{i,j} K_L^{ij} e_i \otimes e_j\quad\text{and}\quad L^{ij} \coloneqq \Pair(L (e^i),e^j).\]
From \eqref{Eq:KernelEquation} we have
\begin{equation} \label{Eq:KernelCoordinates}
\Kern_L^{ij} = (-1)^{(\Abs{L}+1)(\Abs{\Pair}+\Abs{e_i})} L^{ij}\quad \text{for all }i,j.
\end{equation} 

From now on, we will be in the situation of (A) and (B) in the Overview; in particular, we put $V[1]$ in place of $V$ in Definition~\ref{Def:LinSchw}. Let $\Kern_{\Id} \in V[1]^{\otimes 2}$ be the Schwartz kernel of the identity $\Id: V[1]\rightarrow V[1]$ and $\Kern_{\Htp} \in V[1]^{\otimes 2}$ the Schwartz kernel of the cochain homotopy $\Htp: V[1] \rightarrow V[1]$. From~\eqref{Eq:KernelCoordinates}, we get
\begin{equation*} 
\Kern_{\Htp}^{ij} = \Htp^{ij} \quad \text{and} \quad {\Kern_{\Id}}^{ij} = (-1)^{\Abs{e_i} + \Abs{\Pair}} \Pair(e^i,e^j)\quad\text{for all }i,j. 
\end{equation*}
We see that the tensor $\TKer = \sum_{i,j} \TKer^{ij} e_i \otimes e_j$ from~\eqref{Eq:PropagatorT} can be expressed as
\begin{equation*}
\TKer = (-1)^{n-2} \Kern_{\Id}.
\end{equation*}
This is the invariant meaning of $\TKer$. Note that the degrees satisfy 
\[ \Abs{\TKer} = n-2\qquad\text{and}\qquad\Abs{\Kern_{\Htp}} = n - 3. \]

The assumption~\eqref{Eq:ConditionOnG} on $\Htp$ is equivalent to graded antisymmetry of the bilinear form $\Htp^+\coloneqq \Pair\circ (\Htp\otimes \Id): V[1]^{\otimes 2} \rightarrow \R$. This is further equivalent to
\[ \tau(\Kern_{\Htp}) = (-1)^{\Abs{\Kern_{\Htp}}}\Kern_{\Htp}. \]
Therefore, $\Kern_{\Htp}$ satisfies~\eqref{Eq:SymmetryCondition}, and hence it defines an admissible propagator for the construction of $f_{klg}$ for every $k$, $l\ge 1$, $g\ge 0$. We have from~\eqref{Eq:SymmetryForm} that the degree shift $\HTP_{klg}: (\DBCyc V[3-n])^{\otimes k} \rightarrow (\DBCyc V[3-n])^{\otimes l}$ is symmetric. Moreover, using~\eqref{Eq:DegreeForm},~\eqref{Eq:FiltrDegreeForm} and \eqref{Eq:DegDegShift}, we obtain
\[\begin{aligned}
\Abs{\HTP_{klg}} &= - 2d(k+g-1), \\
\Norm{\HTP_{klg}}&\ge \gamma (2-2g-k-l),
\end{aligned}\]
where $(d,\gamma) = (n-3,2)$. These are the degree and filtration conditions on an $\IBLInfty$-morphism from~\cite[Definition~2.8 and (8.3)]{Cieliebak2015}. In fact, our $\HTP=(\HTP_{klg})_{k,l\ge 1, g\ge 0}$ is precisely the $\IBLInfty$-homotopy from \cite[Theorem 11.3]{Cieliebak2015}.

Graded antisymmetry of $\Pair$ is equivalent to 
\[ \tau(\TKer)= (-1)^{\Abs{\TKer}+1}\TKer. \]
Visibly, $\TKer$ does not satisfy~\eqref{Eq:SymmetryCondition}, and hence is not an admissible propagator. Nevertheless, we can still use it to define $f_{210}$ and~$f_{120}$ since the corresponding graphs $\Gamma$ (see Figure~\ref{Fig:GammasGraphs}) have only one internal edge $\mathrm{e}$, and, for a given $L_1$, there is a unique compatible~$L_2$ determined by the orientation of $\mathrm{e}$ (see Example~\ref{Ex:Canon} for the compatibility condition). As for the symmetry of the resulting maps, a transposition of internal vertices or boundary components in~\eqref{Eq:OrientationComplex} can be compensated only by $\mathrm{e}\mapsto-\mathrm{e}$, which produces $(-1)^{\Abs{\TKer}+1}$ (c.f.~the proof of Proposition~\ref{Prop:GraphPairing} (a)). Therefore, if we shift the degrees by $A= - \Abs{\TKer} +  1 = n- 3$, we obtain symmetric maps $\OPQ_{210}: (\DBCyc V[A])^{\otimes 2} \rightarrow \DBCyc V[A]$ and $\OPQ_{120}: \DBCyc V[A] \rightarrow (\DBCyc V[A])^{\otimes 2}$. We show in Example~\ref{Ex:Canon} below that these operations agree with those defined in Definition~\ref{Def:CanonicaldIBL}.

\begin{Example}[The canonical $\dIBL$-operations]\label{Ex:Canon}
We have
\begin{equation}\label{Eq:DefByGraphs}
\begin{aligned}
f_{210}(\psi_1 \otimes \psi_2)(w) & = \frac{1}{1!} \sum_{[\Gamma]\in \RRG_{210}} \frac{1}{\Abs{\Aut(\Gamma)}} \langle \psi_1 \otimes \psi_2 \mid w \rangle_\Gamma^{\Prpg}\quad\text{and} \\
f_{120}(\psi)(w_1 \otimes w_2) & = \frac{1}{2!} \sum_{[\Gamma]\in \RRG_{120}} \frac{1}{\Abs{\Aut(\Gamma)}} \langle \psi \mid w_1 \otimes w_2 \rangle_\Gamma^{\Prpg}.
\end{aligned}
\end{equation}
We parametrize $\RG_{210}$ by the ribbon graphs $\Gamma_{k_1, k_2}$ with $1\le k_1\le k_2$ and $\RG_{120}$ by the ribbon graphs $\Gamma^{s_1, s_2}$ with $0\le s_1 \le s_2$; these graphs are depicted in Figure~\ref{Fig:GammasGraphs}. We have 
$\RRG_{210} = \RG_{210}\backslash \{[\Gamma_{1,1}]\}$ and $\RRG_{120} = \RG_{120}\backslash \{[\Gamma^{0,0}], [\Gamma^{0,1}]\}$. We also have 
\[\Abs{\Aut(\Gamma_{k_1,k_2})} = \begin{cases} 1 & \text{if }k_1 \neq k_2, \\ 
              2 & \text{if }k_1 = k_2, \end{cases}\]
and likewise for $\Gamma^{s_1,s_2}$. We fix labelings~$L_{3}^v$ and parametrize $L_{3}^b$ by $c=1$, $\dotsc$, $k_1 + k_2 - 2$ for $\Gamma_{k_1,k_2}$ and by $c_1 = 1$, $\dotsc$, $s_1$ and $c_2 = 1$, $\dotsc$, $s_2$ for $\Gamma^{s_1,s_2}$ as it is indicated in Figure~\ref{Fig:GammasGraphs}.

There are two possible labelings $L_{1}^v$ for $\Gamma_{k_1, k_2}$ and two possible labelings $L_{1}^b$ for $\Gamma^{s_1, s_2}$; this is the only freedom in choosing a full labeling $L$ because $L_3$ is fixed and $L_2$ is just the orientation of the single internal edge, which is uniquely determined by $L_1$. For both $\Gamma_{k_1, k_2}$ and $\Gamma^{s_1, s_2}$, we will denote the two possible full labelings by $L^1$ and $L^2$. They can be depicted as follows:
\begin{figure}[t]
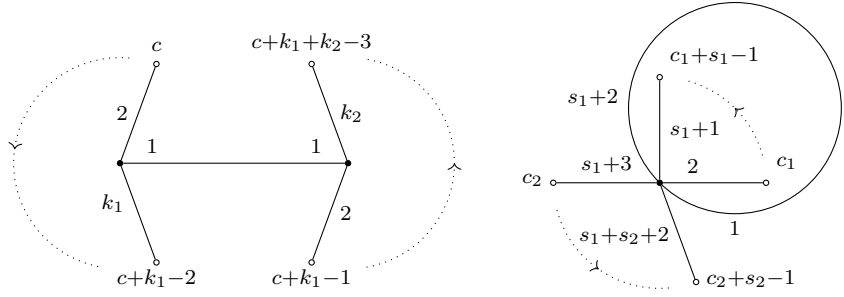

\centering
\input{\GraphicsFolder/gammad1d2.tex}
\quad
\input{\GraphicsFolder/gammas1s2.tex}
\caption[Ribbon graphs for $\OPQ_{210}$ and $\OPQ_{120}$.]{Graphs $\Gamma_{k_1,k_2}$ and $\Gamma^{s_1, s_2}$ with fixed labelings $L_3$.}
\label{Fig:GammasGraphs}
\end{figure}
\begin{equation}\label{Eq:OrientLabel}
\begin{tabular}{c|m{1.5cm} m{1.5cm}}
 & \makebox[1.5cm]{$\Gamma_{k_1,k_2}$} & \makebox[1.5cm]{$\Gamma^{s_1,s_2}$} \\\hline
\rule{0pt}{4ex}$L^1$ &\centering{\begin{tikzpicture}
\tikzset{->-/.style={decoration={markings,mark=at position #1 with {\arrow{>}}},postaction={decorate}}} 
\node (V1) at (0,0) {1};
\node(V2) at (1,0) {2};
\draw[->-={0.5}] (V1) -- (V2);
\end{tikzpicture}}
& \begin{tikzpicture}
\tikzset{->-/.style={decoration={markings,mark=at position #1 with {\arrow{>}}},postaction={decorate}}} 
\node (V1) at (1,0) {1};
\node(V2) at (0,0) {2};
\draw[->-={0.5}] (V1) -- (V2);
\end{tikzpicture}\\[1ex]
$L^2$ & \begin{tikzpicture}
\tikzset{->-/.style={decoration={markings,mark=at position #1 with {\arrow{>}}},postaction={decorate}}} 
\def\Rad{0.4}
\node (B1) at (0,0) {1};
\node (B2) at (0.8,0) {2};
\draw[->-={0.5}] ($([shift=(0:\Rad)]B1)$) arc (0:360:\Rad);
\end{tikzpicture} & \begin{tikzpicture}
\tikzset{->-/.style={decoration={markings,mark=at position #1 with {\arrow{>}}},postaction={decorate}}} 
\def\Rad{0.4}
\node (B1) at (0,0) {2};
\node (B2) at (0.8,0) {1};
\draw[->-={0.5}] ($([shift=(360:\Rad)]B1)$) arc (360:0:\Rad);
\end{tikzpicture}
\end{tabular}
\end{equation}
Let us check that the indicated $L_1$ and $L_2$ are compatible. For the complexes $C_2 \rightarrow C_1 \rightarrow C_0$ from \eqref{Eq:OrientationComplex}, we have the following:
\[\begin{aligned}
\Gamma_{k_1,k_2}:\qquad &\langle \mathrm{b} \rangle \xrightarrow{\Bdd_2 = 0}  \langle \mathrm{e} \rangle \xrightarrow{\Bdd_1} \langle \mathrm{v}_2 - \mathrm{v}_1 \rangle \oplus \langle \mathrm{v}_1 + \mathrm{v}_2 \rangle,  \\
\Gamma^{s_1, s_2}:\qquad &\langle \mathrm{b}_1 - \mathrm{b}_2 \rangle \oplus \langle \mathrm{b}_1 + \mathrm{b}_2 \rangle  \xrightarrow{\Bdd_2} \langle \mathrm{e} \rangle \xrightarrow{\Bdd_1 = 0} \langle \mathrm{v} \rangle.  
\end{aligned}\]
As for $\Gamma_{k_1,k_2}$, the basis $\mathrm{v}_2 - \mathrm{v}_1$, $\mathrm{v}_1 + \mathrm{v}_2$ of $C_0$ is positively oriented with respect to the basis $\mathrm{v}_2$, $\mathrm{v}_1$. Therefore, $\mathrm{e}$ has to be oriented such that $\Bdd_1 \mathrm{e} = \mathrm{v}_2 - \mathrm{v}_1$; i.e., it is a path from $\mathrm{v}_1$ to $\mathrm{v}_2$. As for $\Gamma^{s_1,s_2}$, the basis $\mathrm{b}_1-\mathrm{b}_2$, $\mathrm{b}_1 + \mathrm{b}_2$ of $C_2$ is positively oriented with respect to $\mathrm{b}_1$, $\mathrm{b}_2$. Therefore, $\mathrm{e}$ has to be oriented such that $\mathrm{e} = \Bdd_2 (\mathrm{b}_1 - \mathrm{b}_2)$. Recall that we orient the boundary of a $2$-simplex by the ``outer normal first'' convention. We conclude that the labelings from~\eqref{Eq:OrientLabel} are indeed compatible.

As for $f_{210}$, the permutations $\sigma_1\coloneqq\sigma_{L^1}$ and $\sigma_2\coloneqq \sigma_{L^2}$ corresponding to the labelings $L^1$ and $L^2$, respectively, read
\[\begin{aligned}
\sigma_{1} &= \biggl(\begin{array}{cc|ccc}
 1 & 2     & \dots & c+2 & \dots \\
 1 & k_1+1 & \undermat{k_1+k_1-2}{\dots& c+2 & \dots}\dots & 2& \dots 
\end{array}\biggr)\quad\text{and}\\[4ex]
\sigma_{2} &= \biggl(\begin{array}{cc|ccc}
 1 & 2     & \dots & c+2 & \dots  \\
 1 & k_2+1 & \undermat{k_1+k_2-2}{\dots & k_2 + 2 & \dots}\dots & k_2 + 2 & \dots
\end{array}\biggr).\\[3ex]
\end{aligned}\]
The underbracket marks the block which represents a cyclic permutation of the remaining indices. We see that 
\[\begin{aligned}
\sigma_{1}&:  V^{\otimes 2} \otimes V^{\otimes s} \longrightarrow V^{\otimes k_1}\otimes V^{\otimes k_2},\quad e_i e_j w \longmapsto e_i w^1 e_j w^2,  \\
\sigma_{2}&: V^{\otimes 2}\otimes V^{\otimes s} \longrightarrow V^{\otimes k_2}\otimes V^{\otimes k_1},\quad e_i e_j w  \longmapsto e_i w^2 e_j w^1,
\end{aligned}\]
where $w^1 = w_c \dots w_{c+k_1-2}$, $w^2 = w_{c+k_1-1}\dots w_{c+k_1+k_2-3}$ and $s \coloneqq k_1 + k_2 - 2$. Defining $\tilde{w}^1 \coloneqq w^2$ and $\tilde{w}^2 \coloneqq w^1$, The Koszul sign of $\sigma_2$ can be written as
\[ \varepsilon(w\mapsto w^1 w^2) (-1)^{\Abs{w^2}\Abs{e_j} + \Abs{w^1}\Abs{w^2}} = \varepsilon(w\mapsto \tilde{w}^1 \tilde{w}^2 ) (-1)^{\Abs{\tilde{w}^1} \Abs{e_j}}. \]
We use these facts to rewrite \eqref{Eq:DefByGraphs} as follows:
\begin{align*}
& f_{210}(\psi_1 \otimes \psi_2)(w) \\
&\qquad=\begin{aligned}[t]
& \sum_{\substack{1\le k_1 < k_2}} \sum_{i,j} \TKer^{ij} \Bigl(\sum_{k(w^1) = k_1 - 1} \varepsilon(w\mapsto w^1 w^2)(-1)^{\Abs{w^1}\Abs{e_j}} \psi_1(e_i w^1) \psi_2(e_j w^2) \\
&{}+ \sum_{k(w^1) = k_2 - 1} \varepsilon(w\mapsto w^1 w^2) (-1)^{\Abs{w^2}\Abs{e_j} + \Abs{w^1}\Abs{w^2}} \psi_1(e_i w^2) \psi_2(e_j w^1)\Bigr) \\
&{}+ \sum_{1< k_1 = k_2} \frac{1}{2} \Bigl(\sum_{k(w^1) = k_1 - 1} \varepsilon(w\mapsto w^1 w^2)(-1)^{\Abs{w^1}\Abs{e_j}} \psi_1(e_i w^1) \psi_2(e_j w^2) \\ 
&{}+\sum_{k(w^1) = k_2 - 1} \varepsilon(w\mapsto w^1 w^2) (-1)^{\Abs{w^2}\Abs{e_j} + \Abs{w^1}\Abs{w^2}} \psi_1(e_i w^2) \psi_2(e_j w^1)\Bigr)\end{aligned}\\
&\qquad=\sum_{\substack{k_1, k_2 \ge 1 \\ k_1 + k_2 > 2}} \sum_{\substack{ k(w^1) = k_1 - 1 \\ k(w^2) = k_2 - 1}}  \TKer^{ij}\varepsilon(w \mapsto w^1 w^2) (-1)^{\Abs{w^1}\Abs{e_j}} \psi_1(e_i w^1) \psi_2(e_j w^2).
\end{align*}
This coincides with the formula from Definition~\ref{Def:CanonicaldIBL}.

As for $f_{120}$, the permutations $\sigma_1\coloneqq\sigma_{L^1}$ and $\sigma_2\coloneqq \sigma_{L^2}$ corresponding to the labelings $L^1$ and $L^2$, respectively, read
\[\begin{aligned}
\sigma_{1} &= \biggl(\begin{array}{cc|ccc|ccc}
1 & 2      & \dots & c_1+2 & \dots & \dots & c_2 + s_1 + 2 & \dots \\
 1 & s_1 + 2& \undermat{s_1}{\dots & c_1+2 & \dots}\dots & 2 & \dots &\undermat{s_2}{\dots & c_2 + s_1 + 2 & \dots}\dots & s_1 + 3 & \dots
\end{array}\biggr)\quad\text{and} \\[4ex]
\sigma_{2} &= \biggl(\begin{array}{cc|ccc|ccc}
 1       & 2 &  \dots  & c_2 + 2 & \dots & \dots & c_1 + s_2 + 2 & \dots  \\
 s_1 + 2 & 1 & \undermat{s_2}{\dots & c_2+2 & \dots} \dots  & s_1 + 3 & \dots & \undermat{s_1}{\dots & c_1 + s_2 + 2 & \dots} \dots &  2     & \dots
\end{array}\biggr), \\[3ex]
\end{aligned}\]
where the underbracketed blocks denote cyclic permutations of consecutive indices on the corresponding boundary component. We see that 
\[\begin{aligned}
\sigma_1 &: V^{\otimes 2}\otimes V^{\otimes s_1}\otimes V^{\otimes s_2} \longrightarrow V^{\otimes k}, \quad e_i e_j w_1 w_2 \longmapsto e_i w_1^1 e_j w_2^1, \\
\sigma_2 &:  V^{\otimes 2}\otimes V^{\otimes s_2} \otimes V^{\otimes s_1} \longrightarrow V^{\otimes k}, \quad  e_i e_j w_1 w_2 \longmapsto e_j w_2^1 e_i w_1^1,
\end{aligned}\]
where $w_i^1$ denotes a cyclic permutation and $k\coloneqq s_1 + s_2 +2$. The Koszul sign of~$\sigma_2$ can be written as
\[\begin{aligned}&(-1)^{\Abs{e_i}\Abs{e_j} + \Abs{w_1}\Abs{w_2} + \Abs{e_i}\Abs{w_2}} \varepsilon(w_1 \mapsto w_1^1)\varepsilon(w_2\mapsto w_2^1)\\
&\qquad= (-1)^{(\Abs{e_i} + \Abs{w_1})(\Abs{e_j} + \Abs{w_2}) + \Abs{w_1} \Abs{e_j}}\varepsilon(w_1 \mapsto w_1^1)\varepsilon(w_2\mapsto w_2^1).\end{aligned}\]
We use this fact and the cyclic symmetry of $\psi$ to rewrite \eqref{Eq:DefByGraphs} as follows:
\[\begin{aligned}
&f_{120}(\psi)(w_1 \otimes w_2)
\\&\; = \begin{aligned}[t]
& \begin{multlined}[t]\sum_{0\le s_1 < s_2} \Bigl(\delta\bigl(\substack{k(w_1) = s_1 \\ k(w_2) = s_2}\bigr)\sum \TKer^{ij} \varepsilon(w_1\mapsto w_1^1) \varepsilon(w_2\mapsto w_2^1)(-1)^{\Abs{w_1} \Abs{e_j}} \\
\psi(e_i w_1^1 e_j w_2^1) +\delta\bigl(\substack{k(w_1) = s_2 \\ k(w_2) = s_1}\bigr) \sum \TKer^{ij} \varepsilon(w_1\mapsto w_1^1)\varepsilon(w_2\mapsto w_2^1)\\ (-1)^{\Abs{e_i} \Abs{e_j} + \Abs{w_2} \Abs{w_1} + \Abs{e_i}\Abs{w_2}} \psi(e_j w_2^1 e_i w_1^1) \Bigr)
\end{multlined}\\
&\begin{multlined}{}+\sum_{0< s_1 = s_2} \delta\bigl(\substack{k(w_1) = k(w_1) = s_1 \\ k(w_2) = k(w_2) = s_2}\bigr) 
\frac{1}{2} \Bigl(\sum \TKer^{ij} \varepsilon(w_1\mapsto w_1^1) \varepsilon(w_2\mapsto w_2^1)\\ (-1)^{\Abs{w_1} \Abs{e_j}} \psi(e_i w_1^1 e_j w_2^1) + \sum \TKer^{ij} \varepsilon(w_1\mapsto w_1^1)\varepsilon(w_2\mapsto w_2^1) \\ (-1)^{\Abs{e_i} \Abs{e_j} + \Abs{w_2} \Abs{w_1} + \Abs{e_i}\Abs{w_2}} \psi(e_j w_2^1 e_i w_1^1) \Bigr) \end{multlined}
\end{aligned} \\
&\; = \begin{multlined}[t] \smash{\sum_{\substack{s_1, s_2 \ge 0 \\ s_1 + s_2 > 0}}} \delta\bigl(\substack{k(w_1) = s_1 \\ k(w_2) = s_2}\bigr) \sum \TKer^{ij} \varepsilon(w_1 \mapsto w_1^1)\varepsilon(w_2 \mapsto w_2^1)(-1)^{\Abs{w_1}\Abs{e_j}} \\ \psi(e_i w_1^1 e_j w_2^1).\end{multlined}
\end{aligned}\]
This coincides with the formula from Definition~\ref{Def:CanonicaldIBL}.
\end{Example}

\section{De Rham case}

We will now establish a formal analogy between the finite-dimensional and the de Rham case, which will explain the signs in Definition~\ref{Def:PushforwardMCdeRham}.

\textbf{The finite-dimensional case.} Consider the situation of (A) -- (D) in the Overview. To recall briefly, we have a finite-dimensional cyclic dga $(V,\Pair,m_1, m_2)$ and a subcomplex $\Harm \subset V$ such that there is a projection $\pi: V[1] \rightarrow \Harm[1]$ chain homotopic to $\Id$ via a chain homotopy $\Htp: V[1] \rightarrow V[1]$. Using~$m_2$, one constructs the canonical Maurer-Cartan element $\MC$ for $\dIBL(\CycC(V))$. The algebraic Schwartz kernel $\Kern_{\Htp}$ of $\Htp$ is an admissible propagator used to construct the $\IBLInfty$-quasi-isomorphism $\HTP = (\HTP_{klg}): \dIBL(\CycC(V)) \rightarrow \dIBL(\CycC(\Harm))$. The Maurer-Cartan element $\MC$ is then pushed forward along $\HTP$ to obtain the Maurer-Cartan element $\PMC \coloneqq \HTP_* \MC$  for $\dIBL(\CycC(\Harm))$ (see \cite[Lemma 9.5]{Cieliebak2015}). The formula for~$\PMC$ given in~\cite[Remark~12.10]{Cieliebak2015} reads
\begin{equation} \label{Eq:PushforwardMC}
\begin{aligned} &\PMC_{lg}(\Susp^l w_1 \otimes \dotsb \otimes  w_l) \\
&\qquad=\frac{1}{l!}\sum_{[\Gamma]\in \TRRG_{klg}} \frac{1}{\Abs{\Aut(\Gamma)}} (-1)^{k(n-2)}\langle (m_2^+)^{\otimes k}, w_1 \otimes \dotsb \otimes w_l \rangle_{\Gamma}^{\Kern_{\Htp}}.
\end{aligned}\end{equation}
Here the artificial sign $(-1)^{k(n-2)}$ is added because our sign conventions for $m_2^+$ differ (see Remark~\ref{Rem:mukplus}).

\textbf{The de Rham case.}
We are in the setting of Definition~\ref{Def:PushforwardMCdeRham}. To recall briefly, we have the cyclic dga $(\DR(M), \Pair, m_1, m_2)$, the subspace of harmonic forms $\Harm\subset \DR$, the harmonic projection $\pi_\Harm: \DR\rightarrow \Harm$ and a Hodge propagator $\Prpg\in \DR(\Bl_\Diag(M\times M))$, which is the Schwartz kernel of a chain homotopy $\Htp: \DR \rightarrow \DR$ between $\pi_\Harm$ and~$\Id$. In analogy with the finite-dimensional case, the canonical Maurer-Cartan element~\eqref{Eq:CanonMC} for $\dIBL(\Harm)$ satisfies $\MC_{10} = (-1)^{n-2}m_2^+$ with $m_2^+ = \Pair(m_2 \otimes \Id)$. Because $\dim(\DR)$ is not of finite type, Definition~\ref{Def:CanonicaldIBL} does not give the canonical $\dIBL$-structure on $\CycC(\DR)$, and hence we have neither $\HTP$ nor~$\PMC$ in the standard sense.

In order to deduce the formal analogy, we embed $\DR(M)^{\otimes 2}$ into $\DR(\Bl_{\Diag}(M\times M))$ using the external wedge product $(\eta_1,\eta_2)\mapsto \tilde{\pi}_1^*\eta_1 \wedge \tilde{\pi}_2^*\eta_2$ and suppose that the Hodge propagator $\Prpg$ satisfies $\Prpg \in \DR^{\otimes 2}$. This never happens, so what follows is just a formal computation whose purpose is to deduce candidates for signs.%

\begin{Proposition}\label{Prop:FinDimAnalog}
In the de Rham case, suppose that $\Prpg\in \DR(M)^{\otimes 2}$. Then \eqref{Eq:PushforwardMC} reduces to \eqref{Eq:PushforwardMCdeRham}.
\end{Proposition}
\begin{proof}
Consider the intersection pairing $\tilde{\Pair}$ and its degree shift $\Pair$ (see Proposition~\ref{Prop:DGAs}). According to Definition~\ref{Def:LinSchw}, they extend to pairings on $\DR(M)^{\otimes k}$ and $\DR(M)[1]^{\otimes k}$ for all $k\ge 1$, respectively. For all $\eta_{11}$, $\eta_{12}$, $\eta_{21}$, $\eta_{22}\in \DR(M)$, we have:
\begin{equation}\label{Eq:PairComp}
\begin{aligned}
&\Pair(\SuspU^2 \eta_{11}\otimes \eta_{12},\SuspU^2 \eta_{21}\otimes \eta_{22}) \\ 
&\qquad = (-1)^{\eta_{11} + \eta_{21}} \Pair(\SuspU\eta_{11}\otimes \SuspU\eta_{12}, \SuspU \eta_{21}\otimes \SuspU\eta_{22}) \\
&\qquad = (-1)^{\eta_{11} + \eta_{21} + (1+\eta_{12})(1+\eta_{21})} \Pair(\SuspU \eta_{11}, \SuspU \eta_{21}) \Pair(\SuspU \eta_{12},\SuspU \eta_{22})\\
&\qquad =(-1)^{1+\eta_{12}\eta_{21}} \tilde{\Prpg}(\eta_{11},\eta_{21})\tilde{\Pair}(\eta_{12},\eta_{22}) \\
&\qquad = - \tilde{\Pair}(\eta_{11}\otimes \eta_{12}, \eta_{21}\otimes \eta_{22}).
\end{aligned}
\end{equation}
One can also check that
\[ \tilde{\Pair}(\eta_{11} \otimes \eta_{12}, \eta_{21}\otimes\eta_{22}) = \int_{x,y} \eta_{11}(x)\eta_{12}(y)\eta_{21}(x)\eta_{22}(y). \]

For the Hodge homotopy $\Htp: \DR(M) \rightarrow \DR(M)$ and its Hodge propagator $\Prpg \in \DR(M)^{\otimes 2}$, we have the following:
\[ \forall \eta_1, \eta_2\in \DR(M): \quad \tilde{\Pair}(\Htp(\eta_1),\eta_2) = \int_{x,y} \Prpg(x,y)\eta_1(x)\eta_2(y) = \tilde{\Pair}(\Prpg, \eta_1\otimes \eta_2). \]
From this and \eqref{Eq:PairComp}, we obtain
\[\begin{aligned}
\Pair(\Htp(\SuspU \eta_1),\eta_2) &= \Pair(\SuspU \Htp(\eta_1),\SuspU \eta_2) = (-1)^{1+\eta_1} \tilde{\Pair}(\Htp(\eta_1),\eta_2) \\ &= (-1)^{1+\eta_1} \tilde{\Pair}(\Prpg,\eta_1\otimes \eta_2) = (-1)^{\eta_1}\Pair(\SuspU^2 \Prpg, \SuspU^2 \eta_1 \otimes \eta_2)  \\ &= \Pair(\SuspU^2 \Prpg, \SuspU\eta_1 \otimes \Susp\eta_2).
\end{aligned}\]
Therefore, the element $\SuspU^{2} \Prpg \in V[1]^{\otimes 2}$ corresponds to the Schwartz kernel $\Kern_{\Htp}$ of $\Htp: V[1] \rightarrow V[1]$. We write this correspondence as
\[ \Kern_{\Htp}\in V[1]^{\otimes 2}\ \sim\ \SuspU^2 \Prpg \in \Bl_\Diag(M\times M)[2]. \]

Let us check that $\SuspU^{2}\Prpg$ satisfies~\eqref{Eq:SymmetryCondition}. First of all, if we embed $\DR(M)^{\otimes k}$ into $\DR(M^{\times k})$ using the external wedge product $\eta_1 \otimes \dotsb \otimes \eta_k \mapsto \pi_1^*\eta_1 \wedge \dotsm \wedge \pi_k^* \eta_k \eqqcolon \eta_1(x_1)\wedge \dotsm \wedge \eta_k(x_k)$, then for all $\eta_1$, $\dotsc$, $\eta_k \in \DR(M)$ we have
\[\sigma(\eta_1 \otimes \dotsb \otimes \eta_k)(x_1,\dotsc,x_k) = \eta_1(x_{\sigma_1})\wedge \dotsc \wedge \eta_k(x_{\sigma_k}), \]
where the action on the left-hand side is given by~\eqref{Eq:Perm}. Now, the symmetry property \eqref{Eq:SymProp} implies
\[ \tau(\SuspU^{2}\Prpg) = - \SuspU^{2}\tau^*(\Prpg) = (-1)^{n+1} \SuspU^{2} \Prpg = (-1)^{\Abs{\SuspU^2 \Prpg}} \SuspU^2 \Prpg. \]
Therefore, the symmetry condition \eqref{Eq:SymmetryCondition} is indeed satisfied.

Let $\Gamma\in \TRRG_{klg}$, and let $L$ be a labeling of $\Gamma$. We abbreviate $\sigma\coloneqq \sigma_L \in \Perm_{3k} $. Given $\eta_{ij}\in \DR(M)$ for $j=1$, $\dotsc$, $s_i$ and $i=1$, $\dotsc$, $l$, where $s_i$ is the valency of the $i$-th boundary component, we set $\eta_i = \eta_{i1}\otimes \dotsb \otimes \eta_{is_i}$, $\eta =\eta_1 \otimes \dotsb \otimes \eta_l$, $\alpha_{ij}=\SuspU \eta_{ij}$, $\omega_i = \alpha_{i1}\otimes \dotsb \otimes \alpha_{i s_i}$ and $\omega = \omega_1\otimes \dotsb \otimes \omega_l$. We denote $s \coloneqq s_1 + \dotsb + s_l$, so that $3k = 2e + s$, where $e$ is the number of internal edges. We have
\begin{align*}(m_2^+)^{\otimes k} \bigl(\sigma((\SuspU^{ 2} \Prpg)^{\otimes e} \otimes \omega)\bigr) &= \varepsilon(\SuspU,\eta)(m_2^+)^{\otimes k} \bigl(\sigma((\SuspU^{ 2} \Prpg)^{\otimes e} \otimes \SuspU^{s} \eta)\bigr) \\
& = (-1)^{s e (n-1)}\varepsilon(\SuspU,\eta)(m_2^+)^{\otimes k} \bigl(\sigma(\SuspU^{2e+s}\Prpg^{\otimes e} \otimes \eta)\bigr) \\
&= \underbrace{(-1)^{\sigma + s e (n-1)} \varepsilon(\SuspU,\eta) }_{\eqqcolon \varepsilon_1}(m_2^+)^{\otimes k} \bigl(\SuspU^{2e+s} \sigma( \Prpg^{\otimes e} \otimes \eta)\bigr),
\end{align*}
where $\varepsilon(\SuspU,\eta)$ is the Koszul sign to order $\SuspU^s \eta_{11}\dots\eta_{l s_l} \mapsto \SuspU\eta_{11} \dots \SuspU \eta_{l s_l}$ and the operation $m_2^+: \DR(M)[1]^{\otimes 3} \rightarrow \R$ is given by $m_2^+ = \Pair(m_2 \otimes \Id)$. We denote $\kappa \coloneqq \Prpg^{\otimes e} \otimes \eta = \kappa_1 \otimes \dotsb \otimes \kappa_{3k}$, $\kappa_i \in \DR(M)[1]$ and compute
\begin{align*}
 & (m_2^+)^{\otimes k} (\SuspU^{3k} \sigma(\kappa)) \\
 &\quad \underset{\mathclap{\substack{\uparrow\rule{0pt}{1.7ex} \\ \ \ \ \Abs{m_2^+} = 3 - n}}}{=} \varepsilon(\sigma, \kappa) (m_2^+)^{\otimes k} (\SuspU^{3k} \kappa_{\sigma_1^{-1}} \otimes \dotsb \otimes \kappa_{\sigma_{3k}^{-1}})  \\
 & \quad = \begin{multlined}[t] (-1)^{\frac{1}{2}k(k-1)n}\varepsilon(\sigma,\kappa) (m_2^+)^{\otimes k}\bigl(\SuspU^3(\kappa_{\sigma_1^{-1}} \otimes \kappa_{\sigma_2^{-1}} \otimes \kappa_{\sigma_3^{-1}}) \otimes \dotsb \\ \otimes \SuspU^3(\kappa_{\sigma_{3k-2}^{-1}} \otimes \kappa_{\sigma_{3k-1}^{-1}} \otimes \kappa_{\sigma_{3k}^{-1}}) \bigr)\end{multlined}\\
 &\quad = \begin{multlined}[t]\overbrace{(-1)^{\frac{1}{2} k(k-1) n  + \kappa_{\sigma^{-1}_{2}} + \dotsb + \kappa_{\sigma^{-1}_{3k-1}}} \varepsilon(\sigma,\kappa)}^{\eqqcolon\varepsilon_2} (m_2^+)^{\otimes k} \bigl((\SuspU \kappa_{\sigma^{-1}_1} \otimes \SuspU \kappa_{\sigma^{-1}_2} \\ \otimes \SuspU \kappa_{\sigma^{-1}_3}) \otimes \dotsb \otimes (\SuspU \kappa_{\sigma^{-1}_{3k-2}} \otimes \SuspU \kappa_{\sigma^{-1}_{3k-1}} \otimes \SuspU \kappa_{\sigma^{-1}_{3k}})\bigr). \end{multlined}
\end{align*}
Next, using the formula \eqref{Eq:ChernSimons} for $m_2^+$, we get
\begin{align*}
&(m_2^+)^{\otimes k}\bigl((\SuspU \kappa_{\sigma^{-1}_{1}} \otimes \SuspU \kappa_{\sigma^{-1}_{2}} \otimes \SuspU \kappa_{\sigma^{-1}_{3}}) \otimes \dotsb \otimes (\SuspU \kappa_{\sigma^{-1}_{3k-2}} \otimes \SuspU \kappa_{\sigma^{-1}_{3k-1}} \otimes \SuspU \kappa_{\sigma^{-1}_{3k}})\bigr) \\
&\quad= \begin{multlined}[t](-1)^{\kappa_{\sigma^{-1}_{2}} + \dotsb + \kappa_{\sigma^{-1}_{3k-1}}} \Bigl(\int_{x_1} \kappa_{\sigma^{-1}_{1}}(x_1)\kappa_{\sigma^{-1}_{2}}(x_1)\kappa_{\sigma^{-1}_{3}}(x_1) \Bigr) \dotsm \\ \Bigl(\int_{x_k} \kappa_{\sigma^{-1}_{3k-2}}(x_k)\kappa_{\sigma^{-1}_{3k-1}}(x_k) \kappa_{\sigma^{-1}_{3k}}(x_k) \Bigr)\end{multlined} \\
&\quad= \begin{multlined}[t](-1)^{\kappa_{\sigma^{-1}_{2}} + \dotsb + \kappa_{\sigma^{-1}_{3k-1}}} \int_{x_1,\dotsc,x_k} \kappa_{\sigma_{1}^{-1}}(x_1)\kappa_{\sigma_{2}^{-1}}(x_1)\kappa_{\sigma_{3}^{-1}}(x_1) \dotsm \\ \kappa_{\sigma_{3k-2}^{-1}}(x_k) \kappa_{\sigma_{3k-1}^{-1}}(x_k) \kappa_{\sigma_{3k}^{-1}}(x_k) \end{multlined}\\
&\quad = \begin{multlined}[t]\overbrace{(-1)^{\kappa_{\sigma^{-1}_{2}} + \dotsb + \kappa_{\sigma^{-1}_{3k-1}}} \varepsilon(\sigma,\kappa)}^{\eqqcolon\varepsilon_3} \int_{x_1,\dotsc,x_k} \kappa_{1}(x_{\xi(\sigma_{1})})\kappa_{2}(x_{\xi(\sigma_{2})}) \kappa_{3}(x_{\xi(\sigma_{3})}) \dotsm \\ \kappa_{3k-2}(x_{\xi(\sigma_{3k-2})}) \kappa_{3k-1}(x_{\xi(\sigma_{3k-1})})\kappa_{3k}(x_{\xi(\sigma_{3k})}), \end{multlined}
\end{align*}
where $\xi(3j-2) = \xi(3j-1) = \xi(3j) = j$ for $j=1$, $\dotsc$, $k$ (see Definition~\ref{Def:PushforwardMCdeRham}). In total, we have 
\begin{align*} &(m_2^+)^{\otimes k} \bigl(\sigma((\SuspU^{ 2} \Prpg)^{\otimes e} \otimes \omega)\bigr) \\ 
&\quad = \begin{multlined}[t]\varepsilon_1 \varepsilon_2 \varepsilon_3  \int_{x_1,\dotsc,x_k} \Prpg(x_{\xi(\sigma_1)},x_{\xi(\sigma_2)}) \dotsm  \Prpg(x_{\xi(\sigma_{2e-1})},x_{\xi(\sigma_{2e})}) \\ \alpha_{11}(x_{\xi(\sigma_{2e+1})}) \dotsm \alpha_{ls_{l}}(x_{\xi(\sigma_{2e+s})}),\end{multlined}
\end{align*}
where
\[ \varepsilon_1 \varepsilon_2 \varepsilon_3 = (-1)^{\sigma + s e (n-1)+ \frac{1}{2}k(k-1)n} \varepsilon(\SuspU,\eta). \]
Using~\eqref{Eq:EulerFormula} and~\eqref{Eq:TrivalentFormula} and $\varepsilon(\SuspU,\eta) = (-1)^{\Prpg(\omega)}$, we get the total sign
\[ (-1)^{k(n-2)}\varepsilon_1 \varepsilon_2 \varepsilon_3  = (-1)^{s(k,l) + \sigma + \Prpg(\omega)},  \]
where $(-1)^{k(n-2)}$ is the artificial sign from~\eqref{Eq:PushforwardMC}. This proves the proposition.
\end{proof}

\Add[noline,caption={Other de Rham formulas}]{What are the de Rham formulas for $\HTP_{klg}$, $\HTP^\MC_{klg}$, and so on? What kind of Feynman graphs? See Page 99 in Diary I.}

\begin{Remark}[Signs for the Fr\'echet $\dIBL$-structure on $\DR(M)$]\label{Rem:Frechet}
In \cite[Section~13]{Cieliebak2015}, they consider the weight-graded nuclear Fr\'echet space $\DBCyc \DR(M)_{\infty} \subset \DBCyc \DR(M)$ generated by $\varphi\in \DBCyc \DR(M)$ which have a smooth Schwartz kernel $k_\varphi\in \DR(M^{\times k})$; they showed that there is a canonical Fr\'echet $\dIBL$-structure on $\DBCyc \DR(M)_{\infty}[2-n]$. In order to deduce the signs, we can consider the subspace $\DBCyc \DR(M)_{\mathrm{alg}} \subset \DBCyc \DR(M)_\infty$ generated by $\varphi\in \DBCyc \DR(M)$ with an algebraic Schwartz kernel $\Kern_\varphi \in \DR(M)[1]^{\otimes k}$, rewrite \eqref{Eq:DefByGraphs} in terms of $\Kern_\varphi$ and extend the obtained formulas to $\DBCyc \DR(M)_\infty$.
\end{Remark}

\chapter{Reduced cyclic homology of A-infinity-algebras}
\label{App:AInfty}
In this appendix, we prove Proposition~\ref{Prop:Reduced}. The idea from \cite{LodayCyclic} is to resolve cyclic (co)invariants degree-wise and obtain certain bicomplexes with better properties.

In Section~\ref{Sec:FF}, given a strictly unital $\AInfty$-algebra on a graded vector space $V$, we define the normalized and reduced Hochschild (co)chain complexes (Definition~\ref{Def:NormRedHoch}) and prove the computational prerequisites CP1--CP4 (Lemmas~\ref{Lem:CP1}, \ref{Lem:CP2}, \ref{Lem:CP3} and \ref{Lem:CP4}); they are necessary for the development of the cyclic homology theory in the upcoming section. These prerequisites, like squaring to zero of the Hochschild differential, seem to be much harder computationally for $\AInfty$-algebras than for $\DGA$'s. Proofs of some of these relations in different formalisms appeared already in \cite{Mescher2016} and \cite{Lazarev2003}.

In Section~\ref{Sec:HomBi}, we define Loday's and Connes' cyclic half-plane bicomplexes for $\AInfty$-algebras together with their normalized and reduced versions (Definition~\ref{Def:CycBico}). We then summarize some convergence results for spectral sequences associated to horizontal, vertical and diagonal filtrations (Proposition~\ref{Prop:ConvOfSpSeq}). In a series of lemmas (Lemmas~\ref{Lem:LodCycBiCycHom}, \ref{Lem:LodConCycBi}, \ref{Lem:ConNormVer} and~\ref{Lem:ReducedCyclic}), we prove that some of the (co)homologies are isomorphic. These lemmas copy results for $\DGA$'s from \cite{LodayCyclic}; we just do them carefully for half-plane bicomplexes and more explicitly. There is a new phenomenon of long chains coming from completing the direct sum total complex; these long chains seem to disappear in homology if the degrees of $V$ are bounded (Lemma~\ref{Lem:BddDegrees}). Additionally, we point out some differences between first-quadrant and half-plane bicomplexes (Remark~\ref{Rem:SpecSeq}) and mention the relation to mixed complexes (Remark~\ref{Rem:MixedCompl}).

In Section~\ref{Sec:FinRem}, we obtain short exact sequences for reduced Connes' bicomplexes (Lemma~\ref{Lem:ConBiRed}), which replace, up to quasi-isomorphisms, the non-exact sequences for reduced cyclic Hochschild (co)chains. We summarize the isomorphisms of (co)homologies from Section~\ref{Sec:HomBi} (Figure~\ref{Fig:FinalPictureHom}), finish the proof of Proposition~\ref{Prop:Reduced} and formulate a few open question (Questions~\ref{Q:OpenProbAInftx}).
\section{Computational prerequisites} \label{Sec:FF}

The heart of cyclic (co)homology theory, following \cite{LodayCyclic}, are the following five \emph{computational prerequisites} (CP):
\begin{description}
\item[\quad CP0\,{\normalfont (horizontal relations)}:] $\ker \CountOp = \im (\Id-\CycPermOp)$, $\ker(\Id-\CycPermOp) = \im \CountOp$,
\item[\quad CP1\,{\normalfont (vertical relations)}:] $\Hd\circ\Hd = 0$, $\Hd'\circ \Hd' = 0$,
\item[\quad CP2\,{\normalfont (vertical-horizontal relations)}:] $\Hd'\circ \CountOp = \CountOp\circ \Hd$, $(\Id-\CycPermOp)\circ \Hd' = \Hd\circ (\Id-\CycPermOp)$,
\end{description}
and in the strictly unital case
\begin{description}[resume]
\item[\quad CP3\,{\normalfont (null-homotopy of the bar resolution)}:] $\Hd'\circ \InsOneOp_1 + \InsOneOp_1\circ \Hd' = \Id$ and
\item[\quad CP4\,{\normalfont (contraction onto normalized chains)}:] $\NormProj: \HC V \rightarrow \HNC V$ is a quasi-isomorphisms.
\end{description}
The definitions of $\CycPermOp$ (cyclic permutation), $\Hd$ ($\AInfty$-Hochschild differential), $\Hd'$ (acyclic $\AInfty$-Hochschild differential) and $\HC V$ ((reduced) bar complex with reversed grading shifted by one) can be found in Section~\ref{Sec:Alg2}; in particular, consult Definition~\ref{Def:CycHom}. The new players are the \emph{counting operator}
\[ \CountOp \coloneqq \sum_{k=1}^\infty \underbrace{\sum_{i=0}^{k-1} t_k^i}_{\displaystyle{\eqqcolon\CountOp_k}} : \HC V \longrightarrow \HC V \]
and the projection $\bar{p}: \HC V \rightarrow \HNC V$ to normalized Hochschild chains --- this we define below.

\begin{Definition}[Normalized and reduced Hochschild complex]\label{Def:NormRedHoch}
\Correct[noline,caption={DONE Wrong def of normalized}]{The normalized complex is not properly defined!! The projection does not work.}
Let $(V,(\mu_j),\NOne)$ be a strictly unital $\AInfty$-algebra. Let $\bar{V}[1]\coloneqq V[1]/\langle\NOne\rangle$. We define the \emph{normalized Hochschild chain complex} by
\[ \HNC V \coloneqq \bigoplus_{l=0}^\infty V[1]\otimes\bar{V}[1]^{\otimes l}. \]
We consider the canonical projection $\NormProj: V[1]\rightarrow \bar{V}[1]$ and define $\NormProj: \HC V \rightarrow \HNC V$ by
\[ \Restr{\NormProj}{V[1]^{\otimes l}} \coloneqq \Id \otimes \underbrace{\NormProj \otimes \dotsb \otimes \NormProj}_{l-\text{times}}.\]
For every $l\ge 1$, we define the operator $\InsOneOp_l : \HC V \rightarrow \HC V$ by inserting $\NOne$ at the $l$-th position of a tensor product, where the position $l=1$ is in front; i.e., we have
\[ \InsOneOp_1(v_1\otimes\dotsb\otimes v_i) = \NOne\otimes v_1\otimes\dotsb\otimes v_i\quad\text{for all }v_j\in V[1]\text{ and }i\ge j\ge 1.\]
We define the \emph{normalized Hochschild cochain complex} by 
\[ \HNC^* V \coloneqq \{\varphi\in\HC^*V \mid \varphi \circ \InsOneOp_l = 0\text{ for all }l\ge 2\}. \]

If $u: \HC \R \rightarrow \HC V$ is the unit map (in the strictly unital case, $\Restr{u}{\R[1]^{\otimes k}}\coloneqq u^{\otimes k}$ where $u: \R[1]\to V[1]$ is the canonical injection; see also Definition~\ref{Def:AugUnit} and the discussion below) and $\NormIncl: \HNC^*V \rightarrow \HC^* V$ the inclusion, we define the \emph{reduced Hochschild chain and cochain complexes} $\HC^{\RedMRM} V$ and $\HC_{\RedMRM}^* V$ by
\[ \HC^{\RedMRM} V \coloneqq \coker(\NormProj\circ u)\quad\text{and}\quad\HC_{\RedMRM}^*
 \coloneqq \ker(u^* \circ \NormIncl),\quad\text{respectively}.\]
We denote by $p^{\RedMRM}: \HC V \rightarrow \HC^{\RedMRM} V$ and $\iota_{\RedMRM}: \HC^*_{\RedMRM} V \rightarrow \HC^* V$ the canonical projection and inclusion, respectively.

All chain complexes above are graded by degree and equipped with a boundary operator induced naturally from $\Hd$ (see the remark below).
\end{Definition}

\begin{Remark}[Some details on normalized and reduced complexes]
Since $\NOne$ is a unit for~$\mu_2$, we have
\begin{align*}
0 & = (-1)^{\Abs{v_1} + \dotsb + \Abs{v_{k-1}}}\bigl(v_1 \dotsb \mu_2(v_k, \NOne) + (-1)^{\Abs{v_k}}\mu_2(\NOne,v_1)\dotsb v_k \bigr)\quad\text{and}\\
0 & = (-1)^{\Abs{v}_1 + \dotsb + \Abs{v_{i-2}}}\bigl(v_1 \dotsb \mu_2(v_{i-1}, \NOne) v_i \dotsb v_k + (-1)^{\Abs{v_{i-1}}} v_1 \dotsb  v_{i-1}\mu_2(\NOne, v_i) \dotsb v_k \bigr)
\end{align*}
for all $i=2$, $\dotsc$, $k$. This fact and strict unitality implies
\[ \Hd\bigl(\sum_{i\ge 2}\im \InsOneOp_i\bigr)\subset\sum_{i\ge 2}\im \InsOneOp_i = \ker \NormProj.\]
Therefore, $\Hd$ induces a differential on $\HNC V$. Since $\HNC^* V = \{\varphi\in \HC^*V \mid \varphi(\sum_{i\ge 2}\im\InsOneOp_i) = 0\}$, the dual $\Hd^*$ restricts to $\HNC^* V$. Clearly, both $\NormProj$ and $\NormIncl$ are chain maps, and they are compatible under the dualization from Definition~\ref{Def:Pairings}; i.e., $\NormIncl \simeq \NormProj^*$ under $\HNC^* V \simeq (\HNC V)^{\GD}$ and $\HC^* V \simeq (\HC V)^{\GD}$, where $^{\GD}$ denotes the graded dual.
 
As for the reduced complexes, $u$ is a chain map, and thus $\ker$ and $\coker$ are chain complexes. Again, it holds $\HC_{\RedMRM}^* V \simeq (\HC^{\RedMRM} V)^{\GD}$ and $\iota_\RedMRM \simeq p^{\RedMRM,*}$ under the dualization.
\end{Remark}

We will now prove CP1, CP2, CP3 and CP4 for strictly unital $\AInfty$-algebras. We do not prove CP0 because it is a standard fact which does not depend on the algebra we work with (see \cite{LodayCyclic}). A proof of CP1 in a slightly different notation and in a more general setting (coefficients in a bimodule) can also be found in \cite{Mescher2016}. The proofs of CP2 and CP3 work in the same way as the proofs for $\DGA$'s from \cite{LodayCyclic}. The computation is just a little longer. As for CP4, we can not use the proof for $\DGA$'s from \cite[Proposition~1.6.5]{LodayCyclic} anymore because we do not have a simplicial module; instead of this, we consider an explicit homotopy inspired by~\cite{Lazarev2003}, where CP4 is also proven in a slightly different notation.

We first introduce some notation which simplifies computations:
\begin{Definition}[Notation]
For the cyclic permutation $\CycPermOp_k^i$ ($\coloneqq i$-times $t_k$), we define $c\coloneqq k - i + 1$ and write
\[ v_c \dotsb v_{c-1} \coloneqq \CycPermOp_k^i(v_1\otimes \dotsb \otimes v_k). \] 
We compute indices modulo $k$ and often omit writing the tensor product.

For every $i = 1$,~$\dotsc$, $k$ and $j = 1$,~$\dotsc$, $k-i+ 1$, we define the \emph{closed bracket} by
\begin{equation} \label{Eq:Inclmu} \begin{aligned} 
&v_1 \dotsb \MuII{v_{i} \dotsb v_{i + j - 1}} \dotsb v_k \\
&\qquad \coloneqq (-1)^{\Abs{v_1} + \dotsb + \Abs{v_i-1}} v_1 \otimes \dotsb \otimes \mu_j(v_i \otimes \dotsb \otimes v_{i+j-1})\otimes \dotsb \otimes v_k.
\end{aligned} \end{equation}
If we apply the closed bracket two-times, we write the first application as an \emph{underbracket} and the second as an \emph{overbracket}; for instance, we have
\[ \begin{aligned} & v_1 \dotsb \MuII{v_{i_1} \dotsb v_{i_2}} \dotsb \MuI{v_{i_3}\dotsb v_{i_4}} \dotsb v_k \\ 
& \quad =  \begin{multlined}[t] (-1)^{\Abs{v_{i_1}} + \dotsb + \Abs{v_{i_3-1}}}
 v_1 \otimes \dotsb \otimes \mu_{j_2}(v_{i_1}\otimes \dotsb \otimes v_{i_2})\otimes \dotsb \\ \otimes \mu_{j_1}(v_{i_3}\otimes \dotsb \otimes v_{i_4})\otimes \dotsb \otimes v_k,
\end{multlined}\end{aligned}\]
where $j_1 = i_4 - i_3 + 1$, $j_2 = i_2 - i_1 + 1$. Clearly, the difference is only in the sign. We denote
 \[ v_1 \dotsb v_{i-1} \NOneII v_{i} \dotsb v_k \coloneqq \InsOneOp_i(v_1 \dotsb v_k) = (-1)^{\Abs{v_1}+\dotsb+\Abs{v_{i-1}}} v_1\dotsb v_{i-1} \NOne v_i \dotsb v_k. \]
 If $\InsOneOp_i$ is composed with an other operation, we write $\NOneI$ if the corresponding $\NOne$ was inserted first and $\NOneII$ if it was inserted second. For example, we have
 \[ v_1 \NOneI v_2 v_3 \NOneII v_4 = \InsOneOp_5(\InsOneOp_2(v_1 v_2 v_3 v_4)). \]

For $j\ge 1$ and $1\le i_1 \le i_2 \le k$ with $i_2 - i_1 \ge j$, we define the \emph{open bracket} as follows:
 \[ v_1 \dotsb \OMuIIO[j]{v_{i_1} \dotsb v_{i_2}} \dotsb v_k \coloneqq \sum_{\substack{i_1\le i_3 \le i_4 \le i_2 \\ i_4 - i_3 = j}} v_1 \dotsb v_{i_1} \dotsb \MuII{v_{i_3}\dotsb v_{i_4}} \dotsb v_{i_2} \dotsb v_k. \]
\end{Definition}

Using the notation above, it holds
\[ \begin{aligned}
b'(v_1\dotsb v_k) &= \sum_{1\le i_1 \le i_2 \le k} v_1 \dotsb \MuII{v_{i_1} \dotsb v_{i_2}} \dotsb v_k = \sum_{j=1}^k \OMuIIO[j]{v_1\dotsb v_k}, \\
R(v_1\dotsb v_k) & = \sum_{\substack{2 \le c \le k}} \MuII{v_{c} \dotsb v_{1}} \dotsb v_{c-1},
\end{aligned} \]
and the $\AInfty$-relations simplify to
\begin{equation} \label{Eq:AInftyCyclic}
\sum_{1 \le i_1 \le  i_2 \le k} \MuII{v_{1}\dotsb \MuI{v_{i_1} \dotsb v_{i_2} } \dotsb v_{k}} = \sum_{j=1}^k \MuII{\OMuIO[j]{v_1\dotsb v_k}} = 0. 
\end{equation}
\noindent Because all signs are, in fact, Koszul signs for the symbols $\mu_{j_1}$, $\mu_{j_2}$, $v_1$, $\dotsc $, $v_k$, and because $\mu$'s have odd degree, we have for every $1 \le i_1 \le i_2 \le i_3 \le i_4 \le k$ the following relation:
\begin{equation} \label{Eq:OddDeg}
v_{1}\dotsb \MuI{v_{i_1} \dotsb v_{i_2}} \dotsb \MuII{v_{i_3} \dotsb v_{i_4}} \dotsb v_{k} +  v_{1}\dotsb \MuII{v_{i_1} \dotsb v_{i_2}} \dotsb \MuI{v_{i_3} \dotsb v_{i_4}} \dotsb v_{k} = 0.
\end{equation}

\begin{Lemma}[CP1] \label{Lem:CP1}
For an $\AInfty$-algebra $(V,(\mu_j))$, it holds
\[ \Hd'\circ \Hd' = 0\quad\text{and}\quad \Hd \circ \Hd = 0. \]
\end{Lemma}
\begin{proof}
We write
\[ \Hd\circ\Hd  = (\Hd' + R)\circ(\Hd' + R) = \Hd'\circ \Hd' + \Hd' \circ R + R \circ \Hd' + R \circ R \]
and evaluate it on a tensor $v_1\dotsb v_k \in \HC V$. We claim that a summand of $\Hd(\Hd(v_1\dotsb v_k))$ coming from the subsequent application of the operations can be uniquely determined by the following data: 
\begin{itemize}
\item the information whether it comes from $\Hd'\circ \Hd'$, $\Hd'\circ R$, $R\circ \Hd'$ or $R\circ R$;
\item a cyclic permutation $c$ of $v_1$, $\dotsc$, $v_k$;
\item positions of the under- and upperbracket.
\end{itemize}
The reason for this is that both $\Hd'$ and $R$ produce only Koszul signs, and hence the total sign of a summand in $\Hd(\Hd(v_1\dots v_k))$ is the Koszul sign for the symbols $\mu_{j_1}$, $\mu_{j_2}$, $v_1$,~$\dotsc$,~$v_k$, which depends only on the start and final position of the symbols; this is precisely encoded in the data above.

For $c=1$, only $\Hd'\circ\Hd'$ contributes; however, using \eqref{Eq:AInftyCyclic} and \eqref{Eq:OddDeg}, we get
\begin{align*}
(\Hd'\circ \Hd')(v_{1}\dotsb v_{k}) &= \begin{aligned}[t] &\sum_{1\le i_1 \le i_2 \le i_3 \le i_4\le k} v_{1}\dotsb \MuII{v_{i_1} \dotsb \MuI{v_{i_2} \dotsb v_{i_3}} \dotsb v_{i_4}} \dotsb v_{k} \\ {}+ &\sum_{\substack{1\le i_1 \le i_2 \le {k-1} \\ i_2+1 \le i_3 \le i_4 \le {k} }} v_{1}\dotsb \MuI{v_{i_1} \dotsb v_{i_2}} \dotsb \MuII{v_{i_3} \dotsb v_{i_4}} \dotsb v_{k} \\ {}+&\sum_{\substack{2\le i_3 \le i_4 \le {k} \\ 1 \le i_1 \le i_2 \le i_3 - 1 }} v_{1}\dotsb \MuII{v_{i_1} \dotsb v_{i_2}} \dotsb \MuI{v_{i_3} \dotsb v_{i_4}} \dotsb v_{k} \end{aligned} \\
 & = 0.
\end{align*}
Let $c \ge 2$ and consider the summands from $\Hd'\circ R$, $R\circ \Hd'$ and $R\circ R$ based on $v_{c} \dots v_{c-1}$. The contribution of $R\circ \Hd'$ consists of the following three parts:
\begin{EqnList}
\item $\displaystyle \sum_{\substack{c \le i_1 \le i_2 \le k \\ 1 \le i_4 \le c-1}}\MuII{v_{c} \dotsb \MuI{v_{i_1} \dotsb v_{i_2}} \dotsb v_k v_1 \dotsb v_{i_4}} v_{i_4+1} \dotsb v_{c-1}$,
\item $\displaystyle\sum_{\substack{1\le i_1 \le i_2 \le i_4 \le c-1}}\MuII{v_{c} \dotsb v_k v_1 \dotsb  \MuI{v_{i_1} \dotsb v_{i_2}} \dotsb v_{i_4}} v_{i_4+1} \dotsb v_{c-1}$,
\item $\displaystyle \sum_{1\le i_4 \prec i_1 \le i_2 \le c-1} \MuII{v_{c} \dotsb v_k v_1 \dotsb v_{i_4}} v_{i_4+1} \dotsb \MuI{v_{i_1} \dotsb v_{i_2}} \dotsb v_{c-1}$.
\end{EqnList}
The contribution of $\Hd'\circ R$ consists of the following two parts:
\begin{EqnList}[resume]
\item $\displaystyle\sum_{1 \le i_2 \le i_4 \le c-1}\MuII{\MuI{v_{c} \dotsb v_k v_1 \dotsb v_{i_2}} \dotsb v_{i_4}} v_{i_4+1}\dotsb v_{c-1}$,
\item $\displaystyle\sum_{1\le i_2 \prec i_3 \le i_4 \le c-1}\MuI{v_{c} \dotsb v_k v_1 \dotsb v_{i_2}} v_{i_2+1} \dotsb \MuII{v_{i_3} \dotsb v_{i_4}}\dotsb v_{c-1}$.
\end{EqnList}
The contribution of $R \circ R$ is:
\begin{EqnList}[resume]
\item $\displaystyle\sum_{\substack{c \prec i_1 \le k\\1\le i_2 \le i_4 \le c-1}}\MuII{v_{c} \dotsb \MuI{v_{i_1}\dotsb v_k v_1\dotsb v_{i_2}} \dotsb v_{i_4}} v_{i_4+1} \dotsb v_{c-1}$.
\end{EqnList}
Using \eqref{Eq:OddDeg}, it is easy to see that III cancels with V. The sum of the other terms I, II, IV, VI vanishes for fixed $1 \le i_4 \le c-1$ due to \eqref{Eq:AInftyCyclic}.
\end{proof}

\begin{Lemma}[CP2] \label{Lem:CP2}
For an $\AInfty$-algebra $(V,(\mu_j))$, the following relations hold:
\begin{ClaimList}
\item $\Hd'\circ \CountOp = \CountOp\circ \Hd$,
\item $(\Id-\CycPermOp)\circ \Hd' = \Hd\circ (\Id-\CycPermOp)$.
\end{ClaimList}
\end{Lemma}
\begin{proof}
\begin{ProofList}
\item We denote $z_j \coloneqq \mu_j \otimes \Id^{k-j}$ and omit writing the composition $\circ$. We consider the components
\[ {\Hd'}_k^j \coloneqq \sum_{i=0}^{k-j} \CycPermOp^i_{k-j+1}z_j\CycPermOp_k^{-i} \qquad\text{and}\qquad R_k^j \coloneqq \sum_{i=1}^{j-1} z_j\CycPermOp_k^i. \]
It holds
\begin{align*} 
{\Hd'}_k^j \CountOp_k &= \sum_{l=0}^{k-1}\sum_{i=0}^{k-j} \CycPermOp^i_{k-j+1} z_j \CycPermOp^{-i + l}_k \\
& = \sum_{l=0}^{k-1} \sum_{i=0}^{k-j} \CycPermOp^l_{k-j+1} \CycPermOp^{i-l}_{k-j+1} z_j \CycPermOp^{-(i-l)}_k \\
& = \sum_{u=1-k}^{k-j} \bigl(\sum_{l\in L_u} \CycPermOp^l_{k-j+1}\bigr) \CycPermOp^u_{k-j+1} z_j \CycPermOp^{-u}_k,
\end{align*}
where $u \coloneqq i - l$ and 
\[L_u \coloneqq \{ l\in \{0, \dotsc, k-1\} \mid \exists i\in \{0,\dotsc, k-j\} : u = i-l \}. \] 
We distinguish the cases
\[ L_u = \begin{cases}
         \{0,\dotsc, k-j-u\} & \text{for }0\le u \le k-j, \\
         \{-u, \dotsc, k-j-u\} & \text{for } 1-j \le u \le -1\quad\text{and} \\
         \{-u, \dotsc, k -1\} & \text{for }1-k \le u \le -j
         \end{cases}\]
and denote the corresponding sums by $\mathrm{I}$, $\mathrm{II}$ and $\mathrm{III}$, respectively. It holds         
\[ \mathrm{I} = \sum_{u=0}^{k-j} \bigr(\sum_{l=0}^{k-j-u} \CycPermOp^l_{k-j+1} \bigl) \CycPermOp^u_{k-j+1} z_j \CycPermOp^{-u}_k \]
and  
\begin{align*}
\mathrm{III} &= \sum_{u = 1- k}^{-j} \sum_{l = -u}^{k-1} \CycPermOp^l_{k-j+1} \CycPermOp^u_{k-j+1} z_j \CycPermOp^{-u}_k \\
& = \sum_{u = 1- k}^{-j} \sum_{l = -u}^{k-1} \CycPermOp^l_{k-j+1} \CycPermOp^{u+k-j+1}_{k-j+1} z_j \CycPermOp^{-u-k}_k  \\
&= \sum_{u=1}^{k-j} \sum_{l = k - u}^{k-1} \CycPermOp^{l-j+1}_{k-j+1} \CycPermOp^u_{k-j+1} z_j \CycPermOp^{-u}_k \\ 
& = \sum_{u=1}^{k-j} \bigl(\sum_{l =  k - j -u + 1}^{k-j} \CycPermOp^l_{k-j+1}\bigr) \CycPermOp^u_{k-j+1} z_j \CycPermOp^{-u}_k.
\end{align*}
Therefore, we have
\[ \mathrm{I} + \mathrm{III} = \sum_{u=0}^{k-j} \bigl( \sum_{l=0}^{k-j} \CycPermOp^l_{k-j+1} \bigr) \CycPermOp^u_{k-j+1} z_j \CycPermOp^{-u}_k = \CountOp_{k-j+1} {b'}_k^j \]
Next, we have
\[ \mathrm{II} = \sum_{u = 1-j}^{-1} \sum_{l=-u}^{k-j-u} \CycPermOp^l_{k-j+1} \CycPermOp^u_{k-j+1} z_j \CycPermOp^{-u}_k = \sum_{u = 1}^{j-1} \bigl(\sum_{l=0}^{k-j} \CycPermOp^l_{k-j+1}\bigr) z_j \CycPermOp^u_k = \CountOp_{k-j+1} R_k^j. \]
We conclude that
\[ {\Hd'}_k^j \CountOp_k = \CountOp_{k-j+1}{\Hd'}_k^j + \CountOp_{k-j + 1} R_k^j = \CountOp_{k-j+1}\Hd_k. \]
This proves the claim.
\item For every $k\ge 1$ and $1\le j \le k$ we compute
\begin{align*}
(\Id-\CycPermOp){\Hd}^j_k &= {\Hd'}_k^j - \sum_{i=0}^{k-j} \CycPermOp^{i+1}_{k-j+1}z_j \CycPermOp^{-(i+1)}_k\CycPermOp_k\\
& = {\Hd'}_k^j - \sum_{i=1}^{k-j + 1} \CycPermOp^{i}_{k-j+1}z_j \CycPermOp^{-i}_k\CycPermOp_k \\
& = {\Hd'}_k^j - \sum_{i=0}^{k-j} \CycPermOp^{i}_{k-j+1}z_j \CycPermOp^{-i}_k \CycPermOp_k + \CycPermOp^0_{k-j+1} z_j \CycPermOp^{-0}_k \CycPermOp_k - \CycPermOp^{k-j+1}_{k-j+1} z_j \CycPermOp^{-k+j-1}_k \CycPermOp_k \\
& = {\Hd'}_k^j(\Id-\CycPermOp_k)  + z_j \CycPermOp_k - z_j \CycPermOp^{j}_k  \\ 
&= {\Hd'}_k^j(\Id-\CycPermOp_k) + \sum_{i=1}^{j-1} z_j \CycPermOp^i_k (\Id-\CycPermOp_k) \\
& = ({\Hd'}_k^j + {R}_k^j)(\Id-\CycPermOp_k). 
\end{align*} 
This proves the claim. \qedhere
\end{ProofList}
\end{proof}

\begin{Lemma}[CP3] \label{Lem:CP3}
For a strictly unital $\AInfty$-algebra $(V,(\mu_k),\NOne)$, it holds
\[ \Hd'\circ \InsOneOp_1 + \InsOneOp_1 \circ \Hd' = \Id. \]
\end{Lemma}
\begin{proof}
For any $k\ge 1$ and $v_1$, $\dotsc$, $v_k \in V[1]$, we compute
\[ \begin{aligned} 
\Hd' \InsOneOp_1(v_1\dots v_k) + \InsOneOp_1 \Hd' (v_1\dots v_k) &= \MuII{\NOneI v_1} v_2 \dots v_k  + \NOneI \OMuIIO{v_1 \dots v_k} + \NOneII \OMuIO{v_1 \dots v_k} \\ &= v_1 \dots v_k. 
\end{aligned} \]
This proves the claim.
\end{proof}

\begin{Lemma}[CP4] \label{Lem:CP4}
Let $\mathcal{A} = (V,(\mu_k),\NOne)$ be a strictly unital $\AInfty$-algebra. For all $k\ge 2$, we define $h_k: \HC V \rightarrow \HC V$ by
\[ h_k\coloneqq \InsOneOp_k \circ \Hd + \Hd\circ \InsOneOp_k + \Id. \]
Then the formulas
\begin{align*}
    s^* &\coloneqq  s^*_2 + s^*_3 \circ h^*_2 + s^*_4 \circ h^*_3 \circ h_2^* + \dotsb\quad\text{and} \\
    h^* &\coloneqq \dotsb\circ h^*_k\circ\dotsb\circ h^*_2
\end{align*}
define homogenous linear maps~$s^*$ and $h^*: \HC^*V \rightarrow \HC^* V$ of degrees~$1$ and~$0$, respectively. The map $h^*$ is a projection onto $\HNC^* V$, and the following homotopy relation holds:
\begin{equation} \label{Eq:Homot}
s^* \circ \Hd^* + \Hd^* \circ s^* = h^* - \Id.
\end{equation}
It implies that $\NormIncl$ and $\NormProj$ are quasi-isomorphisms.
\end{Lemma}

\begin{proof}
We set $\HNC^*_{(1)} V\coloneqq \HC^* V$, and for all $k\ge 2$, we define
\[ \HNC^*_{(k)} V \coloneqq \{\psi\in \HC^* V \mid \psi \circ \InsOneOp_i = 0\text{ for all }i=2, \dotsc, k\}. \]
We will show first that $h_k^*$ restricts to a projection $\HNC^*_{(k-1)}V \rightarrow \HNC^*_{(k)} V$.  Let $i \ge 1$ and $v_1$,~$\dotsc$, $v_i\in V[1]$. We make the following computations:
\begin{PlainList}
\item For $i<k-1$, we have
\[(\InsOneOp_i \Hd + \Hd \InsOneOp_i)(v_1 \dots v_i) = 0 \]
by the definition of $\InsOneOp_ik$ and by the fact that $\Hd$ does not increase weights.
\item For $i = k-1$, we have
\begin{align*}
& (\InsOneOp_i \Hd + \Hd \InsOneOp_i)(v_1 \dots v_i) \\
&\quad = \OMuIO[$1$]{v_1 \dots v_i}\NOneII  + \OMuIIO[$1$]{v_1\dots v_i}\NOneI + \sum_{j=2}^{i} \OMuIIO[$j$]{v_1\dots v_i}\NOneI + v_1\dots \MuII{v_i \NOneI} + \MuII{\NOneI v_1} \dots v_i \\
& \quad = \sum_{j=2}^{i} \OMuIIO[$j$]{v_1\dots v_i}\NOneI.
\end{align*}
Notice that $\NOne$ in the result is at positions $<k$.
\item For $i> k-1$, we have
\begin{align*} 
&(\InsOneOp_i \Hd + \Hd \InsOneOp_i)(v_1 \dots v_i) \\
& \quad  =  \begin{aligned}[t] &\hphantom{+}\sum_{j = 1}^{i-k+2} \OMuIO[$j$]{v_1 \dots v_{k+j-2}} \NOneII v_{k+j-1} \dots v_i + \sum_{j=1}^{i-k+1} v_1 \dots v_{k-1} \NOneII \OMuIO[$j$]{v_{k}\dots v_i} \\
&{}+ \sum_{m=1}^{i-k+1} \sum_{c=m+k-1}^i \MuI{v_c \dots v_i v_1 \dots v_{m}}v_{m+1}\dots v_{m+k-2} \NOneII v_{m+k-1} \dots v_{c-1}  \\
&{}+ \sum_{j=1}^{k-1} \OMuIIO[j]{v_1\dots v_{k-1}} \NOneI v_{k}\dots v_i + \sum_{j=1}^{i-k+1} v_1 \dots v_{k-1} \NOneI \OMuIIO[j]{v_{k}\dots v_i} \\
&{}+v_1\dots \MuII{v_{k-1} \NOneI} v_{k} \dots v_i + v_1 \dots v_{k-1} \MuII{\NOneI v_{k}}\dots v_i 
\\
&{}+ \sum_{m=1}^{k-1} \sum_{c=k}^i \MuII{v_c \dots v_i v_1 \dots v_{m}}v_{m+1}\dots v_{k-1} \NOneI v_{k} \dots v_{c-1} \end{aligned} \\
& \quad= \begin{aligned}[t] &\hphantom{+}\sum_{j = 1}^{i-k+2} \OMuIO[$j$]{v_1 \dots v_{k+j-2}} \NOneII v_{k+j-1} \dots v_i + \sum_{j=1}^{k-1} \OMuIIO[j]{v_1\dots v_{k-1}} \NOneI v_{k}\dots v_i \\
&{}+ \sum_{m=1}^{i-k+1} \sum_{c=m+k-1}^i \MuI{v_c \dots v_i v_1 \dots v_{m}}v_{m+1}\dots v_{m+k-2} \NOneII v_{m+k-1} \dots v_{c-1} \\
&{}+\sum_{m=1}^{k-1} \sum_{c=k}^i \MuII{v_c \dots v_{i} v_1 \dots v_{m}}v_{m+1}\dots v_{k-1} \NOneI v_{k} \dots v_{c-1}
\end{aligned} \\
&\quad = \begin{aligned}[t]
&\hphantom{+}\overbrace{\sum_{j = 2}^{i-k+2} \OMuIO[$j$]{v_1 \dots v_{k+j-2}} \NOneII v_{k+j-1} \dots v_i}^{\eqqcolon\mathrm{I}} + \overbrace{\sum_{j=2}^{k-1} \OMuIIO[j]{v_1\dots v_{k-1}} \NOneI v_{k}\dots v_i}^{\eqqcolon\mathrm{II}} \\
&{}+ \overbrace{\sum_{m=2}^{i-k+1} \sum_{c=m+k-1}^i \MuI{v_c \dots v_i v_1 \dots v_{m}}v_{m+1}\dots v_{m+k-2} \NOneII v_{m+k-1} \dots v_{c-1}}^{\eqqcolon\mathrm{III}} \\
&{}+\overbrace{\sum_{m=2}^{k-1} \sum_{c=k}^i \MuII{v_c \dots v_i v_1 \dots v_{m}}v_{m+1}\dots v_{k-1} \NOneI v_{k} \dots v_{c-1}}^{\eqqcolon\mathrm{IV}}
\end{aligned}
\end{align*}
Notice that $\NOne$ is at the $k$-th position in $\mathrm{I}$ and $\mathrm{III}$, whereas at positions $<k$ in $\mathrm{II}$ and $\mathrm{IV}$.
\end{PlainList}
Let $k\ge 2$, and let $\psi\in \HNC_{(k-1)}^* V$. In order to show that $\InsOneOp_j^* h_k^* \psi = 0$ for $2 \le j \le k$, let $i\ge j$, and let $v_1$,~$\dotsc$, $v_{i}\in V[1]$ be such that $v_{j} = \NOne$.
Clearly, $\psi(\mathrm{II}) = \psi(\mathrm{IV}) = 0$ for any~$v$'s. As for $\mathrm{III}$, the vector $v_j=\NOne$ lies either inside $\mu_j$ with $j\ge 3$ or at a position $<k$. It follows that $\psi(\mathrm{III})=0$. As for $\mathrm{I}$, we write 
\[ \mathrm{I} = \overbrace{\OMuIO[2]{v_1 \dotsb v_k} \NOneII v_{k+1} \dotsb v_i}^{\mathrm{Ia}} + \overbrace{\sum_{j\ge 3}^{i-k+2} \OMuIO[j]{v_1\dotsb v_{k+j-2}}\NOneII v_{k+j-1}\dotsb v_i}^{\mathrm{Ib}}. \]
It holds $\psi(\mathrm{Ib})=0$. For $2\le j<k$, it holds  
\begin{align*}
\psi(\mathrm{Ia}) &= \begin{multlined}[t]\psi(v_1 \dots \MuI{v_{j-1}\NOne} v_{j+1} \dots v_k \NOneII v_{k+1} \dots v_i) \\{}+ \psi(v_1 \dots v_{j-1}\MuI{\NOne v_{j+1}} \dots v_k \NOneII v_{k+1} \dots v_i) \end{multlined} \\ 
&= 0,
\end{align*}
whereas for $j=k$, we have
\[ \psi(v_1 \dotsb \MuI{v_{k-1} \NOne} \NOneII v_{k+1} \dotsb v_i) = - \psi(v_1 \dots v_{i}). \]
It follows that
\begin{equation}
 h_k^*\psi(v_1 \dotsb v_i) = \psi(v_1\dotsb v_i) + \psi\bigl((\InsOneOp_i \Hd + \Hd \InsOneOp_i)(v_1\dotsb v_i)\bigr) = 0.
\end{equation}
Therefore, we have $h^*_k \psi \in \HNC_{(k)}^* V$. If $\psi\in \HNC_{(k)}^* V$, then clearly $h_{k}^*(\psi) = \psi$. Consequently,~$h_k^*$ is a projection $h^*_k: \HNC_{(k-1)}^*V \rightarrow \HNC_{(k)}^* V$.

For $k\ge 2$, we define
\begin{align*}
 \leftidx{^k}{s}{^*} &\coloneqq \InsOneOp_2^* +  \InsOneOp_3^* \circ h_2^* + \dotsb + \InsOneOp_k^* \circ h_{k-1}^* \circ \dotsb \circ h_2^*, \\
 \leftidx{^k}{h}{^*} &\coloneqq h_k^* \circ \dotsb \circ h_2^*.
\end{align*}
Let $\Filtr^n_{\WeightMRM} \HC V = \bigoplus_{k=1}^n \HC_k V$ be the filtration of $\HC V$ by weights. For all $k\ge n+1$ it holds
\[ \Restr{\leftidx{^{k}}{h}{^*}\psi}{\Filtr^n_{\WeightMRM} \HC V} = \leftidx{^{n+1}}{h}{^*}\psi\quad\text{and}\quad \Restr{\leftidx{^k}{s}{^*}\psi}{\Filtr^n_{\WeightMRM} \HC V} = \leftidx{^{n+1}}{s}{^*}\psi. \]
It follows that $h^*$, $s^*: \HC^* V \rightarrow \HC^* V$ are well-defined and that $h^*$ is a projection onto~$\HNC^* V$. Also, it suffices to prove  \eqref{Eq:Homot} with~$\leftidx{^k}{s}{^*}$ and~$\leftidx{^k}{h}{^*}$ instead of~$s^*$ and~$h^*$ for each~$k\ge 2$. For~$k=2$, it holds by definition. Suppose that \eqref{Eq:Homot} holds for some~$k\ge 2$. Then
\begin{align*}
&\Hd^*\circ (\leftidx{^{k+1}}{s}{^*}) + (\leftidx{^{k+1}}{s}{^*})\circ \Id \\
&\quad = \Hd^*\circ (\leftidx{^{k}}{s}{^*}) + (\leftidx{^{k}}{s}{^*}) \circ \Hd^* + \Hd^*\circ \InsOneOp_{k+1}^*\circ h_k^*\circ \dotsb \circ h_2^* + \InsOneOp_{k+1}^*\circ h_k^*  \circ \dotsb \circ h_2^* \circ \Hd^* \\
&\quad = \leftidx{^k}{h}{^*}- \Id + (\Hd^*\circ \InsOneOp_{k+1}^* + \InsOneOp_{k+1}^*\circ \Hd^*)\circ h_k^* \circ \dotsb \circ h_2^*  \\
&\quad = \leftidx{^k}{h}{^*}- \Id + (h_{k+1}^* - \Id)\circ h_k^* \circ \dotsb \circ h_2^* \\
&\quad = \leftidx{^{k+1}}{h}{^*} - \Id.
\end{align*}
The lemma is finally proven.\qedhere
\end{proof}

\section{Homological algebra of bicomplexes}\label{Sec:HomBi}
We will consider homological and cohomological half-plane bicomplexes, which we depict~as
\Correct[caption={DONE Switsch p q},noline]{Switch p and q! There is more mistakes. It shoudl be just $B_{q,p} \mapsto B_{q,p}^*$ in the picture, not a change of numbering }
\Add[caption={DONE Why completion arises for half-plane},noline]{Give here the example with the snake of $\R$'s.}
\begin{center}
$B:\ $\begin{tikzcd}[column sep=scriptsize, row sep=scriptsize]
{}\arrow[d]&{}\arrow[d]&{}\arrow[d]&{} \\
B_{q,p+2} \arrow[d] & B_{q+1,p+2} \arrow[l] \arrow[d]  & B_{q+2,p+2} \arrow[l] \arrow[d]  &{}\arrow[l]\\
B_{q,p+1} \arrow[d] & B_{q+1,p+1} \arrow[d] \arrow[l] & \arrow[d] \arrow[l] B_{q+2,p+1}  &{} \arrow[l] \\
B_{q, p} \arrow[d] & B_{q+1,p}\arrow[d] \arrow[l] & B_{q+2,p}\arrow[d]  \arrow[l]  & {}\arrow[l] \\
{}&{}&{} &{}
\end{tikzcd}
\end{center}
and 
\begin{center}
$\quad B^*:\ $\begin{tikzcd}[column sep=scriptsize, row sep=scriptsize]
{}&{}&{} &{}\\
B^{q,p+2} \arrow[u] \arrow[r,]& B^{q+1,p+2} \arrow[u] \arrow[r] & B^{q+2,p+2} \arrow[u,] \arrow[r]  &{}\\
B^{q,p+1} \arrow[u] \arrow[r] & B^{q+1,p+1} \arrow[u] \arrow[r]  & \arrow[u] \arrow[r]  B^{q+2,p+1} &{} \\
B^{q,p} \arrow[u] \arrow[r]  & B^{q+1,p}\arrow[u] \arrow[r]  & B^{q+2,p}\arrow[u]  \arrow[r]  & {} \\
{}\arrow[u]&{}\arrow[u]&{}\arrow[u]&{}
\end{tikzcd},
\end{center}
respectively. The standard convention is that the squares anticommute (see \cite{LodayCyclic})!
 \ToDo[noline,caption={DONE Do bicomplexes commute or anticommute?}]{According to our convention, the squares anti-commute. Boardmann says they should anticommute!!! }
 
We consider the \emph{total complexes} $(\TotI(B), \Bdd)$ and $(\TotII(B),\Bdd)$, where for all $q\in \Z$, the chain groups are defined by
\[ (\TotI B)_q \coloneqq \bigoplus_{i + j =q} B_{i,j},\quad\text{and}\quad (\TotII B)_q \coloneqq \prod_{i + j =q} B_{i,j}, \]
respectively, and where $\Bdd = \Bdd_v + \Bdd_h$ is the total boundary operator consisting of the vertical and horizontal boundary operators $\Bdd_v$ and $\Bdd_h$, respectively. 
The homologies of $\TotI$ and $\TotII$ are denoted by $\H B$ and $\H \hat{B}$, respectively. We proceed similarly in cohomology.\Add[caption={DONE Remark about half-plane bi},noline]{Make a remark why half-plane happen in A-infinity, example of the snake where the spectral sequence does not compute the homology properly, make a remark about dualizations. These are the main problems.}

For each $B$ and $B^*$, we consider the vertical and horizontal filtrations which are defined in such a way that they are preserved by all the arrows and that the $k$-th group contains the $k$-th column and the $k$-th row, respectively. More precisely, the vertical filtration~$\Filtr_{\VertMRM}^k B$ of $B$ consists of the columns $0$,~$\dotsc$, $k$, whereas the vertical filtration~$\Filtr_{\VertMRM}^k B^*$ of $B^*$ consists of the columns $k+1$, $k+2$,~$\dotsc$; the horizontal filtration~$\Filtr^\HorMRM_k B$ of $B$ consists of the rows $k$, $k-1$,~$\dotsc$, whereas the horizontal filtration~$\Filtr_\HorMRM^k B^*$ of $B^*$ consists of the rows $k$, $k+1$,~$\dotsc$, and so~on.

In order to check that morphisms of bicomplexes induce quasi-isomorphisms of total complexes, we will use the techniques of spectral sequences. Because we work with half-plane bicomplexes, our spectral sequences do not lie in the first quadrant, as in~\cite{Weibel1994}, and the notion of conditional convergence from~\cite{Boardmann1999} comes in handy. In the following, we recall some basic theory and formulate a proposition about the convergence of some unbounded spectral sequences.

A \emph{cohomological spectral sequence} is a collection $\SSPage_r$ of bigraded vector spaces and differentials $\SSDiff_{r}: \SSPage_r^{\bullet\bullet} \rightarrow \SSPage_r^{\bullet+r,\bullet-r+1}$ for $r\in \N$ such that $\SSPage_{r+1} = \H(\SSPage_r,\SSDiff_r)$. Let $(C^*,\Dd)$ be a cochain complex with a decreasing filtration $\Filtr^s C^*$ (it has to be graded and preserved by~$\Dd$). For every $s\in \Z$, the short exact sequence
\[ 0\longrightarrow \Filtr^{s+1} C^* \overset{i}{\longrightarrow} \Filtr^s C^* \overset{j}{\longrightarrow} \Gr_s(C^*) \coloneqq \Filtr^{s} C^* / \Filtr^{s+1} C^* \longrightarrow 0 \]
induces the long exact sequence\Modify[caption={DONE Labels on arrows},noline]{Add labels $i$, $j$, $\delta$ to arrows!}
\[\dotsb \longrightarrow \H^\bullet(\Filtr^{s+1} C^*) \overset{i}{\longrightarrow}  \H^\bullet(\Filtr^{s} C^*) \overset{j}{\longrightarrow} \H^\bullet(\Gr_s C^*) \overset{\delta}{\longrightarrow} \H^{\bullet+1}(\Filtr^{s+1} C^*) \longrightarrow \dotsb, \]
which wraps into the exact couple of bigraded vector spaces
\[\begin{tikzcd}
A_1 \coloneqq \bigoplus_{s\in \Z} \H(\Filtr^s C^*)[s]  \arrow{rr}{i} && A_1 \arrow{dl}{j}\\
 & \SSPage_1 \coloneqq \bigoplus_{s\in \Z} \H(\Gr_s C^*)[s] \arrow{ul}{\delta}.
\end{tikzcd}\]
This is the so called \emph{geometric grading} convention.\footnote{It is chosen such that $\SSPage_1^{sd} = \H(B^{sd},\Dd_\VertMRM)$ for the vertical filtration.} We also define $A_1^s \coloneqq \H(\Filtr^s C^*)$ and $\SSPage_1^s \coloneqq \H(\Gr_s C^*)$. By deriving this triangle (see, e.g., \cite{Cieliebak2013}), one obtains a spectral sequence $\SSPage_r$ associated to the filtration. One defines the $\SSPage_\infty$ page (see~\cite{Boardmann1999}) and studies the convergence of $\SSPage_r$ to a filtered group $G$. In order to formulate this, one considers the limit $A^\infty \coloneqq \lim_s A^s$, the colimit $A^{-\infty} \coloneqq \colim_s A^s$ and the right derived module for the limit $RA^\infty \coloneqq \lim_s^1 A^s$. We will use the following notions of convergence:
\begin{enumerate}
\item\emph{Strong convergence to a filtered group $G$}\ $:\Equiv$\ For each $s\in \Z$, we have $\Gr_s G \simeq \SSPage_\infty^s/\SSPage_\infty^{s+1}$ and the filtration on $G$ is exhaustive, Hausdorff and complete (i.e., $G^\infty = 0$, $G^{-\infty} = G$ and $RG^\infty = 0$ for $G^s \coloneqq \Filtr^s G$).
\item\emph{Conditional convergence to the colimit $G\coloneqq A^{-\infty}$}\ $:\Equiv$\ It holds $A^\infty = 0$ and $RA^\infty = 0$.
\end{enumerate}
Note that neither notion implies, in general, the other (see (b) of Remark~\ref{Rem:SpecSeq}).

\begin{Proposition}[Convergence of certain unbounded spectral sequences]\label{Prop:ConvOfSpSeq}
The following statements about convergence of spectral sequences hold:
\begin{ClaimList}
\item For any $\Z$-graded cochain complex $(C^*,\Dd)$ with the canonical filtration $\Filtr^k_{\CanMRM} C^* \coloneqq \bigoplus_{i\ge k} C^i$, the associated spectral sequence converges strongly and conditionally to the colimit $\H(C^*)$.
\item The spectral sequence associated to the total complex $\TotI B^*$ of a cohomological half-plane bicomplex $B^*$ with the filtration induced from the horizontal filtration $\Filtr^k_{\HorMRM} B^* = \bigoplus_{i\in\Z}\bigoplus_{j\ge k} B^{ij}$ converges strongly to the colimit $\H(\TotI B^*)$.
\item The spectral sequence associated to the total complex $\TotI B^*$ of a cohomological half-plane bicomplex $B^{*}$ with the filtration induced from the diagonal filtration 
\[ \Filtr^k B^* = \bigoplus_{j-i>k} B^{ij} \oplus \bigoplus_{i\in \N_0} Z^{i,k+i}, \]
where $Z^{ij} \subset B^{i j}$ is such that $\Dd_{\HorMRM} Z^{ij} = 0$ and $\Dd_{\HorMRM} B^{i-1 j} \subset Z^{ij}$, converges strongly to the colimit $\H(\TotI B^*)$.
\item The spectral sequence associated the total complex $\TotII B^*$ of a cohomological half-plane bicomplex $B^{*}$ with the filtration induced from the vertical filtration $\Filtr^k_{\VertMRM} B^* = \bigoplus_{i\ge k}\bigoplus_{j\in\Z} B^{ij}$ converges conditionally to the colimit $\H(\TotII B^*)$.
\end{ClaimList}
The following statements about morphisms of spectral sequences hold:
\begin{ClaimList}[resume]
\item Let $f$ be a morphism of filtered complexes of types (a), (b) or (c). If it induces an isomorphism of $\SSPage_r$ for some $r$, then it induces an isomorphism of the target groups.
\item Let $f$ be a morphism of filtered complexes of type (d). If it induces an isomorphism of $\SSPage_r$ for some $r$, then it induces an isomorphism of the target groups.
\end{ClaimList}
\end{Proposition}

\begin{proof}
\begin{ProofList}
\item Strong convergence follows from (b) by embedding $C^*$ as the first column of~$B^*$. Also, the computation there show that $A^{-\infty} = \H(C^*)$ and $A^\infty = 0$. The condition $RA^\infty = 0$ is equivalent to the degreewise completeness of the filtration, which is true as the filtration is degreewise trivial. This shows the conditional convergence.
\item Let us compute the first page with the geometrical bigrading:
\begin{align*}
\SSPage_1^{sd} & = \H(\Gr_s \TotI B^*,\Dd_\HorMRM)[s]^d \\
 &= \H^{d+s}(\TotI(s\text{-th row of }B^*),\Dd_\HorMRM) \\
 &= \H(B^{d,s},\Dd_\HorMRM).
\end{align*}
We want to use the following theorem:
\begin{ProofTheorem}[{\cite[Theorem~6.1]{Boardmann1999}}]\label{Thm:Boardman61}
Suppose that $\SSPage^s = 0$ for all $s>0$ and $A^{-\infty} = 0$. Then the spectral sequence converges strongly to the colimit $A^{\infty}$.
\end{ProofTheorem}
The proof can be done degreewise (see \cite{MO336781}), and it can be shown that Theorem~\ref{Thm:Boardman61} generalizes appropriately under the weaker assumption of ``exiting differentials''. This means that the pages occupy a half-plane and for any fixed $(s,d)$, all but finitely many differentials $\Dd_r: \SSPage_r^{sd} \rightarrow \SSPage_{r}^{s+r,d-r+1}$  leave the half-plane (and thus vanish). In our case,~$\SSPage_r^{sd}$ occupy the half-plane $\{(s,d)\mid d\ge 0\}$, and because $d-r+1 \to -\infty$ as $r\to \infty$, the condition of exiting differentials is satisfied.

We still have to check that $A^{\infty}=0$ and compute $A^{-\infty}$. Because the colimit is an exact functor, it commutes with $\H$, and we have
\[ A^{-\infty} = \colim_s \H(\Filtr^s\TotI B^*) = \H(\colim_s\Filtr^s\TotI B^*) = \H\Bigl(\bigcup_s \Filtr^s \TotI B^*\Bigr) = \H(\TotI B^*). \]
We used here that $\Filtr$ is exhaustive. The limit $A^\infty$ can be represented as
\[ A^\infty \simeq \Bigl\{([a_s]) \in \prod_{s\in \Z} A^s \mid [a_{s+1}]\mapsto [a_s]\Bigr\}, \]
where $\H(\Filtr^{s+1}\TotI B^*) \rightarrow \H(\Filtr^s \TotI B)$ is induced by the inclusion $\Filtr^{s+1}\TotI B^* \xhookrightarrow{} \Filtr^s \TotI B^*$. Pick $s_0\in \Z$ and consider $[a_{s_0}] \in A^{s_0}$ with a fixed representative $a_{s_0}\in \Filtr^{s_0} B^*$. Because the cohomological degrees of $a_{s_0}$ are bounded, let's say that $d_0\in \Z$ is an upper bound, and the filtration is degreewise bounded from below, there is an $s_1 \ge s_0$ such that $\Filtr^{s} \TotI B^* \cap (\TotI B^*)^{d} = 0$ for all $s\ge s_1$ and $d\le d_0$. Now, we have $[a_{s_1}] \mapsto [a_{s_0}]$, and hence $[a_{s_0}] = 0$. Because $s_0$ was arbitrary, we get $([a_s]) = 0$. Therefore, it holds $A^\infty=0$.

Alternatively, a direct proof of (b) can be found in \cite{Cencelj1998}.

\item The first page reads
\[\SSPage_1^{sd} = \H^{s+d}(\Gr_s \TotI B^*) = \H(B^{\lfloor\frac{d}{2}\rfloor,s+\lceil\frac{d}{2}\rceil},\Dd'), \]
where $\Dd'$ is the differential on $\Gr \TotI B^*$. We see that $\SSPage_r^{sd}$ occupy the half-plane $\{(s,d)\mid d\ge 0\}$, and hence the condition of exiting differentials is satisfied. The groups~$A^\infty$ and~$A^{-\infty}$ are computed as in (b). The strong convergence is again implied by a generalization of Theorem~\ref{Thm:Boardman61}.

Alternatively, one can modify the direct proof for the horizontal filtration from~\cite{Cencelj1998}.

\item We want to use the following theorem:
\begin{ProofTheorem}[{\cite[Theorem~9.2]{Boardmann1999}}]\label{Thm:Thm92}
Let $C^*$ be a cochain complex filtered by an exhaustive, Hausdorff and complete filtration. Then the spectral sequence converges conditionally to $\H(C^*)$.
\end{ProofTheorem}
We have
\[ \Filtr^k_{\VertMRM} (\TotI B^*)^d = \bigoplus_{i\ge k} B^{i,d-i}, \]
and hence $\TotII B^*$ is the completion of $\TotI B^*$ with respect to $\Filtr_{\VertMRM}$. Hence, it is complete. Clearly, it is also Hausdorff and exhaustive, and we can apply Theorem~\ref{Thm:Thm92}.

\item This follows from (a), (b), (c) and the following theorem:
\begin{ProofTheorem}[{\cite[Theorem~5.3]{Boardmann1999}}]
Let $f: C^* \rightarrow \bar{C}^*$ be a morphism of filtered cochain complexes. Suppose that $\SSPage_r$ converges strongly to a filtered group~$G$ and that~$\bar{\SSPage}_r$ converges (strongly) to a filtered group~$\bar{G}$. If~$f$ induces an isomorphism~$\SSPage_r \simeq \bar{\SSPage}_r$ for some~$r$, then it induces an isomorphism~$G\simeq\bar{G}$.
\end{ProofTheorem}

\item We want to use the following theorem:
\begin{ProofTheorem}[{\cite[Theorem~7.2]{Boardmann1999}}]\label{Thm:Thm72}
Let $f: C^* \rightarrow \bar{C}^*$ be a morphism of filtered cochain complexes. Suppose that $\SSPage^s = \bar{\SSPage}^s = 0$ for all $s<0$ and that the spectral sequences converge conditionally to the colimits $A^{-\infty}$ and $\bar{A}^{-\infty}$, respectively. If $f$ induces isomorphisms $\SSPage_\infty \simeq \bar{\SSPage}_\infty$ and $R\SSPage_\infty \simeq R\bar{\SSPage}_\infty$, then it induces an isomorphism $A^\infty\simeq\bar{A}^\infty$. 
\end{ProofTheorem}
The first page for the vertical filtration reads:
\begin{align*}
 \SSPage_1^{sd} & = \H(\Gr_s \TotII B^*, \Dd_{\VertMRM})[s]^d \\
                & = \H^{d+s}(\TotII(s\text{-th column of }B^*),\Dd_{\VertMRM}) \\
                & = \H(B^{s d},\Dd_{\VertMRM}).
\end{align*} 
Therefore, the condition $\SSPage^s = \bar{\SSPage}^s = 0$ for all $s<0$ is satisfied.\footnote{It is again possible to relax this assumption and prove \ref{Thm:Thm72} when the condition of ``entering differentials'' is satisfied. This means that the pages occupy a half-plane and for any fixed $(s,d)$, all but finitely many differentials $\Dd_r: \SSPage_r^{s-r,d+r-1} \rightarrow \SSPage_{r}^{s,d}$ start outside of the half-plane (and thus vanish). See \cite{MO336781}.} By (d), conditional convergence is guaranteed. Since both $\SSPage_\infty$ and $R\SSPage_\infty$ depend only on $\SSPage_r$ for $r\ge r_0$ and any $r_0$ (see \cite[p.~16]{Boardmann1999}), the rest of the assumptions of Theorem~\ref{Thm:Thm72} is fulfilled.\qedhere
\end{ProofList}
\end{proof}

\begin{Remark}[Differences to first-quadrant bicomplexes]\phantomsection\label{Rem:SpecSeq}
\begin{RemarkList}
\item Given a half-plane bicomplex $B$, we have
\[ (\TotI B)^{\GD} \simeq \TotII B^*, \]
where $B^* = \bigoplus_{i,j} B_{ij}^*$ is the ``pointwise dual'' to $B$. This is why we have to consider homology and cohomology separately and can not just dualize the results.
\item The vertical filtration of $B^*$ might not converge strongly to $\H(\TotI B^*)$. Indeed, let
\begin{center}
$B^*: \quad \begin{tikzcd}
 \R & 0 & 0  \\
 \R\arrow{u}{\Id}\arrow{r}{\Id} & \R & 0   \\
 0 & \R\arrow{u}{\Id}\arrow{r}{\Id} & \dotsb
\end{tikzcd}$.
\end{center}
Then $\H(\TotI B^*) = \R$ (the $\R$ in the first column), but $\SSPage_1 = 0$ because every column is exact. Notice that $\H(\TotII B^*) = 0$. Taking $0$'s instead of $\Id$'s in the definition of $B^*$, we see that the horizontal filtration does not converge conditionally to $\H(\TotI B^*)$ because its filtration by $A^s$ is incomplete.
\qedhere
\end{RemarkList}
\end{Remark}

We will work with the following bicomplexes.

\begin{Definition}[Bicomplexes for cyclic (co)homology]\label{Def:CycBico}
Let $\mathcal{A} = (V, (\mu_k))$ be an $\AInfty$-algebra. \emph{Loday's cyclic bicomplexes} are defined by
\begin{center}
$\LodCycBi(\mathcal{A}):\ $\begin{tikzcd}[column sep=scriptsize, row sep=scriptsize]
\arrow[d] & \arrow[d] & \arrow[d] & \arrow[d] &{} \\
\HC_2 V \arrow[d]{}{\Hd} & \HC_2V \arrow[l]{}{\Id - \CycPermOp} \arrow[d]{}{-\Hd'} & \HC_2V \arrow[d]{}{\Hd} \arrow[l]{}{\CountOp} & \HC_2V \arrow[d]{}{-\Hd'} \arrow[l]{}{\Id - \CycPermOp} &{}\arrow[l]{}{\CountOp}\\ 
\HC_1V \arrow[d]{}{\Hd} & \HC_1V \arrow[l]{}{\Id - \CycPermOp} \arrow[d]{}{-\Hd'} & \HC_1V \arrow[d]{}{\Hd} \arrow[l]{}{\CountOp} & \HC_1V \arrow[d]{}{-\Hd'} \arrow[l]{}{\Id - \CycPermOp} &{}\arrow[l]{}{\CountOp}\\
\HC_0V \arrow[d]{}{\Hd} & \HC_0V \arrow[l]{}{\Id - \CycPermOp} \arrow[d]{}{-\Hd'} & \HC_0V \arrow[d]{}{\Hd} \arrow[l]{}{\CountOp} & \HC_0V \arrow[d]{}{-\Hd'} \arrow[l]{}{\Id - \CycPermOp} &{}\arrow[l]{}{\CountOp}\\
 {} & {} & {} & {} & {}
\end{tikzcd}
\end{center}
and
\begin{center}
$\LodCycBi^*(\mathcal{A}):\ $\begin{tikzcd}[column sep=scriptsize, row sep=scriptsize]
{} & {} & {} & {} & {}\\
 \HC^2V \arrow[u]{}{\Hd^*} \arrow[r]{}{\Id - \CycPermOp^*} & \HC^2V \arrow[u]{}{-{\Hd'}^*} \arrow[r]{}{\CountOp^*} & \HC^2V \arrow[u]{}{\Hd^*} \arrow[r]{}{\Id - \CycPermOp^*} & \HC^2V \arrow[u]{}{-{\Hd'}^*} \arrow[r]{}{\CountOp^*} &{}\\ 
 \HC^1V \arrow[u]{}{\Hd^*} \arrow[r]{}{\Id - \CycPermOp^*} & \HC^1V \arrow[u]{}{-{\Hd'}^*} \arrow[r]{}{\CountOp^*} & \HC^1V \arrow[u]{}{\Hd^*} \arrow[r]{}{\Id - \CycPermOp^*} & \HC^1V \arrow[u]{}{-{\Hd'}^*} \arrow[r]{}{\CountOp^*} &{}\\
 \HC^0V \arrow[u]{}{\Hd^*}\arrow[r]{}{\Id - \CycPermOp^*} & \HC^0V \arrow[u]{}{-{\Hd'}^*} \arrow[r]{}{\CountOp^*} & \HC^0V \arrow[u]{}{\Hd^*} \arrow[r]{}{\Id - \CycPermOp^*} & \HC^0V \arrow[u]{}{-{\Hd'}^*}\arrow[r]{}{\CountOp^*} &{} \\
\arrow[u]{}{\Hd^*} & \arrow[u]{}{-{\Hd'}^*} & \arrow[u]{}{\Hd^*} & \arrow[u]{}{-{\Hd'}^*} &{} 
\end{tikzcd}.
\end{center}
Clearly, $\LodCycBi^*$ is the ``pointwise'' graded dual to $\LodCycBi$ and analogously for other bicomplexes we are going to define.

Let $\NOne$ be a strict unit for $\mathcal{A}$. We define the \emph{Connes' operator} $\Cd: \HC V \rightarrow \HC V$ by 
\begin{equation}\label{Eq:ConnesOperator}
\Cd\coloneqq (\Id-\CycPermOp)\circ\InsOneOp_1\circ\CountOp.
\end{equation}
\emph{Connes' cyclic bicomplexes} are defined by
\begin{center}
$\ConCycBi(\mathcal{A}):$
\begin{tikzcd}[column sep=scriptsize, row sep=scriptsize]
{}\arrow[d]{}{\Hd}&{}\arrow[d]{}{\Hd}&{}\arrow[d]{}{\Hd}&{} \\
\HC_2V \arrow[d]{}{\Hd} & \HC_1V \arrow[d]{}{\Hd} \arrow[l]{}{\Cd}& \HC_0V \arrow[d]{}{\Hd} \arrow[l]{}{\Cd}  &{}\arrow[l]{}{\Cd}\\
\HC_1V \arrow[d]{}{\Hd} & \HC_0V \arrow[d]{}{\Hd} \arrow[l]{}{\Cd}  & \arrow[d]{}{\Hd} \arrow[l]{}{\Cd}  \HC_{-1}V  &{}\arrow[l]{}{\Cd} \\
\HC_0V \arrow[d]{}{\Hd} & \HC_{-1}V \arrow[d]{}{\Hd} \arrow[l]{}{\Cd}  & \HC_{-2}V \arrow[d]{}{\Hd}  \arrow[l]{}{\Cd}  & {} \arrow[l]{}{\Cd} \\
{}&{}&{} &{}
\end{tikzcd} 
\end{center}
and
\begin{center}
$\ConCycBi^{*}(\mathcal{A}):$
\begin{tikzcd}[column sep=scriptsize, row sep=scriptsize]
{}&{}&{} &{}\\
\HC^2V \arrow[u]{}{\Hd^*} \arrow[r]{}{\Cd^*}& \HC^1V \arrow[u]{}{\Hd^*} \arrow[r]{}{\Cd^*} & \HC^0V \arrow[u]{}{\Hd^*} \arrow[r]{}{\Cd^*}  &{}\\
\HC^1V \arrow[u]{}{\Hd^*} \arrow[r]{}{\Cd^*} & \HC^0V \arrow[u]{}{\Hd^*} \arrow[r]{}{\Cd^*}  & \arrow[u]{}{\Hd^*} \arrow[r]{}{\Cd^*}  \HC^{-1}V  &{} \\
\HC^0V \arrow[u]{}{\Hd^*} \arrow[r]{}{\Cd^*}  & \HC^{-1}V \arrow[u]{}{\Hd^*} \arrow[r]{}{\Cd^*}  & \HC^{-2}V \arrow[u]{}{\Hd^*}  \arrow[r]{}{\Cd^*}  & {} \\
{}\arrow[u]{}{\Hd^*}&{}\arrow[u]{}{\Hd^*}&{}\arrow[u]{}{\Hd^*}&{}
\end{tikzcd} 
\end{center}
We define the \emph{normalized Connes' operator} $\NCd: \HNC V \rightarrow \HNC V$ by\footnote{The definition does not depend on the chosen section of $\bar{p}: \HC V \rightarrow \HNC V$.}
\begin{equation}\label{Eq:NCd}
\NCd \coloneqq \NormProj\circ\Cd = \NormProj\circ \InsOneOp_1 \circ \CountOp.
\end{equation}
The \emph{normalized Connes' cyclic bicomplexes} are defined by
\begin{center}
$\NConCycBi(\mathcal{A}):$
\begin{tikzcd}[column sep=scriptsize, row sep=scriptsize]
{}\arrow[d]{}{\Hd}&{}\arrow[d]{}{\Hd}&{}\arrow[d]{}{\Hd}&{} \\
\HNC_2V \arrow[d]{}{\Hd} & \HNC_1V \arrow[d]{}{\Hd} \arrow[l]{}{\NCd}& \HNC_0V \arrow[d]{}{\Hd} \arrow[l]{}{\NCd}  &{}\arrow[l]{}{\NCd}\\
\HNC_1V \arrow[d]{}{\Hd} & \HNC_0V \arrow[d]{}{\Hd} \arrow[l]{}{\NCd}  & \arrow[d]{}{\Hd} \arrow[l]{}{\NCd}  \HNC_{-1}V  &{}\arrow[l]{}{\NCd} \\
\HNC_0V \arrow[d]{}{\Hd} & \HNC_{-1}V \arrow[d]{}{\Hd} \arrow[l]{}{\NCd}  & \HNC_{-2}V \arrow[d]{}{\Hd}  \arrow[l]{}{\NCd}  & {} \arrow[l]{}{\NCd} \\
{}&{}&{} &{}
\end{tikzcd} 
\end{center}
and
\begin{center}
$\NConCycBi^*(\mathcal{A}):$
\begin{tikzcd}[column sep=scriptsize, row sep=scriptsize]
{}&{}&{} &{}\\
\HNC^2V \arrow[u]{}{\Hd^*} \arrow[r]{}{\NCd^*}& \HNC^1V \arrow[u]{}{\Hd^*} \arrow[r]{}{\NCd^*} & \HNC^0V \arrow[u]{}{\Hd^*} \arrow[r]{}{\NCd^*}  &{}\\
\HNC^1V \arrow[u]{}{\Hd^*} \arrow[r]{}{\NCd^*} & \HNC^0V \arrow[u]{}{\Hd^*} \arrow[r]{}{\NCd^*}  & \arrow[u]{}{\Hd^*} \arrow[r]{}{\NCd^*}  \HNC^{-1}V  &{} \\
\HNC^0V \arrow[u]{}{\Hd^*} \arrow[r]{}{\NCd^*}  & \HNC^{-1}V \arrow[u]{}{\Hd^*} \arrow[r]{}{\NCd^*}  & \HNC^{-2}V \arrow[u]{}{\Hd^*}  \arrow[r]{}{\NCd^*}  & {} \\
{}\arrow[u]{}{\Hd^*}&{}\arrow[u]{}{\Hd^*}&{}\arrow[u]{}{\Hd^*}&{}
\end{tikzcd}.
\end{center}
The \emph{reduced Connes' cyclic bicomplexes} $\ConCycBi^{\RedMRM}$ and $\ConCycBi_{\RedMRM}^*$ are defined by replacing $\HNC V$ and $\HNC^*V$ by $\HC^{\RedMRM} V$ and $\HC_{\RedMRM}^* V$, respectively.

The coordinate $(0,0)$ in the bicomplexes above always correspond to $\HC^0 V$ in the first column (bottom-left position in the figures). 
\end{Definition}
Note that another convention of drawing homological bicomplexes in the left half-plane and cohomological bicomplexes in the right half-plane might be more natural.
\Add[caption={DONE Origin of bicomplexes},noline]{Define where the point $(0,0)$ is for the bicomplexes.}
\begin{Remark}[Mixed complexes]\label{Rem:MixedCompl}
One can equivalently encode the data of a cohomological Connes bicomplex into that of a mixed complex $(\HC^*,\Hd^*,\Cd^*)$. In general, it is a graded vector space $\HC^*$ with a differential $\Hd^*$, $\Abs{\Hd^*}=1$ and a boundary operator $\Cd^*$, $\Abs{\Cd^*}=-1$ which anticommute. One introduces the formal symbol~$\FormU$ of degree $\Abs{\FormU}= 2$ and considers the polynomial ring $\HC^*[\FormU]$ in $\FormU$ with values in~$\HC^*$ and the ring of power series $\HC^*[[\FormU]]$ with the differential $\Hd^* + \Cd^*\FormU$. Clearly, the former is quasi-isomorphic to $\TotI(\ConCycBi^{*})$ and the latter to $\TotII(\ConCycBi^{*})$ (columns of $\ConCycBi^{*}$ are indexed with non-negative powers of $\FormU$).
Altogether, there are seven versions $[\FormU]$, $[\FormU^{-1}]$, $[\FormU,\FormU^{-1}]$, $[[\FormU^{-1},\FormU]$, $[\FormU^{-1},\FormU]]$, $[[\FormU,\FormU^{-1}]]$ whose relation is studied in \cite{Cieliebak2018b}. Some of these are related to periodic and negative versions of cyclic homology (see~\cite{LodayCyclic}).
\end{Remark}
In the following proofs, we might not need the full strength of Proposition~\ref{Prop:ConvOfSpSeq} since the spectral sequences for the bicomplexes from Definition~\ref{Def:CycBico} mostly collapse already on the second page (see (iii) of Questions~\ref{Q:OpenProbAInftx}).
\begin{Lemma}[Loday's cyclic bicomplexes and cyclic homology]\label{Lem:LodCycBiCycHom}
Let $\Alg = (V,(\mu_k))$ be an $\AInfty$-algebra. The projection $p^\lambda: \LodCycBi \rightarrow \HC^\lambda$ to the first column modulo $\im(\Id-\CycPermOp)$ is a chain map and induces an isomorphism $\H(\widehat{\LodCycBi}) \simeq \H^\lambda(\Alg)$. The inclusion $\iota_\lambda: \HC_\lambda^{*} \rightarrow \LodCycBi^{*}$ into the first column is a chain map and induces an isomorphism $\H_\lambda^*(\Alg) \simeq \H(\LodCycBi^*)$.
\end{Lemma}
\begin{proof}
The fact that $p^\lambda$ and $\iota_\lambda$ are chain maps is obvious. We consider the horizontal filtration $\Filtr^\HorMRM$ of $\LodCycBi^*$. Because of CP1, the rows are acyclic, and we see that the only non-zero terms of the first page of the corresponding spectral sequence are 
\[ \SSPage_1^{0 d} =  \HC^d V / \ker(\Id - \CycPermOp^*). \] The differential $\Dd_1$ is easy to check to be $\Hd^*$, and the inclusion $\iota_\lambda$ induces the isomorphism $\HC_{\lambda}^d\simeq \HC^d V / \ker(\Id - \CycPermOp^*)$. Considering the canonical filtration on $\HC^*_\lambda$, claims (a), (b) and~(e) of Proposition~\ref{Prop:ConvOfSpSeq} apply.

As for homology, we consider the degree reversed cochain complex $\DegRev(\TotII \LodCycBi)$ and the reversed filtration $\DegRev(\Filtr^\HorMRM)_s = \Filtr^\HorMRM_{-s}$. For the corresponding cohomological spectral sequence~$\tilde{\SSPage}_r$, we have
\begin{align*}
\tilde{\SSPage}_1^{sd} &= \H^{s+d}(\DegRev(\Filtr)_{s}\DegRev(\TotII \LodCycBi)/\DegRev(\Filtr)_{s+1}\DegRev(\TotII\LodCycBi)) \\
&=\H_{-s-d}(\TotII(-s\text{-th row of }\LodCycBi)) \\
&=\H(B_{-d,-s},\Bdd_\HorMRM).
\end{align*}
Therefore, the only groups are $\tilde{\SSPage}^{s0} = \HC_{-s}/\im(\Id-\CycPermOp)$, and the spectral sequence converges conditionally to $\DegRev(\H(\widehat{\LodCycBi}))$. Clearly, $\iota_\lambda$ induces an isomorphism of the first pages, where on $\DegRev(\HC^\lambda)$ we consider the canonical filtration. Proposition~\ref{Prop:ConvOfSpSeq} and its proof finishes the argument.
\end{proof}

Claim (a) of the following is similar to \cite[Lemma 2.12]{Cieliebak2018b}.

\begin{Lemma}[No long chains in homology for bounded degrees] \label{Lem:BddDegrees}
Let $\Alg=(V,(\mu_k),\NOne)$ be a strictly unital $\AInfty$-algebra. Suppose that $V$ has bounded degrees. Then the canonical inclusion $\TotI \hookrightarrow \TotII$ induces the following isomorphism: 
\begin{ClaimList}
\item $\H\bigl(\LodCycBi(\Alg)\bigr) \simeq \H\bigl(\widehat{\LodCycBi}(\Alg)\bigr)$,
\item $\H\bigl(\NConCycBi(\Alg)\bigr) \simeq \H\bigl(\widehat{\NConCycBi}(\Alg)\bigr)$.
\end{ClaimList} 
\end{Lemma}
\begin{proof}
\begin{ProofList}
\item We will denote $\LodCycBi(\Alg)$, $\NConCycBi(\Alg)$ and $\HC V$ simply by $\LodCycBi$, $\NConCycBi$ and $\HC$, respectively. Consider the (increasing) filtration $\Filtr_{\WeightMRM}$ of $\HC$ by weights. We first prove the following subclaim:
\begin{SubClaim}[Weight normalization]\label{SubClaim:LodCycBi}
Let $c = (c_i)_{i=0}^\infty\in {\TotII}_k(\LodCycBi)$ be a closed chain of degree~$k$; i.e., for all $i\in\N_0$, we have $c_i \in \HC_{k-i}$, and the relations
\[ \Hd c_{2i} + (\Id-\CycPermOp)c_{2i+1} = 0 \quad\text{and}\quad -\Hd' c_{2i+1} + \CountOp c_{2i+2} = 0\]
hold. Suppose that we are given $j\ge 1$ and $n_0\in \N$ such that $c_{j-1}\in \Filtr^{n_0}_{\WeightMRM} \HC_{k-j+1}$. Then we can construct $\tilde{c}_j\in \Filtr^{n_0}_{\WeightMRM}\HC_{k-j}$, $\tilde{c}_{j+1}\in \HC_{k-j-1}$ and $\tilde{z}_{j+1}\in \HC_{k-j}$ such that if we define
\begin{equation}\label{Eq:DefOfChains}
\tilde{c}_i \coloneqq \begin{cases}
    \tilde{c}_j & \text{for } i=j, \\
    \tilde{c}_{j+1} & \text{for } i=j+1, \\
    c_i & \text{otherwise}
\end{cases}\quad\text{and}\quad z_i \coloneqq \begin{cases} 
\tilde{z}_{j+1} & \text{for }u = j+ 1,\\
0 & \text{otherwise},
\end{cases}
\end{equation}
then $\tilde{c}\coloneqq (\tilde{c}_i)_{i=0}^\infty$ is a closed chain and $z\coloneqq(z_i)_{i=0}^\infty$ satisfies $\Bdd z = c - \tilde{c}$. By repeating this procedure inductively, we obtain chains $c'\in\TotII_k(\LodCycBi)$ and $z'\in \TotII_{k+1}(\LodCycBi)$ such that $c'_i \in \Filtr^{n_0}_{\WeightMRM}\HC_{k-i}$ for all $i\ge j$ and $c-c' = \Bdd z'$.
\end{SubClaim}
\begin{proof}[Proof of the Subclaim]
\begin{figure}
\centering
\begin{tikzcd}
 {} & c_j, \tilde{c}_j\in\HC_{k-j} \arrow{l}{\Id-\CycPermOp}\arrow{d}{-\Hd'} & \arrow{l}{\CountOp}\tilde{z}_{j+1}\in \HC_{j-k}\arrow{d}{\Hd} \\
 {} & {} & \arrow{l}{\CountOp}c_{j+1},\tilde{c}_{j+1}\in\HC_{j-k-1} \arrow{d}{\Hd} \\
 {} & {} & {} 
\end{tikzcd}
\caption[Illustration of weight normalization in the Loday's cyclic bicomplex.]{Positions of element in $\LodCycBi$ for $j$ odd.}
\label{Fig:PosOfElLodCycBi}
\end{figure}
We will assume that $j$ is odd; the proof is analogous for $j$ even with the roles of $(\Id-\CycPermOp)$ and $\CountOp$, resp.~$\Hd$ and $-\Hd'$ switched. The situation is depicted in Figure~\ref{Fig:PosOfElLodCycBi}. Because $c_{j-1}\in \Filtr^{n_0}_{\WeightMRM} \HC_{k-j+1}$ and $\Hd$ does not increase weights, we have $\Hd c_{j-1} \in \Filtr^{n_0}_{\WeightMRM} \HC_{k-j}$. Since $c$ is closed, we have $(\Id-\CycPermOp)c_{j} = - \Hd c_{j-1} \in \Filtr^{n_0}_{\WeightMRM} \HC_{k-j}$. Therefore, there is a $\tilde{c}_j \in \Filtr^{n_0}_{\WeightMRM} \HC_{k-j}$ such that $(\Id-\CycPermOp)\tilde{c}_j = (\Id-\CycPermOp)c_j$. As $c_j - \tilde{c}_j \in \ker(\Id-\CycPermOp) = \im \CountOp$, we have $c_j - \tilde{c}_j = \CountOp \tilde{z}_{j+1}$ for some $\tilde{z}_{j+1}\in \HC_{k-j}$. We define $\tilde{c}_{j+1}\coloneqq c_{j+1} - \Hd \tilde{z}_{j+1}$. The following relations hold:
\begin{EqnList}
\item $(\Id-\CycPermOp)\tilde{c}_j = (\Id-\CycPermOp)c_j$,
\item $\begin{aligned}[t]
- \Hd' \tilde{c}_{j} + \CountOp \tilde{c}_{j+1} &= - \Hd' \tilde{c}_{j} + \CountOp c_{j+1} - \CountOp \Hd \tilde{z}_{j+1} \\
 &= - \Hd' \tilde{c}_{j} + \CountOp c_{j+1} - \Hd'\CountOp \tilde{z}_{j+1}  \\ 
&= - \Hd' \tilde{c}_{j} + \CountOp c_{j+1}  - \Hd'(c_j - \tilde{c}_j) \\ & = -\Hd'c_j + \CountOp c_{j+1} \\
& = 0,
\end{aligned}$
\item $\Hd \tilde{c}_{j+1} = \Hd c_{j+1}$,
\item $\CountOp \tilde{z}_{j+1} = c_j - \tilde{c}_j$,
\item $\Hd \tilde{z}_{j+1} = c_{j+1}- \tilde{c}_{j+1}$.
\end{EqnList}
The relations I--III show that $\tilde{c}$ is closed and the relations IV--V that $\Bdd z = c - \tilde{c}$.\footnote{In fact, $\Bdd \tilde{c} = 0$ follows from $\Bdd c = 0$ and $\Bdd z = c - \tilde{c}$.}
\renewcommand{\qed}{\hfill\textit{(Subclaim) }$\square$}

Starting with $c^{1} \coloneqq \tilde{c}$ and $z^{1} \coloneqq z$, we repeat the construction above to produce the telescopic sequence of homotopies
\begin{align*}
c - c^{1} & = \Bdd z^{1} \\
c^{1} - c^{2} & = \Bdd z^{2} \\
\dotsb &= \dotsb 
\end{align*}
such that $c^l_i\in \Filtr^{n_0}_{\WeightMRM} \HC_{k-i}$ for all $j\le i\le j+l$. The limit chain $c'\coloneqq \sum_{k=0}^{\infty} c^{k} \in \TotII\LodCycBi$ has the property that $c'_i \in \Filtr^{n_0}_{\WeightMRM}\HC_{k-i}$ for all $i\ge j$, and the limit homotopy $z' \coloneqq \sum_{k=1}^\infty z^k \in \TotII \LodCycBi$ converges and satisfies $\Bdd z' = c-c'$.
\end{proof}

We will now prove surjectivity of the map on homology induced by the inclusion $\TotI \hookrightarrow \TotII$. Given $[c]\in \H_k(\widehat{\LodCycBi})$, using the Subclaim, we can assume that there is an $n_0\in\N$ such that $c_i\in \Filtr^{n_0}_{\WeightMRM}\HC_{k-i}$ for all $i\in\N_0$. However, we have $\Abs{c_i} = -k+i-1$ for the degrees in $\B V$, and because the degrees of $V$ are bounded, $c_i$ eventually, as $i\to \infty$, reach degrees which can not be produced by $n_0$ vectors. Therefore, there is an $i_0\in \N_0$ such that $c_i = 0$ for all $i\ge i_0$; this means that $c \in \TotI \LodCycBi$.

To show injectivity of the induced map on homology, suppose that $c\in \TotI \LodCycBi$ satisfies $c=\Bdd z$ for some $z\in \TotII \LodCycBi$. Let $i_0\in \CountOp$ be such that $c_i = 0$ for all $i\ge i_0$. We use the Subclaim to alter $z$ and obtain a chain $\tilde{z}\in\TotI \LodCycBi$ such that $\tilde{z}_{i} = z_i$ for $i\le i_0$ and $\Bdd \tilde{z} = c$. This shows injectivity.

\item We will prove an analogy of the Subclaim from (a):

Given a closed $c = (c_i)_{i=0}^\infty \in \TotII_k \NConCycBi$, every $c_i \in \HNC_{k-2i} V$ can be written as 
\[ c_i = \tilde{c}_i + \NOne \hat{c}_i \]
for unique $\tilde{c}_i\in \HC_{q-2i} \bar{V}$ and $\hat{c}_i \in \HC_{q-2i-1} \bar{V}$. Using strict unitality, we have $\Hd(\NOne\hat{c}_i) = (\Id-\CycPermOp)\hat{c}_i - \NOne \Hd'\hat{c}_i$, and hence
\begin{align*}
   \Hd c_i &=  \Hd\bar{c}_i + (\Id-\CycPermOp)\hat{c}_i - \NOne \Hd'\hat{c}_i, \\
   \NCd c_i &= \NOne \CountOp \bar{c}_i.
\end{align*}
For the second equality, recall the definition \eqref{Eq:NCd} and note that the $\bar{p}$ in front ``kills'' any input of $\NCd$ containing at least one $\NOne$. We see that $\Bdd c = 0$ is equivalent to $\Hd c_i = - \NCd c_{i+1}$ which is equivalent to
\[ \begin{aligned}
\Hd \bar{c}_i + (\Id-\CycPermOp)\hat{c}_i & = 0\quad\text{and}  \\
\Hd'\hat{c}_i &= \CountOp \bar{c}_{i+1}
\end{aligned} \] 
for all $i\in \N_0$.
\begin{figure}
\centering
\begin{tikzcd}
c_{j-1}\in\HNC_{k-2j+2}\arrow{d}{\Hd} & \arrow{l}{\NCd} z_j\in \HNC_{k-2j+1} \arrow{d}{\Hd} & \\
{} & \arrow{l}{\NCd} c_j, \tilde{c}_j \in \HNC_{k-2j} \arrow{d}{\Hd} & \arrow{l}{\NCd} z_{j+1}\in\HNC_{k-2j-1}\arrow{d}{\Hd} \\
{} & {} & \arrow{l}{\NCd} c_{j+1}, \tilde{c}_{j+1} \in \HNC_{k-2j-2}
\end{tikzcd}
\caption[Illustration of weight normalization in the Connes' bicomplex.]{Positions of element in $\ConCycBi$ for $j$ odd.}
\label{Fig:PosOfElConCycBi}
\end{figure}

Suppose that $c_{j-1} \in \Filtr^{n_0}_{\WeightMRM} \bar{\HC}_{k-2 j + 2}$ for some $j\ge 1$ and $n_0\in \N_0$. Then $\bar{c}_{j-1}\in\Filtr^{n_0}_{\WeightMRM}\HNC_{k-2 j + 2}$ and $\hat{c}_{j-1}\in \Filtr^{n_0-1}_{\WeightMRM}\HNC_{k-2j+1}$. Because $\Hd'\hat{c}_{j-1} = \CountOp \bar{c}_j$, we can find $\bar{d}_{j} \in \Filtr^{n_0-1}_{\WeightMRM} \HC_{k-2j}\bar{V}$ such that $\CountOp \bar{d}_{j} = \Hd' \hat{c}_{j-1}$. Because $\bar{d}_j - \bar{c}_j \in \ker \CountOp = \im (\Id-\CycPermOp)$, we can find $\hat{z}_j \in \HNC_{k-2j}V$ such that $(\Id-\CycPermOp)\hat{z}_j = \bar{d}_{j} - \bar{c}_{j}$. We compute
\begin{align*}
 \Hd \bar{d}_j & = \Hd \bigl( \bar{c}_j + (\Id-\CycPermOp)\hat{z}_j)\bigr) \\
 & = - (\Id-\CycPermOp)\hat{c}_j + \Hd (\Id-\CycPermOp)\hat{z}_j \\
 & = - (\Id-\CycPermOp)\hat{c}_j + (\Id-\CycPermOp) \Hd' \hat{z}_j \\
 & = - (\Id-\CycPermOp)\bigl(\hat{c}_j - \Hd'\hat{z}_j\bigr).
\end{align*}
Because $\bar{d}_j \in \Filtr^{n_0-1}_{\WeightMRM}\HC_{k-2j}\bar{V}$ and $\Hd$ does not increase the filtration, we can find $\hat{d}_j \in \Filtr^{n_0-1}_{\WeightMRM}\HC_{k-2j-1}\bar{V}$ such that $(\Id-\CycPermOp)\hat{d}_j = - \Hd \bar{d}_j$. Now, $\hat{d}_j - (\hat{c}_j - \Hd'\hat{z}_j) \in \ker (\Id-\CycPermOp) = \im \CountOp$, and hence there is a $\bar{z}_{j+1}\in \HNC_{k-2j-1}$ such that $\CountOp \bar{z}_{j+1} = \hat{d}_j - (\hat{c}_j - \Hd'\hat{z}_j)$. We define the following elements:\Correct[caption={DONE : in definition},noline]{Take care of alignment of $\coloneqq$ and $=$ in one column!}
\begin{align*}
 \tilde{c}_j &\coloneqq c_j + \NCd \bar{z}_{j+1} + \Hd(\NOne \hat{z}_j) \\
 & = c_j + \NCd \bar{z}_{j+1} + (\Id-\CycPermOp)\hat{z}_j - \NOne \Hd'\hat{z}_j \\
 & = c_j + \NOne \CountOp \bar{z}_{j+1} + (\Id-\CycPermOp)\hat{z}_j - \NOne\Hd'\hat{z}_j \\
 & =  c_j + \NOne \hat{d}_j -\NOne \hat{c}_j + \NOne \Hd'\hat{z}_j + \bar{d}_j - \bar{c}_j - \NOne\Hd'\hat{z}_j\\
 & = \bar{d}_j + \NOne \hat{d}_j, \\
 \tilde{c}_{j+1} & \coloneqq c_{j+1} + \Hd \bar{z}_{j+1}, \\
 \tilde{z}_j & \coloneqq \NOne \hat{z}_j, \\
 \tilde{z}_{j+1} &\coloneqq \bar{z}_{j+1}.
\end{align*}
The following relations hold:
\begin{EqnList}
\item $\NCd \tilde{c}_j = \NCd c_j$,
\item $\Hd \tilde{c}_j = \Hd c_j + \Hd \NCd \bar{z}_{j+1} = - \NCd c_{j+1} - \NCd  \Hd \bar{z}_{j+1} = - \NCd \tilde{c}_{j+1}$,
\item $\Hd \tilde{c}_{j+1} = \Hd c_{j+1}$,
\item $\NCd \tilde{z}_j = \NCd \NOne \hat{z}_j = 0$,
\item $\Hd \tilde{z}_j = \tilde{c}_j - c_j - \NCd \tilde{z}_{j+1}$,
\item $\Hd \tilde{z}_{j+1} = \tilde{c}_{j+1} - c_{j+1}$.
\end{EqnList}
Relations I--III show that $\tilde{c}$ is closed, and relations IV--VI show that $\Bdd z = \tilde{c} - c$. Here $\tilde{c}$ is defined as in \eqref{Eq:DefOfChains} and $z$ has one more term:
\[ z_i \coloneqq \begin{cases} \tilde{z}_j & \text{for } i = j, \\ \tilde{z}_{j+1} & \text{for } i = j+1, \\
0 & \text{otherwise}. \end{cases} \]
Since $\tilde{c}_j = \bar{d}_j + \NOne \hat{d}_j$, $\bar{d}_j\in \Filtr^{n_0-1}_{\WeightMRM}\HC_{k-2j}\bar{V}$ and $\hat{d}_j\in \Filtr^{n_0-1}_{\WeightMRM}\HC_{k-2j-1}\bar{V}$, we have $\tilde{c}_j \in \Filtr^{n_0}_{\WeightMRM} \HNC_{k-2j}$.

Having the recursive step, the rest can be done in the same way as in (a). \qedhere
\end{ProofList}
\end{proof}

\begin{Lemma}[Loday's and Connes' bicomplexes are quasi-isomorphic]\label{Lem:LodConCycBi}
Let $\Alg=(V,(\mu_k),\NOne)$ be a strictly unital $\AInfty$-algebra. The map
\begin{align*}
I: \TotII \ConCycBi &\longrightarrow \TotII\LodCycBi\\
(c_0, c_1, c_2, \dotsc) &\longmapsto (c_0, \InsOneOp_1 \CountOp c_1, c_1, \InsOneOp_1 \CountOp c_2, c_2, \dotsc )
\end{align*}
is a chain map inducing the isomorphisms 
\[\H( \widehat{\ConCycBi})\simeq \H(\widehat{\LodCycBi})\quad\text{and}\quad\H(\ConCycBi)\simeq\H(\LodCycBi).\] Analogously, the map 
\begin{align*}
P: \TotII\LodCycBi^* &\longrightarrow \TotII \ConCycBi^* \\
(\psi_0, \psi_1,\psi_2, \dotsc )&\longmapsto (\psi_0, \psi_2 + \CountOp^* \InsOneOp_1^* \psi_1, \psi_4 + \CountOp^* \InsOneOp_1^* \psi_3, \dotsc )
\end{align*}
is a chain map inducing the isomorphisms 
\[\H(\widehat{\ConCycBi}^*) \simeq \H(\widehat{\LodCycBi}^*)\quad\text{and}\quad\H(\ConCycBi^*) \simeq \H(\LodCycBi^*).\]
\end{Lemma}
\begin{proof}
The following computation shows that $\iota$ is a chain map:
\begin{align*}
\Bdd_{\LodCycBi}I(c_0,c_1,c_2\dotsc)&= \bigl(\Hd c_0 + (\Id-\CycPermOp)\InsOneOp_1\CountOp c_1, - \Hd'\InsOneOp_1\CountOp c_1 + \CountOp c_1, \Hd c_1 + (\Id-\CycPermOp)\InsOneOp_1 \CountOp c_2, \dotsc \bigr)\\
&= \bigl(\Hd c_0 + \Cd c_1, -\CountOp c_1 + \InsOneOp_1 \Hd' \CountOp c_1 + \CountOp c_1 ,\Hd c_1 + \Cd c_2,\dotsc\bigr) \\
&= \bigl(\Hd c_0 + \Cd c_1, \InsOneOp_1 \CountOp \Hd c_1,\Hd c_1 + \Cd c_2,\dotsc\bigr) \\
&= I\bigl(\Hd c_0 + \Cd c_1, \Hd c_1 + \Cd c_2, \dotsc \bigr) \\ 
&= I\Bdd_{\ConCycBi}\bigl(c_0,c_1,c_2,\dotsc\bigr).
\end{align*}
Clearly, $I$ is injective, and hence it induces an isomorphisms of chain complexes 
\[ \bigl(\TotII\ConCycBi,\Bdd_{\ConCycBi}\bigr) \simeq \bigl(\im(I),\Restr{\Bdd_{\LodCycBi}}{\im(I)}\bigr)\subset (\TotII \LodCycBi, \Bdd_{\LodCycBi}). \]
Consider the subcomplex $(\TotII \LodCycBi_{\mathrm{odd},\bullet},-{\Hd'})\subset (\TotII\LodCycBi,\Bdd_{\LodCycBi})$ which consists of odd columns of $\LodCycBi$. It is a direct complement of $\im(I)$ in $\TotII \LodCycBi$. Indeed, $(0,c_1,0,c_2,\dotsc) \in \im(I)$ implies $c_i = 0$ for all $i\in\N$, which gives $\TotII(\LodCycBi_{\mathrm{odd},\bullet}) \cap \im(I) = 0$; also, for any $(c_i)\in \TotII \LodCycBi$, we have
\[ (c_0, c_1, c_2, c_3, c_4 \dotsc ) = I\bigl((c_0, c_2, c_4, \dotsc )\bigr) - (0,\InsOneOp_1\CountOp c_2 - c_1 , 0, \InsOneOp_1\CountOp c_4 - c_3, 0, \dotsc), \]
which gives $\im(I) + \TotII(\LodCycBi_{\mathrm{odd},\bullet}) = \TotII \LodCycBi$. Now, $\TotII(\LodCycBi_{\mathrm{odd},\bullet})$ is contractible by CP3, and hence $\H(\TotII \ConCycBi) \simeq \H(\TotII \LodCycBi)$ (using an argument with the long exact sequence in homology). Clearly, $I$ restricts to short chains $\TotI$, and thus $\H(\TotI \ConCycBi) \simeq \H(\TotI \LodCycBi)$ holds too.

A similar discussion applies in cohomology. The following computation shows that $P$ is a chain map:
\begin{align*}
 &P \Dd_{\LodCycBi}(\psi_0,\psi_1,\psi_2, \psi_3, \psi_4,\dotsc) \\
 &\quad = P(\Hd^*\psi_0,(\Id-\CycPermOp^*)\psi_0 - \Hd'\psi_1,\CountOp^*\psi_1+ \Hd^*\psi_2,(\Id-\CycPermOp^*)\psi_2 - {\Hd'}^*\psi_3, \CountOp^*\psi_4 + \Hd^*\psi_3, \dotsc ) \\
 &\quad = \begin{aligned}[t]
  \bigl(\Hd^*\psi_0, & \CountOp^*\psi_1 + \Hd^*\psi_2 + \CountOp^*\InsOneOp_1^*((\Id-\CycPermOp^*)\psi_0 - {\Hd'}^*\psi_1), \\
  & \CountOp^*\psi_4 + \Hd^*\psi_3+ \CountOp^*\InsOneOp_1^*((\Id-\CycPermOp^*)\psi_2 - {\Hd'}^*\psi_3),\dotsc\bigr)
 \end{aligned} \\
 &\quad=\begin{aligned}[t]
 \bigl(\Hd^*\psi_0, \Cd^* \psi_0 + \Hd^* \psi_2 + \CountOp^*(\psi_1 - \InsOneOp_1^*{\Hd'}^* \psi_1), \Cd^* \psi_2 + \Hd^* \psi_4 + \CountOp^*(\psi_3 - \InsOneOp_1^*{\Hd'}^* \psi_3),\dotsc  \bigr)
\end{aligned}\\
&\quad = \bigl(\Hd^*\psi_0, \Cd^*\psi_0 + \Hd^*(\psi_2 + \CountOp^*\InsOneOp_1^* \psi_1), \Cd^*\psi_2 + \Hd^*(\psi_4 + \CountOp^*\InsOneOp_1^* \psi_3), \dotsc \bigr) \\
&\quad = \Dd_{\ConCycBi}\bigl(\psi_0,\psi_2 + \CountOp^*\InsOneOp_1^*\psi_1, \psi_2, \psi_4 + \CountOp^*\InsOneOp_1^*\psi_3,\dotsc\bigr)\\
&\quad = \Dd_{\ConCycBi} P (\psi_0,\psi_1,\psi_2,\psi_3,\psi_4,\dotsc)
\end{align*}
The fourth equality uses that $\InsOneOp_1 \Hd' + \Hd' \InsOneOp_1 = \Id$ and $\Hd'\CountOp = \CountOp\Hd$. Because $(\psi_0,\psi_1,\dotsc) = P(\psi_0,0,\psi_1,0,\dotsc )$, $P$ is surjective, and hence it induces an isomorphism of cochain complexes $\TotII \LodCycBi^*/\ker(P) \simeq \TotII \ConCycBi^*$.  It is easy to see that 
\[ \ker(P) = \bigl\{(0,\psi_1,-\CountOp^*\InsOneOp_1^*\psi_1,\psi_3,-\CountOp^*\InsOneOp_1^*\psi_3,\dotsc)\bigr\} \]
and that the map
\begin{align*}
Z: \TotII(\LodCycBi^{\mathrm{odd},\bullet}) &\longrightarrow \ker(p) \\
(\psi_1, \psi_3, \dotsc) &\longmapsto (0 ,\psi_1, -\CountOp^* \InsOneOp_1^* \psi_1, \psi_3, -\CountOp^* \InsOneOp_1^* \psi_3, \dotsc )
\end{align*}
is an isomorphism of the cochain complexes
\[ \bigl(\TotII(\LodCycBi^{\mathrm{odd},\bullet}),-{\Hd'}^*\bigr) \simeq \bigl(\ker(P),\Restr{\Bdd_{\LodCycBi}}{\ker(P)}\bigr) \subset (\TotII\LodCycBi,\Bdd_{\LodCycBi}). \]
Indeed, we have 
\begin{align*}
 \Bdd_{\LodCycBi}Z(\psi_1,\psi_3,\dotsc) &= (0,-{\Hd'}^*\psi_1, \CountOp^*\psi_1 - \Hd^*\CountOp^*\InsOneOp_1^*\psi_1, -(\Id-\CycPermOp^*)\CountOp^*\InsOneOp_1^*\psi_1 - {\Hd'}^*\psi_3, \dotsc ) \\
  &= \bigl(0,-{\Hd'}^*\psi_1,-\CountOp^*\InsOneOp_1^*(-{\Hd'}^*\psi_1), -{\Hd'}^*\psi_3,\dotsc\bigr) \\
  & = Z(-{\Hd'}^*)(\psi_1,\psi_3,\dotsc).
\end{align*}
Therefore, $\ker P$ is contractible, and the statement is implied by an argument with the long exact sequence in homology.
\end{proof}

\begin{Lemma}[Connes' cyclic bicomplexes are quasi-iso to their normalized versions]\label{Lem:ConNormVer}
Let $\Alg=(V,(\mu_k),\NOne)$ be a strictly unital $\AInfty$-algebra. The projection $\NormProj$ and the inclusion $\NormIncl$ (see Definition~\ref{Def:CycBico}) induce the isomorphisms $\H(\ConCycBi) \simeq \H(\NConCycBi)$ and $\H(\widehat{\ConCycBi}^*)\simeq\H(\widehat{\NConCycBi}^*)$, respectively.
\end{Lemma}
\begin{proof}
It follows from CP4 using the spectral sequence associated to the vertical filtration. In cohomology, we use (d) and (f) of Proposition~\ref{Prop:ConvOfSpSeq}.

In homology, we have $\tilde{\SSPage}^{sd} = \H(\ConCycBi_{-s,-d},\Bdd_\VertMRM)$ for the reversed spectral sequence (see the proof of Lemma~\ref{Lem:LodCycBiCycHom}), and hence strong convergence is implied by Theorem~\ref{Thm:Boardman61} from the proof of Proposition~\ref{Prop:ConvOfSpSeq}. Claim (e) of Proposition~\ref{Prop:ConvOfSpSeq} finishes the proof.
\end{proof}

The following is based on \cite[Proposition 2.2.14]{LodayCyclic} and its proof.

\begin{Lemma}[Reduced Connes' cyclic bicomplexes and cyclic homology are quas-iso] \label{Lem:ReducedCyclic}
Let~$\Alg=(V,(\mu_k),\NOne)$ be a strictly unital $\AInfty$-algebra. The projection $p^\lambda: \ConCycBi^{\RedMRM} \rightarrow \HNC^\lambda$ to the first column modulo $\im(\Id-\CycPermOp)$ is a chain map and induces an isomorphism $\H(\widehat{\ConCycBi}^{\RedMRM})\simeq\H(\HNC^\lambda)$ ($\eqqcolon\H^{\lambda,\RedMRM}(\Alg)$). The inclusion $\iota_\lambda: \HNC_\lambda^* \rightarrow \ConCycBi_{\RedMRM}^*$ into the first column of $\ConCycBi_{\RedMRM}^*$ is a chain map and induces an isomorphism $\H(\HNC_\lambda^*)\simeq\H(\ConCycBi_{\RedMRM}^*)$. 
\end{Lemma}

\begin{proof}
We start with the cohomology. It is easy to see that $\InsOneOp_1$ induces $\InsOneOp_1^*: \HC_{\RedMRM}^* V \rightarrow \HC_{\RedMRM}^* V$ and that for this map, we have
\[ Z\coloneqq \ker \InsOneOp_1^* = \im \InsOneOp_1^* \simeq \HC^* \bar{V}. \]
We consider the following diagonal filtration of $\ConCycBi^*_{\RedMRM}$:
 \begin{center}
$\Filtr^s\ConCycBi^*_{\RedMRM}: \quad$\begin{tikzcd}[column sep=scriptsize, row sep=scriptsize]
{}&{}&{} &{}\\
\HC_{\mathrm{red}}^{s+2} \arrow[u]{}{\Hd^*} \arrow[r]{}{\NCd^*}& \HC_{\mathrm{red}}^{s+1} \arrow[u]{}{\Hd^*} \arrow[r]{}{\NCd^*} & Z^s \arrow[u]{}{\Hd^*} \arrow[r]{}{\NCd^*}  &{}\\
\HC_{\mathrm{red}}^{s+1} \arrow[u]{}{\Hd^*} \arrow[r]{}{\NCd^*} & Z^s \arrow[u]{}{\Hd^*} \arrow[r]{}{\NCd^*}  & \arrow[u] \arrow[r]  0 &{} \\
Z^s \arrow[u]{}{\Hd^*} \arrow[r]{}{\NCd^*}  & 0 \arrow[u] \arrow[r]  & 0 \arrow[u]  \arrow[r]  & {} \\
{}\arrow[u]&{}\arrow[u]&{}\arrow[u]&{}
\end{tikzcd} 
\end{center}
The first page of the corresponding spectral sequence consists of the columns
\[ \SSPage_1^s = \H(\Gr_s \TotI) = \H\bigl(Z^s \xrightarrow{\Hd^*} \HC_{\RedMRM}^{s+1}/Z^{s+1} \xrightarrow{\NCd^*} Z^s \xrightarrow{\Hd^*} \dotsb \bigr), \]
where the cochain complex starts in degree $s$. Because $\InsOneOp_1^* \Hd^* = - {\Hd'}^* \InsOneOp_1^* + (\Id - \CycPermOp^*)$ and $\NCd^* = \CountOp^* \InsOneOp_1^*$ on $\HC_{\RedMRM}$, we have the commutative diagram
\[\begin{tikzcd}
 Z^s \arrow{r}{\Hd^*}\arrow{d}{\Id} & \HC_{\RedMRM}^{s+1}/Z^{s+1} \arrow{r}{\NCd^*}\arrow{d}{\InsOneOp_1^*} & Z^s \arrow{r}{\Hd^*}\arrow{d}{\Id} & \dotsb \\
 Z^s \arrow{r}{\Id-\CycPermOp^*} & Z^s \arrow{r}{\CountOp^*} & Z^s \arrow{r}{\Id-\CycPermOp^*} & \dotsb.
\end{tikzcd}\]
Therefore, the only non-zero terms of the first page are
\[ \SSPage_1^{s,0} = Z_s / \ker(\Id-\CycPermOp^*) = \bar{\HC}_\lambda^s \]
with the differential $\Dd_1 = \Hd^*$. This is precisely the first page of the canonical filtration of~$\bar{\HC}_\lambda^*$. The map induced by $\iota_\lambda$ on the first page is the identity, and Proposition~\ref{Prop:ConvOfSpSeq} implies the rest.

The situation in homology is analogous. We consider the restriction $\InsOneOp_1: \HC^{\RedMRM} \rightarrow \HC^{\RedMRM}$ and the subspace
\[ B\coloneqq \ker \InsOneOp_1 = \im \InsOneOp_1. \]
The diagonal filtration is now
\begin{center}
 $\Filtr^s \ConCycBi^{\RedMRM}: \quad$\begin{tikzcd}[column sep=scriptsize, row sep=scriptsize]
{}\arrow[d] & {}\arrow[d] & {}\arrow[d] & {} \\
0\arrow[d] & 0 \arrow[d]\arrow[l] & B_{s}\arrow[d]{}{\Hd}\arrow[l]{}{\NCd} & {}\arrow[l] \\
0\arrow[d] & B_{s}\arrow[d]{}{\Hd}\arrow[l]{}{\NCd} & \HC^{\RedMRM}_{s-1}\arrow[d]{}{\Hd}\arrow[l]{}{\NCd} & {}\arrow[l] \\
B_{s}\arrow[d] & \HC^{\RedMRM}_{s-1}\arrow[d]\arrow[l]{}{\NCd} & \HC^{\RedMRM}_{s-2}\arrow[d]\arrow[l]{}{\NCd} & {}\arrow[l] \\
{} & {} & {} & {}
\end{tikzcd} 
\end{center}
The reversed filtration $\DegRev(\Filtr)_s = \Filtr^{-s}$ of the degree reversed cochain comples $\DegRev(\TotI)$ satisfies
\begin{align*}
\tilde{\SSPage}_1^{s} & = \H(\DegRev(\Filtr)_{s}\DegRev(\TotI)/\DegRev(\Filtr)_{s+1}\DegRev(\TotI) ) \\
& = \H(\HC^{\RedMRM}_{-s-1}/B_{-s-1} \xleftarrow{\Hd} B_{-s} \xleftarrow{\NCd} \HC^{\RedMRM}_{-s-1}/B_{-s-1} \xleftarrow{\Hd}\dotsb)
\end{align*}
where the first group has degree $s$. We have the commutative diagram
\[\begin{tikzcd}
\HC^{\RedMRM}_{-s-1}/B_{-s-1}\arrow{d}{\InsOneOp_1} & \arrow{l}{\Hd}\arrow{d}{\Id} B_{-s} & \arrow{l}{\NCd} \HC^{\RedMRM}_{-s}/B_{-s}\arrow{d}{\InsOneOp_1} & \arrow{l}{\Hd}\dotsb \\
B^{-s}&\arrow{l}{\Id-\CycPermOp}B^{-s}&\arrow{l}{\CountOp}B^{-s}&\arrow{l}{\Id-\CycPermOp}\dotsb,
\end{tikzcd}\]
and thus $\tilde{\SSPage}_1^{s,0} = B^{-s}/\im(\Id-\CycPermOp) = \HC^{\lambda,\RedMRM}_{-s}$. The rest is as in the case of cohomology.
\end{proof}

\section{Final argument and remarks}\label{Sec:FinRem}

We are finally in position to prove Proposition~\ref{Prop:Reduced}. Precisely as in the proof sketch, we replace the unit-augmentation sequence \eqref{Eq:UnitAugSS}, up to a quasi-isomorphism, by a short exact sequence of normalized Connes' cyclic bicomplexes.

\begin{Lemma}[Short exact sequence of normalized Connes' cyclic bicomplexes]\label{Lem:ConBiRed}
Let $\mathcal{A}=(V,(\mu_j),\NOne,\varepsilon)$ be a strictly augmented strictly unital $\AInfty$-algebra. The short exact sequences of bicomplexes
\[\begin{tikzcd}
 0 \arrow{r} &  \NConCycBi(\R) \arrow{r}{u} & \NConCycBi(V)\arrow{r}{p^{\RedMRM}} & \ConCycBi^{\RedMRM} \arrow{r} & 0
\end{tikzcd}\]
and
\[\begin{tikzcd}
 0 \arrow{r} & \ConCycBi_{\RedMRM}^* \arrow{r}{\iota_{\RedMRM}} & \NConCycBi^*(V) \arrow{r}{u^*} & \NConCycBi(\R) \arrow{r} & 0
\end{tikzcd}\]
split. From this, we obtain the following isomorphisms:
\begin{equation}\label{Eq:ConRedBiIso}
\begin{aligned}
\H(\NConCycBi)&\simeq  \H(\ConCycBi^{\RedMRM}) \oplus \H^\lambda(\R), & \H( \widehat{\NConCycBi}) &\simeq  \H(\widehat{\ConCycBi}^{\RedMRM})\oplus H^\lambda(\R), \\
\H(\NConCycBi^*) &\simeq \H(\ConCycBi^*_{\RedMRM}) \oplus H_\lambda^*(\R), & \H(\widehat{\NConCycBi}^*) &\simeq  \H(\widehat{\ConCycBi}^*_{\RedMRM})\oplus H_\lambda^*(\R).
\end{aligned}
\end{equation}
\end{Lemma}
\begin{proof}
Because we have a strict unit and a strict augmentation, the maps $u: \HC \R \rightarrow \HC V$ and $\varepsilon: \HC V \rightarrow \HC \R$ satisfy $\varepsilon \circ u = \Id$. It follows that $\HC V = \ker\varepsilon\oplus\im u$. The same holds for the induced maps $\bar{u}: \HNC \R \rightarrow \HNC V$ and $\bar{\varepsilon}: \HNC V \rightarrow \HNC \R$. We can now define the splitting $r: \NConCycBi(V) \rightarrow \NConCycBi(\R)$ of the homological short exact sequence by projecting to $\ker \bar{\varepsilon}$ along $\im \bar{u}$. It is easy to check that it is a morphism of bicomplexes. This dualizes ``pointwisely'' to cohomology.

Because 
\[ (\HNC \R)_i = \begin{cases} 0 & \text{for }i \neq 0, \\ \R & \text{for } i = 0, \end{cases} \]
the bicomplex $\NConCycBi(\R)$ is diagonal, and hence $\TotI$ and $\TotII$ are the same; both compute $\H^\lambda(\R)$ (using results from the previous sections). The same is true in cohomology. This shows~\eqref{Eq:ConRedBiIso}.
\end{proof} 

We summarize our results in Figure~\ref{Fig:FinalPictureHom}. Having $\H^\lambda(\Alg) \simeq \H^{\lambda,\RedMRM}(\Alg) \oplus \H^{\lambda}(\R)$, Proposition~\ref{Prop:Reduced} in Section~\ref{Sec:Alg2} follows by dualization. 

\begin{Questions}\phantomsection\label{Q:OpenProbAInftx}
\begin{RemarkList}
\item Suppose that $V$ has bounded degrees and look at Figure~\ref{Fig:FinalPictureHom}. Does $\H(\widehat{\LodCycBi}^*)\simeq \H(\LodCycBi^*)$ hold? Does $\H(\ConCycBi^*)\simeq\H(\NConCycBi^*)$ hold?
\item Does Proposition~\ref{Prop:Reduced} hold for homological unital and homological augmented $\AInfty$-algebras? A strategy would be to construct a quasi-isomorphic strictly unital and strictly augmented $\AInfty$-algebra. Does it exist?
\item How is it with the conditional and strong convergence of spectral sequences associated to different filtrations of bicomplexes from Definition~\ref{Def:CycBico}? Because of the simple internal data, lots of them collapse.\qedhere
\end{RemarkList}
\end{Questions}
\begin{figure}[t]
\centering
 \begin{tikzcd}  
  \H^\lambda(\Alg) \arrow[dash]{r}{Lem.~\ref{Lem:LodCycBiCycHom}}  & \H(\widehat{\LodCycBi}) \arrow[dash]{d}{Lem.~\ref{Lem:LodConCycBi}} \arrow[dash,dashed]{r}{Lem.~\ref{Lem:BddDegrees}}  & \H(\LodCycBi) \arrow[dash]{d}{Lem.~\ref{Lem:LodConCycBi}} \\
   {} &  \H(\widehat{\ConCycBi}) & \H(\ConCycBi) \arrow[dash]{d}{Lem.~\ref{Lem:ConNormVer}} \\
    {} &  \H(\widehat{\NConCycBi}) \arrow[dash]{d}{Lem.~\ref{Lem:ConBiRed}} \arrow[dash,dashed]{r}{Lem.~\ref{Lem:BddDegrees}} &   \H(\NConCycBi) \arrow[dash]{d}{Lem.~\ref{Lem:ConBiRed}} \\
{} & \H(\widehat{\ConCycBi}^{\RedMRM}) \oplus \H^\lambda(\R) \arrow[dash]{d}{Lem.~\ref{Lem:ReducedCyclic}} & \H(\ConCycBi^{\RedMRM}) \oplus \H^\lambda(\R) \\
{}& \H^{\lambda,\RedMRM}(\Alg) \oplus \H^\lambda(\R)  & {}
 \end{tikzcd}
\\[1cm]
 \begin{tikzcd}  
   \H(\widehat{\LodCycBi}^*) \arrow[dash]{d}{Lem.~\ref{Lem:LodConCycBi}} & \H(\LodCycBi^*) \arrow[dash]{d}{Lem.~\ref{Lem:LodConCycBi}} & \H_\lambda^* \arrow[dash]{l}{Lem.~\ref{Lem:LodCycBiCycHom}} \arrow[bend left = 60,dotted,dash]{ldddd}\\
   \H(\widehat{\ConCycBi}^*) \arrow[dash]{d}{Lem.~\ref{Lem:ConNormVer}} & \H(\ConCycBi^*) & {} \\
   \H(\widehat{\NConCycBi}^*) \arrow[dash]{d}{Lem.~\ref{Lem:ConBiRed}} &   \H(\NConCycBi^*) \arrow[dash]{d}{Lem.~\ref{Lem:ConBiRed}} & {} \\
  \H(\widehat{\ConCycBi}^*_{\RedMRM})\oplus\H_\lambda^*(\R) &  \H(\ConCycBi^*_{\RedMRM})\oplus\H_\lambda^*(\R) \arrow[dash]{d}{Lem.~\ref{Lem:ReducedCyclic}} & {} \\
  {} & \H_{\lambda,\RedMRM}^*(\Alg) \oplus \H_{\lambda}^*(\R) & {}
 \end{tikzcd}
\caption[Isomorphisms of various versions of cyclic (co)homologies of an $\AInfty$-algebra.]{Isomorphisms of (co)homologies for a strictly unital strictly augmented $\AInfty$-algebra $\Alg$ on a graded vector space $V$. A solid line denotes an isomorphism which is always valid and a dashed line an isomorphism which is valid provided that the degrees of $V$ are bounded. The dotted line denotes the isomorphism obtained by dualizing the corresponding isomorphism in homology under the assumptions that the degrees of $V$ are bounded.}
\label{Fig:FinalPictureHom}
\end{figure}

\chapter{Towards an invariant definition of canonical IBL-operations}
\label{Section:AppEqDefPrCoPr}
\allowdisplaybreaks
\Correct[caption={Change notation},noline]{Change the notation for the pre-Lie algebra product and for the $D$-space!!! I propose $\triangle$ and }
The canonical $\IBL$-operations $\OPQ_{210}$ and $\OPQ_{120}$ on cyclic cochains of an odd symplectic vector space $V$ have been defined in coordinates in \cite{Cieliebak2015}. In Part~I, we used Definition~\ref{Def:CanonicaldIBL}; it takes and gives invariant objects but requires a choice of basis to describe the inner ``trace'' mechanism. Example~\ref{Ex:Canon} shows that an invariant definition can be obtained from the invariant formalism for evaluation of ribbon graphs developed in Appendix~\ref{Section:Appendix} by plugging in the algebraic Schwarz kernel of the identity as a propagator. In this appendix, we ask the following question.
\begin{Question}\label{Q:CanOp}
Do $\OPQ_{210}$ and $\OPQ_{120}$ arise as a combination of some natural operations coming from the structure of Hochschild cochains on $V$ or from odd symplectic geometry? 
\end{Question}
\Add[caption={DONE Gerstenhaber bracket},noline]{Mention that it is, in fact, the Gerstenhaber bracket!}
In Section~\ref{Sec:LieBrHoch}, we define a canonical Lie bracket $[\cdot,\cdot]$ on Hochschild cochains on~$V$ with values in~$V$ (Definition~\ref{Def:CanonLA} and Proposition~\ref{Prop:CanonLA}), which comes from a natural pre-Lie algebra structure which is visualized as grafting of trees (Equation~\eqref{Eq:PreLie}, Figure~\ref{Fig:GraftTrees} and Lemma~\ref{Lem:PreLie}). This is, in fact, the Gerstenhaber bracket from \cite{Gerstenhaber1963}. Next, we use the symplectic form to define the operator~${}^+$ which ``lowers indices'' (Equation~\ref{Eq:ff}) and show that it takes $[\cdot,\cdot]$ to a Lie bracket on cyclic cochains on~$V$ with values in~$\R$ (Definition~\ref{Def:NewProduct} and Proposition~\ref{Prop:NewLieBr}). By writing down everything in coordinates, we show that this Lie bracket is a degree shift of $\OPQ_{210}$ (Proposition~\ref{Prop:EqOfDefCool}). This answers Question~\ref{Q:CanOp} for $\OPQ_{210}$.

In Section~\ref{Sec:CoprNic}, we relate the cobracket $\OPQ_{120}$ to the canonical $\BV$-operator $\BVOpSym$ on ``functions'' on an odd symplectic vector space (Proposition~\ref{Prop:HochCycSym}). We do it by rewriting~$\OPQ_{120}$ in terms of certain double derivative operators on the tensor algebra, which are associated to a given basis (Definition~\ref{Def:DefOfHochOp}). We prove that both definitions are equivalent (Proposition~\ref{Prop:EqCoprod}). The $\BV$-operator $\BVOpSym$ can be defined geometrically (see~\cite{Doubek2018}), and the cobracket $\OPQ_{120}$ is a factorization of an extension of~$\BVOpSym$ to cyclic invariants with respect to the cyclic shuffle product. We do not know whether this characterization determines $\OPQ_{120}$ uniquely, and hence Question~\ref{Q:CanOp} for~$\OPQ_{120}$ remains open (see Questions~\ref{Q:OpenProbBrCo} for a list of related questions). This section was stimulated by a discussion with J.~Pullman and L.~Peksov\'a at a winter school in Srn\'i, 2019.%

\section{Bracket and grafting of trees}\label{Sec:LieBrHoch}

Let $V$ be a $\Z$-graded vector space. We define the weight-graded vector spaces
\begin{equation}\label{Eq:DefOfD}
\DSp V \coloneqq \bigoplus_{k=0}^\infty \Hom(V^{\otimes k},V)\quad\text{and}\quad \DSpRed V \coloneqq \bigoplus_{k=1}^\infty \Hom(V^{\otimes k},V),
\end{equation}
where $\Hom$ denotes the graded vector space generated by homogenous morphisms. The latter space is the weight-reduced version of the former (see Definition~\ref{Def:Grading}).

\begin{figure}
\centering
\input{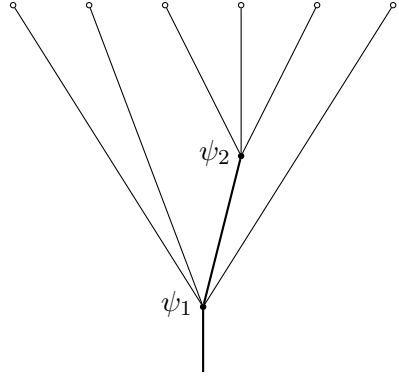}
\caption{The Gerstenhaber bracket as grafting of trees.}
\label{Fig:GraftTrees}
\end{figure}

For homogenous $\psi_1 \in \Hom(V^{\otimes k_1},V)$, $\psi_2 \in \Hom(V^{\otimes k_2},V)$ with $k_1$, $k_2\in\N$ and vectors $v_1$, $\dotsc$, $v_{k_1+k_2-1}\in V$, we define\footnote{If $k_1 = 0$ or $k_2 = 0$, we define $\Star$ to be $0$.}
\begin{equation}\label{Eq:PreLie}
\begin{aligned}
&(\psi_1 \OpPre \psi_2)(v_1 \dotsb v_{k_1 + k_2 -1}) \\
&\ \coloneqq \begin{multlined}[t]\sum_{i=1}^{k_1} (-1)^{\psi_2(v_1 + \dotsb + v_{i-1})}\psi_1(v_1\dotsb v_{i-1} \psi_2(v_i \dotsb v_{k_2 + i -1}) v_{k_2+i} \dotsb v_{k_1 + k_2 - 1}).\end{multlined}
\end{aligned}
\end{equation}
Here and almost everywhere in this section, we suppress writing the tensor product. We recall that we use the same symbol $v$ to denote both a homogenous vector and its degree in the exponent. From \eqref{Eq:PreLie}, we get a bilinear operation 
\[ \OpPre: \DSp V\otimes \DSp V \longrightarrow \DSp V, \]
which preserves the degree and decreases the weight by $1$. It can be visualized as grafting of trees (see Figure~\ref{Fig:GraftTrees}).
\begin{Lemma}[Pre-Lie algebra]\label{Lem:PreLie}\Reminder[noline,caption={Non-reduced pre-Lie algebra.}]{Can one define $\Star$ differently when there are no inputs and get something non-trivial?}
Let $V$ be a graded vector space. The pair $(\DSp V,\OpPre)$ is a graded pre-Lie algebra with respect to the degree, i.e., it holds
\[ (\psi_1 \OpPre \psi_2)\OpPre\psi_3 -  \psi_1 \OpPre (\psi_2\OpPre\psi_3) = (-1)^{\psi_2 \psi_3}\bigl[(\psi_1 \OpPre \psi_3)\OpPre\psi_2 - \psi_1 \OpPre (\psi_3\OpPre\psi_2)\bigr] \]
for all homogenous $\psi_1$, $\psi_2$, $\psi_3 \in \DSp V$. 
\end{Lemma}
\begin{proof}
In the following computations, $U$ denotes a tensor product of homogenous vectors from the vector space~$V$ and $U_i$ denotes a part of a decomposition of $V$, i.e., $U = U_1 \dotsb U_k$. We compute schematically
\begin{align*}
&\bigl((\psi_1 \OpPre \psi_2)\OpPre \psi_3\bigr)(U)  =\begin{aligned}[t]
&\sum (-1)^{\psi_3(U_1 + U_2 + U_3) + \psi_2 U_1}\psi_1(U_1\psi_2(U_2)U_3\psi_3(U_4)U_5) \\
{}+ &\sum (-1)^{\psi_3 U_1 + \psi_2(U_1 + U_2 + U_3+\psi_3)}\psi_1(U_1\psi_3(U_2)U_3\psi_2(U_4)U_5) \\
{}+ &\sum (-1)^{\psi_3(U_1 + U_2)+ \psi_2 U_1}\psi_1(U_1\psi_2(U_2\psi_3(U_3)U_4)U_5)\end{aligned}
\end{align*}
and
\begin{align*}
\bigl(\psi_1 \OpPre (\psi_2\OpPre \psi_3)\bigr)(U)=\sum (-1)^{(\psi_2 - \psi_3) U_1 + \psi_3 U_2}\psi_1(U_1\psi_2(U_2\psi_3(U_3)U_4)U_5).
\end{align*}
Therefore, we have
\begin{align*}
&\bigl[\bigl((\psi_1 \OpPre \psi_2)\OpPre \psi_3\bigr) - \bigl(\psi_1 \OpPre (\psi_2\OpPre \psi_3)\bigr)\bigr](U) \\
&\qquad\qquad=\begin{aligned}[t]&\sum (-1)^{\psi_3(U_1 + U_2 + U_3) + \psi_2 U_1}\psi_1(U_1\psi_2(U_2)U_3\psi_3(U_4)U_5) \\
{}+ &\sum (-1)^{\psi_3 U_1 + \psi_2(U_1 + U_2 + U_3+\psi_3)}\psi_1(U_1\psi_3(U_2)U_3\psi_2(U_4)U_5)
\end{aligned}\\
&\qquad\qquad=\begin{aligned}[t]
(-1)^{\psi_2 \psi_3}\Bigl[&\sum (-1)^{\psi_2(U_1 + U_2 + U_3) + \psi_3 U_1 }\psi_1(U_1\psi_3(U_2)U_3\psi_2(U_4)U_5) \\
{}+&\sum (-1)^{\psi_2 U_1 + \psi_3(U_1 + U_2 + U_3 + \psi_2)}\psi_1(U_1\psi_2(U_2)U_3\psi_3(U_4)U_5)\Bigr]
\end{aligned}\\
&\qquad\qquad=(-1)^{\psi_2 \psi_3}\bigl[\bigl((\psi_1 \OpPre \psi_3)\OpPre \psi_2\bigr) - \bigl(\psi_1 \OpPre (\psi_3\OpPre \psi_2)\bigr)\bigr](U).
\end{align*}
This finishes the proof.
\end{proof}

\begin{Definition}[Canonical Lie algebra on Hochschild cochains]\label{Def:CanonLA}
Let $V$ be a graded vector space. For all $\psi_1$, $\psi_2 \in \DSp V$, we define the bracket by
\[ [\psi_1, \psi_2] \coloneqq \psi_1 \OpPre \psi_2 - (-1)^{\psi_1 \psi_2} \psi_2 \OpPre \psi_1. \]
It is called the \emph{Gerstenhaber bracket}.
\end{Definition}
\begin{Proposition}[Canonical Lie algebra on Hochschild cochains]\label{Prop:CanonLA}
In the situation of Definition~\ref{Def:CanonLA}, the pair $(\DSp V,[\cdot,\cdot])$ is a graded Lie algebra with bracket $[\cdot,\cdot]$ of degree $0$ and weight $-1$.
\end{Proposition}
\begin{proof}
The proof is standard.
\end{proof}

Let $\langle\cdot,\cdot\rangle: V \otimes V \rightarrow \R$ be a homogenous bilinear form of degree $-n$. This means that
\[ \langle v_1, v_2 \rangle \neq 0\quad\Implies\quad v_1 + v_2 = n. \]
To every $\psi\in \Hom(V^{\otimes k},V)$, we associate $\psi^+\in \Hom(V^{\otimes k + 1},\R)$ defined for all $v_1$, $\dotsc$, $v_{k + 1} \in V$ by the formula
\begin{equation}\label{Eq:ff}
\psi^+(v_1\dotsb v_{k+1}) \coloneqq \langle \psi(v_1 \dotsb v_{k}),v_{k+1}\rangle.
\end{equation}
We define the weight-graded vector spaces
\begin{equation}\label{Eq:DSpPlusDef}
\DSpPlus V \coloneqq \bigoplus_{k=0}^\infty \Hom(V^{\otimes k},\R)\quad\text{and}\quad\DSpPlusRed V \coloneqq \bigoplus_{k=1}^\infty \Hom(V^{\otimes k},\R).
\end{equation}
We use here the cohomological grading convention; i.e., $\psi\in \DSpPlus V$ has degree~$d$ and weight~$k$ if and only if for all homogenous $v_1$, $\dotsc$, $v_k\in V$, the following implication holds:
\[ \psi(v_1 \dotsb v_k) \neq 0 \quad\Implies \quad v_1 + \dotsb + v_k = d. \] 
We say that $\varphi\in \DSp V$ is cyclically symmetric if 
\[ \varphi(v_1 \dotsb v_{k}) = \underbrace{(-1)^{v_{k}(v_1 + \dotsb + v_{k-1})}}_{\eqqcolon\varepsilon(v_1\dotsb v_{k},v_{k} v_1 \dotsb v_{k-1})} \varphi(v_{k} v_1 \dotsb v_{k-1}) \]
for all homogenous $v_1$, $\dotsc$, $v_{k}\in V$. We define the weight-graded vector spaces
\begin{align*}
\DSpPlusCyc V &\coloneqq \{\varphi\in \DSpPlus V \mid \varphi\text{ is cyclically symmetric}\}\quad\text{and} \\
\DSpCyc V &\coloneqq \{\psi \in \DSp V \mid \psi^+ \in \DSpPlusCyc V \},
\end{align*}
and similarly for the reduced versions. Formula \eqref{Eq:ff} defines the following linear maps of weight $+1$:
\begin{equation}\label{Eq:CycDiag}
\begin{tikzcd}
 {}^+: & \DSp V \arrow{r} & \DSpPlusRed V \\
 {}^+\coloneqq \Restr{{}^+}{\DSpCyc V}: & \DSpCyc V \arrow{r}\arrow[hook]{u} & \DSpPlusCycRed V\arrow[hook]{u}
\end{tikzcd}
\end{equation}
This might be related to Remark~\ref{Rem:NWG} about weight-reduced bar complexes.

Because $\DSpPlus V$ has the cohomological grading and~$\DSp V$ the standard grading for homogenous maps, ${}^+$ is not homogenous; instead, it satisfies 
\[\Abs{\psi^+} = n - \Abs{\psi}. \]
It is also useful to note that if $\psi^+(v_1\dotsb v_{k+1})\neq 0$, then
\[ \varepsilon(v_1\dotsb v_{k+1},v_{k+1} v_1 \dotsb v_k) = (-1)^{v_{k+1}(\psi^+ - 1)} = (-1)^{v_{k+1}(\psi - n - 1)}. \]

\begin{Lemma}[Restriction of Lie bracket to cyclic invariants]
Let $V$ be a graded vector space and $\langle \cdot,\cdot\rangle: V\otimes V \rightarrow \R$ a homogenous bilinear form of degree $-n$ which is graded antisymmetric. This means that for all homogenous $v_1$, $v_2\in V$, we have
\[ \langle v_1, v_2 \rangle = (-1)^{1+v_1 v_2}\langle v_2, v_1\rangle. \]
Then the Lie bracket $[\cdot,\cdot]$ on $\DSp V$ satisfies $[\DSpCyc V, \DSpCyc V]\subset\DSpCyc V$, and thus it restricts to the Lie bracket (which we denote by the same symbol) 
\[ [\cdot,\cdot]: \DSpCyc V \otimes \DSpCyc V \longrightarrow \DSpCyc V. \]
\end{Lemma}
\begin{proof}
For any homogenous $v_1$, $\dotsc$, $v_{k_1+k_2} \in V$, we compute
\begin{align*}
&\langle[\psi_1,\psi_2](v_1,\dotsc,v_{k_1+k_2-1}),v_{k_1+k_2}\rangle \\
& = \begin{aligned}[t]
&\sum_{i=1}^{k_1} (-1)^{\psi_2(v_1+\dotsb+v_{i-1})}\langle \psi_1(v_1 \dotsb v_{i-1}\psi_2(v_i \dotsb v_{i+k_2-1}) v_{i+k_2}\dotsb \\ &v_{k_1+k_2-1}), v_{k_1+k_2}\rangle - (-1)^{\psi_1 \psi_2}\sum_{j=1}^{k_2}(-1)^{\psi_1(v_1+\dotsb+v_{j-1})}\langle \psi_2(v_1\dotsb \\
&v_{j-1}\psi_1(v_j\dotsb v_{j+k_1-1})v_{j+k_1}\dotsb v_{k_1+k_2-1})), v_{k_1+k_2}\rangle\end{aligned}\\
& = \begin{aligned}[t]
& (-1)^{v_{k_1+k_2}(v_1 + \dotsb v_{k_1+k_2-1})} \\
& \Bigl[\sum_{i=1}^{k_1-1} (-1)^{\psi_2(v_{k_1+k_2}+v_1+\dotsb+v_{i-1})} \langle \psi_1(v_{k_1+k_2} v_1 \dotsb \psi_2(v_i \dotsb v_{i+k_2-1})\\
& v_{i+k_2}\dotsb v_{k_1+k_2-2}), v_{k_1+k_2-1}\rangle+(-1)^{\psi_2(v_1+\dotsb+v_{k_1-1})+v_{k_1+k_2}\psi_2}\\
&\langle \psi_1(v_{k_1+k_2} v_1\dotsb v_{k_1-1}),\psi_2(v_{k_1}\dotsb v_{k_1+k_2-1})\rangle\\
&- (-1)^{\psi_1 \psi_2}\Bigl(\sum_{j=1}^{k_2-1}(-1)^{\psi_1(v_{k_1+k_2}+v_1+\dotsb+v_{j-1})}\langle \psi_2(v_1\dotsb\psi_1(v_j\dotsb\\
& v_{j+k_1-1})v_{j+k_1}\dotsb v_{k_1+k_2-1})), v_{k_1+k_2}\rangle + (-1)^{\psi_1(v_1+\dotsb+v_{k_2-1})+v_{k_1+k_2}\psi_1} \\
& \langle \psi_2(v_{k_1+k_2} v_1\dotsb v_{k_2-1}),\psi_1(v_{k_2},\dotsb,v_{k_1+k_2-1})\rangle\Bigr)\Bigr].
\end{aligned}
\end{align*}
Using graded antisymmetry, we have for the last summand in square brackets
\begin{align*}
&(-1)^{\psi_1(v_1+\dotsb+v_{k_2-1} + v_{k_1 + k_2})}\langle \psi_2(v_{k_1+k_2} v_1\dotsb v_{k_2-1}),\psi_1(v_{k_2},\dotsb,v_{k_1+k_2-1})\rangle \\
&= \begin{multlined}[t](-1)^{1 + \psi_1 \psi_2 + (\psi_2 + v_{k_1 + k_2}+v_1+\dotsb+v_{k_2-1})(v_{k_2}+ \dotsb + v_{k_1+k_2-1})} \\ 
\langle \psi_1(v_{k_2}\dotsb v_{k_1+k_2-1}),\psi_2(v_{k_1+k_2} v_1 \dotsb v_{k_2-1})\rangle\end{multlined} \\
& = (-1)^{1+\psi_1 \psi_2}\langle\psi_1(\psi_2(v_{k_1 + k_2} v_1 \dotsb v_{k_2-1}) v_{k_2}\dotsb v_{k_1 + k_2 - 2}),v_{k_1 + k_2 - 1}\rangle
\end{align*}
and similarly for the second summand
\begin{align*}
&(-1)^{\psi_2(v_1+\dotsb+v_{k_1-1} + v_{k_1+k_2})}\langle\psi_1(v_{k_1+k_2}v_1\dotsb v_{k_1-1}),\psi_2(v_{k_1}\dotsb v_{k_1+k_2-1})\rangle\\
&=\begin{multlined}
(-1)^{1+\psi_1 \psi_2} \langle \psi_2(\psi_1(v_{k_1+k_2} v_1 \dotsb v_{k_1-1}) v_{k_1} \dotsb v_{k_1+k_2-2}),v_{k_1 + k_2 - 1}\rangle.
\end{multlined}
\end{align*}
Therefore, the square bracket equals
\[ \langle[\psi_1,\psi_2](v_{k_1+k_2} v_1 \dotsb v_{k_1 + k_2 - 2}), v_{k_1 + k_2 -1} \rangle \]
and the lemma is proven.
\end{proof} 

The horizontal arrows in \eqref{Eq:CycDiag} become isomorphisms provided that $\langle \cdot,\cdot \rangle$ is non-degenerate (injectivity) and $V$ is of finite type (surjectivity). Therefore, the following definition is possible.
\Reminder[noline,caption={DONE Nondeg, fintype and $+$}]{Be careful, I used to have here in addition degrees bounded from below and now I don't see why. NO PROBLEM, THE ISSUE DISCUSSED IN A FOOTNOTE LATER.}

\newcommand{\NDeg}{\mathbb{\nu}}
\begin{Definition}[Candidate for $\IBL$-product]\label{Def:NewProduct}
Let $V$ be a graded vector space of finite type, and let $\langle \cdot,\cdot \rangle: V\otimes V \rightarrow \R$ be a non-degenerate homogenous graded antisymmetric bilinear form of degree $-n$. Let $\NDeg$ be a formal symbol of degree $n$ realizing the degree shift by $-n$. We define the operation 
\[ \OPQ^*_{210}: (\DSpPlusCycRed V)[-n] \otimes (\DSpPlusCycRed V)[-n] \longrightarrow (\DSpPlusCycRed V)[-n] \] 
for all $\psi_1$, $\psi_2 \in \DSpCyc V$ by
\[ \OPQ^*_{210}( \NDeg\psi_1^+\otimes \NDeg\psi_2^+) \coloneqq \NDeg[\psi_2,\psi_1]^+. \]
\end{Definition}

\begin{Proposition}[Candidate for $\IBL$-product]\label{Prop:NewLieBr}
In the situation of Definition~\ref{Def:NewProduct}, the pair $((\DSpPlusCycRed V)[-n],\OPQ_{210}^*)$ is a graded Lie algebra with the bracket $\OPQ^*_{210}$ of degree $-2n$ and weight $-2$.
\end{Proposition}
\begin{proof}
As for the degree, we have
\begin{align*}
\Abs{\NDeg [\psi_2,\psi_1]^+} &= n+ \Abs{[\psi_2,\psi_1]^+} \\
&= n - \Abs{[\psi_2,\psi_1]}  + n \\
&= n  - \Abs{\psi_1}  - \Abs{\psi_2} + n\\
&= n + \Abs{\psi_1^+} + \Abs{\psi_2^+} - n \\
& = -n + \Abs{\NDeg \psi_1^+} + \Abs{\NDeg \psi_2^+} - n \\
& = \Abs{\NDeg \psi_1^+} + \Abs{\NDeg \psi_2^+} -2n.
\end{align*}
The weights are clear. As for the graded anticommutativity, we have 
\begin{align*}
\OPQ_{210}^*(\NDeg \psi_2^+ \otimes \NDeg\psi_1^+) &= \NDeg[\psi_1,\psi_2]^+ \\
&= -(-1)^{\psi_1 \psi_2} \NDeg[\psi_2,\psi_1]^+ \\ 
&= -(-1)^{\Abs{\NDeg\psi_1}\Abs{\NDeg\psi_2}} \OPQ_{210}^*(\NDeg \psi_1^+ \otimes \NDeg\psi_2^+),
\end{align*}
where we used that 
\[ \Abs{\NDeg \psi^+} = n + \Abs{\psi^+} = n + n - \Abs{\psi} = 2n - \Abs{\psi}. \] 
From the same reason, the graded Jacobi identity for $\OPQ_{210}^*$ is implied by the graded Jacobi identity for $[\cdot,\cdot]$.
\end{proof}

In the situation of Definition~\ref{Def:NewProduct}, let $(e_i)$ be a basis of $V$, and let $(e^i)$ be the dual basis of $V$ such that $\langle e_i, e^j \rangle= \delta_i^j$. To an index $i$, we will assign the degree of $e_i$, so that, e.g., we can write $(-1)^i$ instead of $(-1)^{e_i}$. We will also use the Einstein summation convention.

We define the coordinates
\begin{align*}
g_{ij} &= \langle e_i,e_j\rangle, \\
g^{ij} &= \langle e^i, e^j\rangle, \\
\psi(e_{i_1}\dotsb e_{i_{k_1}}) &= \psi_{i_1 \dotsb i_{k_1}}^i e_i, \\
\psi^+(e_{i_1}\dotsb e_{i_{k_1+1}}) & = \psi_{i_1 \dotsb i_{k_1 + 1}},
\end{align*}
where $\psi\in\DSp V$ and $\psi^+\in\DSpPlus V$. It holds
\[ g_{ij} = (-1)^{i j + 1} g_{ji}, \quad g^{ij} = (-1)^{ij + 1} g^{ji} \quad \text{and}\quad g_{ik}g^{jk} = \delta_i^j. \] 
Equation \eqref{Eq:ff} is now equivalent to 
\begin{equation}\label{Eq:TransfForm}
\psi^i_{i_1 \dotsb i_{k_1}} g_{i i_{k_1+1}}  = \psi_{i_1\dotsb i_{k_1 + 1}},\quad\text{or to}\quad \psi^i_{i_1\dotsb i_{k_1}} = \psi_{i_1 \dotsb i_{k_1} k} g^{ik}.
\end{equation}
We define the operation $\mu: (\DSpPlusCycRed V)^{\otimes 2} \rightarrow \DSpPlusCycRed V$ for $\varphi_1\in \CycHom(V^{k_1 + 1},\R)$, $\varphi_2\in \CycHom(V^{\otimes k_2 + 1},\R)$ with $k_1$, $k_2\in \N_0$ such that $k_1+k_2 \ge 1$ in coordinates by the formula 
\begin{equation}\label{Eq:CoordFormula}
\mu(\varphi_1 \otimes \varphi_2)_{i_1 \dotsb i_{k_1 + k_2}} \coloneqq\begin{multlined}[t] 
\sum_{i,j}\sum_{c=1}^{k_1 + k_2} \varepsilon(i_1 \dotsb i_{k_1 + k_2},i_c \dotsb i_{c-1})(-1)^{(\varphi_1 - i)j} \\ (-1)^i g^{ij} (\varphi_1)_{i i_c \dotsb i_{c+k_1-1}} (\varphi_2)_{j i_{c+k_1} \dotsb i_{c-1}}.
\end{multlined}
\end{equation}
\Correct[noline, caption={DONE possibly wrong sign}]{There might be a sign mistake in $(-1)^a$ because what is needed is probabaly $(-1)^{e^a}$ which differs by $n$. IT IS CORRECT AS IT IS.}
\Modify[caption={DONE finite type},noline]{One can replace finite dimensional by finite type.}
\begin{Remark}[Comparison to \cite{Cieliebak2015}]\label{Rem:CompToCFL}
In order to compare~$\mu$ from \eqref{Eq:CoordFormula} to~$\mu$ from~\cite[Section~10]{Cieliebak2015} (and to the definition of~$\OPQ_{210}$ from Section~\ref{Sec:Alg3}), we must make the replacements
\[ V\mapsto V[1], \quad  n \mapsto n - 2\quad\text{and}\quad\DSpPlusCycRed V\mapsto \DBCyc V. \]
Then $\OPQ_{210}$ and $\OPQ^*_{210}: (\DBCyc V)[2-n]^{\otimes 2} \rightarrow (\DBCyc V)[2-n]$ clearly have the same degree $2(2-n)$.\footnote{The degree of $\OPQ_{210}$ on tensor powers of $C = \DBCyc V [2-n]$, is obtained from the degree on $\Ext C = \Sym(C[1])$ in Definition~\ref{Def:CanonicaldIBL} by substracting $1$ (and by adding $1$ for the cobracket).} Also, $\varphi_1 j + i j$ is precisely the sign needed to move $j$ over $i_{c}\dotsb i_{c+k_1-1}$. We also recall the identification of $\DBCyc V$ with the weight-graded dual $(\BCyc V)^{\WGD}$ from Remark~\ref{Rem:Identifications} and remind that $\BCyc V$ and $\DBCyc V$ were defined to be weight-reduced in order to ease the terminology and notation (see Definition~\ref{Def:BarComplex}).
\Correct[noline,caption={DONE wrong degree}]{The degrees seem wrong. $\OPQ_{210}$ has $-2(n-3)-1$ and... NO PROBLEM.}
\end{Remark}

The equality $\OPQ_{210}^*= \OPQ_{210}$ is obtained from the next proposition (c.f., Remark~\ref{Rem:CompToCFL}).

\begin{Proposition}[Equivalence of definitions of $\IBL$-product]\label{Prop:EqOfDefCool}
Let $V$ be a graded vector space of finite type, and let $\langle \cdot,\cdot\rangle: V\otimes V \rightarrow \R$ be a non-degenerate homogenous graded antisymmetric bilinear form of degree $-n$. Then the operation $\OPQ_{210}^*: (\DSpPlusCyc V)[-n]^{\otimes 2} \rightarrow (\DSpPlusCyc V)[-n]$ defined invariantly in Definition~\ref{Def:NewProduct} is a degree shift of the operation $\mu: (\DSpPlusCycRed V)^{\otimes 2} \rightarrow \DSpPlusCycRed V$ defined in coordinates above. More precisely, for all $\varphi_1$, $\varphi_2 \in \DSpPlusCycRed V$, it holds
\[ \OPQ_{210}^*(\NDeg^2 \varphi_1 \otimes \varphi_2) = \NDeg \mu(\varphi_1 \otimes \varphi_2)\qquad\text{in }(\DSpPlusCycRed V)[-n].\]
\end{Proposition}
\begin{proof}
Consider $\psi_1$, $\psi_2 \in \DSpCyc V$ with $k_1$, $k_2$ inputs, respectively. Writing the definition of $[\cdot,\cdot]$ in coordinates, we get 
\begin{align*}
&[\psi_1,\psi_2]^i_{i_1 \dotsb i_{k_1 + k_2-1}} \\
&\qquad=\begin{aligned}[t]
&\sum_{j=1}^{k_1} (-1)^{\psi_2(i_1 + \dotsb + i_{j-1})}(\psi_1)^i_{i_{1} \dotsb i_{j-1} k i_{j+k_2} \dotsb i_{k_1 + k_2 -1}} (\psi_2)^k_{i_{j} \dotsb i_{j+k_2 - 1}} \\ {}+&\sum_{j=1}^{k_2} (-1)^{\psi_1(i_1 + \dotsb + i_{j-1} + \psi_2) + 1}(\psi_2)^i_{i_{1} \dotsb i_{j-1} k i_{j+k_1} \dotsb i_{k_1 + k_2 -1}} (\psi_1)^k_{i_{j} \dotsb i_{j+k_1 - 1}}.
\end{aligned}
\end{align*}
Using \eqref{Eq:TransfForm} and the cyclic symmetry of $\psi_1^+$ and $\psi_2^+$, we compute
\begin{align*}
&[\psi_1,\psi_2]_{i_1 \dotsb i_{k_1 + k_2}} \\ 
&= [\psi_1,\psi_2]^i_{i_1 \dotsb i_{k_1 + k_2 -1}}g_{i i_{k_1 + k_2}}  \\
&=\begin{aligned}[t]
&\sum_{j=1}^{k_1} (-1)^{\psi_2(i_1 + \dotsb + i_{j-1})} (\psi_1)_{i_1 \dotsc i_{j-1} k i_{j+k_2} \dotsc i_{k_1 + k_2}} (\psi_2)^k_{i_j \dotsc i_{j+k_2-1}} \\
+ &\sum_{j=1}^{k_2} (-1)^{\psi_1(i_1 + \dotsb + i_{j-1}+\psi_2)+1} (\psi_2)_{i_1 \dotsc i_{j-1} k i_{j+k_1} \dotsc i_{k_1 + k_2}} (\psi_1)^k_{i_j \dotsc i_{j+k_1-1}}
\end{aligned}\\
& = \begin{aligned}[t] &\sum_{j=1}^{k_1} (-1)^{\psi_2(i_1 + \dotsb + i_{j-1})} g^{ab}(\psi_1)_{i_1 \dotsc i_{j-1} a i_{j+k_2} \dotsc i_{k_1 + k_2}} (\psi_2)_{i_j \dotsc i_{j+k_2-1} b} \\
+&\sum_{j=1}^{k_2} (-1)^{\psi_1(i_1 + \dotsb + i_{j-1} + \psi_2)+1} g^{ab} (\psi_2)_{i_1 \dotsc i_{j-1} a i_{j+k_1} \dotsc i_{k_1 + k_2}} (\psi_1)_{i_j \dotsc i_{j+k_1-1} b}
\end{aligned}\\
& = \begin{aligned}[t]
&\sum_{j=1}^{k_1} \underbrace{(-1)^{\psi_2(i_1+\dotsb+i_{j-1}) + (i_1 + \dotsb + i_{j-1})(\psi_1 - n -1) + b(\psi_2 - n - 1)}}_{\eqqcolon\varepsilon_1} \\
&\qquad g^{ab} (\psi_1)_{a i_{j+k_2}\dotsb i_{k_1+k_2} i_1 \dotsb i_{j-1}}(\psi_2)_{b i_j \dotsb i_{j+k_2-1}}  \\
{}+ & \sum_{j=1}^{k_2} \underbrace{(-1)^{\psi_1(i_1+\dotsb+i_{j-1} + \psi_2) + 1 + (i_1+\dotsb +i_{j-1})(\psi_2 - n - 1) + a(\psi_1-n - 1) + ab + 1}}_{\eqqcolon\varepsilon_2} \\ 
&\qquad g^{ab} (\psi_1)_{a i_j\dotsb i_{j+k_1-1}}
(\psi_2)_{b i_{j+k_1}\dotsb i_{k_1+k_2} i_1 \dotsb i_{j-1}}.
\end{aligned} 
\end{align*}
Notice that the lower indices $i_1$, $\dotsc$, $i_{k_1+k_2}$ appear in cyclic permutations. The first sum consists of cyclic permutation starting with $i_{k_2 + 1}$ and going up to $i_{k_1+k_2+1}$, and the second sum consists of cyclic permutations starting with $i_{1}$ and going up to $i_{k_2}$; hence, all cyclic permutations appear, and it just remains to check the signs. We see that
\begin{align*}
\varepsilon_1 &= (-1)^{\psi_2(i_1+\dotsb+i_{j-1}) + (i_1 + \dotsb + i_{j-1})(\psi_1-n-1) + b(\psi_2 -n- 1)}\\
&=\begin{multlined}[t]\underbrace{(-1)^{(i_1 + \dotsb + i_{j-1} + \psi_2 - n - b)(\psi_1 - n - a + \psi_2 - n - b - 1)}}_{=\varepsilon(i_1\dotsb i_{k_1+k_2}, i_{j+k_2} \dotsb i_{k_1+k_2} i_1 \dotsb i_{j+k_2-1})} (-1)^{b(\psi_1 -n- a)}(-1)^a \\ (-1)^{n(\psi_1 - n) + \psi_2 \psi_1}
\end{multlined}
\end{align*}
and
\begin{align*}
\varepsilon_2 &= (-1)^{\psi_1(i_1+\dotsb+i_{j-1} + \psi_2) + (i_1+\dotsb +i_{j-1})(\psi_2 - n - 1) + a(\psi_1-n-1) + a b} \\
& = \begin{multlined}[t]\underbrace{(-1)^{(i_1 + \dotsb + i_{j-1})(\psi_1 -n - a + \psi_2 -n- b - 1)}}_{=\varepsilon(i_1 \dotsb i_{k_1 + k_2}, i_j \dotsb i_{k_1+k_2} i_1 \dotsb i_{j-1})} (-1)^{b(\psi_1-n-a)}(-1)^a \\
(-1)^{n(\psi_1 - n) + \psi_1 \psi_2 }.
\end{multlined}
\end{align*}
This finishes the proof.
\end{proof}

\section{Cobracket and odd symplectic geometry}\label{Sec:CoprNic}

We start with a remark about the naive dualization of the bracket on the dual.

\begin{Remark}[Naive dualization]
If we ignore the fact that the dualization of spaces like~$\DSp V$ is problematic and we might not get $(\DSp V^{\GD})^{\GD}\simeq \DSp V$,\footnote{It holds $V\simeq (V^{\GD})^{\GD}$ provided that $V$ is of finite type. The tensor product of graded vector spaces of finite type is also of finite type provided that the grading is non-negative. Moreover, one has to require that $W^0 =W^1=0$ for $W$ for which $V=W[1]$ or take suitable completions with respect to weights in order to deal with arbitrary long tensor products in the same degree in the dual of $\DSp V$.} a naive way to obtain a cobracket $\bar{\delta}: \DSp V\rightarrow\DSp V \otimes\DSp V$ would be to dualize the  canonical bracket $[\cdot,\cdot]^*: \DSp V^{\GD} \otimes \DSp V^{\GD} \rightarrow \DSp V^{\GD}$ from the previous section (replacing $V$ by $V^{\GD}$). Because $w([\cdot,\cdot]^*) = -1$, the dual would have $w(\bar{\delta}) = 1$, where $w$ denotes the weights. However, the $\IBL$-cobracket has $w(\OPQ_{120})=-2$ (see Definition~\ref{Def:CanonicaldIBL}), and hence 
\[w\bigl(\bigl(({}^+)^{-1}\otimes({}^+\bigr)^{-1}\bigr)\circ\OPQ_{120}\circ{}^+\bigr) = w(\OPQ_{120}) - 1 = - 3 \neq 1 = w(\bar{\delta}), \]
where the map ${}^+: \DSp V\rightarrow\DSpPlus V$ of weight $+1$ was defined in \eqref{Eq:CycDiag}. It might be interesting to write down the coordinate expression for such $\bar{\delta}$ which comes from~$\OPQ_{120}$ and study its origin and properties on~$\DSp V$.
\end{Remark}

In the following definition, we will use the terms ``cobracket'' and ``$\BV$-operator'' to denote some operations without actually knowing that they satisfy the corresponding relations (see Questions~\ref{Q:OpenProbBrCo} at the end of this section).
 
\begin{Definition}[Hochschild cobracket and $\BV$-operator]\label{Def:DefOfHochOp}
Let $V$ be a graded vector space, let $(e_i)\in V$ be its homogenous basis, and let $(\eta^i)\subset V^{\GD}$ be the dual basis, i.e., it holds $\eta^i(e_j) = \delta^i_j$. We denote by~$\Ten(V^{\GD})$ the tensor algebra over~$V^{\GD}$ and consider its basis $\eta^{\otimes I} \coloneqq \eta^{i_1} \otimes \dotsb \otimes \eta^{i_k}$ for all multiindices $I = (i_1,\dotsc,i_k)$.

For each $i$, $j$, we define the \emph{cyclic double derivative} $\Der_{ij}:\Ten(V^{\GD}) \longrightarrow \Ten(V^{\GD}) \otimes \Ten(V^{\GD})$ on the basis by the following formula, which we explain below:
\begin{equation}\label{Eq:DoubleDeriv}
\Der_{ij}(\eta^{\otimes I}) \coloneqq \sum_{(i,j)\in I} \sum_{\substack{c_1 \in \CycPerm_{k_1}\\ c_2 \in \CycPerm_{k_2}}}\frac{1}{k_1}\frac{1}{k_2}\varepsilon(I,ijI_1^{c_1}I_2^{c_2}) \bigl(\eta^{\otimes I_1^{c_1}} \otimes \eta^{\otimes I_2^{c_2}}\bigr)
\end{equation}
The first sum is over all pairs of indices $(i,j)$ at two distinct positions in $I$. We consider the cyclic order on positions in $I$ (of length $k$) starting from the left and define~$I_1$ and~$I_2$ as the substrings between~$i$ and~$j$ (of length~$k_1$) and between~$j$ and~$i$ (of length~$k_2$), respectively. We include neither $i$ nor $j$ in $I_1$ or $I_2$, i.e., $k= k_1 + k_2 + 2$. The second sum is an average over all cyclic permutations $c_1$ and $c_2$ of $I_1$ and $I_2$, respectively. As usual, $\varepsilon$ stands for the Koszul sign and $i$ has the degree of $e_i$. If $k_1$ or $k_2$ vanishes, we replace the average over $c_1$ or $c_2$, respectively, by $1$ and set $\eta^{\emptyset} = 1$.

On $\Ten(V^{\GD})$, we consider the \emph{shuffle product} $\ProdSh: \Ten(V^{\GD})\otimes\Ten(V^{\GD}) \rightarrow \Ten(V^{\GD})$ which is defined by
\[ \ProdSh(\eta^{\otimes I_1},\eta^{\otimes I_2}) \coloneqq  \sum_{\mu\in\Perm_{k_1,k_2}} \frac{k_1! k_2!}{(k_1 + k_2)!} \varepsilon(I_1 I_2, \mu(I_1 I_2)) \eta^{\otimes\mu(I_1 I_2)}, \]
where $k_1$ and $k_2$ are the lengths of $I_1$ and $I_2$, respectively, and $\Perm_{k_1,k_2}$ denotes the shuffle permutations.

If $V$ is of finite type and $\langle\cdot,\cdot\rangle$ is a non-degenerate homogenous bilinear form on~$V$, we consider the coordinates $g^{ij} = \langle e^i,e^j\rangle$, where $(e^i)\subset V$ is the basis such that $\langle e^i,e_j\rangle = \delta^i_j$. We define the \emph{Hochschild cobracket} $\CoProdHoch: \Ten(V^{\GD}) \rightarrow \Ten(V^{\GD}) \otimes \Ten(V^{\GD})$ by
\begin{equation}
\CoProdHoch = \frac{1}{2}\sum_{i,j} (-1)^{e_i} g^{ij} \Der_{ij}
\end{equation}
and the \emph{Hochschild $\BV$-operator} $\BVOpHoch: \Ten(V^{\GD}) \rightarrow \Ten(V^{\GD})$ by
\begin{equation}
\BVOpHoch \coloneqq \ProdSh \circ \CoProdHoch.
\end{equation}
\end{Definition}

It is easy to check that the definitions of $\ProdSh$, $\CoProdHoch$ and $\BVOpHoch$ do not depend on the choice of the basis $(e_i)$ of $V$.

We denote by~$\Ten(V^{\GD})_{\CycMRM}$ the subspace of cyclic invariants of $\Ten(V^{\GD})$  and by $\Ten(V^{\GD})_{\SymMRM}$ the subspace of symmetric invariants. We consider the cyclization map $\pi_{\CycMRM}: \Ten(V^{\GD}) \rightarrow \Ten(V^{\GD})_{\CycMRM}$ and the symmetrization map $\pi_{\SymMRM}: \Ten(V^{\GD}) \rightarrow \Ten(V^{\GD})_{\SymMRM}$; they are defined by the averages
\begin{align*}
\pi_{\CycMRM}(\eta^{\otimes I})&= \sum_{c\in\CycPerm_k}\frac{1}{k} \varepsilon(I,I^c) \eta^{\otimes I^c}\qquad\text{and}\\
\pi_{\SymMRM}(\eta^{\otimes I}) &= \sum_{\sigma\in\Perm_k}\frac{1}{k!} \varepsilon(I,I^\sigma) \eta^{\otimes I^\sigma},
\end{align*}
respectively, where $k$ is the length of $I$ (if~$k=0$, we have~$\eta^{\emptyset}=1$ and~$\pi_{\CycMRM}(1) \coloneqq 1$). We denote by $\CycCoProj$ the projection to cyclic coinvariants and by $\SymCoProj$ the projection to symmetric coinvariants.

\begin{Definition}[Cyclic and symmetric versions]\label{Def:CycSymVersions}
In the situation of Definition~\ref{Def:DefOfHochOp}, we define the cyclic and symmetric versions of the operations $\Der_{ij}$, $\CoProdHoch$, $\ProdSh$ and $\BVOpHoch$ by precomposing with the inclusions $\CycInc: \Ten(V^{\GD})_{\CycMRM} \rightarrow \Ten(V^{\GD})$ and $\SymInc: \Ten(V^{\GD})_{\SymMRM} \rightarrow \Ten(V^{\GD})$ and postcomposing with the projections $\CycProj: \Ten(V^{\GD}) \rightarrow \Ten(V^{\GD})_{\CycMRM}$ and $\SymProj: \Ten(V^{\GD}) \rightarrow \Ten(V^{\GD})_{\SymMRM}$, respectively. In formulas, we have
\begin{equation}\label{Eq:CycSymVersions}\begin{aligned}
 \DerCyc_{ij} & \coloneqq (\CycProj\otimes\CycProj)\circ\Der_{ij}\circ\CycInc, & \DerSym_{ij} & \coloneqq (\SymProj\otimes\SymProj)\circ\Der_{ij}\circ\SymInc, \\
 \CoProdCyc & \coloneqq (\CycProj\otimes\CycProj)\circ\CoProdHoch\circ\CycInc, &  \CoProdSym & \coloneqq (\SymProj\otimes\SymProj)\circ\CoProdHoch\circ\SymInc, \\ 
 \ProdCyc & \coloneqq \CycProj\circ\ProdSh\circ(\CycInc\otimes\CycInc), & \ProdSym & \coloneqq \SymProj\circ\ProdSh\circ(\SymInc\otimes\SymInc), \\ 
 \BVOpCyc & \coloneqq \CycProj\circ\BVOpHoch\circ\CycInc, & \BVOpSym & \coloneqq \SymProj\circ\BVOpHoch\circ\SymInc.
\end{aligned}\end{equation}
\end{Definition}

\begin{Proposition}[Hochschild, cyclic and symmetric operations]\label{Prop:HochCycSym}
In the situation of Definition~\ref{Def:CycSymVersions}, we consider for every multiindex $I$ the linear functional $\alpha^I: \Ten V \rightarrow\R$ given with respect to the basis $e_{\otimes J}$ by 
\[\alpha^I(e_{\otimes J}) \coloneqq \varepsilon(I,J). \]
We set
\begin{equation*}
\alpha^I_{\mathrm{sym}} = \sum_{\sigma\in \Perm_{k}} \frac{1}{k!} \varepsilon(I,I^\sigma) \alpha^{I^\sigma}\quad\text{and}\quad
\alpha^I_{\mathrm{cyc}} = \sum_{c\in\CycPerm_{k}}\frac{1}{k} \varepsilon(I,I^c)\alpha^{I^c}.
\end{equation*}
We define the monomorphisms $I$, $I_{\CycMRM}$ and $I_{\SymMRM}$ by
\begin{equation}\label{Eq:IdentDiaCycSym}
\begin{tikzcd}[ampersand replacement=\&]
 TV^{\GD} \arrow[hook]{r}{I} \& \DSpPlus V: \& \eta^{\otimes I} \arrow[mapsto]{r} \&  \alpha^I, \\
 (TV^{\GD})_{\CycMRM} \arrow[hook]{r}{I_{\CycMRM}}\arrow[hook]{u} \& \DSpPlusCyc V: \arrow[hook]{u} \& \CycProj(\eta^{\otimes I}) \arrow[mapsto]{r} \& \alpha_{\CycMRM}^I, \\
(TV^{\GD})_{\SymMRM} \arrow[hook]{r}{I_{\SymMRM}}\arrow[hook]{u} \& \DSpPlusSym V: \arrow[hook]{u} \& \SymProj(\eta^{\otimes I}) \arrow[mapsto]{r} \& \alpha_{\SymMRM}^I.
\end{tikzcd}
\end{equation}
The vertical arrows are inclusions and the diagram commutes. Moreover, $I$, $I_{\CycMRM}$ and $I_{\SymMRM}$ become isomorphisms provided that $V$ is of finite type with non-negative degrees. 
In this case, we transfer the operations and obtain the following commutative diagram: 
\begin{equation}\label{Eq:ComDiagram}
\begin{tikzcd}
\BVOpHoch: &\DSpPlus V \arrow{r}{\CoProdHoch} & \DSpPlus V \otimes \DSpPlus V \arrow{r}{\ProdSh} & \DSpPlus V \\
\BVOpCyc: &\DSpPlusCyc V \arrow{r}{\CoProdCyc} \arrow[hook]{u} & \DSpPlusCyc V \otimes \DSpPlusCyc V \arrow{r}{\ProdCyc} \arrow[hook]{u} & \DSpPlusCyc V \arrow[hook]{u} \\
\BVOpSym: & \DSpPlusSym V \arrow{r}{\CoProdSym}\arrow[hook]{u} & \DSpPlusSym V \otimes \DSpPlusSym V \arrow{r}{\ProdSym} \arrow[hook]{u} & \DSpPlusSym V. \arrow[hook]{u}
\end{tikzcd}
\end{equation}
The following formulas hold:
\begin{equation}\label{Eq:CycSymExplicit}
\begin{aligned}
 \CoProdCyc(\alpha^I_{\CycMRM}) &= \frac{1}{2}\sum_{i,j} (-1)^{e_i} g^{ij} \sum_{(i,j)\in I} \varepsilon(I,i j I_1 I_2)(\alpha^{I_1}_{\CycMRM}\otimes \alpha^{I_2}_{\CycMRM}),\\
 \ProdSym(\alpha^{I_1}_{\SymMRM},\alpha^{I_2}_{\SymMRM}) &= \alpha^{I_1 I_2}_{\SymMRM},\\
 \BVOpSym(\alpha^I_{\SymMRM}) &= \frac{1}{2}\sum_{i,j} (-1)^{e_i} g^{ij} \sum_{(i,j)\in I} \varepsilon(I,i j I_1 I_2)\alpha^{I_1 I_2}_{\SymMRM}.
\end{aligned} 
\end{equation}
\end{Proposition}

\begin{proof}
The commutativity of \eqref{Eq:IdentDiaCycSym} is clear from the definitions. The commutativity of~\eqref{Eq:ComDiagram} is equivalent to the commutativity of the same diagram with $\DSpPlus V$, $\DSpPlusCyc V$, $\DSpPlusSym V$ replaced by $\Ten(V^{\GD})$, $\Ten(V^{\GD})_{\CycMRM}$, $\Ten(V^{\GD})_{\SymMRM}$, respectively. This commutativity then follows from the defining relations \eqref{Eq:CycSymVersions} if we show the following inclusions:
\begin{enumerate}[label=(\arabic*)]
\item $\Der_{ij}(\Ten(V^{\GD})_{\CycMRM})\subset\Ten(V^{\GD})_{\CycMRM}\otimes\Ten(V^{\GD})_{\CycMRM}$, 
\item $\Der_{ij}(\Ten(V^{\GD})_{\SymMRM})\subset\Ten(V^{\GD})_{\SymMRM}\otimes\Ten(V^{\GD})_{\SymMRM}$,
\item $\ProdSh(\Ten(V^{\GD})_{\CycMRM}\otimes\Ten(V^{\GD})_{\CycMRM})\subset \Ten(V^{\GD})_{\CycMRM}$,
\item $\ProdSh(\Ten(V^{\GD})_{\SymMRM}\otimes\Ten(V^{\GD})_{\SymMRM})\subset \Ten(V^{\GD})_{\SymMRM}$. 
\end{enumerate}
Inclusion (1) is clear from \eqref{Eq:DoubleDeriv}. In fact, it holds even $\Der_{ij}(\Ten(V^{\GD})) \subset  \Ten(V^{\GD})_{\CycMRM} \otimes \Ten(V^{\GD})_{\CycMRM}$. As for inclusion (2), we have
\[ \Der_{ij}\bigl(\pi_{\SymMRM}(\eta^{\otimes I})\bigr) = \sum_{\sigma\in \Perm_k}\frac{1}{k!}\sum_{(i,j)\in I^\sigma}\sum_{\substack{c_1\in\CycPerm_{k_1}\\c_2\in\CycPerm_{k_2}}}\frac{1}{k_1}\frac{1}{k_2}\varepsilon(I,i j I_1^{\sigma c_1}I_2^{\sigma c_2})\eta^{\otimes I_1^{\sigma c_1}}\otimes\eta^{\otimes I_2^{\sigma c_2}}. \]
A pair $(\sigma_1,\sigma_2)\in\Perm_{k_1}\times\Perm_{k_2}$ induces a bijection of the domain of summation, which consists of the elements $\sigma$, $(i,j)$, $c_1$, $c_2$, by keeping $(i,j)$, $c_1$, $c_2$ fixed and defining a new $\tilde{\sigma}$ by altering $\sigma$ to compensate for the effect of $(\sigma_1, \sigma_2)$. The signs fit since we only consider the Koszul sign. Therefore, we have
\[ (\sigma_1\otimes \sigma_2)\Der_{ij}\bigl(\pi_{\SymMRM}(\eta^{\otimes I})\bigr) = \Der_{ij}\bigl(\pi_{\SymMRM}(\eta^{\otimes I})\bigr), \]
and (2) follows. As for (3), we have
\begin{align*}
& \ProdSh\bigl(\CycProj(\eta^{\otimes I_1}),\CycProj(\eta^{\otimes I_2})\bigr)\\
&\qquad= \sum_{\mu\in\Perm_{k_1,k_2}}\frac{k_1!k_2!}{(k_1+k_2)!} \sum_{\substack{c_1\in \CycPerm_{k_1}\\ c_2\in\CycPerm_{k_2}}}\frac{1}{k_1}\frac{1}{k_2}\varepsilon(I_1 I_2,\mu(I_1^{c_1},I_2^{c_2})) \eta^{\otimes \mu(I_1^{c_1},I_2^{c_2})}.
\end{align*}
A permutation $c\in \Perm_k$ again induces a bijection of the domain of summation, which consists of the elements $\mu$, $c_1$, $c_2$. A similar argument holds for (4).

We now derive \eqref{Eq:CycSymExplicit}. Because
\[ \Der_{ij}(\alpha^I_{\CycMRM}) = \sum_{(i,j)\in I} \varepsilon(I,ijI_1I_2) \alpha^{I_1}_{\CycMRM}\otimes\alpha^{I_2}_{\CycMRM}, \]
the formula for $\CoProdCyc$ is clear. As for $\BVOpSym$, because $\ProdSym = \pi_{\SymMRM} \circ \ProdConcat$ for the concatenation product $\ProdConcat(\eta^{\otimes I_1},\eta^{\otimes I_2}) = \eta^{\otimes I_1 I_2}$, we have
\begin{align*}
(\ProdSym\circ \Der_{ij})(\pi_{\SymMRM}\bigl(\eta^{\otimes I})\bigr) & = \sum_{\sigma\in \Perm_k}\frac{1}{k!}\sum_{(i,j)\in I^\sigma}\sum_{\substack{c_1\in\CycPerm_{k_1}\\c_2\in\CycPerm_{k_2}}}\frac{1}{k_1}\frac{1}{k_2}\pi_{\SymMRM}\bigl(\varepsilon(I,i j I_1^{\sigma c_1}I_2^{\sigma c_2})\eta^{\otimes I_1^{\sigma c_1}I_2^{\sigma c_2}}\bigr) \\
& = \sum_{(i,j)\in I} \varepsilon(I,ijI_1 I_2) \pi_{\SymMRM}(\eta^{\otimes I_1 I_2}),
\end{align*}
where we used that $\pi_{\SymMRM} \circ \sigma = \pi_{\SymMRM}$ for every permutation $\sigma$. The formula for $\BVOpSym$ follows. The formula for $\ProdSym$ is also clear.
\end{proof}

In the formula \eqref{Eq:CycSymExplicit} for $\BVOpSym$, we recognize the canonical $\BV$-operator on $\Sym V^{\GD}$ from~\cite[Definition~4]{Doubek2018}.\footnote{In fact, they work on a bigger ``algebra of functions'' on $V$, which is a completion of $\Sym V^{\GD}$ with respect to a suitable filtration. Their statement is that the algebra of functions on an odd symplectic vector space carries a canonical $\BV$-operator. It is given by the divergence of the Hamiltonian vector field.}
We will now relate $\zeta_{\text{cyc}}$ to $\OPQ_{120}$. The steps are similar to Section~\ref{Sec:LieBrHoch}. We define the coproduct $\OPQ_{120}^*: (\DSpPlusCyc V)[-n] \rightarrow (\DSpPlusCyc V)[-n]^{\otimes 2}$, where $-n$ is the degree of $\langle\cdot,\cdot\rangle$, by 
\begin{equation}\label{Eq:CoprCandid}
\OPQ_{120}^*(\nu\varphi) = \nu^2\zeta_{\text{cyc}}(\varphi)\qquad\text{for all }\varphi\in \DSpPlusCyc V,
\end{equation}
where $\nu$ is a formal symbol of degree $n$.

For $\varphi\in \CycHom(V^{\otimes k},\R)$ with $k\ge 4$, we define $(\delta\varphi)\in(\DSpPlusCycRed V)^{\otimes 2}$ using the coordinates
\begin{equation}\label{Eq:CoefCoord}
\varphi_{I} \coloneqq \varphi(e_{\otimes I})\quad\text{and}\quad (\delta\varphi)_{I_1;I_2} = (\delta\varphi)(e_{\otimes I_1} \otimes e_{\otimes I_2})
\end{equation}
by
\begin{equation}\label{Eq:StdCoprCoord}
(\delta\varphi)_{I_1;I_2} \coloneqq \sum_{\substack{c_1 \in\CycPerm_{k_1} \\ c_2\in\CycPerm_{k_2}}} \sum_{i,j} (-1)^{e_i}g^{ij} \varepsilon(I_1,I_1^c)\varepsilon(I_2,I_2^c)(-1)^{e_j I_1} \varphi_{i I_1^{c_1} j I_2^{c_2}}.
\end{equation}
We set $\delta \varphi = 0$ for $k\le 3$. In order to obtain $(\delta\varphi)\in(\DSpPlusCycRed V)^{\otimes 2}$, we use the embedding $(\DSpPlusCyc V)^{\otimes k} \subset (\Ten V)^{\otimes k*}$ from Remark~\ref{Rem:Identifications}. Note that we can also use the canonical identification $(\Ten V/\CycMRM)^{\WGD}\simeq \DSpPlusCyc V$ and work with the cyclic words $e_I \coloneqq \CycCoProj(e_{\otimes I})$ instead. We see that~\eqref{Eq:StdCoprCoord} is precisely the defining equation \cite[Equation~(10.6)]{Cieliebak2015} for the $\IBL$-cobracket $\delta: \DSpPlusCycRed V \rightarrow (\DSpPlusCycRed V)^{\otimes 2}$, and $\OPQ_{120}:(\DSpPlusCycRed V)[-n]\rightarrow(\DSpPlusCycRed V)[-n]^{\otimes 2}$ is the degree shift of $\delta$.

The equality $\OPQ_{120}^*= \OPQ_{120}$ on the reduced part
is obtained from the next proposition (c.f., Remark~\ref{Rem:CompToCFL} for the translation from the $\DSpPlusCycRed$- to the  $\DBCyc$-notation).\Modify[caption={DONE finite type},noline]{One can replace finite dimensional by finite type.}
\begin{Proposition}[Equivalence of definitions of $\IBL$-cobracket]\label{Prop:EqCoprod}
Let $V$ be a graded vector space of finite type, and let $\langle\cdot,\cdot\rangle$ be a non-degenerate homogenous bilinear form of degree~$-n$. For every $\varphi\in \DSpPlusCycRed V$, it holds
\[ \OPQ_{120}^*(\nu \varphi) = \nu^2 \delta \varphi\qquad\text{in}\quad(\DSpPlusCycRed V)[-n]^{\otimes 2}. \]
\end{Proposition}
\begin{proof}
We consider the basis $(\alpha^I_{\CycMRM})$ of $\DSpPlusCyc V$ from Proposition~\ref{Prop:HochCycSym}. Given $\varphi\in\DSpPlusCyc V$, we can write $\varphi=\sum_{I}\frac{1}{\Abs{I}}\varphi_I\alpha^I_{\CycMRM}$, where we sum over all multiindices $I$ and $\varphi_I\in \R$ are coefficients cyclically symmetric in $I$. For every $I$ of length $k$, we have
\[ \varphi(e_{\otimes I}) = \sum_{c\in \CycPerm_k}\frac{1}{k}\varphi_{I^c} \alpha^{I^c}_{\CycMRM}(e_{\otimes I}) = \sum_{c\in\CycPerm_k}\frac{1}{k}\varphi_{I^c} \varepsilon(I^c,I) = \varphi_I, \]
and hence $\varphi_I$ correspond to the coefficients \eqref{Eq:CoefCoord}. Using \eqref{Eq:CycSymExplicit}, we compute
\begin{align*}
 \CoProdCyc(\varphi)&= \frac{1}{2}\sum_I \sum_{(i,j)\in I} \frac{1}{\Abs{I}}\varphi_I  (-1)^{e_i} g^{ij}\varepsilon(I,i j I_1 I_2) \alpha^{I_1}_{\CycMRM} \otimes \alpha^{I_2}_{\CycMRM} \\
 &=\frac{1}{2}\sum_J \sum_{c\in \CycPerm_{\Abs{J}}} \sum_{j \in J\backslash\{j_1\}} \frac{1}{\Abs{J}}\varphi_{J^c} (-1)^{e_{j_1}} g^{{j_1} j}(-1)^{j J_1}\varepsilon(J,J^c) \alpha^{J_1}_{\CycMRM} \otimes \alpha^{J_2}_{\CycMRM} \\
 &=\frac{1}{2}\sum_J\sum_{j\in J\backslash\{j_1\}} \varphi_J(-1)^{e_{j_1}}g^{j_1 j}(-1)^{j J_1} \alpha_{\CycMRM}^{J_1}\otimes \alpha_{\CycMRM}^{J_2}  \\
 & = \sum_{J_1, J_2}\underbrace{\Bigl(\frac{1}{2}\sum_{i,j} (-1)^{e_i} g^{ij}(-1)^{j J_1} \varphi_{i J_1 j J_2} \Bigr)}_{\eqqcolon(*)_{J_1;J_2}}\alpha^{J_1}_{\CycMRM}\otimes\alpha^{J_2}_{\CycMRM}.
\end{align*}
On the second line, we used the bijection of the summation domains consisting of $I$, $(i,j)$, resp.~$J$, $c$, $b$, which assigns to~$I$ its cyclic permutation~$J$ such that the position of~$i$ corresponds to the first position of~$J$, i.e., it holds~$j_1 = i$. On the third line, we used that~$\varphi_I$ are cyclic symmetric, and on the fourth line, we just rewrote the summation using $J = i J_1 j J_2$. We now relate $(*)_{J_1;J_2}$ to $(\delta\varphi)_{J_1;J_2}$. We compute (notice that there is no Koszul sign from the application of the tensor product, c.f., Remark~\ref{Rem:Identifications})
\begin{align*}
\zeta_{\text{cyc}}(\varphi)(e_{I_1}\otimes e_{I_2}) & = \Bigl(\sum_{\substack{c_1\in\CycPerm_{k_1}\\c_2\in\CycPerm_{k_2}}}(*)_{I_1^{c_1};I_2^{c_2}}\alpha_{\CycMRM}^{I_1^{c_1}}\otimes\alpha_{\CycMRM}^{I_2^{c_2}}\Bigr)(e_{I_1}\otimes e_{I_2}) \\
& = \sum_{\substack{c_1\in\CycPerm_{k_1}\\c_2\in\CycPerm_{k_2}}}\varepsilon(I_1^{c_1},I_1)\varepsilon(I_2^{c_2},I_2)(*)_{I_1^{c_1};I_2^{c_2}} \\
& = (\delta\varphi)_{J_1;J_2}.
\end{align*}
This shows the proposition.
\end{proof}

\begin{Questions}\phantomsection\label{Q:OpenProbBrCo}
\begin{RemarkList}
\item Are $(\DSpPlusCyc V, \BVOpCyc, \ProdCyc)$ and $(\DSpPlus V, \BVOpHoch, \ProdSh)$ also $\BV$-algebras like $(\DSpPlusSym V, \BVOpSym, \ProdSym)$? Recall the 2nd-derivative identity (or the ''seven term identity'') which is used to define a $\BV$-algebra:
\begin{equation}\label{Eq:secondderivative}\begin{aligned}
\BVOp(abc) & = \BVOp(ab)c + (-1)^{a(b+c)}\BVOp(bc)a + (-1)^{bc}\BVOp(ac)b \\
&\hphantom{=}{}- \BVOp(a)bc - (-1)^{ab}\BVOp(b)ac - (-1)^{c(a+b)}\BVOp(c)ab.
\end{aligned}\end{equation}
It clearly uses associativity (and in this form also commutativity). It is a known fact that~$\ProdSh$ is both associative and commutative (see~\cite{LodayCyclic}). Therefore, the question for~$\BVOpHoch$ makes sense. However,~$\ProdCyc$ might not be associative (is it?), and hence~$\BVOp_{\text{cyc}}$ might not be a $\BV$-operator in the classical sense. Is it possible to define a second order derivative, and hence the notion of a $\BV$-operator, on a non-associative algebra?
\item Do the factorizations $\BVOpCyc = \ProdCyc \circ \CoProdCyc$, resp.~$\BVOpSym= \ProdSym \circ \CoProdSym$ define $\CoProdCyc$, resp.~$\CoProdSym$ uniquely? Do other $\BV$-operators factorize like this? Is $\CoProdSym$ a cobracket and is there a bracket on ``symmetric Hochschild cochains'' inducing a bialgebra structure on~$\DSpPlusSym V$? 
\item Are the new operations chain maps with respect to $\Hd$ and~$\Hd + \Dd$? 
\qedhere
\end{RemarkList}
\end{Questions}

\allowdisplaybreaks

\chapter{Filtered IBL-infinity-algebras as filtered MV-algebras}\label{App:IBLMV}
\Correct[caption={$BV$ and $BVinfty$}]{Change $\BV$ to $\BVInfty$.}
\Correct[caption={DONE Degree of BV},noline]{Probably $|\BVOp| = -1$ in our formalism. Indeed, it is $-1$. Correct!}
In this appendix, we want to understand the $\BV$-formalism for $\IBLInfty$-algebras and the equivalence
\[\IBLInfty:\quad \begin{gathered}
\underline{\text{Surface calculus}} \\
(\OPQ_{klg}), (f_{klg}), \text{gluing relations}
\end{gathered}
\;\Longleftrightarrow\;
\begin{gathered}
\underline{\BV\text{-formalismus}} \\
 \BVOp, \MVMorF,\text{\,relations }\BVOp^2 = 0, e^\MVMorF \BVOp^+ = \BVOp^- e^\MVMorF.
\end{gathered}\]
This was originally done in \cite{Cieliebak2015} using filtrations and formal power series in $\hbar$ as a ``bookkeeping''. Nevertheless, the author thinks that some details were not fully addressed and was also curious about extending the $\MV$-formalism from \cite{Markl2015} to the filtered setting.

As a matter of fact, the main motivation for better understanding of these details was a brief discussion with Kyler Siegel at a workshop about $\BV$-quantization in Stony Brook, 2019. \Add[noline,caption={DONE Kyler's surname}]{Look up Kyler's surname.}It was pointed out that a twisted $\infty$-algebra is, in some sense, ``isomorphic'' to the untwisted one because the twisting is just conjugation with the exponential of the Maurer-Cartan element. The case of $\AInfty$-algebras from~\cite{FOOOI} was mentioned. The author of this text was confused because, by Part~I, the twisting with the Chern-Simons Maurer-Cartan element ``adds'' the information about string topology to the canonical structure and one expects to obtain a non-isomorphic structure. The precise notions of isomorphisms needed to be clarified.

In Section~\ref{Sec:DetailsOnFiltr}, we start by dealing with filtrations and completions in terms of series in more details. We formulate the Resummation Lemma (Lemma~\ref{Lem:TechLem}), define (complete) filtered algebras (Definition~\ref{Def:FiltAlg}) and comment on units and augmentations (Remark~\ref{Rem:FiltrUnitAug}). We then consider combinations of two filtrations (Equation~\eqref{Eq:CombinedFiltr}) and show that the symmetric bialgebra with the filtration which is the union of the induced filtration and the filtration by weights might not be a filtered bialgebra (Example~\ref{Ex:CombinedOnSymetric}); it is, however, under a boundedness assumption on the filtration (Lemma~\ref{Lem:BoundCondOnFiltr}). We then consider the exponential and the logarithm on a complete filtered algebra (Lemma~\ref{Lem:Exponential}) and generalize it for the convolution product of morphisms with codomain a complete filtered algebra and domain a complete filtered coalgebra satisfying the limit conilpotency property (Proposition~\ref{Prop:ConvPwrSer}). This is the case of completions of conilpotent coalgebras, e.g., the symmetric bialgebra (Lemma~\ref{Lem:SymAlgLimConilp}). We remark that the exponential of a morphism might not be invertible as a map (Remark~\ref{Rem:ExpLogStar}).

In Section~\ref{Sec:FilteredMV}, we recall $\MV$-algebras from \cite{Markl2015} and introduce their complete filtered version (Definition~\ref{Def:FilteredMV}). Based on this, in analogy to \cite{Markl2015}, we define the notion of a complete filtered $\IBLInfty$-algebra in $\MV$-formalism and its morphisms (Definition~\ref{Def:ComplFiltrIBL}). We study the equivalent formulation in components via the surface calculus (Proposition~\ref{Prop:EqCharOfMVIBL}). We show that the notion of a filtered $\IBLInfty$-algebra in $\MV$-formalism is equivalent to the notion of a filtered $\IBLInfty$-algebra from \cite{Cieliebak2015} in some cases, e.g., for the dual cyclic bar complex from Part~I (Proposition~\ref{Prop:BVforIBL}). The formalism of \cite{Cieliebak2015} is symmetric on exchanging inputs and outputs whereas the $\MV$-formalism is not; we illustrate how bubblings at inputs and outputs are handled differently (Example~\ref{Ex:AsymOfMV}). Finally, we consider twisting with a Maurer-Cartan element (Proposition~\ref{Eq:TwistingProp}).

In Section~\ref{Sec:BVCompl}, we define the $\BV$-chain complexes associated to a complete filtered $\IBLInfty$-algebra in $\MV$-formalism (Definition~\ref{Def:BVCompl}) and observe that morphisms and the twisting with the Maurer-Cartan element induce chain maps of these chain complexes (Proposition~\ref{Prop:ObservationsMor}). We formulate some open questions (Questions~\ref{Q:SomeQuestionsFilter}).

In Section~\ref{Sec:CompConvA}, we study the composition of maps which are convolutions of other maps (Lemma~\ref{Lem:ItCompCond}). This leads to the definition of compositions controlling the number of ``veins'' between the individual maps (Definition~\ref{Def:ConComp}). This can be used to formulate the surface calculus algebraically (Proposition~\ref{Prop:PartCompAComp}). We give proofs of some formulas for partial compositions from Section~\ref{Sec:Alg1} (Proposition~\ref{Prop:PartCompositions}).

\section{Filtered bialgebras and exponential in convolution product}\label{Sec:DetailsOnFiltr}

We use the definitions of a filtration, completion and the induced filtration on the completion from Section~\ref{Sec:Alg1a}. We formulate the statements in the category of vector spaces, but similar statements hold in the category of graded vector spaces (writing ``homogenous'' and ``graded'' everywhere). All filtrations are decreasing, i.e., $\Filtr^{\lambda_1} \supseteq \Filtr^{\lambda_2}$ whenever $\lambda_1 \le \lambda_2$.
We work over a field $\K$ of characteristic zero.

We start with the following technical lemma, which shows that convergence in completion is very much like absolute convergence in analysis.\footnote{In fact, for complex numbers $v_k\in \C$ for $k\in\N_0$, property (b) of Lemma~\ref{Lem:TechLem} is equivalent to the absolute convergence of $\sum_{k=0}^\infty v_k$ in $\C$.
Thanks to Ji\v r\'i Zeman for noting that.}

\begin{Lemma}[Resummation Lemma]\label{Lem:TechLem}
Let $V$ be a vector space filtered by a decreasing filtration~$\Filtr$, and let~$\hat{V}$ be its completion. 
\begin{ClaimList}
\item Let $v_{ij}\in V$ for all $i$, $j\in\N_0$ be vectors such that $\sum_{j=0}^\infty v_{ij}$ converges for every $i\ge 0$ (i.e., $\Norm{v_{ij}}\to\infty$ as $j\to\infty$, so that $\sum_{j=0}^\infty v_{ij}\in\hat{V}$) and $\Norm{\sum_{j=0}^\infty v_{ij}} \to \infty$ as $i\to \infty$. Then for any bijection $r: \N_0\rightarrow \N_0 \times \N_0$, the sum $\sum_{i=0}^\infty v_{r(i)}$ converges and the limit does not depend on $r$; we denote it by $\sum_{i=0}^\infty\sum_{j=0}^\infty v_{ij} \in \hat{V}$.\footnote{It is an exercise to prove a similar statement for nested infinite sums $\sum_{i_1=0}^\infty \dotsb \sum_{i_n=0}^\infty v_{i_1,\dotsc,i_n}$.}
\item Suppose that $\sum_{k=0}^\infty v_k$ for $v_k\in V$ converges, and let $s: Z\subset \N_0\times \N_0 \rightarrow \N_0$ be a bijection. If we define $w_{ij} \coloneqq v_{s(i,j)}$ for $(i,j)\in Z$ and $w_{ij} \coloneqq 0$ otherwise, then $\sum_{i=0}^\infty \sum_{j=0}^\infty w_{ij}$ converges (in the sense of (a)) and equals $\sum_{k=0}^\infty v_k$.
\item Let $V$ and $V'$ be vector spaces filtered by exhaustive filtrations, and let $\sum_{i=0}^\infty v_{i}$ for $v_i\in V$ and $\sum_{i=0}^\infty v_{i}'$ for $v_i'\in V'$ converge. Then for any bijection $r: \N_0 \rightarrow \N_0 \times \N_0$, $i\to (r_1(i),r_2(i))$, the sum $\sum_{i=0}^\infty v_{r_1(i)}\otimes v_{r_2(i)}'$ converges and the limit does not depend on $r$; we denote it by $\sum_{i=0}^\infty v_i \otimes \sum_{j=0}^\infty v_j' \in V\COtimes V'$.
\end{ClaimList}
Moreover, we have the following statements about maps of filtered vector spaces:
\begin{ClaimList}[resume]
\item If $f: \hat{V}\rightarrow\hat{V}$ is a linear map of finite filtration degree, i.e., $\Norm{f}>-\infty$, and $\sum_{i=0}^\infty v_i$ with $v_i\in V$ a convergent series, then $\sum_{i=0}^\infty f(v_i)$ converges to $f(\sum_{i=0}^\infty v_i)$.
\item Given linear maps $f_i: \hat{V}\to \hat{V}$ for $i\in \N_0$, we say that the sum $\sum_{i=0}^\infty f_i$ converges if $\sum_{i=0}^\infty f_i(v)$ converges for all $v\in \hat{V}$. In this case, $f(v) \coloneqq \sum_{i=0}^\infty f_i(v)$ defines a linear map $\hat{V}\rightarrow\hat{V}$ which satisfies $\Norm{f}\ge \inf_{i\in\N_0} \Norm{f_i}$.
\end{ClaimList}
\end{Lemma}
\begin{proof}
\begin{ProofList}
\item Let $r:\N_0\rightarrow\N_0\times\N_0$ be a bijection. We first prove the convergence of $\sum_{i=0}^\infty v_{r(i)}$. Given $K>0$, let $i_0\in\N_0$ be such that $\Norm{v_{ij}}\ge K$ for $(i,j)\in \{i\ge i_0, j\ge 0\}$. Such $i_0$ exists by the assumption that $\Norm{\sum_{j=0}^\infty v_{ij}}\to \infty$ as $i\to \infty$. Let $j_0\in \N_0$ be such that $\Norm{v_{ij}}\ge K$ for $(i,j)\in \{i \le i_0, j\ge j_0\}$. Such $j_0$ exists from the convergence of $\sum_{j=0}^\infty v_{0j}$, $\dotsc$, $\sum_{j=0}^\infty v_{i_0 j}$. We can now pick $k_0\in \N$ such that $r(\{k< k_0\}) \supset \{i\le i_0, j\le j_0\}$, and hence $\Norm{v_{r(k)}}\ge K$ for all $k\ge k_0$. This implies the convergence of $\sum_{i=0}^\infty v_{r(i)}$.

Let now $r': \N_0 \rightarrow \N_0 \times \N_0$ be another bijection. Given $K>0$, we find $i_0$, $j_0\in \N_0$ so that $\Norm{v_{ij}}\ge K$ for all $(i,j)\not\in\{i\le i_0, j\le j_0\}$ (like in the previous paragraph). There exits $k_0\in \N$ such that $r(\{k < k_0\}), r'(\{k \le k_0\}) \supset \{i\le i_0, j\le j_0\}$. Hence, for any $k\ge k_0$, the contribution of $v_{ij}$ with $(i,j)\in\{i\le i_0, j\le j_0\}$ cancels in $\Delta_k \coloneqq \sum_{i=0}^k v_{r(i)} - \sum_{i=0}^k v_{r'(i)}$, and we get $\Norm{\Delta_k} \ge K$. It follows that the sums are equal in the limit.
\item It is easy to see that $w_{ij}$ satisfy the assumptions of (a). The claim follows from~(a) by defining $r\coloneqq s^{-1}\sqcup \Id: \N_0 = \N_0 \sqcup (\N_0\times \N_0\backslash Z) \rightarrow \N_0\times \N_0$, where the notation $\N_0 = \N_0 \sqcup (\N_0\times \N_0\backslash Z)$ means an infinite shuffle permutation. Clearly, $\sum_{i=0}^\infty w_{r(i)}$ is precisely $\sum_{i=0}^\infty v_i$ after crossing out possibly infinitely many zeros.

\item We start with convergence. Let $K>0$. Pick indices $i_0\in \N_0$ and $i_0'\in\N_0$ such that $\Norm{v_i} \ge K - \inf_j \Norm{v_j'}$ and $\Norm{v_i'}\ge K - \inf_j \Norm{v_j}$ for all $i\ge i_0$ and $i\ge i_0'$, respectively. Note that the right-hand sides of the inequalities are finite because the filtrations are exhaustive. Pick $k_0\in\N$ such that $r(\{k<k_0\})\supset\{i\le i_0, j\le j_0\}$. Then, for $k\ge k_0$, we have 
\begin{align*}
\Norm{v_{r_1(k)} \otimes v_{r_2(k)}'} &\ge \Norm{v_{r_1(k)}} + \Norm{v_{r_1(k)}'}\\
&\ge \begin{cases}
(K - \inf_j \Norm{v_j'}) + \inf_j \Norm{v_j'} = K & \text{if }r_1(k)\ge i_0, \\
 \inf_j \Norm{v_j} + (K - \inf_j \Norm{v_j}) = K & \text{if }r_2(k)\ge i_0'.
\end{cases}
\end{align*}
It follows that $\sum_{i=0}^\infty v_{r_1(i)}\otimes v_{r_2(i)}'$ converges.

If $r': \N_0 \rightarrow \N_0\times\N_0$ is another bijection, we write $r' = r \circ s$ for a bijection $s: \N_0 \rightarrow \N_0$ and apply (b) to $V\otimes V'$ with the bijection $s: Z\coloneqq \N_0 \times\{0\}\simeq \N_0\rightarrow \N_0$. It follows that $\sum_{i=0}^\infty v_{r(i)} = \sum_{i=0}^\infty v_{r'(i)}$.
\item This is clear because $f(\sum_{i=0}^\infty v_i) = \sum_{i=0}^n f(v_i) + f(\sum_{i=n+1}^\infty v_i)$ for any $n\in \N_0$ and $\Norm{f(\sum_{i=n+1}^\infty v_i)} \ge \Norm{f} + \inf_{i>n}\Norm{v_i} \to \infty$ as $n\to\infty$.
\item Linearity is clear. The inequality $\Norm{f}\ge \inf_i\Norm{f_i}$ follows immediately from $\Norm{f(v)} \ge \inf_i \Norm{f_i(v)} \ge \inf_i \Norm{f_i} + \Norm{v}$ for all $v\in \hat{V}$.\qedhere
\end{ProofList}
\end{proof}

Note that (a) and (c) of the Resummation Lemma are, in fact, reformulations of the facts that the canonical inclusions induce the isomorphisms $\hat{V}\simeq\hat{\hat{V}}$ and $V\COtimes V\simeq\hat{V}\COtimes \hat{V}$, respectively.

We will work with (complete) filtered algebras, coalgebras and bialgebras, which we now define schematically. They are basically algebras over the corresponding (pr)operad in the category of (complete) filtered vector spaces. 

\begin{Definition}[(Complete) filtered algebras, coalgebras, bialgebras]\label{Def:FiltAlg}
A \emph{filtered algebra, coalgebra or bialgebra} is a filtered vector space $V$ with linear operations $\mu_{klg}: V^{\otimes k} \rightarrow V^{\otimes l}$  of non-negative filtration degree for $k$, $l$, $g\in \N_0$ such that $(V,(\mu_{klg}))$ is an algebra, coalgebra or bialgebra, respectively.

A \emph{complete filtered algebra, coalgebra or bialgebra} is a complete filtered vector space~$V$ with linear operations $\mu_{klg}: V^{\COtimes k} \rightarrow V^{\COtimes l}$ of non-negative filtration degree for $k$, $l$, $g\in \N_0$ such that $\mu_{klg}$ satisfy the relations of an algebra, coalgebra or bialgebra with $\otimes$ replaced by $\COtimes$, respectively.

A morphism of (complete) filtered algebras, coalgebras or bialgebras is a linear map $f:V_1 \rightarrow V_2$ with non-negative filtration degree intertwining $\mu_{klg}$.
\end{Definition}

\begin{Remark}[On (complete) filtered algebras]
\begin{RemarkList}
\item Clearly, completion is a functor from filtered algebras to complete filtered algebras. In deformation theory, we may deal with complete filtered algebras which are not in the image of this functor.
\item  The $\IBLInfty$-algebra on $C$ from Definition~\ref{Def:IBLInfty} for $\gamma = 0$ is, in fact, a ``complete filtered $\IBLInfty$-algebra on~$\hat{C}$'' in the sense of Definition~\ref{Def:FiltAlg} plus the data of the filtered vector space~$C$. It might not be the completion of a ``filtered $\IBLInfty$-algebra on~$C$''; in fact, we called an $\IBLInfty$-algebra ``completion-free'' if the operations on~$\hat{C}$ arose as continuous extensions of operations on~$C$. Definition~\ref{Def:IBLInfty} seemed natural when we were trying to compute examples of the twisted structure, hoping that we do not encounter infinite sums. After dealing with completions, however, Definition~\ref{Def:FiltAlg} is more logical as an abstract definition.

Notice that for $\gamma > 0$, the norm $\Norm{\OPQ_{klg}} \ge \gamma(2-2g-k-l)$ is allowed to be negative; in particular, $\Norm{\OPQ_{210}}\ge -\gamma$ and $\Norm{\OPQ_{120}}\ge -\gamma$. Therefore, one has to allow finite filtration degree in Definition~\ref{Def:FiltAlg} in order to accommodate it to $\IBLInfty$-algebras with $\gamma \ge 0$.
\end{RemarkList}
\end{Remark}

\begin{Remark}[Units and augmentations]\label{Rem:FiltrUnitAug}
The ground field $\K$ is filtered by the trivial complete filtration $\Filtr^{\lambda\le 0}\K = \K$ and $\Filtr^{\lambda>0}\K = 0$ (see also \eqref{Eq:TrivFiltr}). Suppose that $\MVUnit: \K \rightarrow V$ is a unit and $\MVAug: V \rightarrow \K$ an augmentation of a complete filtered algebra~$V$; in particular, we have $\MVAug \circ \MVUnit = \Id$ and $\Norm{\MVUnit}$, $\Norm{\MVAug}\ge 0$. From this, we deduce the following implications for~$v\in V$:
\begin{align*}
\Norm{v}>0 &\quad\Implies\quad v\in \ker \MVAug =: \bar{V},\\
v\in \im \MVUnit &\quad\Implies\quad \Norm{v} = 0.
\end{align*}
Recall that $V = \im \MVUnit \oplus \bar{V}$.
\end{Remark}

Given a vector space $U$ over $\K$, we will work with the \emph{symmetric bialgebra} $\Sym U = \bigoplus_{k=0}^\infty \Sym_k U$ from Definition~\ref{Def:SymAlgebra}. It has the concatenation product $\MVProd: \Sym U \otimes \Sym U \rightarrow \Sym U$ and the shuffle coproduct $\MVCoProd: \Sym U \rightarrow \Sym U \otimes \Sym U$ which are given for all $u_{ij}$, $u_i \in U$ and $k$, $k'\in \N_0$ by
\begin{align*}
\MVProd(u_{11}\dotsb u_{1k} \otimes u_{21} \dotsb u_{2k'}) &= u_{11} \dotsb u_{1k} u_{21} \dotsb u_{2k'}\quad \text{and}\\
\MVCoProd(u_1 \dotsb u_k) &= \sum_{\substack{k_1,\,k_2 \ge 0\\ k_1 + k_2 = k}} \sum_{\sigma\in \Perm_{k_1, k_2}} \varepsilon(\sigma,u) u_{\sigma^{-1}_1} \dotsb u_{\sigma_{k_1}^{-1}}\otimes u_{\sigma_{k_1+1}^{-1}}\dotsb u_{\sigma_{k_1 + k_2}^{-1}},
\end{align*}
respectively. Here $\Perm_{k_1,k_2}$ denotes the shuffle permutations and $\varepsilon(\sigma,u)$ the Koszul sign. The unit $\MVUnit: \K \rightarrow \Sym U$ is given by $\MVUnit(1) = 1 \in \Sym_0 U = \K$, and the augmentation $\MVAug: \Sym U \rightarrow \K$ is determined by its reduced algebra $\RedSym U = \bigoplus_{k=1}^\infty \Sym_k U$. If $U$ is filtered, then $\MVProd$, $\MVCoProd$, $\MVUnit$ and~$\MVAug$ preserve the induced filtration on $\Sym U$ (see Definition~\ref{Def:Filtrations}), and hence $\Sym U$ is a filtered bialgebra. We will also use the canonical projection $\pi_k: \Sym U \rightarrow \Sym_k U$ and the canonical inclusion $\iota_l: \Sym_l U \rightarrow \Sym U$ for $k\in\N_0$ and $l\in \N_0$. They too preserve the filtration.

The symmetric bialgebra of a filtered vector space has naturally two filtrations --- the induced filtration and the filtration by weights $\Filtr_\WeightMRM^\lambda \Sym U \coloneqq \bigoplus_{k\ge \lambda} \Sym_k U$.\footnote{Note that the filtration by weights can be also viewed as the induced filtration from the filtration on $U$ defined by $\Filtr^{\le 1}U \coloneqq U$, $\Filtr^{>1}U\coloneqq 0$.} In general, having two filtrations~$\Filtr_1$ and~$\Filtr_2$ on a vector space $V$, we define the filtrations
\begin{equation}\label{Eq:CombinedFiltr}
\Filtr_\cup^\lambda \coloneqq \Filtr_1^\lambda + \Filtr_2^\lambda\quad\text{and}\quad\Filtr_\cap^{\lambda} \coloneqq \Filtr_1^\lambda \cap \Filtr_2^\lambda
\end{equation}
and call them the \emph{combined filtrations} --- the union and the intersection. It is easy to see that \Correct[caption={DONE Inequality for combined fitlrations},noline]{There has to be an inequality!!}
\[ \Norm{\cdot}_\cup \ge \max(\Norm{\cdot}_1,\Norm{\cdot}_{2})\quad\text{and}\quad\Norm{\cdot}_\cap \le \min(\Norm{\cdot}_1,\Norm{\cdot}_{2}) \]
hold for the corresponding filtration degrees.

\begin{figure}[t]\label{Fig:FiltrationTypes}
\centering
\input{\GraphicsFolder/completion2.tex}
\input{\GraphicsFolder/completion1.tex}
\input{\GraphicsFolder/completion3.tex}
\input{\GraphicsFolder/completion4.tex}
\caption[Natural filtrations of a bigraded vector space.]{For a bigraded vector space $V = \bigoplus_{i} \bigoplus_{j} V_{ij}$, let $\Filtr^{\lambda_1}_{\VertMRM}\coloneqq\bigoplus_{i\ge \lambda_1}\bigoplus_{j} V_{ij}$ be the vertical and $\smash{\Filtr^{\lambda_2}_{\HorMRM}}\coloneqq\bigoplus_{i}\bigoplus_{j\ge \lambda_2} V_{ij}$ the horizontal filtration. The illustrations above depict $\Filtr_{\VertMRM}$, $\Filtr_{\HorMRM}$, $\Filtr_{\VertMRM}\cup\Filtr_{\HorMRM}$ and $\Filtr_{\VertMRM}\cap \Filtr_{\HorMRM}$, respectively. We imagine the component $V_{ij}$ sitting at the position $(i,j)$ in the plane. Given $v_{ij}\in V_{ij}$, the sum $\sum_{i,j} v_{ij}$ converges with respect to the given filtration if and only if for every $\lambda$, only finitely many $v_{ij}$'s in the white region are non-zero.}
\end{figure}

The following example shows that $\Sym U$ might not be a filtered bialgebra with respect to combined filtrations.

\begin{Example}[Combined filtrations of symmetric bialgebra $\Sym U$]\label{Ex:CombinedOnSymetric}
Let $U$ be a vector space filtered by a filtration $\Filtr$. Consider the symmetric bialgebra $\Sym U$ with the induced filtration $\Filtr$ and the filtration by weights $\Filtr_{\WeightMRM}^\lambda$. It is easy to see that the operations $\MVProd$, $\MVCoProd$, $\MVUnit$ and $\MVAug$ preserve both $\Filtr$ and $\Filtr_\WeightMRM$. Consider the combined filtrations $\Filtr_\cup$ and $\Filtr_\cap$. Because
\begin{align*}
\min(\lambda_1 + \lambda_2,k_1+k_2) &\ge \min(\lambda_1,k_1) + \min(\lambda_2,k_2)\quad\text{and}\\
\max(\lambda_1 + \lambda_2,k_1+k_2) &\le \max(\lambda_1,k_1) + \max(\lambda_2,k_2),
\end{align*}
it holds $\Norm{\MVProd}_\cap \ge 0$ and $\Norm{\MVCoProd}_\cup\ge 0$, respectively.

We now demonstrate that it might happen that $\Norm{\MVProd}_\cup = - \infty$ and $\Norm{\MVCoProd}_\cap = -\infty$. Let $U = \Ten W$ be the tensor algebra over a non-zero vector space $W$. On $U$, consider the filtration by weights $\Filtr^\lambda U = \bigoplus_{k\ge \lambda} W^{\otimes k}$. Let $0\neq w\in W$, and define 
\[ v_k \coloneqq \underbrace{w\smallotimes \dotsb \smallotimes w}_{k\text{-times}} \in \Filtr^{k} \Sym_1 U\quad\text{for all }k\in\N, \]
where $\smallotimes$ denotes the tensor product on $\Ten W$. For $0\neq u \in \Ten_0 W = \K$, define 
\[ v_k' \coloneqq \underbrace{u\dotsb u}_{k\text{-times}} \in \Filtr^0 \Sym_k U\quad\text{for all }k\in \N, \]
where $\cdot$ denotes the concatenation product on $\Sym U$. It follows that 
\[ \Norm{v_k}_{\cup} = k,\quad\Norm{v_k'}_{\cup} = k\quad\text{and}\quad\Norm{v_k v_k'}_\cup = k+1. \]
Therefore, for all $k\in \N$, it holds
\begin{align*}
\Norm{\MVProd}_\cup &\le \Norm{v_k v_k'}_\cup - \Norm{v_k \otimes v_k'}_{\cup} \\
&\le k + 1 - \Norm{v_k}_\cup - \Norm{v_k'}_\cup \\
&=  1 - k.
\end{align*}
Consequently $\Norm{\mu}_\cup = -\infty$. Next, it holds 
\[ \Norm{v_k v_k'}_\cap = k\quad\text{and}\quad\Norm{v_k\otimes v_k'}_\cap = 1, \]
and we compute
\begin{align*}
\Norm{\MVCoProd}_\cap &\le \Norm{\MVCoProd(v_k v_k')}_\cap - \Norm{v_k v_k'}_\cap \\ 
&=  1 - k.
\end{align*}
We used here that the summands of $\delta(v_k v_k')$ are tensor products of $v_1''=u^{i} v_k \in \Filtr^1 \Sym_{i+1} U$ and $v_2''=u^{k-i} \in \Filtr^0 \Sym_{k-i} U$ for $i=0$, $\dotsc$, $k$, so that $\Norm{v_1'' \otimes v_2''}_\cap = 1$. It follows that $\Norm{\MVCoProd}_\cap = -\infty$. A heuristic explanation of these phenomenons is that the tensor product of combined filtrations is not the combined filtration of the tensor product.

In the next paragraph, we will argue that it might still be possible to extend $\MVProd$ to $\Sym U\COtimes_\cup\Sym U$ --- the completion of $\Sym U \otimes\Sym U$ with respect to the tensor product of the union filtrations --- even though $\Norm{\MVProd}=-\infty$. A similar discussion applies for $\MVCoProd$.

Suppose that the filtration on $U$ is bounded from above. Let $v_i\in\Filtr^{\lambda_i}\Sym_{k_i}U$ and $v_i'\in \Filtr^{\lambda_i'}\Sym_{k_i'}U$ for all $i\in \N_0$ be such that $\max(\lambda_i,k_i) + \max(\lambda_i',k_i') \to \infty$ as $i\to \infty$, so that $\sum_{i=0}^\infty v_i \otimes v_i'$ converges in $\Sym U\COtimes_\cup\Sym U$ (in fact, any element of $\Sym U\COtimes_\cup \Sym U$ can be written in this way). We have $\mu(v_i,v_i') \in\Filtr^{\lambda_i + \lambda_i'}\Sym_{k_i+k_i'}U$ for all $i\in \N_0$. Suppose that $\sum_{i=0}^\infty v_i v_i'$ does not converge with respect to $\Filtr_\cup$. Thus, $\max(\lambda_i + \lambda_i',k_i+k_i')$ is bounded, hence $\lambda_i + \lambda_i'$ and $k_i + k_i'$ are bounded, and since $k_i$, $k_i'\ge 0$, also~$k_i$ and~$k_i'$ are bounded. Nevertheless, in order to comply with the assumption on the convergence of $\sum_{i=0}^\infty v_i \otimes v_i'$, we need one of $\lambda_i$ or $\lambda_i'$, let's say $\lambda_i$, to diverge to $\infty$. Because $\lambda_i + \lambda_i'$ is bounded, $\lambda_i'$ has to diverge to~$-\infty$. But this is not allowed by the boundedness assumption.
\end{Example}

It turns out that the union filtration is often identical with the induced filtration.

\begin{Lemma}[Combined filtration and boundedness condition]\label{Lem:BoundCondOnFiltr}
Let $U$ be a vector space filtered by a decreasing filtration $\Filtr$. Suppose that there is $\gamma > 0$ such that
\begin{equation}\label{Eq:BoundedFromAbove}
\Filtr^\gamma U = U. 
\end{equation}
Then the following holds:
\begin{align*}
 \gamma\ge 1\quad&\Implies\quad\forall\lambda\in\R:\ \Filtr^\lambda_\cup \Sym U = \Filtr^\lambda \Sym U,\\
 0<\gamma < 1\quad&\Implies\quad\forall v\in \Sym U:\ \Norm{v}_\cup \ge \Norm{v}\ge \gamma\Norm{v}_\cup.
\end{align*}
\end{Lemma}
\begin{proof}
Under the assumption \eqref{Eq:BoundedFromAbove}, we have for any $k\in \N$ and $\lambda\in\R$ the following:
\begin{equation}\label{Eq:Condition}
\lambda\le k \gamma\quad \Implies\quad \Filtr^\lambda \Sym_k U = \sum_{\lambda_1 + \dotsb + \lambda_k = \lambda} \Filtr^{\lambda_1}U \otimes \dotsb \otimes \Filtr^{\lambda_k} U \Bigl/ \Perm_k = \Sym_k U.
\end{equation}
We compute
\begin{align*}
\Filtr_\cup^\lambda \Sym U &= \bigoplus_{k<\lambda}\Filtr^{\lambda}\Sym_k U \oplus  \bigoplus_{k\ge \lambda} \Sym_k U \\
&= \begin{cases}
     \displaystyle\bigoplus_{k=0}^\infty \Filtr^\lambda \Sym_k U = \Filtr^\lambda \Sym U & \text{for } \gamma \ge 1, \\
    \displaystyle\bigoplus_{\substack{k<\lambda\text{ or }k\ge\frac{\lambda}{\gamma}}} \Filtr^\lambda \Sym_k U \oplus \bigoplus_{\lambda \le k < \frac{\lambda}{\gamma}}\Sym_k U\subset \Filtr^{\lambda\gamma} \Sym U& \text{for }0<\gamma< 1.
   \end{cases}
\end{align*}
The first case holds because if $\gamma\ge 1$, then $k\ge \lambda$ implies $\gamma k \ge \lambda$, and thus $\Sym_k U = \Filtr^\lambda \Sym_k U$ by \eqref{Eq:Condition}. The second case holds from the following reasons. If $0<\gamma<1$ and $k\ge \frac{\lambda}{\gamma}$, so that $\Sym_k U = \Filtr^\lambda \Sym_k U$ by \eqref{Eq:Condition}, then $k>\lambda$. Now, $\Filtr^\lambda \Sym_k U \subset \Filtr^{\lambda \gamma} \Sym_k U$ because the filtration is decreasing and because $\lambda \ge \lambda \gamma$, and if $v\in \Sym_k U$ with $\lambda \le k < \frac{\lambda}{\gamma}$, then $k\gamma \ge \lambda \gamma$, and hence $\Sym_k U = \Filtr^{\lambda\gamma}\Sym_k U$ by \eqref{Eq:Condition}. The claim follows.
\end{proof}

We consider the exponential and the logarithm on filtered algebras.

\begin{Lemma}[Exponential and logarithm on filtered algebras]\label{Lem:Exponential}
Let $V$ be a complete filtered associative algebra with unit~$1$. Given $v\in V$ with $\Norm{v}>0$, we define the exponential
\[ e^v \coloneqq 1 + v + \frac{1}{2!}v^2 + \frac{1}{3!} v^3 + \dotsb \in V. \]
It holds $e^v e^{-v} = 1$. Given $v\in V$ with $\Norm{v-1}>0$, we define the logarithm
\[ \log(v) = \sum_{r=1}^\infty \frac{(-1)^{r-1}}{r} (v-1)^{r}\in V. \]
It holds $\log(e^v) = v$ for $v\in V$ with $\Norm{v}>0$ and $e^{\log v} = v$ for $v\in V$ with $\Norm{v-1}>0$.
\end{Lemma}
\begin{proof}
For any $v\in V$, we have
\begin{align*}
e^{-v}e^v &= e^{-v}\Bigl(\sum_{i=0}^\infty \frac{1}{i!} v^i\Bigr)\\
  &=\sum_{i=0}^\infty \frac{1}{i!} e^{-v}v^i \\
  & = \sum_{i=0}^\infty \sum_{j=0}^\infty \frac{1}{i! j!}(-1)^j v^{i+j} \\
  &= \sum_{n=0}^\infty \frac{1}{n!}\underbrace{\sum_{k=0}^n (-1)^k \binom{n}{k}}_{=0\text{ for }n>0} v^n.
\end{align*}
On the second line, we used that the multiplication with $e^{-v}$ from the left has finite filtration degree, and hence it commutes with infinite sums; on the third line, we used that the multiplication with $v^i$ from the right has finite filtration degree; on the fourth line, we used the Resummation Lemma (Lemma~\ref{Lem:TechLem}). The facts about $\log$ are proven similarly, the equations reducing to known combinatorial identities.
\end{proof}

Let $(V,\MVCoProd)$ be a coalgebra and $(V',\MVProd')$ an algebra. The \emph{convolution product} $\Star$ on $\K$-linear maps $f_1$, $f_2: V \rightarrow V'$ is defined by
\[ f_1\Star f_2 \coloneqq \MVProd'\circ(f_1\otimes f_2)\circ\MVCoProd. \]
See also \cite[Section~1.6]{Loday2012}.
If~$\MVAug: V \rightarrow \K$ is a counit for~$V$ and~$\MVUnit': \K \rightarrow V'$ a unit for~$V'$, then
\begin{equation}\label{Eq:ConvUnit}
\StarProdOne \coloneqq \MVUnit'\circ\MVAug: V \rightarrow V'
\end{equation}
is the unit for $\Star$. The convolution product is associative and commutative provided that the algebra and coalgebra are.

Let us recall the conilpotency property of a coaugmented counital coassociative coalgebra $(V,\MVCoProd,\MVAug,\MVUnit)$, where $\MVAug: V \rightarrow \K$ is the counit and $\MVUnit: \K\rightarrow V$ the coaugmentation. We write
\[ \MVCoProd =\bar{\MVCoProd}+ \MVCoProd_0, \]
where $\MVCoProd_0: V \rightarrow V\otimes V$ is defined by
\begin{equation}\label{Eq:CoProdOne}
\MVCoProd_0(v) \coloneqq \begin{cases} 1 \otimes 1 & \text{if }v=1, \\
1\otimes v + v\otimes 1 & \text{if }v\in \bar{V},\end{cases}
\end{equation}
with respect to the decomposition $V = \langle 1 \rangle \oplus \bar{V}$. The other map is then defined simply as the difference 
\[ \bar{\MVCoProd} \coloneqq \MVCoProd - \MVCoProd_0: V \longrightarrow V\otimes V. \]
We call~$\MVCoProd_0$ the \emph{trivial coproduct} and $\bar{\MVCoProd}$ the \emph{reduced diagonal.} They are both coassociative, and $\MVCoProd_0$ is even cocommutative. The \emph{conilpotency property} reads: for every $v\in V$, there exists an $n\in \N$ such that $\bar{\MVCoProd}^{(n)}(v) = 0$, where $\bar{\MVCoProd}^{(n)}\coloneqq(\bar{\MVCoProd}\otimes\Id^{\otimes n-2})\circ \dotsb\circ\bar{\MVCoProd}$ is the iterated reduced diagonal with $n$ outputs.

Using the conilpotency property, it is possible to weaken the condition $\Norm{f}>0$ on the convergence of a power series in $f\in\Hom(V,V')$ in the convolution algebra $(\Hom(V,V'),\Star,\StarProdOne)$ to $\Norm{f}\ge 0$ and $\Norm{f(1)}>0$. The latter condition can not be weakened because $e^f(1) = e^{f(1)}$ is the exponential in $(V',\MVProd',\MVUnit')$.

\begin{Proposition}[Power series in $\Star$ with coefficients in $\K$]\label{Prop:ConvPwrSer}
Let $(V,\MVCoProd,\MVAug, \MVUnit)$ be a complete filtered coaugmented counital cocommutative coassociative coalgebra such that the following \emph{limit conilpotency property} holds:
\begin{equation}\label{Eq:LimConilp}
\bar{\MVCoProd}^{(n)}(v) \to 0\quad\text{as}\quad n\to\infty\quad\text{for all }v\in V.
\end{equation}
Let $(V',\MVProd',\MVUnit',\MVAug')$ be a complete filtered augmented unital commutative associative algebra. Let~$R$ be a complete filtered $\K$-algebra. Then the following holds:
\begin{ClaimList}
\item The convolution product $\Star$ extends naturally to $R$-linear maps $\MVMorF_i: V\COtimes R\rightarrow V'\COtimes R$ ($\otimes = \otimes_\K$) of finite filtration degrees, and for its iterations, we have
\begin{equation}\label{Eq:ItStarProd}
\MVMorF_1\Star\dotsb\Star\MVMorF_n = {\MVProd'}^{(n)}\circ(\MVMorF_1\COtimes_R\dotsb\COtimes_R\MVMorF_n)\circ \MVCoProd^{(n)},
\end{equation}
where ${\MVProd'}^{(n)}$ and $\MVCoProd^{(n)}$ are the continuous $R$-linear extensions of the iterated product and coproduct, respectively. (Recall that the continuous extension of a linear map~$f$ of finite filtration degree on $V$ to the completion $\hat{V}$ is defined by $f(\sum_{i=0}^\infty v_i) \coloneqq \sum_{i=0}^\infty f(v_i)$ for all $v_i\in V$ with $\Norm{v_i}\to\infty$.)\Modify[caption={DONE Fin filtr degree},noline]{This holds for maps with finite filtration degree.}
\item Let $\MVMorF: V\COtimes R\rightarrow V'\COtimes R$ be an $R$-linear map of finite filtration degree such that 
\begin{equation}\label{Eq:ConditionsPowerSeries}
\Norm{\MVMorF}\ge 0\quad\text{and}\quad\Norm{\MVMorF(1)}>0.
\end{equation}
Then any power series $\sum_{k=0}^\infty \alpha_k \MVMorF^{\Star k}$ with coefficients $\alpha_k\in \K$, where $\MVMorF^{\Star 0}\coloneqq \StarProdOne$ is the unit of the convolution product, converges to an $R$-linear map $V\COtimes R \rightarrow V'\COtimes R$ of non-negative filtration degree, and it holds
\[ \bigl(\sum_{k=0}^\infty \alpha_k \MVMorF^{\Star k}\bigr)(1) = \sum_{k=0}^\infty \alpha_k \MVMorF(1)^k. \]
\end{ClaimList}
\end{Proposition}
\begin{proof}
\begin{ProofList}
\item 
Given a $\K$-linear map $u: V^{\COtimes k} \rightarrow \hat{V}^{\otimes l}$ of finite filtration degree for some $k$, $l\ge 0$, we consider its $R$-linear extension $u\otimes \Id: V^{\COtimes k}\otimes R \rightarrow V^{\COtimes l}\otimes R$ and extend it continuously to $V^{\COtimes k}\COtimes R\rightarrow V^{\COtimes l}\COtimes R$. Because $V^{\COtimes n}\COtimes R \simeq (V\COtimes R)^{\COtimes_R n}$ via canonical maps (this can be proven explicitly using Proposition~\ref{Prop:IsoCrit}), we get the desired continuous $R$-linear extension $(V\COtimes R)^{\COtimes_R k}\rightarrow (V\COtimes R)^{\COtimes_R l}$. We apply this construction to $u = {\mu'}^{(n)}$, $\delta^{(n)}$ and define $\MVMorF_1\Star\dotsb\Star\MVMorF_n$ using \eqref{Eq:ItStarProd} for all $n\ge 2$. It is then easy to see that
\begin{align*}
& (\dotsb((\MVMorF_1\Star\MVMorF_2)\Star\MVMorF_3)\Star\dotsb)\Star\MVMorF_n\\
&\quad= \MVProd'(\MVProd'(\dotsb\MVProd'(\MVMorF_1 \COtimes_R \MVMorF_2)\MVCoProd \COtimes_R \MVMorF_3)\MVCoProd\COtimes_R \dotsb \COtimes_R \MVMorF_n)\MVCoProd  \\
&\quad = \underbrace{\MVProd'(\MVProd'\COtimes_R \Id)\dotsb(\MVProd'\COtimes_R\Id^{n-2})}_{={\MVProd'}^{(n)}}(\MVMorF_1\COtimes_R\dotsb\COtimes_R\MVMorF_n)\underbrace{(\MVCoProd\COtimes_R\Id^{n-2})\dotsb(\MVCoProd\COtimes_R\Id)\MVCoProd}_{\MVCoProd^{(n)}} \\
&\quad = \MVMorF_1\Star\dotsb\Star\MVMorF_n.
\end{align*}
\item \begin{figure}
\centering
\begin{tikzpicture}[point/.style = {draw, circle, fill=black, minimum size=2pt,inner sep=0pt},
mylabel/.style args={at #1 #2  with #3}{
    postaction={decorate,
    decoration={
      markings,
      mark= at position #1
      with  \node [#2] {#3};
 } } } 
]
\def\maxy{4}
\def\maxx{8}
\draw[->,name path=xaxis](-.5,0) -- (\maxx,0) node[below] {$n$};
\draw[->,name path=yaxis](0,-.5) -- (0,\maxy) node[left] {$\lambda$};
\coordinate[point,label={[left]$n_1 \Norm{\MVMorF(1)}$}] (f1) at (0,3);
\coordinate[point,label={[below]$n_1$}] (n1) at (7,0);
\path[name path=f1--n1] (f1) -- (n1);
\draw[name path=plot,mylabel=at 0.9 above left with {$\Norm{\bar{\delta}^{(n)}(w)}$}] (0,0)..controls (3,0.5) and (4,1)..(\maxx-.5,\maxy-.5);
\path[name intersections={of=f1--n1 and plot,by=inter}];
\path[name path=inter--xaxis] (inter) --+(0,-2);
\path[name intersections={of=inter--xaxis and xaxis,by=n0int}];
\coordinate[point,label={[below]$n_0$}] (n0) at (n0int);
\draw[dashed] (n0) -- (inter);
\draw[dashed] (f1) -- (n1);
\path let \p1 = (inter) in coordinate (K0) at (-.5,\y1);
\path let \p1 = (inter) in coordinate (K1) at (\maxx,\y1);
\path let \p1 = (inter) in coordinate[label={[above left]K}] (Kmid) at (0,\y1);
\draw[dotted] (K0) -- (K1);
\end{tikzpicture}
\caption[Bound for the filtration degree in the iterated convolution product.]{Given $K>0$, pick $n_0\in\N$ such that $\Norm{\bar{\delta}^{(n)}(w)}\ge K$ for $n\ge n_0$. Find $n_1\ge n_0$ such that $n_1 \Norm{\MVMorF(1)} \ge K + n_0 \Norm{\MVMorF(1)}$. Then $(n - i)\Norm{\MVMorF(1)} + \Norm{\bar{\delta}^{(i)}(w)} \ge K$ for $i=0$, $\dotsc$, $n$ for any $n\ge n_1$.}
\label{Fig:Convergence}
\end{figure}
Let $w\in V\COtimes R$ and $n\ge 2$. For the map $\MVCoProd^{(n)}: V\COtimes R \rightarrow (V\COtimes R)^{\COtimes_R n}$, we write
\begin{equation}\label{Eq:ItCoprod}
\MVCoProd^{(n)}(w) = \sum_{i=0}^n\sum_{(n-i)\times 1} \sum_{\bar{\MVCoProd^{(i)}}} w_{(1)}\COtimes_R 1 \COtimes_R 1 \COtimes_R w_{(2)}\COtimes_R 1 \COtimes_R \dotsb \COtimes_R w_{(i)}\COtimes_R 1,
\end{equation}
where the second sum is over all insertions of $n-i$ units $1$ into the tensor product and the third sum denotes the sum from the Sweedler's notation
\[ \bar{\delta}^{(i)}(w) = \sum w_{(1)}\COtimes_R \dotsb \COtimes_R w_{(i)}. \]
Formula \eqref{Eq:ItCoprod} is easy to see from $\MVCoProd = \MVCoProd_0 + \bar{\MVCoProd}$.

Before we continue, let us check that $\bar{\MVCoProd}: V\COtimes R \rightarrow (V\COtimes R)^{\COtimes_R 2}$ also satisfies the limit conilpotency property (the proof is similar to the proof of Lemma~\ref{Lem:SymAlgLimConilp} below). Every element of $V\COtimes R$ can be written as $\sum_{i=0}^\infty r_i v_i$ for $r_i\in R$ and $v_i\in V$ such that $\Norm{r_i v_i} \to \infty$ as $i\to \infty$; this follows from the Resummation Lemma (Lemma~\ref{Lem:TechLem}). Given $K>0$, we pick $i_0\in \N$ such that $\Norm{r_iv_i}\ge K$ for all $i\ge i_0$. We pick $n_0\in \N$ such that $\Norm{\bar{\MVCoProd}^{(n)}(v_0)}+\Norm{r_0}$, $\dotsc$, $\Norm{\bar{\MVCoProd}^{(n)}(v_{i_0-1})}+\Norm{r_{i_0-1}} \ge K$ for all $n\ge n_0$. According to the construction in (a), we have 
\[ \bar{\MVCoProd}^{(n)}(\sum_{i=0}^\infty r_i v_i) = \sum_{i=0}^\infty r_i \bar{\MVCoProd}^{(n)}(v_i). \]
For $n\ge n_0$, it holds $\Norm{r_i\bar{\MVCoProd}^{(n)}(v_i)}\ge \Norm{r_i} + \Norm{\bar{\MVCoProd}^{(n)}(v_i)}\ge K$ for $i<i_0$ and $\Norm{r_i\bar{\MVCoProd}^{(n)}(v_i)}\ge \Norm{r_i v_i} \ge K$ for $i\ge i_0$. It follows that $\bar{\MVCoProd}^{(n)}(\sum_{i=0}^\infty r_i v_i) \to 0$ as $n\to \infty$.

Going on with the main proof, using~\eqref{Eq:ItStarProd} and~\eqref{Eq:ItCoprod}, we have
\begin{align*}
\Norm{\MVMorF^{\Star n}(w)} & = \Norm{\mu^{(n)}(\MVMorF\COtimes_R \dotsb \COtimes_R \MVMorF)\delta^{(n)}(w)} \\
 & = \begin{multlined}[t] \bigl\|\sum_{i=0}^n \sum_{(n-i)\times 1} \sum_{\bar{\delta}^{(i)}} \MVMorF(w_{n,(1)})\COtimes_R \MVMorF(1)\COtimes_R \MVMorF(1) \COtimes_R \MVMorF(w_{n,(2)}) \COtimes_R \MVMorF(1)\COtimes_R \dotsb \\ \COtimes_R \MVMorF(w_{n,(n-i)})\COtimes_R \MVMorF(1)\bigr\| \end{multlined}\\
 & \ge \min_{i=0,\dotsc,n}\bigl( (n-i) \Norm{\MVMorF(1)} + i \Norm{\MVMorF} + \Norm{\bar{\delta}^{(i)}(w)} \bigr) \\
& \ge \underbrace{\min_{i=0,\dotsc,n}\bigl( (n-i) \Norm{\MVMorF(1)} + \Norm{\bar{\delta}^{(i)}(w)} \bigr)}_{=:(*)_n}.
\end{align*}
Let $K>0$ be arbitrary. Because $\Norm{\MVMorF}\ge 0$, $\Norm{\MVMorF(1)}>0$ and $\bar{\delta}^{i}(w) \to 0$ as $i\to \infty$ by the assumptions, we obtain an $n_1\in \N$ such that $(*)_n\ge K$ for all $n \ge n_1$ (see Figure~\ref{Fig:Convergence}).
This implies the convergence of $\sum_{n=0}^\infty \alpha_n\MVMorF^{\Star n}(w)$ in $V\COtimes R$. Using (e) of the Resummation Lemma, we get a well-defined $\K$-linear map $\sum_{k=0}^\infty \alpha_k \MVMorF^{\Star k}: V\COtimes R \rightarrow V\COtimes R$ with $\Norm{\sum_{k=0}^\infty \alpha_k \MVMorF^{\Star k}} \ge \inf_{k=0,\dotsc,\infty} \Norm{\MVMorF^{\Star k}} \ge  0$. It is easy to see that it is $R$-linear as well. \qedhere
\end{ProofList}
\end{proof}

\begin{Lemma}[Completion of conilpotent coalgebra satisfies limit conilpotency]\label{Lem:SymAlgLimConilp}
Let $(V,\MVCoProd,\MVAug,\MVUnit)$ be a filtered coaugmented counital coassociative coalgebra which is conilpotent. Then the completion $(\hat{V},\MVCoProd,\MVAug,\MVUnit)$ is a complete filtered coaugmented counital coassociative coalgebra which satisfies the limit conilpotency property \eqref{Eq:LimConilp}.
\end{Lemma}
\begin{proof}
Let $v\in \hat{V}$, and write $v = \sum_{i=0}^\infty v_i$ for $v_i\in V$ with $\Norm{v_i} \to \infty$. Given $K>0$, find $i_0\in \N$ such that $\Norm{v_i}\ge K$ for all $i\ge i_0$. From the conilpotency, we can find $n_0\in \N$ such that $\bar{\MVCoProd}^{(n)}(v_0) = \dotsc = \Norm{\bar{\MVCoProd}^{(n)}(v_{i_0-1})} = 0$ for all $n\ge n_0$. Because $\bar{\MVCoProd}$ has finite filtration degree, $\bar{\MVCoProd}^{(n)}$ has finite filtration degree too, and so we can permute it with the infinite sum and write 
\[ \bar{\MVCoProd}^{(n)}(v) = \sum_{i=0}^\infty \bar{\MVCoProd}^{(n)}(v_i) = \sum_{i=i_0}^\infty \bar{\MVCoProd}^{(n)}(v_i). \]
Because $\Norm{\bar{\MVCoProd}} \ge 0$, we have $\Norm{\bar{\MVCoProd}^{(n)}} \ge 0$, and hence 
\[ \Norm{\bar{\MVCoProd}^{(n)}(v)} \ge \inf_{i=i_0,\dotsc,\infty}\Norm{\bar{\MVCoProd}^{(n)}(v_i)} \ge \inf_{i=i_0,\dotsc,\infty} \Norm{v_i} \ge K \]
for all $n\ge n_0$. This shows the limit conilpotency property.
\end{proof}

\begin{Remark}[Exponential and logarithm in $\Star$]\phantomsection\label{Rem:ExpLogStar}
\begin{RemarkList}
\item It is obvious that Lemma~\ref{Lem:Exponential} holds for~$\Star$ with the weakened conditions~\eqref{Eq:ConditionsPowerSeries}. To sum up, the conditions for the existence of $\exp(\MVMorF)$ are $\Norm{\MVMorF}\ge 0$ and $\Norm{\MVMorF(1)}>0$, and the conditions for the existence of $\log(\MVMorF)$ are $\Norm{\MVMorF}\ge 0$ and $\Norm{\MVMorF(1) - 1}>0$.

As for $\log(\MVMorF)$, we will explain how $\Norm{\MVMorF}\ge 0$ and $\Norm{\MVMorF(1)}>0$ together imply $\Norm{\MVMorF - \StarProdOne}\ge 0$. Let $\lambda < 0$, $\mu\in \R$ and $v\in \Filtr_{\mu}(V\COtimes R)$ be arbitrary. If $v$ is not a multiple of $1$, then $(\MVMorF - \StarProdOne)(v) = \MVMorF(v) \in \Filtr_{\mu + \lambda}$ as $\lambda<\Norm{\MVMorF}$. If $v = \tau$ for $0\neq\tau\in\K$, then it must hold $\mu \le 0$, and further because of $\Norm{\MVMorF(1)-1}>0$, there is an $\varepsilon>0$ such that $(\MVMorF - \StarProdOne)(\tau) \in \Filtr_\varepsilon \subset \Filtr_{\mu + \lambda}$. It follows that $\Norm{\MVMorF - \StarProdOne}\ge 0$.

\item Given $\MVMorF: V\COtimes R \rightarrow V\COtimes R$ satisfying~\eqref{Eq:ConditionsPowerSeries}, it holds $e^{-\MVMorF}\Star e^{\MVMorF} = \StarProdOne$ by Lemma~\ref{Lem:Exponential} because $\StarProdOne$ is the unit for the convolution product. However, there might not be any $\MVMorG: V\COtimes R \rightarrow V\COtimes R$ such that $\MVMorG \circ e^{\MVMorF} = \Id$ (or $e^{\MVMorF}\circ\MVMorG = \Id$). To see this, let $V=\hat{\Sym} U$, $R=\K$, and let $\HTP_{110}: U \rightarrow U$ be a linear map with $\Norm{\HTP_{110}}\ge 0$ such that there is $0 \neq v\in U$ with $\HTP_{110}(v)=0$. Consider the trivial extension $\MVMorF\coloneqq \HTP_{110}: \hat{\Sym}U\rightarrow\hat{\Sym}U$ (it equals $\HTP_{110}$ on $\hat{\Sym}_1 U\rightarrow \hat{\Sym}_1 U$ and $0$ otherwise). Because
\[ \HTP^{\Star n}(\underbrace{v \dotsb v}_{k\text{-times}}) = \begin{cases}
 n! \HTP_{110}(v)^n & \text{if } k = n, \\
 0 & \text{otherwise},
\end{cases}\]
for all $k$, $n\in\N_0$, it holds $e^{\MVMorF}(v) = \HTP_{110}(v) = 0$, and so $e^{\MVMorF}: \hat{\Sym}U\rightarrow \hat{\Sym}U$ is not invertible.

\item In the case of $V=\hat{\Sym}U$, one can check that $\log(\Id)$ is the trivial extension of $\Id_{110}: \hat{U}\rightarrow \hat{U}$ (see also \cite[Example~21]{Markl2015} in the non-filtered setting).\qedhere
\end{RemarkList}
\end{Remark}

\section{Filtered IBL-infinity-algebras in filtered MV-formalism}\label{Sec:FilteredMV}

In \cite{Markl2015}, $\MV$-algebras were introduced; they are precisely the algebras governed by the $\BV$-relations $\BVOp^2 = 0$ and $e^\MVMorF \BVOp^+ = \BVOp^- e^\MVMorF$ (following \cite{Cieliebak2015}, we will denote by ${}^+$ the source and by ${}^-$ the target). Schematically, there is the following inclusion of categories (explanations will be given below):
\begin{equation}\label{Eq:InclOfCat}
\begin{gathered}[t]
\underline{\MV\text{-algebras}} \\
\begin{aligned}
&\text{Obj.:}\ \begin{aligned}[t]
&(V,\MVProd,\MVCoProd,\MVUnit,\MVAug), R, \\
&\BVOp : V\COtimes R \rightarrow V\COtimes R,\\
& \BVOp^2 = 0, \Abs{\BVOp}=1, \\
& \BVOp(1) = 0.
\end{aligned}\\
&\text{Mor.:}\ \begin{aligned}[t]
&\MVMorF: V^+\COtimes R\rightarrow V^-\COtimes R, \\
&e^\MVMorF \circ \BVOp^+ = \BVOp^- \circ e^\MVMorF,  \\
&\MVMorF(1) \subset V^-\COtimes\mathfrak{m}^-.
\end{aligned}
\end{aligned}
\end{gathered}
\ \supset\ 
\begin{gathered}[t]
\underline{\text{$\BVInfty$\text{-algebras}}^{*}} \\
\begin{aligned}[t]
&R = \K[[\hbar]], \\
&\Delta= \BVOp_1 + \hbar \BVOp_2 + \hbar^2 \BVOp_3 + \dotsb, \\
& \begin{aligned}[t]\Delta_i: V \rightarrow V \text{ differential}\\\text{operator of order}\le i, \end{aligned}\\
&\MVMorF= \MVMorF_1 + \hbar \MVMorF_2 + \hbar^2 \MVMorF_3 + \dotsb.
\end{aligned}
\end{gathered}
\ \supset\ 
\begin{gathered}[t]
\underline{\text{$\IBLInfty$\text{-algebras}}^{**}} \\
\begin{aligned}
&V = \Sym U,  \\
&\Sym_{k > i}(U) \subset \ker \MVMorF_i.
\end{aligned}
\end{gathered}
\end{equation}
Here, $R$ is a complete local Noetherian ring with residue field~$\K$ of characteristic $0$ and maximal ideal~$\mathfrak{m}$, $(V,\MVProd,\MVUnit,\MVAug)$ is a graded commutative associative algebra over $\K$ with unit $\MVUnit: \K \rightarrow V$ and augmentation $\MVAug: V\rightarrow\K$, $(V,\MVCoProd,\MVAug,\MVUnit)$ is a \emph{conilpotent} graded cocommutative coassociative coalgebra with counit~$\MVAug$ and coaugmentation~$\MVUnit$ (the same maps as $\MVAug$ and $\MVUnit$ for $(V,\mu)$), $V\COtimes R \coloneqq \varprojlim_n (V\otimes R/V\otimes \mathfrak{m}^n)$, where $\mathfrak{m}^n = \mathfrak{m}\dotsb\mathfrak{m}$, is the completed tensor product, the operators~$\BVOp$ and~$\MVMorF$ are $R$-linear and have non-negative filtration degrees with respect to the filtration by $V\COtimes \mathfrak{m}^n$.

If $R$ is a $\K$-algebra,\footnote{The ring $\Z_4$ is a local ring which is not a $\K$-algebra over its residue field $\K=\Z_2$. The ring of polynomials $\R[x]$ in a single variable $x$ is a $\K$-algebra over its residue field $\K=\R$, but it is not a local ring because both $(x)$ and $(x+1)$ are maximal ideals. In contrast to this, the ring of power series $\K((\hbar))$ is both a local ring and a $\K$-algebra. Thanks to Thorsten Hertl for pointing this out.} then this is precisely the same setting as that of Proposition~\ref{Prop:ConvPwrSer} for the filtrations $\Filtr^{\le 0} V = V$, $\Filtr^{>0} V= 0$ and $\Filtr^{\lambda\le 0} R = R$, $\Filtr^{\lambda>0} R = \mathfrak{m}^{\lceil\lambda\rceil}$. Clearly, $\HTP(1)\subset V^- \COtimes \mathfrak{m}^-$ is equivalent to $\Norm{\HTP(1)}>0$. Therefore, $e^\MVMorF$ exists by Proposition~\ref{Prop:ConvPwrSer}.

Recall from \cite[p.\,5]{Markl2015} that a linear operator $D: V\rightarrow V$ on a commutative associative algebra~$V$ with unit $1$ is called a \emph{differential operator} of order $\le k$ for $k\in \N_0$ if it holds\label{Page:DiffOp}
\[ \psi^D_{k+1}(v_1, \dotsc, v_{k+1}) \coloneqq [[\dotsb[[D,L_{v_1}],L_{v_2}],\dotsc],L_{v_{k+1}}] = 0 \quad\text{for all }v_1, \dotsc, v_{k+1}\in V, \]
where 
\[ L_v(w) \coloneqq v w\quad\text{for all }w\in V \]
is the left-multiplication with $v\in V$ and $[\cdot,\cdot]$ is the graded commutator.

The categories of $\BVInfty$- and $\IBLInfty$-algebras contained in the $\MV$-formalism are not the most general ones; this is what $*$ and $**$ indicate. Before we explain this, let us agree on calling $(\OPQ_{klg})$ a \emph{strict $\IBLInfty$-algebra} if $\OPQ_{0lg}=\OPQ_{k0g}=0$ for all $k$, $l$, $g\in \N_0$, an \emph{input-strict $\IBLInfty$-algebra} if $\OPQ_{0lg} = 0$ for all $l$, $g\in \N_0$, and a \emph{weak $\IBLInfty$-algebra} if we want to emphasize that operations with no input and output are allowed. The same terminology will be used for morphisms.
\begin{itemize}
\item[*] It is the category of \emph{augmented strictly commutative $\BVInfty$-algebras} (see \cite[Section~5]{Cieliebak2007} for the definition of a commutative $\BVInfty$-algebra). The augmentation $\varepsilon$ is an additional data to get an $\MV$-algebra. The coproduct is defined by $\delta\coloneqq \delta_{0}$ (see~\eqref{Eq:CoProdOne}).
\item[**] It is the category of \emph{(non-filtered) input-strict $\IBLInfty$-algebras} $(\OPQ_{klg})_{k \ge 1, l, g\ge 0}$ and input-strict morphisms~$(\HTP_{klg})_{k\ge 1,l, g\ge 0}$ which satisfy the following \emph{finiteness condition} for all $k\ge 1$ and $g\ge 0$: for any $v\in \Sym U$, we have\Add[caption={DONE Add a reference},noline]{Remark (6) at the beginning of CFL15.}
\begin{equation}\label{Eq:RegCond}
\OPQ_{klg}(v) = 0\quad\text{and}\quad\MVMorF_{klg}(v) = 0 \quad\text{for all but finitely many }l\in \N_0.
\end{equation}
See \cite[Remark~(6), p.\,14]{Cieliebak2015} for the same finiteness condition. This is precisely what symplectic field theory for exact cobordisms gives.

(The transformation formulas between $\OPQ_{klg}$ and $\BVOp_i$ and $\MVMorF_{klg}$ and $\MVMorF_i$ were written down in Remark~\ref{Rem:BVForm}. We will repeat them in Proposition~\ref{Prop:EqCharOfMVIBL} below.)
\end{itemize}

The following filtered version of $\MV$-algebras will allow us to describe more general weak $\IBLInfty$-algebras. In particular, we will be able to remove the restriction~\eqref{Eq:RegCond}.\Add[caption={DONE Resolve strict and norm continuity},noline]{We choose to work ie do proof with $\Norm{}$ because it is more easier. However, it can be redone using $\Filtr_\lambda$. There is some class of filtrations for which $\Norm{c} = \lambda$ is equivalent to $c\in \Filtr_\lambda$ and $c\not\in\Filtr_\mu$ for $\mu>\lambda$. Make a remark.}

\begin{Definition}[Complete filtered $\MV$-algebra]\label{Def:FilteredMV}
A \emph{complete filtered $\MV$-algebra} over a complete filtered graded algebra $R$ over a field $\K$ is a complete filtered graded vector space~$V$ over $\K$, a homogenous $R$-linear map $\BVOp: V\COtimes R \rightarrow V\COtimes R$ of finite filtration degree satisfying 
\[ \Abs{\BVOp} = -1, \quad \Norm{\BVOp}\ge 0,\quad\Norm{\BVOp(1)}>0 \quad\text{and}\quad\BVOp\circ \BVOp = 0,\]
and operations $\MVProd: V\COtimes V \rightarrow V$, $\MVCoProd: V \rightarrow V\COtimes V$, $\MVUnit: \K \rightarrow V$ and $\MVAug: V\rightarrow \K$ such that $(V,\mu,\eta,\varepsilon)$ is a complete filtered augmented unital commutative associative algebra and $(V,\delta,\varepsilon,\eta)$ is a complete filtered coaugmented counital cocommutative coassociative coalgebra satisfying the limit conilpotency property \eqref{Eq:LimConilp}. We often denote a complete filtered $\MV$-algebra simply by $(V,\BVOp)$.

A \emph{morphism of complete filtered $\MV$-algebras} $(V^+,\BVOp^+)$ and $(V^-,\BVOp^-)$ over $R$ is a homogenous $R$-linear map $\MVMorF: V^+\COtimes R\rightarrow V^-\COtimes R$ of finite filtration degree \Add[caption={DONE Degree $0$?},noline]{Add that $\HTP$ is of degree $0$?}such that 
\[ \Abs{\MVMorF} = 0,\quad \Norm{\MVMorF}\ge0, \quad \Norm{\MVMorF(1)}>0\quad\text{and}\quad e^{\MVMorF}\circ\BVOp^+ = \BVOp^- \circ e^\MVMorF. \]

The \emph{composition} $\DiamComp$ of two composable morphisms $\MVMorF_1$ and $\MVMorF_2$ (i.e., the target of $\MVMorF_2$ is the source of $\MVMorF_1$) is defined by
\begin{equation}\label{Eq:CompositionMV}
\MVMorF_1\DiamComp\MVMorF_2 \coloneqq \log(e^{\MVMorF_1}\circ e^{\MVMorF_2}).
\end{equation}
(The exponential and logarithm are well-defined by Proposition~\ref{Prop:ConvPwrSer}.)
\end{Definition}

\begin{Remark}[On complete filtered $\MV$-algebras]\begin{RemarkList}
\item The condition $\Norm{\MVMorF(1)}>0$ is required for the exponential to converge; i.e., it might be seen as a technical condition. On the other hand, the condition $\Norm{\BVOp(1)}>0$ is optional, and we impose it in the manner of \cite{Markl2015} and \cite{Cieliebak2015}.
\item We do not define ``non-complete filtered $\MV$-algebras'' because it is not clear whether $\MVMorF_1\DiamComp\MVMorF_2: V^+\COtimes R \rightarrow V^-\COtimes R$ for $V^+\otimes R \xrightarrow{\MVMorF_2} V^{+-}\otimes R \xrightarrow{\MVMorF_1} V^-\otimes R$ restricts to a map $V^+\otimes R \rightarrow V^-\otimes R$. This forces us to define morphisms as maps of completions $V^+\COtimes R \rightarrow V^-\COtimes R$. In such category, two filtered $\MV$-algebras would be isomorphic if and only if their completions were. This is not desired, and thus we define only ``complete filtered $\MV$-algebras''.\qedhere
\end{RemarkList}\end{Remark}

Inspired by \eqref{Eq:InclOfCat}, we define the following version of weak $\IBLInfty$-algebras.

\begin{Definition}[Complete filtered $\IBLInfty$-algebra in $\MV$-formalism]\phantomsection\label{Def:ComplFiltrIBL}
Let $d\in\Z$ and $\gamma>0$. A \emph{complete filtered $\IBLInfty$-algebra of bidegree $(d,\gamma)$ in $\MV$-formalism} is a complete filtered $\MV$-algebra $(V,\BVOp, R, \MVProd, \MVCoProd, \MVAug, \MVUnit)$ satisfying the following conditions:
\begin{enumerate}[label=(\arabic*)]
\item $R=\K((\hbar))$ is the ring of Laurent series\footnote{Another notation for $\K((\hbar))$ is $\K[[\hbar]][\hbar^{-1}]$ to emphasize that it is the localization of the ring of power series $\K[[\hbar]]$ at the powers of $\hbar$. An element of $\K((\hbar))$ is a formal power series $\sum_{i=-\infty}^\infty a_i \hbar^i$ with only finitely many non-zero $a_i\in \K$ for $i\le 0$.} in a formal variable~$\hbar$ of degree $\Abs{\hbar} = 2 d$ equipped with the complete, Hausdorff and exhaustive filtration\Correct[noline,caption={DONE It is The multiplication is not continuous}]{The multiplication is not continuous}
\begin{equation}\label{Eq:MVFiltr}
\Filtr^\lambda \K((\hbar)) \coloneqq \Bigl\{\sum_{i=-\infty}^\infty a_i \hbar^i \in \K((\hbar)) \mid a_i = 0\text{ for }i<\frac{\lambda}{2\gamma}\Bigr\}\quad\text{for }\lambda\in\R.
\end{equation}
\item There is a complete filtered graded vector space $W$ such that if we define
\begin{equation}\label{Eq:DefOfU}
 U\coloneqq W[1]\quad\text{and}\quad\Filtr^\lambda U \coloneqq (\Filtr^{\lambda - \gamma} W)[1]\quad\text{for all }\lambda\in \R,
 \end{equation}
then $(V=\hat{\Sym} U, \MVProd, \MVCoProd, \MVUnit, \MVAug)$ is the completion of the symmetric bialgebra on $U$ filtered by the induced filtration.
\item The $\BV$-operator $\BVOp: \hat{\Sym}U((\hbar))\rightarrow\hat{\Sym}U((\hbar))$, where $\hat{\Sym}U((\hbar)) \coloneqq \Sym U\COtimes\K((\hbar))$,\footnote{In contrast to $\K((\hbar))$, an element of $\hat{\Sym}U((\hbar))$, when seen as a power series, can have a non-zero coefficient at every power $\hbar^i$.} decomposes as
\begin{equation}\label{Eq:BVOpDecompRel}
 \BVOp = \hbar^{-1}\BVOp_0 + \BVOp_1 + \hbar\BVOp_2 + \hbar^2 \BVOp_3 + \dotsb,
\end{equation}
where for all $i\ge 0$ the operator $\BVOp_i: \hat{\Sym}U \rightarrow \hat{\Sym}U$ is a linear differential operator of order $\le i$.\footnote{The definition of a differential operator on p.\,\pageref{Page:DiffOp} generalizes in a straightforward way to complete filtered algebras.}
\end{enumerate}
We often denote the data of a complete filtered $\IBLInfty$-algebra in $\MV$-formalism simply by~$(W,\BVOp)$. 
 
A \emph{morphism of complete filtered $\IBLInfty$-algebras $(W^+,\BVOp^+)$ and $(W^-,\BVOp^-)$ of bidegree $(d,\gamma)$ in $\MV$-formalism} is a morphism of complete filtered $\MV$-algebras $\MVMorF: \hat{\Sym}U^+((\hbar)) \rightarrow \hat{\Sym}U^-((\hbar))$ which decomposes as 
\begin{equation}\label{Eq:MorDecompRel}
 \MVMorF = \hbar^{-1}\MVMorF_0 + \MVMorF_1 + \hbar\MVMorF_2 + \hbar^2\MVMorF_3 + \dotsb,
\end{equation}
where for all $i\ge 0$ the map $\MVMorF_i : \hat{\Sym}U^+\rightarrow\hat{\Sym}U^-$ is $\K$-linear and satisfies 
\begin{equation}\label{Eq:CondOnMor}
\hat{\Sym}_{j}U^+\subset\ker \MVMorF_i\quad\text{for all }j>i.
\end{equation}
\end{Definition}

\begin{Notation}[Replacing $\Star$ with $\odot$ for $V=\Sym U$]
From now on, we will denote the convolution product~$\Star$ on morphisms of $\Sym U$ by $\odot$. It is namely the same operation on morphisms which is denoted by~$\odot$ in Section~\ref{Sec:Alg1} and in~\cite{Cieliebak2015}.
\end{Notation}
\renewcommand{\Star}{\odot}
For a map $\MVMorF = \sum_{i=-\infty}^\infty \MVMorF_{i+1}\hbar^i: \hat{\Sym}U^+((\hbar)) \rightarrow \hat{\Sym}U^-((\hbar))$, where $\MVMorF_{i}: \hat{\Sym}U^+\rightarrow\hat{\Sym}U^-$ is $\K$-linear, we denote 
\[ \langle \MVMorF\rangle_{klg} \coloneqq \pi_l \circ \MVMorF_{k+g}\circ \iota_k: \hat{\Sym}_k U^+ \rightarrow \hat{\Sym}_l U^-\quad\text{for all }k, l\ge 0, g\in \Z. \]
This is the notation introduced in \cite[Equation~(2.14)]{Cieliebak2015}.
\ToDo[caption={DONE Source and target},noline]{One has to define the product on $\HTP: V \rightarrow V'$, i.e., differenti source and target.}

Based on \cite{Cieliebak2015}, we will now give an equivalent characterization of complete filtered $\IBLInfty$-algebras in $\MV$-formalism and their morphisms in terms of their components $(\OPQ_{klg})$ and $(\HTP_{klg})$ within the surface calculus. We will not repeat the interpretation of the algebraic relations in terms of gluing of surface; for this, see \cite{Cieliebak2015}. We need the following lemma.

\begin{Lemma}[Differential operators on filtered symmetric bialgebras]\label{Lem:ComplSymBialg}
Let $U$ be a complete filtered graded vector space. Then the following holds for the complete filtered symmetric bialgebra $\hat{\Sym} U$ and all $k\in \N_0$:
\begin{ClaimList}
 \item A linear homogenous map $D: \hat{\Sym}U \rightarrow \hat{\Sym}U$ of finite filtration degree is a differential operator of order $\le k$ if and only if it can be written as 
\begin{equation}\label{Eq:DifOpForm}
D = \sum_{i=0}^k D_i \Star \Id
\end{equation}
for linear homogenous maps $D_i : \hat{\Sym}_i U\subset\hat{\Sym}U \rightarrow \hat{\Sym} U$ of finite filtration degrees (here $D_i = 0$ on $\hat{\Sym}_j U$ for $j\neq i$ is the trivial extension). The maps~$D_i$ are uniquely determined by $D$.
 \item A linear homogenous map $\MVMorF:\hat{\Sym}U\rightarrow\hat{\Sym}U$ of finite filtration degree satisfies $\hat{\Sym}_{i>k}U\subset\ker\MVMorF$ if and only if it can be written as $\MVMorF = \MVMorF_0 + \dotsb + \MVMorF_k$ for linear homogenous maps $\MVMorF_i: \hat{\Sym}_iU\rightarrow\hat{\Sym}U$ of finite filtration degrees. The maps $\MVMorF_i$ are uniquely determined by $\MVMorF$.
\end{ClaimList}
\end{Lemma}

\begin{proof}
\begin{ProofList}
\item The claim is implied by the following subclaim and proposition, which generalize in a straightforward way to the filtered case. Recall that an operator $D: \Sym U \rightarrow \Sym U$ is called a \emph{derivative} of order $\le k$ if 
\[ D(1)=0\quad\text{and}\quad\psi_i^D(v_1,\dotsc,v_i)(1)=0\quad\text{for all }i\ge k+1\text{ and }v_1, \dotsc, v_i\in V, \]
where $\psi_i^D$ were defined on page~\pageref{Page:DiffOp}.
\begin{SubClaim}[Derivatives and differential operators]\label{SubClaimFiltr}
A homogenous linear operator $D: \Sym U \rightarrow \Sym U$ is a differential operator of order $\le k$ if and only if the linear operator $D' \coloneqq D - D_0\Star \Id$, where $D_0 = D\circ\iota_0$, is a derivative of order $\le k$ (recall that $\iota_k: \Sym_k U \rightarrow \Sym U$ is the inclusion).
\end{SubClaim}
\begin{proof}
Given a derivative $D'$ of order $\le k$ and a linear map $D_0: \Sym_0 U\subset \Sym U \rightarrow \Sym U$, we will check that $D\coloneqq D' + D_0\Star \Id$ is a differential operator of order $\le k$. We have
\begin{align*}
 \underbrace{\psi_{k+2}^{D'}(v_1,\dotsc,v_{k+2})(1)}_{=0}&=
 [\psi^{D'}_{k+1}(v_1,\dotsc,v_{k+1}),L_{v_{k+2}}](1)\\
 &=\psi_{k+1}^{D'}(v_1,\dotsc,v_{k+1})(v_{k+2})-\underbrace{\psi_{k+1}^{D'}(v_1,\dotsc,v_{k+1})(1)}_{=0} v_{k+2}
\end{align*}
for all $v_1$, $\dotsc$, $v_{k+2}\in V$, and hence $D'$ is also a differential operator of order $\le k$. For all $v\in \Sym U$, it holds
\[ (D_0\Star\Id)(v) = D_0(1)v,\]
and hence for all $v$, $v_1\in \Sym U$, we have
\begin{align*}
\psi^{D_0\Star \Id}_1(v_1)(v) &= (D_0\Star \Id)(v_1 v) - (-1)^{D_0 v_1} v_1 (D_0\Star \Id)(v) \\
& = D_0(1)v_1 v - (-1)^{D_0 v_1} v_1 D_0(1) v \\
& = 0.
\end{align*}
It follows that $D_0 \Star \Id$ is a differential operator of order $0$. Clearly, a sum of differential operators of orders $\le k_1$ and $\le k_2$ is a differential operator of order $\le \max(k_1,k_2)$. Therefore, $D = D' + D_0\Star\Id$ is a differential operator of order $\le k$. Moreover, since $D'(1) = 0$ by definition, we have $D(1) = D_0(1)$, i.e., $D_0 = D \circ \iota_0$.

Given a differential operator $D$ of order $\le k$, we define $D_0\coloneqq D\circ\iota_0$ and $D'\coloneqq D - D_0\Star\Id$. Firstly, it holds 
\[ D'(1) = D(1) - (D_0\Star \Id)(1) = 0. \]
Secondly, we have 
\[ \psi_i^D(v_1,\dotsc,v_i) = \psi^{D'}(v_1,\dotsc,v_i) = 0, \]
and hence $\psi_i^D(v_1,\dotsc,v_i)(1) = 0$ for all $i\ge k+1$ and $v_1$, $\dotsc$, $v_i\in V$. Therefore, $D'$ is a derivative of order $\le k$.
\renewcommand{\qed}{\hfill\textit{(Subclaim) }$\square$}
\end{proof}
\begin{ProofProposition}[{\cite[Proposition~3.2]{Markl1997}}]\label{ProofProp:Filtr}
A linear operator $D: \Sym U \rightarrow \Sym U$ is a derivative of order $\le k$ if and only if it can be written as $D = \sum_{i=1}^k D_i \Star \Id$ for unique $D_i : \Sym_i U \rightarrow \Sym U$. 
\end{ProofProposition}
\item This is clear. We have $\MVMorF_i = \MVMorF \circ \iota_i$ for all $i=0$, $\dotsc$, $k$. \qedhere
\end{ProofList}
\end{proof}

\begin{Proposition}[Filtered $\IBLInfty$-algebras in $\MV$-formalism in components]\label{Prop:EqCharOfMVIBL}
We have the following equivalences of structures:
\begin{ClaimList}
\item The data of a complete filtered $\IBLInfty$-algebra of bidegree $(d,\gamma)$ in $\MV$-formalism $(W,\BVOp)$ is equivalent to the data of a complete filtered graded vector space $W$ and a collection of $\K$-linear maps $\OPQ_{klg}: \hat{\Sym}_k U \rightarrow \hat{\Sym}_l U$ for all $k$, $l$, $g\in \N_0$ which are homogenous, have finite filtration degrees and satisfy the following conditions for all $k$, $l$, $g\ge 0$:
\begin{enumerate}[label=(\arabic*)]
\item  $\Abs{\OPQ_{klg}} = -2d(k+g-1) - 1$.
\item $\Norm{\OPQ_{klg}}\ge -2\gamma(k+g-1)$.
\item The inequality in (2) is strict whenever $k=0$.
\item The sum of maps $Q_{kg} \coloneqq \sum_{l=0}^\infty \OPQ_{klg}: \hat{\Sym}_k U \rightarrow \hat{\Sym}U$ converges (in the sense of Lemma~\ref{Lem:TechLem}).
\item The following equation holds:\footnote{Notice that \eqref{Eq:IBLFormula} does not contain $\OPQ_{00g}$ with $g\ge 0$ for any $k$, $l$, $g\ge 0$ (otherwise all~$\OPQ_{klg}$ appear). In fact, it will be clear from the proof that there is no condition on $\OPQ_{00g}$ coming from $\BVOp^2 = 0$; if we work over a general ring, then $\OPQ_{00g}(1)$ for $g\ge 0$ can be arbitrary odd elements in it. Note that this contradicts \cite[p.\,47, bottom-most paragraph]{Cieliebak2015}.}
\begin{equation}\label{Eq:IBLFormula}
 \sum_{\substack{h \ge 1 \\ k_1, k_2, l_1, l_2, g_1, g_2 \ge 0 \\ k_1 + k_2 - h = k \\ l_1 + l_2 - h = l \\ g_1 + g_2 + h-1 = g}} \OPQ_{k_1 l_1 g_1}\circ_h \OPQ_{k_2 l_2 g_2} = 0.
\end{equation}
\end{enumerate}
\item The data of a strict morphism $\MVMorF: (W^+,\BVOp^+)\rightarrow(W^-,\BVOp^-)$ of complete filtered $\IBLInfty$-algebras of bidegree $(d,\gamma)$ in $\MV$-formalism, strict meaning that $\MVAug^- \circ \MVMorF = \MVMorF \circ \MVUnit^+ = 0$, is equivalent to a collection of $\K$-linear maps $\HTP_{klg}: \hat{\Sym}_k U^+ \rightarrow \hat{\Sym}_l U^-$ for all $k$, $l$, $g\in \N_0$ which are homogenous, have finite filtration degrees and satisfy the following conditions for all $k$, $l$, $g\ge 0$:
\begin{enumerate}[label=(\arabic*)]
\item $\Abs{\HTP_{klg}} = -2d(k+g-1)$.
\item $\Norm{\HTP_{klg}}\ge -2\gamma(k+g-1)$.
\item The inequality in (2) is strict whenever $k=0$.
\item The sum $F_{kg}\coloneqq \sum_{l=0}^\infty \HTP_{klg}: \hat{\Sym}_k U^+ \rightarrow \hat{\Sym}U^-$ converges.
\ToDo[caption={DONE $r=0$ in composition},noline]{Is there $r=0$? Because twisting with $MC$ element has this term, but it does not make much sense in the composition! Do the relations hold also for $k, l =0$ or we do not require that?}
\item It holds $\HTP_{k0g}=\HTP_{0lg}=0$.
\item The following equation holds:
\begin{equation}\label{Eq:WeakIBLMor}\begin{multlined}
\sum_{r=0}^\infty \frac{1}{r!}\sum_{\substack{h_1, \dotsc, h_r \ge 1 \\ k_1, \dotsc, k_r, k^-, l_1, \dotsc, l_r, l^-, g_1, \dotsc, g_r, g^-\ge 0 \\k_1 + \dotsb + k_r = k \\ l_1+ \dotsb + l_r - k^- + l^- = l \\ g_1 + \dotsb + g_r + g^- + k^- - r = g}} \OPQ_{k^- l^- g^-}^-\circ_{h_1,\dotsc,h_r}(\HTP_{k_1 l_1 g_1},\dotsc,\HTP_{k_r l_r g_r}) \\ = \sum_{r=0}^\infty\frac{1}{r!}\sum_{\substack{h_1, \dotsc, h_r \ge 1\\k_1, \dotsc, k_r, k^+, l_1, \dotsc, l_r, l^+, g_1, \dotsc, g_r, g^+ \ge 0 \\k_1 + \dotsb + k_r + k^+ - l^+ = k \\ l_1+ \dotsb + l_r = l \\ g_1 + \dotsb + g_r + g^+ + l^+ - r = g}} (\HTP_{k_1 l_1 g_1},\dotsc,\HTP_{k_r l_r g_r})\circ_{h_1,\dotsc,h_r}\OPQ_{k^+ l^+ g^+}^+.
\end{multlined}\end{equation}
The term $r=0$ on the left- or right-hand side is possible only for $k=0$ or $l=0$ and equals $\OPQ_{0lg}^-\circ\StarProdOne: \hat{\Sym}_0 U^+ \rightarrow \hat{\Sym}_l U^-$ or $\StarProdOne\circ\OPQ_{k0g}^+: \hat{\Sym}_k U^+ \rightarrow \hat{\Sym}_0 U^-$, respectively, where $\StarProdOne$ is the $\hat{\Sym}_0 U^+ \rightarrow \hat{\Sym}_0 U^-$ component of the unit for the convolution product~\eqref{Eq:ConvUnit} (the identity under $\hat{\Sym}_0 U^{\pm} \simeq \K$ via $\MVUnit^{\pm}$). \footnote{If $\HTP$ is not strict, it would be interesting to know whether $e^{\MVMorF} \BVOp^+ = \BVOp^- e^{\MVMorF}$ is still equivalent to the connected calculus \eqref{Eq:WeakIBLMor} or whether there are some other equations possibly involving disconnected gluing. Note that trying to incorporate~$\HTP_{0lg}$ to the right-hand side and~$\HTP_{k0g}$ to the left-hand side of \eqref{Eq:WeakIBLMor} always leads to the disconnected calculus. Also note that $\HTP_{00g}$ for $g\ge 0$ do not appear in \eqref{Eq:WeakIBLMor} for any $k$, $l$, $g\ge 0$; otherwise all~$\HTP_{klg}$ appear. Finally, note that~$\OPQ_{klg}^+$ and~$\OPQ_{klg}^-$ appear for all $k$, $l$, $g\ge 0$ and that it follows from \eqref{Eq:WeakIBLMor} that $\StarProdOne\circ\OPQ_{00g}^+$ and $\OPQ_{00g}^-\circ\StarProdOne$ have to be equal for all $g\ge 0$.}
\end{enumerate}
\end{ClaimList}
The equivalences (a) and (b) are given by the formulas
\begin{equation}\label{Eq:FormulasSums}
\BVOp_i = \sum_{\substack{k, g\ge 0 \\k+g=i}} Q_{kg}\Star\Id \quad\text{and}\quad\MVMorF_i = \sum_{\substack{k,g\ge 0\\k+g=i}} F_{kg}\quad\text{for all }i\ge 0,\text{ respectively.}
\end{equation}
We remark that the operations $\circ_{h_1,\dotsc,h_r}$ are  defined in Definition \ref{Def:ConComp} in the next section or in Definition~\ref{Def:CircS}.

Suppose that $\MVMorF^+$ and $\MVMorF^-$ are strict composable morphisms of complete filtered $\IBLInfty$-algebras in $\MV$-formalism. Then we have the following:
\begin{ClaimList}[resume]
\item The composition $\MVMorF\coloneqq \MVMorF^-\DiamComp\MVMorF^+$ is a strict morphism of complete filtered $\IBLInfty$-algebras in $\MV$-formalism and its components $\HTP_{klg}$ for $k$, $l\ge 1$, $g\ge 0$ are given by
\begin{equation}\label{Eq:CompositionOfMorphisms}
\HTP_{klg} = \hspace{-1.8cm}\sum_{\substack{r^-, r^+ \ge 0 \\ k_{1}^-, l_1^-, \dotsc, k_{r^-}^-, l_{r^-}^- \ge 1 \\ k_{1}^+, l_1^+,\dotsc, k_{r^+}^+, l_{r^+}^+ \ge 1  \\ g_1^+, \dotsc, g_{r^+}^+, g_1^-,\dotsc, g_{r^-}^- \ge 0 \\ k_1^- + \dotsb + k^-_{r^-} = l_1^+ + \dotsb + l_{r^+}^+ \\ k_1^+ + \dotsb + k_{r^+}^+ = k \\ l_1^- + \dotsb + l_{r^-}^- = l \\ g_1^+ + \dotsb + g_{r^+}^+ + g_1^- + \dotsb + g_r^- - r^+ - r^- \\ 
+ k_1^- + \dotsb + k_{r^-}^- + 1 = g}}\hspace{-1.5cm}\frac{1}{r^+! r^-!} (\HTP^-_{k_1^- l_1^- g_1^-},\dotsc,\HTP^-_{k_{r^-}^- l_{r^-}^- g_{r^-}^-})\circ_{\mathrm{con}} (\HTP^+_{k_1^+ l_1^+ g_1^+},\dotsc,\HTP^+_{k_{r^+}^+ l_{r^+}^+ g_{r^+}^+}),
\end{equation}
where $\circ_{\mathrm{con}}$ denotes the connected composition (see Definition~\ref{Def:ConComp} in the next section).
\end{ClaimList}
\end{Proposition}
\begin{proof}
\begin{ProofList}
\item Suppose first that $(W,\BVOp)$ is a complete filtered $\IBLInfty$-algebra of bidegree $(d,\gamma)$ in $\MV$-formalism; i.e., we have $\BVOp = \BVOp_0 \hbar^{-1} + \BVOp_1 + \BVOp_2 \hbar + \dotsb$ for $\BVOp_i: \hat{\Sym}U \rightarrow \hat{\Sym}U$ differential operators of order $\le i$, $\Abs{\BVOp} = -1$, $\Norm{\BVOp}\ge 0$ and $\Norm{\BVOp(1)}>0$. Using Lemma~\ref{Lem:ComplSymBialg}, we write $\BVOp_i = \sum_{j=0}^i D_{i j} \Star \Id$ for $\K$-linear maps $D_{ij}:\hat{\Sym}_j U\subset\hat{\Sym}U\rightarrow\hat{\Sym}U$ of finite filtration degrees which are uniquely determined by $\BVOp_i$, and we define $Q_{kg} \coloneqq  D_{k+g,k}: \hat{\Sym}_{k}U \rightarrow \hat{\Sym}U$ for all $k$, $g\in \N_0$. Next, we define $\OPQ_{klg} \coloneqq \pi_l \circ Q_{kg}: \hat{\Sym}_k U \rightarrow \hat{\Sym}_l U$ for all $k$, $l$, $g\in \N_0$.

In the following computations, we will keep in mind that $\Abs{\hbar} = 2d$ and $\Norm{\hbar} = 2\gamma$. Using the formula $\Abs{f \circ g} = \Abs{f} + \Abs{g}$, which is valid for homogenous linear maps $f$ and $g$, we have 
\begin{align*}
\Abs{\BVOp}=-1 &\quad\Equiv\quad\forall i\ge -1: \Abs{\BVOp_{i+1}} = -1 - d i \\
&\quad\Equiv\quad\forall k, g \ge 0, k + g = i+1: \Abs{Q_{kg}} = - 1 - di \\
&\quad\Equiv\quad\forall k, l, g \ge 0, k+g=i+1: \Abs{\OPQ_{klg}} = - 1 - d(k+g-1).
\end{align*}
Thus (1) follows. Using that $\Norm{f\circ g} \ge \Norm{f} + \Norm{g}$ and $\Norm{\sum_i v_i}\ge \inf_i \Norm{v_i}$, we have
\begin{align*}
\Norm{\BVOp}\ge 0 &\quad\Equiv\quad\forall i\ge -1: \Norm{\BVOp_{i+1}} \ge -2\gamma i \\
 &\quad\Equiv\quad\forall k, g \ge 0, k+g=i+1: \Norm{Q_{kg}} \ge -2\gamma i \\
 &\quad\Equiv\quad\forall k, l, g \ge 0, k+g=i+1: \Norm{\OPQ_{klg}} \ge -2\gamma (k+g-1).
\end{align*}
Thus (2) follows. The argument for ``$\Implies$'' on the second line is inductive using that
\[ Q_{kg} = \Bigl(\BVOp_{k+g} - \sum_{j=0}^{k-1} Q_{j, k+g - j} \Star \Id\Bigr)\circ\iota_k \]
for all $k$, $g\ge 0$. Evaluation of $\BVOp(1)$ gives
\begin{align*}
\BVOp(1)&=\BVOp_0(1)\hbar^{-1}+\BVOp_1(1)+\BVOp_2(1)\hbar+\dotsb \\
        &=Q_{00}(1)\hbar^{-1}+Q_{01}(1)+Q_{02}(1)\hbar+\dotsb \\
& = \Bigl(\sum_{l=0}^\infty \OPQ_{0l0}(1) \Bigr)\hbar^{-1} +  \Bigl(\sum_{l=0}^\infty \OPQ_{0l1}(1)\Bigr) + \hbar\Bigl(\sum_{l=0}^\infty \OPQ_{0l2}(1)\Bigr) + \dotsb,
\end{align*}
and we see that $\Norm{\BVOp(1)}>0$ is equivalent to (3). As for (4), it is implied by the Resummation Lemma (Lemma~\ref{Lem:TechLem}). Namely, we have 
\[ \sum_{l=0}^\infty \OPQ_{klg}(v) = \sum_{l=0}^\infty \pi_l(Q_{kg}(v)) \overset{\ref{Lem:TechLem}}{=} Q_{kg}(v)\quad \text{for all }v\in \hat{\Sym}_k U. \]
Thus $\sum_{l=0}^\infty \OPQ_{klg}$ converges to $Q_{kg}$.

On the other hand, given $\OPQ_{klg}$ satisfying (1), (2), (3) and (4), we clearly get an operator $\BVOp: \hat{\Sym}U((\hbar))\rightarrow\hat{\Sym}U((\hbar))$ which has the decomposition \eqref{Eq:BVOpDecompRel} and which satisfies $\Abs{\BVOp} = 0$, $\Norm{\BVOp}\ge 0$ and $\Norm{\BVOp(1)}>0$.

Assuming (1), (2), (3) and (4), it remains to check that $\BVOp^2 = 0$ is equivalent to (5). This is the same computation as in \cite[Section~2]{Cieliebak2015}, just in the filtered setting and allowing $k=0$ or $l=0$ (it will turn out that it does not change anything). Recall the notation from Section~\ref{Sec:Alg1} that $\hat{D} \coloneqq D \Star \Id = \mu(D\COtimes \Id)\delta$. We will be using formulas from Remark~\ref{Rem:Compositions} in that section, which are proven in Proposition~\ref{Prop:PartCompositions} in the next section. Note that by the Resummation Lemma, elements of the completion can be summed up in any order, even as nested infinite sums, provided one (and hence all) of these sums converges. Therefore, we could, in principle, write just one sum $\sum$ and think of $\BVOp$ as of the sum $\sum \hat{\OPQ}_{klg} \hbar^{k+g-1}$ over all $k$, $l$, $g\ge 0$.

We have
\[ \begin{aligned} \BVOp\circ\BVOp& = \sum_{i=-2}^\infty \Bigl(\sum_{\substack{k_1, l_1, g_1, k_2, l_2, g_2 \ge 0 \\ k_1 + k_2 + g_1 + g_2 = i + 2}} \hat{\OPQ}_{k_1 l_1 g_1}\circ\hat{\OPQ}_{k_2 l_2 g_2}\Bigr) \hbar^i \\
 & = \sum_{i=-2}^\infty \Bigl(\sum_{\substack{k_1, l_1, g_1, k_2, l_2, g_2 \ge 0 \\ k_1 + k_2 + g_1 + g_2 = i + 2\\ }}\sum_{h=0}^{\min(k_1,l_2)}\reallywidehat{\OPQ_{k_1 l_1 g_1}\circ_h \OPQ_{k_2 l_2 g_2}}\Bigr) \hbar^i, \end{aligned}\]
and we see that 
\[ \langle \BVOp^2 \rangle_{klg} = \sum_{\substack{k_1, l_1, g_1, k_2, l_2, g_2 \ge 0 \\ k_1 + k_2 + g_1 + g_2 = k+g + 1\\ }}\sum_{h=0}^{\min(k_1,l_2)}\pi_l\circ\reallywidehat{\OPQ_{k_1 l_1 g_1}\circ_h \OPQ_{k_2 l_2 g_2}}\circ\iota_k\quad\text{for all }k,l\ge 0, g\ge -1.  \]
Because $\OPQ_{k_1 l_1 g_1}\circ_h \OPQ_{k_2 l_2 g_2}: \hat{\Sym}_{k_1+k_2-h}U\rightarrow\hat{\Sym}_{l_1+l_2-h}U$, it follows from the definition $\hat{f} \coloneqq \mu(f\otimes \Id)\delta$ that $\pi_l \circ \reallywidehat{\OPQ_{k_1 l_1 g_1}\circ_h \OPQ_{k_2 l_2 g_2}}\circ\iota_k = 0$ unless $0\le k_1 + k_2 - h \le k$, $0\le l_1 + l_2 - h\le l$ and $l_1 + l_2 + h - (k_1 + k_2 + h) = l-k$. For fixed $k$, $l$, $h\ge 0$, $g\ge -1$, it holds
\begin{align*}
\left\{\begin{aligned}
&k_1, l_1, g_1, k_2, l_2, g_2\ge 0\\
& 0 \le k_1 + k_2 - h \le k \\
& 0 \le l_1 + l_2 - h \le l \\
& l_1 + l_2 - k_1 - k_2 = l - k \\
&k_1 + k_2 + g_1 + g_2 = k + g + 1
\end{aligned}\right\} &= \bigsqcup_{\substack{s=0, \dotsc, k}}\left\{\begin{aligned}
&k_1, k_2, l_1, l_2, g_1, g_2 \ge 0\\
&k_1 + k_2 - h = k-s \\
&l_1 + l_2 - h = l-s \\
&g_1 + g_2 + h - 1 = g+s
\end{aligned}\right\},
\end{align*}
and hence\Correct[caption={DONE This is incorrect!!},noline]{The computation is incorrect--it is either $k'<k$ or $l'<l$ and so on.}
\begin{equation}\label{Eq:ComputingComponentsBVOp}
\begin{aligned}
\langle \BVOp^2 \rangle_{klg} &=
\sum_{s=0}^k \sum_{h = 0}^{g + s +1} \sum_{\substack{k_1, k_2, l_1, l_2, g_1, g_2 \ge 0\\k_1+k_2-h=k-s\\l_1+l_2-h=l-s\\ g_1 + g_2 + h - 1 = g+s}} \pi_l \circ \reallywidehat{\OPQ_{k_1 l_1 g_1}\circ_h\OPQ_{k_2 l_2 g_2}}\circ \iota_k \\
&=\underbrace{\sum_{h = 0}^{g+1} \sum_{\substack{k_1, k_2, l_1, l_2, g_1, g_2 \ge 0\\k_1+k_2-h=k\\l_1+l_2-h=l\\ g_1 + g_2 + h - 1 = g}}\OPQ_{k_1 l_1 g_1}\circ_h\OPQ_{k_2 l_2 g_2}}_{\displaystyle=:(\square)_{klg}} + \sum_{s=1}^k  \pi_l \circ \reallywidehat{(\square)_{k-s,l-s,g+s}}\circ \iota_k.
\end{aligned}
\end{equation}
We see immediately that 
\begin{equation}\label{Eq:ComponentsBlaBla}
\langle\BVOp^2 \rangle_{klg} = 0\quad\text{for all }k, l\ge 0, g\ge -1,
\end{equation}
which is equivalent to $\BVOp^2 = 0$, follows from $(\square)_{klg} = 0$ for all $k$, $l\ge 0$, $g\ge -1$. The reverse implication is proven by induction on the linear order $\prec$ on signatures $(k,l,g)$ defined in \cite[Definition~2.4]{Cieliebak2015}. It holds $(k',l',g')\prec (k,l,g)$, by definition, if one of the following  conditions is satisfied:
\begin{enumerate}[label=(\roman*)]\label{Enum:OrderingOfSignatures}
\item $k' + l' + 2g' < k + l + 2g$,
\item $k'+l'+2g' = k + l + 2g$ and $g' > g$, or
\item $k' + l' + 2g' = k + l + 2g$ and $g' = g$ and $k' < k$.
\end{enumerate}
For the last sum in \eqref{Eq:ComputingComponentsBVOp}, we denote $k_s\coloneqq k-s$, $l_s\coloneqq l-s$ and $g_s\coloneqq g+s$ and compute
\[  (k-k_s) + (l-l_s) + 2(g-g_s) = s + s + 2(-s) = 0. \]
Therefore, case (ii) applies, and so $(k_s,l_s,g_s)\prec (k,l,g)$ for all $s=1$, $\dotsc$, $k$. This shows the equivalence of $\BVOp^2 = 0$  and $(\square)_{klg} = 0$ for all $k$, $l\ge 0$, $g\ge -1$. Finally, $(\square)_{k,l,-1} = 0$ holds automatically because $g_1 + g_2 + h =0$ implies $h=0$, and all terms $\OPQ_{k_1 l_1 g_1}\circ_h \OPQ_{k_2 l_2 g_2}$ with $h=0$ vanish. This is because~$\OPQ_{klg}$ are odd and $f_1 \circ_0 f_2 = (-1)^{\Abs{f_1}\Abs{f_2}} f_2\circ_0 f_1$.
\item Exactly as in (a), we first prove the equivalence of $\Abs{\MVMorF} = 0$, $\Norm{\MVMorF} \ge 0$, $\Norm{\MVMorF(1)}>0$ and the conditions \eqref{Eq:MorDecompRel} and \eqref{Eq:CondOnMor} to (1), (2), (3) and (4) under the correspondence \eqref{Eq:FormulasSums}. Clearly, $\MVAug^- \circ \MVMorF = \MVMorF \circ \MVUnit^+ = 0$ is equivalent to (5). 

Assuming (1), (2), (3) and (4), it remains to check that the equation $e^\MVMorF \BVOp^+ = \BVOp^- e^\MVMorF$ is equivalent to~(6). This is again the same computation as in \cite[Section~2]{Cieliebak2015}. We will do it in the weak case and then restrict to the strict case for the induction. We have
\begin{align*}
& e^\MVMorF \circ \BVOp^+ \\ 
&\quad = \begin{multlined}[t] \Bigl(\sum_{r=0}^\infty \frac{1}{r!} \sum_{i_1, \dotsc, i_r \ge -1} \sum_{\substack{k_1, l_1, g_1, \dotsc, k_r,l_r,g_r \ge 0\\ k_1 + g_1 = i_1 + 1, \dotsc, k_r + g_r = i_r + 1}}\HTP_{k_1 l_1 g_1}\Star \dotsb \Star \HTP_{k_r l_r g_r} \hbar^{i_1 + \dotsb + i_r}\Bigr)\\\circ\Bigl(\sum_{i^+ = -1}^\infty \sum_{\substack{k^+, l^+, g^+ \ge 0\\k^+ + g^+ = i^+ + 1}} \hat{\OPQ}^+_{k^+ l^+ g^+}\hbar^{i^+}\Bigr)\end{multlined}
\\
&\quad =\sum_{i=-\infty}^\infty\Bigl(\sum_{r=0}^\infty\frac{1}{r!}\sum_{{\substack{k^+, l^+, g^+, k_1, l_1, g_1, \dotsc, k_r,l_r,g_r \ge 0 \\
k^+ + k_1 + \dotsb + k_r + g^+ + g_1 + \dotsb + g_r = i + r + 1 }}}(\HTP_{k_1 l_1 g_1}\Star \dotsb \Star \HTP_{k_r l_r g_r})\circ\hat{\OPQ}_{k^+l^+g^+}^+\Bigr)\hbar^i
 \\
&\quad =\sum_{i=-\infty}^\infty\Bigl(\sum_{r=0}^\infty\frac{1}{r!} \!\!\!\!\underbrace{\sum_{\substack{k^+, l^+, g^+\ge 0 \\ k_1, l_1, g_1, \dotsc, k_r,l_r,g_r \ge 0 \\
k^+ + k_1 + \dotsb + k_r + g^+ \\ + g_1 + \dotsb + g_r = i + r + 1 }}\sum_{\substack{h_1, \dotsc, h_r \ge 0 \\ h_1 + \dotsb + h_r = l^+}}(\HTP_{k_1 l_1 g_1},\dotsc,\HTP_{k_r l_r g_r})\circ_{h_1,\dotsc,h_r}\OPQ_{k^+l^+g^+}^+}_{\displaystyle =:(*)^+_{i,r}:\hat{\Sym}U^+ \rightarrow\hat{\Sym}U^-}\Bigr)\hbar^i
\end{align*}
and similarly
\begin{align*}
& \BVOp^-\circ\,e^{\MVMorF} \\
& \quad = \sum_{i=-\infty}^\infty \Bigl(\sum_{r=0}^\infty \frac{1}{r!}\!\!\!\!\underbrace{\sum_{\substack{k^-, l^-, g^-\ge 0 \\ k_1, l_1, g_1, \dotsc, k_r,l_r,g_r \ge 0 \\
k^- + k_1 + \dotsb + k_r + g^- \\ + g_1 + \dotsb + g_r = i + r + 1 }}\sum_{\substack{h_1, \dotsc, h_r \ge 0 \\ h_1 + \dotsb + h_r = k^-}}\OPQ_{k^- l^- g^-}^-\circ_{h_1,\dotsc,h_r}(\HTP_{k_1 l_1 g_1},\dotsc,\HTP_{k_r l_r g_r})}_{\displaystyle =: (*)^-_{i,r}: \hat{\Sym}U^+ \rightarrow\hat{\Sym}U^-}\Bigr)\hbar^i.
\end{align*}
We will consider $\pi_l \circ (*)^+_{i,r} \circ \iota_k$ for fixed $i\in \Z$ and $k$, $l\ge 0$. Because of the definition of $\circ_{h_1,\dotsc,h_r}$, only the terms with $l_1 + \dotsb + l_r = l$ and $k^+ + k_1 + \dotsb + k_r - h_1 - \dotsb - h_r = k^+ + k_1 + \dotsb + k_r - l^+= k$ survive in $\pi_l \circ (*)^+_{i,r} \circ \iota_k$. Denoting $g\coloneqq i - k + 1$, we have
\[\begin{aligned}
k^+ + k_1 + \dotsb + k_r + g^+ + g_1 + \dotsb + g_r &= i +r +1 \\
k^+ + k_1 + \dotsb + k_r  - l^+ &= k \\
l_1 + \dotsb + l_r &= l
\end{aligned}\!\Equiv
\begin{aligned}
g_1 + \dotsb + g_r + g^+ + l^+ -  r &= g \\
k^+ + k_1 + \dotsb + k_r  - l^+ &= k \\
l_1 + \dotsb + l_r &= l.
\end{aligned}\]
Therefore, it holds
\begin{align*}
\pi_l \circ (*)^+_{i,r} \circ \iota_k  = \!\!\sum_{\substack{k^+,l^+,g^+\ge 0\\ 
k_1,l_1,g_1,\dotsc,k_r,l_r,g_r\ge 0 \\
k^+ + k_1 + \dotsb + k_r - l^+ = k \\
l_1 + \dotsb + l_r = l  \\ g_1 + \dotsb + g_r + g^+ + l^+ - r = g }}\sum_{\substack{h_1,\dotsc,h_r\ge 0\\ h_1 + \dotsb + h_r = l^+}}(\HTP_{k_1 l_1 g_1},\dotsc,\HTP_{k_r l_r g_r})\circ_{h_1,\dotsc, h_r} \OPQ_{k^+ l^+ g^+}^+.
\end{align*}\ToDo[caption={DONE Finis the computation of the eponential},noline]{Finish the computation of the exponential!!}
Collecting the terms with $h_i=0$ and using the graded commutativity of $\Star$, we obtain
\begin{align*}
\pi_l \circ (*)^+_{i,r} \circ \iota_k &= \begin{multlined}[t]\underbrace{\sum_{\substack{h_1, \dotsc, h_r \ge 1\\ k^+, l^+, g^+, k_1, l_1, g_1, \dotsc, k_r, l_r, g_r\ge 0 \\  h_1 + \dotsb + h_{r} = l^+ \\k_1 + \dotsb + k_r + k^+ - l^+ = k \\ l_1+ \dotsb + l_r = l \\ g_1 + \dotsb + g_r + g^+ + l^+ - r = g}} (\HTP_{k_1 l_1 g_1},\dotsc,\HTP_{k_r l_r g_r})\circ_{h_1,\dotsc,h_r}\OPQ_{k^+ l^+ g^+}^+}_{\displaystyle=:(\square^+)^r_{klg}}\\
+ \sum_{r'=0}^{r-1} \binom{r}{r'}\hspace{-.5cm}\sum_{\substack{h_1,\dotsc,h_{r'}\ge 1 \\
k^+,l^+,g^+ \ge 0\\
k_1,l_1,g_1,\dotsc,k_r,l_r,g_r\ge 0 \\
h_1 + \dotsb + h_{r'} = l^+ \\
k_1 + \dotsb + k_r + k^+ - l^+ = k \\
l_1 + \dotsb + l_r = l \\
g_1 + \dotsb + g_r + g^+ + l^+ - r = g }}\hspace{-.5cm}\begin{aligned}[t]
(\HTP_{k_1 l_1 g_1},\dotsc,\HTP_{k_{r'} l_{r'} g_{r'}})\circ_{h_1,\dotsc, h_{r'}} \OPQ^+_{k^+ l^+ g^+}&\\
\Star \HTP_{k_{r'+1} l_{r'+1} g_{r' + 1}}\Star\dotsb\Star\HTP_{k_r l_r g_r}&
\end{aligned}
\end{multlined} \\
&= (\square^+)^r_{klg} + \sum_{r'=0}^{r-1}\binom{r}{r'}\begin{aligned}[t]&\sum_{\substack{0\le k' \le k \\ 0\le l'\le l \\ 0\le g'\le g + r - r'}} \Biggl((\square^+)_{k'l'g'}^{r'} \\ &\Star\hspace{-1.21cm}\underbrace{\sum_{\substack{k_{r'+1},l_{r'+1},g_{r'+1},\dotsc,k_r,l_r,g_r\ge 0\\
k_{r'+1} + \dotsb + k_r = k-k' \\
l_{r'+1} + \dotsb + l_r = l-l' \\
g_{r'+1} + \dotsb + g_r - (r - r') = g-g'}}\HTP_{k_{r'+1} l_{r'+1} g_{r' + 1}}\Star\dotsb\Star\HTP_{k_r l_r g_r}}_{\displaystyle =: (\triangle)^{r-r'}_{k-k',l-l',g-g'}}\Biggr). \end{aligned}
\end{align*}
We used here that for fixed $r\ge 0$, $0\le r'\le r-1$, $k$, $l\ge 0$ and $g\in \Z$, we have
\begin{equation*}
\left\{\begin{aligned}
&h_1, \dotsc, h_{r'} \ge 1 \\
&k^+, l^+, g^+ \ge 0 \\
&k_1, l_1, g_1, \dotsc, k_r, l_r, g_r \ge 0 \\\hline
&h_1+\dotsb+h_{r'}=l^+\\
&k_1+\dotsb+k_r+k^+-l^+=k\\
&l_1+\dotsb+l_r=l\\
&g_1+\dotsb+g_r+g^++l^+-r=g
\end{aligned}\right\} = \!\!\!\!\!\!\!\begin{aligned}[t]\bigsqcup_{\substack{0\le k' \le k\\ 0\le l'\le l\\ -r'\le g' \le g+r-r'}}\left\{\begin{aligned}
&h_1, \dotsc, h_r'\ge 1\\
&k^+,l^+,g^+\ge 0\\
&k_1,l_1,g_1,\dotsc,k_{r'},l_{r'},g_{r'} \ge 0\\\hline
&h_1 + \dotsb + h_{r'} = l^+\\
&k_1 + \dotsb + k_{r'} + k^+ - l^+ = k' \\
&l_1 + \dotsb + l_{r'} = l' \\
&g_1 + \dotsb + g_{r'} + g^+ + l^+ - r' = g'
\end{aligned}\right\}\\
\times
\left\{\begin{aligned}
&k_{r'+1}, l_{r'+1}, g_{r'+1}, \dotsc, k_r, l_r, g_r \ge 0 \\\hline
&k_{r'+1} + \dotsb + k_r = k-k' \\
&l_{r'+1} + \dotsb + l_r = l-l' \\
&g_{r'+1} + \dotsb + g_r - (r-r') = g-g'
\end{aligned}\right\},
\end{aligned}
\end{equation*}
where the notation is the vertical version of $\{ \cdot \mid \cdot \}$. In fact, the summation starts from $g'=0$ because if $g'<0$, then $(\square^+)_{k'l'g'}^{r'} = 0$. Summing over $r\in \N_0$, we get
\begin{equation}\label{Eq:PositiveEq}\begin{aligned}
&\sum_{r=0}^\infty \frac{1}{r!} \pi_l \circ (*)_{i,r}^+ \circ \iota_k \\
&\quad = \sum_{r=0}^\infty \frac{1}{r!} (\square^+)_{klg}^r  + \sum_{r=0}^\infty \sum_{r'=0}^{r-1} \frac{1}{r'!} \frac{1}{(r-r')!}\!\!\!\!\!\sum_{\substack{0\le k' \le k \\ 0\le l'\le l \\ 0 \le g'\le g + r - r'}}\!\!\!\!(\square^+)^{r'}_{k'l'g'}\Star (\triangle)^{r-r'}_{k-k', l- l', g - g'} \\
&\quad= \sum_{r=0}^\infty \frac{1}{r!} (\square^+)_{klg}^r + \sum_{\substack{0\le k' \le k \\ 0\le l'\le l \\ g'\ge 0}}\Bigl(\sum_{r' = 0}^\infty \frac{1}{r'!} (\square^+)_{k'l'g'}^{r'}\Bigr)\Star\Bigl(\sum_{t=1}^\infty \frac{1}{t!}(\triangle)_{k-k',l-l',g-g'}^{t}\Bigr),
\end{aligned}\end{equation}
where we used the substitution $t=r-r'$ and the fact that if $g-g' < -t$, then $(\triangle)_{k-k',l-l',g-g'}^t = 0$. Similarly, we obtain
\begin{equation}\label{Eq:NegativeEq}\begin{aligned}
&\sum_{r=0}^\infty \frac{1}{r!} \pi_l \circ (*)_{i,r}^- \circ \iota_k \\
&\quad = \sum_{r=0}^\infty \frac{1}{r!} (\square^-)_{klg}^r + \sum_{\substack{0\le k' \le k \\ 0\le l'\le l \\ g'\ge 0}}\Bigl(\sum_{t=1}^\infty \frac{1}{t!}(\triangle)_{k-k',l-l',g-g'}^{t}\Bigr)\Star\Bigl(\sum_{r' = 0}^\infty \frac{1}{r'!} (\square^-)_{k'l'g'}^{r'}\Bigr).
\end{aligned}\end{equation}
Now, $e^{\MVMorF}\BVOp^+ = \BVOp^- e^{\MVMorF}$ is equivalent to 
\begin{equation}\label{Eq:SequenceOfEquations}
\sum_{r=0}^\infty \frac{1}{r!} \pi_l \circ (*)^+_{i,r} \circ \iota_k = \sum_{r=0}^\infty \frac{1}{r!}\pi_l \circ (*)^-_{i,r} \circ \iota_k\quad\text{for all }k, l \ge 0\text{ and }i\in \Z.
\end{equation}
By our previous computations, this follows from
\[ \sum_{r=0}^\infty \frac{1}{r!} (\square^+)_{klg}^r = \sum_{r=0}^\infty \frac{1}{r!} (\square^-)_{klg}^r\quad \text{for all }k, l, g\ge 0, \]
which are precisely equations \eqref{Eq:WeakIBLMor}. In order to prove the reverse implication, one needs to do induction on signatures $(k,l,g)$ as in (a). It is possible in the strict case by the following lemma.
\begin{SubClaim}[Induction in strict case I]
Suppose that $\HTP_{0lg} = \HTP_{k0g} = 0$ for all $k$, $l$, $g\ge 0$. Then for any $k$, $l\ge 0$ and $i\in\Z$, $g=i-k+1$, only the terms with $(k',l',g')\prec(k,l,g)$ contribute to the sums on the right-hand side of \eqref{Eq:PositiveEq} and \eqref{Eq:NegativeEq}. These sums vanish if $(k,l,g) = (0,0,0)$.
\end{SubClaim}
\begin{proof}
Recall the definition of $\prec$ on p.\,\pageref{Enum:OrderingOfSignatures}. First of all, $(\triangle)_{k-k',l-l',g-g'}^t \neq 0$ implies
\[ D\coloneqq (k+l+2g) - (k' + l' + 2 g') = (k-k') + (l-l') + 2(g-g') \ge t + t + 2(-t) = 0, \]
where $k-k'$, $l-l'\ge t$ holds due to strictness. Suppose that $D=0$. Then $g'\ge g$ must hold because $k-k'\ge 0$ and $l-l'\ge 0$. If $g'=g$, then $l=l'$ and $k=k'$ must hold; this is a contradiction with $k-k'\ge t \ge 1$. Therefore, it holds $g'>g$, and case (ii) of the definition of $\prec$ applies.

If $(k,l,g) = (0,0,0)$, then $(\triangle)^t_{0,0,-g'} = 0$ because $t\ge 1$ and $k-k'\ge t$.\renewcommand{\qed}{\hfill\textit{(Subclaim) }$\square$}
\end{proof}
\item We have
\[ e^\MVMorF = \sum_{k,l,g\ge 0} \langle e^\MVMorF \rangle_{klg} \hbar^{k+g-1}, \]
where
\begin{align}
\langle e^{\MVMorF}\rangle_{klg} &= \sum_{\substack{r\ge 0 \\ k_1, l_1, g_1, \dotsc, k_r, l_r, g_r \ge 0 \\ k_1 + \dotsb + k_r = k \\ l_1 + \dotsb + l_r = l \\ g_1 + \dotsb + g_r - r + 1 = g}} \frac{1}{r!} \HTP_{k_1 l_1 g_1}\Star\dotsb\Star\HTP_{k_r l_r g_r} \nonumber \\
& = \HTP_{klg} + \sum_{\substack{r\ge 2 \\ k_1, l_1, g_1, \dotsc, k_r, l_r, g_r \ge 0 \\ k_1 + \dotsb + k_r = k \\ l_1 + \dotsb + l_r = l \\ g_1 + \dotsb + g_r - r + 1 = g}} \frac{1}{r!} \HTP_{k_1 l_1 g_1}\Star\dotsb\Star\HTP_{k_r l_r g_r} \label{Eq:ComponentOfExpo}
\end{align}
for all $(k,l,g)\neq (0,0,1)$ with $k$, $l\ge 0$, $g\in\Z$. For $(k,l,g) = (0,0,1)$, one has to add the unit $\StarProdOne$ coming from the summand with $r=0$. We find that
\[\langle e^{\MVMorF^-}e^{\MVMorF^+}\rangle_{klg} = \hspace{-2cm}\sum_{\substack{r^-, r^+ \ge 0 \\ k_{1}^-, l_1^-, g_1^-, \dotsc, k_{r^-}^-, l_{r^-}^-, g_{r^-}^- \ge 0 \\ k_{1}^+, l_1^+, g_1^+, \dotsc, k_{r^+}^+, l_{r^+}^+, g_{r^+}^+ \ge 0 \\ k_1^- + \dotsb + k^-_{r^-} = l_1^+ + \dotsb + l_{r^+}^+ \\ k_1^+ + \dotsb + k_{r^+}^+ = k \\ l_1^- + \dotsb + l_{r^-}^- = l \\ g_1^+ + \dotsb + g_{r^+}^+ + g_1^- + \dotsb + g_r^- - r^+ - r^- + k_1^- + \dotsb + k_{r^-}^- + 1 = g}}\hspace{-2.4cm}\frac{1}{r^+! r^-!} (\HTP^-_{k_1^- l_1^- g_1^-} \Star \dotsb \Star \HTP^-_{k_{r^-}^- l_{r^-}^- g_{r^-}^-})\circ (\HTP^+_{k_1^+ l_1^+ g_1^+} \Star \dotsb \Star \HTP^+_{k_{r^+}^+ l_{r^+}^+ g_{r^+}^+})\]
for all $k$, $l\ge 0$, $g\in \Z$. The composition $\MVMorF^-\DiamComp\MVMorF^+ = \log(e^{\MVMorF^-}\circ e^{\MVMorF^+})$ is the unique $\MVMorF$ such that $e^{\MVMorF} = e^{\MVMorF^-}\circ e^{\MVMorF^+}$; this is equivalent to 
\begin{equation}\label{Eq:IndEqComp}
 \langle e^\MVMorF \rangle_{klg} = \langle e^{\MVMorF^-}e^{\MVMorF^+}\rangle_{klg}\quad\text{for all }k, l\ge 0, g\in\Z.
\end{equation}
In the strict case, this can be solved for $(\HTP_{klg})$ by induction on signatures.
\begin{SubClaim}[Induction in strict case II]
Suppose that $\HTP^+$ and $\HTP^-$ are strict. Then it holds $(k_i,l_i,g_i)\prec (k,l,g)$ for all $i=1$,~$\dotsc$, $r$ and $r\ge 2$ in \eqref{Eq:ComponentOfExpo} for all $k$, $l\ge 0$, $g\in\Z$.
\end{SubClaim}
\begin{proof}
We compute
\[ k + l + 2g - 2 = (k_1 + l_1 + 2 g_1 -2) + \dotsb + (k_r + l_r + 2 g_r -2), \]
where $k_i + l_i + 2g_i - 2 \ge 0$ by strictness. It follows that $k+l+2g \ge k_i + l_i + 2 g_i$ holds for all $i=1$,~$\dotsc$, $r$. If, e.g., $k + l + 2g = k_1 + l_1 + 2 g_1$, then $k_2 = l_2 = \dotsb = k_r = l_r = 1$, $g_2 = \dotsb = g_r = 0$, and hence $g = 1-r$. However, this can not happen as $r\ge 2$ and $g\ge 0$. Therefore, case (i) of $\prec$ always occurs.
\renewcommand{\qed}{\hfill\textit{(Subclaim) }$\square$}
\end{proof}
Using this, we can set $\HTP_{110} = \HTP^-_{110}\circ\HTP^+_{110}$ and $\HTP_{0lg}=\HTP_{k0g}\coloneqq 0$ for all $k$, $l$, $g\ge 0$ and solve~\eqref{Eq:IndEqComp} for $(\HTP_{klg})$ for all $k$, $l\ge 1$, $g\ge 0$ by induction over $(k,l,g)$. The solution \eqref{Eq:CompositionOfMorphisms} will solve~\eqref{Eq:IndEqComp} for all $k$, $l\ge 0$, $g\in \Z$ (the equations for $g<0$ consist of disconnected gluings and can be checked by splitting into connected components).\Correct[caption={Negative $g$},noline]{How is it with negative $g$'s? I never really checked.} Conditions on the filtration degree are easy to check as in \cite[Lemma~8.5]{Cieliebak2015}.
\qedhere
\end{ProofList}
\end{proof}

We expect that weak complete filtered $\IBLInfty$-algebras in $\MV$-formalism with weak morphisms form a subcategory of the category of complete filtered $\MV$-algebras (to see this, it remains to prove a version of (c) of Proposition~\ref{Prop:EqCharOfMVIBL} for weak morphisms). The identity morphism in this category is the continuous $\K((\hbar))$-linear extension of the trivial extension of the identity $\Id: \hat{U} \rightarrow \hat{U}$ to $\hat{\Sym}U((\hbar))\rightarrow\hat{\Sym}U((\hbar))$ (see (iii) of Remark~\ref{Rem:ExpLogStar}).

In the following, we will compare Definition~\ref{Def:ComplFiltrIBL} to the definition of filtered $\IBLInfty$-algebras from \cite[Section~8]{Cieliebak2015}.

A \emph{filtered $\IBLInfty$-algebra of bidegree $(d,\gamma)$ over $\K$ on a complete filtered graded vector space $W$ according to \cite{Cieliebak2015}} is a collection of homogenous $\K$-linear maps $\OPQ_{klg}:\hat{\Sym}_k U\rightarrow\hat{\Sym}_l U$ of finite filtration degrees for all $k$, $l$, $g\in \N_0$, where $U$ was defined in \eqref{Eq:DefOfU}, which satisfy, firstly, (1) and (2) of (a) of Proposition~\ref{Prop:EqCharOfMVIBL} with the strict inequality in (2) for all $(k,l,g)$ from the set 
\begin{equation}\label{Eq:UnstableSignatures}
\bigl\{(0,0,0), (1,0,0), (0,1,0), (2,0,0), (0,2,0), (0,0,1)\bigr\}
\end{equation}
and, secondly, the equations \eqref{Eq:ComponentsBlaBla}. A \emph{morphism of filtered $\IBLInfty$-algebras of bidegree $(d,\gamma)$ over~$\K$ on complete filtered graded vector spaces $W^+$ and $W^-$ according to \cite{Cieliebak2015}} 
is a collection of homogenous $\K$-linear maps $\HTP_{klg}: \hat{\Sym}_k U^+ \rightarrow \hat{\Sym}_l U^-$ of finite filtration degrees for all $k$, $l$, $g\in \N_0$ which satisfy, firstly, (1) and (2) of (b) of Proposition~\ref{Prop:EqCharOfMVIBL} with the strict inequality in (2) for $(k,l,g)\in\eqref{Eq:UnstableSignatures}$ and, secondly, the equations~\eqref{Eq:SequenceOfEquations}.

Recall \eqref{Eq:DefOfU} and notice that if we define the filtration $\Filtr_W^\lambda U \coloneqq (\Filtr^\lambda W)[1]$ for all $\lambda\in \R$ and denote the corresponding filtration degree by $\Norm{\cdot}_W$, then it holds $\Norm{\cdot} = \Norm{\cdot}_W + \gamma$ on~$U$, and for a map $\OPQ_{klg}:\hat{\Sym}_k U\rightarrow\hat{\Sym}_l U$, we have
\[ \Norm{\OPQ_{klg}}\ge -2\gamma(k+g-1)\quad\Equiv\quad\Norm{\OPQ_{klg}}_W \ge \gamma(2-2g-k-l). \]

\begin{figure}
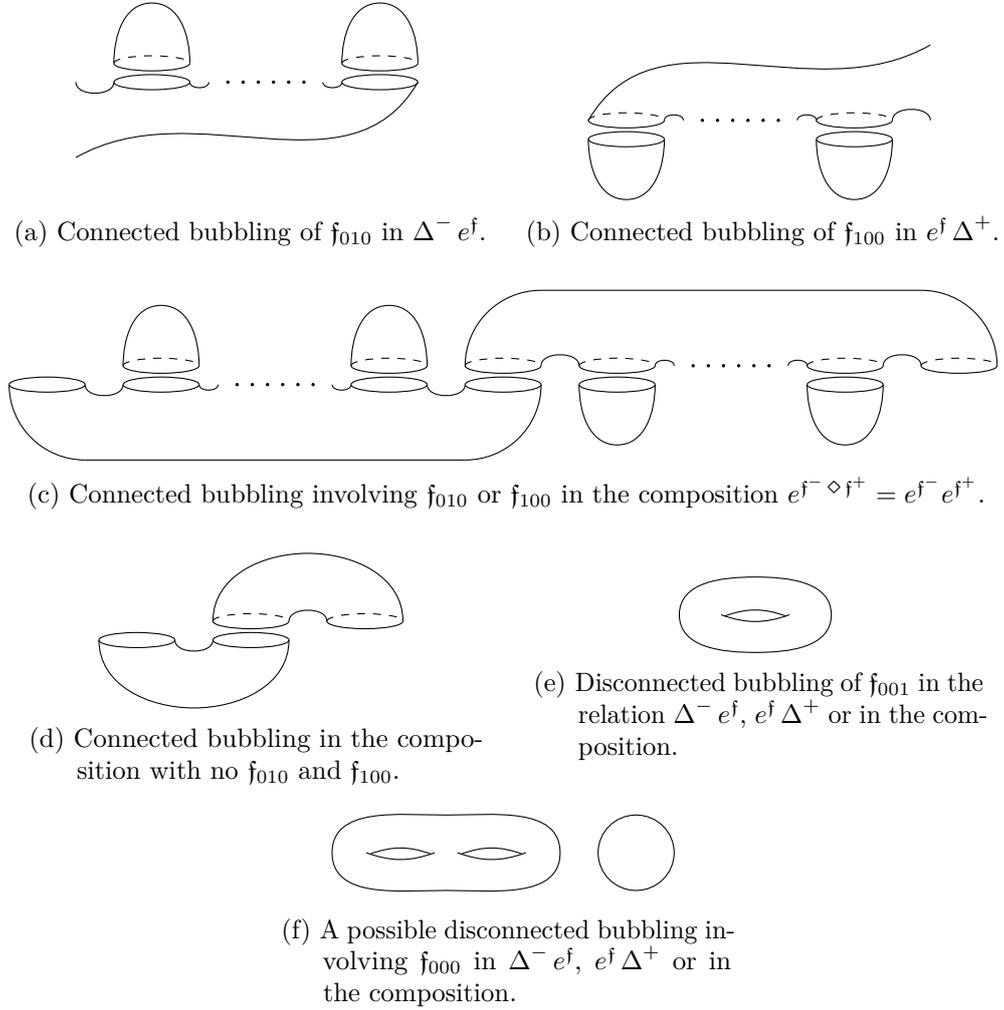

\centering
{ \begingroup \allowdisplaybreaks
\def\dist{0.25} %
  \def\rad{0.5} %
  \def\ecc{0.1} %
  \def\hght{1} %
  \def\dif{1.5} %
  \def\radO{\rad} %
  \def\eccO{\ecc} %
  \def\hghtO{2*\hght+\dist} %
  \def\difO{\dif} %
  \def\gencanc{0.05} %
  \def\genecc{20} %
  \def\genrad{0.45} %
\begin{subfigure}{.45\textwidth}
\centering
\input{\GraphicsFolder/bubbling5.tex}
\caption{Connected bubbling of $\HTP_{010}$ in $\BVOp^- e^{\MVMorF}$.}
\end{subfigure}
\begin{subfigure}{.45\textwidth}
\centering
\input{\GraphicsFolder/bubbling6.tex}
\caption{Connected bubbling of $\HTP_{100}$ in $e^{\MVMorF}\BVOp^+$.}
\end{subfigure}\\[.5cm]
\begin{subfigure}{\textwidth}
\centering
\input{\GraphicsFolder/bubbling1.tex}
\caption{Connected bubbling involving $\HTP_{010}$ or $\HTP_{100}$ in the composition $e^{\MVMorF^-\DiamComp\MVMorF^+} = e^{\MVMorF^-}e^{\MVMorF^+}$.}
\end{subfigure}\\[.5cm]
\begin{subfigure}{.4\textwidth}
\centering
\input{\GraphicsFolder/bubbling2.tex}
\caption{Connected bubbling in the composition with no $\HTP_{010}$ and~$\HTP_{100}$.}
\end{subfigure}
\hspace{.5cm}
\begin{subfigure}{.4\textwidth}
\centering
\input{\GraphicsFolder/bubbling3.tex}
\caption{Disconnected bubbling of $\HTP_{001}$ in the relation $\BVOp^- e^{\MVMorF}$, $e^{\MVMorF}\BVOp^+$ or in the composition.}
\end{subfigure}\\[.2cm]
\begin{subfigure}{.4\textwidth}
\centering
\input{\GraphicsFolder/bubbling4.tex}
\caption{A possible disconnected bubbling involving $\HTP_{000}$ in $\BVOp^- e^{\MVMorF}$, $e^{\MVMorF}\BVOp^+$ or in the composition.}
\end{subfigure}\\[.2cm]
\endgroup}
\caption[Bubbling in $\IBLInfty$-relations.]{
Bubbling in $\IBLInfty$-relations. If we glue any of the components above to a surface of signature $(k,l,g)$, the signature remains the same. Note that since~$g$ is defined via the Euler characteristic, adding a disconnected component without inputs and outputs decreases $g$ by one.}
\label{Fig:Bubbling}
\end{figure}

Signatures \eqref{Eq:UnstableSignatures} together with $(1,1,0)$ correspond to unstable surfaces, i.e., those $(k,l,g)$ for which $\chi_{klg}=2-2g-k-l \ge 0$. Allowing morphisms $\MVMorF$ which have non-zero components of these signatures leads to the appearance of an infinite number of summands in $\langle \BVOp^- e^{\MVMorF} \rangle_{klg}$, $\langle e^{\MVMorF}\BVOp^+\rangle_{klg}$ and $\langle e^{\MVMorF^-}e^{\MVMorF^+}\rangle_{klg}$ (and in the Maurer-Cartan equation \eqref{Eq:TwistingEq} later); the strict inequalities for filtration degrees of these components seem to be the minimal condition to algebraically handle this situation. Again, the author does not see any technical reason for imposing strict filtration degree conditions for~$\OPQ_{klg}$. See Figure~\ref{Fig:Bubbling} for the graphical explanation of the \emph{bubbling}.

We observe the following differences between our approach to weak $\IBLInfty$-algebras and the original approach from \cite{Cieliebak2015}: 

\begin{enumerate}[label=(\roman*)]
\item \emph{Bubbling.} The definition of \cite{Cieliebak2015} is symmetric in inputs and outputs, whereas our definition is not (compare \eqref{Eq:UnstableSignatures} and (3) of Proposition~\ref{Prop:EqCharOfMVIBL}). We will illustrate this in Example~\ref{Ex:AsymOfMV} below.

\item \emph{General ring.} The theory of \cite{Cieliebak2015} is formulated over a filtered graded commutative ring~$R$, i.e., $W$ is a filtered $R$-module and the maps are $R$-linear. We can tweak our formalism to handle this situation if $R$ is a $\K$-algebra (e.g., the Novikov ring) by replacing $\K((\hbar))$ with $R((\hbar))$ in Definition~\ref{Def:ComplFiltrIBL}.

\item \emph{Completions.} The $\BV$-formalism of \cite{Cieliebak2015} uses the completion $\hat{\Sym}U$ with respect to the union filtration $\Filtr_\cup$ of the induced filtration and the filtration by weights (see Section~\ref{Sec:DetailsOnFiltr}); however, this is not necessarily a filtered bialgebra (see Example~\ref{Ex:CombinedOnSymetric}), Proposition~\ref{Prop:ConvPwrSer} might not apply, and it is not clear whether~$e^{\HTP}$ (or the multiplication with the exponential of the Maurer-Cartan element later) are well-defined operators on $\hat{\Sym}U((\hbar))$.
\end{enumerate}

In the following case, our theory and the theory of \cite{Cieliebak2015} agree.

\begin{Proposition}[Equivalence of definitions in bounded case over $\K$]\label{Prop:BVforIBL}
Let $d\in \Z$ and $\gamma>0$. For $\zeta\in \R$, we denote by $\mathcal{W}_\zeta$ the class of complete filtered graded vector spaces $W$ for which there is an $\alpha>\zeta$ such that
\begin{equation}\label{Eq:FiltrCondU}
 \Filtr^{\alpha} W = W.
\end{equation}
For such $W$, we will consider weak $\IBLInfty$-algebras of bidegree $(d,\gamma)$ over $\K$ and their weak $\IBLInfty$-morphisms. We have the following:
\begin{ClaimList}
\item Filtered $\IBLInfty$-algebras and morphisms from \cite{Cieliebak2015} over $\mathcal{W}_{-\gamma}$ are also complete filtered $\MV$-algebras and morphisms from Definition~\ref{Def:ComplFiltrIBL}, respectively.
\item The two definitions agree over $\mathcal{W}_0$. If in addition $\gamma\ge 1$, then also the $\BV$-formalisms are identical.
\item The canonical $\dIBL$-structure on the (reduced) dual cyclic bar complex 
\[W = \CDBCyc V[2-n]\]
for a Poincar\'e duality algebra $V$ of degree $n$ from Section~\ref{Sec:Alg3} satisfies (b).
\end{ClaimList}
\end{Proposition}
\begin{proof}
For $W\in \mathcal{W}_\zeta$, there is an $\alpha>\zeta$ such that
\[ \Norm{\cdot} = \Norm{\cdot}_W + \gamma \ge  \alpha + \gamma \quad\text{on }U, \]
and hence
\begin{equation}\label{Eq:IneqWeight}
\Norm{\OPQ_{klg}(v)}_U \ge l (\alpha + \gamma) \quad\text{for all }v\in\hat{\Sym}_k U
\end{equation}
and any $k\ge 0$, $l\ge 1$, $g\ge 0$. It holds also for $l=0$ because the filtration on $\hat{S}_0 U \simeq \K$ is non-negative. We see that if $\zeta\ge -\gamma$, which is the case of both (a) and (b), then the sums $\sum_{l=0}^\infty \OPQ_{klg}$ converge automatically; i.e.,~(4) of (a) of Proposition~\ref{Prop:EqCharOfMVIBL} holds. It remains to study the relation of the strict filtration degree conditions
\begin{equation}\label{Eq:StrCondI}
\forall l, g\in \N_0:\quad  \Norm{\OPQ_{0lg}} = \Norm{\OPQ_{0lg}(1)}  > - 2\gamma(g-1)
\end{equation}
and
\begin{equation}\label{Eq:StrCondII}
\forall (k,l,g)\in \eqref{Eq:UnstableSignatures}:\quad\Norm{\OPQ_{klg}} > -2\gamma(k+g-1).
\end{equation}
Here, $\Norm{\OPQ_{0lg}} = \Norm{\OPQ_{0lg}(1)}$ holds because the trivial filtration on $\K$ is non-negative and it holds $\Norm{1} = 0$. For morphisms, the story is the same, and the rest is implied by Proposition~\ref{Prop:EqCharOfMVIBL}. 
\begin{ProofList}
\item Assuming \eqref{Eq:StrCondII}, we have to prove \eqref{Eq:StrCondI}. It is easy to see that if $\zeta \ge -\gamma$, then~\eqref{Eq:StrCondI} follows from~\eqref{Eq:IneqWeight} except for $(k,l,g) = (0,0,0)$, $(0,1,0)$, $(0,2,0)$ and $(0,0,1)$. These are precisely the signatures from \eqref{Eq:UnstableSignatures} with no input; in particular, they are implied by \eqref{Eq:StrCondII}.
\item Assuming \eqref{Eq:StrCondI}, we have to prove \eqref{Eq:StrCondII}. Clearly,~\eqref{Eq:StrCondI} implies the strict inequality for all~$(k,l,g)\in$\eqref{Eq:UnstableSignatures} with no input. It remains to check it for $(1,0,0)$ and $(2,0,0)$. Using $\zeta\ge-\gamma$, we obtain
\[ \Norm{\HTP_{100}(v)} \ge \Norm{\HTP_{100}} + \Norm{v} \ge 0 + (\gamma + \alpha) > 0\quad\text{for all }v\in\hat{\Sym}_1 V, \] 
and using $\zeta \ge 0$, we obtain
\[ \Norm{\HTP_{200}(v)} \ge \Norm{\HTP_{200}} + \Norm{v} \ge - 2 \gamma + 2(\gamma + \alpha)>0\quad\text{for all }v\in\hat{\Sym}_2 V. \]
Because $\im \HTP_{100}$, $\im \HTP_{200} \subset \K$, it must hold $\HTP_{100} = \HTP_{200} = 0$, and hence the strict filtration degree condition is automatically satisfied.

The equivalence of $\BV$-formalisms for $\gamma\ge 1$ follows from Lemma~\ref{Lem:BoundCondOnFiltr} because $\Filtr^{1} U = (\Filtr^{1-\gamma} W)[1] = W[1] = U$, where $\alpha>0 \ge 1-\gamma$.
\item For any $k\in \N_0$ and $k \le \lambda < k+1$, we have
\begin{align*}
\Filtr^\lambda_{\WeightMRM} \BCyc V &= \BCyc_1 V \oplus \dotsb \oplus \BCyc_k V, \\
\Filtr^\lambda \DBCyc V &= \{ \psi \in \DBCyc V \mid \Restr{\psi}{\Filtr^\lambda \BCyc V} = 0\} \\
& \simeq (\DBCyc V)_{k+1} \oplus (\DBCyc V)_{k+2} \oplus \dotsb,
\end{align*}
where $\Filtr_{\WeightMRM}$ is the increasing filtration by weights, $\Filtr$ the dual filtration and $(\DBCyc V)_{i}$  the graded dual to $\BCyc_i V$. We also see that $\Filtr^\lambda \DBCyc V = \DBCyc V$ for all $\lambda<1$. Since $\CDBCyc V$ is the completion of $\DBCyc V$ in the graded category, we have $\CDBCyc V\in W_0$. Finally, we know that $\gamma = 2$ for the canonical $\dIBL$-algebra.\Correct[caption={DONE Possible problems},noline]{There might be problem with using \eqref{Eq:IneqWeight} in the way I am using this! Maybe over $\K$ there are even more zero operations.}\qedhere
\end{ProofList}
\end{proof}
\begin{Remark}[Generalization over algebra]\label{Rem:GenOverAlg}
It is easy to see that (a) of Proposition~\ref{Prop:BVforIBL} also holds when we work over a non-negatively filtered augmented unital graded $\K$-algebra~$R$. However, (b) should not generalize as there are no conditions in the filtered $\MV$-formalism implying the strict filtration degree condition for $(1,0,0)$ and $(2,0,0)$.
\end{Remark}

\begin{Example}[Asymmetry of $\MV$-formalism]\label{Ex:AsymOfMV}
How is it possible that there are no strict filtration degree conditions on $(1,0,0)$ and $(2,0,0)$ in the filtered $\MV$-formalism even though these unstable surfaces can obviously bubble off (see Figure~\ref{Fig:Bubbling})?\footnote{Note that this question is relevant only when we work over a general $\K$-algebra $R$, so that $\HTP_{100}$ and $\HTP_{200}$ do not necessarily vanish.}

Heuristically, increasing the number of $(1,0,0)$'s glued to the outputs increases the number of times $\bar{\MVCoProd}$ has to be applied to the input to split it and feed it into the new $(1,0,0)$'s. The limit conilpotency property~\eqref{Eq:LimConilp} steps in, increases the filtration degree and provides convergence of the infinite sum. As for the bubbling of $(2,0,0)$, we look at (c) and (d) of Figure~\ref{Fig:Bubbling} and see that there is always an increasing  number of $(0,2,0)$'s or $(1,0,0)$'s in the adjacent components. Now, $(0,2,0)$ increases the filtration degree due to the strict filtration degree condition, and $(1,0,0)$ increases the filtration degree due to the limit conilpotency property as explained above. The convergence is again established.\Add[caption={Particular bubbling in composition},noline]{How is the bubbling of $\HTP_{200}^{\otimes k}\HTP_{100}^{\otimes k}\HTP_{030}^{\otimes k}$ handled in the composition? Maybe it is because there is a strict filtration degree condition on $\HTP_{0lg}$.}

We will now illustrate the bubbling of $\HTP_{100}$ and~$\HTP_{010}$ in the relations for morphisms with a concrete computation. Let $\HTP_{100}: \hat{\Sym}_1 U \rightarrow\hat{\Sym}_0 U$, $\HTP_{010}: \hat{\Sym}_0 U \rightarrow \hat{\Sym}_1 U$, $\OPQ_{1l0}: \hat{\Sym}_1 U \rightarrow \hat{\Sym}_l U$ and $\OPQ_{k10}: \hat{\Sym}_k U \rightarrow \hat{\Sym}_1 U$ for $k$, $l\ge 0$ be $\K$-linear maps such that the following sums converge to $\K((\hbar))$-linear operators $\hat{\Sym}U((\hbar))\rightarrow\hat{\Sym}U((\hbar))$:
\begin{align*}
\BVOp &\coloneqq \sum_{l=1}^\infty \hat{\OPQ}_{1l0}, & \MVMorF&\coloneqq \HTP_{100}, \\
\BVOp'&\coloneqq \sum_{k=1}^\infty \hat{\OPQ}_{k10}\hbar^{k-1}, & \MVMorF' &\coloneqq\HTP_{010}\hbar^{-1}.
\end{align*}
Assume that $\Norm{\HTP} = \Norm{\HTP_{100}}\ge 0$. Because $\HTP_{100}(1)=0$, convergence of the exponential
\[ e^{\MVMorF} = \sum_{r=0}^\infty \frac{1}{r!}\HTP_{100}^{\Star r}: \hat{\Sym}U((\hbar))\rightarrow\hat{\Sym}U((\hbar)) \]
relies purely on the limit conilpotency property (see the proof of Proposition~\ref{Prop:ConvPwrSer}). Assume that $\Norm{\HTP_{010}} = \Norm{\HTP_{010}(1)}>2\gamma$, i.e., $\Norm{\MVMorF'} = \Norm{\MVMorF'(1)} = \Norm{\HTP_{010}(1)} - 2 \gamma >0$. Clearly, convergence of the exponential
\begin{equation}\label{Eq:ExpII}
e^{\MVMorF'} = \sum_{r=0}^\infty \frac{1}{r!}\HTP_{010}^{\Star r}\hbar^{-r}: \hat{\Sym}U((\hbar))\rightarrow\hat{\Sym}U((\hbar))
\end{equation}
relies purely on the strict filtration degree condition. 

For all $v\in\Sym_1 U$, we have
\begin{equation}\label{Eq:BubbleI}
\begin{aligned}
 (e^{\MVMorF}\BVOp)(v) &= \sum_{l=1}^\infty e^{\MVMorF}\hat{\OPQ}_{1l0}(v) \\
 & = \sum_{l=1}^\infty \frac{1}{l!} \mu^{(l)}\HTP_{100}^{\otimes l}\bar{\MVCoProd}^{(l)}\OPQ_{1l0}(v)\\ 
 & = \sum_{l=1}^\infty \HTP_{100}^{\otimes l}\bigl(\OPQ_{1l0}(v)\bigr).
\end{aligned}
\end{equation}
In this simplest case, the limit conilpotency property takes the form $\bar{\MVCoProd}^{(k)}(\Sym_l U) = 0$ for $k>l$, which bounds the number of $\HTP_{100}$'s applied to individual summands.
\Correct[caption={DONE Combinatorial factor},noline]{Maybe there is no combinatorial factor! Checked and it is correct.}
Next, we have
\begin{equation}\label{Eq:BubbleII}
\begin{aligned}
(\BVOp' e^{\MVMorF'})(1) & = \sum_{r=0}^\infty \frac{1}{r!} \BVOp'\HTP_{010}(1)^r\hbar^{-r} \\
& = \sum_{r=0}^\infty \frac{1}{r!} \sum_{i=1}^\infty \hat{\OPQ}_{i10} \HTP_{010}(1)^r \\
& = \sum_{r=1}^\infty\sum_{i=1}^r \frac{1}{i!(r-i)!} \OPQ_{i10}(\HTP_{010}(1)^i)\HTP_{010}(1)^{r-i} \hbar^{-r + i - 1} \\
& = \Bigl(\sum_{i=1}^\infty\frac{1}{i!}\OPQ_{i10}(\HTP_{010}(1)^i)\Bigr)\Bigl(\underbrace{\sum_{t=0}^\infty\frac{1}{t!}\HTP_{010}(1)^t \hbar^{-t-1}}_{=e^{\HTP'(1)}\hbar^{-1}}\Bigr).
\end{aligned}
\end{equation}
Clearly, the condition $\Norm{\HTP_{010}(1)}>0$ is required for the convergence of the sum in the first bracket.
\end{Example}

Let us now consider the twisting of a complete filtered $\IBLInfty$-algebra in $\MV$-formalism $(\OPQ_{klg})$ on $W$ with a Maurer-Cartan element $\PMC$. The \emph{Maurer-Cartan element} is, by definition, a morphism from the trivial $\IBLInfty$-algebra $0$ to $W$. Because $\hat{\Sym}0 = \hat{\Sym}_0 0 = \K$, only the $(0,l,g)$-components denote by $\PMC_{lg}: \K \rightarrow \hat{\Sym}_l U$ for $l$, $g\ge 0$ might be non-zero, and thus\Add[noline,caption={$l=0$ Maurer Cartan}]{Kai does not have $\PMC_{0g}$. Notice that having $\PMC_{0g}$ causes disconnected twisting!}
\[ \PMC = \sum_{l,g \ge 0} \PMC_{lg} \hbar^{g-1}:\ \hat{\Sym}0((\hbar))\longrightarrow\hat{\Sym}U((\hbar)).\]
Clearly, $e^\PMC(1) = e^{\PMC(1)}\in \hat{\Sym}U((\hbar))$. Consider the left multiplication $L_{e^{\PMC(1)}}: \hat{\Sym}U((\hbar)) \rightarrow \hat{\Sym}U((\hbar))$, and let $\BVOp$ be the $\BV$-operator for $(\OPQ_{klg})$. The $\BV$-operator for the twisted $\IBLInfty$-algebra $(\OPQ_{klg}^\PMC)$ is defined by
\begin{equation}\label{Eq:TwistingEq}
\BVOp^\PMC \coloneqq L_{e^{-\PMC(1)}} \circ \BVOp \circ L_{e^{\PMC(1)}}:\hat{\Sym}U((\hbar))\longrightarrow\hat{\Sym}U((\hbar)).
\end{equation}
The twisting always produces an input-strict $\IBLInfty$-algebra. Indeed, we have
\[ \BVOp \circ e^\PMC = 0\quad\Equiv\quad\BVOp(e^{\PMC(1)})=0\quad\Equiv\quad\BVOp^{\PMC}(1) = 0. \]
This is called the \emph{Maurer-Cartan equation}.

\begin{Proposition}[Twisted $\IBLInfty$-algebra]\label{Eq:TwistingProp}
Let $(W,\BVOp)$ be a complete filtered $\IBLInfty$-algebra in $\MV$-formalism, and let $\PMC$ be a Maurer-Cartan element. Then $\BVOp^\PMC$ defined by~\eqref{Eq:TwistingEq} is an input-strict complete filtered $\IBLInfty$-algebra in $\MV$-formalism of the same bidegree as $(W,\BVOp)$, and for its components $(\OPQ_{klg}^\PMC)$, it holds
\[ \OPQ^\PMC_{klg} = \sum_{\substack{r\ge 0\\ k', l', g', l_1, g_1, \dotsc, l_{r}, g_{r} \ge 0 \\ k' - h_1 - \dotsb - h_{r}  = k \\ l' + l_1 + \dotsb + l_{r} - h_1 - \dotsb - h_r = l \\ g' + h_1 + \dotsb + h_{r} - r + g_1 + \dotsb+ g_{r} = g}}\hspace{-1cm}\OPQ_{k'l'g'}\circ_{h_1,\dotsc,h_{r}}(\PMC_{l_1 g_1},\dotsc,\PMC_{l_{r},g_{r}})\quad\text{for all }k\ge 1, l, g\ge 0, \]
where $(\OPQ_{klg})$ and $(\PMC_{lg})$ are components of $\BVOp$ and $\PMC$, respectively.
\end{Proposition}
\begin{proof}
Using $L_{e^{\PMC(1)}} = e^{\PMC}\Star \Id$, we compute
\begin{align*}
&\BVOp \circ L_{e^{\PMC(1)}} \\
&\quad = \sum_{\substack{r\ge 0 \\ k', l', g', l_1, g_1, \dotsc, l_{r'}, g_{r'} \ge 0}}  \frac{1}{r!} \hat{\OPQ}_{k'l'g'}\circ(\PMC_{l_1 g_1} \Star \dotsb \Star \PMC_{l_r g_r}\Star \Id)\hbar^{g_1 + \dotsb + g_r - r + k' + g' - 1} \\
&\quad = \sum_{\substack{r\ge 0 \\ k, k', l', g', l_1, g_1, \dotsc, l_{r'}, g_{r'} \ge 0 \\ h_1, \dotsc, h_r \ge 0 \\ h_1 + \dotsb + h_r + k = k'}}  \frac{1}{r!} \OPQ_{k'l'g'}\circ_{h_1,\dotsc,h_r,k}(\PMC_{l_1 g_1}, \dotsb, \PMC_{l_r g_r}, \Id)\hbar^{g_1 + \dotsb + g_r - r + k' + g' - 1} \\
&\quad = \sum_{\substack{
r\ge 0 \\ 0 \le r' \le r \\ k, k', l', g', l_1, g_1, \dotsc, l_{r'}, g_{r'} \ge 0 \\ h_1, \dotsc, h_{r'} \ge 1 \\ h_1 + \dotsb + h_{r'} + k = k'}}\hspace{-.2cm}\begin{multlined}[t]\frac{1}{(r-r')!}\frac{1}{r'!} \PMC_{l_{r'+1} g_{r'+1}}\Star \dotsb \Star \PMC_{l_r g_r} \\ \Star \bigl(\OPQ_{k'l'g'}\circ_{h_1,\dotsc,h_{r'},k}(\PMC_{l_1 g_1},\dotsc,\PMC_{l_{r'} g_{r'}},\Id)\bigr) \hbar^{g_1 + \dotsb + g_r - r + k' + g' - 1}\end{multlined}\\
&\quad=\begin{multlined}[t]
\Bigl(\sum_{t\ge 0} \frac{1}{t!} \PMC_{l_1' g_1'}\Star\dotsb\Star\PMC_{l_{t}'g_t'} \hbar^{g_{1}' + \dotsb + g_t' - t}\Bigr) \\
\Star \Bigl(\sum_{k,l,g\ge 0}\hspace{-.5cm}\sum_{\substack{r'\ge 0\\
k,k',l',g',l_1,g_1,\dotsc,l_r,g_r\ge 0\\
h_1,\dotsc, h_{r'}\ge 1 \\
k' - h_1 - \dotsb - h_{r'} =  k\\
l' + l_1 + \dotsb + l_{r'} - h_1 - \dotsb - h_{r'} = l\\
g' + h_1 + \dotsb + h_{r'} - r' + g_1 + \dotsb + g_{r'} = g
}} \hspace{-.5cm} \frac{1}{r'!}\OPQ_{k'l'g'}\circ_{h_1,\dotsc,h_{r'},k}(\PMC_{l_1 g_1},\dotsc,\PMC_{l_{r'} g_{r'}},\Id) \hbar^{k + g -1}\Bigr)
\end{multlined}\\
&\quad=\begin{multlined}[t]
\Bigl(\sum_{t\ge 0} \frac{1}{t!} \bigl(\PMC_{l_1' g_1'}\Star\dotsb\Star\PMC_{l_{t}'g_t'}\Star\Id\bigr) \hbar^{g_{1}' + \dotsb + g_t' - t}\Bigr) \\
\circ \Bigl(\sum_{k,l,g\ge 0}\hspace{-.5cm}\sum_{\substack{r'\ge 0\\
k,k',l',g',l_1,g_1,\dotsc,l_r,g_r\ge 0\\
h_1,\dotsc, h_{r'}\ge 1 \\
k' - h_1 - \dotsb - h_{r'} =  k\\
l' + l_1 + \dotsb + l_{r'} - h_1 - \dotsb - h_{r'} = l\\
g' + h_1 + \dotsb + h_{r'} - r' + g_1 + \dotsb + g_{r'} = g
}} \hspace{-.5cm} \frac{1}{r'!}\OPQ_{k'l'g'}\circ_{h_1,\dotsc,h_{r'}}(\PMC_{l_1 g_1},\dotsc,\PMC_{l_{r'} g_{r'}})\Star\Id \hbar^{k + g -1}\Bigr)
\end{multlined}\\
&\quad=L_{e^{\PMC(1)}}\circ \Bigl(\sum_{k,l,g\ge 0}\hspace{-.5cm}\sum_{\substack{r'\ge 0\\
k,k',l',g',l_1,g_1,\dotsc,l_r,g_r\ge 0\\
h_1,\dotsc, h_{r'}\ge 1 \\
k' - h_1 - \dotsb - h_{r'} =  k\\
l' + l_1 + \dotsb + l_{r'} - h_1 - \dotsb - h_{r'} = l\\
g' + h_1 + \dotsb + h_{r'} - r' + g_1 + \dotsb + g_{r'} = g
}} \hspace{-.5cm} \frac{1}{r'!}\reallywidehat{\OPQ_{k'l'g'}\circ_{h_1,\dotsc,h_{r'}}(\PMC_{l_1 g_1},\dotsc,\PMC_{l_{r'} g_{r'}})}\hbar^{k + g -1}\Bigr).
\end{align*}
The claim follows.
\end{proof}

\section{BV-complexes for IBL-infinity-algebras}\label{Sec:BVCompl}

We plan to use the filtered $\MV$-formalism from the previous section to study chain-maps between the following chain complexes.

\begin{Definition}[Chain complexes for $\IBLInfty$-algebras]\label{Def:BVCompl}
Given a complete filtered $\IBLInfty$-algebra in $\MV$-formalism $(U,\BVOp)$, we call the chain complex $(\hat{\Sym}U((\hbar)),\BVOp)$ the \emph{$\BV$-complex}. If $(U,\BVOp)$ is input-strict, i.e., if $\BVOp(1) = 0$, then $(\hat{\Sym}U[[\hbar]],\BVOp)$ is a chain complex as well, and we call it the \emph{strict $\BV$-complex}. If $(U,\BVOp)$ is strict, we also have the chain complex $(\hat{\Sym}_1 U, \OPQ_{110})$, and we call it the \emph{$\IBLInfty$-chain complex} (c.f., Definition~\ref{Def:HomIBL}).
\end{Definition}

Clearly, the definition works also for filtered $\IBLInfty$-algebras from \cite{Cieliebak2015} (we are just not sure about Proposition~\ref{Prop:ObservationsMor} below, and thus we rather stick to our formalism).

\begin{Remark}[$\BV$-bicomplex for surfaces]\label{Rem:BVBicomp}
Writing $\BVOp = \BVOp_1 + \BVOp_2 \hbar + \BVOp_3 \hbar^2 + \dotsb$ in the input-strict case, we have 
\[ 0 \overset{!}{=} \BVOp^2 = \BVOp_1^2 + \hbar(\BVOp_1 \BVOp_2 + \BVOp_2 \BVOp_1) + \hbar^2(\BVOp_2^2 + \BVOp_1 \BVOp_3 + \BVOp_3 \BVOp_1) + \hbar^3(\dotsb) + \dotsb. \]
We see that $\BVOp_1^2 = 0$ and $[\BVOp_1,\BVOp_2] = 0$, where $[\cdot,\cdot]$ is the graded commutator, always hold. Moreover, if $[\BVOp_1,\BVOp_3] = 0$, then also $\BVOp_2^2 = 0$. Now, $\Abs{\BVOp_1} = -1$ always holds, and if $d=-1$, then $\Abs{\hbar} = -2$ and $\Abs{\BVOp_2} = -1 - 2d = 1$. We see that if $[\BVOp_1,\BVOp_3]=0$ and $d=-1$, then $(\hat{\Sym}U, \BVOp_1, \BVOp_2)$ is a (homological) mixed complex. If $\BVOp_k = 0$ for $k \ge 3$, then the $[\hbar^{-1},\hbar]]$- and $[[\hbar]]$-versions of the total complex (see Remark~\ref{Rem:MixedCompl}) compute the $\BV$- and strict $\BV$-homology, respectively. Note that by \cite[Proposition~2.2]{Cieliebak2018b}, these versions are quasi-isomorphism invariants of mixed complexes.\footnote{A morphism of mixed complexes commutes with both the boundary operator and the differential. A quasi-isomorphism of homological mixed complexes is a morphism of mixed complexes which is a quasi-isomorphism with respect to the boundary operator.}

In the case of the canonical $\dIBL$-algebra on cyclic cochains of a Poincar\'e duality algebra of degree $n$, we have $d=n-3$, and so $d=-1$ corresponds to $n=2$.
\end{Remark}
\begin{Proposition}[Morphisms of $\BV$-complexes]\label{Prop:ObservationsMor}
In the category of complete filtered $\IBLInfty$-algebras in $\MV$-formalism, the following holds:
\begin{ClaimList}
 \item A weak morphism of weak $\IBLInfty$-algebras induces a chain map $e^\HTP$ of $\BV$-complexes.
 \item An input-strict morphism of input-strict $\IBLInfty$-algebras induces a chain map of both the $\BV$-complex and its strict version. 
 \item The $(1,1,0)$-part of a strict $\IBLInfty$-morphism of strict $\IBLInfty$-algebras induces a chain map of $\IBLInfty$-chain complexes.
 \item For a Maurer-Cartan element $\PMC$, the left multiplication $L_{e^{\PMC(1)}}: \hat{\Sym}U((\hbar))\rightarrow \hat{\Sym}U((\hbar))$ is an isomorphism of the $\BV$-complexes $(\hat{\Sym}U((\hbar)),\BVOp)\simeq (\hat{\Sym}U((\hbar)),\BVOp^\PMC)$.
\end{ClaimList}
\end{Proposition}
\begin{proof}
Clear.
\end{proof}

Notice that whereas $L_{e^{\PMC(1)}}: \hat{\Sym}U \rightarrow\hat{\Sym}U$ is an isomorphism of chain complexes and it has a well-defined logarithm (i.e., it holds $\Norm{L_{e^{\PMC(1)}}}\ge 0$ and $\Norm{L_{e^{\PMC(1)}}(1)-1}=\Norm{e^{\PMC(1)}-1}>0$), it is typically not equal to $e^\HTP$ for any (weak) $\IBLInfty$-morphism $\HTP$.

As an example, suppose that $\PMC(1) = \hbar^{-1}\PMC_{10}(1) \in \hat{\Sym}U$ (recall that we interpret $\PMC_{10}$ as a map $\K \rightarrow \hat{\Sym}U$) and assume that $L_{e^{\PMC(1)}} = e^\HTP$ for $\HTP = \HTP_0\hbar^{-1} + \HTP_1 + \HTP_2\hbar + \dotsb$ with $\hat{\Sym}_i U \subset \ker(\HTP_k)$ for all $i>k$. In particular, for $u\in U$, there is no negative power of $\hbar$ in $\HTP(u)$. On the other hand, there is precisely one negative power of $\hbar$ in $\sum_{k=1}^\infty \frac{(-1)^k}{k}(L_{e^{\PMC(1)}} - \StarProdOne)^{\Star k}u$, namely $\PMC_{10}(1)\hbar^{-1}$ for $k=1$. Therefore, $L_{e^{\PMC(1)}}$ does not come from a weak $\IBLInfty$-morphism as long as $\PMC_{10}(1)\neq 0$. This sheds light on the matter of the author's confusion from the preamble about twisting with a Maurer-Cartan element.

\begin{Questions}\phantomsection\label{Q:SomeQuestionsFilter}\begin{RemarkList}
\item What is the geometric meaning of 
\begin{equation}\label{Eq:HomGeom}
\H(\hat{\Sym}U((\hbar)),\BVOp) \simeq \H(\hat{\Sym}U((\hbar)),\BVOp^\PMC), \quad\H(\hat{\Sym}U[[\hbar]],\BVOp)\quad\text{and}\quad\H(\hat{\Sym}U[[\hbar]],\BVOp^\PMC),
\end{equation}
where $U=\DBCyc\HDR(M)[3-n]$ is the (reduced) dual cyclic bar complex of the de Rham cohomology of a closed oriented $n$-manifold $M$ with the filtration dual to the (homological, i.e., increasing) filtration by weights, $\BVOp$ represents the canonical $\dIBL$-structure for the intersection pairing on $\HDR(M)$ and $\PMC$ is the Chern-Simons Maurer-Cartan element? Are some of the homologies~\eqref{Eq:HomGeom} isomorphic? How about the bicomplex from Remark~\ref{Rem:BVBicomp} for surfaces?
\item Are there any implications between the following statements (a)--(c)? A strict $\IBLInfty$-morphism of strict $\IBLInfty$-algebras induces
\begin{enumerate}[label=(\alph*),leftmargin=1.5cm]
\item a quasi-isomorphism of $\IBLInfty$-complexes,
\item a quasi-isomorphism of strict $\BV$-complexes and
\item a quasi-isomorphism of $\BV$-complexes? 
\end{enumerate}
\item For a Maurer-Cartan element $\PMC$, is there a formula for
\[ \log L_{e^{\PMC(1)}}:\ \hat{\Sym}U((\hbar)) \longrightarrow \hat{\Sym}U((\hbar))? \]
\qedhere
\end{RemarkList}
\end{Questions}

\Add[caption={Add formulas},noline]{Add here formulas for $\PMC = \HTP_* \MC$. It should be the composition $\HTP \DiamComp \MC$ but $\MC$ is weak morphism! But very simple weak morphism. Maybe the induction will work. Add the formula for $\HTP^\MC = \log(L_{e^{-\MC(1)}}\circ e^\HTP \circ L_{e^{\MC(1)}})$. See Page 99 in Diary I.}

\section{Composition of polynomials in convolution product}\label{Sec:CompConvA}
\renewcommand{\Star}{*}
We will work on a (non-filtered) $\N_0$-graded bialgebra $V$ over $\K$ and consider $\K$-linear maps $V\rightarrow V$. Nevertheless, the results extend in a straightforward way to weight-graded complete filtered bialgebras $\hat{V}$ and $\K$-linear maps of finite filtration degrees, and ultimately to the formal series $\hat{V}((\hbar))$ and $\K((\hbar))$-linear maps of finite filtration degrees.

The theory is based on the following observation.
\begin{figure}[t]
\centering
\input{\GraphicsFolder/ItCompCon.tex}
\caption[The ``heart'' of the iterated bialgebra compatibility condition.]{The ``heart and veins'' of the iterated bialgebra compatibility condition $\MVCoProd^{(n)}\MVProd^{(m)}$, and the definition of $\sigma_{m,n}$.}
\label{Fig:Spider}
\end{figure}
\begin{Lemma}[Iterated compatibility condition]\label{Lem:ItCompCond}
Let $(V, \MVProd, \MVCoProd, \MVUnit, \MVAug)$ be an $\N_0$-graded bialgebra. Write $V = \bigoplus_{i=0}^\infty V_i$,\footnote{We use the lower index because it should correspond to the weight in the weight-graded setting of $V = \Sym U$.} and for every $i\in \N_0$, let $\iota_i : V_i \rightarrow V$ and $\pi_i: V \rightarrow V_i$ be the canonical inclusion and projection, respectively. Let $m$, $n\in \N$, and let $\sigma_{m,n}\in \Perm_{m n}$ be the permutation of $mn$ elements depicted in Figure~\ref{Fig:Spider}. For $A\in \N_0^{m\times n}$, let
\[ F_{A}\coloneqq \bigotimes_{i=1}^m \bigotimes_{j=1}^n \pi_{A_{ij}}\circ \iota_{A_{ij}} : V^{\otimes m n } \longrightarrow V^{\otimes m n}. \]
Then for the iterated product $\MVProd^{(m)}: V^{\otimes m} \rightarrow V$ and the iterated coproduct $\MVCoProd^{(n)}: V\rightarrow V^{\otimes n}$, the following holds:
\begin{equation}\label{Eq:ItCompRel}
\MVCoProd^{(n)}\MVProd^{(m)} = \MVProd^{(m)\otimes n} \sigma_{m,n} \MVCoProd^{(n)\otimes m} =  \sum_{A\in \N_0^{m\times n}} \MVProd^{(m)\otimes n} \sigma_{m,n}F_A \MVCoProd^{(n)\otimes m}.
\end{equation}
For $n=m=2$, this reduces to the well-known compatibility relation
\begin{equation}\label{Eq:BasicCompCond}
\MVCoProd\circ \MVProd = (\MVProd\otimes\MVProd)\circ (\Id\otimes\tau\otimes\Id)\circ(\MVCoProd\otimes\MVCoProd).
\end{equation}
\end{Lemma}
\begin{proof}
Clearly, the second equality of \eqref{Eq:ItCompRel} holds because $\sum_{A\in\N_0^{m\times n}} F_A = \Id^{mn}$. As for the first equality, it holds for $m=n=2$ from the definition of a bialgebra; the rest is obtained by induction on $m$, $n$, distinguishing the two cases $\MVCoProd^{(n)}\MVProd^{(m)} = (\MVCoProd\otimes\Id^{\otimes n-2})\MVCoProd^{(n-1)}\MVProd^{(m)}$ and $\MVCoProd^{(n)}\MVProd^{(m)} = \MVCoProd^{(n)}\MVProd^{(m-1)}(\mu\otimes\Id^{\otimes n-2})$, respectively. In the first case, the induction step reads
\begin{align*}
\MVCoProd^{(n)}\MVProd^{(m)} & = (\MVCoProd\otimes\Id^{\otimes n-2})\MVCoProd^{(n-1)}\MVProd^{(m)} \\
& = (\MVCoProd\otimes\Id^{\otimes n-2})\MVProd^{(m)\otimes n-1}\sigma_{m,n-1}\MVCoProd^{(n-1)\otimes m} \\
& = \MVProd^{(m)\otimes n}(\sigma_{m,2} \MVCoProd^{\otimes m} \otimes \Id^{\otimes m(n-2)})\sigma_{m,n-1}\MVCoProd^{(n-1)\otimes m} \\
& = \MVProd^{(m)\otimes n} \sigma_{m,n} \delta^{(n)\otimes m}.
\end{align*}
On the third line, we used that $\MVCoProd\MVProd^{(m)}=(\MVProd^{(m)}\otimes\MVProd^{(m)})\sigma_{m,2}\MVCoProd^{\otimes m}$, which equals \eqref{Eq:BasicCompCond} for $m=2$, and for $m\ge 3$, it is proven by induction with the induction step
\begin{align*}
 \MVCoProd\MVProd^{(m)} & = \MVCoProd\MVProd^{(m-1)}(\MVProd\otimes\Id^{\otimes m-2}) \\
 & = (\MVProd^{(m-1)}\otimes\MVProd^{(m-1)})\sigma_{m-1,2}\MVCoProd^{\otimes m-1}(\MVProd\otimes\Id^{\otimes m-2}) \\
 & = (\MVProd^{(m-1)}\otimes\MVProd^{(m-1)})\sigma_{m-1,2}(\MVProd\otimes\MVProd \otimes\Id^{m-2})(\Id \otimes \tau\otimes \Id \otimes \Id^{m-2})\MVCoProd^{\otimes m} \\
 & = (\MVProd^{(m)}\otimes\MVProd^{(m)})\sigma_{m,2}\MVCoProd^{\otimes m}.
\end{align*}
The second case is analogous.
\end{proof}
\begin{Definition}[Compositions of monomials in convolution product]\label{Def:ConComp}
Consider an $\N_0$-graded bialgebra $(V,\MVProd,\MVCoProd,\MVUnit,\MVAug)$. Given linear maps $\MVMorF_1$, $\dotsc$, $\MVMorF_r$, $\MVMorF_1', \dotsc, \MVMorF_{r'}': V \rightarrow V$ and a matrix $A\in \N_0^{r'\times r}$ for $r$, $r'\in \N$, we define the \emph{$A$-composition} $(\MVMorF_1,\dotsc,\MVMorF_r)\SquareComp_A(\MVMorF_1,\dotsc,\MVMorF_{r'}): V \rightarrow V$ by the formula
\begin{align*}
&(\MVMorF_1,\dotsc,\MVMorF_r)\SquareComp_A(\MVMorF_1,\dotsc,\MVMorF_{r'}) \\
&\qquad\coloneqq \MVProd^{(r)}\circ(\MVMorF_1 \otimes \dotsb \otimes \MVMorF_r)\circ \Prod^{(r')\otimes r} \circ  \sigma_{r',r}\circ F_A \circ \MVCoProd^{(r)\otimes r'} \circ (\MVMorF_1' \otimes \dotsb \otimes \MVMorF_{r'}')\circ\MVCoProd^{(r')},
\end{align*}
where $\sigma_{r',r}$ and $F_A$ were defined in Lemma~\ref{Lem:ItCompCond}.

We call a matrix $A\in\N_0^{r'\times r}$ \emph{connected} if for the matrix
\[ B \coloneqq \begin{pmatrix} 0 & A \\ A^T & 0 \end{pmatrix}, \]
the matrix product $B^{r+r'}$ has at least one row with all entries non-zero.\footnote{Such $A$ represents a connected weighted bipartite graph.} In this case, we say that the $A$-composition $(\MVMorF_1,\dotsc,\MVMorF_r)\SquareComp_A(\MVMorF_1,\dotsc,\MVMorF_{r'})$ is \emph{connected.}

Suppose that $V$ is connected, i.e., $V_0 = \langle 1\rangle$. Given linear maps $\MVMorF_1$, $\dotsc$, $\MVMorF_r$, $\MVMorF_1'$, $\dotsc$, $\MVMorF_{r'}': V\rightarrow V$ for $r$, $r'\in \N$, we define the \emph{connected composition} $(\MVMorF_1, \dotsc, \MVMorF_r)\circ_{\mathrm{con}}(\MVMorF_1',\dotsc,\MVMorF_{r'}'): V \rightarrow V$ by
\[ (\MVMorF_1, \dotsc, \MVMorF_r)\circ_{\mathrm{con}}(\MVMorF_1,\dotsc,\MVMorF_{r'}) = \sum_{\substack{A\in \N_0^{(r'+1)\times (r+1)}\\ A\  \mathrm{connected} \\ A_{r'+1,r+1} = 0 }} (\MVMorF_1,\dotsc,\MVMorF_r,\Id)\SquareComp_A (\MVMorF_1',\dotsc,\MVMorF_{r'}',\Id).\]
Given linear maps $\MVMorF$, $\MVMorF_1$, $\dotsc$, $\MVMorF_r: V \rightarrow V$ and parameters $h_1$, $\dotsc$, $h_r\in \N_0$ for $r\in \N$, we define the \emph{partial compositions} $\MVMorF\circ_{h_1, \dotsc, h_r}(\MVMorF_1, \dotsc, \MVMorF_r)$, $(\MVMorF_1, \dotsc, \MVMorF_r)\circ_{h_1, \dotsc, h_r}\MVMorF: V\rightarrow V$ by
\begin{align*}
&\MVMorF\circ_{h_1, \dotsc, h_r}(\MVMorF_1, \dotsc, \MVMorF_r) \\
&\qquad\coloneqq\begin{multlined}[t]\MVProd(\MVMorF\otimes \Id)(\MVProd\otimes\Id)(\Id\otimes \tau)\bigl(\bigl[(\MVProd^{(r)}\otimes \MVProd^{(r)})(F_{h_1,\dotsc,h_r} \otimes \Id^{\otimes r})\sigma_r\MVCoProd^{\otimes r}\bigr] \otimes \Id \bigr)\\(\MVMorF_1\otimes \dotsb \otimes \MVMorF_r\otimes \Id)\MVCoProd^{(r+1)},\end{multlined}\\
&(\MVMorF_1, \dotsc, \MVMorF_r)\circ_{h_1, \dotsc, h_r}\MVMorF\\
&\qquad\coloneqq\begin{multlined}[t]\MVProd^{(r+1)}(\MVMorF_1\otimes\dotsb\otimes\MVMorF_r\otimes\Id) \bigl(\bigl[\MVProd^{\otimes r}\sigma_r^{-1}(F_{h_1,\dotsc,h_r}\otimes\Id^{\otimes r})(\MVCoProd^{(r)}\otimes\MVCoProd^{(r)})\bigr]\otimes\Id\bigr)\\(\Id\otimes\tau)(\MVCoProd\otimes\Id)(\MVMorF\otimes\Id)\MVCoProd,\end{multlined}
\end{align*}
where we set $F_{h_1,\dotsc, h_r} \coloneqq F_{(h_1,\dotsc,h_r)} = F_{(h_1,\dotsc,h_r)^T}=\iota_{h_1}\pi_{h_1} \otimes \dotsb \otimes \iota_{h_r}\pi_{h_r}$ and $\sigma_r \coloneqq \sigma_{r,2}$, and we omit writing the composition $\circ$. For $r=1$, we have $\MVProd^{(1)} = \MVCoProd^{(1)}\coloneqq \Id$ by definition, and both equations above reduce to 
\begin{align*}
\MVMorF \circ_h \MVMorF_1 &= \MVProd(\MVMorF\otimes \Id)(\MVProd\otimes\Id)(\Id\otimes\tau)(F_h\otimes\Id^{\otimes 2})(\MVCoProd\otimes \Id)(\MVMorF_1 \otimes \Id)\MVCoProd \\
\Bigl(\!\!&= \MVProd(\MVMorF\otimes \Id)(\MVProd\otimes\Id)(F_h\otimes\Id^{\otimes 2})(\Id\otimes\tau)(\MVCoProd\otimes \Id)(\MVMorF_1 \otimes \Id)\MVCoProd\Bigr).
\end{align*}
(c.f., Definition~\ref{Def:CircS} in Section~\ref{Sec:Alg1} for the case of $V = \Sym(C[1])$)
\end{Definition}

\begin{Proposition}[Partial compositions]\label{Prop:PartCompAComp}
Let $\MVMorF$, $\MVMorF_1$, $\dotsc$, $\MVMorF_r : V \rightarrow V$ be linear maps on a connected $\N_0$-graded bialgebra $V$, and let $h_1$,~$\dotsc$, $h_r \in \N_0$ for $r\in \N$. Then the following formulas hold:
\begin{align*}
\MVMorF\circ_{h_1, \dotsc, h_r}(\MVMorF_1, \dotsc, \MVMorF_r) &= \sum_{\substack{A \in \N_0^{(r+1)\times 2} \\ A = \left(\begin{smallmatrix}
h_1 & \bullet \\
\vphantom{\int\limits^x}\smash{\vdots} &  \smash{\vdots} \\
h_r & \bullet \\ 
\bullet & 0
\end{smallmatrix}\right)}} (\MVMorF,\Id)\SquareComp_{A}(\MVMorF_1,\dotsc,\MVMorF_r,\Id), \\
(\MVMorF_1, \dotsc, \MVMorF_r)\circ_{h_1, \dotsc, h_r}\MVMorF &= \sum_{\substack{A \in \N_0^{2\times(r+1)} \\ A = \left(\begin{smallmatrix}
h_1 & \dotsb & h_r & \bullet \\
\bullet & \dotsb & \bullet & 0
\end{smallmatrix}\right)
}} (\MVMorF_1,\dotsc,\MVMorF_r,\Id)\SquareComp_A(\MVMorF,\Id).
\end{align*}
\end{Proposition}

\begin{proof}
We compute
\begin{align*}
& \sum_{\substack{A \in \N_0^{(r+1)\times 2} \\ A = \left(\begin{smallmatrix}
h_1 & \bullet \\
\vphantom{\int\limits^x}\smash{\vdots} & \smash{\vdots} \\
h_r & \bullet \\ 
\bullet & 0
\end{smallmatrix}\right)}} (\MVMorF\Star\Id)\SquareComp_A (\MVMorF_1\Star\dotsb\Star\MVMorF_r \Star \Id) \\
&\quad =\sum_{\substack{A \in \N_0^{(r+1)\times 2} \\ A = \left(\begin{smallmatrix}
h_1 & \bullet \\
\vphantom{\int\limits^x}\smash{\vdots} & \smash{\vdots} \\
h_r & \bullet \\ 
\bullet & 0
\end{smallmatrix}\right)}} \MVProd(\MVMorF \otimes \Id)\Prod^{(r+1)\otimes 2}\sigma_{r+1,2}F_A\MVCoProd^{\otimes r+1}(\MVMorF_1 \otimes \dotsb \otimes \MVMorF_{r}\otimes\Id)\MVCoProd^{(r+1)} \\
&\quad = \begin{multlined}[t]\MVProd(\MVMorF \otimes \Id)\Prod^{(r+1)\otimes 2}\sigma_{r+1}(\iota_{h_1}\pi_{h_1}\otimes \Id \otimes\dotsb \otimes\iota_{h_r}\pi_{h_r}\otimes \Id\otimes  \Id \otimes \iota_0\pi_0)\MVCoProd^{\otimes r+1} \\ (\MVMorF_1 \otimes \dotsb \otimes \MVMorF_{r}\otimes\Id)\MVCoProd^{(r+1)}\end{multlined}\\
&\quad = \begin{multlined}[t]\MVProd(\MVMorF \otimes \Id)\Prod^{(r+1)\otimes 2}(F_{h_1,\dotsc,h_r}\otimes \Id\otimes  \Id^{\otimes r} \otimes \iota_0\pi_0)\sigma_{r+1} \MVCoProd^{\otimes r+1} \\ (\MVMorF_1 \otimes \dotsb \otimes \MVMorF_{r}\otimes\Id)\MVCoProd^{(r+1)}\end{multlined} \\
&\quad =\begin{multlined}[t]\MVProd(\MVMorF\otimes \Id)(\MVProd\otimes\Id)(\Id\otimes \tau)\bigl(\bigl[\MVProd^{(r)\otimes 2}(F_{h_1,\dotsc,h_r} \otimes \Id^{\otimes r})\sigma_r\MVCoProd^{\otimes r}\bigr] \otimes \Id \bigr)\\(\MVMorF_1\otimes \dotsb \otimes \MVMorF_r\otimes \Id)\MVCoProd^{(r+1)}\end{multlined} \\
&\quad=\MVMorF\circ_{h_1, \dotsc, h_r}(\MVMorF_1, \dotsc, \MVMorF_r).
\end{align*}
On the line before the last line we used that for a connected bialgebra $V$, it holds
\[ (\Id\otimes\iota_0\pi_0)\MVCoProd(v)=v\otimes 1\quad\text{and}\quad (\iota_0\pi_0\otimes\Id)\MVCoProd(v) = 1 \otimes v\quad\text{for all }v\in V.\]
The case of $(\MVMorF_{1},\dotsc,\MVMorF_r)\circ_{h_1,\dotsc,h_r}\MVMorF$ is treated analogously, using that $\sigma_{m,n} = \sigma_{n,m}^{-1}$ (this is visible by looking at the highlighted square in Figure~\ref{Fig:Spider}, which is symmetric under the rotation by $180^\circ$).
\end{proof}

We see that the operations $\circ_{h_1,\dotsc,h_r}$ and $\circ_{\mathrm{con}}$, which are the cornerstone of the surface calculus for $\Sym U$ in Section~\ref{Sec:FilteredMV}, originate naturally from $\SquareComp_A$. Clearly, we can now replace~$\Sym U$ by any connected weight-graded bialgebra~$V$ and develop a surface calculus for $\BVInfty$-algebras over~$V$.

The following formulas were stated in Remark~\ref{Rem:Compositions} in Section~\ref{Sec:Alg1} and come from \cite{Cieliebak2015}, where they were used in a slightly different form (and without a proof).

\begin{Proposition}[Formulas involving partial compositions] \label{Prop:PartCompositions}
For $r\in \N$, let $\MVMorF$, $\MVMorF_1$, $\dotsc$, $\MVMorF_r : V \rightarrow V$ be linear maps on a connected $\N_0$-graded bialgebra $V$. Then the following formulas hold:
\begin{align}\label{Eq:FormI}
\hat{\MVMorF} \circ \hat{\MVMorF}_1 &= \sum_{h\ge 0} \widehat{\MVMorF\circ_h \MVMorF_1}, \\ 
   \hat{\MVMorF} \circ (\MVMorF_1 \Star \dotsb \Star \MVMorF_r) &= \sum_{k\ge 0}\sum_{\substack{h_1, \dotsc, h_r \ge 0 \\ h_1 + \dotsb + h_r = k}} (\MVMorF \iota_k \pi_k)\circ_{h_1,\dotsc, h_r}(\MVMorF_1,\dotsc, \MVMorF_r), \label{Eq:FormII} \\
   (\MVMorF_1 \Star \dotsb \Star \MVMorF_r) \circ \hat{\MVMorF} &= \sum_{l\ge 0} \sum_{\substack{h_1,\dotsc,h_r \ge 0 \\ h_1 + \dotsb + h_r = l}} (\MVMorF_1,\dotsc,\MVMorF_r) \circ_{h_1,\dotsc,h_r} (\iota_l\pi_l\MVMorF),\label{Eq:FormIII} \\
   \MVMorF\circ_{h_1,\dotsc,h_{r-1},0}(\MVMorF_1,\dotsc, \MVMorF_r) &= \MVMorF\circ_{h_1,\dotsc, h_{r-1}}(\MVMorF_1,\dotsc, \MVMorF_{r-1}) \Star \MVMorF_r, \label{Eq:FormIV} \\ \label{Eq:FormV}
  (\MVMorF_1,\dotsc,\MVMorF_r)\circ_{0,h_2,\dotsc,h_r}\MVMorF &= \MVMorF_1 \Star (\MVMorF_2,\dotsc,\MVMorF_r)\circ_{h_2,\dotsc,h_r} \MVMorF.
\end{align}
(Recall that $\hat{\MVMorF} \coloneqq \MVMorF\Star \Id: V \rightarrow V$.)
\end{Proposition}

\begin{proof}
As for \eqref{Eq:FormI}, we compute
\begin{align*}
\hat{\MVMorF} \circ \hat{\MVMorF}_1 &= \MVProd(\MVMorF\otimes \Id)\MVCoProd\MVProd(\MVMorF_1\otimes\Id)\MVCoProd \\
&= \MVProd(\MVMorF\otimes \Id)(\MVProd\otimes\MVProd)(\Id\otimes\tau\otimes\Id)(\MVCoProd\otimes\MVCoProd)(\MVMorF_1\otimes\Id)\MVCoProd \\
&= 
\MVProd(\MVProd\otimes\Id)(\MVMorF\otimes \Id^{\otimes 2})(\MVProd\otimes\Id^{\otimes 2})(\Id\otimes\tau\otimes \Id)(\MVCoProd\otimes\Id^{\otimes 2}) (\MVMorF_1\otimes\Id^{\otimes 2})(\MVCoProd\otimes\Id)\MVCoProd \\
& = 
 \MVProd\bigl(\bigl[\underbrace{\MVProd(\MVMorF\otimes\Id)(\MVProd\otimes\Id)(\Id\otimes\tau)(\MVCoProd\otimes\Id)(\MVMorF_1\otimes\Id)\MVCoProd}_{=\sum_{h\ge 0}\MVMorF\circ_h \MVMorF_1}\bigr]\otimes\Id\bigr)\MVCoProd \\
& = \sum_{h\ge 0} \reallywidehat{\MVMorF \circ_h \MVMorF_1}.
\end{align*}
On the second line, we used the compatibility condition; on the third line, we used associativity and coassociativity.

In order to see \eqref{Eq:FormII}, the easiest is to use Proposition~\ref{Prop:PartCompAComp}:
\begin{align*}
\hat{\MVMorF}(\MVMorF_1 \Star \dotsb \Star \MVMorF_r)&= \sum_{k\ge 0}\sum_{\substack{h_1,\dotsc,h_r\ge 0 \\ h_1 + \dotsb + h_r = k}}\sum_{\substack{A\in \N_0^{r \times 2} \\ A=\left(\begin{smallmatrix}
h_1 & \bullet \\ 
\vphantom{\int\limits^x}\smash{\vdots} & \smash{\vdots} \\
h_r & \bullet
\end{smallmatrix}\right)}}(\MVMorF \iota_k \pi_k,\Id)\SquareComp_{A}(\MVMorF_1,\dotsc,\MVMorF_r) \\
& = \sum_{k\ge 0}\sum_{\substack{h_1,\dotsc,h_r\ge 0 \\ h_1 + \dotsb + h_r = k}}\sum_{\substack{A'\in \N_0^{(r+1)\times 2} \\ A'=\left(\begin{smallmatrix}
h_1 & \bullet \\
\vphantom{\int\limits^x}\smash{\vdots} & \smash{\vdots} \\
h_r & \bullet \\ 
\bullet & 0
\end{smallmatrix}\right)}}(\MVMorF \iota_k \pi_k,\Id)\SquareComp_{A'}(\MVMorF_1,\dotsc,\MVMorF_r,\Id) \\
& = \sum_{k\ge 0}\sum_{\substack{h_1,\dotsc,h_r\ge 0 \\ h_1 + \dotsb + h_r = k}} (\MVMorF\iota_k\pi_k)\circ_{h_1,\dotsc,h_r}(\MVMorF_1,\dotsc,\MVMorF_r).
\end{align*}
On the second line, the bottom-left $\bullet$, i.e., how many ``veins'' of $\Id$ go into $\MVMorF \iota_k \pi_k$, is forced to be $0$ because $\MVMorF \iota_k \pi_k$ has $k$ inputs and $h_1 + \dotsb + h_r = k$. Equation~\eqref{Eq:FormIII} is proven analogously.

As for \eqref{Eq:FormIV}, we prefer to manipulate the expression with bialgebra operations:
\begin{align*}
&\MVMorF\circ_{h_1,\dotsc,h_{r-1},0}(\MVMorF_1,\dotsc,\MVMorF_r) \\
&\quad=\begin{multlined}[t]\MVProd(\MVMorF\otimes\Id)(\MVProd\otimes\Id)(\Id\otimes\tau)\bigl(\bigl[\MVProd^{(r)\otimes 2}(F_{h_1,\dotsc,h_{r-1},0}\otimes\Id^{\otimes r})\sigma_r\MVCoProd^{\otimes r}\bigr]\otimes\Id\bigr)\\(\MVMorF_1\otimes\dotsb\otimes\MVMorF_{r}\otimes\Id)\MVCoProd^{(r+1)}\end{multlined}\\
&\quad=\begin{multlined}[t]\MVProd(\MVMorF\otimes\Id)(\MVProd\otimes\Id)(\Id\otimes\tau)\bigl(\bigl[(\MVProd^{(r-1)}\otimes\MVProd^{(r)})(F_{h_1,\dotsc,h_{r-1}}\otimes\Id^{\otimes r})(\sigma_{r-1}\otimes\Id)\\ (\MVCoProd^{\otimes r-1}\otimes\Id)\bigr]\otimes\Id\bigr)(\MVMorF_1\otimes\dotsb\otimes\MVMorF_{r}\otimes\Id)\MVCoProd^{(r+1)}\end{multlined}\\
&\quad=\begin{multlined}[t]\MVProd(\MVMorF\otimes\Id)(\MVProd\otimes\Id)(\Id\otimes\tau)(\Id\otimes\MVProd\otimes\Id)\bigl(\bigr[\MVProd^{(r-1)\otimes 2}(F_{h_1,\dotsc,h_{r-1}}\otimes\Id^{\otimes r-1})\sigma_{r-1}\\\MVCoProd^{\otimes r-1}\bigr]\otimes \Id^{\otimes 2}\bigr)(\MVMorF_1\otimes\dotsb\otimes\MVMorF_{r}\otimes\Id)\MVCoProd^{(r+1)}\end{multlined}\\
&\quad=\begin{multlined}[t]\MVProd(\MVMorF\otimes\Id)(\MVProd\otimes\Id)(\Id\otimes\MVProd)(\Id\otimes\tau\otimes\Id)\bigl(\bigr[\MVProd^{(r-1)\otimes 2}(F_{h_1,\dotsc,h_{r-1}}\otimes\Id^{\otimes r-1})\sigma_{r-1}\\\MVCoProd^{\otimes r-1}\bigr]\otimes \Id^{\otimes 2}\bigr)(\MVMorF_1\otimes\dotsb\otimes\MVMorF_{r-1}\otimes\Id\otimes\MVMorF_{r})\MVCoProd^{(r+1)}\end{multlined}\\
&\quad=\begin{multlined}[t]\MVProd\bigl(\bigl[\MVProd(\MVMorF\otimes\Id)(\MVProd\otimes\Id)(\Id\otimes\tau)\bigl(\bigl[\MVProd^{(r-1)\otimes 2}(F_{h_1,\dotsc,h_{r-1}}\otimes\Id^{\otimes r-1})\sigma_{r-1} \\ \MVCoProd^{\otimes r-1}\bigr]\otimes\Id\bigr)(\MVMorF_1\otimes\dotsb\otimes\MVMorF_{r-1}\otimes\Id)\MVCoProd^{(r)}\bigr]\otimes\MVMorF_r\bigr)\MVCoProd\end{multlined}\\
&\quad=\MVProd\bigl(\MVMorF\circ_{h_1,\dotsc,h_{r-1}}(\MVMorF_1,\dotsc,\MVMorF_{r-1})\otimes\MVMorF_r\bigr)\MVCoProd \\
&\quad= \MVMorF\circ_{h_1,\dotsc,h_{r-1}}(\MVMorF_1,\dotsc,\MVMorF_{r-1})\Star \MVMorF_r
\end{align*}
Equation~\eqref{Eq:FormV} is proven analogously.
\end{proof}

\backmatter
\phantomsection
\addtocontents{toc}{\protect\vspace{12pt}}%
\addcontentsline{toc}{part}{Bibliography}

\emergencystretch=1em
\printbibliography

\end{document}